\documentclass[a4paper,12pt,twoside,openright]{report}

\usepackage{a4wide}
\usepackage{xspace}
\usepackage{amsmath}
\usepackage{amssymb}
\usepackage{booktabs}
\usepackage{graphicx}
\usepackage{url}
\usepackage[breaklinks=true]{hyperref}
\usepackage{subfig}
\usepackage[countmax]{subfloat}
\usepackage{epigraph}

\newcommand{\ie}{i.e.\ }

\newcommand{\eg}{e.g.\ }
\newcommand{\Eg}{E.g.\ }
\newcommand{\cf}{{\it cf.}\ }
\newcommand{\wrt}{w.r.t.\ }
\newcommand{\cnf}{\emph{cf.}\xspace}
\newcommand{\eq}{Eq.}
\newcommand{\eqs}{Eqs.}
\newcommand{\Eq}{Eq.} 

\newcommand{\order}[1]{{\cal O}\left(#1\right)}
\newcommand{\as}{\alpha_s}
\newcommand{\cone}{\text{SISCone}}
\newcommand{\cam}{\text{Cam}}
\newcommand{\antikt}{\text{anti-}k_t}
\newcommand{\kt}{k_t}
\newcommand{\cNg}{{\cal N}_g}
\newcommand{\cV}{{\cal V}}
\newcommand{\GeV}{\,\mathrm{GeV}}
\newcommand{\TeV}{\,\mathrm{TeV}}
\newcommand{\mb}{\,\mathrm{mb}}
\newcommand{\JA}{\text{JA}}
\newcommand{\bs}[1]{\boldsymbol{#1}}
\newcommand{\gghosts}{\,|\,\{g_i\}}

\newcommand{\avg}[1]{\left\langle #1 \right\rangle}

\newcommand{\PU}{\mathrm{PU}}
\newcommand{\nPU}{{\ensuremath{N_\PU}}\xspace}
\newcommand{\lumi}{{\ensuremath{\mathcal{L}}}\xspace}

\newcommand{\oPJ}{\mathrm{1PJ}}

\newcommand{\hard}{\text{hard}}

\newcommand{\fastjet}{\texttt{FastJet}\xspace}
\newcommand{\fjcontrib}{\texttt{\href{http://fastjet.hepforge.org/contrib/}{fjcontrib}}\xspace}
\newcommand{\ttt}[1]{{\small\texttt{#1}}}

\newcommand{\PJ}{\ttt{PseudoJet}\xspace}

\newcommand{\rhoest}{\ensuremath{\rho_{\rm est}}\xspace}
\newcommand{\Dpt}{\ensuremath{\Delta p_t}\xspace}

\newcommand{\AuAu}{AuAu\xspace}
\newcommand{\PbPb}{PbPb\xspace}
\newcommand{\jet}{\mathrm{jet}}
\newcommand{\pthard}{\ensuremath{p_t^{pp}}}

\newcommand{\ptfullsub}{\ensuremath{p_t^{AA,{\rm sub}}}}

\newcommand{\zcut}{z_\text{cut}}
\newcommand{\pythia}[1]{Pythia\xspace #1}

\newcommand{\SD}{Soft Drop\xspace}

\newcommand{\ntr}{\text{ntr}}

\newcommand{\NpC}{\text{NpC}}
\newcommand{\cut}{\text{cut}}

\newcommand{\ptcut}{p_t^\cut}

\usepackage{color}
\definecolor{darkgreen}{rgb}{0,0.5,0}
\definecolor{comment}{rgb}{0,0.3,0}
\definecolor{identifier}{rgb}{0.0,0,0.3}
\usepackage{listings}
\makeatletter
\newenvironment{acknowledgements}{%
  \if@twocolumn
    \section*{Acknowledgements}%
  \else
    \small
    \begin{center}%
      {\bfseries Acknowledgements\vspace{-.5em}\vspace{\z@}}%
    \end{center}%
    \quotation
    \fi}
  {\if@twocolumn\else\endquotation\fi}
\makeatother

\hypersetup{colorlinks=true,linkcolor=blue,urlcolor=blue}

\usepackage{fancyhdr}
\newcommand{\helv}{\fontfamily{phv} \fontseries{b}\fontsize{10}{12}\selectfont}

\makeatletter
\renewcommand{\chaptermark}[1]{%
  \markboth{\MakeUppercase{\ifnum\c@secnumdepth>\m@ne \@chapapp \ \thechapter. \ \fi #1}}%
           {\MakeUppercase{\ifnum\c@secnumdepth>\m@ne \@chapapp \ \thechapter. \ \fi #1}}%
}
\makeatother

\makeatletter
\def\cleardoublepage{\clearpage\if@twoside \ifodd\c@page\else
\hbox{}
\vspace*{\fill}
\thispagestyle{empty}
\newpage
\if@twocolumn\hbox{}\newpage\fi\fi\fi}
\makeatother

\fancypagestyle{plain}{%
\fancyhf{} 
\fancyfoot[C]{\helv \thepage} 
}

\begin{document}

\begin{titlepage}
  
  \begin{center}
    \vspace*{5.0cm}
    \Huge \textsf{Pileup mitigation at the LHC}
    
    \vspace*{0.7cm}
    
    \LARGE \textsf{A theorist's view}
  \end{center}
  
  \vspace*{2.0cm}
  
  \begin{center}
    {\large \textsf{Gr\'egory Soyez}}\\
    \vspace*{0.7cm}
    Institut de Physique Th\'eorique, CNRS, CEA Saclay,
    Universit\'e Paris-Saclay,\\
    Orme des Merisiers, B\^at 774, F-91191 Gif-sur-Yvette cedex, France\\
    \vspace*{0.3cm}
    email: \href{mailto:gregory.soyez@ipht.fr}{gregory.soyez@ipht.fr}
  \end{center}

\cleardoublepage

\end{titlepage}


\begin{abstract}
  To maximise the potential for new measurements and discoveries at
  CERN's Large Hadron Collider (LHC), the machine delivers as high as
  possible collision rates.
  As a direct consequence, multiple proton-proton collisions occur
  whenever two bunches of protons cross.
  Interesting high-energy (hard) collisions are therefore contaminated by
  several soft, zero-bias, ones. This effect, known as {\it pileup},
  pollutes the final state of the collision.
  It complicates the reconstruction of the objects in this final
  state, resulting in increased experimental measurement
  uncertainties.
  
  To reduce these uncertainties, and thus improve the quality and
  precision of LHC measurements, techniques are devised to correct for
  the effects of pileup.
  This document provides a theoretical review of the main methods
  employed during Run~I and II of the LHC to mitigate pileup effects.
  I will start with an in-depth presentation of the {\em area--median}
  used for the vast majority of applications, including several
  refinements of the original idea, their practical (numerical)
  implementation and an assessment of their efficiency and robustness.
  I will then focus on several theoretical calculations that can
  provide both quantitative and qualitative information on the
  area--median approach.
  
  In the specific case of boosted jets, a field that has seen a wide
  interest recently, a set of methods, known as {\em grooming}
  techniques has also been used. I will describe these techniques,
  address their performance and briefly show that they are amenable to
  a theoretical, analytic, understanding.

  The last part of this review will focus on ideas oriented towards
  future pileup mitigation techniques. This includes a couple of new
  methods that have recently been proposed as well as a large series
  of alternative ideas. The latter are yet unpublished and have not
  received the same amount of investigation than the former but they
  have the potential to bring new developments and further improvement
  over existing techniques in a future where pileup mitigation will be
  crucial.
\end{abstract}

\tableofcontents

\chapter{Introduction}\label{chap:intro}

Collider physics has played a central role in fundamental research
over the past decades. Today, most earth-based experiments exploring
particle physics at high energies are beam--beam collider experiments
with the Large Hadron Collider (LHC) occupying the front row.
This type of experiment brings into collision two beams of particles,
either leptons or hadrons.
The topic of this review applies to proton-proton collisions, with
applications for the on-going LHC programme as the main focus.
As we will show, it also has consequences for heavy-ion collisions
such as the gold--gold collisions at RHIC --- the Relativistic
Heavy-Ion Collider at the Brookhaven National Laboratory --- or
lead--lead collisions at the LHC.
The physics discussed here is even more relevant for future hadronic
colliders.
From now on, let us consider the collision of two proton beams.

The primary goal of the LHC is to explore physics at high energies ---
or, equivalently, short distances --- typically around the TeV scale
(\ie around $10^{-19}$~m or $10^{-4}$~fm).
The vast majority of the analyses at the LHC involve either
measurements of standard-model processes or searches for new phenomena
involving physics beyond the standard model.
Given the constraints already obtained from earlier colliders such as
the Tevatron or LEP, the LHC analyses explore either energy
(kinematic) domains previously unreachable, or rare phenomena in
extreme corners of the phase-space, or both.

With collisions at a centre-of-mass energy of 13~TeV, the LHC is able
to boldly probe energies that no collider had probed before.
This is however not sufficient to explore rare corners of the phase
space for which we also need to accumulate as much data as possible.
For this one wants to reach the highest possible collision rate.
To understand how it works in practice, it is interesting to look a
bit closer at how the LHC works. 
Each beam is made of trains of bunches ($\sim 2800$ bunches per beam
at the LHC), where each bunch contains billions of protons
($\sim 10^{11}$ at the LHC).
There are several ways to maximise the collision rate: increase the
number of bunches, increase the number of protons per bunch, or
improve the beam optics so that the beams are better focalised at the
interaction points (technically, lowering the $\beta^*$ parameter).
The net effect of increasing the number of protons per bunch or
improving the beam optics is that whenever two bunches cross at one of
the interaction points (\ie in one of the experiments), the rate of
collisions increases, to a point that several proton-proton collisions
occur simultaneously, during one bunch crossing.
It is this effect of simultaneous collisions which is called {\em
  pileup} and which is the main topic of this review.

To be more quantitative, the collision rate of a collider is called
the {\em luminosity} ${\cal {L}}$, and can be measured in
$\text{cm}^{-2}\text{s}^{-1}$.
Towards the end of Run~II of the LHC, the machine delivers a
luminosity around $2\,10^{34}\,\text{cm}^{-2}\text{s}^{-1}$, twice its
nominal value.
For a process with a cross-section $\sigma$, the rate of events for
this process is $\lumi\sigma$.
For example, the total cross-section for the production of a $t\bar t$
pair is around $800$~pb and that for a Higgs boson is around $50$~pb,
resulting in about 15 $t\bar t$ pairs produced per second and one Higgs
boson per second at the current luminosity.
More generally, the LHC typically studies processes with a
cross-section that can go as low a few femtobarns, \ie 11-14 orders of
magnitude lower than the total proton-proton cross-section.

To estimate the {\em pileup multiplicity}, \ie the number of
proton-proton interactions occurring per bunch crossing, one also needs
the number of bunches or, equivalently, the time separation between
two bunches. At the LHC it is 25~ns.\footnote{It was 50~ns during
  Run~I.}
The luminosity $\lumi=2\,10^{34}\,\text{cm}^{-2}\text{s}^{-1}$
therefore corresponds to a luminosity per bunch crossing
$\lumi_\text{crossing}=5\,10^{26}\,\text{cm}^{-2}=0.5\,\text{mb}^{-1}$.
Given a total proton-proton cross-section of order 100~mb, this gives
around 50 proton-proton interactions per bunch crossing, \ie an
average pileup multiplicity $\avg{\nPU}\sim 50$.
Pileup itself is not really a new issue: it was already present at the
Tevatron, albeit with a much lower multiplicity (typically around 5),
so that simple techniques were sufficient to address its effects.
It is with the LHC and its large luminosity that pileup multiplicities
started to crank up. 
The average pileup multiplicity was already around $\avg{\nPU}\sim 20$
during Run~I~\cite{pileup-cms-run1} and 30 at the beginning of Run~II.
For the coming high-luminosity LHC (HL-LHC) one expects an average
pileup multiplicity between 140 and
200~\cite{Bruning:2015dfu,Rossi:2015ulc}.
Note that, in practice, the actual number of interactions in a given
bunch crossing, \nPU, fluctuates, following a Poisson distribution
around its average $\avg{\nPU}$.

Imagine now that during a collision we have one of those high-energy
(hard) rare processes that the LHC experiments are keen to study.
A number of other, simultaneous, zero-bias collisions will occur
during the same bunch crossing.
The overall final-state of the collision, observed in the detectors, is
therefore not only the products of one hard collision we are
interested in, but the superposition of these products with those of
all the simultaneous soft collisions.
In other words, the final-state of the collision includes both the
final-state particles of the hard collision and pileup.
Since zero-bias (pileup) proton-proton collisions typically produce a
few tens of soft hadrons in their final-state, the net result of
pileup is to add hundreds to thousands of soft hadrons to the
final-state of the hard collision one wishes to study (referred to as
``the hard event'' or ``the hard collision'' in what follows).

Now that we have described what pileup is, we can discuss what its
effect are.
In a nutshell, adding a large number of soft particles to the
final-state of the collision complicates the reconstruction of the
properties of the hard event itself.
To better understand this, remember that pileup is essentially made of
a large number of soft hadrons, spread over the whole detector.
The reconstruction of any object involving the hadrons produced in the
final-state of the collision will therefore be affected by this
additional hadronic activity.

Probably the most typical situation where pileup impacts the ability
to properly reconstruct the objects in the final-state of a hard
collision is the case of jets. Jets are the collimated sprays of
hadrons stemming from the showering and hadronisation of the
high-energy partons produced in the hard collision (we refer to
Ref.~\cite{Salam:2009jx} for a relatively recent review and
Appendix~\ref{app:jetdefs} for a quick overview).
The hadrons from pileup will overlap with those of the hard collision
so that when jets are reconstructed, particles of the former will be
clustered together with particles of the latter. In other words, a
given jet will have both particles from the hard event and from
pileup.
The properties of the hard jets --- mostly the jet transverse momentum
but also its rapidity and azimuthal angle --- will be biased by these
extra hadrons coming from pileup.

Jets are not the only objects to be affected by pileup: every object
whose reconstruction involves the hadrons in the final state of the
collision will be.
This covers almost all the objects one uses in LHC analyses.
Reconstructing leptons and photons usually involves a condition on the
hadronic activity in the surroundings of the lepton/photon. This
activity will obviously be affected by the extra hadrons from pileup.
Missing transverse energy is accessed by reconstructing the transverse
momentum of all the measured objects in the event. This will obviously
be affected by the extra hadrons from pileup.
Tagging $B$ hadrons (using displaced vertices or other kinematic
variables), or even reconstructing the particles in the final-state
from the output of the detector, also involves hadronic
information. This will also be affected by the extra hadrons from
pileup.

Generally speaking the practical consequences of pileup are
two-folded.
The first effect is almost straightforward: pileup biases the
quantities measured from the final state. For example, the transverse
momentum of jets is shifted upwards due to the extra hadronic
activity; or the hadronic energy around a lepton/photon is increased,
resulting in a loss of reconstruction efficiency.
The second effect is a little bit more subtle and is related to the
fact that the pileup activity fluctuates. These fluctuations arise
from multiple reasons: the actual number of pileup events, $\nPU$,
fluctuates around its average $\avg{\nPU}$; then, the final-state
hadronic activity fluctuates from one zero-bias collisions to
another. The last point takes several forms: the overall (total) energy
deposited in the final state fluctuates between different zero-bias
collisions but the distribution of these particles across the detector
changes as well.
As a consequence, the energy deposited by the pileup particles will
vary in different regions of the detector as well as other properties
like, for example, the relative fraction of charged and neutral
particles coming from pileup.
All these fluctuations yield a smearing of the quantities reconstructed
from the final state of the collision: fluctuations of the pileup
activity will result in fluctuations of the bias due to pileup, \ie a
smearing of the reconstructed quantities and distributions.
For example, the reconstructed jet transverse momentum will be smeared
by the fluctuating pileup activity, affecting the jet energy
resolution. Similarly, the resolution on the missing transverse energy
will degrade due to the fluctuations of the overall pileup activity.
{\it In summary, pileup biases and smears the quantities reconstructed
  from the final state of the hard collision, with the latter
  resulting in a degradation of the resolution.}
We will discuss this more precisely in
Section~\ref{sec:areamed-idea-characterisation}.

To gain a little more insight on the importance of pileup effects, one
can give a quick quantitative estimate of the above effects (see
chapter~\ref{chap:mcstudy} for a more precise discussion).
For 13~TeV collisions and a pileup multiplicity of 50, which is
representative of the situation towards the end of Run~II of the LHC,
an anti-$k_t$ jet of radius $R=0.4$ will see its energy shifted, on
average, by 15-20~GeV, with a smearing of about 7 GeV.
Now imagine that one wants to reconstruct jets down to a few tens of
GeV, say 20-30 GeV, \eg for the measurement of double Higgs production
(a key element of the LHC programme).
Both the energy shift and smearing are becoming serious issues.
Even if this is an extreme example, it is good to keep in mind that
the energy scales of pileup effects when reconstructing jets is not
negligible, typically in the 10-GeV ballpark (a little above for the
bias, a little below for the smearing) and will increase at the HL-LHC.
This is therefore something one cannot neglect and has to be addressed
in almost all the analysis at the LHC. Pileup often turns out to be one
of the important sources of systematic uncertainties at the LHC.
This affects in particular precision measurements, or measurements
involving lower energy scales.

Given the harmful effects of pileup, it is natural to investigate
how to mitigate them.
This exercise naturally means pursuing two goals: (i) correcting for
its biases and (ii) reducing its damaging effects on resolution.
Addressing pileup mitigation techniques is the main purpose of this
review.
For this, it is important to keep in mind the above two goals,
especially when it will come to assessing the performance of pileup
mitigation tools.

Note that one could always take the attitude that pileup is part of
the detector effects on observables and that one could correct for its
effects using an unfolding procedure (as one typically does for other
detector effects), without applying any specific pileup mitigation
technique.
This can be reinforced by the fact that a proper measurement would
anyway have to unfold for detector effects, as well as for residual
pileup effects after one applies (or not) a pileup mitigation
technique.
There are several reasons to believe that there is an added value to
applying carefully-designed pileup mitigation techniques.
In our opinion, the main argument in favour of pileup mitigation
techniques is that it provides a more generic (observable-independent)
and precise approach than a global unfolding. As we shall discuss in
Section~\ref{sec:areamed-idea-characterisation}, and show repeatedly
throughout this review, pileup mitigation techniques aim at reducing
as much as they can the biases and resolution-degradation effects of
pileup by exploiting event-by-event, and even jet-by-jet,
information. This procedure is, to a very large extent, independent of
the details of a specific  observable or distribution one wants to
measure. After these ``in-situ'' corrections one is therefore left
with much smaller residual corrections to apply and one should
therefore expect that this ultimately leads to smaller systematic
uncertainties compared to a situation where all pileup corrections
would have been left to an a posteriori unfolding procedure.
In other words, by first applying an efficient pileup mitigation
method --- reducing the average biases and improving the resolution
in each event in a generic way independent of the details of
the analysis --- the analysis-dependent unfolding that one is left
with is much lighter, hence most likely resulting in smaller and
better-controlled systematic uncertainties.

That said, we should keep in mind that, for precision measurements, one
would still need an unfolding of the residual pileup effects. On the
one hand this means that one can live with small residual biases (see
also discussions in Chapter~\ref{chap:beyond-motivation}), but on the
other hand, it also means that we want to keep these residual
corrections small and process-independent and that we want, for
example, to avoid fluctuations with long tails which are more delicate
to unfold.

Over the past decade several pileup mitigation techniques have been
proposed. One of the first proposed methods for used at the LHC relied
on the ability to reconstruct quite efficiently pileup vertices using
charged tracks. It applied a correction directly proportional to the
number of reconstructed pileup vertices.
Ultimately, Run~I of the LHC has adopted instead the {\em area--median
  method}, initially proposed~\cite{Cacciari:2007fd} by Matteo
Cacciari and Gavin Salam, and significantly extended by Matteo
Cacciari, Gavin Salam and
myself~\cite{Cacciari:2010te,AlcarazMaestre:2012vp,Cacciari:2012mu,Soyez:2012hv,Cacciari:2014jta}.
this method has still extensively been used during Run~II except in
specific applications to which we will return later.
As we will show at length in this document, one of the benchmarks of
the area--median method is that it provides an efficient and robust
way to correct for pileup, with a sizeable reduction of the resolution
degradation and a very small remaining average bias.
However, with the perspective for yet higher pileup multiplicities in
the future, several groups have started to investigate yet more
powerful techniques, yielding several new methods like
cleansing~\cite{Krohn:2013lba}, the
SoftKiller~\cite{Cacciari:2014gra}, PUPPI~\cite{Bertolini:2014bba} or
a wavelet approach~\cite{Monk:2018clo}, or
machine-learning-based techniques like PUMML\cite{Komiske:2018lor} or
PUPPIML~\cite{Martinez:2018fwc}.

In this context, this document serves several purposes. The first and
main one (part~I of this document) is to provide a review of the
area--median method that is used at the LHC.
This includes an extensive discussion of the method itself, and its
application to several observables, including the ubiquitous jet
transverse momentum, the jet mass and jet shapes, and the jet
fragmentation function. We will also perform an in-depth Monte-Carlo
validation of the approach for applications at the LHC and highlight
our recommendation for practical use of the area--median pileup
subtraction technique.
Since pileup subtraction is most often done numerically, we will also
include a description of the area--median interface provided in
\fastjet~\cite{fastjet,fastjet-manual}.

Another aspect that will be explored in the context of our review of
the area--median method is its application to heavy-ion collisions
either in gold--gold collisions at RHIC or in lead--lead collisions at
the LHC.
Even though in both cases luminosities are sufficiently low to neglect
pileup effects, the case of heavy-ion collisions has a different
effect that behaves similarly to pileup: the quark--gluon plasma
created during a heavy-ion collision decays into a large number
(thousands) of particles in the final-state of the collision,
corresponding to a huge Underlying Event activity.
This is in practice very similar to pileup, with many soft particles
scattered all over the event and compromising the reconstruction of
the hard objects in the event.
The activity of the Underlying Event in a LHC lead--lead event is even
larger than the current pileup activity at the LHC: for the most
central collisions, an anti-$k_t$ jet of radius $R=0.4$ can receive an
additional contamination of $100-150$~GeV, with fluctuations above
15~GeV.
We will therefore review as well how the area--median can be used to
correct for the large Underlying Event contamination to jets produced
in heavy-ion collisions.
This however requires a few words of caution.
First of all, the Underlying Event cannot practically be separated
from the ``hard event'' in a well-defined way (this is also the case
in proton--proton collisions).
In heavy-ion terms, the medium and jet particles cannot be uniquely
separated, the jet can for example back-react on the medium, ...
Additionally, the products of the decay of the quark--gluon plasma have
flow structures, like the elliptic flow~(see
\eg~\cite{Ollitrault:1992bk,Poskanzer:1998yz,Adler:2003kt,Adams:2003am,Alt:2003ab,Back:2004mh,Aamodt:2010cz,Chatrchyan:2012wg,ATLAS:2011ah}),
related to the geometry of the collision, which need to be taken into
account when implementing a correction.

Besides a thorough description of the area--median pileup subtraction
method, we also wish to present a series of other results and
techniques. These will cover two main directions.
The first (part~II of this document) concerns a family of methods used
in the specific context of jet substructure.
Jet substructure, used \eg for the identification of boosted objects,
is an entire topic on its own and we will focus here on the aspects
related to pileup mitigation.
One of the important aspects here is that jet substructure studies
typically use fat jets, \ie large-radius jets, which are specially
affected by pileup (and the Underlying Event).
Our focus here will be centred on {\it grooming techniques} which aim
at limiting the overall sensitivity of jets to pileup.
As we shall argue \eg in Chapter~\ref{chap:grooming-mcstudy}),
grooming techniques take a relatively orthogonal approach to pileup
mitigation compared to the area--median method. They reduce the
sensitivity to pileup --- basically by using an effective jet radius
smaller than the one of the original fat jet --- without explicitly
correcting for pileup effects.
For better results, grooming techniques should therefore be
supplemented by a pileup correction like the area--median method, or
the Constituent Subtractor~\cite{Berta:2014eza} (a particle-based
extension of the area--median approach), or
PUPPI~\cite{Bertolini:2014bba} (developed and used extensively by CMS
for substructure analyses during Run~II of the LHC).
Our goal here is to focus on the grooming aspects, so our studies will
use them in conjunction with the area--median approach.\footnote{We
  briefly discuss the Constituent Subtractor in
  Section~\ref{sec:areamed-shapes} and PUPPI in our conclusion study
  (Chapter~\ref{chap:ccl}).}

Finally, in the third and last part of this document, we will discuss
methods that have been proposed more recently and which aim at
reducing the sensitivity to pileup fluctuations and their impact on
the jet energy resolution. These methods are mostly meant for future
use, typically at the HL-LHC and potential future hadronic colliders,
where the pileup conditions will significantly worsen.

Our aim in this last part is not to provide an exhaustive review of
existing methods. In particular, we will not discuss recent tools such
as PUPPI (except for some comparisons in our conclusive study) or
machine-learning-based approaches such as PUMML or PUPPIML. Instead,
we mainly target two goals. The first is to highlight the challenges
of the exercise. In particular, we will argue that improving on the
resolution front is usually done at the expense of a more limited
control over the bias of the method (cf. the two goals of pileup
mitigation introduced earlier).
This will be contrasted with the area--median approach. We will for
example study the SoftKiller (which makes use of key quantities of
the area--median method).
We shall also discuss the applicability of the area--median
subtraction method to charged-hadron-subtracted (CHS) type of
events\footnote{In an idealised viewpoint, for each charged track one
  identifies the interaction vertex it originates from. One can then
  remove tracks not associated directly to the primary hard
  interaction (or leading vertex).} and compare it to an alternative
approach which directly makes use of the ratio of charged tracks to
neutral energy deposits.

The second goal of this third part is to document a series of yet
unpublished results. These mostly cover ideas and candidates for new
pileup mitigation methods organised in two series of studies: one
concerning the possibility to use grooming techniques as a generic
tool for pileup subtraction (not confined to their applications to fat
jets); the other covering a list of candidates for event-wide pileup
mitigation methods that we have come across when developing the
SoftKiller.
In all these cases, the idea is to explore different methods that can
lead to a reduction of the pileup fluctuation effects on jet energy
resolution. Only a few methods have been proposed so far and our hope
here is that the series of preliminary results we present can help
towards building more powerful tools in the future.

Before getting our hands dirty, one last point needs to be addressed
in this introduction: as stated in the title of this review, we
concentrate on theoretical aspects of pileup mitigation and this calls
for clarifications.
Based on what is said above, one can easily get the feeling that
pileup is a purely experimental issue. After all, reconstructing the
fundamental objects in the final state of a collision seems
exclusively an experimental task connected to a proper understanding
of all aspects of the detector's response.
We want to show here that there is more to the story and that there is
room for theoretical ideas and concepts when it comes to investigating
pileup effects and pileup mitigation.

To highlight this, we will provide analytic insights on our results
whenever possible throughout this document. (See the ``reading
guide'' below to navigate between descriptions, results, Monte Carlo
simulations and analytic details.)
In the first part, related to the review of the area--median approach,
we will discuss pileup effects using simple analytic arguments.
We will also show that the concept of jet
areas~\cite{Cacciari:2008gn}, central to the area--median method, has
lots of interesting features in perturbative Quantum Chromodynamics
(QCD), observed as well in Monte Carlo simulations.
The main features of the other central ingredient of the area--median
method, namely the event-by-event estimate of the pileup activity, can
also be understood from a simple toy-model of pileup for which we can
obtain a series of analytic results.

Furthermore, analytic results can also be obtained for the grooming
techniques discussed in the second part of this document.
The fact that jet substructure techniques can be amenable to an
analytic understanding has been one of the major breakthrough in the
jet substructure community over the past few years.
It has even been shown that they are nice candidates for precision
calculations at the LHC (see
\eg~\cite{Frye:2016aiz,Marzani:2017mva,Marzani:2017kqd,Sirunyan:2018xdh,Aaboud:2017qwh}).
In Chapter~\ref{chap:grooming-analytic}, we will give a few insights
into analytic calculations for grooming techniques, focusing primarily
on quantities related to pileup.

Conversely, since the main purpose of this document is to focus on the
theoretical aspects of pileup mitigation, we will extensively use
Monte Carlo simulations but we will not at all discuss additional
complications related to detector effects\footnote{Except for our
  description of the SoftKiller method in
  Chapter~\ref{chap:soft-killer} where we shall briefly investigate
  the robustness of the method in an experimental setup.} or other
effects like out-of-time pileup which comes from the fact that the
detectors readout often spans over the time of several bunch
crossings.
To be fully addressed, these effects need a dedicated experimental
analysis, with details specific to each experiments, which goes beyond
the scope of this document.
It is however worth mentioning that, as far as the area--median pileup
subtraction technique is concerned, we are confident that the picture
described here is not going to be strongly affected by detector
effects. This is mainly due to that fact that the area--median method
is extremely robust against the details of both the hard and the pileup
events.\footnote{One specific example of a detector effect that we
  shall not discuss is the, usually small, non linearities that will
  arise from calorimeter thresholds.}
For the same reason, we shall concentrate our discussion on the
reconstruction of the properties of the jets. Other quantities like
lepton/photon isolation criteria or missing-$E_T$ reconstruction are
much more dependent on the details of the detectors and will not be
investigated here.
Readers who are interested in experimental aspects of pileup
subtraction can for example consult Ref.~\cite{pileup-atlas} for the
ATLAS experiment of Ref.~\cite{pileup-cms} for CMS.

\paragraph{Reading guide and organisation of this document.}
Even without entering into the details of detector effects, this
document remains rather long. We therefore find helpful to give a
short reading guide through this review.

We first reiterate that this document focuses only on pileup
mitigation. A brief description of the commonly-used jet algorithms is
given in Appendix~\ref{app:jetdefs} and we refer to
Ref.~\cite{Salam:2009jx} for a comprehensive introduction to jet
physics in general.
That said, we have separated the text in three main parts: first the
extensive description of the area--median method, followed by a
discussion of grooming techniques in the second part and, finally,
opening towards more recent techniques in the third part of this
document.
I strongly suggest to all readers to go through the first two Sections
of Chapter~\ref{chap:areamed} which provide a generic description of
pileup effects, generic considerations about pileup mitigation and the
net effects on physical observables. These two Sections also introduce
basic pileup characteristics that will be used throughout the whole
document \eg to provide physical interpretations of our results.

The discussion of the area--median method has been organised in three
different steps: we describe the method and its several extensions in
Chapter~\ref{chap:areamed}, we provide a validation of the method
based on Monte Carlo simulations in Chapters~\ref{chap:mcstudy}
(pileup in proton-proton collisions) and \ref{chap:mcstudy-hi}
(Underlying Event in heavy-ion collisions) and introduce a series of
analytic results in Chapter~\ref{chap:analytics}.
Again, the reader interested in the area--median method is invited to
go first through Chapter~\ref{chap:areamed} and in particular through
Section~\ref{sec:areamed-areamed}, which gives the basic description
of the method, and Section~\ref{sec:areamed-practical}, which lists in
details our practical recommendations for use at the LHC.
Some of the more technical aspects like the definition of jet areas
(Section~\ref{sec:areamed-defareas}), extensions to jet shapes
(Section~\ref{sec:areamed-shapes}) and jet fragmentation function
moments (Section~\ref{sec:areamed-fragmentation-function}), or the
practical details of the \fastjet interface
(Section~\ref{sec:areamed-implementation}) can be skipped if one is
just interested in the main lines.
Chapter~\ref{chap:mcstudy} then demonstrates the performance of the
method, justifying the recommendations made in
Section~\ref{sec:areamed-practical}, with
Chapter~\ref{chap:mcstudy-hi} departing from pure pileup mitigation to
concentrate of the application of the area--median method to the
subtraction of the large Underlying Event in heavy-ion collisions.
Our overview of the area--median method ends with
Chapter~\ref{chap:analytics} where we discuss at length analytic
properties of jet areas, and provide a brief translation of the
analytic results of \cite{Cacciari:2009dp}, initially discussing the
analytic estimation of the Underlying Event activity, to the
estimation of the pileup properties.

The second and third parts of this document cover more specialised
topics. 
For the discussion of grooming techniques applied to boosted jets (the
second part of this document), we have kept the same structure as for
the first part: we first provide a detailed description of a series of
grooming methods (Chapter~\ref{chap:grooming-description}), then move
on to a series of Monte Carlo studies
(Chapter~\ref{chap:grooming-mcstudy}) and finish with a description of
a few analytic results for groomers
(Chapter~\ref{chap:grooming-analytic}).
The reader who is not familiar with grooming is obviously invited to
read first the descriptions in
Chapter~\ref{chap:grooming-description}.
The Monte Carlo studies (Chapter \ref{chap:grooming-mcstudy}) mostly
revisit a study done in the context of the Boost 2012
workshop~\cite{Altheimer:2013yza}, studying how groomers mitigate
pileup for boosted jets. However, the study presented here extends the
one originally done for the Boost proceedings to include a more
extensive and modern set of grooming techniques.
At last, Chapter \ref{chap:grooming-analytic} will provide a
calculation of the grooming radius, directly related to the
sensitivity to pileup, for the
SoftDrop~\cite{Larkoski:2014wba} grooming procedure. 
This is merely an example illustrating how jet substructure methods
are amenable to an analytic treatment. This Chapter is more
appropriate for the reader who is curious about recent developments
aiming towards an analytic understanding of jet substructure.

The third part of this document will discuss several methods going
beyond the area--median approach. We encourage all readers to explore
Chapter~\ref{chap:beyond-motivation} which provides an introductory
discussion of the motivations to seek for more powerful techniques as
well as the associated fundamental challenges.
For readers interested in continuing their journey in this direction,
we have then split the results in four separate chapters that can be
read more-or-less independently.
We start in Chapter~\ref{chap:beyond-grooming} with a preliminary
study of the usage of grooming techniques as a generic tool for pileup
mitigation, beyond the scope of boosted jets.
Then, in Chapter~\ref{chap:charged_tracks}, we discuss the use of CHS
events and charged tracks information to help mitigating pileup
contamination for neutral particles. This includes a discussion of the
area--median method and the ``Neutral-proportional-to-Charged''
method~\cite{Cacciari:2014jta} as well as an in-depth comparison to
the recently proposed cleansing technique~\cite{Krohn:2013lba}.
In Chapter~\ref{chap:soft-killer} we introduce the SoftKiller method
that shows significant improvements over the area--median method in
terms of energy resolution, while keeping limited average biases. We
believe that it can be a powerful method for use in high-pileup
environments, especially in view of its remarkable speed properties.
Finally, in Chapter~\ref{chap:beyond-prelimn} we list a series of
preliminary studies including several event-wide subtraction methods
that we have only briefly investigated when we were developing the
SoftKiller method, as well as possible extensions of the SoftKiller
method itself.
These are still preliminary results but I believe that they can prove
helpful in the development of future pileup mitigation methods.

As a conclusion, we will provide a summary of the key performances of
several methods, published and preliminary, using the common framework
that had been introduced in the context of the PileUp WorkShop held at
CERN in May 2014~\cite{PUWS}.
This part is also appropriate for a generic reader who can then go
back and dig more into the additional details he or she would be
interested in.


\part{The area--median approach}
\chapter{Description of the area--median method}\label{chap:areamed}

\section{Basic characterisation of the pileup effects}\label{sec:areamed-idea-characterisation}


Before discussing how to remove the pileup contamination from a jet it
is helpful to characterise that contamination in terms of a few simple
physical effects.
This simple characterisation has its limitations but provides a
simple physical understanding of many situations we shall discuss
throughout this document.

The most fundamental case we will focus on is the reconstruction of
the jet transverse momentum, $p_t$. This is by far the most used jet
characteristic. Applications to more generic jet properties will be
discussed in Sections \ref{sec:areamed-shapes} and
\ref{sec:areamed-fragmentation-function}.

The main characteristic of pileup that the area--median approach makes
use of is the fact that, in each event, the energy (or, rather, the
transverse momentum) deposit coming from pileup is to a large extent
uniform in rapidity and azimuth. In that approximation the pileup
activity is characterised by a single number: {\bf the pileup
  transverse momentum density, $\rho$}. In this framework, $\rho$ is
defined per unit of area in the rapidity--azimuth plane.
It is crucial to point out that, here, $\rho$ is defined {\it event by
  event}.
In practice, $\rho$ has a additional relatively small rapidity
dependence, smooth on average. For the sake of this discussion, let us
temporarily neglect that dependence. We will come back to it in
Section~\ref{sec:areamed-position}.

Given the pileup transverse momentum density $\rho$, the pileup effect
on a jet will be to shift its $p_t$ by an amount proportional to
$\rho$.
Since $\rho$ is defined per unit area in the rapidity--azimuth plane,
the proportionality constant will be the ``area'' of the jet, $A_{\rm
  jet}$.
If one thinks of a calorimeter with towers of size $a\times a$, this
is equivalent to saying that the $p_t$ of a jet will be shifted by the
average pileup $p_t$ in each tower, $\rho_{\rm tower}$, times the
number of towers in the jet, $N_{\rm towers}$.\footnote{One needs to
  be careful how to define the number of active towers in a jet. Even
  if we count the active towers that include pileup, there is always a
  probability that a tower inside a jet remains empty. We will come
  back to this at the beginning of
  Section~\ref{sec:areamed-defareas}, where we introduce a proper
  definition for jet areas.}
In our language, this corresponds to $\rho_{\rm tower}=\rho a^2$ with
a jet areas $A_{\rm jet}=N_{\rm towers}a^2$.

We will discuss at length the generic definition of the area of a jet
in the next section.
In a nutshell, the idea is to add to a given event a dense and uniform
coverage of infinitely soft particles, {\it ghosts}, each carrying a
quantum of area $a_{\rm ghost}$. We then define the {\bf active area}
of a jet as the number of ghosts in the jet, multiplied by
$a_{\rm ghost}$.
Physically, this active area mimics the effect of pileup in the sense
that pileup particles are uniformly distributed, in the limit where
their $p_t$ becomes infinitely small. This last point ensures that
ghosts do not affect the clustering provided one uses an infrared-safe
jet algorithm.

For a given pileup density $\rho$ and jet area $A_{\rm jet}$, the
average effect due to the clustering of pileup particles in a jet is
to shift the jet $p_t$ by $\rho A_{\rm jet}$.


A series of additional effects come on top of the simple transverse
momentum shift discussed above.
The fact that the pileup transverse momentum deposit is uniform in
rapidity and azimuth is just an approximation only true on average. 
If one takes an event with a given $\rho$ one can imagine looking at
the energy deposit in a square patch of size $1\times 1$ in the
rapidity--azimuth plane.
As this patch moves in the event the energy deposit will vary. This
results in a full distribution which is interesting in
itself\footnote{We shall come back to that later in this document, \eg
  to the fact that this distribution has longer tails towards large
  $p_t$ than a Gaussian distribution.} but, for most of the simple
discussions in this document, it is sufficient to see it as a Gaussian
of average $\rho$ and standard deviation $\sigma$.
This new pileup property, {\bf $\sigma$, characterises the pileup
  fluctuations within an event} and will appear each time we discuss the
uncertainty of pileup subtraction techniques.

The effect of the pileup fluctuations on the jet is to smear its
transverse momentum. Indeed, if the jet sits on an upwards
(resp. downwards) pileup fluctuation, the contamination due to the
capture of pileup particle in the jet will be more (resp. less) than
the average $\rho A_{\rm jet}$.
These fluctuations around $\rho A_{\rm jet}$ are given by $\sigma
\sqrt{A_{\rm jet}}$. The proportionality to $\sigma$ is expected but
the appearance of $\sqrt{A_{\rm jet}}$ may be a bit more of a
surprise.
This is actually the fact that fluctuations, \ie dispersion of
statistical distributions, add in quadrature, which means that it is
$\sigma^2$ which scales like the area of the jet.
Another, perhaps more concrete, way to see this is to come back to the
case of a calorimeter. If the pileup $p_t$ in towers fluctuates by an
amount $\sigma_{\rm tower}$, the $p_t$ fluctuations in a jet of
$N_{\rm towers}$ towers will be $\sigma_{\rm tower}\sqrt{N_{\rm
    towers}}$. If we take a patch of area 1 in which there is $1/a^2$
towers, this gives $\sigma_{\rm tower}=\sigma a$ and, remembering
$A_{\rm jet}=N_{\rm towers}a^2$, we get fluctuations in our jet given
by $\sigma\sqrt{A_{\rm jet}}$.


Before turning to the discussion of how pileup affects the jet $p_t$,
there is a third effect that one has to introduce: {\bf back-reaction}
(BR). This relates to the fact that pileup does not just add extra
soft particles, from pileup interactions, to jets: since pileup
particles are clustered at the same time as particles from the hard
interaction, {\bf the particles from the hard interactions clustered
  in a given jet can be affected by the presence of the pileup
  particles}. Back-reaction is the difference between the clustering
done with and without pileup particles, computed only using the
particles from the hard interactions within a given jet .
Again, the effect of back-reaction on the jet transverse momentum is
both an average shift and a dispersion. We will provide detailed
calculations and simulations of back-reaction effects in Section
\ref{sec:areamed-analytic-back-reaction}, but all that we need to know
at this stage is that if one uses the anti-$k_t$ algorithm
\cite{Cacciari:2008gp} to cluster the jets, the average $p_t$ shift
due to back-reaction is negligible and its smearing effects are
proportional to $\sqrt{\rho p_t}$ with a relatively small coefficient
which make them relevant only at large $p_t$.


The discussion so far can be summarised under the form of a pocket
formula which will be extremely useful for many discussions throughout
this report: the effect of pileup on the jet transverse momentum is
given by
\begin{equation}\label{eq:pueffects-central}
\boxed{\,
p_{t,\rm full} =   p_{t,\rm hard}
             +   \rho A_{\rm jet} 
             \pm \sigma\sqrt{A_{\rm jet}}
             +   {\rm BR},\,
}
\end{equation}
where $p_{t,\rm full}$ is the transverse momentum of the jet when
clustering the event with particles from both the hard interaction and
pileup, while $p_{t,\rm hard}$ is the jet $p_t$ with only the hard
particles, \ie the quantity we really want to reconstruct.
In the rhs of this equation, the second term (the $\rho$ term) is the
average shift, the third term (the $\sigma$ term) is the pileup
fluctuations, responsible for degrading the jet energy resolution, and
the fourth is the back-reaction effects, all introduced above.
We will see these effects directly in Monte-Carlo simulations in
Chapter~\ref{chap:mcstudy}.
Similar considerations also apply to other jet properties, with $\rho$
shifting their average value and $\sigma$ smearing their value (see
\eg Section \ref{sec:areamed-shapes} for additional details).


For the sake of future discussions, it is helpful to
introduce a few additional quantities. When averaging over an
ensemble of events, we will denote the average of $\rho$ by
$\avg{\rho}_{\rm ev}$ and its standard deviation by $\sigma_\rho$. 
It is also customary to see the pileup as a superposition of \nPU\ 
minimum bias events. Each minimum bias event has itself a
corresponding $\rho\equiv \rho_{\rm mb}$ and $\sigma=\sigma_{\rm
  mb}$. We shall denote its average by $\avg{\rho_{\rm mb}}_{\rm ev}$
and its standard deviation by $\sigma_{\rho, \rm mb}$. For a class of
events with the same \nPU, we would have, neglecting the fluctuations
across the event sample of all the various $\sigma$,
\begin{equation}\label{eq:avgrho_mb2NPU}
  \avg{\rho}_{\text{ev, fixed }\nPU}=\nPU\avg{\rho_{\rm mb}}_{\rm ev},
  \qquad
  \sigma_{\text{ev, fixed }\nPU} = \sqrt{\nPU}\,\sigma_{\rm mb}
  \quad\text{ and }\quad
  \sigma_{\rho\text{, fixed }\nPU}=\sqrt{\nPU}\,\sigma_{\rho,{\rm mb}}.
\end{equation}
Of course, if one averages over all events, one also needs to take
into account the variations of \nPU\ across events. The dispersion of
$\rho$ will then pick an extra contribution coming from the variation
of \nPU. This gives
\begin{equation}\label{eq:avgrho_NPU2mu}
  \avg{\rho}_{\rm ev}=\mu\avg{\rho_{\rm mb}}_{\rm ev},
  \qquad
  \sigma = \sqrt{\mu}\,\sigma_{\rm mb}
  \quad\text{ and }\quad
  \sigma_\rho^2\simeq \mu\sigma^2_{\rho,{\rm mb}}+\sigma^2_\nPU\avg{\rho_{\rm mb}}_{\rm ev}^2
\end{equation}
where, we have defined $\mu$ as the average number of pileup
interactions and $\sigma_\nPU$ its standard deviation. For a
Poison-distributed pileup, we have $\sigma_\nPU^2=\mu$.

From these considerations, we can reformulate
\eq~(\ref{eq:pueffects-central}) in several ways:\footnote{We can
  also write a series of similar expressions where the averages are
  done at the level of the jets rather than at the level of transverse
  momentum densities. In that case, we would pick an extra
  contribution to the dispersion coming from the fluctuations of the
  area of the jet.}
\begin{align}
p_{t,\rm full}
& =  p_{t,\rm hard}
  + \avg{\rho_{\rm mb}}_{\rm ev} \nPU A_{\rm jet}
  \pm \sigma_{\rho,\rm mb} \sqrt{\nPU} A_{\rm jet}
  \pm \sigma\sqrt{A_{\rm jet}} + {\rm BR}\label{eq:pueffects-shift_prop_npu}\\
& =  p_{t,\rm hard}
  + \avg{\rho_{\rm mb}}_{\rm ev} \mu A_{\rm jet}
  \pm \avg{\rho_{\rm mb}}_{\rm ev} \sqrt{\mu} A_{\rm jet}
  \pm \sigma_{\rho,\rm mb} \sqrt{\mu} A_{\rm jet}
  \pm \sigma\sqrt{A_{\rm jet}} + {\rm BR}\label{eq:pueffects-shift_prop_mu}
\end{align}
These expressions differ in the level at which we compute the average
shift and fluctuations:
\begin{itemize}
\item in (\ref{eq:pueffects-central}), the average shift is
  characterised by $\rho$, defined event by event, leaving the
  intra-event fluctuations $\sigma$;
\item in (\ref{eq:pueffects-shift_prop_npu}) the average shift is
  proportional to the number of pileup vertices and we therefore get
  an extra source of fluctuations,
  $\sigma_{\rho,\rm mb} \sqrt{\nPU} A_{\rm jet}$, coming the fact that
  the fluctuations in the overall pileup deposit of each minbias event
  will translate in fluctuations in $\rho$;
\item in (\ref{eq:pueffects-shift_prop_mu}), the average shift is
  written as a constant
  $p_t=\avg{\rho_{\rm mb}}_{\rm ev} \mu A_{\rm jet}$ for each jet of
  area $A_{\rm jet}$ and we pick yet an extra source of smearing
  coming from the fluctuations of \nPU,
  $\avg{\rho_{\rm mb}}_{\rm ev} \sqrt{\mu} A_{\rm jet}$.
\end{itemize}
As we shall see in this Chapter and the following ones, various pileup
mitigation techniques subtract from the full jet transverse momentum
and estimate of the average shift. One would then be left with
residual fluctuations depending on the level at which this is done.
These various expressions will prove useful when we discuss the net
bias and residual resolution degradation after we apply a pileup
subtraction technique.

These basic considerations are almost enough to introduce the
area--median subtraction method but a few subtleties arise from our
definition of the jet area. We shall therefore introduce that concept
once and for all in Section~\ref{sec:areamed-defareas} and return to
the area--median subtraction method in
Section~\ref{sec:areamed-areamed}.

Note finally that for all the applications that we will consider in
this document, we can assume some form of ordering between the various
quantities introduced above. Typically, the scale of the hard jet will
be larger than the background contamination, itself larger than the
pileup fluctuations, namely, $p_{t,\rm hard}/A_{\rm
  jet}>\rho>\sigma$. This approximation is valid for the case of
pileup at the LHC.

\section{Effects of average shifts and resolution degradation on
  physical observables}\label{sec:description-convolution}

We have seen that the main effects of pileup are to shift and smear
the transverse momentum of the jets, as well as their other kinematic
properties.
The goal of pileup mitigation is, to a large extent, to correct for
the average shift and to reduce as much as possible the smearing
effects. 
At many stages in this document, we will characterise pileup
subtraction techniques using these properties as quality measures. 

In order to put things in perspective and to facilitate the
understanding of several arguments used when we will discuss our
findings, it is helpful to provide a simple description of the effects
that an average shift and a dispersion would have on physical
observables.

For the sake of the argument, let us consider the differential jet
cross-section as a function of the jet transverse momentum,
$d\sigma/dp_t$.
Imagine this has a truth distribution $d\sigma_{\rm truth}/dp_t$
corresponding to a case without pileup, usually denoted by
$d\sigma_{\rm hard}/dp_t$ throughout this document.
Let us now add pileup to that initial distribution and, optionally,
subtract it, to obtain a ``reconstructed'' jet cross-section
$d\sigma_{\rm reco}/dp_t$. This is the result of the convolution of
the initial, truth, spectrum with the spectrum characterising the
residual pileup effects. We want to understand how this will differ
from the expected result $d\sigma_{\rm truth}/dp_t$.

For the vast majority of the results discussed here, several
simplifications can be made. 
First, the effect of pileup and the optional additional subtraction
can, to a large extent, be approximated by a Gaussian of average
$\Delta$ and of dispersion $\sigma$. The former corresponds to the
average shift, or bias, or offset, while the latter corresponds to the
residual smearing, or resolution degradation, or fluctuations.
As long as the residual effects of pileup do not show long
tails\footnote{In particular, long positive tails.}  the Gaussian
approximation is appropriate to capture the physics.

Then, let us approximate the initial distribution by a decreasing
exponential:
\begin{equation}
\frac{d\sigma_{\rm truth}}{dp_t} = \frac{\sigma_0}{\kappa} e^{-p_t/\kappa}.
\end{equation}
This is representative of a steeply-falling distribution and, as long
as $\Delta$ and $\sigma$ are not too large, it is always a good
approximation, at least locally.

If we convolute that distribution with a Gaussian, we find
\begin{align}
  \frac{d\sigma_{\rm reco}}{dp_t}
  & \approx \int dk_t
    \frac{1}{\sqrt{2\pi}\sigma}e^{-\frac{(p_t-k_t-\Delta)^2}{2\sigma^2}}
    \frac{d\sigma_{\rm truth}}{dk_t}\\
  & \approx \frac{d\sigma_{\rm truth}}{dp_t} \exp\left(\frac{\Delta}{\kappa}+\frac{\sigma^2}{2\kappa^2}\right).
    \label{eq:distrib-effect-convoluted}
\end{align}

The above result shows that both the average shift and the dispersion
have an impact on the final reconstructed distribution.
The effect of the shift is rather trivial: an offset $\Delta$ on $p_t$
translates immediately in a factor $\exp(\Delta/\kappa)$. 
The effect of the dispersion is a bit more subtle. It is related to
the fact than, on a steeply-falling distribution, positive
fluctuations will be given a larger weight than negative fluctuations.

An alternative way to see this is to look, for a given reconstructed
$p_{t,\rm reco}$, at the distribution of the initial, truth,
$p_{t,\rm truth}$. This is given by
\begin{align}
  P(p_{t,\rm truth}|p_{t,\rm reco})
  & = \left[\frac{d\sigma_{\rm reco}}{dp_{t,\rm reco}}\right]^{-1}
      \frac{1}{\sqrt{2\pi}\sigma}
      e^{-\frac{(p_{t,\rm reco}-p_{t,\rm truth}-\Delta)^2}{2\sigma^2}}
      \frac{d\sigma_{\rm truth}}{dp_{t,\rm truth}}\\
  & = \frac{1}{\sqrt{2\pi}\sigma}
      \exp\left[-\frac{1}{2\sigma^2}\left(
      p_{t,\rm truth}-p_{t,\rm reco}+\Delta+\frac{\sigma^2}{\kappa}\right)^2\right].
    \label{eq:distrib-effect-most-probably}
\end{align}
This shows that the most probable truth $p_t$ is given by
\begin{equation}
p_{t,\rm truth}^{\text{(most likely)}} = p_{t,\rm reco}-\Delta-\frac{\sigma^2}{\kappa}.
\end{equation}
The presence of the $\Delta$ shift here is again expected and we see
that we get an extra contribution coming from the dispersion. This
extra contribution increases if the initial distribution becomes
steeper (\ie when $\kappa$ decreases) and disappears for a flat
distribution ($\kappa\to \infty$).

From the above perspective, one sees that if we want to mitigate
pileup, the ideal situation would be to have both $\Delta$ and
$\sigma$ as small as possible. 
From \eq~\eqref{eq:distrib-effect-convoluted}, we
see that having $\Delta+\sigma^2/(2\kappa)=0$ would also translate into
an unbiased reconstructed distribution but this requires some level of
fine-tuning and would be process-dependent.
In practice, we typically target pileup mitigation techniques that
result in $\Delta \approx 0$, and try to reduce $\sigma$ as much as
possible.
In an experimental context, the effect of $\sigma$ (and the residual
$\Delta$) would ultimately have to be unfolded together with the other
detector effects.

The discussion in this Section has only taken the jet $p_t$ and the
jet cross-section as an example. The very same arguments can be made
for other processes and for other jet properties. As long as the
residual effect of pileup are small enough, they can reasonable be
approximated by a Gaussian distribution and the physical observable
can locally be approximated by a decreasing exponential. 
In these conditions, the results (\ref{eq:distrib-effect-convoluted})
and (\ref{eq:distrib-effect-most-probably}) would therefore apply and
capture most of the relevant effects to estimate pileup effects and
to assess the quality of pileup mitigation techniques.

Finally, this discussion also justifies that the average shift and
dispersion are the two basic quality measures to study and, up to very
few exceptions, we will do so in all our studies in this document.
The robustness of a method can be assessed by checking that the
average shift $\Delta$ remains close to 0 regardless of the physical
process under consideration, the scale of the hard process, the pileup
conditions and other jet properties like its rapidity.
Once the robustness of a method is established, we can discuss its
efficiency in terms of how small the residual smearing $\sigma$ is.

\section{The jet catchment areas}\label{sec:areamed-defareas}

The first ingredient we will need for the area--median pileup
subtraction method is the definition of the concept of {\it jet
  areas}. 
This concept can be seen as an additional property of each jet, like
their transverse momentum or mass.
Note that, for hadronic collisions, we work with the transverse
momentum ($p_t$), rapidity ($y$), azimuthal angle ($\phi$) coordinate
system\footnote{One could also use the transverse energy and
  pseudo-rapidity. It is however more natural to keep the
  longitudinal-boost-invariant $p_t$ and $y$ since the clustering
  itself uses that set of coordinates. Also, rapidity differences are
  invariant along longitudinal boosts while pseudo-rapidity
  differences are not, except for massless particles.} and by ``jet
area'' we really mean the area in the $y-\phi$ plane.

In the context of a (simple) calorimeter, it is natural to think of
the area of a jet as the sum of the areas of each of the cells
included in the jet.
In a sense, our goal in this Section is to provide a well-defined
concept of jet areas which would also work at particle-level and
remain well-defined in the case of a calorimeter with empty towers in
a jet. 
Naively, one could just say that the area of a jet of radius $R$
should be $\pi R^2$ which would be biased because not all jet
algorithms produce circular jets and the boundary of a jet can be
influenced by the surrounding jets. 
Another tempting definition would be to take the convex hull of all
the particles constituting the jet. Again, this is not an ideal choice
since convex hulls could overlap and would not pave the
rapidity--azimuth plane.
In a nutshell, our solution is to define the area of a jet as a {\em
  catchment area}, \ie the region of the rapidity--azimuth plane where
the jet would catch infinitely soft particles.
There are several ways to implement this in practice, hence we will
introduce three definitions: the {\em passive area}, the {\em active
  area} and the {\em Voronoi area}.

For reasons that will soon become clear, {\bf the active area is the
  most appropriate for pileup subtraction} but we will start our
discussion with the passive area since it is the most simple one.
Finally, jet areas are amenable to analytic calculations and we shall
discuss this in great detail in Section~\ref{sec:analytics-areas}.

\subsection{Passive Area}\label{sec:areamed-defareas-passive}

Suppose we have an event composed of a set of particles $\{p_i\}$
which are clustered into a set of jets $\{J_i\}$ by some infrared-safe
jet algorithm.

Imagine then adding to the $\{p_i\}$ a single {\sl infinitely soft}
ghost particle $g$ at rapidity $y$ and azimuth $\phi$, and repeating
the jet-finding.
As long as the jet algorithm is infrared safe, the set of jets
$\{J_i\}$ is not changed by this operation: their kinematics and hard
particle content will remain the same, the only possible differences
being either the presence of $g$ in one of the jets or the appearance
of a new jet containing only $g$.

The passive area of the jet $J$ can then either be defined as a scalar
\begin{equation}
\label{eq:passive-area-def}
a(J) \equiv \int dy\,d\phi\; f(g(y,\phi),J)\qquad\qquad  f(g,J) =
\left\{\begin{array}{cc}
1 & g \in J \\
0 & g \notin J \\
\end{array}
\right. \; ,
\end{equation}
which corresponds to the area of the region where $g$ is clustered
with $J$, or as a 4-vector,
\begin{equation}
a_\mu(J) \equiv \int dy\,d\phi\; f_\mu(g(y,\phi),J)\qquad\qquad  f_\mu(g,J) =
\left\{\begin{array}{cc}
g_\mu/g_t & g \in J \\
0 & g \notin J \\
\end{array}
\right. \; ,
\end{equation}
where $g_t$ is the ghost transverse momentum.
For a jet with a small area $a(J)\ll 1$, the 4-vector area has the
properties that its transverse component satisfies $a_t(J) = a(J)$, it
is roughly massless and it points in the direction of $J$.  For larger
jets, $a(J)\sim 1$, the 4-vector area acquires a mass and may not
point in the same direction as $J$. We shall restrict our attention
here to scalar areas because of their greater simplicity. Nearly all
results for scalar areas carry over to 4-vector areas, modulo
corrections suppressed by powers of the jet radius (usually
accompanied by a small coefficient).%
\footnote{The above definitions apply to jet algorithms in which each
  gluon is assigned at most to one jet. For a more general jet
  algorithm (such as the ``Optimal'' jet finder of
  \cite{Grigoriev:2003yc,Grigoriev:2003tn} or those which perform
  $3\to2$ recombination like ARCLUS \cite{Lonnblad:1992qd}), then one
  may define the 4-vector area as
  \begin{equation}
    \label{eq:passive-4vect-area-gen}
    a_\mu(J) = \lim_{g_t \to 0} \frac{1}{g_t} \int dy\, d\phi\;
    (J_{\mu}(\{p_i\},g(y,\phi)) - J_{\mu}(\{p_i\}))\,,
  \end{equation}
  where $J_{\mu}(\{p_i\},g(y,\phi))$ is the 4-momentum of the jet as
  found on the full set of event particles $\{p_i\}$ plus the ghost,
  while $J_{\mu}(\{p_i\})$ is the jet-momentum as found just on the
  event particles.
}

\subsection{From passive to Voronoi area}\label{sec:areamed-defareas-voronoi}

The only algorithm for which one can make any statement about the
passive area for a general $n$-particle configuration is the $k_t$
algorithm~\cite{Catani:1993hr,Ellis:1993tq}.

Because of the $k_t$ distance measure, the single ghost will cluster
with one of the event particles before any other clustering takes
place. 
One can determine the region in which the ghost will cluster with a
given particle, and this is a definition of the area $a_{\kt,R}(p_i)$
of a particle $p_i$.  Since the ghost-particle clustering will occur
before any particle-particle clustering, the jet area will be the sum
of the areas of all its constituent particles:
\begin{equation}
  \label{eq:particle-areas}
  a_{\kt,R}(J) = \sum_{p_i \in J} a_{\kt,R}(p_i)\,.
\end{equation}

Can anything be said about the area of a particle? The ghost will
cluster with the event particle to which it is closest, as long it is
within a distance $R$. This corresponds to a geometrical construction
known as the Voronoi diagram, which subdivides the plane with a set of
vertices into cells around each vertices.\footnote{It is this same
  geometrical construction that was used to obtain a nearest neighbour
  graph that allowed $k_t$ jet clustering to be carried out in
  $N \ln N$ time~\cite{fastjet}.} %
Each cell has the property that all points in the cell have as their
closest vertex the cell's vertex.  Thus the Voronoi cell is equivalent
similar to the region in which a ghost will cluster with a particle up
to the limitation that the ghost should be within a distance $R$ of
the particle.
This leads us to define the Voronoi area of particle $i$,
$a_R^{\cV}(p_i)$, to be the area of its Voronoi cell ${\cal V}_i$
intersected with a circle of radius $R$, ${\cal C}_{i,R}$, centred on
the particle:
\begin{equation}
  \label{eq:voronoi-intersect}
  a_R^{\cV}(p_i) \equiv \text{area}({\cal V}_i \cap {\cal
    C}_{i,R})\,.
\end{equation}
Thus given a set of momenta, the passive area of a $k_t$ jet can be
directly determined from the Voronoi diagram of the
event,\footnote{Strictly speaking it should be the Voronoi diagram on
  a $y-\phi$ cylinder, however this is just a technical detail.} using
\eq~(\ref{eq:particle-areas}) and the relation
\begin{equation}
  \label{eq:kt-voronoi}
  a_{\kt,R}(p_i) = a_R^{\cV}(p_i)\,.
\end{equation}

\begin{figure}
  \centering
  \includegraphics[width=0.49\textwidth]{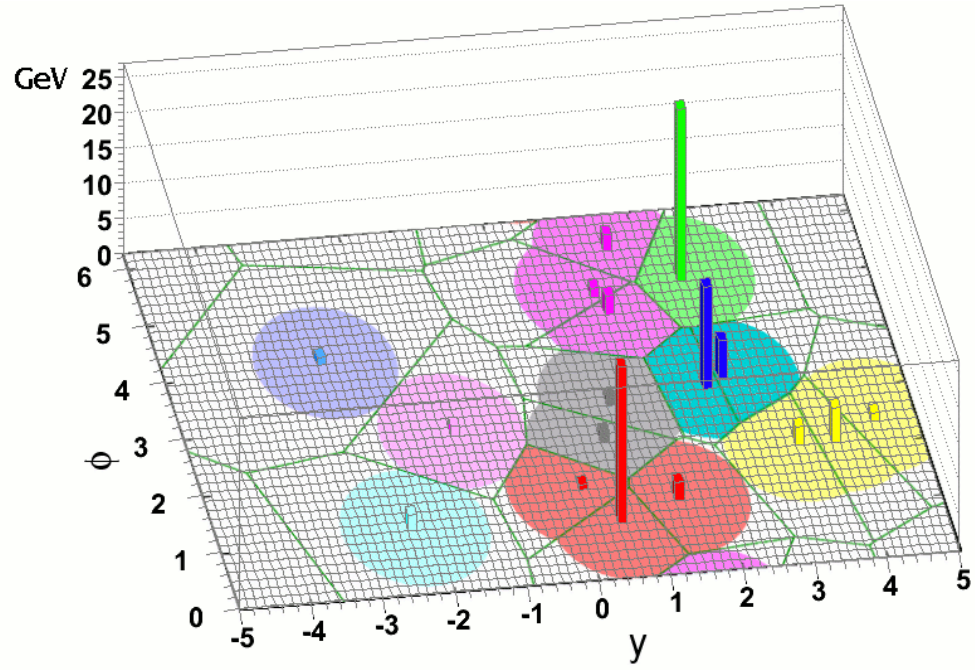}
  \caption{The passive area of jets in a parton-level event generated
    by Herwig and clustered with the $k_t$ algorithm with $R=1$. The
    towers represent calorimeter cells containing the particles, the
    straight (green) lines are the edges of the Voronoi cells and the
    shaded regions are the areas of the jets.}
  \label{fig:event-voronoi-mod}
\end{figure}%

The Voronoi construction of the $k_t$-algorithm passive area is
illustrated in fig.~\ref{fig:event-voronoi-mod}. One sees both the
Voronoi cells and how their intersection with circles of radius $R=1$
gives the area of the particles making up those jets.

Note that it is not possible to write passive areas for jet algorithms
other than $k_t$ in the form \eq~(\ref{eq:particle-areas}). One can
however introduce a new type of area for a generic algorithm, a
{\it Voronoi area}, in the form 
\begin{equation}
  \label{eq:jet-voronoi-area}
  a_{\JA,R}^{\cV}(J) = \sum_{p_i \in J} a_{R}^{\cV}(p_i)\,.
\end{equation}
While for algorithms other than $k_t$ (for which, as we have seen,
$a_{\kt,R}(J) = a_{\kt,R}^{\cV}(J)$), this area is not in general
related to the clustering of any specific kind of background
radiation, it can nevertheless be a useful tool, because its numerical
evaluation is efficient~\cite{Fortune,fastjet} and as we shall discuss
later when we discuss analytic properties of jet areas
(section~\ref{sec:analytics-areas}), for dense
events its value coincides with both passive and active area
definitions.

\subsection{Active Area, catchment area for pileup subtraction}
\label{sec:areamed-defareas-active}

To define an active area, as for the passive area, we start with an
event composed of a set of particles $\{p_i\}$ which are clustered
into a set of jets $\{J_i\}$ by some infrared-safe jet algorithm.
However, instead of adding a single soft ghost particle, we now add a
dense coverage of ghost particles, $\{g_i\}$, randomly distributed in
rapidity and azimuth, and each with an infinitesimal transverse
momentum.\footnote{In most cases the distribution of those transverse
  momenta will be irrelevant, at least in the limit in which the
  density of ghosts is sufficiently high.} %
The clustering is then repeated including the set of particles plus
ghosts. 

During the clustering the ghosts may cluster both with each other and
with the hard particles. This more `active' participation in the
clustering is the origin of the name that we give to the area defined
below. It contrasts with the passive area definition
(section~\ref{sec:areamed-defareas-passive}) in which the single ghost
acted more as a passive spectator.

Because of the infrared safety of any proper jet algorithm, even the addition of
many ghosts does not change the momenta of the final jets $\{J_i\}$.
However these jets do contain extra particles, ghosts, and we use the number
of ghosts in a jet as a measure of its area. Specifically, if the
number of ghosts per unit area (on the rapidity--azimuth cylinder) is
$\nu_g$ and $\cNg(J)$ is the number of ghosts contained in jet $J$,
then the (scalar) active area of a jet, given the specific set of ghosts
$\{g_i\}$ is 
\begin{equation}
  \label{eq:act_area_given_gi}
  \boxed{A (J \,|\, {\{g_i\}})  =   \frac{\cNg(J)}{\nu_g}} \,.
\end{equation}

\begin{figure}
  \centering
  \includegraphics[width=0.49\textwidth]{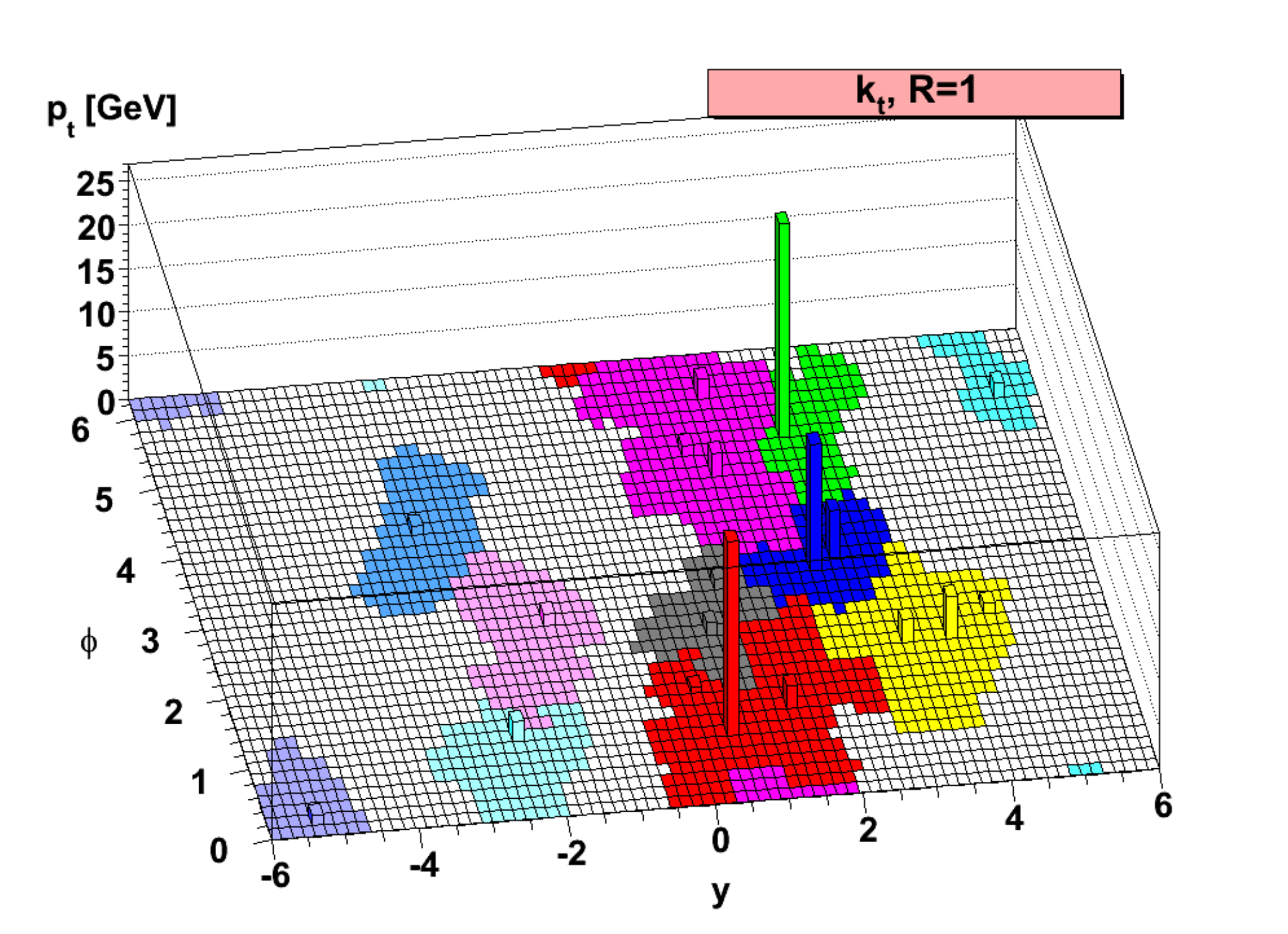}%
  \includegraphics[width=0.49\textwidth]{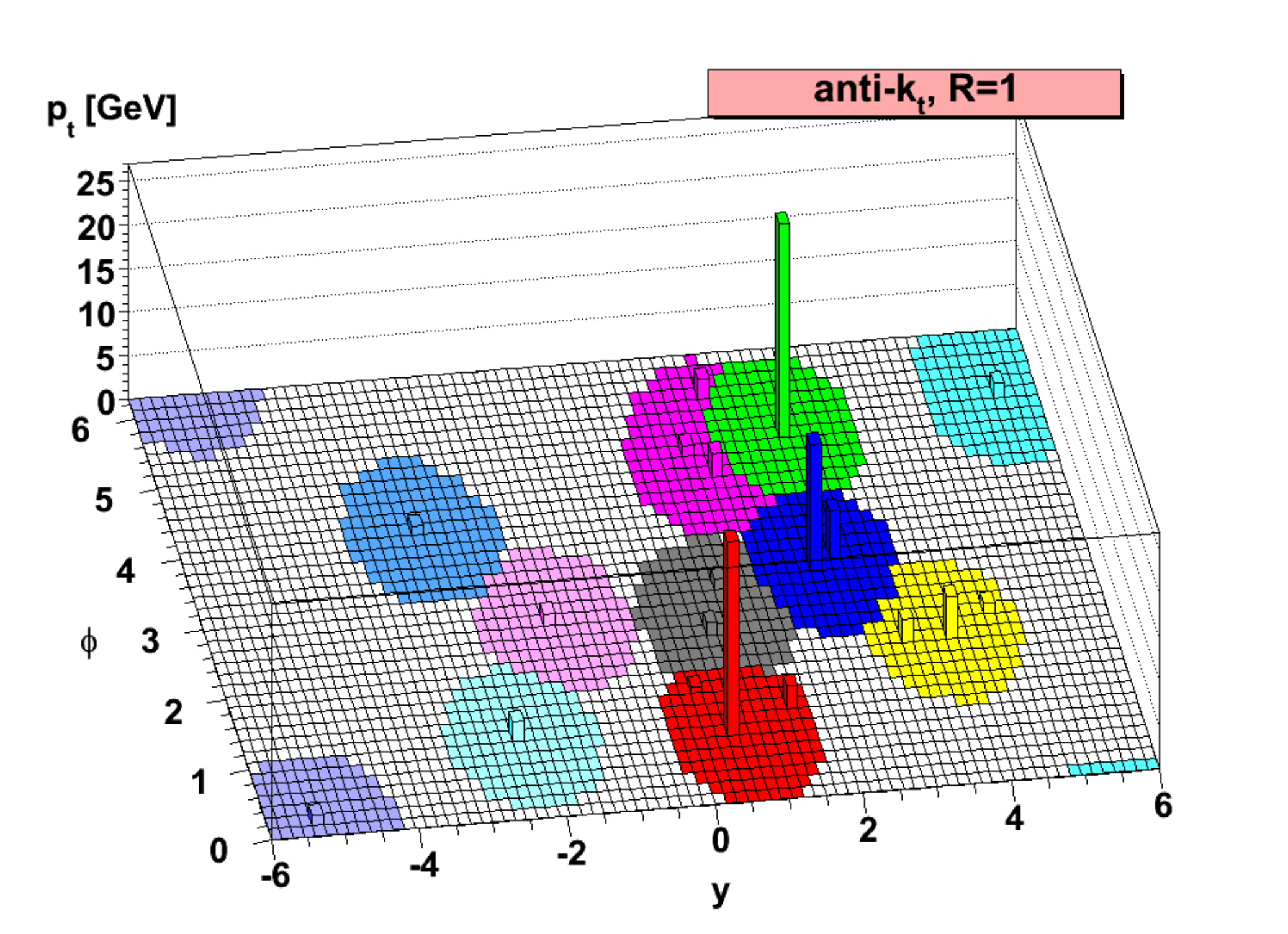}
  \caption{Active area for the same event as in
    figure~\ref{fig:event-voronoi-mod}, once again clustered with the
    $k_t$ algorithm (left) and the anti-$k_t$ algorithm (right) both
    using $R=1$. Only the areas of the hard jets have been shaded ---
    the pure `ghost' jets are not shown.}
  \label{fig:example-active-area}
\end{figure}

An example of jet areas obtained in this way is shown in
figure~\ref{fig:example-active-area}. One notes that the boundaries of
the jets obtained with the $k_t$ algorithm (the left plot) are rather
ragged. Clustering with a different set of ghosts would lead to
different boundaries.
This is because the ghosts can cluster among themselves to form
macroscopic subjets, whose outlines inevitably depend on the specific
set of initial ghosts, and these then subsequently cluster with true
event particles. This can happen for any density of ghosts, and thus
the jet boundaries tend to be sensitive to the randomness of the
initial sets of ghosts.
This is true for most jet algorithms. 
One noticeable exception is the case of the anti-$k_t$ algorithm
(right plot in Fig.~\ref{fig:example-active-area}).
In that case, ghosts will cluster with the particles in the event
before they cluster among themselves.
The boundaries of the jets are thus defined by the structure of the
event, (almost) independently of the initial set of
ghosts.\footnote{Because of the finite ghost density and the
  randomness of the ghosts, the boundaries will slightly fluctuate
  between different sets of ghosts. In a limit of an infinite ghost
  density, the boundary would become independent of the set of ghost.}
In particular, the hard jets will be circular, as seen in
Fig.~\ref{fig:example-active-area}.

This randomness propagates through to the number of ghosts clustered
within a given jet, even in the limit $\nu_g \to \infty$, resulting in
a different area each time. To obtain a unique answer for active area
of a given jet one must therefore average over many sets of ghosts, in
addition to taking the limit of infinite ghost density,\footnote{One
  may wonder if the averaged area (and its dispersion) depends on the
  specific nature of the fluctuations in ghost positions and momenta
  across ensembles of ghosts --- for a range of choices of these
  fluctuations, no significant difference has been observed (except in
  the case of pure ghost jets with SISCone, whose split--merge step
  introduces a strong dependence on the microscopic event structure).
}
\begin{equation}
  \label{eq:act_area}
\boxed{  A(J) = \lim_{\nu_g \to \infty} \left \langle
  A (J \,|\, {\{g_i\}}) \right \rangle_g} \,.
\end{equation}
Note that as one takes $\nu_g \to \infty$, the ghost transverse
momentum density, $\nu_g \langle g_t \rangle$, is to be kept infinitesimal.
For most practical applications, including the case of pileup
mitigation, one has many particles in the event and the boundaries of
the jets are well determined by the particles in the event leaving
only very little dependence on actual set of ghosts.
In practice we can therefore use a unique set of ghost to determine
the active area which is far less time-consuming than having to
average over many sets.
This is what we have done throughout this review.
Note however that for low-multiplicity events the active area can
fluctuate significantly between different sets of ghosts (see for
example Fig.~\ref{fig:ghost-areas} which shows the distribution of
active areas for different sets of ghosts, for events with a single
particle or for jets made purely of ghosts in the absence of any hard
particles).

The active area bears a close resemblance to the average
susceptibility of the jet to a high density of soft radiation like
pileup, since the many soft particles will cluster between each other
and into jets much in the same way as will the ghosts.
In that sense, the ``area of the jet'', $A_{\rm jet}$, referred to in
our introductory description of pileup effects (see
Section~\ref{sec:areamed-idea-characterisation}), has to be seen as
an active area.

A feature that arises when adding many ghosts to an event is that some
of the final jets contain nothing but ghost particles.  They did not
appear in the original list of $\{J_i\}$ and we refer to them as {\em
  pure ghost jets}. These pure ghost jets (not shown in
fig.~\ref{fig:example-active-area}), fill all of the `empty' area, at
least in jet algorithms for which all particles are clustered into
jets.
They will be similar to the jets formed from purely soft radiation in
events with minimum-bias pileup, and so are interesting to study
in their own right and will play a crucial role in the area--median
subtraction method.

In analogy with the 4-vector passive area, we can define a 4-vector
version of the active area. It is given by
\begin{equation}
  \label{eq:act4_area_given_gi}
  A_\mu (J \,|\, {\{g_i\}})  =   
  \frac{1}{\nu_g \langle g_t \rangle } \sum_{g_i \in J} g_{\mu i} \,,
  \qquad\quad
  A_\mu(J) = \lim_{\nu_g \to \infty} \left \langle
  A_\mu (J \,|\, {\{g_i\}}) \right \rangle_g \,.
\end{equation}
The sum of the $g_{\mu i}$ is to be understood as carried out in the same
recombination scheme as used in the jet clustering.%
\footnote{Though we do not give the details it is simple to extend the
  4-vector active area definition to hold also for a general IR safe
  jet algorithm, in analogy with the extension of the passive area
  definition in \eq~(\ref{eq:passive-4vect-area-gen}).}

One may also define the standard deviation $\Sigma(J)$ of the
distribution of a jet's active area across many ghost ensembles,
\begin{equation}
  \label{eq:act_area_stddev}
  \Sigma^2(J) = \lim_{\nu_g \to \infty} \left \langle
  A^2 (J \,|\, {\{g_i\}}) \right \rangle_g  - A^2(J)\,.
\end{equation}
This provides a measure of the variability of a given jet's
contamination from (say) pileup and is closely connected with the
momentum resolution that can be obtained with a given jet algorithm.

\section{Area--median pileup subtraction}\label{sec:areamed-areamed}

\subsection{Basic recipe for pileup subtraction}\label{sec:areamed-basic-recipe}

Now that we have a well-defined concept of the area of a jet in our
toolkit, we can finally introduce the area--median approach.

Coming back to our original discussion of the pileup effects, we have
noticed that pileup introduces an average bias appearing as a shift of
the jet transverse momentum, as well as a smearing of its resolution.
We aim at a method that would remove the $p_t$ shift on average,
\ie be unbiased on average, and limit the residual resolution
degradation effects to a minimum.
Looking at equations (\ref{eq:pueffects-central}),
(\ref{eq:pueffects-shift_prop_npu}) and
(\ref{eq:pueffects-shift_prop_mu}), it is preferable to apply a
correction that avoids making too many averages which would each time
result in additional contributions to the resolution degradation.

The main idea of the area--median pileup subtraction technique is to
subtract from the jet transverse momentum $p_{t,\rm full}$ its average
shift $\rho A_{\rm jet}$ in (\ref{eq:pueffects-central}). This
requires two ingredients: the calculation of the jet area, which we
already have under the form of the {\em active area} introduced in the
previous Section, and a method to obtain an estimate of $\rho$,
$\rho_{\rm est}$.

The estimation of the average pileup activity in an event is based on
the observation that, if we break an event into patches, most of the
patches will not contain particles from the hard jets and for those
patches $p_{t,\rm patch}/A_{\rm patch}$ is, on average, a good
estimate of $\rho$. Averaging the $p_{t,\rm patch}/A_{\rm patch}$ over
all the patches would however not give a good estimate of $\rho$ since
it will be strongly biased by the few patches containing particles
from the hard jets.
In practice, it was found \cite{Cacciari:2007fd} that the median of
$p_{t,\rm patch}/A_{\rm patch}$ provides a good estimate of $\rho$,
not significantly affected by the ``outlying'' patches containing
particles from the hard jets.\footnote{Several alternative approaches
  are can be envisaged, like excluding (recursively or not) patches
  more than $n\sigma$ away from the global average, similarly to what
  ATLAS has used for background subtraction in heavy-ion collisions
  \cite{Aad:2010bu}. We found that the use of a median was equally
  efficient and simpler than alternative methods which require
  additional parameter(s).}

In the end, the area--median subtraction method works as follows:\\

\noindent\centerline{\fbox{\begin{minipage}{0.9\textwidth}
  \begin{itemize}
  \item Find an estimate of $\rho$:
    \begin{enumerate}
    \item Break the event into patches of similar area. Typically, one
      can break the event into grid cells in the $y-\phi$ plane
      ({\it``grid--median'' estimator}). Alternatively, one can use
      $k_t$~\cite{Catani:1993hr,Ellis:1993tq} or
      Cambridge/Aachen~\cite{Dokshitzer:1997in,Wobisch:1998wt}
      jets\footnotemark\ obtained with a jet radius potentially
      different from the one used to cluster the physical jets
      ({\it``jet--median'' estimator}).
    \item Compute the momentum and area of each patch. When patches
      are grid cells, one could use the scalar $p_t$ and scalar,
      geometric, area, while for jet patches, one would rather use
      their 4-vector $p_t$ together with the transverse component of
      the 4-vector active area.
    \item The estimate of $\rho$ is given by
      \begin{equation}\label{eq:rhoest-base}
        \rho_{\rm est}
         =\underset{i \in \rm patches}{{\rm median}}
          \left\{\frac{p_{t,i}}{A_i}\right\}
      \end{equation}
      where the empty cells or the pure-ghost jets are included in the
      calculation.
    \end{enumerate}
  \item Subtract pileup from physical jets:
    \begin{enumerate}
    \item cluster the jets with your favourite jet definition, computing
      their active area 
    \item correct their 4-momentum using
      \begin{equation}\label{eq:subtraction-base}
        p_\mu^{\rm (sub)} = p_\mu - \rho_{\rm est} A_{{\rm jet},\mu}.
      \end{equation}
    \end{enumerate}
  \end{itemize}
\end{minipage}
}} \footnotetext{These algorithms have the advantage that the
resulting jets tend to have reasonably uniform areas. Conversely,
anti-$k_t$ and SISCone suffer from jets with near zero areas or, for SISCone, sometimes huge, ``monster'' jets, biasing the $\rho$ determination. They are therefore not recommended.}

\vspace*{0.3cm}

Modulo possible offsets in the estimation of $\rho$ (see
Section~\ref{sec:analytic-pileup} for details), the area--median
method would subtract from each jet its $\rho A_{\rm jet}$
contamination, leaving a jet with, on average, an unbiased transverse
momentum\footnote{To make things clear, by ``on average'' we mean that
  if we take a jet with a given $p_t$ (or a family of jets with the
  same $p_t$), and repeatedly add pileup to it (or them), the average
  transverse momentum of the jets obtained after applying the
  area--median subtraction procedure would be the original $p_t$.},
with a $p_t$ resolution degradation given by
$\sigma \sqrt{A_{\rm jet}}$. There could also be an additional bias
and resolution degradation coming from back-reaction although, apart
from the resolution degradation at large $p_t$, this can safely be
ignored when subtracting pileup from anti-$k_t$ jets which covers the
vast majority of the applications.\footnote{At large $p_t$ the
  resolution degradation due to back-reaction would grow like
  $\sqrt{\alpha_s\rho p_tA_{\rm jet}}$ (see
  Section~\ref{sec:areamed-analytic-back-reaction}) mostly coming from
  a long negative tail. This would likely have very little effect in
  practical applications, especially when detector effects are also
  taken into account.}
Overall, we therefore expect the area--median approach to provide jets
with better resolution than alternative unbiased approaches like
subtracting an average pileup contamination per pileup vertex or an
overall average pileup contamination, since these would result in
additional sources of resolution degradation.
The key element behind this is the fact that the pileup activity,
$\rho$, is estimated fully dynamically on an event-by-event basis with
the jet area computed individually for each jet. The left-over
resolution degradation due to pileup fluctuations are of the order of
the much-smaller $\sigma$.

When implementing this procedure in a practical context, several
technical details have to be fixed like the rapidity coverage for
particles and ghosts or the size of the patches. We will come back to
these in Section~\ref{sec:areamed-practical}.
Here we rather discuss the physics behind the area--median approach
and carry on with extensions in the next Sections.

The area--median method would work basically with any jet algorithm
giving well-defined active jet areas. Apart from some fancy grooming
techniques, this mostly means that it works with any infrared-safe jet
algorithm.

A potentially non-trivial point is why empty patches (or pure-ghost
jets) have to be included in the calculation of the median. In an
event with a large pileup activity, where the whole event is densely
populated, this would make no difference. However, when the event
becomes sparse, \eg when applying our procedure to parton-level
Monte-Carlo simulations or with events with no or little pileup
activity, most of the non-empty patches would be made of hard
jets. Including empty patches would then avoid having $\rho_{\rm est}$
set to the scale of the hard jets.\footnote{Experimentally, this can
  come with additional complications. For example, since the pileup
  energy deposit is smaller in forward regions, it can be cut by the
  calorimeter noise threshold. This can result in $\rho$ being
  underestimated or even estimated as 0 more often than it should
  \cite{CMS:2010pua, pileup-atlas}. One way around that, used by
  ATLAS \cite{pileup-atlas}, is to estimate the pileup activity
  in the central region and to extrapolate it to forward regions using
  a profile depending on $\nPU$ and $\mu$ (the two are needed because
  of out-of-time pileup).} 

Note that, conceptually, one can view the area--median pileup
subtraction method as part of the jet definition. In that viewpoint,
jets are defined by applying the anti-$k_t$ algorithm with a given
radius (or any other jet definition one wishes to use) and subtracting
the result using the area--median method. It is the combination of
these two steps which can be seen as a jet definition.

Furthermore, the Underlying Event produced in proton-proton collisions
also corresponds to a largely uniform energy deposit. It will
therefore also be subtracted by the area--median method. We will
discuss this in more details in Section~\ref{sec:more}.

\subsection{Positional dependence}\label{sec:areamed-position}

So far, we have assumed that, apart from small up and down
fluctuations of order $\sigma$, the pileup energy deposit is uniform
in rapidity and azimuthal angle. We know however that this is not true
in practice.
The most striking example is that the energy deposit in a minimum bias
collision is, on average, larger in central rapidities than in the
forward regions, translating into an average pileup transverse
momentum $\rho$ larger in the central region than at forward rapidities.
Another example is the case of heavy-ion collisions where
flow~\cite{Ollitrault:1992bk,Poskanzer:1998yz,Adler:2003kt,Adams:2003am,Alt:2003ab,Back:2004mh,Aamodt:2010cz,Chatrchyan:2012wg,ATLAS:2011ah}
would also generate a non-trivial $\phi$ dependence, mostly dominated
by elliptic flow and its $\cos(2\phi)$ dependence.
In both cases, the background density $\rho$ will vary smoothly across an
event.

From the perspective of pileup subtraction, this means that we have to
deal with a situation where $\rho$ depends on the position of the jet
$j$ that one wants to subtract.
The area--median approach proposes two methods to handle the
positional dependence of $\rho$, both based on adaptations of
\eq~(\ref{eq:rhoest-base}).

\begin{figure}
  \includegraphics[width=\textwidth]{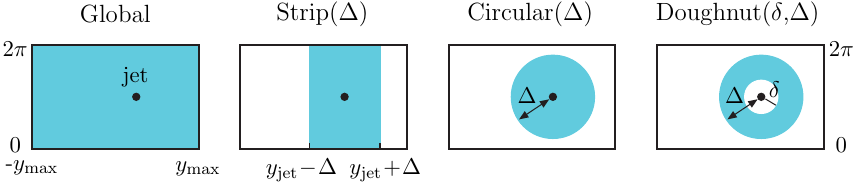}
  \caption{Typical local ranges that can be used for the estimation of
    the positional-dependent pileup density $\rho(j)$, compared to a
    global range (left).}
  \label{fig:local-ranges}
\end{figure}

\paragraph{Local range.} The first method is to estimate $\rho$ only
from the patches which are in the local vicinity of the jet one wishes
to subtract:
\begin{equation}\label{eq:rhoest-local}
\boxed{
  \rho_{{\rm est},{\cal R}}(j)
   =\underset{i \in {\rm patches}\in {\cal R}(j)}{\rm median}
    \left\{\frac{p_{t,i}}{A_i}\right\}}\,,
\end{equation}
where ${\cal R}(j)$ denotes a {\it local range} around jet
$j$. Typical choices for ${\cal R}(j)$ are the following:
\begin{itemize}
\item the {\it Strip} range, ${\cal S}_{\Delta}(j)$, includes the jets
  $j'$ satisfying $|y_{j'}-y_j|<\Delta$,
\item the {\it Circular} range, ${\cal C}_{\Delta}(j)$, includes the
  jets $j'$ satisfying
  $\sqrt{(y_{j'}\!-\!y_j)^2+(\phi_{j'}\!-\!\phi_j)^2} < \Delta$,
\item the {\it Doughnut} range, ${\cal D}_{\delta,\Delta}(j)$, includes
  the jets $j'$ satisfying $\delta <
  \sqrt{(y_{j'}\!-\!y_j)^2+(\phi_{j'}\!-\!\phi_j)^2} < \Delta$.
\end{itemize}
These three options are depicted in Figure~\ref{fig:local-ranges}.

Note that, when using a local range, a compromise has to be found
between taking it large enough so that it contains enough patches for
the estimation of the median to be reliable, and not too large to
capture the positional dependence of $\rho$. Based on analytic
estimates discussed in Section~\ref{sec:analytic-pileup}, we can show
that for jets of radius $R$, if we want the uncertainty on the
determination of $\rho$ to contribute at most for a fraction
$\epsilon$ of the natural pileup fluctuations
$\sigma\sqrt{A_{\rm jet}}$, the range needs to cover at least an area
$A_{{\cal {R}}}$ such that
\begin{equation}
A_{{\cal {R}}} \gtrsim \frac{\pi^2R^2}{4\epsilon}.
\end{equation}
For $\epsilon=0.1$ and $R=0.4$, this gives $A_{{\cal {R}}} \gtrsim
4$. We will discuss this in more details in our Monte Carlo studies of
heavy-ion collisions in Chapter~\ref{chap:mcstudy-hi}.

\paragraph{Rescaling.} The alternative approach is to correct for the
positional dependence before taking the median. In practice, if $f(j)$
captures the positional dependence of the pileup energy deposit as a
function of the position $j$ (arbitrarily normalised), we can use this
as a rescaling function and compute the background density at the
position $j$ using
\begin{equation}\label{eq:rhoest-rescaled}
  \boxed{
    \rho_{{\rm est},f}(j)
    =f(j)\,\underset{i \in {\rm patches}}{\rm median}
    \left\{\frac{p_{t,i}}{A_i\,f(i)}\right\}}\,,
\end{equation}
where $f(i)$ is the positional dependence computed at the position of
``patch $i$''. 

Compared to the local-range approach, this requires the rescaling
function $f$ to be computed before using
\eq~(\ref{eq:rhoest-rescaled}).
In the situation where this is used to correct for the smooth rapidity
dependence of pileup, this function can be taken from an average over
minimum bias events.
For elliptic flow in heavy-ion collisions, $f(j)$ can be parametrised
as a function of the event plane and $v_2$ determined event-by-event. 

We shall discuss in more details these local estimates later, when we
validate the area--median subtraction using Monte-Carlo simulations
(Section~\ref{sec:areamedian-mcstudy:jetpt}). We shall also discuss
the local range approach when we discuss heavy-ion collisions
(Chapter~\ref{chap:mcstudy-hi}).

\subsection{Particle masses}\label{sec:areamed-particle-masses}

In order to obtain a full pileup subtraction of the jet 4-momentum,
there is a last ingredient that needs to be discussed: the effect of
particle masses.

To see this, let us go back to the basic area--median method described
in Section~\ref{sec:areamed-basic-recipe} and to equation
(\ref{eq:subtraction-base}) in particular. The 4-vector jet area is
nothing but a 4-vector sum over the ghosts inside the jets (see \eg
\eq~(\ref{eq:act4_area_given_gi})).
The main formula (\ref{eq:subtraction-base}) can then be seen as
subtracting from the jet 4-vector, $j^\mu$, an amount $a\rho\,g^\mu$ for
each ghost in the jet. In this expression, $a=1/\nu$ is the
fundamental area carried by each ghost and $\rho$ is the appropriate
conversion factor to correct for the jet transverse momentum. 
In other words, (\ref{eq:subtraction-base}) makes a parallel between
the pileup particles contaminating the jet and the ghosts used to
calculate the jet area, using the latter to subtract the former
through an estimate of the proportionality constant $\rho$.

This analogy between the pileup particles and the ghosts is however no
longer exact when pileup particles become massive since the ghosts are
massless. On top of its rapidity and azimuth, a generic 4-vector has
two additional degrees of freedom: its transverse momentum $p_t$ and
its {\it transverse mass} $m_t=\sqrt{p_t^2+m^2}$, with $m$ its mass.
Since $\rho$ is determined from the transverse-momentum component of
the patches, subtracted a 4-vector proportional to a jet area made of
massless ghosts will result in a small mismatch for the components of
the jet sensitive to the mass of the particles.

To put this discussion on a more formal ground, let us take an
infinitesimal patch of the event of size $\delta y\times \delta
\phi$.
The 4-vector corresponding to a uniform pileup deposit in that patch
will take the generic form
\begin{equation}
(p_x,p_y,p_z,E) 
  \equiv (\rho \cos(\phi),
          \rho \sin(\phi),
          (\rho+\rho_m) \sinh(y),
          (\rho+\rho_m) \cosh(y))\, \delta y \, \delta\phi\,
\end{equation}
where we have used the fact that $\rho$ is defined as the pileup
transverse momentum per unit area --- estimated using
\eq~(\ref{eq:rhoest-base}) --- and the longitudinal components, $p_z$
and $E$, can receive an extra contribution coming from the mass of the
pileup particle.
The new term $\rho_m$ is an extra pileup contamination that comes from
the fact that the transverse mass is larger than the transverse
momentum for massive particles.
A uniform pileup deposit is therefore characterised by 2 numbers:
$\rho$ and $\rho_m$.

The correction for the $\rho$ contamination can be carried over using
the method already presented in Section~\ref{sec:areamed-basic-recipe}
and correcting from the new $\rho_m$ term can be straightforwardly
done following the same logic. We first estimate $\rho_m$ using a
median approach:
\begin{equation}\label{eq:rhomest-base}
\boxed{
  \rho_{m,\rm est}
   =\underset{i \in \rm patches}{{\rm median}}
    \left\{\frac{m_{\delta,i}}{A_i}\right\},
}
\end{equation}
with $m_{\delta,{\rm patch}}=\sum_{i\in{\rm
    patch}}\sqrt{p_{t,i}^2+m_i^2}-p_{t,i}$, where the sum runs over
all the particles in the patch. We then correct the jet 4-vector using
\begin{equation}\label{eq:subtraction-with-rhom}
\boxed{
p_\mu^{\rm (sub)} = p_\mu 
  - (\rho_{\rm est} A_{{\rm jet},x},\:
  \rho_{\rm est} A_{{\rm jet},y},\:
  (\rho_{\rm est}+\rho_{m,\rm est}) A_{{\rm jet},z},\:
  (\rho_{\rm est}+\rho_{m,\rm est}) A_{{\rm jet},E}).
}
\end{equation}

These expressions can easily be extended to include positional
dependence either using local ranges or using a rescaling function. In
the second case, note that for all the applications we have seen, a
common rescaling function for $\rho$ and $\rho_m$ is good enough (see
also Section~\ref{sec:area-median:pileup-properties}).

\subsection{Safe subtraction and positivity
  constraints}\label{sec:areamedian-safesubtraction}

There are a few cases where the area--median subtraction formula could
lead to unphysical jets.

The first one is when $\rho A$ is larger than the (unsubtracted) $p_t$
of the jet. This could typically happen for low-$p_t$ jets sitting on
a downwards pileup fluctuation and would lead to jets with
``negative'' $p_t$ after subtraction.\footnote{Strictly speaking, a
  4-vector subtraction would result in a jet pointing in the opposite
  direction in $\phi$ which is not any better than a negative $p_t$.}
In this case, the most reasonable prescription is usually to set the
subtracted jet to a vanishing 4-momentum.

The second case would be a jet with a positive transverse momentum but
a negative squared mass. In this case, a good prescription is to keep
the transverse momentum and azimuth of the jet obtained after
subtraction of the $\rho$ contamination, take the rapidity of the
original jet and set its mass to 0.

We will explicitly refer to these prescriptions as ``safe''
subtraction and discuss them later in our Monte Carlo tests (see
Chapter~\ref{chap:charged_tracks}).

\subsection{Extension to Charged-Hadron Subtracted (CHS)
  events}\label{sec:areamedian-CHSsubtraction}

In several experimental contexts, it is possible to associate charged
tracks either to the leading vertex or to a pileup interaction.
In that context, one builds a {\em Charged-Hadron-Subtracted (CHS)
  event} where one subtracts the tracks associated to pileup
interactions keeping only the tracks associated with the leading
vertex and the neutral particles (or energy deposits) coming either
from the leading vertex or from pileup.
Optionally, the charged tracks associated with a pileup vertex can be
retained as ghosts, \ie with their momentum scaled down by a large
factor, in order to retain information about their presence in the
original event.
Modulo experimental complications\footnote{In an experimental context,
  this comes with the difficulty of having to subtract the energy
  deposit left by charged tracks in the calorimeter. Additional
  complications also come from the smaller rapidity coverage of the
  tracker.}, this considerably reduces the pileup contamination (both
the average shift and the dispersion), since charged tracks account
for about a fraction $f\approx 062$ of the pileup energy deposit.

The area--median method can be straightforwardly applied to CHS
events, with the pileup properties estimated either on the full CHS
event or only on their neutral component.\footnote{The only small
  difference being that for ``safe'' subtraction, one would use the
  rapidity and mass of the charged tracks associated to the leading
  vertex in the cases where the initial area--median subtraction leads
  to squared mass smaller than the squared mass of the charged tracks
  coming from the leading vertex.}
One should expect a reduction of the average pileup contamination
by a factor $1/(1-f)\approx 2.6$ and of the resolution smearing
by a factor $1/\sqrt{1-f}\approx 1.6$.
Although results from Chapter~\ref{chap:mcstudy} directly extend to
CHS events, we will discuss this at length in
Chapter~\ref{chap:charged_tracks} where we compare the area--median
approach to alternative methods that make use of charged track
information.

\section{Jet shapes}\label{sec:areamed-shapes}

So far, we have introduced the area--median method for subtracting
pileup contamination from the jet 4-momentum.
It is now natural to ask if the method can be extended to more general
properties of the jets.
In this Section, we show that it is possible to obtain a generic
area--median subtraction method for jet shapes, \ie
infrared-and-collinear safe functions of the jet
constituents.

Let us therefore consider a generic infrared-and-collinear-safe
function $v_{\rm jet}(\{p_i\})$ of the jet constituents $\{p_i\}$, including a
uniform pileup contamination characterised by the scales $\rho$ and
$\rho_m$. 
We also include in these constituents a set of ghosts each carrying an
area $A_g=1/\nu$ in the $y-\phi$ plane and to which we associate an
infinitesimal transverse momentum $p_{t,g}$ and a (new) infinitesimal
transverse mass $m_{t,g}=p_{t,g}+m_{\delta,g}$ with
$m_{\delta,g}=\sqrt{p_{t.g}^2+m_g^2}-p_{t,g}$.
As discussed in Section~\ref{sec:areamed-particle-masses}, these
ghosts mimic the effect of a uniform pileup deposit and the ghost
scales $p_{t,g}$ and $m_{\delta,g}$ correspond to the pileup
densities $\rho$ and $\rho_m$.

For any given jet we can then consider the derivatives of the jet
shape \wrt to the ghost transverse scales $p_{t,g}$ and $m_{\delta,g}$,
\begin{equation}\label{eq:shape-ghost-derivatives}
v_{\rm jet}^{(m,n)} = A_g^{m+n}
 \partial_{p_{t,g}}^m \partial_{m_{\delta,g}}^n v_{\rm jet}(\{p_i\}).
\end{equation}
In a sense, these derivatives probe the susceptibilities of the shape to
the $\rho$ and $\rho_m$ components of pileup and can be used to
correct the shape $v$ by extrapolating its value down to zero pileup:
\begin{align}
v^{\rm (sub)}
& = \sum_{m=0}^{\infty} \frac{1}{m!}
    \sum_{n=0}^{\infty} \frac{1}{n!}
    (-)^{m+n} \rho^m\rho_m^n v_{\rm jet}^{(m,n)}\\
& \approx v_{\rm jet}
    - \rho v_{\rm jet}^{(1,0)}
    - \rho_m v_{\rm jet}^{(0,1)}
    + \frac{1}{2} \rho^2 v_{\rm jet}^{(2,0)}
    + \rho\rho_m v_{\rm jet}^{(1,1)}
    + \frac{1}{2} \rho_m^2 v_{\rm jet}^{(0,2)}
    + \dots,\label{eq:subtraction-for-shapes}
\end{align}
where the second line is expanded to second order in the pileup properties.
This second-order subtraction is the main formula we shall use to
subtract jet shapes.\footnote{Theoretically, there is no limitation to
  go to higher orders but, in practice, the derivatives will be
  estimated numerically and third-order derivatives become delicate to
  compute for common applications.}

It is illustrative to consider two specific examples of this formula:
the jet transverse momentum and the jet mass squared. 
For the jet transverse momentum, the uniform pileup distribution will
shift the original hard transverse momentum $p_{t,\rm hard}$ (without
pileup), by an amount $A_\perp\rho$, with $A_\perp$ the transverse
momentum component of the jet area 4-vector. On top of that, the
ghosts will add a contribution $p_{t,g}A_\perp/A_g$ to the total
$p_t$, giving successively
\begin{align}
p_t & = p_{t,\rm hard} + (\rho+p_{t,g}/A_g)A_\perp,\\
p_t^{(1,0)} & = A_\perp,
\end{align}
with all other derivatives vanishing, and finally,
\begin{equation}
p_t^{\rm (sub)} 
 = p_t - \rho p_t^{(1,0)}
 = p_t - \rho A_\perp
 = p_{t,\rm hard} + p_{t,g}A_\perp/A_g
 = p_{t,\rm hard}
\end{equation}
where, for the last equality, we have dropped the negligible
contribution coming from the infinitesimal $p_t$ of the ghosts.
This simple exercise shows that (\ref{eq:subtraction-for-shapes})
reduces to the expression $p_t^{\rm (sub)} = p_t - \rho A_\perp$ at the
core of the area--median subtraction.
It also provides an interesting interpretation of the ``jet area'' as
the susceptibility of the jet transverse momentum to variations of the
background density $\rho$.

Albeit a bit more involved technically, the case of the squared jet
mass flows similarly. Let us consider a hard jet of transverse
momentum $p_{t,\rm hard}$ and mass $m_{\rm hard}$ with additional
uniform pileup contamination and ghosts.
We already know that the hard jet transverse momentum is shifted by
$(\rho+p_{t,g}/A_g)A_\perp$ and, in a similar way, the jet transverse
mass will be shifted by
$(\rho+p_{t,g}/A_g)A_\perp+(\rho_m+m_{\delta,g}/A_g)A_{\rm jet}$,
where we have separated the contribution from the transverse momentum
and the one from the particle masses. This gives
\begin{align}\label{eq:expression-for-m2-with-pileup}
m^2 & = \left[\sqrt{p_{t,\rm hard}^2+m_{\rm hard}^2}
             +(\rho+p_{t,g}/A_g)A_\perp
             +(\rho_m+m_{\delta,g}/A_g)A_{\rm jet}\right]^2 
     - \left[p_{t,\rm hard} + (\rho+p_{t,g}/A_g)A_\perp\right]^2,\nonumber\\
m^{2(1,0)} & = 2A_\perp\left[\sqrt{p_{t,\rm hard}^2+m_{\rm hard}^2}-p_{t,\rm hard}
                          +(\rho_m+m_{\delta,g}/A_g)A_{\rm jet}\right],\nonumber\\
m^{2(0,1)} & = 2A_{\rm jet}\left[\sqrt{p_{t,\rm hard}^2+m_{\rm hard}^2}
                          +(\rho+p_{t,g}/A_g)A_\perp
                          +(\rho_m+m_{\delta,g}/A_g)A_{\rm jet}\right],\nonumber\\
m^{2(2,0)} & = 0,\nonumber\\
m^{2(1,1)} & = 2A_\perp A_{\rm jet},\nonumber\\
m^{2(0,2)} & = 2A_{\rm jet}^2,
\end{align}
with higher derivatives vanishing. Substituting this in
(\ref{eq:subtraction-for-shapes}), and neglecting the infinitesimal
ghosts, we find after easy manipulations
\begin{equation}
m^{2\rm (sub)} = m_{\rm hard}^2.
\end{equation}
Note that, in order to recover exactly the hard jet without pileup
contamination, it is important to apply the subtraction procedure to
the jet mass squared. This is mostly because, as can be seen from
(\ref{eq:expression-for-m2-with-pileup}), the squared mass $m^2$ is
explicitly quadratic in $p_{t,g}$ and $m_{\delta,g}$. If we were to
apply the subtraction formula to the jet mass, the extra square root
would translate in non-vanishing higher derivatives and subtracting
pileup contamination up to second-order would leave some
(parametrically small) residual corrections (proportional to
$\rho^3/m^2$ at large $p_t$).

These examples show not too surprisingly that our second-order
expansion would work better if the quadratic approximation works
fine. As we shall see in practice in Monte Carlo simulations, there
are some notorious cases where this approximation tends to break. One
typical example is when a shape has an endpoint and the effect of
pileup is to bring the value of the shape close to that endpoint.

In some situations, it is however possible to work around these
limitations by splitting the shape into components and subtracting
these components individually before reconstructing the final
subtracted shape from the subtracted components. The case of the
$N$-subjettiness ratio $\tau_{21}=\tau_2/\tau_1$ illustrates this very
well: while $\tau_{21}$ itself has a complicated dependence on the
pileup density, (unnormalised) $\tau_1$ and $\tau_2$ are linear in
$\rho$. We should therefore apply the subtraction procedure to $\tau_1$
and $\tau_2$ and reconstruct the subtracted value of $\tau_{21}$ from
those. The only delicate point here is to properly handle the
situation where the subtracted value of $\tau_1$ becomes 0 or
negative, \eg by setting the subtracted value of $\tau_{21}$ to 0.

Generally speaking, subtracting the shape as a single entity or
breaking the shape in components and subtracting these components
individually tend to be affected by different effects. Depending on
the specific details of the shape and how it is used in a given
analysis, one method or the other would give slightly better results
(e.g.\ for $\tau_{21}$, the subtraction as a single entity works
slightly better at smaller $\tau_{21}$ but the splitting into
components is marginally preferred at larger $\tau_{21}$).
For simplicity, we will apply the subtraction to the shape as a single
entity in our later tests in Section~\ref{sec:mcstudy-shapes}.

Finally, let us note that it is possible to rewrite
(\ref{eq:subtraction-for-shapes}) in terms of a single variable. We
can for example introduce a variable $x_g$ such that $p_{t,g}=x_g$ and
$m_{\delta,g}=\rho_m/\rho x_g$ and define
\begin{equation}\label{eq:simplified-shape-ghost-derivatives}
\boxed{v_{\rm jet}^{[k]} = A_g^{k} \partial_{x_g}^k v_{\rm jet}(\{p_i\}).}
\end{equation}
The subtraction formula then takes the simpler form
\begin{equation}\label{eq:simplified-subtraction-for-shapes}
\boxed{v^{\rm (sub)}
 = \sum_{k=0}^{\infty} \frac{1}{k!}
    (-)^{k} \rho^k v_{\rm jet}^{[k]}
 \approx v_{\rm jet}
    - \rho v_{\rm jet}^{[1]}
    + \frac{1}{2} \rho^2 v_{\rm jet}^{[2]}
    + \dots.}
\end{equation}

\section{Jet fragmentation function}\label{sec:areamed-fragmentation-function}

At this stage, we have seen that the area--median pileup subtraction
method can be applied to all infrared and collinear safe jet
observables. In this section, we shall add to that list yet another
observable, the jet fragmentation function, which is collinear unsafe.

Part of the idea behind the area--median subtraction of the jet
fragmentation function is to subtract its moments rather than the
direct distribution. We provide a discussion of the different
representations of a jet fragmentation function in
Appendix~\ref{app:fragmentation-function-representations}, including
some additional physics motivations to consider the moments of the jet
fragmentation function.
Here we shall just recall the basic definitions.

\paragraph{Definition.} The jet fragmentation function (FF) is defined
as the distribution $dN_h/dz$ of the momentum fraction\footnote{Note
  that, in the literature, the ``fragmentation function'' often refers
  to the probability distribution for a high-energy parton to produce
  a given hadron. The scale evolution of these objects can be
  predicted from QCD and the production of a given hard hadron, say a
  $\pi$, can be written as the convolution between the production of a
  hard parton and such a fragmentation function.
  In this review, we instead discuss the ``(jet) fragmentation
  function'' which describes the momentum-fraction distribution of
  hadrons inside a jet.}
\begin{equation}\label{eq:z}
z=\frac{p_{t,h}}{\tilde{p}_t}
\end{equation}
of all the hadrons $h$ in a jet. In the definition of $z$, $p_{t,h}$
is the transverse momentum of the hadron and $\tilde{p}_t$ is the
scalar $p_t$ of the jet, defined as the scalar sum of the transverse
momenta of the constituents of the jet, $\tilde{p}_t=\sum_{i\in{\rm
    jet}} p_{t,i}$. In future discussions it will also be interesting
to define $\xi=\log(1/z)$.
In practice, the distribution $dN_h/dz$ is usually binned in $z$ or
$\xi$ and normalised to the number $N_{\rm jet}$ of jets in the sample.

Instead of applying the pileup subtraction directly to $dN_h/dz$, we
will rather do it to its moments defined as
\begin{equation}\label{eq:def-ff-moments}
M_N = \frac{1}{N_{\rm jet}}\int_0^1 dz\,z^N\,\frac{dN_h}{dz}
    = \frac{1}{N_{\rm jet}}\int_0^\infty d\xi \,e^{-N\xi}\,\frac{dN_h}{d\xi}.
\end{equation}
In practice, one usually computes the $N$th moment for a single jet:
\begin{equation}\label{eq:def-ff-moments-single-jet}
M_N^{\rm jet} = \frac{\sum_{i\in \rm jet}p_{t,i}^N}{\tilde{p}_t^N},
\end{equation}
and average over the jet sample, $M_N=\langle M_N^{\rm jet} \rangle$.

\paragraph{Area--median subtraction.} We can use an adaptation of the
area--median approach to directly subtract the moments $M_N$ of the
fragmentation function.

The procedure to subtract the FF moments is quite similar to the main
area--median approach.
One first determines the expected background contribution per unit
area to a given moment (or rather to its numerator in
\eq~(\ref{eq:def-ff-moments-single-jet})):
\begin{equation}\label{eq:rhoN-from-median}
\boxed{  \rho_N =
  \mathop{\text{median}}_\text{patches}\left\{\frac{\sum_{i\in \text{patch}}
  p_{t,i}^N}{A_\text{patch}}\right\}\,.}
\end{equation}
For example, for $N=0$ this will be the median particle multiplicity
per unit area.
The subtracted FF moment is then obtained by separately taking the
numerator and denominator of \eq~(\ref{eq:def-ff-moments-single-jet}),
as measured in the full event, and respectively subtracting $\rho_N A$
and $\rho A$:
\begin{equation}\label{eq:momsub}
\boxed{  M_N^\text{sub}
    =\frac{\sum_{i\in \text{jet}} p_{t,i}^N - \rho_N A}
          {(p_{t,\text{full}} - \rho A)^N} \, .}
\end{equation}

\paragraph{Improved background-subtracted fragmentation
  function.}

In Section~\ref{sec:mcstudy-hi-ff}, we will apply this subtraction
method to FF measurements in heavy-ion collisions and observe that it
is insufficient to fully correct the jet FF moments. We believe this
can largely be attributed to background fluctuations and we describe
below an improved subtraction method that corrects for the dominant
fluctuation effects.  

Background fluctuations mean that neither the jet's $p_t$ nor the
numerator of the FF moment are perfectly reconstructed for any given
jet.
When selecting jets above some $p_t$ threshold for a process with a
steeply falling jet spectrum, it is favourable to select jets slightly
below the $p_t$ threshold, but which have an upwards background
fluctuation (cf.\ the discussion of
Section~\ref{sec:description-convolution}).
One consequence of this is that the denominator in \eq~(\ref{eq:momsub})
tends to be larger than the actual jet $p_t$.
The larger the value of $N$, the greater the impact of this effect.
One can also understand it as resulting in an underestimate of the $z$
fraction, leading to the FF being spuriously shifted to lower $z$.
Such an effect is already well known: CMS~\cite{Chatrchyan:2012gw},
for example, when comparing with $pp$ FFs, explicitly applies a
correction to the $pp$ FFs to account for the smearing of the
denominator of the $z$ variable that is expected when the jet's
momentum is reconstructed in a HI environment.
ATLAS~\cite{Aad:2014wha} carries out an unfolding to
account for this.
Implementing an additional correction for the dominant background
fluctuations effects will reduce the size of the final unfolding
corrections and hopefully of the associated uncertainties. We note
that a key argument here is that the dominant fluctuation effects are
local in moment space $N$, therefore avoiding more complex (unfolding)
corrections in $z$ space which is additionally affected by
bin-migration effects.

A second consequence of fluctuations affects mostly the low-$z$, or
low-$N$ region of the FF: the upwards fluctuations of the background
$p_t$ that cause a jet to pass the $p_t$ cut even when it is below
threshold also tend to be associated with upwards fluctuations of the
multiplicity of soft background particles.
It is this effect that causes the FF to be overestimated for low
values of $N$.%
\footnote{ATLAS correct for this~\cite{Aad:2014wha} by multiplying the
  expected background contribution by a $z$-independent factor that is
  a function of the jet $p_t$.}

One way of verifying the above interpretation is to consider
$\gamma+$jet events, selecting events based on the photon $p_t$ and
normalising the fragmentation function $z$ also to the photon $p_t$.
In this case, the background-induced fluctuations of the jet's
reconstructed $p_t$ are of no relevance, and in explicit simulations,
we have found that plain subtraction is already quite effective.

For the dijet case, in the limit where the background fluctuations are
reasonably small compared to the transverse momentum of the jet, it is
possible to devise a simple correction in moment space for both kinds
of fluctuation-induced bias.
In order to do so, we start by rewriting the subtracted moments of the
jet FF in \eq~(\ref{eq:momsub}) as
\begin{equation}
M_N^\text{sub} = \frac{\sum_i p_{t,i}^N - \rho_N A}{(p_{t,\text{full}} - \rho
A)^N} \equiv \frac{S_N}{S_1^N}
\end{equation}
where we have introduced the
shorthands $S_N$ for the subtracted numerator and $S_1$ for
$p^\text{sub}_{t,\text{full}}$, as defined by the area--median
subtraction, \eq~(\ref{eq:subtraction-base}).
Given our assumption that background fluctuations are moderate, we can
locally approximate the hard jet cross-section by an exponential (as
already done in Section~\ref{sec:description-convolution})
\begin{equation}
\label{eq:Hspectrum}
  H(p_t)\equiv \frac{d\sigma}{dp_t} = \frac{\sigma_0}{\kappa}\,{\rm exp}(-p_t/\kappa)\,.
\end{equation}
We then take a Gaussian approximation for the spectrum of background
transverse-momentum fluctuations from one jet to the next.
Denoting the fluctuation by $q_t$, the distribution of $q_t$ is then
\begin{equation}
\label{eq:Bspectrum}
  B(q_t)\equiv \frac{dP}{dq_t} 
  = \frac{1}{\sqrt{2\pi A}\sigma}\,
       \exp\left(-\frac{q_t^2}{2\sigma^2 A}\right),
\end{equation}
where $\sigma$ is a parameter that describes the size of
fluctuations from one patch of area $1$ to another. 
For patches of area $A$, the variance of the $q_t$ across the
patches is $\sigma^2 A$. 

We further introduce the variable $Q_N$ to denote the difference
between the actual background contribution to $S_N$ in a specific jet
and the expected contribution, $\rho_N A$, \ie
\begin{equation}
  \label{eq:QN-def}
  Q_N=\left(\sum_{i \, \in\, \text{jet\,(bkgd)}} k_{t,i}^N \right)-\rho_N A \, , 
\end{equation}
where the sum runs just over the background constituents $k_{t,i}$ of the
jet. By construction, $Q_1 = q_t$. 
In practice $Q_N$ cannot be determined for a single jet, since we
don't know which particles are the background ones, but its
statistical properties can be determined by looking at many jets, and
we denote the standard deviation of its fluctuations from one patch of area $1$ to another by $\sigma_N$,
in direct analogy to $\sigma$.
The fluctuations of $Q_N$ are not independent of the
fluctuations $q_t$ of the background's transverse momentum: there is a correlation
coefficient $r_N$ between them, defined as
\begin{equation}
  \label{eq:r_N-definition}
  r_N = \frac{\mathrm{Cov}(q_t,
    Q_N)}{\sqrt{\mathrm{Var}(q_t)\mathrm{Var}(Q_N)}}\,,
\end{equation}
where $\mathrm{Var}(X)$ is the variance of the variable $X$ and
$\mathrm{Cov}(X,Y)$ the covariance of $X$ and $Y$.
Using the fact that $\mathrm{Var}(q_t) = \sigma^2 A$,
$\mathrm{Var}(Q_N) = \sigma_N^2 A$, we have that the average value for
$Q_N$ as a function of $q_t$ is
\begin{equation}
  \label{eq:expected-QN}
  \langle Q_N \rangle(q_t) 
  = \frac{\mathrm{Cov}(q_t, Q_N)}{\mathrm{Var}(q_t)} q_t
  = r_N \frac{\sigma_N}{\sigma} q_t\,.
\end{equation}

With these ingredients we can now correct for the
fluctuation effects as follows.
If we measure a certain value $S_N$ in a jet, then as a function of the $q_t$
fluctuation, the expected true hard contribution to it is
\begin{equation}
  \label{eq:SN-hard-v-qt}
  S_N^\text{hard} = S_N - \langle Q_N\rangle(q_t)
   = S_N - r_N\frac{\sigma_N}{\sigma} q_t\,,
\end{equation}
where we have averaged over possible $Q_N$ values, given the
$q_t$ fluctuation.
To obtain our estimate for $M_N^\text{hard}$, this should be normalised by the
$N^\text{th}$ power of the true hard $p_t$ of the jet, $(S_1 -
q_t)^{N}$:
\begin{equation}
  \label{eq:MN-hard-v-qt}
  M_N^\text{hard} = \frac{S_N^{\hard}}{(S_1 - q_t)^{N}}
  = \frac{S_N}{S_1^N} + N \frac{S_N q_t}{S_1^{N+1}} 
         - r_N\frac{\sigma_N q_t}{\sigma S_1^N} + \order{q_t^2}\,.
\end{equation}
One subtlety here is that this is an estimate for $M_N^\text{hard}$ in
hard jets with true $p_t = S_1 - q_t$.
However because $M_N$ is a slowly varying function of $p_t$, by
taking the result as contributing to $M_N^\text{hard}(S_1)$ rather than
$M_N^\text{hard}(S_1-q_t)$ we make only a small mistake, of the same order as
other terms that we shall neglect.\footnote{This can be seen by
  observing that $M_N$ satisfies a DGLAP-style equation for its
  evolution $dM_N/d\ln p_t \sim \alpha_s M_N$. Given that $q_t$ is
  itself small compared to $p_t$, $M_N$ for jets with $p_t = S_1 -
  q_t$ differs from that for jets with $p_t = S_1$ by a relative
  amount $\sim \alpha_s q_t/S_1$, which we can neglect in the same way
  that we neglect $\order{q_t^2}$ terms.
  If we wanted to improve on this approximation, then one approach
  might be, for each event, to use the middle expression in
  \eq~(\ref{eq:MN-hard-v-qt}) and assign it to $M_N$ at $p_t = S_1 -
  \langle q_t \rangle$.
  We have not, however, investigated this option in detail and other
  improvements would probably also be necessary at a similar accuracy,
  \eg taking into account deviations from the simple exponential and
  Gaussian approximations that we have used for $H(p_t)$ and $B(q_t)$.
}

Following \eq~(\ref{eq:distrib-effect-most-probably}), the average
fluctuation $q_t$ for a given reconstructed $S_1$ is
\begin{equation}\label{eq:av-qt}
\langle q_t \rangle = \frac{\sigma^2 A}{\kappa},
\end{equation}
therefore, retaining the terms linear in $q_t$ in
\eq~(\ref{eq:MN-hard-v-qt}) and averaging now over possible $q_t$
values, leads to the following prescription for an ``improved''
subtracted $M_N(S_1)$, corrected for fluctuations effects up to first
order in $q_t/S_1$:
\begin{equation}
  \label{eq:subimp-v3}
  \boxed{
  M_N^\text{sub,imp} =  M_N^\text{sub} \times \left(1 +
    N \frac{\sigma^2 A}{S_1 \kappa}\right)
   - r_N\frac{ \sigma \sigma_N A}{\kappa S_1^N}\,.}
\end{equation}
This is simpler than the corresponding correction would be
directly in $z$ space, in particular because in $z$ space the
correction to one bin of the fragmentation function depends in a
non-trivial way on the contents of nearby bins.
One can think of the advantages of moment space as being that the
correction to a given $N$ value does not depend on $M_N$ at all other
values of $N$, and that it is straightforward to account for
correlations between fluctuations in the jet-$p_t$ and in the moments.

Note that in a real experimental context, calorimeter fluctuations of
the reconstructed jet and background $p_t$'s would have an effect akin
to increasing $\sigma$ and decreasing the correlation coefficient
$r_N$.
The in-situ methods that we use for the determination of $\sigma$,
$\sigma_N$ and $r_N$ would automatically take this into account.
Noise-reduction methods in the reconstruction of the jet $p_t$, as
used by CMS~\cite{Chatrchyan:2011sx}, would have the effect of
reducing $\sigma$ (and probably also $r_N$).
However noise reduction is likely to complicate the meaningful
determination of $r_N$, since it acts differently on pure background
jets as compared to jets with a hard fragmenting component.

The correction in \eq~(\ref{eq:subimp-v3}) can be applied jet-by-jet to
correct for the fluctuation effects.
We shall discuss the practical application of these techniques in the
case of heavy-ion collisions in Section~\ref{sec:mcstudy-hi-ff}.

\section{Numerical Implementation}\label{sec:areamed-implementation}

All of the methods described above are made numerically available
either directly in \fastjet (for the standard application to $p_t$ and
4-vector subtraction) or in various contributions to \fjcontrib (for
the more advanced applications).
We describe in this Section some details of their implementation and
leave to the next Section the recommendations on how to use the
area--median pileup subtraction technique in practical applications.

Following the same logic as above, the first ingredient that is needed
is a computation of the jet areas and the second ingredient is an
implementation of the estimation of the background density $\rho$.

In what follows, we assume that the reader is familiar with the basic
usage of \fastjet for jet clustering, \ie is familiar with the usage
of the three basic \fastjet classes: \ttt{PseudoJet},
\ttt{JetDefinition}, \ttt{ClusterSequence}. We refer to the \fastjet
manual \cite{fastjet-manual} for more details.

A complete description of the \fastjet classes related to jet areas
and background subtraction can be found in the \fastjet manual. Here
we just give a quick overview of the relevant classes and their
typical usage.

\subsection{Clustering with jet areas}\label{sec:areas}

All three main types of jet areas (active, passive and Voronoi) are
directly available from \fastjet.
Compared to standard clustering, the user is exposed to two new
classes:
\begin{lstlisting}
  class fastjet::AreaDefinition;
  class fastjet::ClusterSequenceArea;
\end{lstlisting}
with input particles, a jet definition and an area definition being
supplied to a \ttt{ClusterSequenceArea} in order to obtain jets with
area information.
Typical usage would be as follows:
\begin{lstlisting}
  // include the appropriate header
  #include "fastjet/ClusterSequenceArea.hh"

  // define an area with ghosts up to the particle acceptance
  double ghost_maxrap = 5.0; // e.g. if particles go up to y=5
  AreaDefinition area_def(active_area_explicit_ghosts, GhostedAreaSpec(ghost_maxrap));

  // cluster using that definition of jet areas
  ClusterSequenceArea clust_seq(input_particles, jet_def, area_def);

  // retrieve the jets and show how to get their area
  vector<PseudoJet> jets = sorted_by_pt(clust_seq.inclusive_jets());
  double area_hardest_jet = jets[0].area();
\end{lstlisting}

\paragraph{Defining areas.} For active areas, pertinent for pileup
subtraction, one would use
\begin{lstlisting}
  AreaDefinition(fastjet::AreaType area_type, fastjet::GhostedAreaSpec ghost_spec);
\end{lstlisting}
with an \ttt{area\_type} being one of the two active area choices:
\begin{lstlisting}
  enum AreaType{ [...], active_area, active_area_explicit_ghosts, [...]};
\end{lstlisting}
The two variants give identical results for the areas of hard jets.
The second one explicitly includes the ghosts when the user requests
the constituents of a jet and also leads to the presence of ``pure
ghost'' jets.

The second argument used to define an \ttt{AreaDefinition} specifies
how to distribute the ghosts.
This is done via the class \ttt{GhostedAreaSpec} whose constructor is
\begin{lstlisting}
  GhostedAreaSpec(double ghost_maxrap, 
                  int    repeat        = 1,   double ghost_area    = 0.01, 
                  double grid_scatter  = 1.0, double pt_scatter    = 0.1,
                  double mean_ghost_pt = 1e-100);
\end{lstlisting}
The ghosts are distributed on a uniform grid in $y$ and $\phi$, with
small random fluctuations to avoid clustering degeneracies.

\begin{figure}
  \centering
  \includegraphics[width=0.48\textwidth]{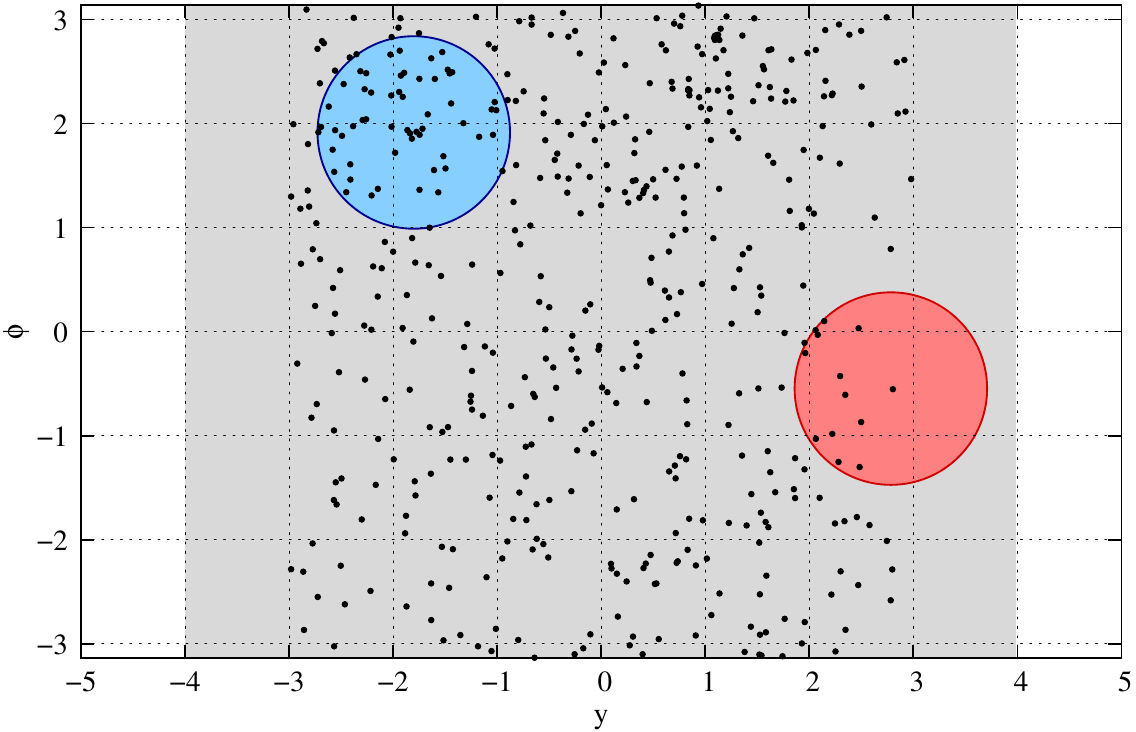}\hfill
  \includegraphics[width=0.48\textwidth]{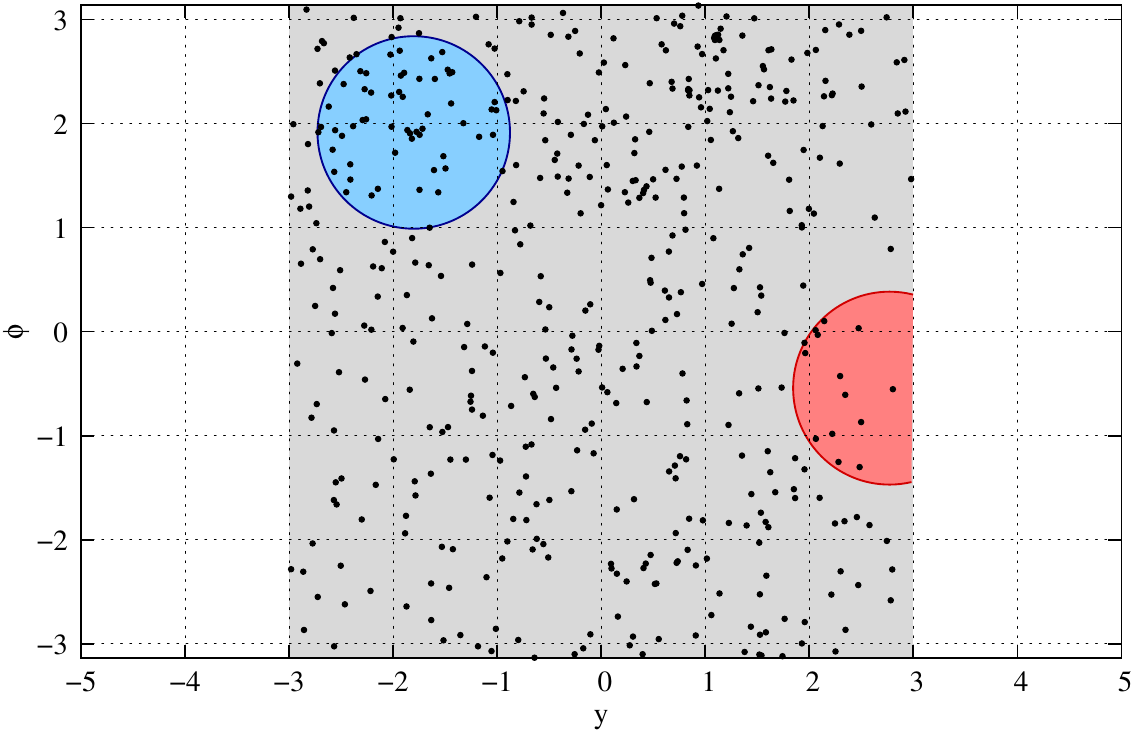}
  \caption{Two choices for ghost placement. The grey area in each plot
    indicates the region containing ghosts, while the dots are
    particles, here accepted up to $|y|<3$. 
    The circular regions indicate the areas that will be found for two
    particular jets.  
    In the left-hand case, with ghosts that extend well beyond the
    acceptance for particles, jet areas are unaffected by the particle
    acceptance; 
    in the right-hand case, with ghosts placed only up to the
    acceptance limit, jet areas are clipped at the edge of the
    acceptance.  }
  \label{fig:ghost-placement}
\end{figure}

The \ttt{ghost\_maxrap} variable defines the maximum rapidity up to
which ghosts are generated.
If one places ghosts well beyond the particle acceptance (at least $R$
beyond it), then jet areas also stretch beyond the acceptance, giving
a measure of the jet's full extent in rapidity and azimuth.
If ghosts are placed only up to the particle acceptance, then the jet
areas are clipped at that acceptance and correspond more closely
to a measure of the jet's susceptibility to contamination from
accepted soft particles.
The two choices are illustrated in fig.~\ref{fig:ghost-placement}.
To define more complicated ghost acceptances it is possible to replace
\ttt{ghost\_maxrap} with a \ttt{Selector}, which must be purely
geometrical and have finite rapidity extent.

The two other relevant arguments to construct a \ttt{GhostedAreaSpec}
object are \ttt{ghost\_area} and \ttt{n\_repeat}.
The \ttt{ghost\_area} parameter is the area associated with a single
ghost. 
Smaller values will mean more accurate determinations of the jet area
at the cost of a longer CPU-time for clustering.
We have found the default ghost area given above to be adequate for
most applications, with smaller ghost areas beneficial mainly for
high-precision determinations of areas of jets with small $R$.

Since the ghost positions and $p_t$'s include small random
fluctuations, clustering with different sets of ghosts will give
different jet areas. The \ttt{n\_repeat} parameter controls how many
sets of ghosts will be used. With $\ttt{repeat} > 1$, the final area
is averaged over the different sets.\footnote{And a pseudo-Monte Carlo
  estimate of the error is available through the
  \ttt{jet.area\_error()} call.}
For most applications, including those related to pileup mitigation,
keeping the default value $\ttt{n\_repeat}=1$ is a good default.
Finally, the other parameters of \ttt{GhostedAreaSpec} control the
size of the position and $p_t$ fluctuations and can safely be kept to
their default values.

Note that, for the sake of reproducibility, facilities are also
available to set and retrieve the seeds for the random-number
generator. This is done through the \texttt{set\_random\_status(...)}
and \texttt{get\_random\_status(...)} members of
\texttt{GhostedAreaSpec}.

\paragraph{Clustering with areas.} As mentioned above, once an
AreaDefinition has been constructed, the clustering with jet-area
calculation is obtained using the \ttt{ClusterSequenceArea} class
which is constructed using
\begin{lstlisting}
  template<class L> ClusterSequenceArea(const std::vector<L> & input_particles, 
                                        const JetDefinition & jet_def,
	                                const AreaDefinition & area_def);
\end{lstlisting}

As long as an instance of this class is in scope, a user can access
information about the area of its jets using the following methods of
\ttt{PseudoJet}:
\begin{lstlisting}
  /// Returns the scalar area associated with the given jet
  double area = jet.area();

  /// Returns the error (uncertainty) associated with the determination of the
  /// scalar area of the jet; gives zero when the repeat=1 and for passive/Voronoi areas
  double area_error = jet.area_error();

  /// Returns the 4-vector area associated with the given jet
  PseudoJet area_4vector = jet.area_4vector();

  /// When using active_area_explicit_ghosts, returns true for jets made 
  /// exclusively of ghosts and for ghost constituents.
  bool is_pure_ghost = jet.is_pure_ghost();
\end{lstlisting}

It is also possible to specify explicitly the set of ghosts to be
used provided one directly uses the more basic
\ttt{ClusterSequenceActiveAreaExplicitGhosts} class:
\begin{lstlisting}
  template<class L> ClusterSequenceActiveAreaExplicitGhosts
         (const std::vector<L> & pseudojets, 
          const JetDefinition & jet_def_in,
          const std::vector<L> & ghosts,
          double ghost_area,
	  const bool & writeout_combinations = false);
\end{lstlisting}
This is useful \eg to keep area information in tools which require a
re-clustering.

\subsection{Background estimation and subtraction}
\label{sec:BackgroundEstimator}

\fastjet natively provide implementations of the two main methods to
estimate the background density $\rho$ using
equation~(\ref{eq:rhoest-base}). 

\subsubsection{Background estimation.}
The simplest workflow for background estimation is first, typically
outside the event loop, to create a background estimator.
For the grid based method one creates a
\ttt{fastjet::GridMedianBackgroundEstimator},
\begin{lstlisting}
  #include "fastjet/tools/GridMedianBackgroundEstimator.hh"
  // ...
  GridMedianBackgroundEstimator bge(double max_rapidity,
                                    double requested_grid_spacing);
\end{lstlisting}
where \ttt{max\_rapidity} is the maximal $|y|$ covered by the grid and
\ttt{requested\_grid\_spacing} the (requested) grid-cell size.

For the jet-based method, one creates a
\ttt{JetMedianBackgroundEstimator},
\begin{lstlisting}
  #include "fastjet/tools/JetMedianBackgroundEstimator.hh"
  // ...
  JetMedianBackgroundEstimator bge(const Selector & selector,
                                   const JetDefinition & jet_def,
                                   const AreaDefinition & area_def);
\end{lstlisting}
where the selector is used to indicate which jets are used for
background estimation (for most use cases, one just specifies a
rapidity range, \eg \ttt{SelectorAbsRapMax(4.5)} to use all jets
with $|y|<4.5$), together with a jet definition and an area
definition.\footnote{It is also possible to construct a
  \ttt{JetMedianBackgroundEstimator} from a
  \ttt{ClusterSequenceAreaBase} object and a \ttt{Selector}. The
  latter is useful \eg when one uses several estimators sharing the
  same clustering to define their patches.}
An active area with explicit ghosts is recommended.\footnote{With the
  $k_t$ algorithm one may also use a Voronoi area
  (\texttt{effective\_Rfact = 0.9} is recommended), which has the
  advantage of being deterministic and faster than ghosted areas. In
  this case however one must use a selector that is geometrical and
  selects only jets well within the range of event particles.
  \Eg if particles are present up to $|y| = y_{\max}$ one should only
  use jets with $|y| \lesssim y_{\max} - R$.
  When using ghosts, the selector can instead go right up
  to the edge of the acceptance, if the ghosts also only go right up
  to the edge, as in the right-hand plot of
  fig.~\ref{fig:ghost-placement}.}

Both of the above background estimators derive from a
\ttt{fastjet::BackgroundEstimatorBase} class and the remaining
functionality is common to both.
In particular, for each event, one tells the background estimator
about the event particles,
\begin{lstlisting}
  bge.set_particles(event_particles);
\end{lstlisting}
where \ttt{event\_particles} is a vector of \PJ, and then extracts the
background density and a measure of its fluctuations with the two
following calls 
\begin{lstlisting}
  // the median of ($p_t$/area) for grid cells, or for jets that pass the selection cut, 
  // making use also of information on empty area in the event (in the jets case)
  rho = bge.rho(); 

  // an estimate of the fluctuations in the $p_t$ density per unit $\smash{\sqrt{A}}$,
  sigma = bge.sigma(); 
\end{lstlisting}

\subsubsection{Background subtraction.}
The \ttt{Subtractor} class, defined in
\ttt{include/tools/Subtractor.hh}, directly implements the subtraction
method (\ref{eq:subtraction-base}). Its constructor takes a pointer to
a background estimator:\footnote{Alternatively, it is also possible to
  construct the Subtractor with a fixed value for $\rho$.}
\begin{lstlisting}
  GridMedianBackgroundEstimator bge(....); // or a jet-based estimator
  Subtractor subtractor(&bge);
\end{lstlisting}
The subtractor can then be used as follows:
\begin{lstlisting}
  PseudoJet jet;
  PseudoJet subtracted_jet = subtractor(jet);
  vector<PseudoJet> jets;
  vector<PseudoJet> subtracted_jets = subtractor(jets);
\end{lstlisting}
The subtractor normally returns \ttt{jet - bge.rho(jet)*jet.area\_4vector()}.
If \ttt{jet.pt()} is smaller than
\ttt{bge.rho(jet)*jet.area\_4vector().pt()}, then the subtractor
instead returns a jet with zero 4-momentum (so that
\ttt{(subtracted\_jet==0)} returns \ttt{true}).
In both cases, the returned jet retains the user and structural
information of the original jet.

Note that the \ttt{Subtractor} has an option
\begin{lstlisting}
  void set_known_selectors(const Selector & sel_known_vertex,
                           const Selector & sel_leading_vertex);
\end{lstlisting}
The idea here is that there are contexts where it is possible, for
some of a jet's constituents, to identify which vertex they come from.
In that case it is possible to provide a user-defined a \ttt{Selector}
that indicates whether a particle comes from an identified vertex or
not and a second user-defined \ttt{Selector} that indicates whether
that vertex was the leading vertex.
The 4-momentum from the non-leading vertex is then discarded, that
from the leading vertex is kept, and subtraction is applied to
component that is not from identified vertices.
It follows that $\rho$ must correspond only to the momentum flow from
non-identified vertices.
With this setup, the jet $p_t$ is bounded to be at least equal to that
from the leading vertex, as is the mass if the ``safe'' subtraction
option is enabled (see below).
The typical application would be CHS events where charged particles
would have an identified vertex (see Chapter~\ref{chap:charged_tracks}
for more details).

\subsubsection{Positional dependence of background}

The positional dependence can be treated using one of the two
techniques introduced in Section~\ref{sec:areamed-position} where,
now, the properties $\rho$ and $\sigma$ depend on the position of the
jet.

In \fastjet, the properties of the background are to be obtained
through the methods (available for both
\ttt{JetMedianBackgroundEstimator} and
\ttt{GridMedianBackgroundEstimator})
\begin{lstlisting}
  double rho  (const PseudoJet & jet); // $p_t$ density per unit area $A$ near jet
  double sigma(const PseudoJet & jet); // fluctuations in the $p_t$ density near jet
\end{lstlisting}

\paragraph{Rescaling method.}
To account for positional dependence of the background using the
``rescaling'' technique, one first need to parametrise the average
shape of the rapidity dependence from some number of pileup events.
Then for subsequent event-by-event background determinations, one
carries out a global $\rho$ determination and then applies the
previously determined average rescaling function to that global
determination to obtain an estimate for $\rho$ in the neighbourhood of
a specific jet.

The rescaling approach is available for both grid and jet-based
methods.
To encode the background shape, the user defines a rescaling function
derived from \ttt{FunctionOfPseudoJet<double>} (see the \fastjet
manual for details). \fastjet provides a 4th order polynomial in $y$
as a simple illustrative example:
\begin{lstlisting}
  // gives rescaling$(y) = 1.16 + 0\cdot y -0.023 \cdot y^2 + 0\cdot y^3 + 0.000041 \cdot y^4$
  fastjet::BackgroundRescalingYPolynomial rescaling(1.16, 0, -0.023, 0, 0.000041);
\end{lstlisting}
Then one tells the background estimator (whether jet or grid based)
about the rescaling with the call
\begin{lstlisting}
  bge.set_rescaling_class(&rescaling);
\end{lstlisting}
Subsequent calls to \ttt{rho()} will return the median of the
distribution $p_t/A / \ttt{rescaling}(y)$ (rather than $p_t/A$).
Any calls to \ttt{rho(jet)} and \ttt{sigma(jet)} will include an
additional factor of \ttt{rescaling}$(y_\ttt{jet})$.
Note that any overall factor in the rescaling function cancels out for
\ttt{rho(jet)} and \ttt{sigma(jet)}, but not for calls to \ttt{rho()}
and \ttt{sigma()} (which are in any case less meaningful when a
rapidity dependence is being assumed for the background).

\paragraph{Local estimation (jet based).}
The ``local estimation'' technique is available for now only in the
case of the jet-based estimator, involves constructing a
\ttt{JetMedianBackgroundEstimator} with a \ttt{Selector} retaining
only jets in a local range. Such a \ttt{Selector} should be able to
take a reference, like \ttt{SelectorStrip(}$\Delta y$\ttt{)} which
keeps jets in a strip of half-width $\Delta y$ centred on the
reference jet.
When the user calls either \ttt{rho(jet)} or \ttt{sigma(jet)},
\ttt{jet} will be used as the reference for the \ttt{Selector} and only
the jets in the event that pass the cut specified by this newly
positioned \ttt{selector} are used to estimate $\rho$ or
$\sigma$.
This method is adequate if the number of jets that pass the selector
is much larger than the number of hard jets in the range (otherwise
the median $p_t/A$ will be noticeably biased by the hard jets).
One can attempt to remove some given number of hard jets before
carrying out the median estimation, \eg with a \ttt{selector} such
as
\begin{lstlisting}
  selector = SelectorStrip($\Delta y$) * (!SelectorNHardest(2))
\end{lstlisting}
which removes the 2 hardest jets globally and then, of the remainder,
takes the ones within the strip.\footnote{If you use non-geometric
  selectors such as this in determining $\rho$, the area must have
  explicit ghosts in order to simplify the determination of the empty
  area. If it does not, an error will be thrown.}
This is however not always very effective, because one may not know
how many hard jets to remove.

\paragraph{Estimation in regions (grid based).}
The grid-based estimator does not currently provide for local
estimates, in the sense that the set of tiles used for a given
instance of the grid-based estimator is always fixed.
However, as of \fastjet 3.1, it is possible to obtain relatively fine
control over which fixed set of tiles a grid-based estimator uses.
This is done with the help of the \ttt{RectangularGrid} class
\begin{lstlisting}
  RectangularGrid grid(rap_min, rap_max, rap_spacing, phi_spacing, selector);
  GridMedianBackgroundEstimator bge(grid);
\end{lstlisting}
A given grid tile will be used only if a massless \ttt{PseudoJet}
placed at the centre of the tile passes the \ttt{selector}.
So, for example, to obtain an estimate for $\rho$ based on the activity
in the two forward regions of a detector, one might set \ttt{rap\_min}
and \ttt{rap\_min} to cover the whole detector and then supply a
\ttt{SelectorAbsRapRange(rap\_central\_max, rap\_max)} to select just
the two forward regions.

\subsubsection{Handling masses and safe subtraction}\label{sec:BGE-masses}

There are several subtleties in handling masses in
subtraction. 
The first is related to the hadrons masses which, as we have seen in
Section~\ref{sec:areamed-particle-masses}, requires the introduction
of the $\rho_m$ property of the background.  
Since \fastjet 3.1, the grid and jet-median background
estimators automatically determine $\rho_m$ by default. 
It is returned from a call to \ttt{rho\_m()} or \ttt{rho\_m(jet)},
with fluctuations accessible through \ttt{sigma\_m()} or
\ttt{sigma\_m(jet)}.
The determination of $\rho_m$ involves a small speed penalty and can
be disabled with a call to \ttt{set\_compute\_rho\_m(false)} for
both background estimators.

To avoid changes in results relative to version 3.0, by default
\fastjet 3.1 and 3.2 do not use $\rho_m$ in the \ttt{Subtractor},
\ie it uses \eq~(\ref{eq:subtraction-base}) instead of the more
complete \eq~(\ref{eq:subtraction-with-rhom}).
However, for a given subtractor, a call to
\ttt{set\_use\_rho\_m(true)}, will cause it to instead use
\eq~(\ref{eq:subtraction-with-rhom}) for all subsequent
subtractions.
We \emph{strongly recommend} switching this on if your input particles
have masses, and in future versions of \fastjet we may change the
default so that it is automatically switched on.\footnote{It is also
  possible to construct a \texttt{Subtractor subtractor($\rho$,
    $\rho_m$)} with explicit $\rho$ and $\rho_m$ values; if this is
  done, then $\rho_m$ use \emph{is enabled} by default.}
An alternative is to make the input particles massless.

A second issue, relevant both for \eqs~(\ref{eq:subtraction-base}) and
(\ref{eq:subtraction-with-rhom}), is that sometimes the resulting
squared jet mass is negative.\footnote{If $m^2<0$, \ttt{m()} returns
  $-\sqrt{-m^2}$, to avoid having to introduce complex numbers just
  for this special case.}
This is obviously unphysical.
By default the 4-vector returned by the subtractor is left in that
unphysical state, so that the user can decide what to do with it.
For most applications a sensible behaviour is to adjust the
4-vector so as to maintain its $p_t$ and azimuth, while setting the
mass to zero.
This behaviour can be enabled for a given subtractor by a call to
its \ttt{set\_safe\_mass(true)} function (available since v.3.1).
In this case the rapidity is then taken to be that of the original
unsubtracted jet.

For situations where ``known selectors'' have been specified,
\ttt{set\_safe\_mass(true)} will impose stronger constraints: if the
transverse momentum of the jet after subtraction is smaller than the
transverse momentum from all the known particles from the leading
vertex, we set the subtracted 4-momentum to the sum of the known
particles coming from the leading vertex.
If the subtracted jet has larger transverse momentum but smaller mass,
we set the transverse momentum and azimuthal angle of the subtracted
jet to the result of the pileup subtraction and set the rapidity and
mass from the sum of the known particles coming from the leading
vertex. 
For jets with no known particles coming from the leading vertex,
whenever the subtracted 4-vector has an ill-defined rapidity or
azimuthal angle, we use those of the original, unsubtracted, jet.

In future versions of \fastjet, the default behaviour may be changed
so that ``safe'' subtraction is automatically enabled.

\subsection{Subtraction for jet shapes}

The area--median subtraction procedure for jet shapes introduced in
Section~\ref{sec:areamed-shapes} is not directly included in \fastjet
but is made available as \ttt{GenericSubtractor} in \fjcontrib.

This introduces a class named \ttt{GenericSubtractor}. In practice, the
code varies the ghost transverse momentum in order to estimate the
derivatives (\ref{eq:shape-ghost-derivatives}). Note that to do that,
it brings the ghost $p_t$ scale from $10^{-100}$ to a value where its
effect on the shape is tractable within machine precision.

The basic usage of this class is as follows:
\begin{lstlisting}
  GenericSubtractor gensub(&some_background_estimator);
  FunctionOfPseudoJet<double> shape;  // whatever shape you are using
  double subtracted_shape_value = gensub(shape, jet);
\end{lstlisting}

By default, this only subtracts the transverse momentum density $\rho$
(obtained from the background estimator provided to the class). Two
options allow also for the inclusion of the $\rho_m$ term: one can
either instruct the \ttt{GenericSubtractor} object to compute $\rho_m$
from the same background estimator using\footnote{Since
  \fastjet 3.1, this option would work with any background
  estimator that internally computes $\rho_m$ (and has that
  computation enabled). For \fastjet 3.0, it is only
  available for \ttt{JetMedianBackgroundEstimator}. }
\begin{lstlisting}
  gensub.use_common_bge_for_rho_and_rhom();
\end{lstlisting}
or explicitly construct \ttt{gensub} with two
background estimators\footnote{In that case, $\rho_m$ will be estimated
  using the \ttt{rho()} method of \ttt{\_bge\_for\_rhom} --- and not
  \ttt{rho\_m()} --- unless one explicitly calls
  \ttt{use\_rhom\_from\_bge\_rhom(bool value=true);}. Also, if $\rho$
  and, optionally, $\rho_m$ are known independently, the subtractor
  can be constructed directly as \ttt{GenericSubtractor
    gensub($\rho$,$\rho_m$)}.}
\begin{lstlisting}
  GenericSubtractor gensub(& bge_for_rho, & bge_for_rhom);
\end{lstlisting}

In all cases, extra information (first order subtraction, derivatives,
...) can be retrieved using
\begin{lstlisting}
  GenericSubtractorInfo info;
  double subtracted_shape_value = gensub(shape, jet, info);
\end{lstlisting}
(see directly the class definition in \ttt{GenericSubtractor.hh} for
details)

Finally, there are some cases where it is possible to use clever
tricks to improve the subtraction or its efficiency.
The first case is the one when a shape is made of several components
and one wishes to subtract each of these components independently
rather than subtracting the shape as a whole. (A typical case would be
ratios of shapes.) To do that, one derives the shape from
\ttt{ShapeWithComponenents}.
The second case is the one where the calculation of a shape first
partitions the jet into subjets before proceeding to an actual
calculation of the shape. To avoid recomputing that partition
every time \ttt{GenericSubtraction} varies the ghost scale, one can
derive such a shape from the \ttt{ShapeWithPartition}.
We refer to the complete documentation of the \ttt{GenericSubtractor}
contrib for more details.

\subsection{Subtraction for fragmentation function moments}
\label{sec:areamed-description-fragmentation}

The technique to subtract jet fragmentation function moments
introduced in Section~\ref{sec:areamed-fragmentation-function} has
been implemented in the \ttt{JetFFMoments} contribution to
\fjcontrib. 

It introduces a class, \ttt{JetFFMoments}, which computes the
subtracted FF moments for given values of $N$. It is constructed in
one of the two following ways:
\begin{lstlisting}
  /// constructor from a vector of N values, with an optional background estimator
  JetFFMoments(const std::vector<double> & ns, JetMedianBackgroundEstimator *bge=0);

  /// constructor using nn regularly-spaced values of N between nmin and nmax.
  JetFFMoments(double nmin, double nmax, unsigned int nn, 
               JetMedianBackgroundEstimator *bge=0);
\end{lstlisting}
specifying the values of $N$ either explicitly or as a number of
values between a minimum and a maximum value.
The class can be used to compute just the FF moments, without any form
of background subtraction, by not providing a
\ttt{JetMedianBackgroundEstimator} to the constructor.

To obtain the FF moments corresponding to a given jet, one uses
\begin{lstlisting}
  virtual vector<double> operator()(const PseudoJet &jet) const;
  virtual vector<double> operator()(const PseudoJet &jet, Info & info) const;
  virtual vector<vector<double> > operator()(const vector<PseudoJet> &jets) const;
\end{lstlisting}
The first version just returns a vector with the FF moments computed
at the requested values of $N$. The third version does the same for a
vector of jets. The second version requires an extra argument of the
type \ttt{JetFFMoments::Info} which stores potentially useful
intermediate information used during the calculation (see the internal
contrib's documentation for details).

We can control how FF moments are normalised using
\begin{lstlisting}
  void use_scalar_sum(bool use_scalar_sum = true);
  void set_return_numerator(bool return_numerator);
  void set_denominator(double norm);
\end{lstlisting}
The first decides whether the normalisation uses the scalar $p_t$ of
the jet, $\tilde{p}_t$ of the jet $p_t$ itself, with the former being
the default; the second only computes the numerator (\ie sets the
denominator to 1); and the third sets the denominator to a fixed
value, as one would do in, say, $\gamma+$jet events when one wants to
use the $p_t$ of the photon to normalise the jet FF moments.

Finally, improved subtraction is turned on by calling\footnote{This
  requires at least version 3.1 of \fastjet. For the \fastjet 3.0
  series, additional arguments have to be provided.
With improved subtraction switched on, the computation of FF moments
automatically includes the correction (\ref{eq:subimp-v3}).
Note also that the actual \fastjet code uses {\ttt {mu}} to denote the
argument. Here we have used {\ttt {kappa}} simply for internal coherence
with the rest of this document.  }
\begin{lstlisting}
  /// kappa is the slope $\kappa$ of $d\sigma/dpt$ parametrised as $\sigma_0/\kappa \exp(-p_t/\kappa)$
  /// in the vicinity of the jet
  void set_improved_subtraction(double kappa);
\end{lstlisting}

\section{Recommendations for practical usage}\label{sec:areamed-practical}

We have introduced all the concepts and ingredients needed to apply
the area--median pileup subtraction method. However, at the end of the
day, several details have to be specified for a practical use.
We give below our recommendations for applications at the LHC and we
will provide extensive studies and tests of these settings in the next
chapters.
For simplicity, we assume that we have events where particles are kept
up to a rapidity $|y|=y_{\rm max}$.

\paragraph{Jet area type.} To define the area of the jets, an active
area has to be used. We strongly advise the use of explicit ghosts
(\ie an area of the \ttt{active\_area\_explicit\_ghosts} type) which
allows for an explicit computation of the contribution from the ``empty
patches'' to (\ref{eq:rhoest-base}).
Furthermore, having ghosts explicitly kept makes possible the
calculation of jet areas in more elaborate jet processing like
substructure tools.
Note that \fastjet provides a \ttt{PseudoJet::is\_pure\_ghost()}
method which tells whether a jet is only made of ghosts or not, and a
\ttt{SelectorIsPureGhosts} selector which can be used \eg to eliminate
the ghosts when asking for the constituents of a jet.

\paragraph{Ghosts used for jet areas.} Using active areas means that one
has to specify the placement of the ghosts. With particles up to
$|y|=y_{\rm max}$, we advise to place the ghosts up to a rapidity
$y_{\rm max}$ too (see the right of Figure~\ref{fig:ghost-placement}).
We have found that in practice, using a unique set of ghosts, \ie
setting \ttt{n\_repeat=1}, is sufficient. It also allows for the set
of explicit ghosts to be manipulated in jet tools without
ambiguity.\footnote{When using explicit ghosts with \ttt{n\_repeat}
  larger than 1, the last set of ghosts is retained. For full usage
  with \ttt{n\_repeat} larger than 1 one should retain explicitly all
  the sets of ghosts, a feature which is not implemented in \fastjet.}
For many applications, the default fundamental ghost area,
\ttt{ghost\_area}, of 0.01 is sufficient. For the production of the
final results, it might nevertheless be useful to set this to a
smaller value. Although this has a potentially large impact on
timings, it also tends to improve the resolution we obtain at the end
of the subtraction procedure (see \eg
Fig.\ref{fig:ghost-area-impact}).
Finally, if one wants to ensure reproducibility, one can always
retrieve and set the random seed using
\ttt{GhostedAreaSpec::get\_random\_status(...)} and
\ttt{GhostedAreaSpec::set\_random\_status(...)}.

\paragraph{$\mathbf \rho$ determination.} We recommend the use of the
\ttt{GridMedianBackgroundEstimator}, \ie the estimation of the
background properties based on a grid-based computation of the
patches, mainly for speed reasons.\footnote{The
  \ttt{JetMedianBackgroundEstimator}, with jets of radius in the
  0.4-0.6 range, performs equally well but tends to be slower due to
  the extra clustering it involves.} We have found that, in practice,
a grid size in the range 0.5-0.7 leads to good estimates of $\rho$,
with 0.55 (our recommended default) providing a good default across a
wide range of processes and pileup conditions.

\paragraph{Positional dependence.} To handle the non-negligible
rapidity dependence of pileup (and in the case of heavy-ion
collisions, also the azimuthal dependence of the underlying event), we
tend to prefer the rescaling method, with local estimation also a
useful option \eg in the case of heavy-ion collisions.
Note that, for the sake of convenience, \fastjet provides a
parametrisation of a 4th order polynomial which has been sufficient
for our own studies. This specific form is not meant to be a
recommendation and user-defined rescaling functions should be used if
the simple function in \fastjet fails to describe the positional
dependence correctly.

\paragraph{Particle masses.} If you are interested in jet masses and
your input particles have non-zero masses, make sure you use the
$\rho_m$ component in \eq~(\ref{eq:subtraction-with-rhom}) by calling
your subtractor's \ttt{set\_use\_rho\_m()} method.
You should also pay attention to what happens with negative squared
masses, and consider calling the subtractor's \ttt{set\_safe\_mass()}
option.
For backwards compatibility reasons, both of these options are
disabled by default (this may change in versions of \fastjet larger
than 3.2).

\paragraph{CHS events.} The area--median pileup subtraction technique
can also be used in conjunction with Charged-Hadron-Subtracted (CHS)
types of events.
In that case, all the pileup properties have to be estimated from the
neutral particles or from the CHS events directly.
Note that, optionally, charged tracks from pileup vertices can be kept
as ghosts, \ie with their momentum scaled down by a large factor.
Alternatively, one can cluster the event including the (unmodified)
charged tracks from pileup interactions and use something like
\ttt{subtractor.set\_known\_selectors(SelectorIsCharged(),
  SelectorIsLeadingVertex())} so that the \ttt{subtractor} itself
discards the charged tracks from pileup vertices.\footnote{This is not
  our recommendations because keeping the charged tracks from pileup
  interactions in the initial clustering would slightly increase the
  back-reaction effects.}
CHS events will be discussed at length in
Chapter~\ref{chap:charged_tracks}.

\paragraph{Subtraction for groomed jets.} For pileup mitigation in jet
substructure studies, one usually works with fat jets --- \ie jets of
large radius --- for which one applies a grooming procedure to clean
the soft contamination to the jets.
Whenever possible, we strongly recommend to use also the area--median
pileup subtractions technique. The main physical reason behind this is
that grooming mostly works by reducing the jet area, replacing, say
the original $A_{\rm jet}$ into a groomed area $A_{\rm groomed}$. The
contamination (\ref{eq:pueffects-central}) still applies with
$A=A_{\rm groomed}$ and the area--median pileup subtraction would
correct for it.
Many of the common grooming tools have an implementation that supports
attaching a \ttt{Subtractor} to them and provide automatic internal
pileup subtraction.
We will come back to this in our studies in
Chapter~\ref{chap:grooming-mcstudy}.


\chapter{Monte-Carlo validation of approach}\label{chap:mcstudy}

In this Chapter, we provide an extensive validation of the
area--median pileup subtraction technique based on Monte-Carlo
studies. 
Our goal is to validate two key aspects of the area--median method: (i)
that it indeed corrects for pileup contamination and does it
for a wide range of jet observables, (ii) that it is a robust method,
fairly insensitive to variations of the pileup conditions and details
of the hard event over wide kinematic ranges.
We also want to justify our recommendations outlined in
Section~\ref{sec:areamed-practical}.

Before entering into this vast territory, we shall first briefly
discuss the basic pileup properties $\rho$ and $\sigma$. This is meant
to build a cleaner quantitative picture of pileup effects and is done
in Section~\ref{sec:area-median:pileup-properties}.
We shall then perform Monte Carlo studies of the area--median
performance for several observables and several environments.
\begin{itemize}
\item {\it The jet $p_t$} (Section~\ref{sec:areamedian-mcstudy:jetpt}).
  This is by far the most common application, relevant for the vast
  majority of LHC analyses.
\item {\it Jet shapes} (Section~\ref{sec:mcstudy-shapes}).
  This includes the jet mass as well as a series of other
  (infrared-and-collinear safe) jet observables, used in more specific
  LHC analyses.
\item {\it Heavy-ion Underlying Events} (Chapter~\ref{chap:mcstudy-hi}).
  We will show that the area--median method is also suited to subtract
  the overwhelming Underlying Event created in heavy-ion collisions at
  RHIC and at the LHC.
  This also includes a series of interesting analytic estimates in
  Section~\ref{sec:hi-analytic-estimates} which could also be relevant
  for applications to pileup subtraction in proton-proton collisions.
\item {\it Jet fragmentation function} (Section~\ref{sec:mcstudy-hi-ff}).
  Still using heavy-ion collisions, we validate our subtraction method
  for fragmentation function moments.
\end{itemize}

For all these studies, we will explore various kinematic dependencies
so as to assess the robustness of the area--median subtraction
method. 
This includes varying the pileup multiplicity, the jet transverse
momentum and rapidity, but also checking that the jet multiplicity in
the hard event does not introduce large biases.
On top of that, we also want to make sure that our findings do not
depend on the specific details of Monte-Carlo generators.
In the end, this translates into rather CPU-expensive simulations.
Even though some of the simulations performed in the context of the
original publications are slightly out of date (\eg use old versions
of Monte Carlo generators and tunes, or use different collision
energies), we have decided to keep the original results and not
update them to a more modern setup {\it except} for our study of the
jet transverse momentum. 
There are several reasons justifying this approach. First, as already
mentioned, the jet transverse momentum is by far the most important
use-case and so we consider that it is important to revisit the
original study done in \cite{AlcarazMaestre:2012vp} with a more modern
setup.\footnote{The interested reader can also consult the original
  write-up with our initial findings and check that our conclusions are
  unaffected.}  This is also the study that provides most of the bases
for our recommendations in Section~\ref{sec:areamed-practical}, again
justifying an up-to-date series of Monte-Carlo studies.
Once this important baseline has been established, we will have a
clear indication that the area--median subtraction procedure is robust
and insensitive to the details of the simulation. We have therefore
estimated that it was not critical to update the other Monte Carlo
studies.
We will discuss more at the beginning of each section the differences
between the original simulations, the current (experimental) situation
and the results presented in this document.

\section{Properties of the pileup density}\label{sec:area-median:pileup-properties}

\begin{figure}[t]
\begin{center}
\includegraphics[width=0.4\textwidth]{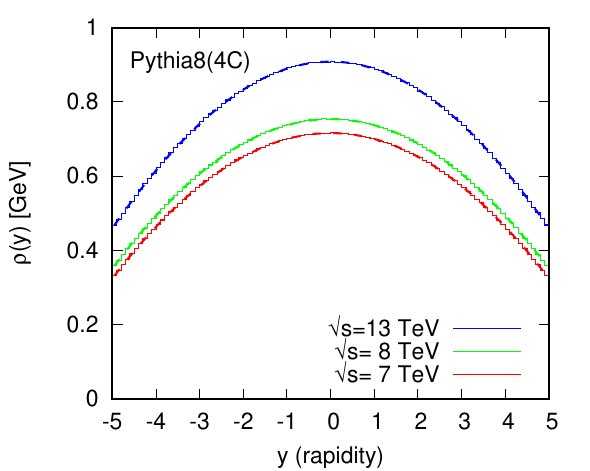}%
\hspace*{0.8cm}%
\includegraphics[width=0.4\textwidth]{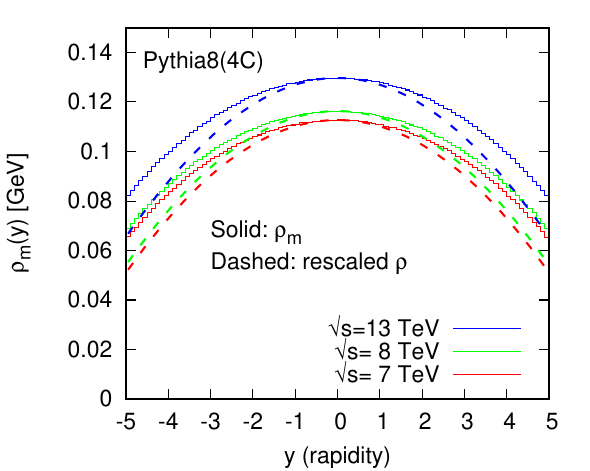}
\end{center}
\caption{Rapidity dependence of the background density $\rho$
  calculated on minimum bias events as obtained by Pythia~8 with the
  tune 4C. Left: $\rho$ profile; right: $\rho_m$ profile. For all
  plots, we show the results for the 3 LHC energies, $\sqrt{s}=7,8$
  and $13$ TeV. The dashed lines on the rightmost plots represent the
  result of fitting $\rho$ with a fourth order polynomial as in
  \eq~\eqref{eq:rho-rapidity-profile} and rescaling it so that it
  matches at $y=0$.}\label{fig:rho-rapidity-dependence}
\end{figure}

Before digging into the systematic validation of the area--median
pileup subtraction technique, it is useful to have a quantitative idea
of how large we should expect the pileup contamination to be.

In the case of pileup in $pp$ collisions, we can estimate directly the
pileup contamination from minimum bias collisions. 

The first result of such a study is presented in
Fig.~\ref{fig:rho-rapidity-dependence}. We have used
Pythia~8.186~\cite{Pythia8}, with tune 4C~\cite{Corke:2010yf}, to
generate minimum bias events for the three main LHC centre-of-mass
energies: 7, 8, and 13 TeV.\footnote{The 4C tune gives reasonable
  agreement with a wide range of minimum-bias data, as can be seen by
  consulting MCPlots~\cite{Karneyeu:2013aha}.}
Let us first look at the leftmost plot, which shows the average
transverse momentum density $\rho$ in the event as a function of
rapidity. In this case, it has been computed exactly, in each rapidity
bin of width 0.1, from the minimum bias events. We see that, in
central rapidity, each pileup event would, on average bring around
700~MeV at 7-8 TeV and around 900~MeV at 13 TeV. This drops down by
about a factor of 2 if we go to forward rapidities, $|y|=5$.
The next average property of pileup that we have discussed is
$\rho_m$, the contribution from particle masses, or, more precisely,
from $m_t-p_t$. This is shown on the right plot of
Fig.~\ref{fig:rho-rapidity-dependence} and it indicates that it is
only about 15\% of the transverse component $\rho$.

\begin{table}
\begin{center}
\begin{tabular}{c|ccc|}
$\sqrt{s}$ & $a$ & $b$ & $c$ \\
\hline
 7 TeV & 0.7161921 & -0.0159483 & 1.01133$\times10^{-5}$ \\
 8 TeV & 0.7538506 & -0.0164948 & 1.44286$\times10^{-5}$ \\ 
13 TeV & 0.9084412 & -0.0189777 & 4.05981$\times10^{-5}$ \\
\hline
\end{tabular}
\end{center}
\caption{Coefficients of the fit of the rapidity profile $\rho(y)$,
  \eq~(\ref{eq:rho-rapidity-profile}) for the three centre-of-mass
  energies under consideration.}\label{tab:rapidity-profiles}
\end{table}

To handle the rapidity dependence of pileup when using the
area--median subtraction method, we have seen that one approach is to
use a rescaling function. The rescaling function itself can be readily
extracted from minimum bias data. 
The profiles we obtain on Fig.~\ref{fig:rho-rapidity-dependence} can
very well be approximated by a fourth degree polynomial of the form
\begin{equation}\label{eq:rho-rapidity-profile}
\rho(y) = (a + by^2+cy^4)\:\text{GeV}.
\end{equation}
The fitted parameters for the different centre-of-mass energies are
given in Table~\ref{tab:rapidity-profiles}.

On the rightmost plot of Fig.~\ref{fig:rho-rapidity-dependence}, we
have shown, using dashed lines, the result of rescaling the $\rho$
rapidity profile so that it matches $\rho_m$ at $y=0$. We see that it
reproduces the $\rho_m$ rapidity dependence within about 15\%.

\begin{figure}[t]
\begin{center}
\includegraphics[width=0.4\textwidth]{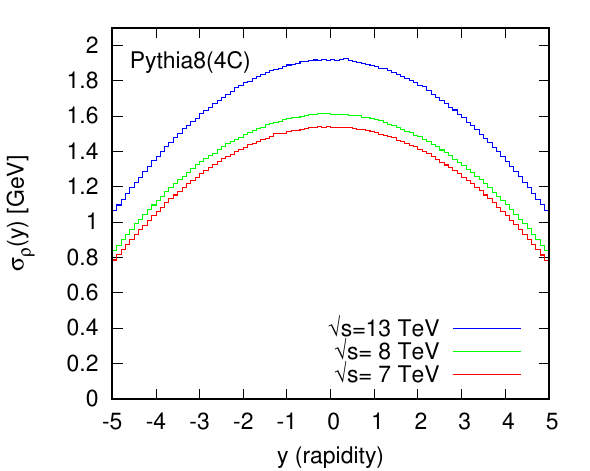}%
\hspace*{0.8cm}%
\includegraphics[width=0.4\textwidth]{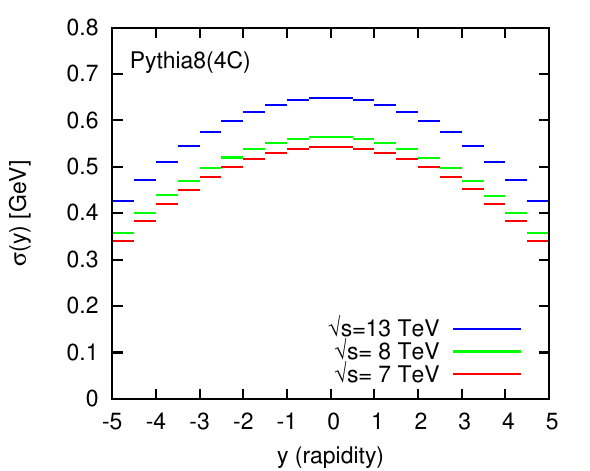}%
\end{center}
\caption{Rapidity dependence of the background fluctuations between
  different events ($\sigma_\rho$, left plot) and inside an event
  ($\sigma$, right plot).}\label{fig:sigma-rapidity-dependence}
\end{figure}

Let us now move to the other fundamental properties of pileup, namely
measures of its fluctuations: $\sigma_\rho$ (between events) and
$\sigma$ (inside an event).
The former is the simplest to measure: we just look at the dispersion
of $\rho(y)$ across our event sample. The result is plotted on the
left panel of Fig.~\ref{fig:sigma-rapidity-dependence} and it shows
large fluctuations, reaching about 1.5~GeV for Run I energies and
nearly 2 GeV at 13~TeV.
Next, to estimate $\sigma$ in a given rapidity range $y_1<y<y_2$, we
split the azimuthal angle range in several (25) cells and compute the
dispersion of the cell scalar $p_t$ divided by the square root of the
cell area is an estimate of $\sigma$.
The resulting rapidity distribution for $\sigma$ is plotted on the
right panel of Fig.~\ref{fig:sigma-rapidity-dependence}. It shows a
similar shape than $\rho$, and reaches $\sim$500 (resp. $\sim$650) MeV
at central rapidities for 7~TeV (resp. 13~TeV) collisions.

\begin{figure}
  \centerline{\includegraphics[width=0.4\textwidth]{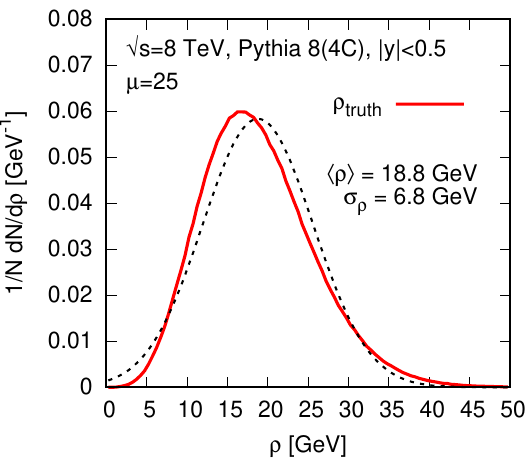}}
  \caption{Distribution of the true pileup density $\rho$ computed for
    $|y|<0.5$ with Poisson-distributed events with
    $\mu=\langle\nPU\rangle=25$. The dotted line represents a Gaussian
    fit with the average and dispersion indicated on the
    plot.}\label{fig:rho-distribution}
\end{figure}

\begin{figure}
\includegraphics[width=0.33\textwidth]{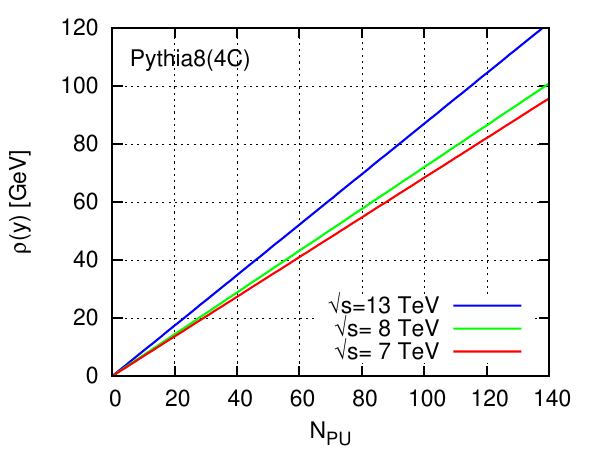}%
\hfill%
\includegraphics[width=0.33\textwidth]{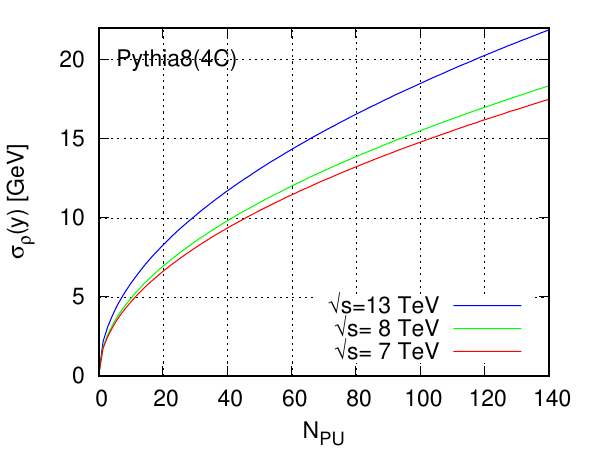}%
\hfill%
\includegraphics[width=0.33\textwidth]{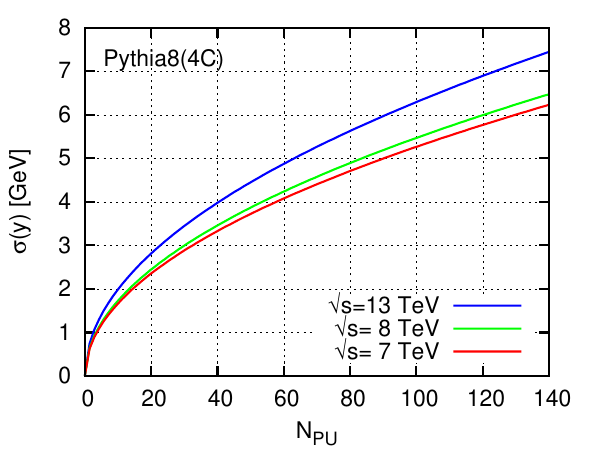}
  \caption{Evolution of the pileup fundamental properties as a
    function of \nPU. The properties are averaged for
    $|y|<2.5$.}\label{fig:rhosigma-npu-dependence}
\end{figure}

One might also wonder how precisely the distribution of the pileup
density $\rho$ in a given event is characterised by its average
density $\langle\rho\rangle$ and its fluctuations $\sigma_\rho$. 
This is illustrated in Fig.~\ref{fig:rho-distribution} where we looked
at the $\rho$ distribution for a sample of Poisson-distributed pileup
events. The dotted line, corresponding to a Gaussian fit, shows a
reasonable description of the Monte-Carlo results. The Gaussian
approximation appears therefore good enough for the simple analytic
estimates we do in this document. 

It is a trivial exercise to infer from the above numbers the pileup
dependence of $\rho$, $\sigma_\rho$ and $\sigma$.
Based on our introductory discussion, we know that $\rho$ will scale
like \nPU\ while $\sigma_\rho$ and $\sigma$ scale like $\sqrt{\nPU}$.
To illustrate this, we have averaged the minimum bias properties
in the central region, between $|y|=0$ and $|y|=2.5$.\footnote{For the
  fluctuations, we have actually averaged the square.}
The expected \nPU\ scaling is plotted on
Fig.~\ref{fig:rhosigma-npu-dependence} for $\rho$, $\sigma_\rho$ and
$\sigma$ respectively.
If, for the sake of the argument, we assume that a jet of radius $R$
has an area $\pi R^2$, which is a good approximation for the standard
anti-$k_t$ jets, the pileup contamination into a jet can be inferred
from these curves by multiplying $\rho$ and $\sigma_\rho$ by
$A=\pi R^2$ and $\sigma$ by $\sqrt{A}=\sqrt{\pi} R$. For a typical jet
obtained with $R=0.4$, this gives $A\approx 0.5$ and
$\sqrt{A}\approx 0.7$.

\section{Widest application: the jet transverse momentum}\label{sec:areamedian-mcstudy:jetpt}

By far the most widely used property of a jet for which pileup
subtraction is mandatory is its transverse momentum. It is used in
a majority of analyses at the LHC. We shall therefore begin our
systematic study of the area--median method by a study of its
performance for the jet $p_t$.

This study serves different purposes. First, it is meant to be a
validation of the area--median approach. To that aim, we shall also
consider an alternative approach, based on the number of observed
pileup vertices.
Next, we want to study the various approaches presented in the
previous Chapter to handle the positional dependence of $\rho$.
Finally, we want to assess the robustness of the method so we shall
use a wide range of hard processes --- this is crucial since it is
supposed to be used across the whole range of analyses at the LHC ---
and test different Monte-Carlo generators in order to estimate the
uncertainty on our conclusions.

Most of the study presented below comes from a contribution to the
proceedings of the 2011 Physics at TeV colliders workshop in Les
Houches. 
However, since the reconstruction of the jet transverse momentum is
used in such a wide variety of LHC analysis, we have renewed our
studies with more realistic conditions for the run I of the
LHC. Compared to the original study, we have moved from 7 to 8~TeV for
the centre-of-mass collision energy, extended our range of \nPU to 50
(instead of 30) and used $\mu=25$ as an average pileup multiplicity
instead of 10.
Note that, while conducted at 8 TeV, the features observed below are
not expected to be significantly altered at 13 TeV. We will see some
example studies carried at 13 TeV in the last part of this document
where we shall discuss more elaborate techniques for Run II and beyond.

\subsection{Subtraction techniques}\label{ssec:pusub_formula}

Since we want to validate the various aspects of the area--median
method, we shall test both the jet-based and the grid-based
approaches, as well as several approaches to handling the positional
dependence.
In order to provide a baseline for comparison and discussion, we shall
also test a simpler alternative based on the number of (seen) pileup
vertices.

In all cases, the subtraction of the jet transverse momentum will be
done through the central formula, \eq~(\ref{eq:subtraction-base}),
and define the {\em subtracted jet} transverse momentum using
\begin{equation}\label{eq:subtract}
  p_{t,\rm sub} = p_t - \rhoest A
\end{equation}
where \rhoest is the estimated value for the background density per
unit area. 
The various methods we put under scrutiny here only depend in how they
estimate \rhoest.

\paragraph{Using {\em seen vertices}.} Since experimentally it might
be possible --- within some level of accuracy that goes beyond the
scope of this discussion --- to count the number of pileup vertices
using charged tracks, one appealing way to estimate the background
density in a given event would be to count these vertices and subtract
a pre-determined number for each of them:
\begin{equation}\label{eq:npuseen}
\rhoest^{(\nPU)}(y) = f(y)\,N_{\rm PU,seen},
\end{equation}
where we have made explicit the fact that the proportionality constant
$f(y)$ can carry a rapidity dependence. $f(y)$ can be studied from
minimum bias collisions (see the previous Section) and can take into
account the fact that not all the PU vertices will be reconstructed.

\paragraph{Area--median subtraction.} This estimates the background
using the central formula \eqref{eq:rhoest-base}:
\begin{equation}\label{eq:median}
\rhoest^{\rm (global)} = \underset{i\in\, \rm patches}{\rm median}
  \left\{\frac{p_{t,i}}{A_i}\right\}
\end{equation}
We shall test both jets and grid cells as patches.

\paragraph{Area--median with a local range.} This uses a local estimate
of the area--median background-density as obtained from
\eq~\eqref{eq:rhoest-local}. A typical example, that we shall study later
on, is the case of a {\em strip range} where only the jets with
$|y-y_j|<\Delta$ are included.

\paragraph{Area--median with rescaling.} This is the other option to
correct for the positional dependence. It uses
\eq~\eqref{eq:rhoest-rescaled}.
The rescaling $f(y)$ can be taken directly from
Section~\ref{sec:area-median:pileup-properties}.

\subsection{Testing framework}\label{sec:testing-framework}

\paragraph{Embedding.} The remainder of this Section is devoted to an
in-depth comparison of the subtraction methods proposed above. Our
testing framework goes as follows. First, we embed a hard event into a
pileup background, either with a fixed number \nPU of minimum-bias
events or with a Poisson-distributed number of pileup interactions,
with $\mu$ pileup events on average.
Then, we reconstruct and subtract the jets in both the hard and full
events.\footnote{One may argue whether or not one should subtract the
  jets in the hard event. We decided to do so to cover the case where
  the hard event contains Underlying Event which, as a relatively
  uniform background, will also be subtracted together with the
  pileup.} Jets in the hard (resp. full) event will be referred to as
hard jets (resp. full jets) and the kinematic cuts are applied on the
hard jets.

\paragraph{Matching.} Testing \eq~(\ref{eq:deltapt}) requires that we
match the jets in the full event (optionally subtracted) with the jets
in the initial hard event.
For each of the full jets, we do that by finding a hard jet such that
their common constituents contribute for at least 50\% of the
transverse momentum of the hard jet. Compared to a matching where we
use distances between the jet axes, this has the advantage that a
given hard jet is matched to at most one jet in the full event, and it
avoids the arbitrariness of choosing a threshold for the matching
distance.
Also, this matching is independent of the subtraction procedure.

\paragraph{Quality measures.} For a matched pair of jets, we shall use
the notation $p_t^{\rm hard}$ and $p_t^{\rm full}$ for the transverse
momentum of the jet in the hard and full event respectively, and
$p_t^{\rm hard,sub}$ and $p_t^{\rm full,sub}$ for their subtracted
equivalents. We can then compute the shift
\begin{equation}\label{eq:deltapt}
  \Dpt = p_t^{\rm full,sub} - p_t^{\rm hard,sub},
\end{equation}
\ie the difference between the reconstructed-and-subtracted jet with
and without pileup. A positive (resp. negative) \Dpt would mean that
the PU contamination has been underestimated (resp. overestimated).
Though in principle there is some genuine information in the complete
\Dpt distribution --- \eg it could be useful to deconvolute the extra
smearing brought by the pileup, see \eg \cite{Cacciari:2010te} and
\cite{Jacobs:2010wq} and the discussion in
Chapter~\ref{chap:mcstudy-hi} below --- we shall mainly focus on two
simpler quantities: the average shift $\langle\Dpt\rangle$ and the
dispersion $\sigma_{\Dpt}$. While the first one is a direct measure of
how well one succeeds at subtracting the pileup contamination on
average, the second quantifies the remaining effects on the
resolution. One thus wishes to have $\langle\Dpt\rangle$ close to 0
and $\sigma_{\Dpt}$ as small as possible. Note that these two
quantities can be studies as a function of variables like the rapidity
and transverse momentum of the jets or the number of pileup
interactions. In all cases, a flat behaviour would indicate a robust
subtraction method.

Instead of these quality measures, some studies in the literature (see
\eg \cite{Krohn:2013lba}), prefer to use the correlation coefficient
between $p_t^{\rm full,sub}$ and $p_t^{\rm hard,sub}$.\footnote{Or
  between ${\cal O}^{\rm full,sub}$ and ${\cal O}^{\rm hard,sub}$ for
  any observable ${\cal O}$.}
We find that quantitative interpretations of correlation coefficients
can sometimes be delicate, essentially because they combine the
covariance of two observables with the two observables' individual
variances. this is further discussed in
Appendix~\ref{app:correlation-coefs}).
To the exception of Chapter~\ref{chap:charged_tracks}, we will thus
use $\langle\Dpt\rangle$ and $\sigma_{\Dpt}$ as indicators of the
performance of pileup mitigation techniques.

\paragraph{Robustness.} The robustness of our conclusions can be
checked by varying many parameters:
\begin{itemize}
\item one can study various hard processes to check that the PU
  subtraction is not biased by the hard event. In what follows we
  shall study dijets with $p_t$ ranging from 20 GeV to 1 TeV, as well
  as fully hadronic $t\bar t$ events as a representative of busier
  final states.
\item The Monte-Carlo used to generate the hard event and PU can be
  varied. For the hard event, we have used Pythia~6.4.28 with the
  Perugia 2011 tune~\cite{TunePerugia}, Pythia 8.186\footnote{The
    $Z$+jet studies in Section \ref{sec:more} have used version
    8.150.}~\cite{Pythia8} with tune 4C~\cite{Corke:2010yf} and Herwig
  6.5.10~\cite{Corcella:2000bw,Corcella:2002jc} with the ATLAS tune
  and we have switched multiple interactions on (our default) or
  off. For the minimum bias sample used to generate PU, we have used
  Pythia 8, tune 4C.\footnote{In the context of the original study,
    carried at 7~TeV in the context of the Les-Houches Physics at TeV
    Colliders workshop, we had also checked that our conclusions
    remain unchanged when using Herwig++~\cite{Gieseke:2011na} (tune
    LHC-UE7-2). Since we have no good reason to believe that it would
    be any different in this slightly different case, we have not
    reiterated that part of the study.}
\end{itemize}

\paragraph{Additional details of the analysis.} For the sake of
completeness, we list here the many other details of how the \Dpt
analysis has been conducted.
We have considered particles with $|y|\le 5$ with no $p_t$ cut or
detector effect; jets have been reconstructed with the anti-$k_t$
algorithm with $R=0.5$ keeping jets with $|y|\le 4$; for area
computations, we have use active areas with explicit ghosts with
ghosts placed\footnote{Note that we have used the ghost placement of
  FastJet 3 which differs slightly from the one in v2.4.}  up to
$|y|=5$.
For jet-based background estimations, we have used the $k_t$ algorithm
with $R=0.4$ though other options will be discussed (and the 2 hardest
jets in the set have been excluded from the median computation to
reduce the bias from the hard event); for grid-based estimations the
grid extends up to $|y|=5$ with cells of edge-size 0.55 (other sizes
will be investigated).
For estimations using a local range, a strip range of half-width 1.5
has been used (again excluding the two hardest jets in the
strip\footnote{It would actually be more appropriate to exclude
  instead the 2 hardest jets in the event.})  and for estimations
using rapidity rescaling, we have used the results obtained in the
previous Section (see \eg Table.~\ref{tab:rapidity-profiles}).
Jet reconstruction, area computation and background estimation have
all been carried out using FastJet (v3.1)
\cite{fastjet,fastjet-manual}.
Whenever we quote an average pileup multiplicity, pileup is generated
as a superposition of a Poisson-distributed number of minimum bias
events and we will vary the average number of pileup interactions. We
shall always assume $pp$ collisions with $\sqrt{s}= 8$~TeV.

For the ``seen vertices'' pileup subtraction, we need to provide a
proper definition of what we call ``seen vertices''. In what follows,
we shall define that as a minimum bias interaction that has at least 2
charged tracks with $|y|\le 2.5$ and $p_t\ge 100$ MeV, which
corresponds to 69.7\% of the events.\footnote{This is a bit optimistic
  but does not affect in any way our discussion.}

\subsection{Generic performance and rapidity dependence}\label{sec:rapdep}

\begin{figure}
\centerline{
\includegraphics[width=0.4\textwidth]{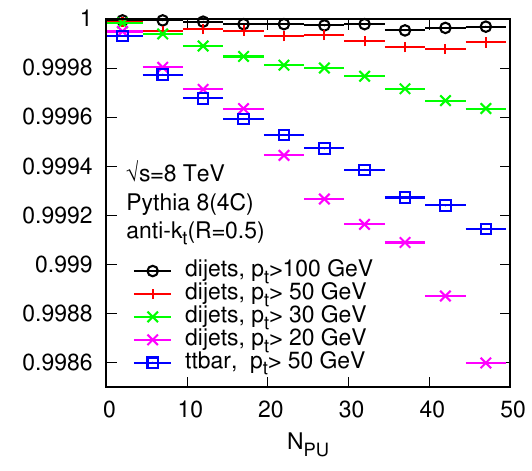}%
}
\caption{fraction of the jets in the hard event, passing the $p_t$
  cut, which have been matched to a jet in the full event. Different
  curves correspond to different processes. Efficiencies are shown as
  a function of the number of pileup interactions.
}\label{fig:matching}
\end{figure}

\begin{figure}
\centerline{
\includegraphics[width=0.4\textwidth]{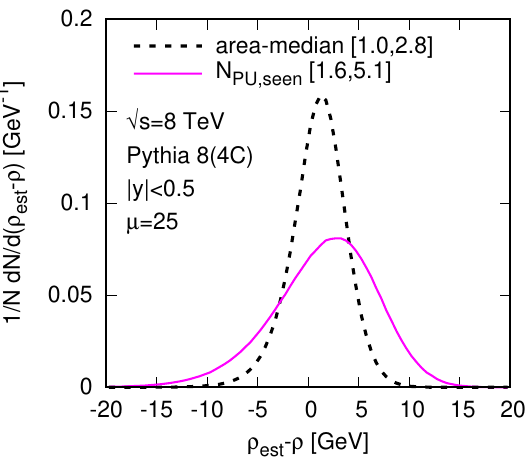}%
\hspace*{0.8cm}%
\includegraphics[width=0.4\textwidth]{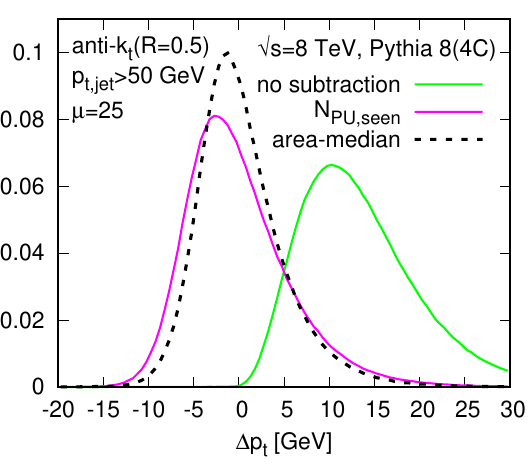}
}
\caption{Left: distribution of difference between the estimated
  $\rho$, $\rhoest$, and its true value in the $|y|<0.5$ rapidity
  interval. The calculation is done using an average of 25
  Poisson-distributed pileup events for a dijet sample generated with
  a generator $p_t$ cut of 40~GeV. The dashed (black) line
  corresponds to the area--median estimation (using a
  rapidity-rescaled grid estimator) while the solid (magenta) line
  corresponds to the ``seen vertices'' estimate. The numbers indicated
  inside brackets in the legend correspond to the average and
  dispersion of the corresponding distributions. Right: the
  corresponding \Dpt distribution for anti-$k_t$($R=0.5$) jets above
  50 GeV. We have also added the distribution obtained when no
  subtraction is performed (solid, green, line).}\label{fig:dpt-distributions}
\end{figure}

The first thing we want to look at is the matching efficiency between
the hard and full jets. This is shown in Fig.~\ref{fig:matching} where
we plot, as a function of \nPU and for different processes, the
fraction of hard jets which have been matched to a jet in the full
event. We see that even with busy events and high pileup
multiplicities, the matching efficiencies remain very high, always
above 99.85\% and often much higher.
As for the $p_t$ dependence, the matching efficiency quickly goes
close to 1 when $p_t$ increase, with again large matching efficiencies
all the way down to jets of 20~GeV.

Before we focus on the quality measured, $\langle\Dpt\rangle$ and
$\sigma_{\Dpt}$, averaged over event samples, let us first have a
brief look at global distributions. 
A few illustrative examples are presented in
Fig.~\ref{fig:dpt-distributions}. On the left plot, we directly plot
the quality of the background estimation at central rapidity. We see
that both the area--median estimate and the ``seen vertices''
estimate are close to the actual, true, $\rho$. The width of
these distributions show that the area--median tends to give a better
estimate than the ``seen vertices'' approach.
On the right plot, we show how this translates in a more practical
situation where we subtract pileup contamination from a sample of jets
above 50~GeV. In this case, we have also included the unsubtracted
distribution which is clearly biased and wider than both subtracted
ones.
Overall, all these distributions can be well-approximated by Gaussian
distributions and we can safely use the averaged quantities
$\langle\Dpt\rangle$ and $\sigma_{\Dpt}$ to assess the performance and
robustness of the various subtractions methods.

\begin{figure}
\centerline{
\includegraphics[width=\textwidth]{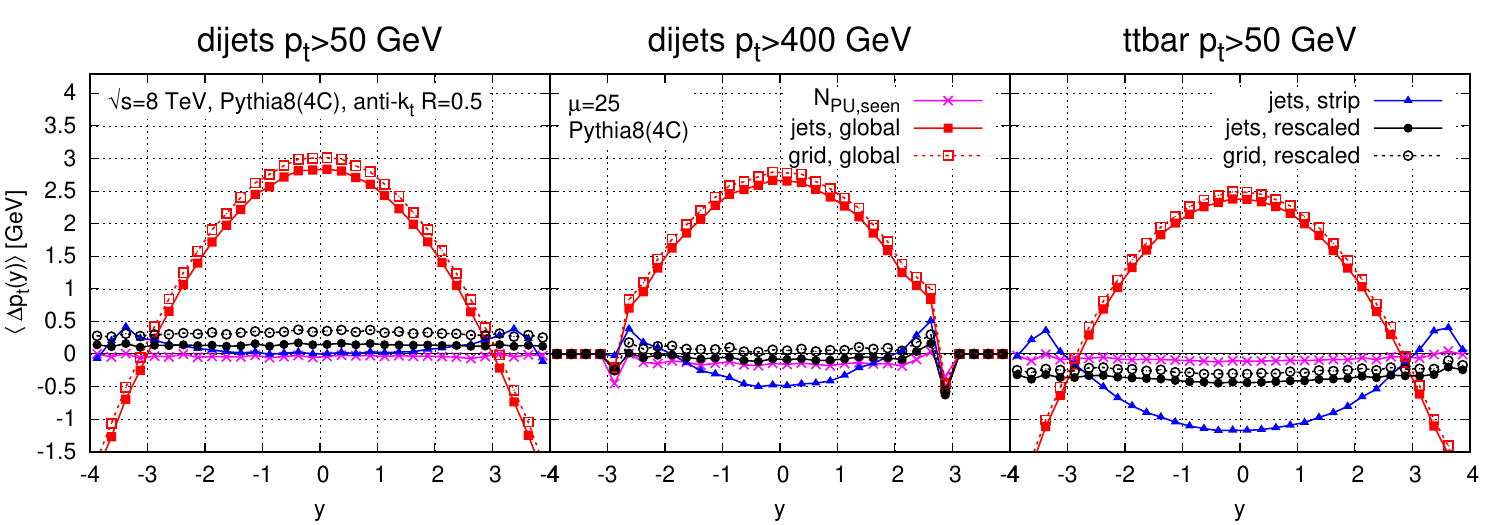}
}
\caption{Residual average shift as a function of the jet rapidity for
  all the considered subtraction methods. For the left (resp.  centre,
  right) plot, the hard event sample consists of dijets with $p_t\ge
  50$~GeV (resp. dijets with $p_t\ge 400$~GeV, and jets above $p_t\ge
  50$~GeV in $t\bar t$ events), generated with Pythia 8 (tune 4C) in
  all cases. The typical PU contamination (for unsubtracted jets) is
  around 14~GeV at central rapidities.}\label{fig:pusub_rapdep}
\end{figure}

Let us therefore carry on with more detailed performance
benchmarks, starting with the study of the rapidity dependence of PU
subtraction. Fig.~\ref{fig:pusub_rapdep} shows the residual average
shift ($\langle\Dpt\rangle$) as a function of the rapidity of the hard
jet. These results are presented for different hard processes,
generated with Pythia 8 (v8.186) and using Poisson-distributed
pileup with $\mu=25$. Robustness \wrt that choice will be discussed in
the next Section but does not play any significant role for the
moment.

The first observation is that the subtraction based on the number of
seen PU vertices does a very good job in all 3 cases. Then, global
area--median (using jets or grid cells) estimations of $\rho$, \ie the
(red) square symbols, do a fair job on average but, as expected, fail
to correct for the rapidity dependence of the PU contamination. If one
now restricts the median to a rapidity strip around the jet, the
(blue) triangles, or if one uses rapidity rescaling, the (black)
circles, the residual shift is very close to 0, typically a few
hundreds of MeV, and flat in rapidity.

Note that the strip-range approach seems to have a small residual
rapidity dependence and overall offset for high-$p_t$ processes or
multi-jet situations. That last point may be due to the fact that
smaller ranges tend to be more affected by the presence of the hard
jets (see also Section~\ref{sec:analytic-pileup}), an effect which is
reinforced for multi-jet events.
The fact that the residual shift seems a bit smaller for grid-based
estimates will be discussed more extensively in the next Section.

\begin{figure}
\centerline{
\includegraphics[width=\textwidth]{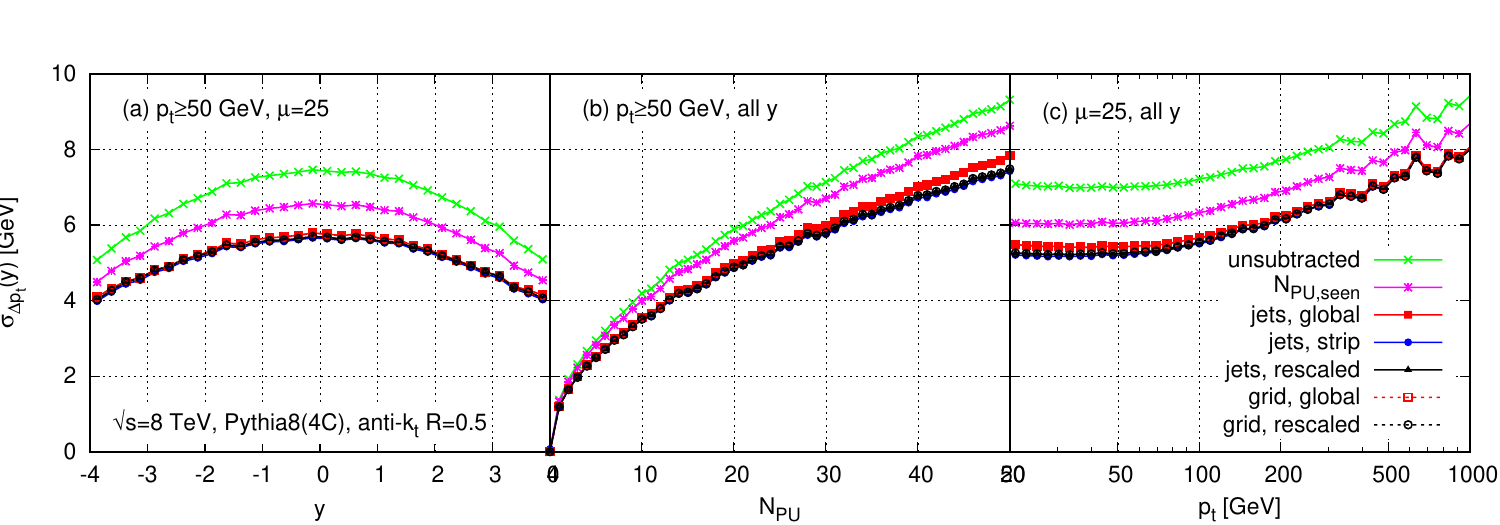}
}
\caption{Dispersion $\sigma_{\Dpt}$. Each curve corresponds to a
  different subtraction method and the results are presented as a
  function of different kinematic variables: left, as a function of
  the rapidity of the hard jet for a sample of jets with
  $p_t\ge 100$~GeV and assuming an average of 25 PU
  (Poisson-distributed) events; centre: as a function of the number of
  PU events for a sample of jets with $p_t\ge 100$; right: as a
  function of the $p_t$ of the hard jet, assuming again an average of
  25 PU events}\label{fig:pusub_resolution}
\end{figure}

Next, we turn to the dispersion of \Dpt, a direct measure of the
impact of PU fluctuations on the $p_t$ resolution of the jets. Our
results are plotted in Fig. \ref{fig:pusub_resolution} as a function
of the rapidity of the hard jet (left panel), the number of PU
vertices (central panel) and the transverse momentum of the hard jet
(right panel). All subtraction methods have been included as well as
the dispersion one would observe if no subtraction were performed.

The results show a clear trend: first, a subtraction based on the
number of seen PU vertices brings an improvement compared to not doing
any subtraction; second, area--median estimations of $\rho$ give a
more significant improvement; and third, all area--median approaches
perform similarly well (with a little penalty for global approaches).

The reason why the area--median estimations of $\rho$ outperform the
estimation based on the number of seen PU vertices is simply because
minimum bias events do not all yield the same energy deposit and this
leads to an additional source of fluctuations in the ``seen
vertices'' estimation compared to all area--median ones.
This is the main motivation for using an event-by-event determination
of $\rho$ based on the energy deposited in the event. This motivation
is further strengthened by the fact that additional issues like vertex
resolution or out-of-time PU would affect both $\langle\Dpt\rangle$
and $\sigma_{\Dpt}$ if estimated simply from the number of seen
vertices while area--median approaches are more robust.

Note that even though local ranges and rapidity rescaling do correct
for the rapidity dependence of the PU on average, the dispersion still
depends on rapidity. The increase with the number of PU vertices is in
agreement with the expected $\sqrt{\nPU}$ behaviour and the increase
with the $p_t$ of the hard process can be associated with {\em
  back-reaction}, see the next Section as well as
Section~\ref{sec:hidetails-back-reaction}. These number can also be
compared to the typical detector resolutions which would be
$\sim$10~GeV for 100~GeV jets and $\sim$20~GeV at $p_t =400$~GeV
\cite{ATLAS-resolution,CMS-resolution}.

\subsection{Digression: back-reaction effects}\label{sec:aside-back-reaction}

\begin{figure}
\centerline{
\includegraphics[width=0.9\textwidth]{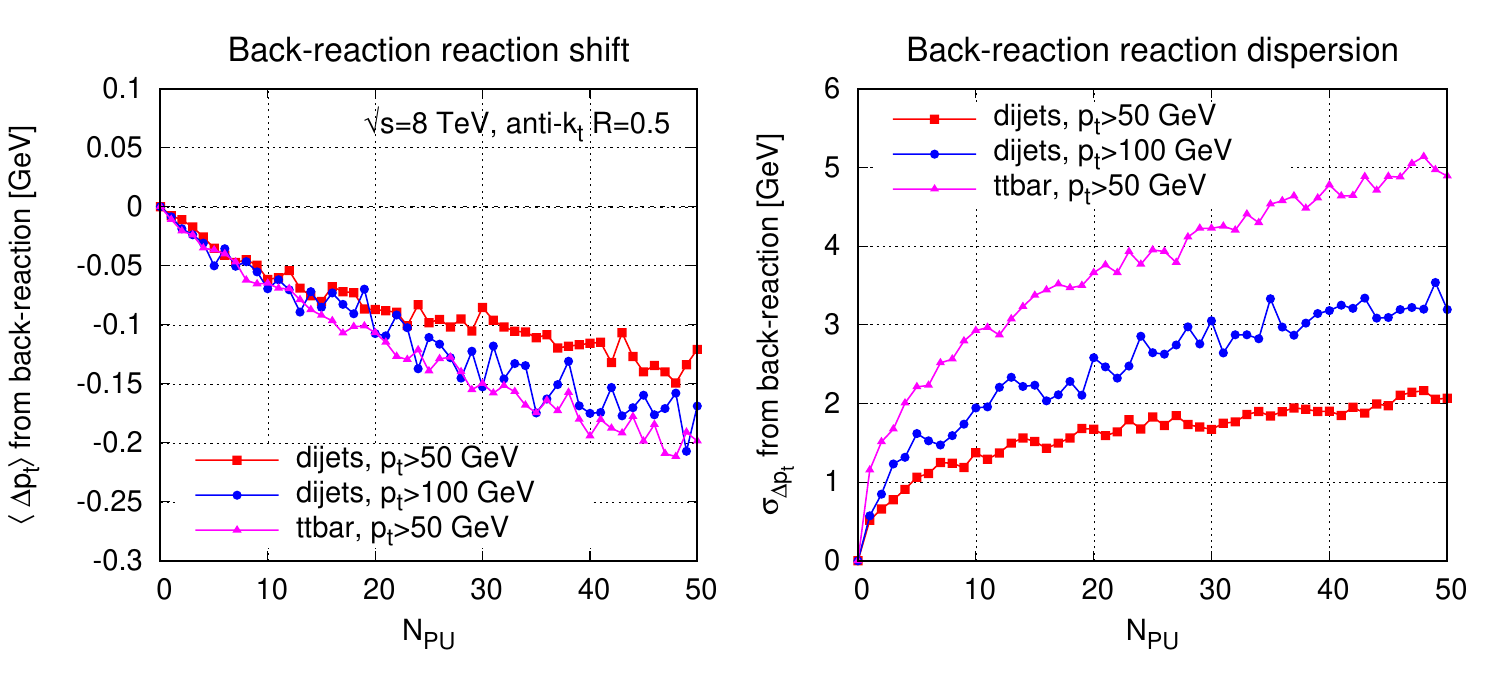}
}
\caption{Contribution of the back-reaction to the pileup average shift
  and dispersion.}\label{fig:pusub_br}
\end{figure}

\begin{figure}
\centerline{
\includegraphics[width=0.9\textwidth]{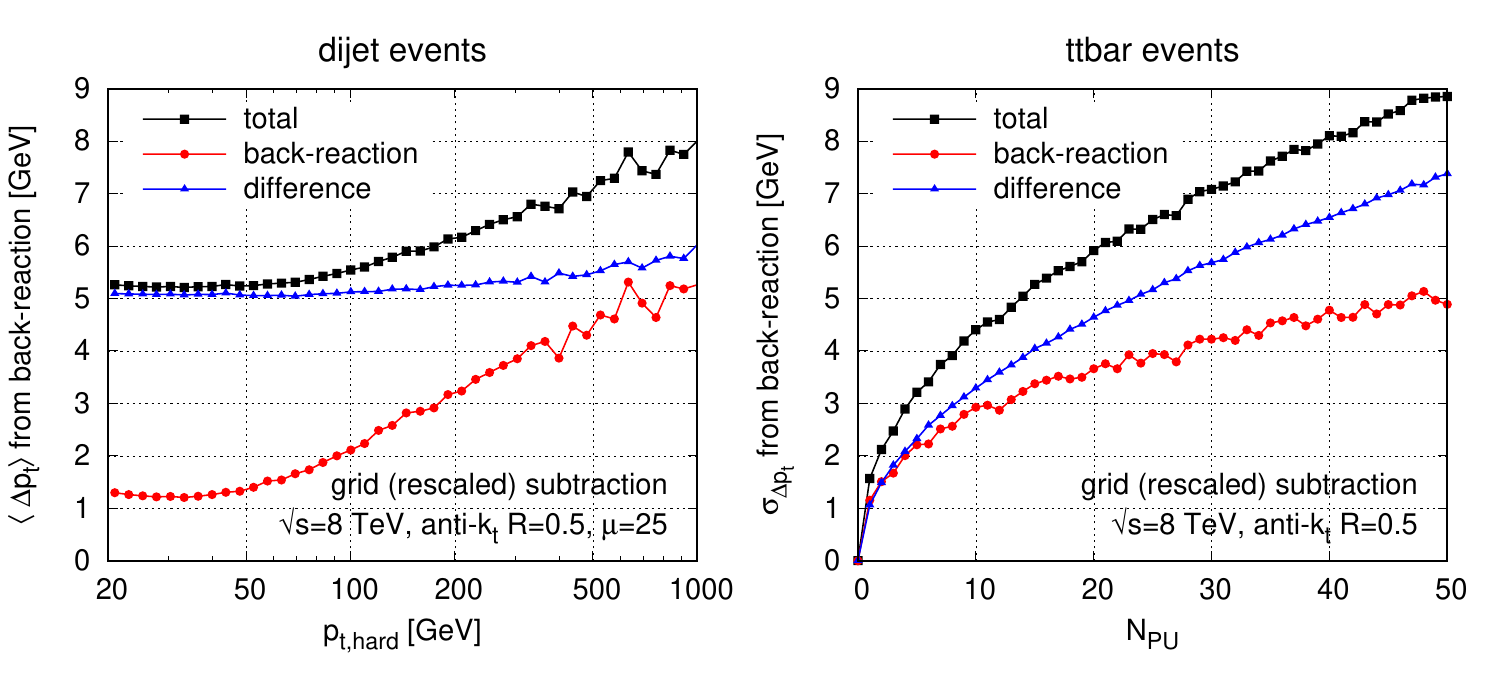}
}
\caption{Dispersion $\sigma_{\Delta p_t}$ broken into two
  contributions: the back-reaction contribution and the remaining
  contribution due intrinsically to residual pileup fluctuations. The
  left plot shown dijet events as a function of the jet $p_t$, the
  right plot focuses on $t\bar t$ events as a function of the number
  of pileup interactions.}\label{fig:pusub_br_split}
\end{figure}

Remember that back-reaction is the effect of pileup on the clustering
of the particles from the hard interaction themselves.
This effect is independent of the subtraction method.
In this short Section, we discuss a few of the quantitative effects of
back-reaction on the jet transverse momentum and compare them to the
average shift and dispersion quoted in the previous Section.

The determination of the back reaction works as follows: for each pair
of matching hard and full jets, we look at the particles from the full
jet that come from the hard interaction, say
$p_t^{\text{hard-in-full}}$.\footnote{This is doable in our Monte
  Carlo setup where we have access to the true origin of each particle
  in the full event.}  The back-reaction is then computed for each jet
as
\begin{equation}\label{eq:back-reaction-one-jet}
  \Delta p_t^{\text{back-reaction}} = p_t^{\text{hard-in-full}} -
  p_t^{\rm hard}.
\end{equation}
We can then, as before, look at the average shift and dispersion due
to back-reaction. These are shown in Figure~\ref{fig:pusub_br}.
We see that, as expected, the effect of back-reaction on the average
shift is very small (less than 200 MeV). We also see that it comes
with a non-negligible dispersion effect, in particular for busy events
($t\bar t$) or high-$p_t$ jets.

To investigate this a little further, it is interesting to look at the
contribution of back-reaction to the overall dispersion observed in
Fig.~\ref{fig:pusub_resolution}. This is shown in
Fig.~\ref{fig:pusub_br_split} for dijet events as a function of the jet
$p_t$ and for $t\bar t$ events as a function of the number of pileup
interactions. In both cases, the total dispersion (the black curve) is
as obtained with the rescaled grid subtraction in
Fig.~\ref{fig:pusub_resolution}. The back reaction is as shown in
Fig.~\ref{fig:pusub_br}. The difference, computed as
$[\sigma_{\Delta p_t}^2 - (\sigma_{\Delta
  p_t}^\mathrm{BR})^2]^{\frac12}$,
can be directly associated with background fluctuations and
misestimation of $\rho$.
One sees that the background-fluctuation component dominates over the
back-reaction component although the latter is certainly
non-negligible. In particular, it is the back-reaction component which
is mainly responsible for the rise of the dispersion with the jet
$p_t$, with a mostly-$p_t$-independent background-fluctuation contribution.

This is the case because the back-reaction dispersion is dominated by
rare events in which two similarly hard subjets are separated by a
distance close to $R$ (specifically by $R\pm\epsilon$ with
$\epsilon \ll 1$). In such a configuration, the background's
contribution to the two subjets can affect whether they recombine and
so lead to a large, $\order{p_t}$, change to the jet's momentum.
In the limit of a uniform background, $\sigma/\rho \ll 1$, this can be
shown (see Section~\ref{sec:areamed-analytic-back-reaction} for
details) to occur with a probability of order $\as \rho R^2/p_t$. Thus
the contribution to the average shift $\avg{\Delta p_t}$ is
proportional to $\as \rho R^2$, while the contribution to
$\avg{\Delta p_t^2}$ goes as $\as \rho R^2 p_t$, and so leads to a
dispersion that should grow asymptotically as $\sqrt{p_t}$
(fig.~\ref{fig:pusub_br_split} is, however, probably not yet in the
asymptotic regime).

It is worth keeping in mind that even though rare but large
back-reaction dominates the overall dispersion, it will probably not
be the main contributor in distorting the reconstructed jet spectrum.
Such distortions come from upwards $\Delta p_t$ fluctuations, whereas
large back-reaction tends to be dominated by downwards fluctuations. The
reason is simple: in order to have an upwards fluctuation from
back-reaction, there must be extra $p_t$ near the jet in the original
$pp$ event. This implies the presence of a harder underlying $2\to 2$
scattering than would be deduced from the jet $p_t$, with a
corresponding significant price to pay in terms of more suppressed
matrix elements and PDFs.
We shall show a similar behaviour for heavy-ion collisions and discuss
analytic results in more details in
Section~\ref{sec:areamed-analytic-back-reaction}.

\subsection{Robustness and Monte-Carlo dependence}\label{sec:robustness}

\begin{figure}
\centerline{
\includegraphics[width=\textwidth]{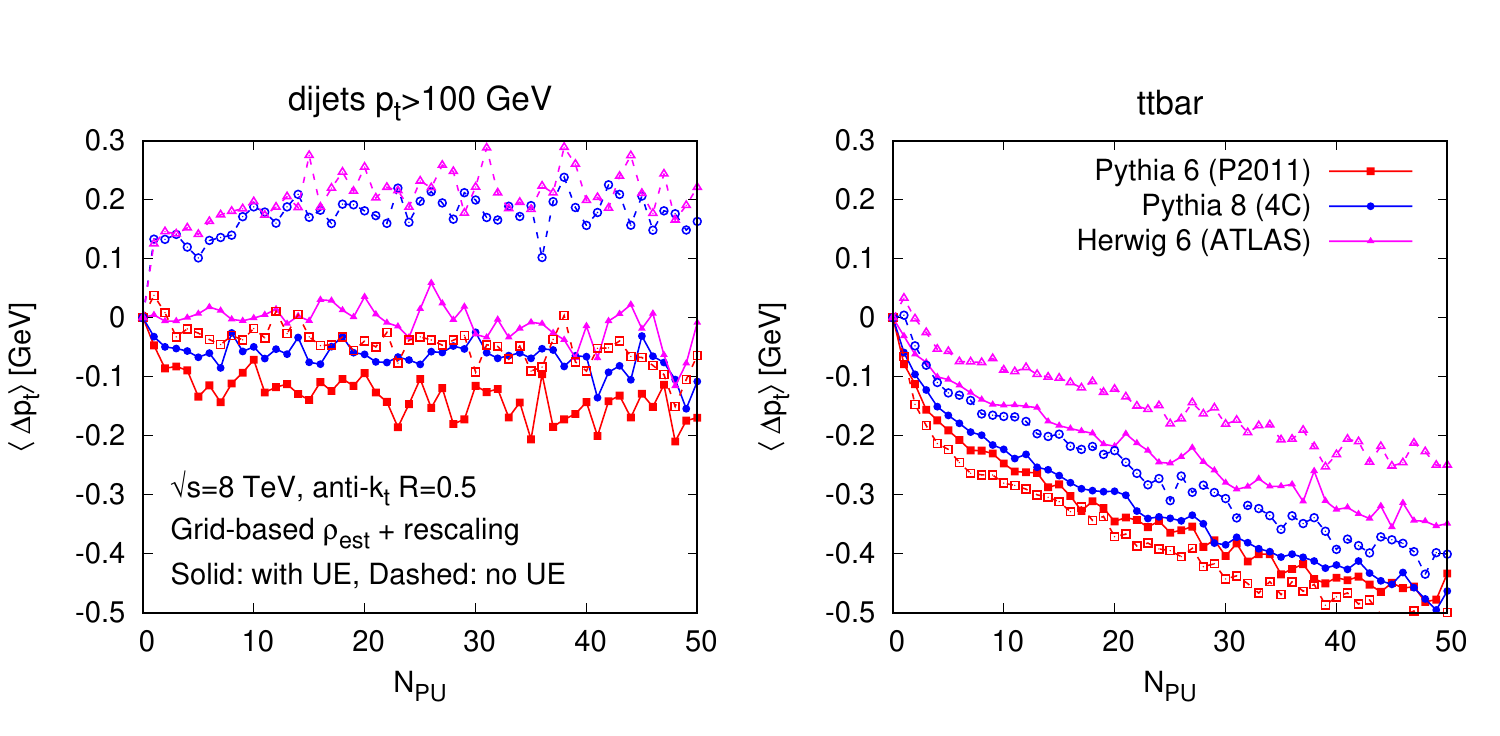}
}
\caption{Dependence of the average $p_t$ shift as a function of the
  number of PU vertices for various Monte-Carlo generators. For the
  left plot, the hard sample is made of dijets with $p_t\ge 100$ GeV
  while for the right plot, we have used a hadronic $t\bar t$
  sample. For each generator, we have considered both the case with
  the Underlying Event switched on (filled symbols) and off (open
  symbols). All results have been obtained using a grid-based median
  estimation of $\rho$ using rapidity rescaling.
}\label{fig:pusub_allmc}
\end{figure}

The last series of results we want to present addresses the
stability and robustness of the area--median estimation of the PU
density per unit area. 

To do that, the first thing we shall discuss is the Monte-Carlo
dependence of our results. In Fig. \ref{fig:pusub_allmc} we compare
the different Monte-Carlo predictions for the $\langle\Dpt\rangle$
dependence on the number of PU vertices in the case of a grid-based
median estimate of $\rho$ with rapidity rescaling. For each of the
three considered Monte-Carlos, we have repeated the analysis with and
without Underlying Event (UE) in the hard event. 
The first observation is that all the results span a range of 300-400
MeV in \Dpt and have a similar dependence on the number of PU
vertices. The dependence on $\nPU$ is flat for dijet events but
shows a small decrease for busier events. The 300-400 MeV shift splits
into a 100-200 MeV effect when changing the generator, which is likely
due to the small but non-zero effect of the hard event on the median
computation, and a 100-200 MeV effect coming from the switching on/off
of the UE.

\begin{figure}
\centerline{
\includegraphics[width=0.995\textwidth]{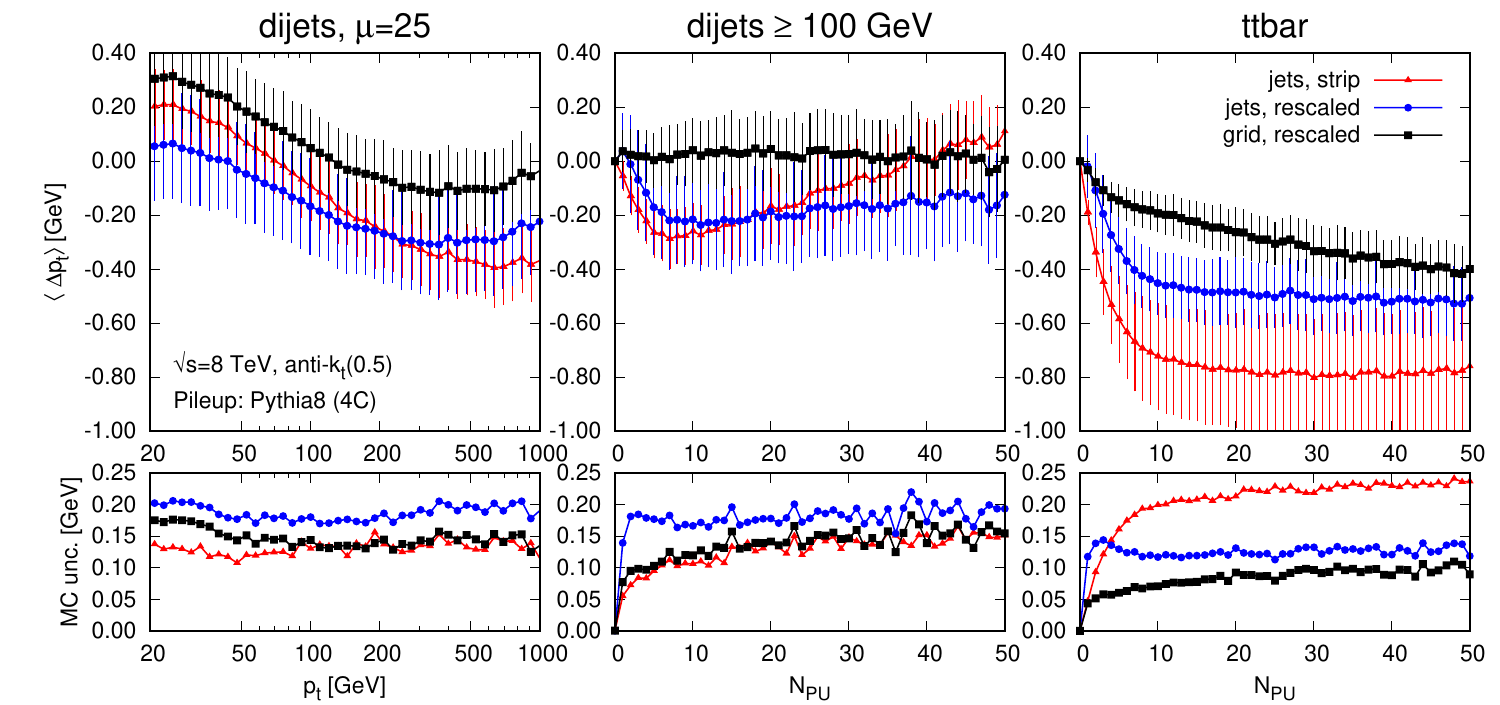}
}
\caption{Average shift, $\langle \Dpt\rangle$, plotted as a function
  of the jet $p_t$ for an average of 25 PU events (left), or as
  a function of the number of PU vertices for dijets with $p_t\ge 100$
  GeV (centre) and for $t\bar t$ events (right). In all cases, we
  compare 3 methods: the rapidity-strip range, (red) triangles, the
  jet-based approach with $y$ rescaling, (blue) circles, and the
  grid-based approach with $y$ rescaling, (black) squares. Each curve
  is the result of averaging over the various Monte-Carlo generator
  options. Their dispersion is represented both as error
  bars on the top row and directly on the bottom
  row.}\label{fig:pusub_summary}
\end{figure}

This question of subtracting the UE deserves a discussion: since the
UE is also a soft background which is relatively uniform, it
contributes to the median estimate and, therefore, one expects the UE,
or at least a part of it, to be subtracted together with the
PU. Precisely for that reason, when we compute \Dpt, our subtraction
procedure is not applied only on the ``full jet'' (hard jet+PU) but
also on the hard jet, see \eq (\ref{eq:deltapt}). The 100-200 MeV
negative shift observed in Fig. \ref{fig:pusub_allmc} thus means that,
when switching on the UE, one subtracts a bit more of the UE in the
full event (with PU) than in the hard event alone (without PU).
This could be due to the fact (see Section~\ref{sec:analytic-pileup}
for details) that for sparse events, as is typically the case with UE
but no PU, the median tends to slightly underestimate the ``real''
$\rho$, \eg if half of the event is empty, the median estimate would
be 0. This is in agreement with the fact that for $t\bar t$ events,
where the hard event is busier, switching on the UE tends to have a
smaller effect.
Note finally that as far as the size of the effect is concerned, this
100-200 MeV shift has to be compared with the $\sim$1 GeV
contamination of the UE in the hard jets.

Finally, we wish to compare the robustness of our various subtraction
methods for various processes \ie hard events and PU conditions. In
order to avoid multiplying the number of plots, we shall treat the
Monte-Carlo (including the switching on/off of the UE) as an error
estimate. That is, an average measure and an uncertainty will be
extracted by taking the average and dispersion of the 6 Monte-Carlo
setups. The results of this combination are presented on
Fig. \ref{fig:pusub_summary} for various situations and subtraction
methods. For example, the 6 curves from the left plot of
Fig. \ref{fig:pusub_allmc} have been combined into the (black) squares
of the central panel in Fig. \ref{fig:pusub_summary}.

Two pieces of information can be extracted from these results.
First of all, for dijets, the average bias of PU subtraction remains
very small and, to a large extent, flat as a function of the $p_t$ of
the jets and the number of PU vertices. When moving to multi-jet
situations, we observe an additional residual shift in the 100-300 MeV
range, extending to $\sim$500 MeV for the rapidity-strip-range
method. This slightly increased sensitivity of the
rapidity-strip-range method also depends on the Monte-Carlo. While in
all other cases, our estimates vary by $\sim$ 100 MeV when changing
the details of the generator, for multi-jet events and the
rapidity-strip-range approach this is increased to $\sim$200 MeV.

Overall, the quality of the subtraction is globally very good. Methods
involving rapidity rescaling tends to perform a bit better than the
estimate using a rapidity strip range, mainly a consequence of the
latter's greater sensitivity to multi-jet events. In comparing
grid-based to jet-based estimations of $\rho$, one sees that the
former gives slightly better results, though the differences remain
small.

Since the grid-based approach is considerably faster than the
jet-based one, as it does not require an additional clustering of the
event\footnote{Note that the clustering of the main event still needs
  to include the computation of jet areas since they are needed in
  \eq~(\ref{eq:subtract}).}, the estimation of $\rho$ using a
grid-based median with rapidity rescaling comes out as a very good
default for PU subtraction. One should however keep in mind
local-range approaches for the case where the rapidity rescaling
function cannot easily be obtained.

\subsection{PU v. UE subtraction: an analysis on $Z$+jet events}\label{sec:more}

To give further insight on the question of what fraction of the
Underlying Event gets subtracted together with the pileup, we have
performed an additional study of $Z+$jet events. We look at events
where the $Z$ boson decays into a pair of muons.
We have considered 5 different situations: events without PU or UE,
events with UE but no PU with and without subtraction, and events with
both UE and PU again with and without subtraction.
Except for the study of events without UE, this analysis could also be
carried out directly on data.

\begin{figure}
\centerline{
\includegraphics[width=\textwidth]{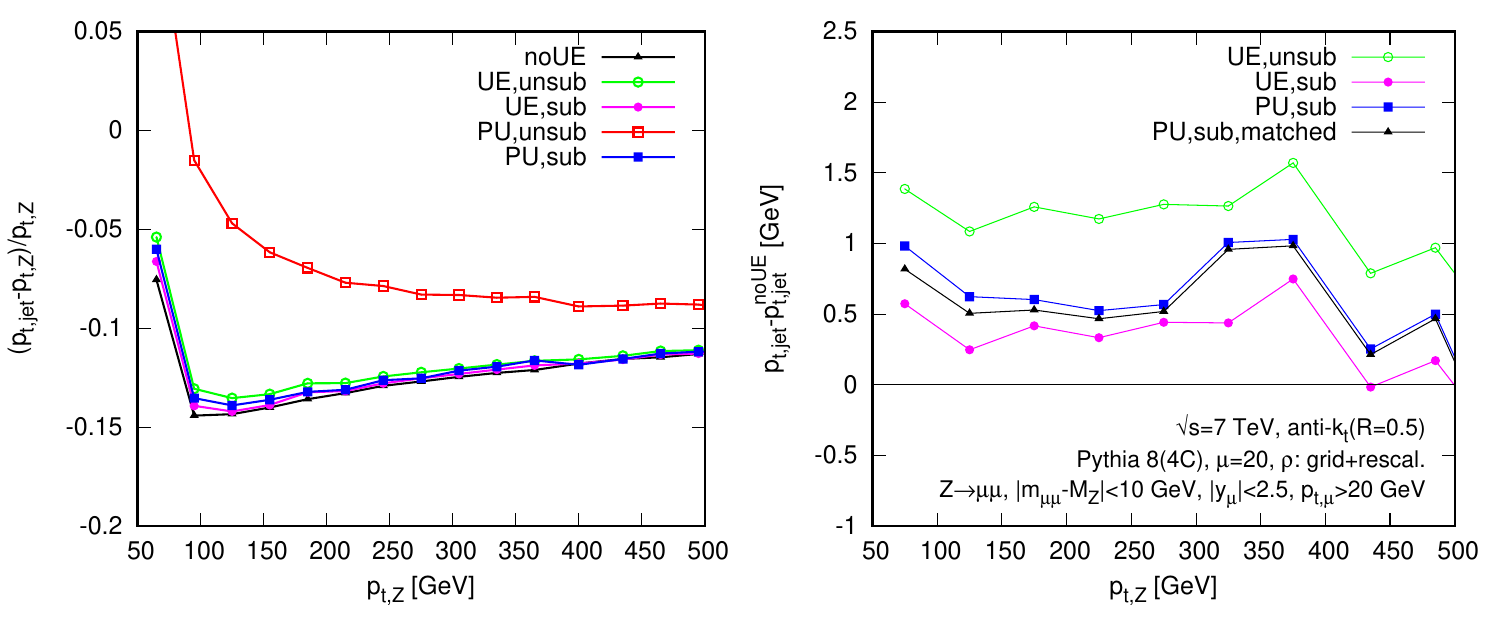}
}
\caption{Left: relative difference between the reconstructed jet and
  the reconstructed $Z$ boson transverse momenta. Right: at a given
  $p_t$ of the reconstructed $Z$ boson, difference between the
  reconstructed $p_t$ of the jet and the ideal $p_t$ with no UE or PU,
  \ie $p_t$ shift \wrt the ``noUE'' curve, the (black) triangles,
  on the left panel. See the text for the details of the
  analysis.}\label{fig:zjet}
\end{figure}

Practically, we impose that both muons have a transverse momentum of
at least 20 GeV and have $|y|\le 2.5$, and we require that their
reconstructed invariant mass is within 10 GeV of the nominal $Z$
mass. As previously, jets are reconstructed using the anti-$k_t$ jet
algorithm with $R=0.5$ and the pileup subtraction is performed using
the grid-based-median approach with rapidity rescaling and a grid size
of 0.55. All events have been generated at $\sqrt{s}=7$~TeV with Pythia
8 (tune 4C) and we have assumed an average PU multiplicity of 20
events.\footnote{Compared to the in-depth validation of the
  subtraction quality for the jet transverse momentum presented
  earlier in this Section, the simulations in this Section have not
  been updated to 8~TeV and correspond to the original results from
  the Les Houches studies.}
Note that for this study, no matching between the hard and full events
are required.
In practice, we only focus on the hardest jet in the event, with the
cut applied after the area--median subtraction for the subtracted
cases.

In Fig. \ref{fig:zjet}, we have plotted the ratio $p_{t,\rm
  jet}/p_{t,Z}-1$, with $p_{t,\rm jet}$ the transverse momentum of the
leading jet, for the various situations under considerations. Compared
to the ideal situation with no PU and no UE, the (black) triangles,
one clearly sees the expected effect of switching on the UE, the empty
(green) circles, or adding PU, the empty (red) squares: the UE and PU
add to the jet $\sim$1.2 and 13 GeV respectively. 

We now turn to the cases where the soft background is subtracted, \ie
the filled (blue) squares and (magenta) circles, for the cases with and
without PU respectively. There are two main observations:
\begin{itemize}
\item with or without PU, the UE is never fully subtracted: from the
  original 1-1.5 GeV shift, we do subtract about 800 MeV to be left
  with a $\sim$500 MeV effect from the UE. That effect decreases when
  going to large $p_t$.
\item in the presence of PU, the subtraction produces results very
  close to the corresponding results without PU and where only the UE
  is subtracted. 
  This nearly perfect agreement at large $p_{t,\rm jet}$ slightly
  degrades into an additional offset of a few hundreds of MeV when
  going to smaller scales.
  This comes about for the following reason: the non-zero $p_t$
  resolution induced by pileup (even after subtraction) means that in
  events in which the two hardest jets have similar $p_t$, the one
  that is hardest in the event with pileup may not correspond to the
  one that is hardest in the event without pileup.
  This introduces a positive bias on the hardest jet $p_t$ (a similar
  bias would be present in real data even without pileup, simply due
  to detector resolution).
  The ``matched'' curve in Fig.~\ref{fig:zjet} (right) shows that if,
  in a given hard event supplemented with pileup, we explicitly use
  the jet that is closest to the hardest jet in that same event
  without pileup, then the offset disappears, confirming its origin as
  due to resolution-related jet mismatching.
\end{itemize}

\subsection{Summary and discussion}

In this Section, we have validated the area--median approach as a
powerful and robust method for pileup mitigation at the LHC.
We have compared it to an alternative approach which applies a
correction based on the observed number of pileup vertices.
We have also studied different methods to correct for the positional
dependence of the pileup energy deposit.

The subtraction efficiency has been studied by embedding hard events
into PU backgrounds and investigate how jet reconstruction was
affected by measuring the remaining $p_t$ shift after subtraction
($\langle\Delta p_t\rangle$) as well as the impact on resolution
($\sigma_{\Delta p_t}$).
We have investigated the robustness of the area--median approach by
varying several kinematic parameters and studying different
Monte-Carlo generators and setups.

The first important message is that, though all methods give a very
good overall subtraction ($\langle\Delta p_t\rangle\approx 0$),
event-by-event methods should be preferred because their smaller PU
impact on the $p_t$ resolution (see
Fig.~\ref{fig:pusub_resolution}). This is mostly because the ``seen
vertices'' method has an additional smearing coming from the
fluctuations between different minimum bias collisions as expected
from our discussion in
Section~\ref{sec:areamed-idea-characterisation}.
This does not happen in event-by-event methods that are only affected
by point-to-point fluctuations in an event. Note also that
event-by-event methods are very likely more robust than methods based
on identifying secondary vertices when effects like vertex
identification and out-of-time PU are taken into account.

The next observation is that the two techniques to correct for the
positional dependence of the pileup energy deposit work (see
Fig.~\ref{fig:pusub_rapdep}). The median approach using a local range
(with jets as patches) or rapidity rescaling (using jets or grid cells
as patches) all give an average offset below 500 MeV, independently of
the rapidity of the jet, its $p_t$ or the number of PU vertices (see
Fig.~\ref{fig:pusub_summary}) and are thus very suitable methods for
PU subtraction at the LHC. Pushing the analysis a bit further one may
argue that the local-range method has a slightly larger offset when
applied to situations with large jet multiplicity like $t\bar t$
events (the right panel of Fig.~\ref{fig:pusub_summary}) though this
argument seems to depend on the Monte-Carlo used to generate the
hard-event sample. Also, since it avoids clustering the event a second
time, the grid-based method has the advantage of being faster than the
jet-based approach.

Our observations are to a large extent independent of the details of
the simulations, with the average bias remaining small when we vary
the kinematic properties of the jet (its transverse momentum and
rapidity), the pileup activity \nPU, or the Monte-Carlo generator used
for the simulation.

At the end of the day, this study is the base for our recommendations
in Section~\ref{sec:areamed-practical}: {\em we can recommend the
  area--median subtraction method with rapidity rescaling and using
  grid cells as patches} as a powerful default PU subtraction method
at the LHC. One should nonetheless keep in mind that the use of jets
instead of grid cells also does a very good job and that local-ranges
can be a good alternative to rapidity rescaling if the rescaling
function cannot be computed.
Also, though we have not discussed that into details, a grid cell size of
0.55 is a good default as is the use of $k_t$ jets with $R=0.4$.

To conclude, let us make a few general remarks.
First, our suggested method involves relatively few assumptions,
which helps ensure its robustness.
Effects like in-time v.\ out-of-time PU or detector response should not
have a big impact.
Many of the studies performed here can be repeated with ``real data''
rather than Monte-Carlo simulations. The best example is certainly the
$Z$+jet study of Section~\ref{sec:more} which could be done using LHC
data samples with different PU activity.  Also, the rapidity rescaling
function can likely be obtained from minimum bias collision data and
the embedding of a hard event into pure PU events could help
quantifying the remaining ${\cal O}$(100~MeV) bias.

Finally, we could also investigate hybrid techniques where one would
discard the charged tracks that do not point to the primary vertex
(CHS events) and apply the subtraction technique described here to the
rest of the event. This would have the advantage to further reduce
fluctuation effects (roughly by a factor
$\sim \sqrt{1/(1-f_{\rm chg})}\approx 1.6$, where
$f_{\rm chg}\approx 0.62$ is the fraction of charged particles in an
event). We will discuss this option at length in
Chapter~\ref{chap:charged_tracks}, together with other possible
approaches using information from charged tracks.

\section{Jet mass and jet shapes}\label{sec:mcstudy-shapes}

\begin{figure}
 \begin{minipage}[c]{0.55\textwidth}
    \includegraphics[width=\textwidth]{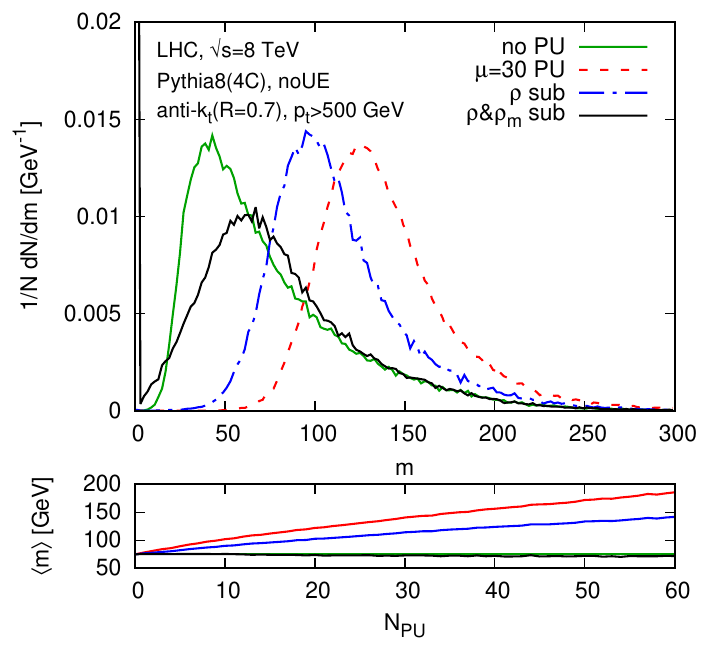}
  \end{minipage}\hfill
  \begin{minipage}[l]{0.4\textwidth}
    \caption{ Area--median subtraction for the jet mass. Top panel:
      mass distribution plotted for the hard events with no pileup
      (solid green line), the full events including (unsubtracted)
      pileup with $\mu=30$ (dashed red line), the subtracted full
      events including only the $\rho$ term in the correction
      (dash-dotted blue line) and the subtracted events including both
      the $\rho$ and $\rho_m$ terms (solid black line. Bottom panel:
      the corresponding average mass shift as a function of the number
      of pileup vertices.
    }\label{fig:area-median-subtraction-mass}
  \end{minipage}
\end{figure}

Let us first briefly discuss the subtraction of the jet mass and the
necessity to include the $\rho_m$ term in
(\ref{eq:subtraction-with-rhom}) when particles are massive.
This is illustrated on Fig.~\ref{fig:area-median-subtraction-mass}
where we have measured the jet-mass spectrum for anti-$k_t(R=0.7)$
jets with $p_t>500$~GeV, together with the shift of the average mass as
a function of \nPU on the lower panel. Compared to the distribution
observed without pileup (the solid green line), we see that adding
pileup (dashed red line) significantly shifts the distribution to
larger masses.
If we apply the area--median subtraction
\eq~(\ref{eq:subtraction-base}), the dash-dotted line in
Fig.~\ref{fig:area-median-subtraction-mass}, we still observe a large
mass shift. 
Only once we also include the $\rho_m$ correction do we recover the
right average mass, explicitly proving the importance of that
correction with massive particles.

It is also interesting to notice that the mass peak is not quite well
reproduced at small jet masses. In that region, the jet mass is rather
sensitive to pileup fluctuations resulting in a smeared peak.

We now turn to the validation of the pileup subtraction method
for jet shapes introduced in Section~\ref{sec:areamed-shapes}.
To investigate the performance of our correction procedure, we
consider a number of jet shapes:
\begin{itemize}
\item Angularities~\cite{Berger:2003iw,Almeida:2008yp}, adapted to
  hadron-collider jets as $\theta^{(\beta)} = \sum_i p_{ti} \Delta
  R_{i,\jet}^{\beta}/ \sum_i p_{ti}$, for $\beta = 0.5,1,2,3$;
  $\theta^{(1)}$, the ``girth'', ``width'' or ``broadening'' of the
  jet, has been found to be particularly useful for quark/gluon
  discrimination~\cite{Gallicchio:2011xq,ATLASqgdiscr}.
\item Energy-energy-correlation (EEC) moments, advocated for their
  resummation simplicity in~\cite{Banfi:2004yd}, $E^{(\beta)} =
  \sum_{i,j} p_{ti} p_{tj} \Delta R_{i,j}^{\beta}/ 
  (\sum_i p_{ti})^2$, using the same set of $\beta$
  values. EEC-related variables have been studied recently also
  in~Refs.\cite{Jankowiak:2011qa,Larkoski:2013eya,Dasgupta:2015lxh}. 
\item ``Subjettiness'' ratios, designed for characterising
  multi-pronged jets~\cite{Kim:2010uj,Thaler:2010tr,Thaler:2011gf}:
  one defines the subjettiness $\tau_N^{(\text{axes},\beta)} = $
  $\sum_{i} p_{ti} \min(\Delta R_{i1},\ldots,\Delta
  R_{iN})^\beta/\sum_i p_{ti}$, where $\Delta R_{ia}$ is the distance
  between particle $i$ and axis $a$, where $a$ runs from $1$ to $N$.
  One typically considers ratios such as $\tau_{21} \equiv
  \tau_2/\tau_1$ and $\tau_{32} \equiv \tau_3/\tau_2$ (the latter used
  \eg in a recent search for R-parity violating gluino
  decays~\cite{ATLAS:2012dp}); we consider $\beta=1$ and $\beta=2$, as
  well as two choices for determining the axes: 
  ``$\text{kt}$'', which exploits the $k_t$ algorithm to decluster the
  jet to $N$ subjets and then uses their axes;
  and ``$\text{1kt}$'', which adjusts the ``$\text{kt}$'' axes so as
  to obtain a single-pass approximate minimisation of
  $\tau_N$~\cite{Thaler:2011gf}.
\item A longitudinally invariant version of the planar
  flow~\cite{Thaler:2008ju,Almeida:2008yp}, involving a $2\times2$
  matrix $M_{\alpha\beta} = \sum_{i} p_{ti} (\alpha_i -
  \alpha_{\text{jet}})(\beta_i - \beta_{\text{jet}})$, where $\alpha$
  and $\beta$ correspond either to the rapidity $y$ or azimuth
  $\phi$; the planar flow is then given by $\text{Pf} =
  4\lambda_1\lambda_2/(\lambda_1 + \lambda_2)^2$, where
  $\lambda_{1,2}$ are the two eigenvalues of the matrix.
\end{itemize}

One should be aware that observables constructed from ratios of
shapes, such as $\tau_{n,n-1}$ and planar flow, are not infrared and
collinear (IRC) safe for generic jets.
In particular Pf and $\tau_{21}$ are IRC safe only when applied to jets
with a structure of at least two hard prongs, usually guaranteed by
requiring the jets to have significant mass;
$\tau_{32}$ requires a hard three-pronged structure,\footnote{%
  Consider a jet consisting instead of just two hard particles with
  $p_t = 1000\GeV$, with $\phi=0,0.5$ and two further soft particles
  with $p_t = \epsilon$, at $\phi=0.05,0.1$, all particles having
  $y=0$.
  It is straightforward to see that $\tau_{32}$ is finite and
  independent of $\epsilon$ for $\epsilon \to 0$, which results in an
  infinite leading-order perturbative distribution for $\tau_{32}$.
  Note however that some level of calculability can be reached for
  these, so-called Sudakov-safe, observables \cite{Larkoski:2013paa}
  (see also \cite{Dasgupta:2015lxh}).
} a condition not imposed in
previous work, and that we will apply here through a cut on
$\tau_{21}$.

For the angularities and EEC moments we have verified that
the first two numerically-obtained derivatives agree with analytical
calculations in the case of a jet consisting of a single hard
particle.
For variables like $\tau_N$ that involve a partition of a
jet, one subtlety is that the partitioning can change as
the ghost momenta are varied to evaluate the numerical derivative.
The resulting discontinuities (or non-smoothness) in the observable's
value would then result in nonsensical estimates of the derivatives.
We find no such issue in our numerical method to evaluate the
derivatives, but were it to arise, one could choose to force a
fixed partitioning.

To test the method in simulated events with pileup, we use
Pythia~8.165, tune 4C.
We consider 3 hard event samples: dijet, $WW$ and $t\bar t$
production, with hadronic $W$ decays, all with underlying event (UE)
turned off (were it turned on, the subtraction procedure would remove
it too).
We use anti-$k_t$ jets with $R=0.7$, taking only those with
$p_t > 500\GeV$ (before addition of pileup).
All jet-finding is performed with \fastjet~3.0.
The determination of $\rho$ and $\rho_m$ for each event follows the
area--median approach using \eqs~\eqref{eq:rhoest-base} and
\eqref{eq:rhomest-base} to estimate the background properties and
\eqref{eq:subtraction-with-rhom} to correct the jet 4-momentum. Here,
the estimation of $\rho$ and $\rho_m$ is done using
jets\footnote{These results were obtained with version 3.0 of \fastjet
  for which $\rho_m$ estimation was only available for jet-based
  methods. Since \fastjet v3.1, it is now also possible to use
  grid-based methods for $\rho_m$.}  obtained with the $k_t$ algorithm
with $R=0.4$ and includes rapidity rescaling.

\begin{figure}[t]
  %
  \includegraphics[width=0.32\textwidth]{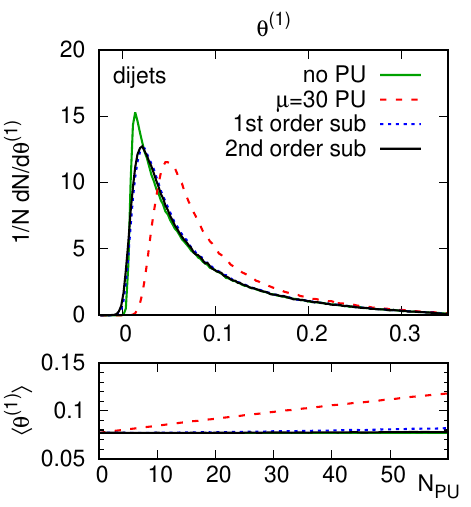}\hfill
  \includegraphics[width=0.32\textwidth]{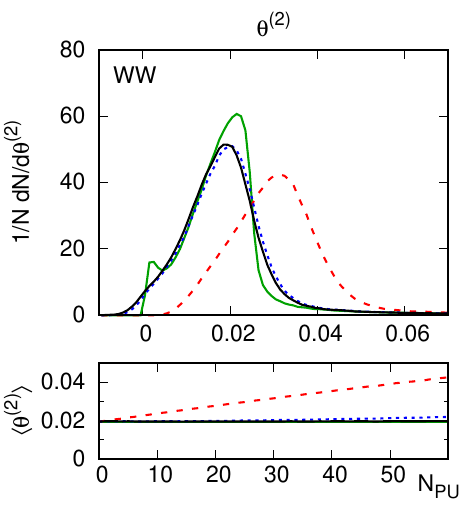}\hfill
  \includegraphics[width=0.32\textwidth]{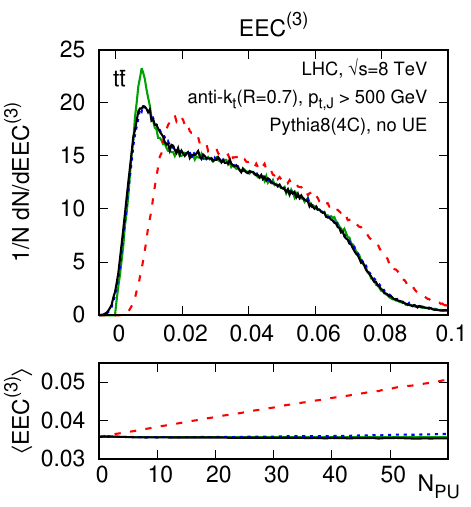}\hfill
  \includegraphics[width=0.32\textwidth]{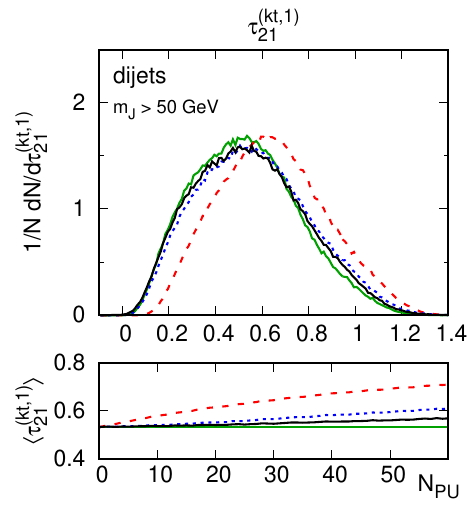}\hfill
  \includegraphics[width=0.32\textwidth]{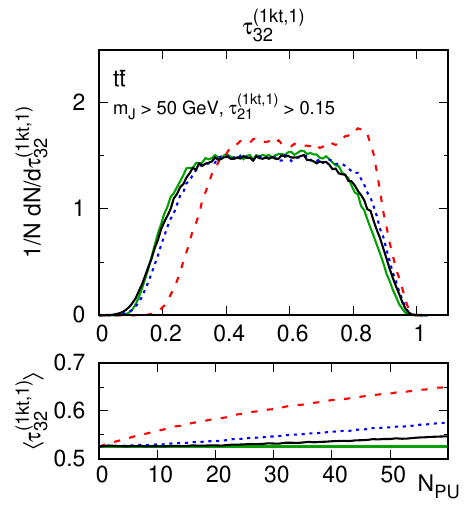}\hfill
  \includegraphics[width=0.32\textwidth]{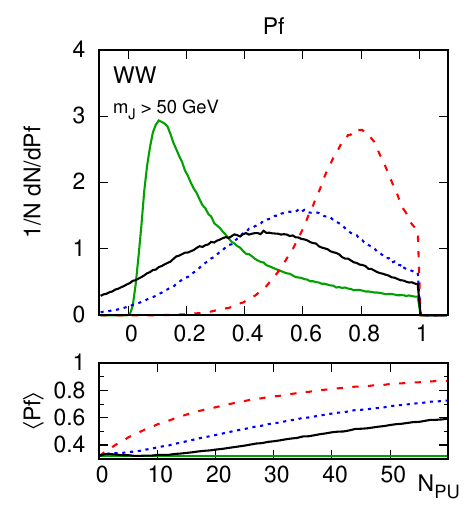}
  \centering
  \caption{Impact of pileup and subtraction on various jet-shape
    distributions and their averages, in dijet, $WW$ and $t\bar t$
    production processes.
    The distributions are shown for Poisson 
    distributed pileup (with an average of $30$ pileup events) and
    the averages are shown as a function of the number of pileup
    events, $\nPU$.
    The shapes are calculated for jets with $p_t > 500\GeV$ (the cut
    is applied before adding pileup, as are the cuts on the jet mass
    $m_J$ and subjettiness ratio $\tau_{21}$ where relevant).  }
  \label{fig:shapes}
\end{figure}

\begin{figure*}[t]
  \includegraphics[width=0.32\textwidth]{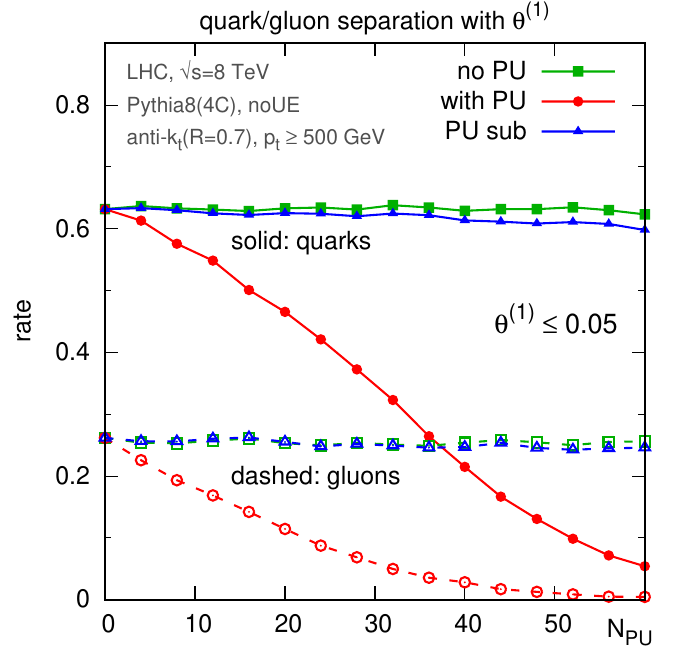}\hfill
  \includegraphics[width=0.32\textwidth]{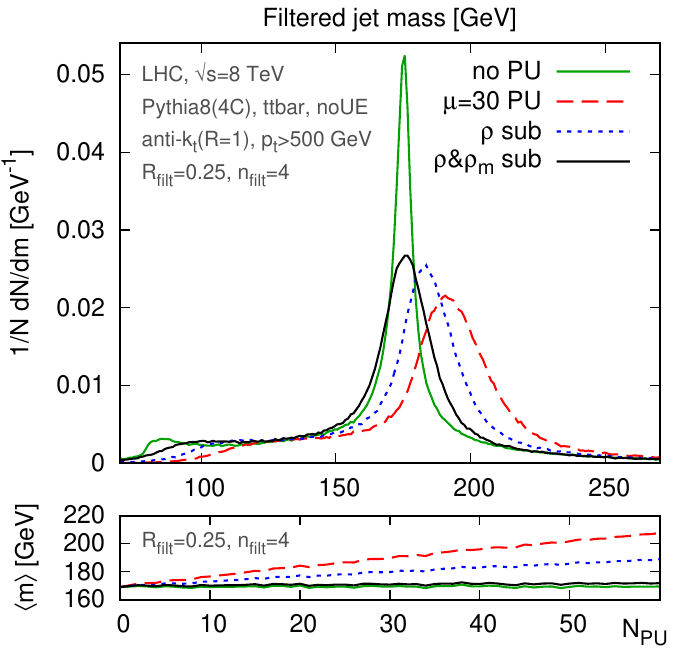}\hfill
  \includegraphics[width=0.32\textwidth]{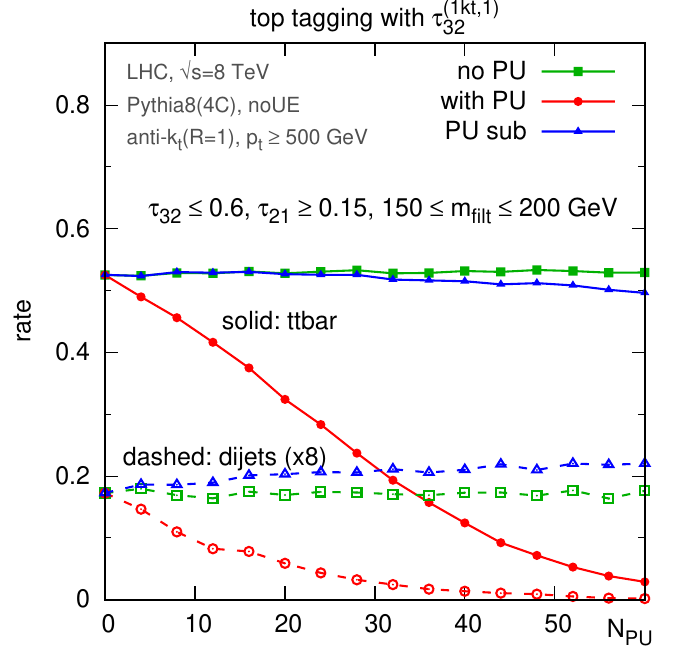}
  \caption{Left: rate for tagging quark and gluon jets using a fixed cut on
    the jet width, shown as a function of the number of pileup
    vertices.
    Middle: filtered jet-mass distribution 
    for fat jets in $t\bar t$ events, showing the impact of 
    the $\rho$ and $\rho_m$ components of the subtraction.
    Right: tagging rate of an $N$-subjettiness top tagger for $t\bar
    t$ signal and dijet background as a function of the number of
    pileup vertices.
    All cuts are applied after addition (and possible subtraction) of
    pileup. Subtraction acts on $\tau_1$, $\tau_2$ and $\tau_3$
    individually. See text for further details. }
  \label{fig:qg}
\end{figure*}

We have 17 observables and 3 event samples.
Fig.~\ref{fig:shapes} gives a representative subset of the resulting
$51$ distributions, showing in each case both the distribution (top
panel) and average shift (bottom panel, plotted as function of \nPU)
for the shape without pileup (solid green line), the result with
pileup (dashed line) and the impact of subtracting first and second
derivatives (dotted and solid black lines respectively).
The plots for the distributions have been generated using a Poisson
distribution of pileup events with an average of $\mu=30$ events.

For nearly all the jet shapes, the pileup has a substantial impact,
shifting the average values by up to $50-100\%$ (as compared to a $5-10\%$
effect on the jet $p_t$).
The subtraction performs adequately: the averaged
subtracted results for the shapes usually return very close to their
original values, with the second derivative playing a small but
sometimes relevant role.
For the distributions, tails of the distributions are generally well
recovered; however intrajet pileup fluctuations cause sharp peaks to
be somewhat broadened. These cannot be corrected for without applying
some form of noise reduction, which would however also tend to
introduce a bias.\footnote{See \eg Chapter~\ref{chap:soft-killer}.}
Of the 51 combinations of observable and process that we examined,
most were of similar quality to those illustrated in
Fig.~\ref{fig:shapes}, with the broadening of narrow peaks found to be
more extreme for larger $\beta$ values.
The one case where the subtraction procedure failed was the
planar flow for (hadronic) $WW$ events:
here the impact of pileup is dramatic, transforming a peak near
the lower boundary of the shape's range, $\text{Pf}=0$, into a peak
near its upper boundary, $\text{Pf}=1$ (bottom-right plot of
Fig.~\ref{fig:shapes}).
This is an example where one cannot view the pileup as simply
``perturbing'' the jet shape, in part because of intrinsic large
non-linearities in the shape's behaviour; with our particular set of
$p_t$ cuts and jet definition, the use of the small-$\rho$ expansion
of \eq~(\ref{eq:subtraction-for-shapes}) fails to
adequately correct the planar flow for more than about $15$ pileup
events.

Next, we want to show the behaviour of the area--median subtraction for
jet shapes in more practical applications. 

We first consider the use of the subtraction approach in the context
of quark/gluon discrimination.
In a study of a large number of shapes, Ref.~\cite{Gallicchio:2011xq}
found the jet girth or broadening, $\theta^{(1)}$, to be the most
effective single infrared and collinear safe quark/gluon
discriminator.
Fig.~\ref{fig:qg} (left) shows the fraction of quark and gluon-induced
jets that pass a fixed cut on $\theta^{(1)} \le 0.05$ as a function of
the level of pileup --- pileup radically changes the impact of the cut,
while after subtraction the q/g discrimination returns to its original
behaviour.

Our last test involves top tagging, which we illustrate on $R=1$,
anti-$k_t$ jets using cuts on the ``filtered'' jet mass and on the
$\tau_{32}$ subjettiness ratio.
The filtering selects the 4 hardest $R_\text{filt}=0.25$,
Cambridge/Aachen subjets after pileup subtraction.
The distribution of filtered jet mass is shown in Fig.~\ref{fig:qg}
(middle), illustrating that the subtraction mostly recovers the
original distribution and that $\rho_m$ is as important as $\rho$
(specific treatments of hadron masses, \eg setting them to zero, may
limit the impact of $\rho_m$ in an experimental context).
The tagger itself consists of cuts on $\tau_{32} < 0.6$, $\tau_{21}\ge
0.15 $ and a requirement that the filtered~\cite{Butterworth:2008iy}
jet mass be between $150$ and $200$ GeV.
The rightmost plot of Fig.~\ref{fig:qg} shows the final tagging
efficiencies for hadronic top quarks and for generic dijets as a
function of the number of pileup events.
Pileup has a huge impact on the tagging, but most of the original
performance is restored after subtraction.


\chapter{Applications to heavy-ion collisions}\label{chap:mcstudy-hi}

In this Chapter, we shall review the performance of the area--median
subtraction method applied to heavy-ion collisions.

Prior to any physics discussion, we should mention that the Monte
Carlo studies presented in this Chapter were made before any heavy-ion
collision actually happened at the LHC, at a time where jet
reconstruction in such a busy environment was still in its early
days.

Our studies mostly correspond to $\sqrt{s}=5.5$~TeV, except for the
jet fragmentation studies presented in Section~\ref{sec:mcstudy-hi-ff}
where we have used $\sqrt{s}=2.76$~TeV.
Since no LHC data were available at the time of our study, the Monte
Carlo generators we used to simulate $PbPb$ collision had to be
extrapolated from RHIC energies.
In practice, it seems that the tunes we have adopted are doing a
decent job at capturing what is now known from LHC data. This is at
the very least good enough for the purpose of the indicative studies
in this Chapter and we have not deemed necessary to update our results
to the lower centre-of-mass energy or to more modern generators and
tunes.
A discussion comparing the tune we used for $\sqrt{s}=2.76$~TeV
collisions and first LHC measurements can be found in
Ref.~\cite{Cacciari:2011tm}.

\section{Challenges of jet reconstruction in heavy-ion collisions}

Jets in HI collisions are produced in an environment that is far from
conducive to their detection and accurate measurement. 
Monte Carlo simulations (and real RHIC data) for gold--gold collisions
at $\sqrt{s_{NN}} = 200$~GeV (per nucleon--nucleon collision) show that
the transverse momentum density $\rho$ of final-state particles is
about 100~GeV per unit area (in the rapidity--azimuth plane).
For lead--lead collisions at $\sqrt{s_{NN}} = 2.36$~TeV or $5.5$~TeV at
the LHC this figure was expected to increase by some factor $\sim 2{-}3$.
This means that jets returned by jet definitions with a radius
parameter of, \eg, $R=0.4$, will contain background contamination of
the order of $\pi R^2 \rho \simeq$ 50 and $100-150 \GeV$ respectively.

A related, and perhaps more challenging obstacle to accurate jet
reconstruction\footnote{Note that, as for the $pp$ studies carried out
  earlier in this chapter, we are considering the reconstruction of
  the jets exclusively at the {\sl particle level}. Detector effects
  can of course be relevant, and need to be considered in detail, but
  are beyond the scope of our analysis. Note also that we shall be
  using the terms `(background-)subtracted momentum' and
  `reconstructed momentum' equivalently.} is due to the fluctuations
both of the background level (from event to event, but also from point
to point in a single event) and of the jet area itself: knowing the
contamination level `on average' is not sufficient to accurately
reconstruct each individual jet.

Our framework has the following characteristics:
\begin{itemize}
\item We shall consider a wider range of jet algorithms, still
  restricting ourselves to IRC safe algorithms. In particular, the jet
  algorithms we use will not be limited to those yielding jets of
  regular shape. This point probably has to be put in the historical
  context: at the time we conducted this study, the LHC had not taken
  any $PbPb$ data and jet reconstruction in heavy-ion collisions was
  still in its infancy. Opening our study to a variety of jet
  algorithms is meant to provide a broader perspective on the
  topic. We shall see that it indeed comes with a few interesting
  observations.
\item Our analysis will avoid excluding small transverse momentum
  particles from the clustering as some original approaches were
  leaning towards. Doing so is collinear-unsafe and inevitably biases
  the reconstructed jet momenta, which must then be corrected using
  Monte Carlo simulations, which can have substantial uncertainties in
  their modelling of jet quenching and energy loss.\footnote{It would
    still be interesting to test the SoftKiller method (see
    Chapter~\ref{chap:soft-killer}) in the context of heavy-ion
    collisions.}
  Instead, we shall try to achieve a bias-free reconstructed jet,
  working with all the particles in the event.\footnote{Our framework
    can, of course, also accommodate the elimination of particles with
    low transverse momenta, and this might help reduce the dispersion
    of the reconstructed momentum, albeit at the expense of biasing
    it.
    Whether one prefers to reduce the dispersion or instead the bias
    depends on the specific physics analysis that one is undertaking.
    Note also that detectors may effectively introduce low-$p_t$
    cutoffs of their own. 
    These detector artefacts should not have too large an impact on
    collinear safety as long as they appear at momenta of the order of
    the hadronisation scale of QCD, \ie a few hundred MeV.
  }
\item The analysis presented here was achieved before the grid-based
  approach to estimating $\rho$ was introduced and before rescaling
  was suggested to correct for positional dependence of $\rho$. 
  The study presented below will therefore be limited to jet-based
  estimates of $\rho$, with local ranges to control the positional
  dependence.
  This is however perfectly sufficient to illustrate our the points we
  want to make.
  Unpublished preliminary studies show that similar performance can be
  reached with rapidity rescaling. In particular a rescaling in $\phi$
  can be used to handle the elliptic flow. We will not include these
  results here.
\end{itemize}

Some of the features observed here are also closely connected to some
analytic considerations discussed in the next Chapter of this
document.

%
\section{Details of the study} \label{sec:hi-study-details}

\subsection{Simulation and analysis framework} \label{sec:simulation}

%
\paragraph{Embedding.}
As for the $pp$ study from Section~\ref{sec:areamedian-mcstudy:jetpt}
above, we shall proceed by embedding a ``truth'' hard jet in a
heavy-ion background and study how precisely area--median subtraction
corrects the jets back to their truth.

This approach is however less motivated in the case of heavy-ion
collisions where the physical separation between the jets and the
heavy-ion Underlying Event (HI-UE) is not well-defined as it was the
case for pileup in $pp$ collisions. This is due to the intrinsic
interaction between the products of the ``hard'' collision and the
dense heavy-ion medium it travels through. 
In today's Monte-Carlo event generators for heavy-ion collisions, such
a separation is however possible and we shall use that to gain some
knowledge of how efficient the area--median subtraction subtracts that
unwanted contribution.
In that context, it is important to check that our conclusions are not
affected neither by the details of the HI-UE, nor by the details of
the embedded hard jets, including the fact that they are quenched of
not.
In that context, the subtraction procedure can be seen as directly
belonging to our definition of jets. 

In practice, if the Monte Carlo program used to simulate heavy-ion
events explicitly provides a separation between hard events and a soft
part --- \eg as does Hydjet \cite{pyquen,hydjet} --- one can extract
one of the former as the single hard event and take the complete event
as the full one.
Alternatively, one can generate a hard event independently and embed
it in a heavy-ion event (obtained from a Monte Carlo or from real
collisions) to obtain the full event.
This second approach, which we have adopted for the bulk of results
presented here because of its greater computational efficiency, is
sensible as long as the embedded hard event is much harder than any of
the semi-hard events that tend to be present in the background.
For the transverse momenta that we consider, this condition is
typically fulfilled.
Note however that for studies in which the presence of a hard
collision is not guaranteed, \eg the evaluation of fake-jet rates
discussed below, one should use the first approach.

\paragraph{Matching and quality measures.}
We shall use the same procedure as for our earlier study of the jet
$p_t$ in $pp$ collisions, Namely, we will reconstruct the jets in both
the hard and the full events, --- yielding the ``hard'' and ``full''
jets --- apply the kinematic cuts on the hard jets to guarantee a
clean sample, and consider a full jet as matched to a hard jet if
their shared constituents make up at least 50\% of the transverse
momentum of the constituents of the hard jet.\footnote{%
  Actually, in our implementation, the condition was that the common
  part should be greater than $50\%$ of the $p_t$ of the hard jet
  after UE subtraction as in section~\ref{sec:subtraction}. In
  practice, this detail has negligible impact on matching efficiencies
  and other results.}
Events without any matched jets are simply discarded and contribute to
the evaluation of matching inefficiencies (see
section~\ref{sec:efficiency}).
For what follows, we shall keep only the pairs of jets which have been
matched to one of the two hardest (subtracted) jets in the hard event.

For a matched pair of jets we can compute the corresponding shift
$\Delta p_t$ as introduced in \eq~(\ref{eq:deltapt}).
We shall mostly focus our study on same two quality measures used in
Section.~\ref{sec:areamedian-mcstudy:jetpt}: the average $p_t$ shift,
$\langle\Delta p_t\rangle$, and its dispersion,
$\sigma_{\Delta p_t} \equiv \sqrt{\avg{\Delta p_t^2}-\avg{\Delta
    p_t}^2}$, where the average is made over the events.
Small (absolute) values of both $\avg{\Delta p_t}$ and
$\sigma_{\Delta p_t}$ will be the sign of a good subtraction.

\paragraph{Monte Carlo simulations.}
Quantifying the quality of background subtraction using Monte Carlo
simulations has several advantages. Besides providing a practical way
of generating the hard ``signal'' separately from the soft background,
one can easily check the robustness of one's conclusions by changing
the hard jet or the background sample.

One difficulty that arises in gauging the quality of jet
reconstruction in heavy-ion collisions comes from the expectation that
parton fragmentation in a hot medium will differ from that in a
vacuum.
This difference is often referred to as \emph{jet
  quenching}~\cite{Baier:1996sk,Zakharov:1997uu} (for reviews,
see~\eg~\cite{CasalderreySolana:2007pr,dEnterria:2009xfs,Wiedemann:2009sh,dEnterria:2010ubj}).
The details of jet quenching are far less well established than those
of vacuum fragmentation and can have an effect on the quality of jet
reconstruction.
Here we shall examine the reconstruction of both unquenched and
quenched jets.
For the latter it will be particularly important to be able to test
more than one quenching model, in order to help build confidence in
our conclusions about any reconstruction bias that may additionally
exist in the presence of quenching.

In practice, for this study we have used both the Fortran
(v1.6)~\cite{pyquen} and the C++ (v2.1)~\cite{hydjet} versions of
Hydjet to generate the background. Hard jets have been generated with
Pythia~6.4~\cite{Sjostrand:2000wi,Sjostrand:2003wg,Sjostrand:2006za},
either running it standalone, or using the version embedded in Hydjet
v1.6.\footnote{The events have been generated using the following
  Hydjet~v1.6 program parameters: for RHIC, $\texttt{nh}=9000$,
  $\texttt{ylfl}=3.5$, $\texttt{ytfl}=1.3$ and
  \texttt{ptmin}$=2.6\GeV$;
  for LHC, $\texttt{nh}=30000$, $\texttt{ylfl}=4$, $\texttt{ytfl}=1.5$
  and $\texttt{ptmin}=10\GeV$. In both cases quenching effects are
  turned on in Hydjet, $\texttt{nhsel}=2$, even when they are not
  included for the embedded $pp$ event. The corresponding Pyquen
  parameters we have 
  used are $\texttt{ienglu}=0$,  $\texttt{ianglu}=0$,
  $\texttt{T0}=1.0\GeV$, $\texttt{tau0}=0.1$~fm and $\texttt{nf}=0$.
  \label{foot:pyquen}}. The quenching effects have been studied using
both QPythia~\cite{Armesto:2009fj,Armesto:2007dt} and
Pyquen~\cite{pyquen,pyquen_tune}.

\subsection{Jet reconstruction: definition and
  subtraction} \label{sec:subtraction}

%
\paragraph{Jet definitions.} 
Since jet reconstruction in heavy-ion collisions is far more complex
than in the $pp$ case, we shall widen a bit our perspective and
consider a series of jet definitions. 
This will also allow us to highlight some of the key differences
between the algorithms, in particular in their response to soft
backgrounds.

We shall therefore use the $k_t$, Cambridge/Aachen (C/A), and
anti-$k_t$ algorithms. In an attempt to further reduce the sensitivity
of the jets to soft backgrounds, we have also investigated grooming
techniques (see the next part of this document for more on this
topic). In practice, we have considered ``filtering''
\cite{Butterworth:2008iy} which works by reclustering each jet with a
radius $R_{\rm filt}$ smaller than the original radius $R$ and keeping
only the $n_{\rm filt}$ hardest subjets (see also
Chapter~\ref{chap:grooming-description}).\footnote{Note that the
  filtering that we use here is unrelated to the Gaussian filter
  approach of \cite{Lai:2008zp} (not used in this paper as no public
  code is currently available).}
Background subtraction is applied to each of the subjets before
deciding which ones to keep.
In this study, we have used filtering applied to Cambridge/Aachen
(C/A(filt)) with $R_{\rm filt}=R/2$ and $n_{\rm filt}=2$. We have not
used SISCone, as its relatively slower speed compared to the
sequential recombination algorithms makes it less suitable for a HI
environment.

In all cases, we have taken the radius parameter $R=0.4$. We have
adopted this value as it is the largest used at RHIC and a reasonable
choice for the LHC.\footnote{The LHC experiments tend however to adopt
  smaller values in the 0.2-0.3 range.} Note that the effect of the
background fluctuations of the jet energy resolution increases
linearly with $R$, disfavouring significantly larger choices.
On the other hand, too small a choice of $R$ may lead to excessive
sensitivity to the details of parton fragmentation, hadronisation and
detector granularity.

\paragraph{Background determination and subtraction.}
This will be carried using the now familiar area--median approach. 
The background density $\rho$ will be obtained from a jet-based
estimation and the choice of an algorithm and a jet radius $R_\rho$
will be discussed below. 

To handle the positional dependence, we shall use the local range
technique. 
Besides the {\it global} estimate of $\rho$, independent of the jet
position, we shall consider three options for a local range introduced
in Section~\ref{sec:areamed-position}: the strip range ${\cal
  S}_{\Delta}(j)$, the circular range ${\cal C}_{\Delta}(j)$, and the
doughnut range ${\cal D}_{\delta,\Delta}(j)$.
Whenever this is necessary, we shall denote by ${\cal R}(j)$ a range
around jet $j$.

When using a local range, a compromise needs to be found between
choosing a range small enough to get a valid local estimation, but
also large enough to contain a sufficiently large number of background
(soft) jets for the estimation of the median to be reliable.
Two effects need to be considered: (1) statistical fluctuations in the
estimation of the background and (2) biases due to the presence of a hard
jet in the region used to estimate the background.
\begin{enumerate}
\item If we require that the dispersion in the reconstructed jet $p_t$
  coming from the statistical fluctuations in the estimation of the
  background (\eq~\eqref{eq:rhoest-purepu-dispersion} multiplied by
  the area of the jet) does not amount to more than a fraction
  $\epsilon$ of the overall
  $\sigma_{\Delta p_t}\approx \sigma\sqrt{A_{\rm jet}}$, then --- see
  Section~\ref{app:minrange-fluct} for details --- the range needs to
  cover an area $A_{\cal R}$ such that
  \begin{equation}
    \label{eq:Ar-lower-lim}
    A_{\cal R} \gtrsim \frac{\pi^2 R^2}{4\epsilon}\,.
  \end{equation}
  Taking $\epsilon = 0.1$ and $R=0.4$ this corresponds to an area
  $A_{\cal R} \gtrsim 25 R^2 \simeq 4$.
  For the applications below, when clustering with radius $R$, we have
  used the rapidity-strip ranges ${\cal S}_{2R}$ and ${\cal S}_{3R}$,
  the circular range ${\cal C}_{3R}$ and the doughnut range
  ${\cal D}_{R,3R}$. One can check that their areas are compatible
  with the $25 R^2$ lower-limit estimated above.

\item The bias in the estimate of $\rho$ due to the presence of $n_b$
  hard jets in the range is given roughly by
  \begin{equation}
    \label{eq:hard_median_offset}
    \langle \Delta \rho \rangle \simeq 1.8 \sigma R_\rho
    \frac{n_b}{A_{\cal R}}\,,
  \end{equation}
  as discussed in Section~\ref{app:minrange-bias}.
  The ensuing bias on the $p_t$ can be estimated as $\langle \Delta
  \rho \rangle \pi R^2$ (for anti-$k_t$ jets). 
  For $R=0.4$, $R_\rho = 0.5$, $A_{\cal R} \simeq 4$ and $n_b=1$, the
  bias in the reconstructed jet $p_t$ is $\simeq 0.1 \sigma$. 
  Given $\sigma$ in the $10-20\GeV$ range (as we will find in
  section~\ref{sec:results}) this corresponds to a $1-2\GeV$ bias.
  In order to eliminate this small bias, we will often choose to
  exclude the two hardest jets in each event when determining of
  $\rho$.\footnote{We deliberately choose to exclude the two hardest
    jets in the event, not simply the two hardest in the range. Note,
    however, that for realistic situations with limited acceptance,
    only one jet may be within the acceptance, in which case the
    exclusion of a single jet might be more appropriate. Excluding a
    second one does not affect the result significantly, so it is
    perhaps a good idea to use the same procedure regardless of any
    acceptance-related considerations.}
  
\end{enumerate}
A third potential bias discussed in ref.~\cite{Cacciari:2009dp} is
that of underestimating the background when using too small a value
for $R_\rho$.
Given the high density of particles in HI collisions, this will 
generally not be an issue as long as $R_\rho\sim 0.5$.

%
\section{Results} \label{sec:results}

\begin{figure}
\centerline{
\includegraphics[width=0.5\textwidth]{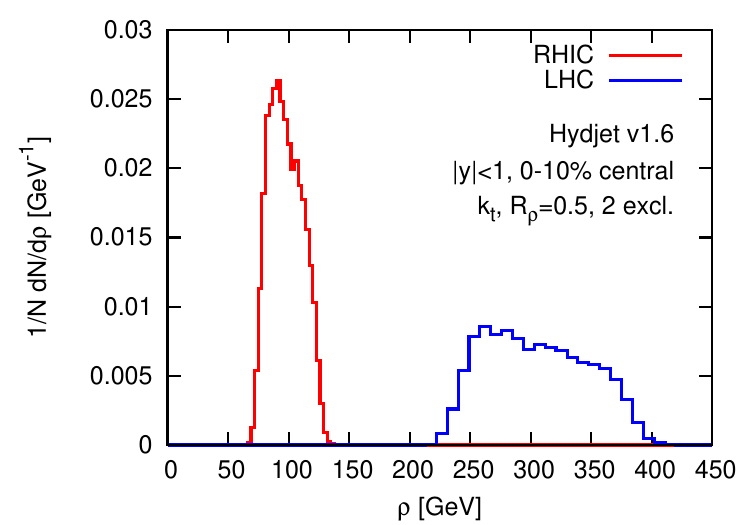}
\includegraphics[width=0.5\textwidth]{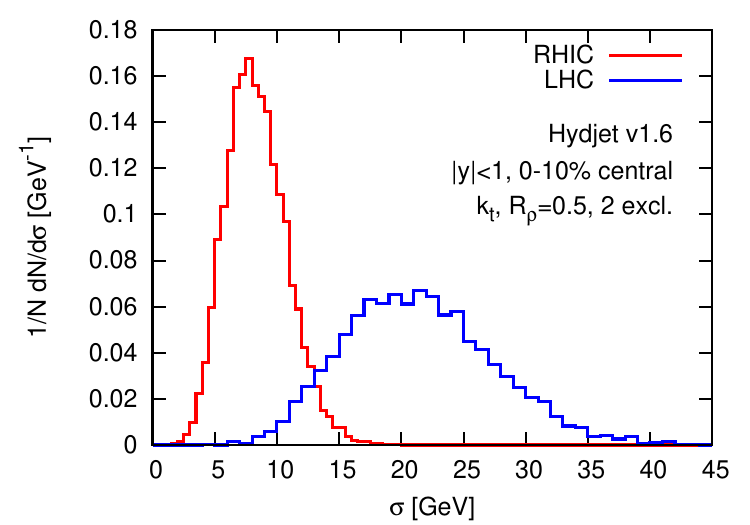}
}
\caption{\label{fig:bkg_distrib}
  Distribution of the background density $\rho$ per unit area (left)
  and its intra-event fluctuations $\sigma$ (right). It has been
  obtained from 5000 Hydjet
  events with RHIC (\AuAu, $\sqrt{s_{NN}} = 200\GeV$) and LHC
  (\PbPb, $\sqrt{s_{NN}} = 5.5\TeV$) kinematics. The background properties
  have been estimated using the techniques presented in section
  \ref{sec:subtraction}, using the $k_t$ algorithm with $R_\rho=0.5$, and
  keeping only the jets with $|y|<1$ (excluding the two hardest).}
\end{figure}

Let us first discuss the main characteristics of the background
properties.
Our setup for the background, using Hydjet v1.6 with 0-10\%
centrality leads, for RHIC (\AuAu, $\sqrt{s_{NN}} =
200\GeV$), to an average background density per unit area at central
rapidity of $\langle\rho\rangle \simeq 99$~GeV with average
fluctuations in a single event of $\langle\sigma\rangle \simeq 8$~GeV
and event-to-event fluctuations $\sigma_\rho \equiv \sqrt{\langle
  \rho^2\rangle - \langle \rho\rangle^2} \simeq 14$~GeV.  
For the LHC (\PbPb, $\sqrt{s_{NN}} = 5.5\TeV$) the corresponding
values are $\langle\rho\rangle \simeq 310$~GeV, $\langle\sigma\rangle
\simeq 20$~GeV and event-to-event fluctuations $\sigma_\rho \simeq
45$~GeV.
Fig.~\ref{fig:bkg_distrib} shows the distributions obtained from the
simulations for $\rho$ and $\sigma$.

In the case of the RHIC simulation, the result for $\avg{\rho}$ is
somewhat higher than the experimental value of $75\GeV$ quoted by the
STAR collaboration~\cite{Salur:2009vz,Ploskon:2009zd}, however this is
probably in part due to limited tracking efficiencies at low $p_t$ at
STAR,\footnote{We thank Helen Caines for discussions on this point.}
and explicit STAR~\cite{Adams:2004cb} and PHENIX~\cite{Adler:2004zn}
results for $dE_t/d\eta$ correspond to somewhat higher $\rho$ values,
about $90\GeV$.
The multiplicity of charged particles ($dN_{ch}/d\eta \simeq 660$ for
$\eta=0$ and 0--6\% centrality) and the pion $p_t$ spectrum in our
simulation are sensible compared to experimental measurements at RHIC
\cite{dnchdy,dnchdpt,dn0dpt}.
For the LHC our charged particle multiplicity is $dN_{ch}/d\eta \simeq
1600$ for $\eta=0$ and 0--10\% centrality, which is comparable to many
of the predictions reviewed in fig.~7 of \cite{Armesto:2009ug}.%
\footnote{For reference, at $2.76\TeV$ with $0-5\%$ centrality, our
  Hydjet simulation gives $dN_{ch}/d\eta \simeq 1520$ for $|\eta|<
  0.5$, which can be compared to the recent ALICE result of
  $dN_{ch}/d\eta \simeq 1584\pm4\pm76$~\cite{Aamodt:2010pb}, which
  appeared subsequent to version~1 of our article. The two numbers are
  in agreement within systematic errors, and we note that our
  conclusions are in any case largely independent of fine details of
  the background.}

An independent control analysis has also been performed with
Hydjet++~2.1~\cite{hydjet} (with default parameters) for the
background. The results at RHIC are similar, while for LHC the
comparison is difficult because the default tune of Hydjet++~2.1
predicts a much higher multiplicity, $dN_{ch}/d\eta \simeq 2800$ for
$\eta=0$ and 0--10\% centrality.

Most of the results of this section will be obtained without
quenching, though in section~\ref{sec:quenching} we will also consider
the impact on our conclusions of the Pyquen and QPythia simulations of
quenching effects.

For the results presented below, we have employed a selection cut of
$|y|\le y_{\rm max}$ on the jets with
$y_{\rm max}=1$ for RHIC and $y_{\rm max}=2.4$ for the
LHC.\footnote{This corresponds roughly to the central region for the
  ATLAS and CMS detectors. 
  Note that we also kept particles beyond $y_{\max}$ in the jet
  clustering and apply the acceptance cut only to the resulting jets.

  For ALICE, the acceptance is more
  limited~\cite{AliceHija,Alice0910}. Some adaptation of our method
  will be needed for estimating $\rho$ in that case, in order for
  information to be derived from jets near the edge of the acceptance
  and thus bring the available area close to the ideal requirements
  set out in Section~\ref{app:minrange-fluct}. One option that we have
  recommended based on preliminary quick tests is to limit the ghosts
  to the detector acceptance and to include all the resulting patches
  in the determination of $\rho$.}
We only consider full jets that are matched to one of the two hardest
jets in the hard event.
The computation of the jet areas in \fastjet, needed both for the
subtraction and the background estimation, has been performed using
active areas (including explicit ghosts), with ghosts up to
$y_{\rm max}+1.8$, a single repetition and a ghost area of 0.01.%
The determination of the background density $\rho$ has been performed
using the $k_t$ algorithm with $R_\rho=0.5$. Though the estimate of
the background depends on $R_{\rho}$ \cite{Cacciari:2009dp} (see also
Section~\ref{sec:analytic-pileup}), we have observed that choices
between 0.3 and 0.5 lead to very similar results (\eg differing by at
most a few hundred MeV at RHIC).

%
\subsection{Matching efficiency}\label{sec:efficiency}

\begin{figure}
\centerline{
\includegraphics[width=0.5\textwidth]{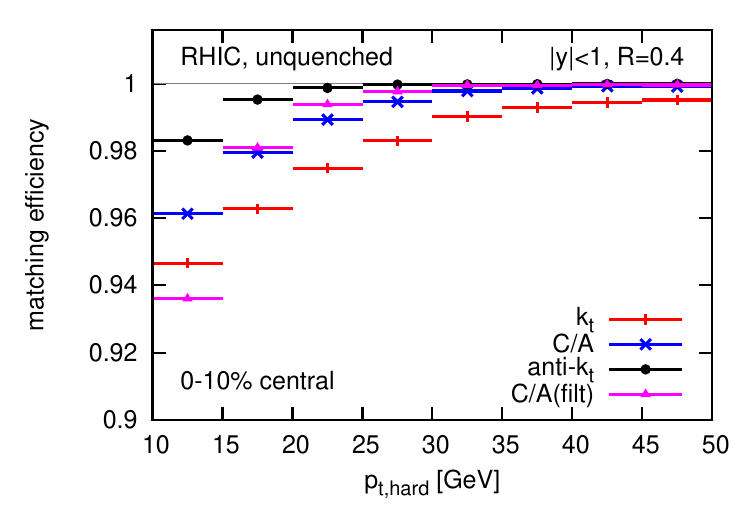}
\includegraphics[width=0.5\textwidth]{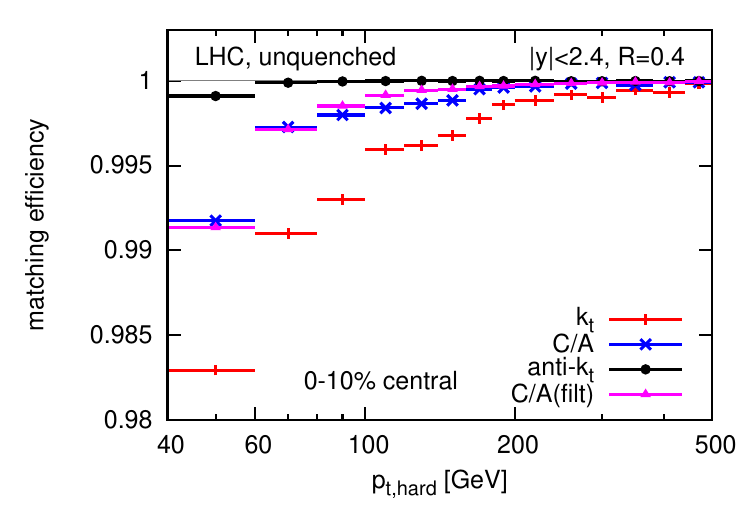}
}
\caption{\label{fig:efficiency} Matching efficiency for reconstructed
  jets as a function of the jet $p_t$. Left: RHIC, right: LHC. These
  results are independent of the choice of background-subtraction
  range in the heavy-ion events, since background subtraction does not
  enter into the matching criterion. 
  Here and in later figures, the label ``unquenched'' refers to
  the embedded $pp$ event; the background is always simulated
  including quenching.
}
\end{figure}

Let us start the presentation of our results with a brief discussion
of the efficiency of reconstructing jets in the medium. As explained
in Section~\ref{sec:simulation}, the jets in the medium are matched to
a ``bare'' hard jet when their common particle content accounts for at
least 50\% of the latter's transverse momentum.

The matching efficiencies we observe depend to some extent on the
details of the Monte Carlo used for the background so our intention is
just to illustrate the typical behaviour we observe and highlight that
these efficiencies tend to be large. We observe from
fig. \ref{fig:efficiency} that we successfully match at least 95\% of
the jets above $p_t \simeq 15$ GeV at RHIC, and at least 99\% of the
jets above $p_t \simeq 60$ GeV at the LHC. It is also interesting to
notice that the anti-$k_t$ algorithm performs best, likely as a
consequence of its `rigidity', namely the fact that anti-$k_t$ jets tend 
to have the same (circular) shape, independently of the soft-particles
that are present.

%
\subsection{Choice of background-estimation range}\label{sec:choice_range}

\begin{figure}[t]
\centerline{
\includegraphics[width=0.5\textwidth]{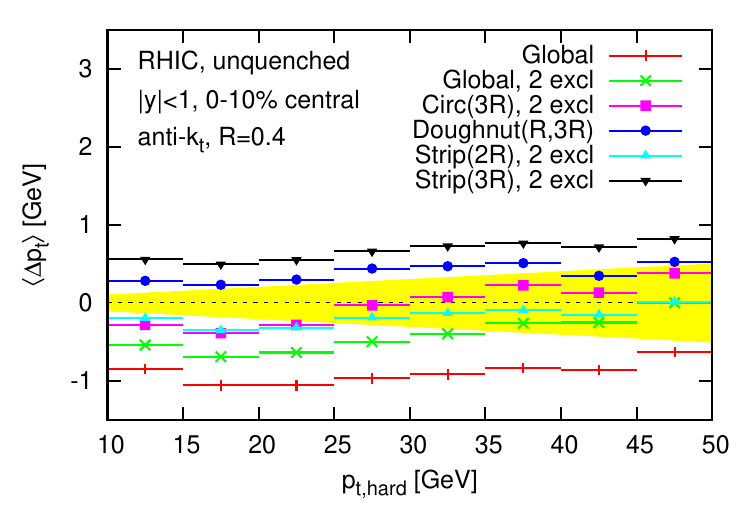}
\includegraphics[width=0.5\textwidth]{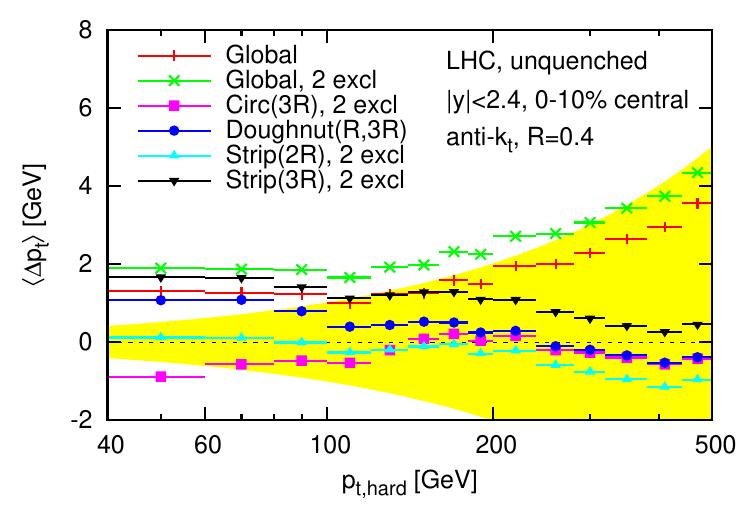}
}
\caption{\label{fig:ptshift_ranges} Effect of the choice of range on
  the average $p_t$ shift, $\Delta p_t$, as defined in
  \eq~(\ref{eq:deltapt}). Left: RHIC, right: LHC. In this figure and
  those that follow, the yellow band corresponds to 1\% of the $p_t$
  of the hard jet.}
\end{figure}

We now turn to the results concerning the measurement of the
background density and the reconstruction of the jet transverse
momentum. We first concentrate on the impact of the choice of a local
range and/or of the exclusion of the two hardest jets when determining
$\rho$.\footnote{Independently of the choice made for the full event,
  we always use a global range up to $|y| = y_{\max}$ for the
  determination of $\rho$ in the hard event, without exclusion of any
  jets. This ensures that the reference jet $p_t$ is always kept the
  same. The impact of subtraction in the hard event is in any case
  small, so the particular choice of range is not critical.}

In fig.~\ref{fig:ptshift_ranges} we show the average shift
$\avg{\Delta p_t}$ for the list of ranges mentioned in section
\ref{sec:subtraction}. The results presented here have been obtained
with the anti-$k_t$ algorithm with $R=0.4$, but the differences among
the various range choices have been seen to be similar with other jet
definitions.  The label ``2 excl'' means that the two hardest jets in
the event have been excluded from the estimation of the background. 
We have found that this improves the precision of the subtractions
whenever expected, \ie for all choices of range except the doughnut
range, where its central hole already acts similarly to the exclusion
of the hardest jets.
To keep the figure reasonably
readable, we have only explicitly shown the effect of removing the two
hardest jets for the global range.
The change of 0.4--0.6~GeV (both for RHIC and the LHC) is
in reasonable agreement with the analytic estimate of about 0.6 GeV
for RHIC and the LHC obtained from $\avg{\Delta p_t}=\pi R^2
\avg{\Delta\rho}$ with $\avg{\Delta\rho}$ calculated using
\eq~(\ref{eq:hard_median_offset}).
Note that at LHC the exclusion of the two hardest jets for the global
range appears to worsen the subtraction, however what is really
happening is that the removal of the two hardest jets exacerbates a
deficiency of the global range, namely the fact that its broad
rapidity coverage causes it to underestimate $\rho$, leading to a
positive net $\avg{\Delta\rho}$.

Other features that can be understood qualitatively include for
example the differences between the two strip and the global (2~excl)
range for RHIC: while the rapidity width of the global range lies
in between that of the two strip ranges, the global range gives a
lower $\avg{\Delta p_t}$ than both, corresponding to a larger $\rho$
estimate, which is reasonable because the global range is centred on
$y=0$, whereas the strip ranges are mostly centred at larger
rapidities where the background is lower.

The main result of the analysis of fig.~\ref{fig:ptshift_ranges} is
the observation that all choices of a local range lead to a small
residual $\Delta p_t$ offset: the background subtraction typically
leaves a $|\avg{\Delta p_t}| \lesssim 1\GeV$ at both RHIC and LHC, \ie
better than 1-2\% accuracy over much of the $p_t$ range of interest.
It is not clear, within this level of accuracy, if one range is to be
preferred to another, nor is it always easy to identify the precise
origins of the observed differences between various
ranges.\footnote{Furthermore, the differences may also be modified by
  jet-medium interactions. }
Another way of viewing this is that the observed differences between
the various choices give an estimate of the residual subtraction error
due to possible misestimation of $\rho$.
For our particular analysis, at RHIC this comment also applies to the
choice of the global range (with the exclusion of the two hardest jets
in the event). This is a consequence of the limited rapidity
acceptance, which effectively turns the global range into a local one,
a situation that does not hold for larger rapidity acceptances, as we
have seen for the LHC results.
In what follows we will use the Doughnut($R,3R$) choice, since it
provides a good compromise between simplicity and effectiveness.

%
%
%
%
%

%
\subsection{Choice of algorithm}\label{sec:choice_alg}

\begin{figure}
  \centering
  \includegraphics[width=0.9\textwidth]{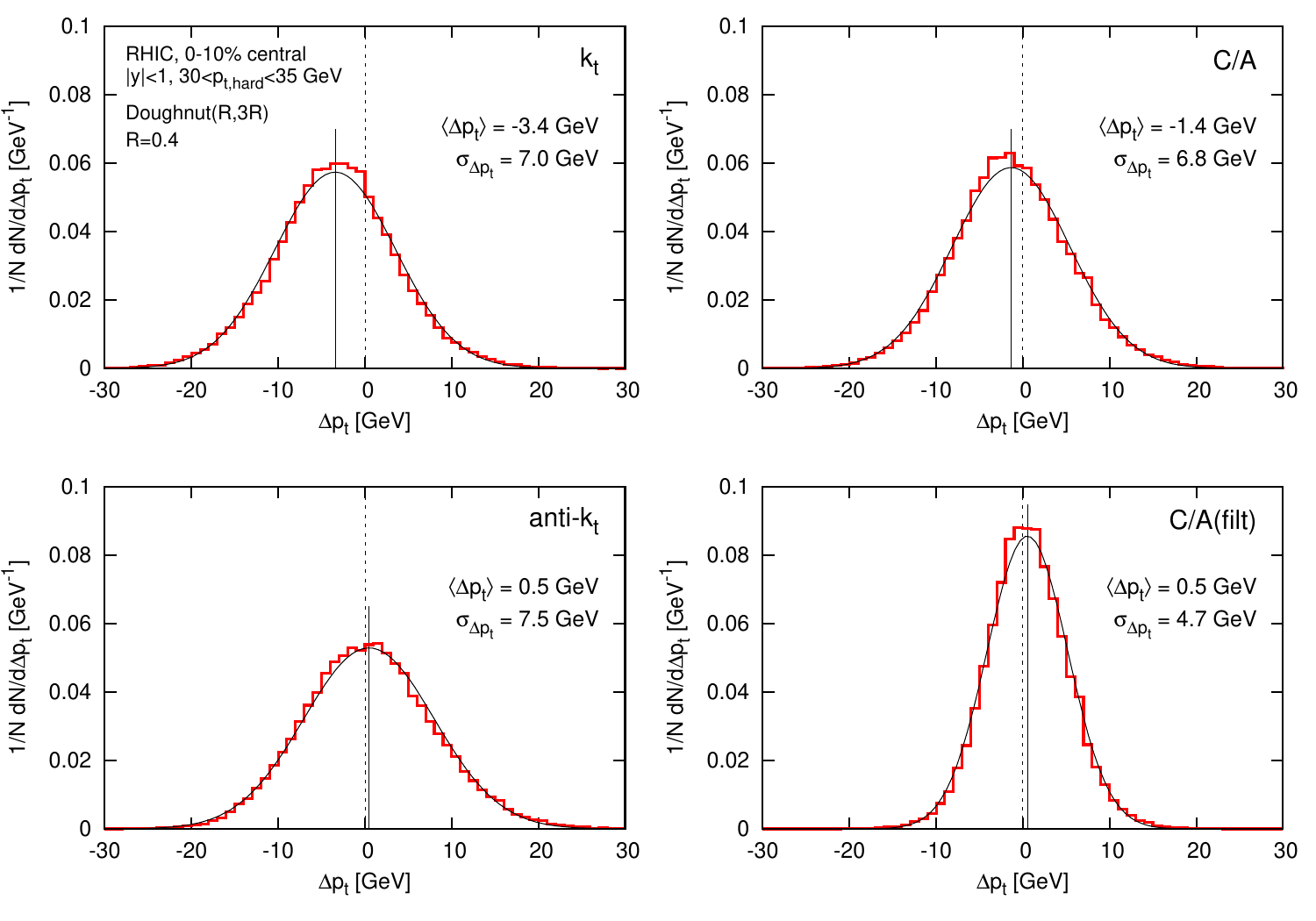}
  \caption{Distribution of $\Delta p_t$ (red histograms) for each of
    our 4 jet algorithms, together with a Gaussian (black curve) whose
    mean (solid vertical line) and dispersion are equal to
    $\avg{\Delta p_t}$ and $\sigma_{\Delta p_t}$ respectively.}
  \label{fig:deltapt-dist}
\end{figure}

The next potential systematic effect that we consider is the choice of
the jet algorithm used for the clustering.\footnote{Recall that in all
  cases, the $k_t$ algorithm with $R_\rho=0.5$ is used for the
  estimation of the background.}
Fig.~\ref{fig:deltapt-dist} shows the distribution of $\Delta p_t$ for each
of our four choices of jet algorithm, $k_t$, C/A, anti-$k_t$ and
C/A(filt), given for RHIC collisions and a specific bin of the hard
jets' transverse momenta, $30 < p_{t,\hard} < 35 \GeV$.
One sees significant differences between the different algorithms.
One also observes that Gaussians with mean and dispersion set equal to
$\avg{\Delta p_t}$ and $\sigma_{\Delta p_t}$ provide a fair
description of the full histograms. 
This validates our decision to concentrate on $\avg{\Delta p_t}$ and
$\sigma_{\Delta p_t}$ as quality measures. 
One should nevertheless be aware that in the region of high $|\Delta
p_t|$ there are deviations from perfect Gaussianity, which are more
visible if one replicates fig.~\ref{fig:deltapt-dist} with a
logarithmic vertical scale (not shown, for brevity).

\paragraph{Average $\Delta p_t$.}

\begin{figure}
\centerline{
\includegraphics[width=0.5\textwidth]{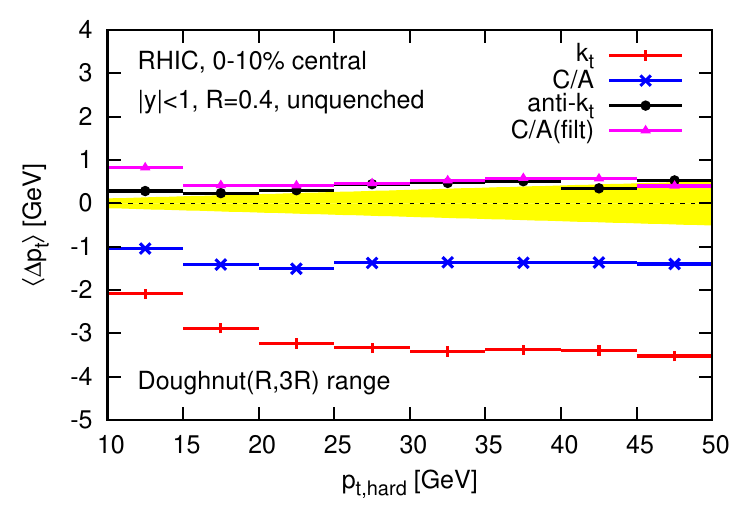}
\includegraphics[width=0.5\textwidth]{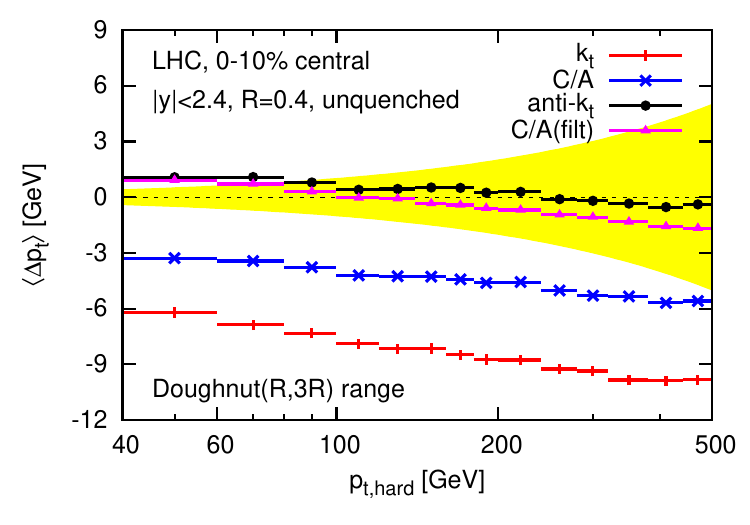}
}
\caption{\label{fig:ptshift} Average shift $\avg{\Delta p_t}$, as a
  function of $p_{t,\hard}$, shown for RHIC (left) and the LHC
  (right).}
\end{figure}

The first observable we analyse is the average $p_t$ shift. We show in
fig.~\ref{fig:ptshift} the $\langle\Delta p_t\rangle$ results for the
four algorithms listed in Section~\ref{sec:subtraction}, as a function
of $p_{t,\hard}$.
We use the doughnut range to estimate the background.
The first observation is that, while the anti-$k_t$ and
C/A(filt) algorithms have a small residual $\avg{\Delta p_t}$, the C/A
and $k_t$ algorithms display significant offsets.
The reason for the large offsets of $k_t$ and C/A is well understood,
related to the effect of {\em back-reaction}. This is the fact that
the addition of a soft background can alter the clustering of the
particles of the hard event: some of the constituents of a jet in the
hard event can be gained by or lost from the jet when clustering the
event with the additional background of the full event.
This happens, of course, on top of the simple background contamination
that adds background particles to the hard jet. Even if this latter
contamination is subtracted exactly, the reconstructed $p_t$ will
still differ from that of the original hard jet as a consequence of
the back-reaction.

\begin{figure}
\centerline{
\includegraphics[width=0.5\textwidth]{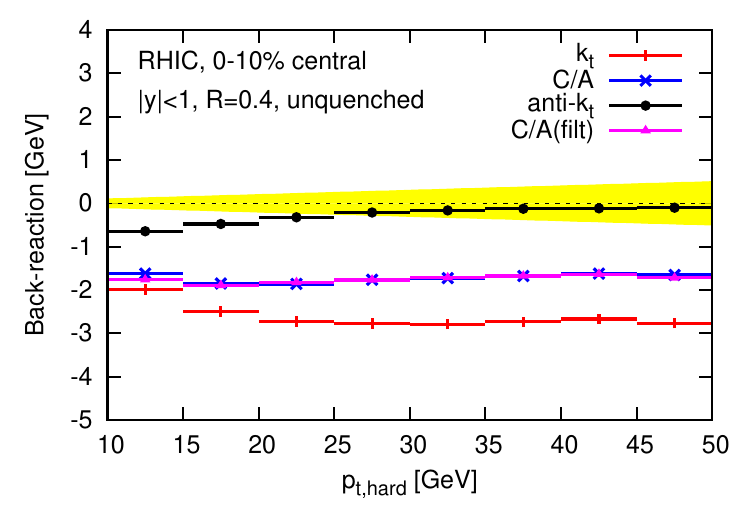}
\includegraphics[width=0.5\textwidth]{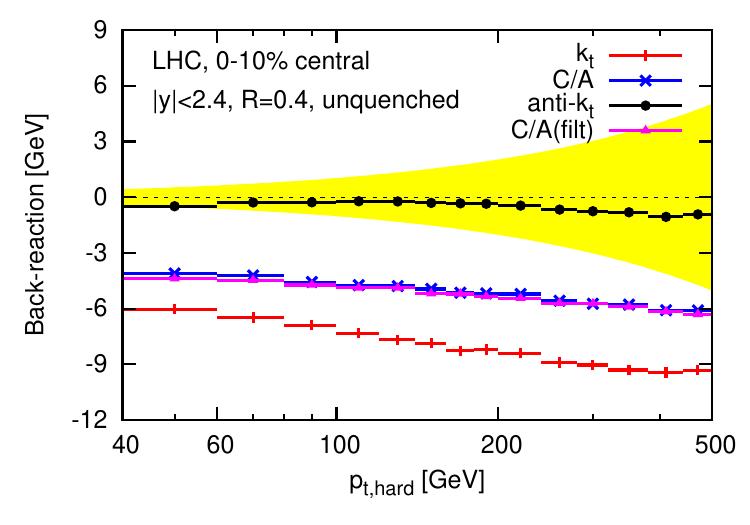}
}
\caption{\label{fig:ptshift_br} Contribution to $\avg{\Delta p_t}$ due
  to back-reaction. Note that these results are independent of the
  range used for estimating $\rho$ in the heavy-ion event. Left: RHIC,
  right: LHC.}
\end{figure}

The effect of the back-reaction can be studied in detail, since in
Monte Carlo simulations it is possible to identify which hard-event
constituents are present in a given jet before and after inclusion of
the background particles in the clustering.
The average $p_t$ shift due to back-reaction can be seen in
fig.~\ref{fig:ptshift_br} for the different jet algorithms.  As
expected (see Section~\ref{sec:areamed-analytic-back-reaction} or
Refs.~\cite{Cacciari:2008gp,Cacciari:2008gn}, it is largest for
  $k_t$, and smallest (almost zero, in fact) for anti-$k_t$.
By comparing fig.~\ref{fig:ptshift_br} and fig.~\ref{fig:ptshift} one
can readily explain the difference between the $\avg{\Delta p_t}$
offsets of the various algorithms in terms of their back-reaction.
The rigidity (and hence small back-reaction) of the anti-$k_t$ jets
manifestly gives almost bias-free reconstructed jets, while the large
back-reaction effects of the $k_t$ algorithm and, to a smaller extent,
of the C/A algorithm translates into a worse performance in terms of
average shift.
The $p_t$ dependence of the back-reaction is weak. This is expected
based on the interplay between the $\ln \ln p_t$ dependence found in
Section~\ref{sec:areamed-analytic-back-reaction}
\cite{Cacciari:2008gn} and the evolution with $p_t$ of the relative
fractions of quark and gluon jets.

The case of the C/A(filt) algorithm is more complex: its small net
offset, comparable to that of the anti-$k_t$ algorithm, appears to be
due to a fortuitous compensation between an under-subtraction of the
background and a negative back-reaction.
The negative back-reaction is very similar to that of C/A without
filtering, while the under-subtraction is related to the fact that the
selection of the hardest subjets introduces a bias towards
positive fluctuations of the background. This effect is discussed in
Section~\ref{app:filtbias}, where we obtain the following estimate
for the average $p_t$ shift (specifically for $R_{\rm filt}=R/2$):
\begin{equation}
  \label{eq:filt-bias-main-text}
\left\langle (\Delta p_t)_{\rm filt}\right\rangle
  \simeq 0.56\,R \sigma,
\end{equation}
yielding an average bias of 2~GeV for RHIC and 4.5~GeV at the LHC,
which are both in good agreement with the differences observed between
C/A with and without filtering in fig.~\ref{fig:ptshift}.
Note that while the bias in \eq~(\ref{eq:filt-bias-main-text}) is
proportional to $\sigma$, the back-reaction bias is instead mainly
proportional to $\rho$.
It is because of these different proportionalities that the
cancellation between the two effects should be considered as
fortuitous.
Since it also depends on the substructure of the jet, one may also
expect that the cancellation that we see here could break down in the
presence of quenching.

\paragraph{Dispersion of $\Delta p_t$.}

\begin{figure}
\centerline{
\includegraphics[width=0.5\textwidth]{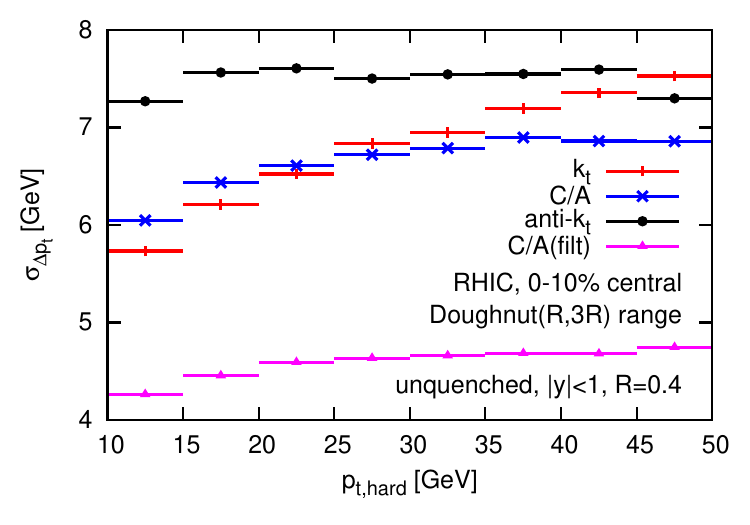}
\includegraphics[width=0.5\textwidth]{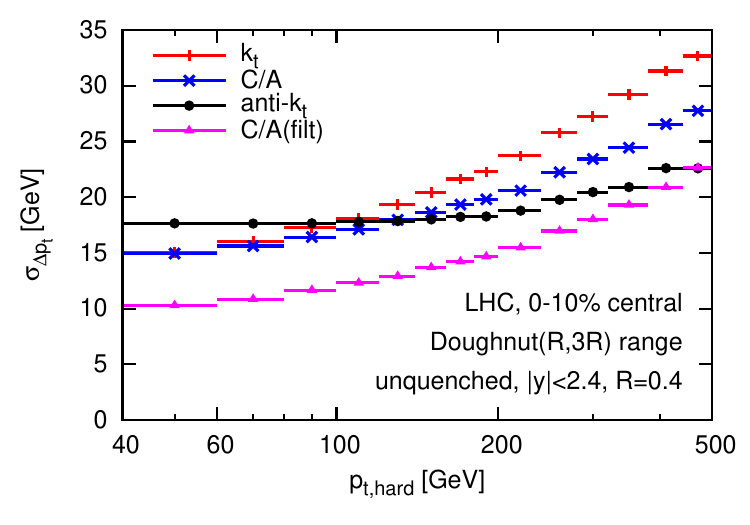}
}
\caption{\label{fig:dispersion} Dispersion $\sigma_{\Delta p_t}$. Left: RHIC, right: LHC.}
\end{figure}

Our results for the $\Delta p_t$ dispersion, $\sigma_{\Delta p_t}$,
are shown in fig.~\ref{fig:dispersion}, again using the doughnut
range. 
(Our conclusions are essentially independent of the particular choice
of range.)

We first discuss the case of RHIC kinematics. For $k_t$ and
anti-$k_t$, the observed dispersions are similar to the result of
$6.8\GeV$ quoted by STAR~\cite{Salur:2009vz,Ploskon:2009zd} (though
the number from STAR includes detector resolution effects, so that the
true physical $\sigma_{\Delta p_t}$ may actually be somewhat lower).
Of note, the advantage enjoyed
by anti-$k_t$ in terms of smallest $\avg{\Delta p_t}$ does not hold at
the level of the dispersion: C/A and $k_t$ tend to behave slightly
better at small transverse momentum. The algorithm which performs best
in terms of dispersion over all the $p_t$ range is now C/A with
filtering, for which the result is smaller than that of the other algorithms
by a factor of about $1/\sqrt{2}$.
This reduction factor can be explained because the dispersion
$\sigma_{\Delta p_t}$ is expected to be proportional to the
square-root of the jet area:
the C/A(filt) algorithm with $R_{\rm filt}=R/2$ and $n_{\rm filt}=2$
produces jets with an area of, roughly, half that obtained with
C/A;
hence the observed reduction of $\sigma_{\Delta p_t}$.

In the LHC setup, the conclusions are quite similar at the lowest
transverse momenta shown.
As $p_t$ increases, the dispersion of the anti-$k_t$ algorithm grows
slowly, while that of the others grows more rapidly, so that at the
highest $p_t$'s shown, the $k_t$ and C/A algorithms have noticeably
larger dispersions than anti-$k_t$, and C/A(filt) becomes similar to
anti-$k_t$.
The growth of the dispersions can be attributed to an increase of the
back-reaction dispersion.
The latter is dominated by rare occurrences, where a large fraction of
the jet's $p_t$ is gained or lost to back-reaction, hence the
noticeable $p_t$ dependence (\cnf
Section~\ref{sec:hidetails-back-reaction}).
An additional effect, especially for the $k_t$ algorithm, might come
from the anomalous dimension of the jet areas, \ie the growth with
$p_t$ of the average jet area.

Note that the dispersion has some limited dependence on the choice of
ghost area --- for example, reducing it from 0.01 to 0.0025 lowers
the dispersions by about $0.2-0.4\GeV$ at RHIC. This is discussed
further in Section~\ref{sec:hidetails-ghost-area-impact}.

%
\subsection{Centrality dependence}\label{sec:centrality}

\begin{figure}
\centerline{\includegraphics[width=0.6\textwidth]{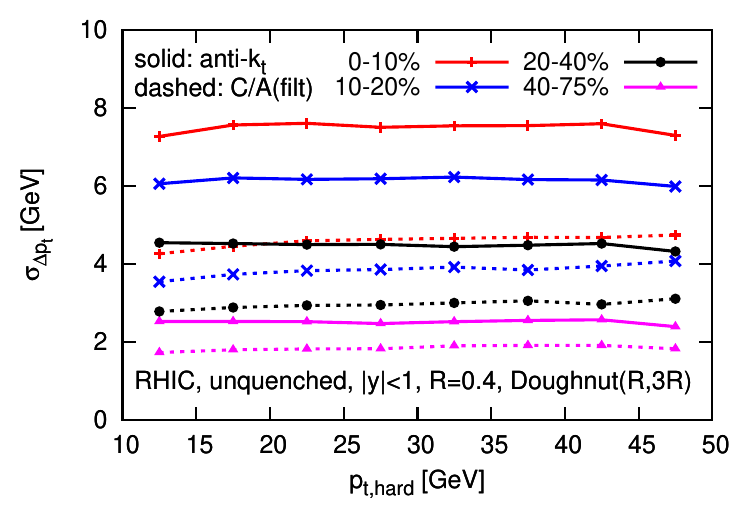}}
\caption{\label{fig:dispersion_centrality}
  $p_t$ dependence of the $\Delta p_t$ dispersion at RHIC,
  for different centrality classes.}
\end{figure}

So far, we have only considered central collisions. Since it is known
that non-central collisions give rise to elliptic
flow~\cite{Ollitrault:1992bk,Poskanzer:1998yz,Adler:2003kt,Adams:2003am,Alt:2003ab,Back:2004mh,Aamodt:2010cz,Chatrchyan:2012wg,ATLAS:2011ah},
one might worry that this leads to an extra source of background
fluctuations and/or non-uniformities, potentially spoiling the
subtraction picture discussed so far.
One can study this on azimuthally averaged jet samples (as we have
been doing so far) or as a function of the azimuthal angle, $\Delta
\phi$, between the jet and the reaction plane.
As above, we use Hydjet~v1.6, whose underlying HYDRO component
includes a simulation of elliptic flow~\cite{Lokhtin:2003ru}.

We have generated heavy-ion background events for RHIC in four
different centrality bins: 0-10\% (as above), 10-20\%, 20-40\% and
40-75\%, with $v_2$ values respectively of $1.7\%$, $3.3\%$, $5.0\%$
and $5.3\%$.%
\footnote{$v_2$ was determined as the average of $\cos2\phi$ over all
  all particles with $|\eta| < 1$ (excluding the additional hard $pp$
  event).  }

We first examine azimuthally averaged results, repeating the studies
of the previous sections for each of the centrality bins.
We find that the results for the average shift, $\avg{\Delta p_t}$,
are largely independent of centrality, as expected if the elliptic
flow effects disappear when averaged over $\phi$.
The results for the dispersion are shown in
fig.~\ref{fig:dispersion_centrality}.
We observe that the dispersion decreases with increasing
non-centrality. Even though one might expect adverse effects from
elliptic flow, the heavy-ion background decreases rapidly when one
moves from central to peripheral collisions, and this directly
translates into a decrease of $\sigma_{\Delta p_t}$.

The first conclusion from this centrality-dependence study is
therefore that the subtraction methods presented in this paper appear
to be applicable also for azimuthally averaged observables in
non-central collisions.

\begin{figure}
  \centerline{
    \includegraphics[width=0.5\linewidth]{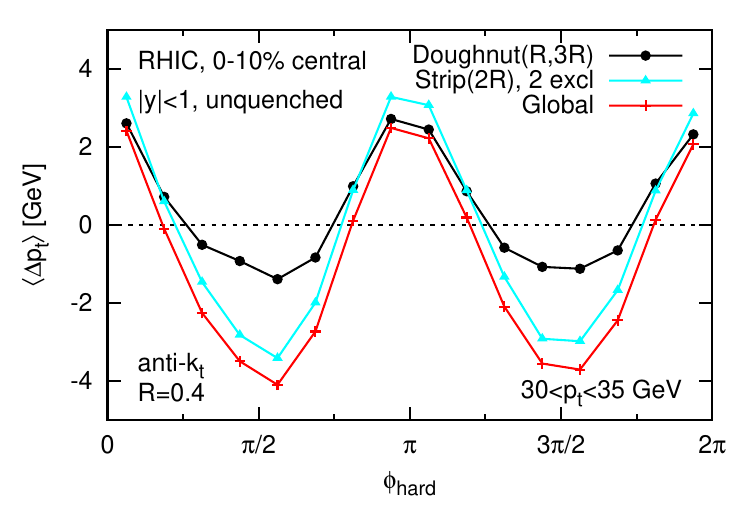}
    \includegraphics[width=0.5\linewidth]{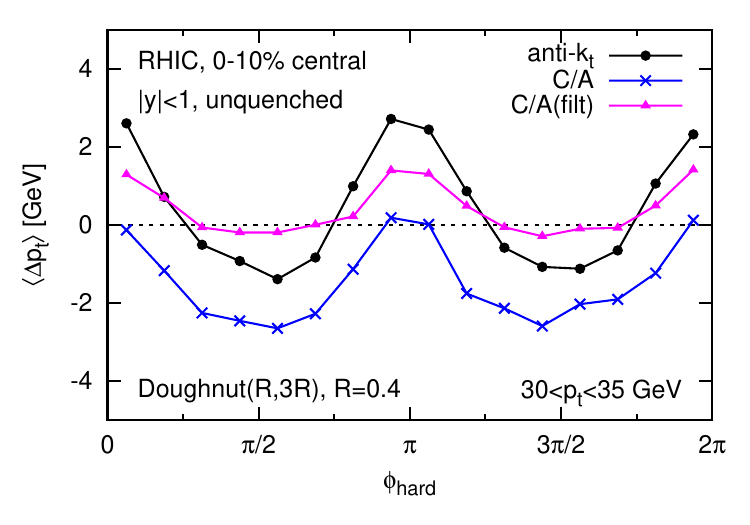}
  }
  \caption{\label{fig:ptshift-v-deltaphi}
    $\phi$ dependence of the $\Delta p_t$ shift at RHIC,
    for the $0-10\%$ centrality bin. Left: for three different ranges
    for the anti-$k_t$ algorithm;
    right: for three different jet algorithms for the doughnut($R,3R$) range.
    The absolute size of the $\phi$ dependence is similar for
    centralities up to $40\%$ and then decreases beyond.  }
\end{figure}

We next consider results as a function of $\Delta \phi$, which is
relevant if one wishes to examine the correlation between jet
quenching and the reaction plane.
An issue in real experimental studies is the determination of the
reaction plane, and the extent to which it is affected by the
presence of hard jets.
In the Hydjet simulations, this problem does not arise because the
reaction plane always corresponds to $\phi=0$.
Figure~\ref{fig:ptshift-v-deltaphi} (left) shows the average $\Delta p_t$ as
a function of $\Delta \phi$ for the anti-$k_t$ algorithm and several
different background-estimation ranges, for the $0-10\%$
centrality bin for RHIC.
The strip range shows significant $\Delta \phi$ dependence, which is
because a determination of $\rho$ averaged over all $\phi$ cannot
possibly account for the local $\phi$-dependence induced by the
elliptic flow.
Other ranges, such as the doughnut range, instead cover a more limited
region in $\phi$. They should therefore be able to provide information
on the $\phi$-dependence of the background.%
\footnote{At the expense of being more strongly affect by jet-medium
  interactions that could manifest themselves as broad enhancement of
  the energy flow in the vicinity of the jet.}
However, since their extent in $\phi$ tends to be significantly larger
than that of the jet, and $\rho$ varies relevantly over that extent,
some residual $\phi$ dependence remains in $\avg{\Delta p_t}$ after
subtraction.
The right-hand plot of figure~\ref{fig:ptshift-v-deltaphi} shows that
the effect is reduced with filtering, as is to be expected since its
initial background contamination is smaller.
The conclusion from this part of the study is that residual
$\phi$-dependent offsets may need to be corrected for explicitly in
any studies of jets and their correlations with the reaction plane.
The investigation of extensions to our background subtraction
procedure to address this issue will be the subject of future work.

%
\subsection{Quenching effects}\label{sec:quenching}

An important issue we wish to investigate is how the phenomenon of jet
quenching (\ie medium effects on parton fragmentation) may affect the
picture developed so far.  The precise nature of jet quenching beyond
its basic analytic properties (see
\eg~\cite{Baier:1996sk,Zakharov:1997uu}) is certainly hard to estimate
in detail, especially at the LHC, where experimental data from, say,
flow or particle spectra measurements are not yet available for the
tuning of the Monte Carlo simulations. Additionally, the
implementation of Monte-Carlo generators that incorporate the analytic
features of jet quenching models is currently a very active
field~\cite{Armesto:2009fj,jewel,martini}. We may therefore expect a
more robust and complete picture of jet quenching in the near future,
related to existing and forthcoming \PbPb data at the LHC.

In this section, we examine the robustness of our HI background
subtraction in the presence of (simulated) jet quenching.\footnote{
  Our focus here is therefore not the study of quenching itself, but
  merely how it may affect our subtraction procedure.}
For this purpose we have used two available models which allow one to
simulate quenched hard jets, Pyquen, which is used by Hydjet~v1.6, and
QPythia. Pyquen has been run with the parameters listed in footnote
\ref{foot:pyquen} for the LHC, and with $\texttt{T0}=0.5\GeV$,
$\texttt{tau0}=0.4$~fm and $\texttt{nf}=2$ for RHIC.\footnote{The
  parameters for RHIC are taken from Ref.~\cite{pyquen}.  The
  difference with the default parameters does not appear to be large
  for the purpose of our investigations and, in any case, a systematic
  study of quenching effects is not among the goals of this paper.}
For QPythia we have tested two options for the values of the transport
coefficient and the medium length ($\hat q = 3$~GeV$^2$/fm, $L = 5$~fm
and $\hat q = 1$~GeV$^2$/fm, $L = 6$~fm), with similar results.
No serious attempt has been made to tune the two codes with each other
or with the experimental data, beyond what is already suggested by the
code defaults: in the absence of strong experimental constraints
on the details of the quenching effects, this allows us to verify the
robustness of our results for a range of conditions.

As in sections \ref{sec:choice_range} and \ref{sec:choice_alg}, we
have embedded the hard Pyquen or QPythia events in a Hydjet~v1.6
background and tested the effectiveness of the background subtraction
for different choices of algorithm.\footnote{The effect of the choice
  of range remains as in section \ref{sec:choice_range} for the
  unquenched case. We will therefore keep employing the doughnut
  range.}
We shall restrict our attention to the anti-$k_t$ and C/A(filt)
algorithms, as they appear to be the optimal choices from our analysis
so far.

\begin{figure}
\centerline{
\includegraphics[width=0.5\textwidth]{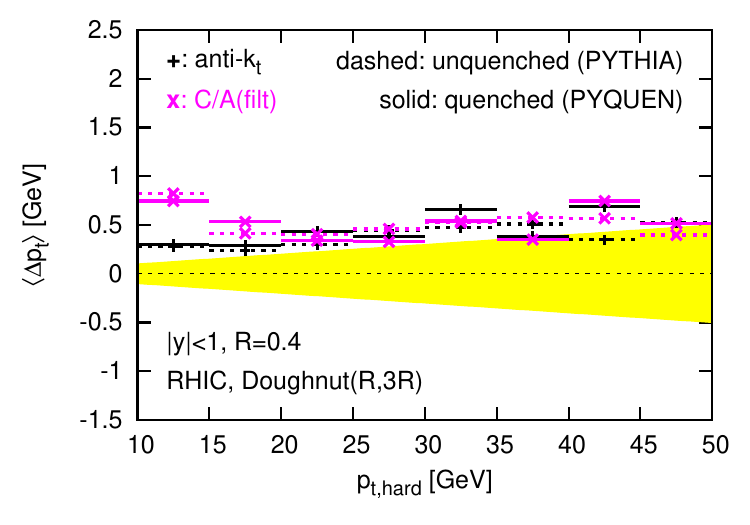}
\includegraphics[width=0.5\textwidth]{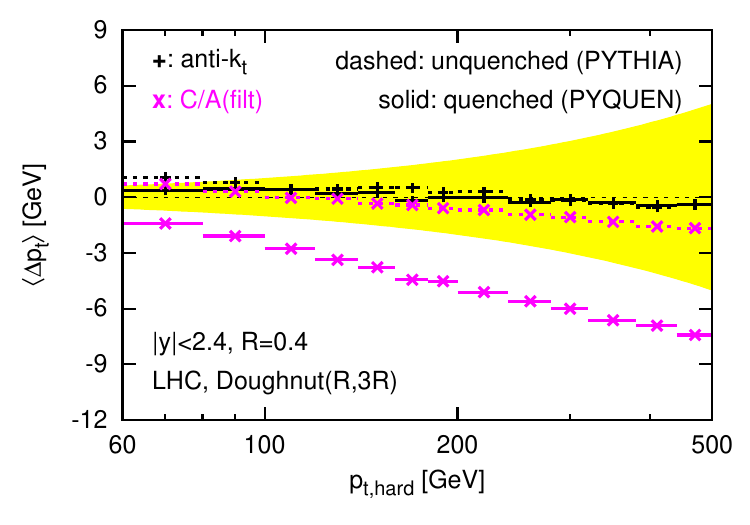}
}
\caption{\label{fig:ptshift_quenched}
  Average $p_t$ shift for background-subtracted jets with the
  anti-$k_t$ and C/A(filt) jet algorithms. The dashed lines correspond
  to unquenched hard jets (Pythia) and the solid ones to quenched hard
  jets (Pyquen). Results are shown for RHIC kinematics on the left
  plot and for the LHC on the right one.}
\end{figure}

We have found that the jet-matching efficiencies are still high, with
essentially no changes at RHIC, and at LHC a doubling of the (small)
inefficiencies that we saw in fig.~\ref{fig:efficiency}(right).
The dispersions $\sigma_{\Delta p_t}$ are also not significantly
affected within our sample of jet-quenching simulations.
We therefore concentrate
on the $\avg{\Delta p_t}$ offset, which is plotted in
fig.~\ref{fig:ptshift_quenched} for Pyquen.
The results are shown for both RHIC and the LHC.
In the case of RHIC, and for the whole $p_t$ range up to about 50 GeV,
quenching can be seen not to significantly affect the subtraction
offset $\langle \Delta p_t\rangle$ (within the usual uncertainty
related to the choice of range, which was shown in
fig.~\ref{fig:ptshift_ranges}). In the LHC case, instead, while the
shift obtained using the anti-$k_t$ algorithm is largely similar to
the unquenched case, the C/A(filt) algorithm performance can be seen
to deteriorate slightly when quenching is turned on, all the more so
at very large transverse momentum.
The $k_t$ and the C/A algorithms are not shown for clarity, but they
share the behaviour of C/A(filt).  This
deterioration of the quality of the subtraction can be traced back to
an increased back-reaction compared to the unquenched jets.
Anti-$k_t$ jets do not suffer from this effect as a consequence of the
usual rigidity of this algorithm.
In the case of C/A(filt), one should nevertheless emphasise that an
error of (at most) 10 GeV on the reconstruction of a 500 GeV jet is
still only a 2\% effect. This is modest, both relative to the likely
experimental precision and to the expected effect of quenching on the
overall jet $p_t$, predicted by Pyquen to be at the level of $10\%$ at
this $p_t$.

Though for brevity we have not explicitly shown them, the results
with QPythia are very similar.

Before closing this section, we reiterate that we have only
investigated simple models for quenching and that our results are
meant just to give a first estimate of the effects that one might have
to deal with in the case of quenched jets. One will be able to address
this question more extensively as more data is collected at the LHC
and the dynamics of jet quenching in QCD is better understood, together
with the development of better ``quenched'' Monte Carlo generators.

\section{Further details and analytic estimates}\label{sec:hi-analytic-estimates}

\subsection{Back reaction contribution to the dispersion}\label{sec:hidetails-back-reaction}

In the context of heavy-ion collisions, where the residual
resolution-degradation effects on the jets can be relatively large, it
is interesting to look a bit deeper in the various sources that
contribute to the dispersion. The natural candidates are the UE
fluctuations themselves and residual uncertainties on the estimation
of $\rho$ (which have been made small by choosing a large enough
range). Here we discuss two additional sources: the dispersion in the
back-reaction which starts kicking in at large $p_t$ and the
uncertainty in our calculation of the jet area.

\begin{figure}
  \centering
  \includegraphics[width=0.5\textwidth]{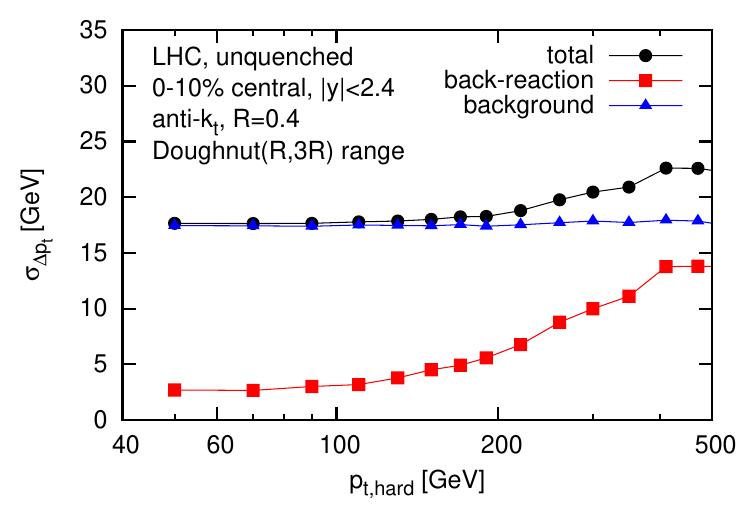}\hfill
  \includegraphics[width=0.5\textwidth]{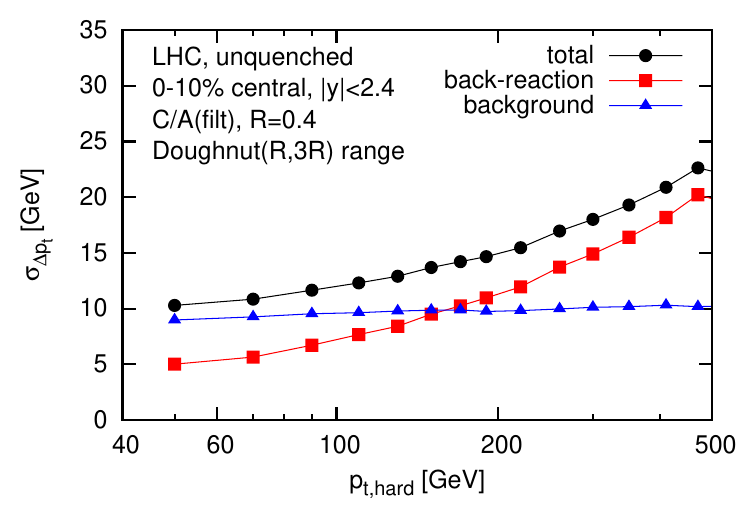}
  \caption{The decomposition of the dispersion into back-reaction and
    ``background'' components (including misestimation of
    $\rho$). The left-hand plot is for the anti-$k_t$ algorithm and
    the right-hand one for C/A(filt). Both correspond to LHC
    collisions at $\sqrt{s_{NN}} = 5.5\TeV$.}
  \label{fig:BR-v-full-dispersion}
\end{figure}

We stated, in section~\ref{sec:choice_alg}, that the increase of the
dispersion at high $p_t$ seen in fig.~\ref{fig:dispersion} was mainly
due to back-reaction. This is made explicit in
fig.~\ref{fig:BR-v-full-dispersion}, which decomposes the dispersion
into its two components: that associated with the back-reaction,
$\sigma_{\Delta p_t}^\mathrm{BR}$ and that associated with background
fluctuations and misestimation of $\rho$ (defined as $[\sigma_{\Delta
  p_t}^2 - (\sigma_{\Delta p_t}^\mathrm{BR})^2]^{\frac12}$).
As already observed in $pp$ collisions (see
Section~\ref{sec:aside-back-reaction}), one sees that the
background-fluctuation component is essentially independent of $p_t$,
while the back-reaction dispersion has a noticeable $p_t$ dependence.
The flatness of the residual, non-back-reaction, component is again
observed and is expected since the anomalous dimension of the jet area
is zero for anti-$k_t$ and small for C/A (with or without filtering),
and in any case leads to a weak scaling with $p_t$, as $\ln \ln p_t$.
Furthermore, there is roughly a factor of $1/\sqrt{2}$ between
anti-$k_t$ and C/A(filt), as expected based on a proportionality of
the dispersion to the square-root of the jet area.


\begin{figure}
  \centering
  \includegraphics[width=0.6\textwidth]{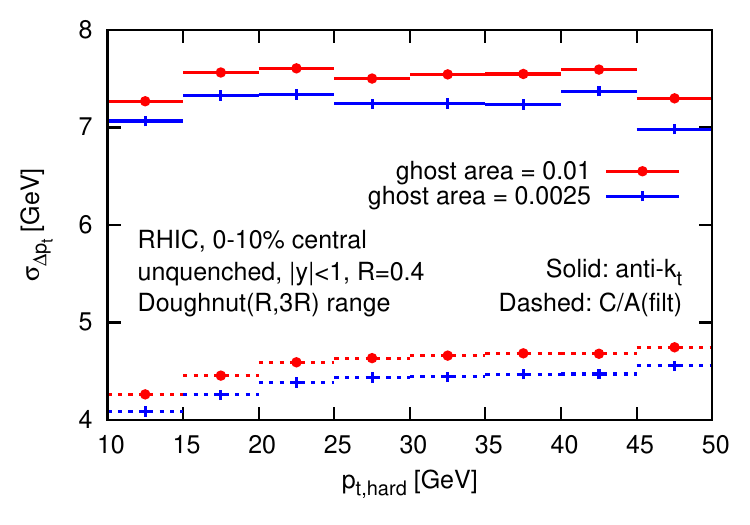}
  \caption{Dispersion, $\sigma_{\Delta p_t}$, for RHIC, as a function
    of the jet $p_t$, with two different choices for the ghost area,
    $0.01$ and $0.0025$. }
  \label{fig:ghost-area-impact}
\end{figure}

\subsection{Quality of area determination}\label{sec:hidetails-ghost-area-impact}

One further source of $\Delta p_t$ fluctuations can come from
imperfect estimation of the area of the jets. 
We recall that throughout this article we have used soft ghosts, each
with area of $0.01$, in order to establish the jet area. That implies
a corresponding finite resolution on the jet area 
and related poor estimation of the exact
edges of the jets, which can have an impact on the amount of
background that one subtracts from each jet, and, consequently, on
the final dispersion.
It is therefore interesting to see,
figure~\ref{fig:ghost-area-impact}, that the dispersion
$\sigma_{\Delta p_t}$ is reduced by about $0.2-0.4\GeV$ if one lowers
the ghost area to $0.0025$.

While this does not affect any of the conclusions of our paper, it does
suggest that for a full experimental analysis there are benefits to be
had from using a ghost area that is smaller than the default FastJet
setting of $0.01$.
These observations are behind one the points made in our final
recommendations, Section~\ref{sec:areamed-practical}.


\subsection{Relative importance of average shift and dispersion}
\label{eq:shift-dispersion-on-spectrum}

To close this section, we examine the relative importance of the
average shift and its
dispersion, taking the illustrative example of their impact on the
inclusive jet cross-section as a function of $p_t$. 
This follows our simple description introduced in
Section~\ref{sec:description-convolution} where we have seen that the
true $p_t$ spectrum decays exponentially \ie
\begin{equation} 
  \label{eq:hard-spectrum}
  \frac{d\sigma^{pp}}{dp_t} = \frac{\sigma_0}{\kappa}\,e^{-p_t/\kappa}.
\end{equation}
The reconstructed spectrum, after embedding in a heavy-ion background and applying
subtraction will be multiplied by a factor
\begin{equation}
  \label{eq:smeared-hard-spectrum}
  \exp\left(\frac{\avg{\Delta p_t}}{\kappa}+\frac{\sigma_{\Delta p_t}^2}{2\kappa^2}\right),
\end{equation}
as obtained in \eq~\eqref{eq:distrib-effect-convoluted}. For a given
reconstructed $p_t$, the most likely original true transverse momentum
is given by \eq~\eqref{eq:distrib-effect-most-probably} and reads
\begin{equation}
  \label{eq:most-likely}
  \text{most likely}\; \pthard \simeq 
   \ptfullsub - \avg{\Delta p_t} - \frac{\sigma_{\Delta p_t}^2}{\kappa}\,,
\end{equation}
where we have neglected the small impact of subtraction on the $pp$
jets.

To illustrate these effects quantitatively, let us first take the
example of RHIC, where between 10 and 60 GeV, the cross-section is
well approximated by \eq~(\ref{eq:hard-spectrum}) with $\kappa=3.3$
GeV. Both the anti-$k_t$ and C/A(filt) have $\avg{\Delta
  p_t}\simeq 0$, leaving only the dispersion effect. In the case of
the anti-$k_t$ (respectively C/A(filt)) algorithm, we see from
fig.~\ref{fig:dispersion} that $\sigma_{\Delta p_t}\simeq 7.5 \GeV$
($4.8\GeV$), which gives a multiplicative factor of about 12 (3).
For a given reconstructed $p_t$, the most likely true $p_t$ is about
$17\GeV$ ($7\GeV$) smaller.
In comparison, for C/A ($k_t$), with $\avg{\Delta p_t}\simeq -1.5\GeV$
($-3.5\GeV$) (fig.~\ref{fig:ptshift}) and $\sigma_{\Delta p_t}\simeq
6.5 \GeV$ (similar for $k_t$) there is a partial compensation between
factors of $0.64$ ($0.35$) and $6.7$ coming respectively from the
shift and dispersion, yielding an overall factor of about $4$ ($2.3$),
while the most likely true $p_t$ is about $14\GeV$ ($12\GeV$) smaller
than the reconstructed $p_t$.

At the LHC ($\sqrt{s_{NN}} = 5.5\TeV$), \eq~(\ref{eq:hard-spectrum})
is a less accurate approximation. Nevertheless, for $p_t \sim
100-150\GeV$, it is not too unreasonable to take $\kappa=20\GeV$
and examine the consequences. For anti-$k_t$ (respectively C/A(filt)),
we have $\avg{\Delta p_t} \simeq 0$ (also for C/A(filt)) and
$\sigma_{\Delta p_t} \simeq 18\GeV$ ($13\GeV$), giving a
multiplicative factor of $1.5$ ($1.2$), \ie far smaller corrections
than at RHIC.
For a given reconstructed $p_t$, the most likely true $p_t$ is about
$16\GeV$ ($8\GeV$) smaller, rather similar to the values we found at
RHIC (though smaller in relative terms, since the $p_t$'s are higher),
with the increase in $\sigma$ being compensated by the increase in
$\kappa$.

In the LHC case, it is also worth commenting on the results for the
$k_t$ algorithm, since this is what was used in the original work from
Ref.~\cite{Cacciari:2007fd}: we have $\avg{\Delta p_t} \simeq -8\GeV$ and
$\sigma_{\Delta p_t} \simeq 18\GeV$, giving a multiplicative factor of
$1.05$, which is consistent with the near perfect agreement that was
seen there between the $pp$ and subtracted $AA$ spectra.
That agreement does not however imply perfect reconstruction, since
the most likely $\pthard$ is about $7\GeV$ lower than $\ptfullsub$.

Though the above numbers give an idea of the relative difficulties of
using different algorithms at RHIC and LHC, 
experimentally what matters most will be the systematic errors on the
correction factors (for example due to poorly understood non-Gaussian
tails of the $\Delta p_t$ distribution). Note also that a compensation
between shift and dispersion factors, as happens for example with the
C/A algorithm, is unlikely to reduce the overall systematic errors.

\subsection{Fluctuations in extracted $\rho$}\label{app:minrange-fluct}

In section \ref{sec:subtraction}, we gave an estimate of the minimum
size of a range one should require for determining $\rho$, given a
requirement that fluctuations in the determination of $\rho$ should be
moderate.
We give the details of the computation in this Section.

We start from the fact that the error made on the estimation of the
background density $\rho$ will translate into an increase of the
dispersion $\sigma_{\Delta p_t}$: on one hand, the dominant
contribution to $\sigma_{\Delta p_t}$ comes from the intra-event
fluctuations of the background \ie is of order
$\sigma \sqrt{A_{\rm jet}}$ with $A_{\rm jet}$ the jet area; on the
other hand, the dispersion $S_{\Delta \rho}$ of the misestimation of
$\rho$ leads to an additional dispersion on the reconstructed jet
$p_t$ of $S_{\Delta \rho} A_{\rm jet}$. As we will show in the next
Chapter, $S_{\Delta \rho}$ can be estimated analytically, see
\eq~(\ref{eq:rhoest-purepu-dispersion}). Adding these two sources of
dispersion in quadrature and using
\eqref{eq:rhoest-purepu-dispersion}, with $A_{\rm tot}=A_{\cal R}$ the
area of the range under consideration, we get
\begin{equation}
\sigma_{\Delta p_t} \simeq \sigma \left(A_{\rm jet} 
  + \frac{\pi A_{\rm jet}^2}{2 A_{\cal R}}\right)^{\frac12}.\label{eq:1}
\end{equation}
If we ask \eg that the contribution to the total dispersion coming
from the misestimation of the background be no more than a fraction
$\epsilon$ of the total $\sigma_{\Delta p_t}$, then we obtain the
requirement
\begin{equation}
  \label{eq:AR-min}
  A_{\cal R} \gtrsim  A_{\min} \simeq \frac{\pi}{4} \frac{A_{\rm jet}}{\epsilon}\,.
\end{equation}
For anti-$k_t$ jets of radius $R$, with $A_{\rm jet} \simeq \pi R^2$,
this translates to
\begin{equation}
  \label{eq:AR-min-in-terms-of-R}
  A_{\cal R} \gtrsim \frac{\pi^2}{4} \frac{R^2}{\epsilon} \simeq 25 R^2\,,
\end{equation}
where the numerical result has been given for $\epsilon = 0.1$. For
$R=0.4$, it becomes $A_{\cal R} \gtrsim 4$.

We can also cast this result in terms of the number of jets that must
be present in ${\cal R}$. Assuming that the jets used to estimate
$\rho$ have a mean area of $0.55 \pi R_\rho^2$,\footnote{This is the
  typical area one would obtain using a (strongly recommended) jet
  definition like the $k_t$ or C/A algorithms for the background
  estimation, see Section~\ref{sec:areamed-areaanalytics-active}.} we
find a minimal number of jets\,,
\begin{equation}
  \label{eq:njets-min}
  n_{\min} \simeq \frac{A_{\min}}{0.55 \pi R_{\rho}^2} \simeq
  \frac{1.4}{\epsilon}\frac{R^2}{R_\rho^2}\,,
\end{equation}
where, as before, we have taken $A_{\jet}\simeq \pi R^2$.
Taking the numbers quoted above and $R_\rho = 0.5$, as used in the
main body of the article, this gives $n_{\min}\simeq 9$.

\subsection{Hard-jet bias in extracted $\rho$}\label{app:minrange-bias}

As we will show in our detailed analytic studies, see
Section~\ref{sec:analytic-rhoest-hard}, the presence of hard jets and
initial-state radiation leads to a bias in the extraction of $\rho$ of
\begin{equation}
  \label{eq:rho-bias}
  \avg{\Delta \rho} \simeq \sigma R_\rho \sqrt{\frac{\pi c_J}{2}}
  \frac{\avg{n_h}}{A_{\cal R}}\,
\end{equation}
where $c_J\simeq 2$ is a numerical constant and $\avg{n_h}$ is the
average number of ``hard'' jets (those above the scale of the
background fluctuations, including initial-state radiation);
$\avg{n_h}$ is given by
\begin{equation}
  \label{eq:nhard}
  \frac{\avg{n_h}}{A_{\cal R}} \simeq \frac{n_b}{A_{\cal R}} +
    \frac{C_i}{\pi^2}\frac{L}{b_0}\,,
    \qquad\quad
    L = \ln \frac{\as\left(\sqrt{c_J} \sigma R_\rho\right)}{\as(p_t)}\,,
\end{equation}
where $n_b$ is the number of ``Born'' partons from the underlying $2\to 2$
scattering that enter the region $\cal R$, while $b_0=(11 C_A -
2n_f)/(12 \pi)$ is the first coefficient of the QCD $\beta$-function.
Contrarily to the original case of UE studies in $pp$ collisions
\cite{Cacciari:2009dp}, in the case of a large background like the UE
in $AA$ collisions, one can safely ignore the impact of the second
term in (\ref{eq:nhard}).
We thus arrive at the result 
\begin{equation}
  \label{eq:rho-bias-result}
  \avg{\Delta \rho} \simeq \sigma R_\rho \sqrt{\frac{\pi c_J}{2}}
  \frac{n_b}{A_{\cal R}} \simeq 1.8 \,\sigma R_\rho
  \frac{n_b}{A_{\cal R}}\,.
\end{equation}

Note that the presence of hard jets and initial-state radiation also
affects the fluctuations in the misestimation of $\rho$ and this
should in principle have been included in the estimates of
Section~\ref{app:minrange-fluct}. 
However, while the effect is not completely negligible, to within the
accuracy that is relevant for us (a few tens of percent in the
estimation of a minimal $A_{\cal R}$) it does not significantly alter
the picture outlined there.

\subsection{Subtraction bias due to filtering}\label{app:filtbias}

We have seen from fig.~\ref{fig:ptshift} in section
\ref{sec:choice_alg} that the subtraction differs when we use the C/A
algorithm with and without filtering. Since this difference is not due
to back-reaction (see fig.~\ref{fig:ptshift_br}), it has to be due to
the subtraction itself.

The difference comes from a bias introduced by the selection
of the two hardest subjets during filtering. The dominant contribution
comes when only one subjet, that we shall assume harder than all the
others, contains the hard radiation, all the other subjets being pure
background. In that case, the selection of the hardest of these
pure-background subjets as the second subjet to be kept tends to
pick positive fluctuations of the background. This in turn results in
a positive offset compared to pure C/A clustering, as observed in
section \ref{sec:choice_alg}.

To compute the effect analytically, let us thus assume that we have
one hard subjet and $n_{\rm bkg}$ pure-background subjets of area
${\cal A}_g = 0.55 \pi R_{\rm filt}^2$ (see
Section.~\ref{sec:areamed-areaanalytics-active}). After subtraction,
the momentum of each of the pure-background subjet can be approximated
as having a Gaussian distribution of average zero and dispersion
$\sigma \sqrt{{\cal A}_g}$.
Assuming that the ``hard'' subjet's transverse momentum remains larger
than that of all the background jets, the 2 subjets that will be kept
by the filter are the hard subjet (subtracted) and the hardest of
all the subtracted background jets. The momentum distribution of the
latter is given by the maximum of the $n_{\rm bkg}$ Gaussian
distributions.\footnote{%
  Within \fastjet's filtering tools, when the subtracted transverse
  momentum of a subjet is negative, the subjet is assumed to be pure
  noise and so discarded.
  This means that the momentum distribution of the hardest subtracted
  background jet is really given by the distribution of the maximum of
  the $n$ Gaussian-distributed random numbers, but with the result
  replaced by zero if all of them are negative.
  In the calculations here we ignore this subtlety, since we will have
  $n=3$ and only $1/8^\mathrm{th}$ of the time are three Gaussian-distributed
  random numbers all negative.}
We are only interested here in computing the average bias introduced
by the filtering procedure, which is then given by
\[
\left\langle (\Delta p_t)_{\rm filt}\right\rangle
 \simeq \int \prod_{k=1}^{n_{\rm bkg}} \left(dp_{t,k}\,\frac{1}{\sqrt{2\pi{\cal
         A}_g}\sigma}e^{-\frac{p_{t,k}^2}{2{\cal A}_g
       \sigma^2}}\right)\:{\rm max}\left(p_{t,1},\dots,p_{t,n_{\rm bkg}}\right)
\]
For the typical case $R_{\rm filt}=R/2$ and $n_{\rm bkg}=3$, one finds
\begin{equation}
\left\langle (\Delta p_t)_{\rm filt}\right\rangle
  \simeq \frac{3\sqrt{{\cal A}_g}\sigma}{2 \sqrt{\pi}} 
  \simeq 0.56\,R\sigma.
\end{equation}
If we insert in that expression the typical values for the
fluctuations quoted in section \ref{sec:results} and $R=0.4$, we find
average biases of 2~GeV for RHIC and 4.5~GeV at the LHC, which are
in good agreement with the differences observed between C/A with and
without filtering in fig.~\ref{fig:ptshift}.

\section{The issue of fakes}\label{sec:fakes}

While the goal of this paper is not to discuss the issue of
``fake-jets'' in detail, it is a question that has been the subject of
substantial debate recently (see for example
\cite{Lai:2008zp,JacobsPragueTalk}). Here, therefore, we wish to
devote a few words to it and discuss how it relates to our
background-subtraction results so far.

In a picture in which the soft background and the hard jets are
independent of each other, one way of thinking about a fake jet is
that it is a reconstructed jet (with significant $p_t$) that is due
not to the presence of an actual hard jet, but rather due to an
upwards fluctuation of the soft background.
The difficulty with this definition is that there is no
uniquely-defined separation between ``hard'' jets and soft background.
This can be illustrated with the example of how Hydjet
simulates RHIC collisions: one event typically consists of a soft
HYDRO background supplemented with $\sim 60$ $pp$ collisions, each simulated
with a minimum $p_t$ cut of $2.6\GeV$ on the $2\to2$ scattering.
To some approximation, the properties of the full heavy-ion events
remain relatively unchanged if one modifies the number of $pp$
collisions and corresponding $p_t$ cut and also retunes the soft
background. 
The fact that this changes the number of hard jets provides one
illustration of the issue that the soft/hard separation is
ill-defined.
Additionally, while there are $\sim 60$ semi-hard $pp$ collisions
($\sim 120$ mostly central semi-hard jets\footnote{Specifically,
  keeping in mind the Hydjet simulation, one can cluster each $pp$ event
  separately to obtain a long list of $pp$ jets from all the separate
  hard events.}) in an event, there is only space within (say) the
acceptance of RHIC for $\order{40}$ jets. Thus there is essentially no
region in an event which does not have a semi-hard jet.
From this point of view, every reconstructed jet corresponds to a
\mbox{(semi-)}hard $pp$ jet and there are no fake jets at all.

\subsection{Inclusive analyses}
\label{sec:fakes:incl}

For inclusive analyses, such as a measurement of the inclusive jet
spectrum, this last point is particularly relevant, because every jet
in the event contributes to the measurement. 
Then, the issue of fakes can be viewed as one of unfolding. 
In that respect it becomes instructive, for a given bin of the 
reconstructed heavy-ion $p_t$, to ask what the corresponding matched
$pp$ jet transverse momenta were.
Specifically, we define a quantity $O(\ptfullsub, p_{t}^{pp})$,
the distribution of the $pp$ ``origin'', $p_{t}^{pp}$, of a heavy-ion
jet with subtracted transverse momentum $\ptfullsub$.
If the origin $O(\ptfullsub, p_{t}^{pp})$ is dominated by a
region of $p_{t}^{pp}$ of the same order as $\ptfullsub$, then
that tells us that the jets being reconstructed are truly hard.
If, on the other hand, it is dominated by $p_{t}^{pp}$ near zero, then
that is a sign that apparently hard heavy-ion jets are mostly due to
upwards fluctuations of the background superimposed on low-$p_t$ $pp$
jets, making the unfolding more delicate.

\begin{figure}
  \includegraphics[width=\textwidth]{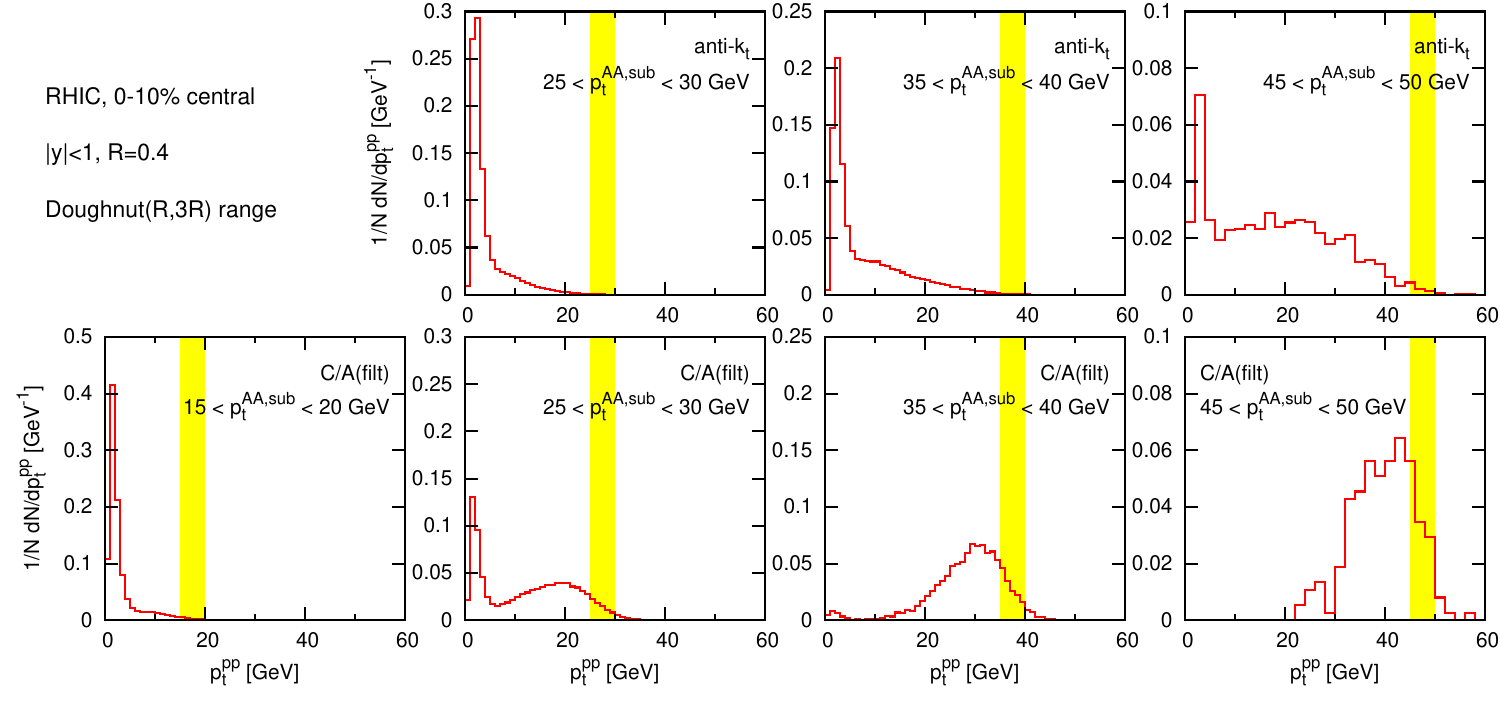}
  \caption{The $p_t$ distribution of the $pp$ jet corresponding to a
    given bin of reconstructed heavy-ion jet $\ptfullsub$ at RHIC, \ie 
    $O(\ptfullsub, p_{t}^{pp})$ as a function of $p_{t}^{pp}$
    for a given bin of $\ptfullsub$.
    The upper row is for the anti-$k_t$ algorithm, while the lower row
    is for C/A(filt). 
    Each column corresponds to a different $\ptfullsub$ bin, as
    indicated by the vertical band in each plot. 
    Cases in which the histogram is broad or peaked near $0$ are
    indicative of the need for special care in the unfolding
    procedure. 
    These plots were generated using approximately 90 million events.
    Each plot has been normalised to the number of events in the
    corresponding $\ptfullsub$ bin.
  }
  \label{fig:origin}
\end{figure}

\begin{figure}
  \includegraphics[width=\textwidth]{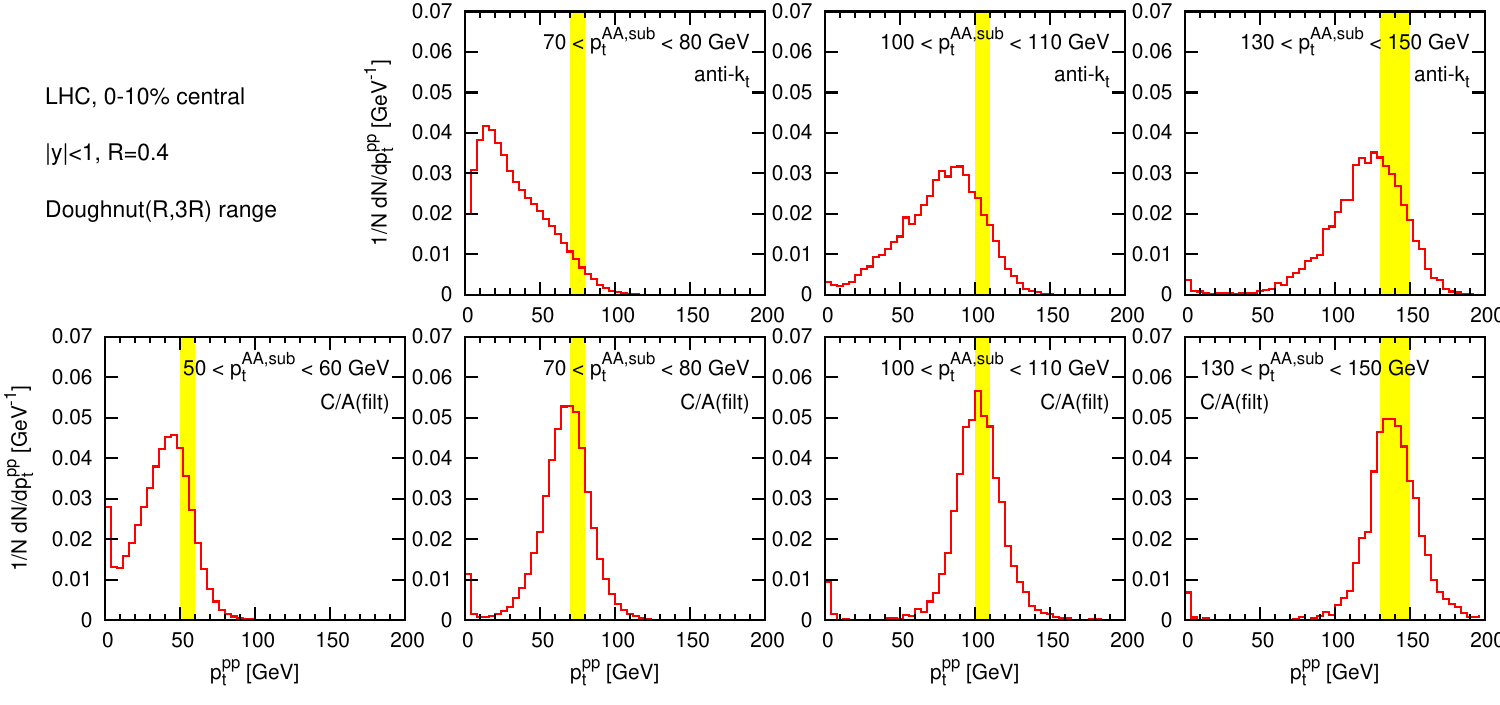}
  \caption{Same as fig.~\ref{fig:origin} for LHC kinematics (PbPb,
    $\sqrt{s} = 5.5\TeV$), generated with approximately 16 million
    events.
    Note the use of a smaller rapidity range here, $|y|<1$, compared to
    the earlier LHC plots. 
  }
  \label{fig:origin_lhc}
\end{figure}

In figure~\ref{fig:origin} we show the origins of heavy-ion jets as
determined in our Hydjet simulations.\footnote{%
A point to be aware of is that multiple $pp$ jets can match a single
heavy-ion jet, \ie have at least half their $p_t$ contained in the
heavy-ion jet.
In evaluating $O(\ptfullsub, p_{t}^{pp})$ we take only the
highest-$p_t$ matched jet. If there is no matched jet (this occurs
only rarely) then we fill the bin at $p_{t}^{pp} = 0$.
Note also that since we are not explicitly embedding hard jets, all
$pp$ jets in the events have undergone Hydjet's quenching.
}
The upper row provides the origin plots for anti-$k_t$ jets at
RHIC. Each plot corresponds to one bin of $\ptfullsub$, and
shows $O(\ptfullsub, p_{t}^{pp})$ as a function of $p_{t}^{pp}$.
At moderate $\ptfullsub$, the $25-30\GeV$ bin, the origin is
dominated by low $\pthard$.
This is perhaps not surprising, given the result in
subsection~\ref{eq:shift-dispersion-on-spectrum} that the $\pthard$
origin is expected to be $\sim 17\GeV$ lower than $\ptfullsub$
for anti-$k_t$ jets --- additionally, that result assumed an
exponential spectrum for the inclusive jet distribution, whereas the
distribution rises substantially faster towards low $\pthard$.
As $\ptfullsub$ increases one sees that the contribution of
high $\pthard$ jets increases, in a manner not too inconsistent with
the expected $\sim 17\GeV$ shift, though the $\pthard$ distribution
remains rather broad and a peak persists at small $\pthard$.
These plots suggest that an inclusive jet distribution measurement
with the anti-$k_t$ algorithm at RHIC is not completely
trivial 
since, up to rather large $\ptfullsub$, one is still sensitive to the
jet distribution at small values of $\pthard$ where the separation
between ``hard'' jets and the soft medium is less clear.
Nevertheless, two points should be kept in mind: firstly, the upper
row of fig.~\ref{fig:origin} shows that different $\ptfullsub$
have complementary sensitivities to different parts of the $\pthard$
spectrum. Thus it should still be possible to ``unfold'' the
$\ptfullsub$ distribution to obtain information about the
$\pthard$, unfolding being in any case a standard part of the
experimental correction procedure.
Secondly STAR quotes~\cite{Salur:2009vz,Ploskon:2009zd} a $10\%$
smaller value for $\sigma_{\Delta p_t}$ than the $7.5\GeV$ that we
find in Hydjet. Such a reduction can make it noticeably easier to
perform the unfolding.

The impact of a reduction in $\sigma_{\Delta p_t}$ is illustrated in
the lower row of fig.~\ref{fig:origin}, which shows the result for
C/A(filt).
Here, even the $25-30\GeV$ bin for $\ptfullsub$ shows a
moderate-$p_t$ peak in the distribution of $\pthard$, and in the
$35-40\GeV$ bin the low-$p_t$ ``fake'' peak has disappeared almost
entirely. 
Furthermore, the $\pthard$ peak is centred about $7\GeV$ lower than
the centre of the $\ptfullsub$ bin, remarkably consistent with
the calculations of section~\ref{eq:shift-dispersion-on-spectrum}.
Overall, therefore, unfolding with C/A(filt) will be easier than
with anti-$k_t$.

Corresponding plots for the LHC are shown in
fig.~\ref{fig:origin_lhc}.
While C/A(filt)'s lower dispersion still gives it an advantage over
anti-$k_t$, for $\smash{\ptfullsub} \gtrsim 80\GeV$, anti-$k_t$ does
now reach a domain where the original $pp$ jets are themselves always
hard.

Procedures to reject fake jets have been proposed,
in~\cite{Grau:2008ed,Lai:2009ai}. They are based on a cut on
(collinear unsafe) jet shape properties and it is thus unclear how
they will be affected by quenching and in particular whether the
expected benefit of cutting the low-$\pthard$ peak in
Figs.~\ref{fig:origin} and \ref{fig:origin_lhc} outweighs the
disadvantage of potentially introducing extra sources of systematic
uncertainty at moderate $p_t$.

One final comment is that experimental unfolding should provide enough
information to produce origin plots like those shown here. As part of
the broader discussion about fakes it would probably be instructive
for such plots to be shown together with the inclusive-jet results.
%

\subsection{Exclusive analyses}
\label{sec:fakes:excl}

An example of an exclusive analysis might be a dijet study, in which
one selects the two hardest jets in the event, with transverse momenta
$p_{t1}$ and $p_{t2}$, and plots the distribution of $\frac12 H_{T,2}
\equiv \frac12(p_{t1} + p_{t2})$.
Here one can define ``fakes'' as corresponding to cases where one or
other of the jets fails to match to one of the two hardest among all
the jets from the individual $pp$ events.
This definition is 
insensitive to the soft/hard boundary in a
simulation such as Hydjet, because it naturally picks out hard $pp$ 
jets that are far above that boundary. And, by concentrating on just
two jets, it also evades the problem of high occupancy from the large
multiplicity of semi-hard $pp$ collisions.
This simplification of the definition of fakes is common to many
exclusive analyses, because they tend to share the feature of
identifying just one or two hard reference jets.

The specific case of the exclusive dijet analysis has the added
advantage that it is amenable to a data-driven estimation of
fakes. 
One divides the events into two groups, those for which the two
hardest jets are on the same side (in azimuth) of the event and those
in which they are on opposite sides (a related analysis was presented
by STAR in ref.~\cite{BrunaPrague}).
For events in which one of the two jets is ``fake,'' the two jets are
just as likely to be on the same side as on the opposite side. This is
not the case for non-fake jets, given that the two hardest ``true''
jets nearly always come from the same $pp$ event and so have to be on
opposite sides.\footnote{%
  At RHIC energies, above $p_t\sim 10-15\GeV$, it is nearly always the
  case that the two hardest jets come from the same $pp$
  event. At the LHC, this happens above $20-30\GeV$.}
Thus by counting the number of same-side versus opposite-side dijets
in a given $\frac12 H_{T,2}$ bin, one immediately has an estimate of
the fake rate.\footnote{Note that for plain $pp$ events, if one has
  only limited rapidity acceptance then the same-side/opposite-side
  separation is not infrared safe, because of events in which only one
  hard jet is within the acceptance and the other ``jet'' is given by
  a soft gluon emission.
  Thus to examine the same-side/opposite-side separation in plain $pp$
  events with limited acceptance, one would need to impose a $p_t$ cut
  on the second jet, say $p_{t,2} > \frac12 p_{t,1}$.  }

\begin{figure}[t]
  \centering
  \includegraphics[width=0.5\textwidth]{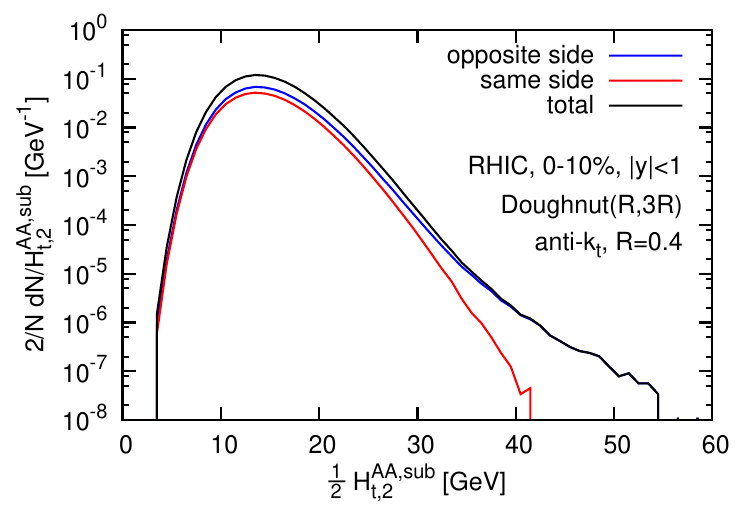}%
  \includegraphics[width=0.5\textwidth]{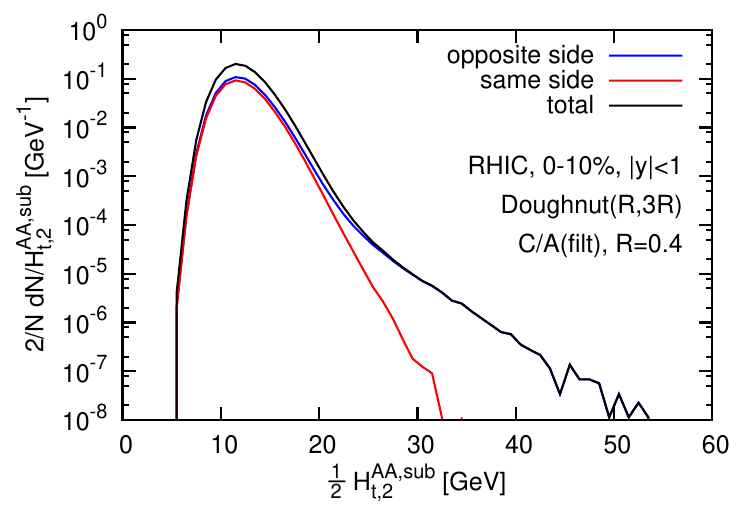}%
  \caption{The distribution of $\frac12 H_{T,2}$ obtained from the two
    hardest full, subtracted heavy-ion jets in each event at RHIC, as
    obtained from simulations with Hydjet~1.6. The left-hand plot is
    for anti-$k_t$ and the right-hand plot for C/A(filt).
    The same-side curves give an approximate measure of (half of) the
    contribution of ``fake'' jets to the dijet spectrum.  }
  \label{fig:dijets}
  \centering
  \includegraphics[width=0.5\textwidth]{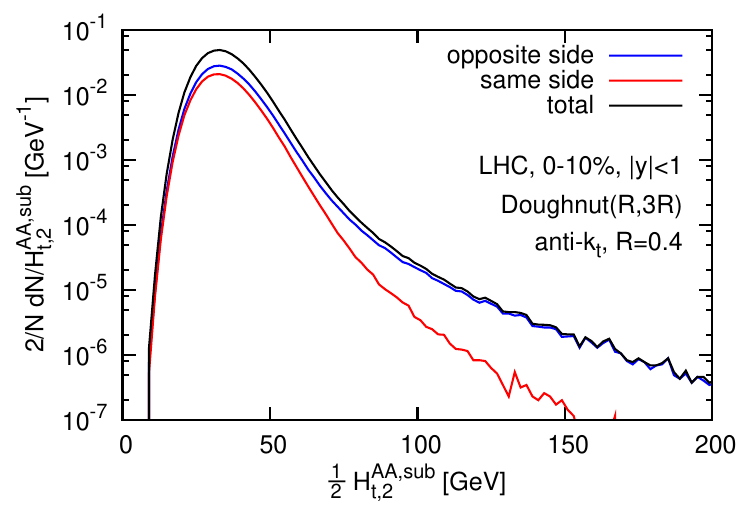}%
  \includegraphics[width=0.5\textwidth]{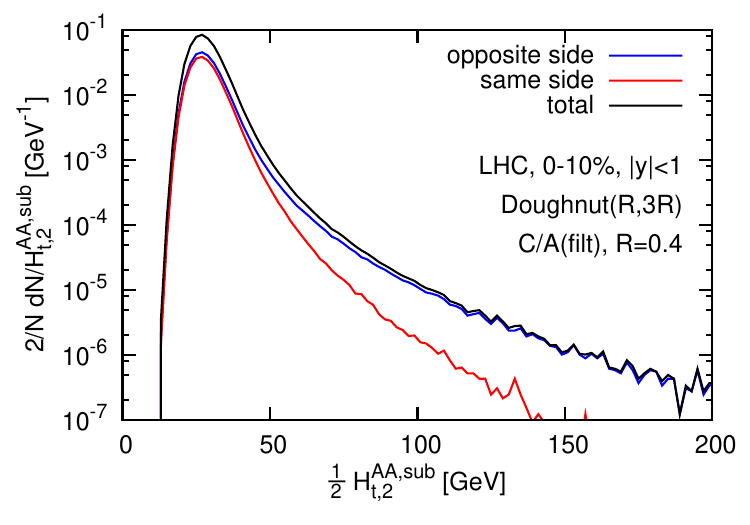}%
  \caption{Same as fig.~\ref{fig:dijets} for LHC kinematics.}
  \label{fig:dijets_lhc}
\end{figure}

This is illustrated in fig.~\ref{fig:dijets}, which shows the
distribution at RHIC of the full, subtracted $\frac12 H_{T,2}$ result,
together with its separation into opposite-side and same-side
components.
One sees that in the peak region, the opposite and same-side
distributions are very similar, indicating a predominantly ``fake''
origin for at least one of two hardest jets (they are not quite
identical, because there is less phase-space on the same side for a
second jet than there is on the away side).
However above a certain full, reconstructed $\frac12 H_{T,2}$ value,
about $30\GeV$ for anti-$k_t$ and $20\GeV$ for C/A(filt) the
same-side distribution starts to fall far more rapidly than the
opposite-side one, indicating that the measurement is now dominated by
``true'' pairs of jets. 

The LHC results, fig.~\ref{fig:dijets_lhc}, are qualitatively similar,
with the same-side spectrum starting to fall off more steeply than the
opposite-side one around $70\GeV$ for the anti-$k_t$ algorithm and $50\GeV$ for
C/A(filt).

One can also examine origin plots for $H_{T,2}$, in analogy with
the Monte Carlo analysis of section~\ref{sec:fakes:incl}. For
brevity, we refrain from showing them here, and restrict ourselves to
the comment that in the region of $H_{T,2}$ where the result is
dominated by opposite-side pairs, the origin plots are consistent with
a purely hard origin for the dijets.

%
\section{Summary and discussions}\label{sec:ccl}

We have shown that the area--median subtraction method gives good
performance to subtract the large heavy-ion UE.
We have tested different choices of range to estimate the rapidity
dependence and observed little difference between various ranges,
as long as they are chosen to be localised in the vicinity of the jet
of interest and of sufficient size (at least 4 units of area for jets
with $R=0.4$).

As for the quality measures, the offset can be brought close to zero
by using the anti-$k_t$ algorithm, while the $k_t$ algorithm has the
largest offset; the Cambridge/Aachen (C/A) algorithm with filtering
also gives a small offset, however this seems to have been due to a
fortuitous cancellation between two only partially related effects.
The dispersion is comparable for anti-$k_t$, C/A and $k_t$, but
significantly smaller for C/A(filt) (except at high transverse momenta
for LHC), as a consequence of its smaller jet area.
Among the different algorithms, anti-$k_t$ is the most robust with
respect to quenching effects, and C/A(filt) seems reasonably robust at
RHIC, though a little less so at the LHC.
The precise numerical results for offset and dispersion can depend a
little on the details of the simulation and the analysis, however the
general pattern remains.

Overall our results indicate that the area-based subtraction method
seems well suited for jet reconstruction in heavy-ion collisions.
Two jet-algorithm choices were found to perform particularly well:
anti-$k_t$, which has small offsets but larger fluctuations, and C/A
with filtering, for which the offsets may be harder to control, but
for which the fluctuations are significantly reduced, with consequent
advantages for the unfolding of experimentally measured jet
spectra.
Ultimately, we suspect that carrying out parallel analyses with these
two choices may help maximise the reliability of jet results in HI
collisions.

\section{Jet fragmentation function}\label{sec:mcstudy-hi-ff}

The last extension of the area--median approach we want to validate is
its application to the subtraction of pileup contamination to the
fragmentation function (moments), as described in
Section~\ref{sec:areamed-description-fragmentation}. We shall apply it
here to heavy-ion collisions, where it is considered as one of the
first observables that one would measure to study jet-quenching
effects~\cite{Borghini:2005em,Guo:2000nz,Armesto:2007dt,Arleo:2008dn,Sapeta:2007ad,Majumder:2009zu,Beraudo:2011bh}.\footnote{The
  generic underlying idea is that jet quenching generates additional
  medium-induced radiation~\cite{Baier:1996sk,Zakharov:1997uu}. One
  therefore expects a reduction of the fragmentation function at large
  $z$ and an increase at small $z$. In terms of the moments of the
  fragmentation, this means an increase for $N<1$, including \eg for
  multiplicity ($N=0$), and a decrease for $N>1$.}

\paragraph{Details of the simulation.} Before proceeding with the
results, it is useful to detail the terminology and simulation tools
that we have used.

What we call `hard' jets in QCD are simulated in proton-proton ($pp$)
collisions using Pythia~6.425 in dijet mode with the DW
tune~\cite{Albrow:2006rt}.\footnote{This tune is not the most up to
  date; however it is not unrealistic for the LHC and has the
  characteristic that it is based on Pythia's virtuality ordered
  shower, which is a prerequisite for use with Pyquen.}
We have not included any $pp$ underlying event (UE), its effect being
minimal in this context anyway.
Jets including quenching effects have been generated using
Pyquen, v1.5.1, as included in Hydjet~\cite{Lokhtin:2003ru,hydjet}.
We consider all hadrons, not just charged tracks, and take $\pi_0$'s
to be stable, so that we are still considering genuine hadron
distributions.

The heavy-ion background (also called underlying event) is simulated
using Hydjet v1.6 for 0-10\% centrality and, where needed, it is
superimposed to the hard event generated by Pythia.
This has been found \cite{Cacciari:2011tm} to be in good agreement with 
the experimental measurements of background fluctuations by 
ALICE~\cite{Abelev:2012ej}.
The jets observed in this combined event will be denoted as `full'
jets. 

All events are generated for either $pp$ or lead--lead (PbPb)
collisions at the LHC, with a centre-of-mass energy of 2.76 TeV per
nucleon-nucleon collision.
Jets are reconstructed using the anti-$k_t$ algorithm with $R=0.4$, as
implemented in FastJet.
In estimating the jet's $p_t$, the HI background is subtracted using
the area--median method using $k_t$ jets with $R=0.4$ to estimate
$\rho$.
At most the two hardest jets passing a hard cut on the subtracted
transverse momentum (see below) are subsequently used for the fragmentation
function analysis.
The cuts that we shall use are $100\GeV$ and $200\GeV$.
%

\paragraph{Effect of the background.}
The addition of the heavy-ion background has the potential to modify a
measured fragmentation function in two ways.
Firstly, the jet's $p_t$ is modified, affecting the normalisation of
$z$ in \eq~(\ref{eq:z}).
Secondly, the heavy-ion background adds many extra particles to the
jet, predominantly at low momenta.

\begin{figure}[t]
  \begin{center}
    \includegraphics[width=0.90\textwidth]{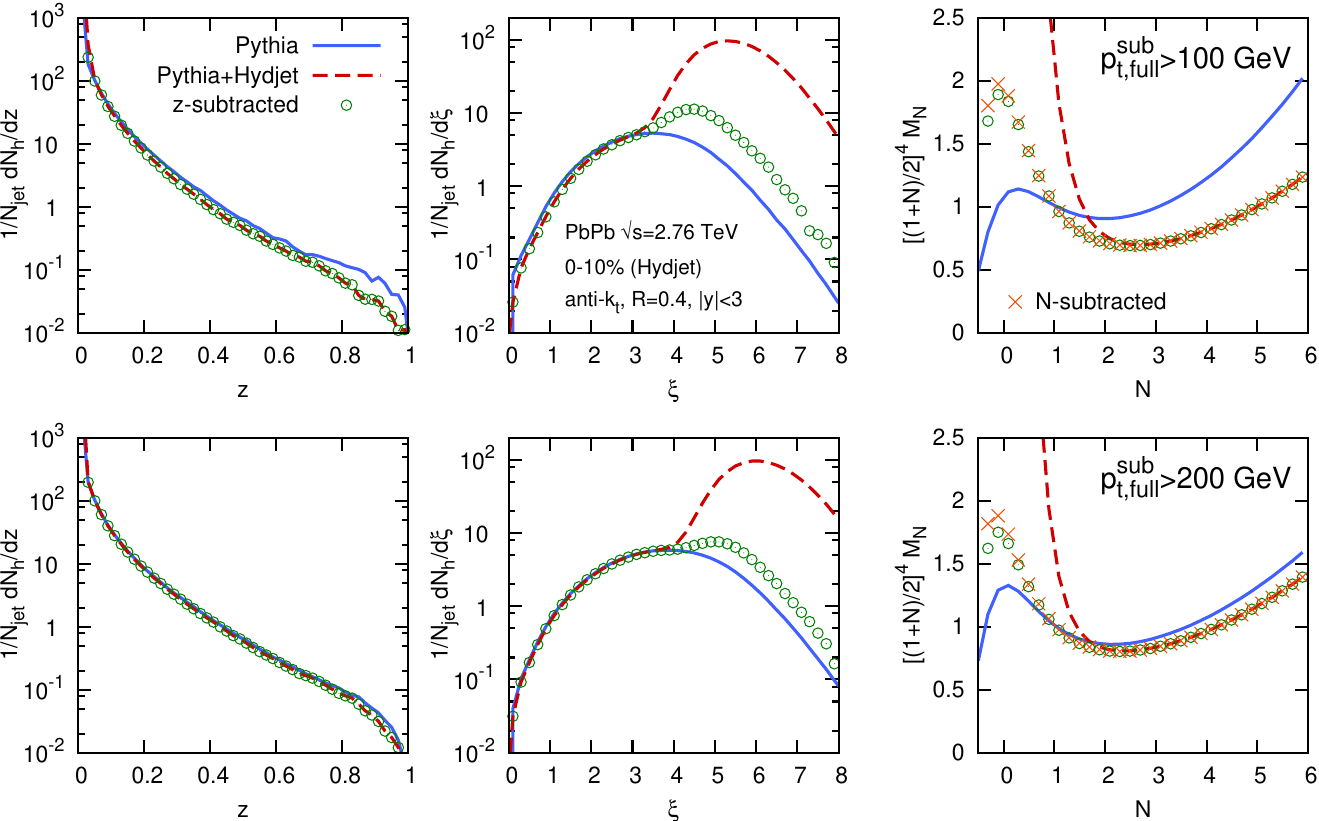}
    \caption{Jet fragmentation functions shown for plain Pythia, with
      the addition of the heavy-ion background (Pythia+Hydjet) and
      after subtraction of the heavy-ion background ($z$-subtracted
      and $N$-subtracted).
      For the results including the heavy-ion background, the jet
      $p_t$ used to define $z$ is always that after subtraction of the
      heavy-ion background. 
      As in Fig.~\ref{fig:hard}, we show the results as a function of
      $z$ (left), $\xi$ (middle) and for the moments versus $N$ (right).
      The upper (lower) row has a jet $p_t$ threshold of $100\GeV$
      ($200\GeV$). 
      \label{fig:hydjet}}
  \end{center}
\end{figure}

To appreciate the impact of the extra particles in the jet from the
heavy-ion background, it is instructive to first examine the 
``Pythia+Hydjet'' dashed red curves of Fig.~\ref{fig:hydjet}.
These show the FF extracted in heavy-ion collisions, without any
subtraction of the background contribution to the FF,
but always using a $z$ value defined such that the jet's $p_t$ has
been corrected for the expected HI background contamination (as is
standard in the experimental measurements).
At this stage we will not perform any unfolding to account for
fluctuations in the HI background.

One sees how the FF acquires a `bump' in the soft
region, which lies at larger $\xi$ (smaller $z$) than the maximum in the original
$pp$ result (blue solid line) and is up to two orders of magnitude
higher. 
For the moments, this presence of the background is seen as a steep
increase in the small-$N$ region, taking the curve far off the scale.

At large $z$ (small $\xi$), the impact of the addition of the
background is only barely visible, as might be expected given that it
is dominated by soft particles.
However, this visually small effect is partially an artefact of the
logarithmic scale used to show the FF versus $z$.
Considering instead the moments, one sees that there is a
non-negligible \emph{reduction} in the FF at large $N$.
This is perhaps surprising given that the background adds particles.
It is a feature related to the interplay between background
fluctuations and the steeply falling jet spectrum. It is well known by
the experiments and is related to the effects discussed at the end of
Section.~\ref{sec:areamed-fragmentation-function}.

\paragraph{$z$-space subtraction.}

The traditional approach to background subtraction from a jet FF
involves the construction of a distribution (in $z$ or $\xi$ space) intended
to approximate that of background-only particles, and the
subtraction of this distribution from the measured one. 
The way the background-only distribution is determined can vary,
the simplest one probably being to consider a region of the event
that is expected to be little affected by the hard jets, and
measure it there. 

To illustrate $z$-space subtraction here, we measure the distribution
of hadrons in two regions transverse in azimuth with respect to the
axis defined by the dijet event.
Event by event, and jet by jet, we subtract those distributions
(measured in a patch of phase space with an area equal to that of the
jet that one is considering) from the jet fragmentation function.
While exact experimental procedures differ in the details, most
choices lead to similar results here.
Perhaps the main distinction of the experimental procedures is that
they sometimes address issues related to flow, which for simplicity we
neglect here in our $z$-space subtraction.
%


The results of applying a $z$-space subtraction are shown in
figure~\ref{fig:hydjet} as green open circles, labelled
`$z$-subtracted' (the moments of these $z$-subtracted results are also
shown).
One sees how these curves come closer to the $pp$ results at small $z$
and small $N$ than do the unsubtracted (dashed red) ones.
The agreement improves as the jet $p_t$ threshold is increased.
However, even with jets of $p_t \gtrsim 200$~GeV, this procedure still
falls short of an accurate reconstruction of the hard FF. 
In fact, the region in $z$ where the subtracted FF works well barely
extends beyond that selected by simply truncating the unsubtracted FF
at low $p_{t,\text{h}}$ so as to avoid the region dominated by the
background.

\begin{figure}[tp]
  \centering
  \includegraphics[width=0.7\textwidth]{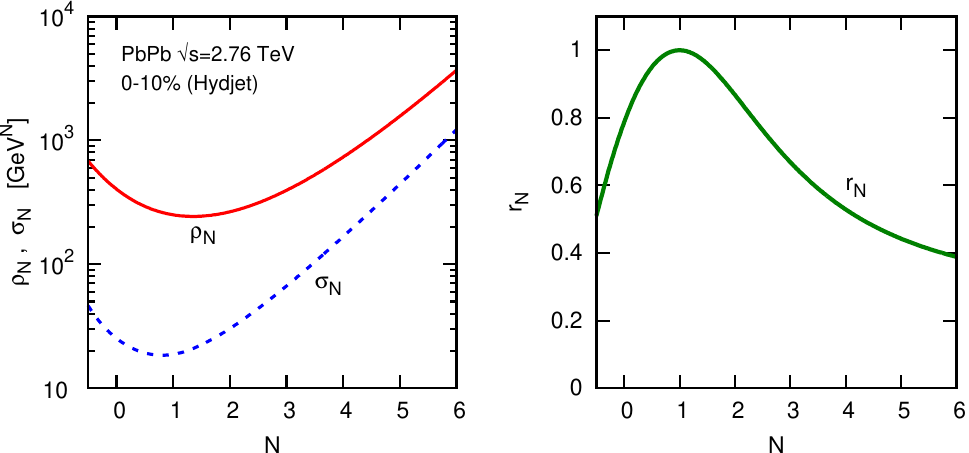}
  \caption{The quantities $\rho_N$, $\sigma_N$ and the correlation
    coefficient $r_N$, shown as a function of $N$ for $0-10\%$ central
    PbPb collisions $\sqrt{s_{NN}} = 2.76 \TeV$ as obtained from
      simulations with Hydjet.}
  \label{fig:correlNrhoNsigmaN}
\end{figure}

\begin{figure}[tp]
\centerline{\includegraphics[width=0.8\textwidth]{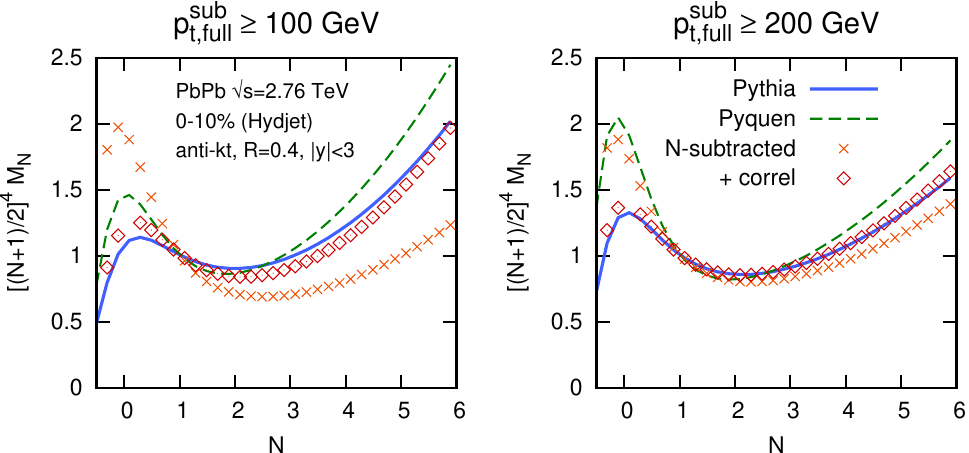}}
\caption{ %
  Jet fragmentation function moments, showing the plain Pythia result,
  the result after embedding in Hydjet and applying plain subtraction
  moment-space subtraction (``N-subtracted'') and after the additional
  improvement to account for correlations (``+ correl''),
  \eq~(\ref{eq:subimp-v3}).  %
  A quenched result (``Pyquen'') is also shown, to help give an
  indication of the order of magnitude of quenching effects as
  compared to residual misreconstruction effects.
}\label{fig:hydjet+correl}
\end{figure}


\paragraph{Area--median moments subtraction.} The results for
$M_N^\text{sub}$ are shown in figure~\ref{fig:hydjet} in the
right-most plots, orange crosses labelled `$N$-subtracted'. One can
see that they are neither better nor worse than the corresponding
$z$-subtracted ones, which have been translated to $N$-space and drawn
in the same plots for direct comparison (green circles).

In order to provide a fragmentation function reconstruction that is
better than that given by the standard $z$-space method, we
introduce in the next section an improved background subtraction
method that is most straightforwardly applied in moment space.

\paragraph{Area--median improved moments subtraction.}  The correction
\eqref{eq:subimp-v3} can be applied jet by jet to correct for the
effects of fluctuations. It requires the prior knowledge of the slope
$\kappa$ of the jet cross-section, which can be obtained from $pp$ data,
or from simulations.\footnote{In practice, $\kappa$ depends on $p_t$ and
  should be taken at the scale $S_1-q_t$ in the integrand. However,
  $\kappa$ varies slowly with $p_t$ and can easily be taken at the fixed
  scale $p_t$ in our small-fluctuations limit. In our analysis $\kappa$
  ranged from $\sim 9$ GeV at $p_t \simeq 50$~GeV to $\sim 28$~GeV at
  $p_t \simeq 200$~GeV. At $p_t \simeq 100$~GeV we had
  $\kappa \sim 16$~GeV.} All the other ingredients that enter this
equation ($\sigma$, $A$, $S_N$, $\sigma_N$, $r_N$) can instead be
determined event-by event or jet-by-jet.
In practice we determine $\sigma$, $\sigma_N$ and $r_N$ from the
ensemble of jets contained in an annulus (or ``doughnut'') of outer
radius $3R$ and inner radius $R$, centred on the jet of interest.
Typical values of $\rho_N$, $\sigma_N$ and $r_N$ are presented as a
function of $N$ in Fig.~\ref{fig:correlNrhoNsigmaN}.%
\footnote{The full specification of our procedure is as follows:
  $\rho$, $\rho_N$ are evaluated using
  \eq~\eqref{eq:rhoN-from-median}, where the ``patches'' correspond to
  the jets in the annulus; $\sigma_N$ (and $\sigma \equiv \sigma_1$) is as returned by
  FastJet on the same
  set of jets, \ie derived from the difference between the 16th percentile
  and the median of the set of $\frac{\sum_{i\in \text{jet}}
    p_{t,i}^N}{A_\text{jet}}$.
  For the last term in \eq~\eqref{eq:subimp-v3} (and also for
  Fig.~\ref{fig:correlNrhoNsigmaN}), $r_N$ is evaluated using
  \eq~\eqref{eq:r_N-definition} as determined on this set of jets,
  with $Q_N$ for a given jet evaluated as $\sum_{i\in \text{jet}}
  p_{t,i}^N - \rho_N A_\text{jet}$ (and $q_t \equiv Q_1$), which
  corresponds to \eq~(\ref{eq:QN-def}) in the approximation that the
  jets in the annulus are predominantly composed of background
  particles.
  We also investigated an alternative procedure, in which the last
  term in \eq~\eqref{eq:subimp-v3} was evaluated as $
  \frac{\mathrm{Cov}(q_t, Q_N)}{\mathrm{Var}(q_t)} \times
  \frac{\sigma^2 A}{\kappa S_1^N}$, \ie directly combining
  \eqs~\eqref{eq:expected-QN} and \eqref{eq:av-qt}, with the ratio of
  covariance and variance evaluated using the usual set of jets in the
  annulus.
  In the region of moderate $N$, where the last term of
  \eq~\eqref{eq:subimp-v3} contributes significantly to the correction
  for $M_N$, this alternative procedure gives nearly identical results
  to our main procedure.
  However, for large $N$ values, we have found that it gives a poor
  evaluation of the last term itself, though overall that term
  contributes negligibly there.
}

We show in figure \ref{fig:hydjet+correl} the result of applying
\eq~(\ref{eq:subimp-v3}) to our subtraction in moment space, for two
jet $p_t$ thresholds. 
The solid blue curve ($pp$ reference) and orange crosses
(N-subtracted) are identical to the results in the rightmost plots in
figure~\ref{fig:hydjet}.
In addition, figure \ref{fig:hydjet+correl} also displays results
obtained using \eq~(\ref{eq:subimp-v3}), shown as red diamonds.
One sees how the quality of the agreement with the `hard' blue curve
is markedly improved.
At low $N$ it is the last, additive, term in 
\eq~(\ref{eq:subimp-v3}) that dominates this improvement, accounting for
the correlation between a jet's reconstructed $p_t$ fluctuations and
the fluctuations in the background's contribution to the moment;
at high $N$ it is the multiplicative $N\sigma^2 A/S_1\kappa$ term that
dominates, correcting for the fact that the more common upwards
fluctuations in the jet's reconstructed $p_t$ cause the fragmentation
$z$ value to be underestimated.

Figure \ref{fig:hydjet+correl} also shows (green dashed curve) the jet
fragmentation function predicted by the quenching model used in
Pyquen. 
One observes that the remaining deficiencies of the reconstruction
are significantly smaller than the difference between the unquenched
(solid blue) and the quenched (dashed green) FFs, pointing to a
potential discriminating power.
This is to be contrasted with subtraction without improvement (orange
crosses), which especially in the soft region, $N<1$, fails to describe
the blue curve sufficiently well to tell whether quenching (as
modelled in Pyquen) is present or simply if an imperfect
reconstruction is taking place.
This serves as an illustration that the improved subtraction
may now be sufficiently good to allow one to discriminate a quenched
FF from an unquenched one.  

An implementation of the tools needed for the jet fragmentation function
subtraction in moments space, as well as for the fluctuations unfolding
improvement, is available in \fjcontrib.



\chapter{Analytic insight}\label{chap:analytics}

In this chapter we shall review a series of analytic properties
related to the area--median pileup subtraction method. These are meant
to serve two main purposes: first, jets being QCD objects, there is a
clear interest to see if their properties are under analytic control
in terms of perturbative QCD; second, we will see that simple analytic
estimates allow to grasp a large fraction of the features observed in
Monte Carlo simulations.

As far as we are willing to understand QCD aspects of jets and their
properties, a large part of this chapter can be seen as a turf for
theoretical QCD aficionados with little phenomenological
consequences. Nevertheless, the results we obtain often allow for a
deeper understanding of the dynamics of jets and can be beneficial in
various respects.\footnote{For example, the analytic understanding of
  jet areas has been a crucial element in devising the anti-$k_t$
  algorithm.}
In the following pages, we shall thus give an extended discussion of
the two cornerstones of the area--median method by discussing the
analytic properties of jet areas (Section~\ref{sec:analytics-areas})
and an analytic understanding of the estimation of the pileup energy
density $\rho$ (see Section~\ref{sec:analytic-pileup}).

On the second front, there is a long list of cases where simple
analytic understanding can help understanding the underlying physics
and build a coherent picture of pileup mitigation, as we have already
seen a few times in the previous chapters. These often rely on a
Gaussian approximation for the pileup $p_t$ distribution, simplified
distributions for the underlying hard processes and basic
understanding of jet areas. We shall provide strong bases for these
arguments throughout this chapter.

Finally, as we shall see for example in
Section~\ref{sec:area-analytics-future}, our analytic investigations
open a series of doors towards future developments.

\section{Properties of jet areas}\label{sec:analytics-areas}

In the context of pileup subtraction, we are mostly interested in the
active jet area. However, passive areas are much simpler to handle
since they do not involve the clustering of ghosts --- it is tempting,
here, to draw an analogy with QED v. QCD where, perturbatively, the
latter is very often simpler to manage than the latter because photons
do not interact while gluons do.
We shall therefore begin with a study of the properties of passive
areas.

That said, our goal is to understand the properties of jet areas in
perturbative QCD (pQCD). To do that, we start by considering jets made
of a single particle, carry on with cases with 2 particles and try,
when possible, to extrapolate to multi-particle configurations.

Before going into detailed calculations, we should mention that jet
areas are infrared-unsafe quantities, as we will see explicitly in the
calculations below.
This should not come as a surprise since they are directly constructed
from infinitely soft particles (ghosts).
In practice, it means that jet areas are sensitive to IR details such
as hadronisation. The main goal of this Section is to show that it is
nevertheless possible to deduce interesting pieces of information
about jet areas from perturbative QCD.
Furthermore, since the jet areas are meant to capture the sensitivity
of jets to soft radiation (and pileup in particular), we believe that
any sensible (i.e.\ related to the jet's sensitivity to UE/pileup type
contamination) definition of area would actually be infrared unsafe.

\subsection{Passive areas}\label{sec:areamed-areaanalytics-passive}

\subsubsection{Areas for 1-particle configurations}\label{sec:areamed-areaanalytics-passive-area-1particle}

We consider passive areas in the context of four different jet
algorithms: (inclusive) anti-$k_t$~\cite{Cacciari:2008gp}, (inclusive)
$k_t$~\cite{Catani:1993hr,Ellis:1993tq}, (inclusive)
Cambridge/Aachen~\cite{Dokshitzer:1997in,Wobisch:1998wt}, and the
seedless infrared-safe cone algorithm, SISCone~\cite{siscone}.
%
%
Each of these jet algorithms contains a parameter $R$ which
essentially controls the reach of the jet algorithm in $y$ and $\phi$.
Given an event made of a single particle $p_1$, the passive area of the jet
$J_1$ containing it is $a(J_1) = \pi R^2$ for all four algorithms.

\subsubsection{Areas for 2-particle configurations}
\label{sec:areamed-areaanalytics-passive-area-2particle}

Let us now consider what happens to the passive jet areas in the presence of
additional soft perturbative radiation. We add a particle $p_2$ such
that the transverse momenta are strongly ordered,
\begin{equation}
\label{eq:strord}
p_{t1} \gg p_{t2} \gg \Lambda_{QCD} \gg g_t\; ,
\end{equation}
and $p_1$ and $p_2$ are separated by a geometrical distance
$\Delta_{12}^2 = (y_1-y_2)^2 + (\phi_1-\phi_2)^2$ in the $y$-$\phi$
plane. Later, we shall integrate over $\Delta_{12}$ and
$p_{t2}$. Note that $g_t$ has been taken to be infinitesimal compared
to all physical scales to ensure that the presence of the ghost
particle does not affect the real jets.

For $\Delta_{12} = 0$ collinear safety ensures that the passive area  
is still equal to $\pi R^2$ for all three algorithms. However, as one increases
$\Delta_{12}$, each algorithm behaves differently.

\paragraph{\boldmath Anti-$k_t$.} Let us first consider the anti-$k_t$
jet algorithm which has the 2-particle distance measure
$d_{ij}=\min(k_{ti}^{-2},k_{tj}^{-2}) \Delta_{ij}^2/R^2$ and beam--particle
distance $d_{iB} = k_{ti}^{-2}$.
Taking $\Delta_{12} \sim \Delta_{1g} \sim \Delta_{2g} \sim R$ and
exploiting the strong ordering~(\ref{eq:strord}) one has
\begin{equation}
d_{1B} \sim d_{12} \sim d_{1g} \ll d_{2B} \sim d_{2g} \ll d_{gB} \; .
\end{equation}
From this ordering of the distances, one sees that the first step of
the clustering will first combine $p_1$ with the closest of $p_2$ or
the ghost, provided it is within a distance $R$ of $p_1$. $p_1$
will then cluster with the other of $p_2$ or the ghost, again provided
it is within a distance $R$ of $p_1$,
The ghost will therefore be clustered with $p_1$ if and only if they
are within a distance $R$ of $p_1$, independently of the presence of
the soft particle $p_2$. The passive area of the jet which will thus
remain $\pi R^2$.

\paragraph{\boldmath $k_t$.} Let us now turn to the $k_t$ jet
algorithm which has the distance measures
$d_{ij} = \min(k_{ti}^2,k_{tj}^2) \Delta_{ij}^2/R^2$ and
$d_{iB} = k_{ti}^2$. This time, we have
\begin{equation}
d_{1B} \gg d_{2B} \sim d_{12} \gg d_{g1} \sim d_{g2} \sim d_{gB} \; .
\end{equation}
One sees that the ghost always undergoes its clustering before either
of the perturbative particles $p_1$ and $p_2$. Specifically, if at
least one of $\Delta_{1g}$ and $\Delta_{2g}$ is smaller than $R$, the
ghost clusters with the particle that is geometrically closer.
If both $\Delta_{1g}$ and $\Delta_{2g}$ are greater than $R$ the ghost
clusters with the beam and will not belong to any of the perturbative
jets.

\begin{figure}[th]
\includegraphics[width=\textwidth]{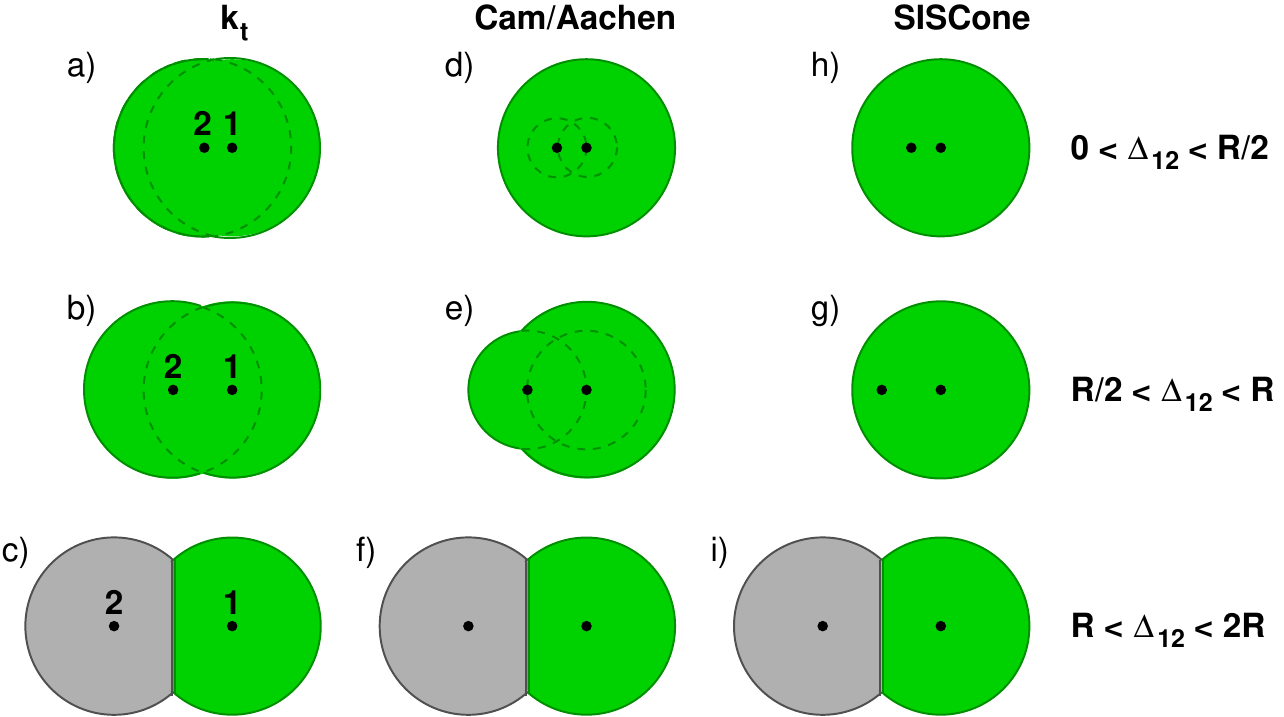}
\caption{Schematic representation of the passive area of a
jet containing one hard particle ``1'' and a softer one ``2'' for various
separations between them and different jet algorithms. Different shadings represent
distinct jets. We have not represented the anti-$k_t$ algorithm for
which the area remains $\pi R^2$ irrespective of the position of the
soft particle $p_2$.}
\label{fig:overlaps} 
\end{figure}

Let us now consider various cases. If $\Delta_{12} < R$,
(Fig.~\ref{fig:overlaps}a,b)
the particles $p_1$ and $p_2$ will eventually end up in the same
jet. The ghost will therefore belong to the jet irrespectively of
having been clustered first with $p_1$ or $p_2$. The area of the jet
will then be given by union of two circles of radius $R$, one
centred on each of the two perturbative particles,
\begin{equation}
  \label{eq:a_kt_Dlt1}
a_{k_t,R}(\Delta_{12}) =  u\!\left(\frac{\Delta_{12}}{R}\right) \, {\pi R^2}\,,
\qquad\qquad
\mathrm{for}~\Delta_{12} < R\,,
\end{equation}
where
\begin{equation}
\label{eq:u}
u(x) = \frac{1}{\pi}\left[x\sqrt{1-\frac{x^2}{4}} + 2\left(\pi -
\arccos\left(\frac{x}{2}\right)\right)\right],
\end{equation}
represents the area, divided by $\pi$, of the union of two circles of
radius one whose centres are separated by $x$.

The next case we consider is $R < \Delta_{12} < 2R $,
Fig.~\ref{fig:overlaps}c. In this case $p_1$ and $p_2$ will never be
able to cluster together. Hence, they form different jets, and the
ghost will belong to the jet of the closer of $p_1$ or $p_2$. The two
jets will each have area
\begin{equation}
  \label{eq:a_kt_Dgt1}
a_{k_t,R}(\Delta_{12}) =   \frac{u\!\left(\Delta_{12}/R\right)}{2}\, \pi R^2\,,
\qquad\qquad
\mathrm{for}~R<\Delta_{12} < 2R\,.
\end{equation}
Finally, for $\Delta_{12} > 2R$ the two jets formed by $p_1$ and $p_2$
each have area $\pi R^2$.

\begin{figure}[t]
  \begin{center}
    \includegraphics[width=0.5\textwidth]{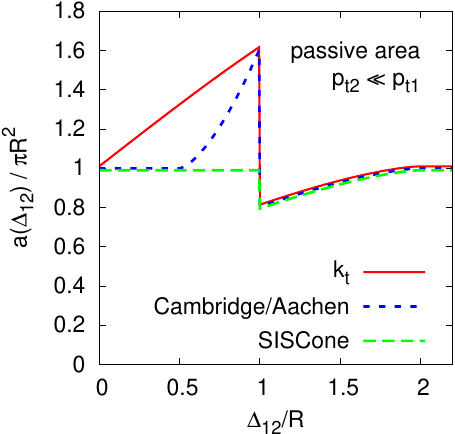}
    \caption{\label{fig:2point} Plot of the passive areas of the hard
      jet as a function of the distance between the hard and the soft
      particle, as given in table \protect\ref{tab:passive-areas}. The
      anti-$k_t$ case is not shown but corresponds to a constant line
      with $a(\Delta_{12})/(\pi R^2)=1$.}
  \end{center}
\end{figure}

The three cases derived above are summarised in
table~\ref{tab:passive-areas} and plotted in Fig.~\ref{fig:2point}.

It is interesting to note that the above results also illustrate the
fact that for the $k_t$ algorithm, the passive area is equal to the
Voronoi area. 
In this 2-particle configuration, the Voronoi diagram consists of a
single line, equidistant between the two points, dividing the plane
into two half-planes, each of which is the Voronoi cell of one of the
particles (this is best seen in Fig.~\ref{fig:overlaps}c). The
intersection of the half-plane with the circle of radius $R$ centred
on the particle has area
${\scriptstyle{\frac{1}{2}}}\pi R^2 u(\Delta_{12}/R)$, and this
immediately gives us the results \eqs~(\ref{eq:a_kt_Dlt1}),
(\ref{eq:a_kt_Dgt1}) according to whether the particles cluster into a
single jet or not.

\paragraph{Cambridge/Aachen.} Now we consider the behaviour of the
Cambridge/Aachen jet
algorithm~\cite{Dokshitzer:1997in,Wobisch:1998wt}, also a sequential
recombination algorithm, defined by the 2-particle distance measure
$d_{ij} = \Delta_{ij}^2/R^2$ and beam-particle distance $d_{iB} = 1$.
Because in this case the distance measure does not involve the
transverse momentum of the particles, the ghost only clusters first if
$\min(\Delta_{1g},\Delta_{2g}) < \Delta_{12}$. Otherwise, $p_1$ and
$p_2$ cluster first into the jet $J$, and then $J$ captures the ghost
if $\Delta_{Jg} \simeq \Delta_{1g} < R$.

The region 
$\Delta_{12} < R$ now needs to be separated into two parts, $\Delta_{12} < R/2$
and $R/2 < \Delta_{12} < R$.

If the ghost clusters first, then it must have been contained in either of the
dashed circles depicted in Figs.~\ref{fig:overlaps}d,e. If $\Delta_{12} < R/2$
both these circles are contained in a circle of radius $R$ centred on $p_1$
(Fig.~\ref{fig:overlaps}d), and so the jet area is $\pi R^2$.

If $R/2 < \Delta_{12} < R$ (Fig.~\ref{fig:overlaps}e)
the circle of radius $\Delta_{12}$ centred on $p_2$ protrudes, and adds to the
area of the final jet, so that
\begin{equation}
{a_{\cam,R}(\Delta_{12})} 
           =  w\!\left(\frac{\Delta_{12}}{R}\right)\, {\pi R^2}\,,
\qquad\qquad
\mathrm{for}~R/2 < \Delta_{12} < R\,,
\end{equation}
where
\begin{equation}
\label{eq:w}
w(x) = \frac{1}{\pi}\left[
\pi-\arccos\left(\frac{1}{2x}\right)+\sqrt{x^2-\frac{1}{4}}+x^2\arccos\left(\frac{1}{2x^2}-1\right)
\right].
\end{equation}
For $\Delta_{12} > R$ a Cambridge/Aachen jet has the same area as
the $k_t$ jet, \cf Fig.~\ref{fig:overlaps}f. As with the $k_t$ algorithm, the
above results are summarised in table~\ref{tab:passive-areas} and
Fig.~\ref{fig:2point}. The latter in particular illustrates the
significant difference between the $k_t$ and Cambridge/Aachen areas
for $\Delta_{12} \sim R/2$, caused by the different order of
recombinations in the two algorithms.

\paragraph{SISCone.}
Modern cone jet algorithms like SISCone identify stable cones and then
apply a split/merge procedure to overlapping stable cones.\footnote{In
  the 2-particle case, the arguments that follow would be identical
  for the midpoint jet algorithm with a Tevatron run~II type split--merge
  procedure~\cite{Blazey}.
  For higher orders, or more realistic events, it is mandatory to use
  an infrared-safe seedless variant, like SISCone.}

For $\Delta_{12} < R$ only a single stable cone is found, centred on
$p_1$. Any ghost within distance $R$ of $p_1$ will therefore belong to
this jet, so its area will be $\pi R^2$, \cf Fig.~\ref{fig:overlaps}h,g.

For $R <\Delta_{12} < 2R$ two stable cones are found, centred on
$p_1$ and $p_2$. The split/merge procedure will then always split
them, because the fraction of overlapping energy is zero. Any ghost
falling within either of the two cones will be assigned to the closer
of $p_1$ and $p_2$ (see Fig.~\ref{fig:overlaps}i). The area of the
hard jet will therefore be the same as for the $k_t$ and Cambridge
algorithms.

Again these results are summarised in table~\ref{tab:passive-areas}
and Fig.~\ref{fig:2point} and one notices the large differences
relative to the other algorithms at $R\lesssim 1$ and the striking
feature that the cone algorithm only ever has negative corrections to
the passive area for these energy-ordered two-particle configurations.

\begin{table}
  \begin{center}
    \begin{tabular}{|c||c|c|c|c|}
      \hline
      &\multicolumn{4}{c|}{$a_{\JA,R}(\Delta_{12})/\pi R^2$}\\
\cline{2-5}
      & anti-$k_t$ & $k_t$ & Cambridge/Aachen & SISCone \\
      \hline
$0 < \Delta_{12} < R/2$ & $1$ 
                        & $ u(\Delta_{12}/R) $ 
                        & $1$ 
			& $1$	 \\		 
$R/2 < \Delta_{12} < R$& $1$ 
                        & $u(\Delta_{12}/R) $ 
                        & $w(\Delta_{12}/R) $
			& $1$ \\
$R < \Delta_{12} < 2R$  & $1$ 
                        & $u(\Delta_{12}/R)\; / \; 2 $
                        & $u(\Delta_{12}/R)\; / \; 2 $
			& $u(\Delta_{12}/R)\; / \; 2 $\\
$\Delta_{12} > 2R$      & $1$ 
                        & $1$
                        & $1$
			& $1$ \\
      \hline
\end{tabular}
\end{center}
\caption{\label{tab:passive-areas} Summary of the passive areas for
  the three jet algorithms for the hard jet in an event containing one
  hard and one soft particle, separated by a $y$-$\phi$  distance
  $\Delta_{12}$. The functions $u$ and $w$ are defined in 
  \eqs~(\ref{eq:u}) and (\ref{eq:w}).}
\end{table}
 
\subsubsection{Area scaling violation}
\label{sec:areamed-areaanalytics-area-scal-viol-passive}

Given that jet passive areas are modified by the presence of a soft
particle in the neighbourhood of a jet, one expects the average jet
area to acquire a logarithmic dependence on the jet transverse
momentum when the jets acquire a sub-structure as a consequence of radiative
emission of gluons.
To determine its coefficient we shall work in the approximation of
small jet radii, motivated by the observation that corrections for
finite $R$, proportional to powers of $R$, are usually accompanied by
a small coefficient~\cite{deFlorian:2007fv,Dasgupta:2007wa}.

In the small-angle limit, the QCD matrix element for the emission of
the perturbative soft gluon, $p_2$, is
\begin{equation}
  \label{eq:softcoll}
  \frac{dP}{dp_{t2} \, d\Delta_{12}} = 
  C_1 \frac{2\as(p_{t2} \Delta_{12})}{\pi}
  \frac{1}{\Delta_{12}} \frac{1}{p_{t2}}\,,
\end{equation}
where $C_1$ is $C_F$ or $C_A$ according to whether the hard particle
$p_1$ is a quark or a gluon. We make use of the fact that
$\Delta_{12}$ is just the angle between the two particles (to within a
longitudinal boost-dependent factor), and will assume that $R$ is sufficiently
small that the small-angle approximation is legitimate. The scale of
the coupling is taken to be the transverse momentum of $p_{2}$
relative to $p_1$.

At order $\as$ the mean jet area with a given jet algorithm (\text{JA}) can be
written
\begin{equation}
  \label{eq:ajfr-base}
  \langle a_{\text{JA},R}\rangle 
  = a_{\text{JA},R}(0) + \langle \Delta a_{\text{JA},R}\rangle
  = \pi R^2 + \langle \Delta a_{\text{JA},R}\rangle\,,
\end{equation}
where we explicitly isolate the $\order{\as}$ higher-order contribution,
\begin{equation}
  \label{eq:ajfr}
  \langle \Delta  a_{\text{JA},R}\rangle \simeq \int_0^{2R} d\Delta_{12} 
  \int_{{Q_0}/\Delta_{12}}^{p_{t1}} d p_{t2} 
  \frac{dP}{dp_{t2} \, d\Delta_{12}} (a_{\text{JA},R}(\Delta_{12}) - a_{\text{JA},R}(0))\,,
\end{equation}
with the $- a_{\text{JA},R}(0)$ term accounting for virtual
corrections. $\avg{\cdots}$ represents therefore an average over perturbative
emission. Note that because of the $1/p_{t2}$ soft divergence for
the emission of $p_{2}$, \eq~(\ref{eq:ajfr}) diverges unless one
specifies a lower limit on $p_{t2}$ --- accordingly we have to
introduce an infrared cutoff ${Q_0}/\Delta_{12}$ on the $p_t$ of the emitted
gluon. This value results from requiring that the transverse momentum
of $p_2$ relative to $p_1$, \ie $p_{t2} \Delta_{12}$, be larger than
${Q_0}$. 
The fact that we need to place a lower limit on $p_{t2}$ is an
explicit manifestation of the fact that jet areas are infrared unsafe
--- they cannot be calculated order by order in perturbative QCD and
for real-life jets they will depend on the details of non-perturbative
effects (hadronisation).
The case of the anti-$k_t$ jet algorithm is an exception since the
area remains $\pi R^2$ independently of $\Delta_{12}$, hence the rhs
of \eqref{eq:ajfr} vanishes independently of $Q_0$.\footnote{Another
  exception is for the SISCone algorithm with a cut, $p_{t,min}$, on
  the minimum transverse momentum of protojets that enter the
  split--merge procedure procedure --- in that situation ${Q_0}$ is
  effectively replaced by $p_{t,min}$.}
In all other cases, one can account for this to some extent by leaving
${Q_0}$ as a free parameter and examining the dependence of the
perturbative result on ${Q_0}$.\footnote{One should of course bear in
  mind that the multi-particle structure of the hadron level is such
  that a single-gluon result cannot contain all the relevant physics
  --- this implies that one should, in future work, examine multi-soft
  gluon radiation as well, perhaps along the lines of the calculation
  of non-global
  logarithms~\cite{Dasgupta:2001sh,Banfi:2002hw,Appleby:2002ke,Appleby:2003sj,Banfi:2005gj},
  though it is not currently clear how most meaningfully to carry out
  the matching with the non-perturbative regime.
  Despite these various issues, we shall see in
  Section~\ref{sec:areamed-areaanalytics-real-life} that the
  single-gluon results work remarkably well in comparisons to Monte
  Carlo predictions. In that section, we shall also argue that in
  cases with pileup, the pileup introduces a natural semi-hard (\ie
  perturbative) cutoff scale that replaces $Q_0$.}

As concerns the finiteness of the $\Delta_{12}$ integration, all the
jet algorithms we consider have the property that 
\begin{subequations}
  \begin{align}
    \label{eq:lim-delta2zero}
    &\lim_{\Delta_{12}\to 0} a_{\text{JA},R}(\Delta_{12}) = 
     a_{\text{JA},R}(0) = 
    \pi R^2\,;
    \\
    &a_{\text{JA},R}(\Delta_{12}) = \pi R^2\quad \mbox{for $\Delta_{12} >
      2R$}
  \end{align}
\end{subequations}
so that the integral converges for $\Delta_{12} \to 0$, and we can
place the upper limit at $\Delta_{12} = 2R$.

After evaluating \eq~(\ref{eq:ajfr}), with the replacement $\Delta_{12}
\to R$, both in the lower limit of the $p_{t2}$ integral and the
argument of the coupling,
we obtain
\begin{equation}
  \label{eq:delta-ajf-res}
  \langle \Delta a_{\text{JA},R}\rangle = d_{\text{JA},R} \frac{2\as C_1}{\pi} \ln
  \frac{R p_{t1}}{{Q_0}} \;, \qquad
  d_{\text{JA},R} = \int_0^{2R} \frac{d\theta}{\theta} ( a_{\text{JA},R}(\theta) -
  \pi R^2)\,,
\end{equation}
in a fixed coupling approximation, and 
\begin{equation}
  \label{eq:delta-ajf-res-running}
  \langle \Delta a_{\text{JA},R}\rangle = 
  d_{\text{JA},R} 
  \frac{C_1}{\pi b_0}
   \ln \frac{\as({Q_0})}{\as(R p_{t1})}\,,
\end{equation}
with a one-loop running coupling, where $b_0 = \frac{11C_A -
  2n_f}{12\pi}$. The approximation 
$\Delta_{12} \sim R$ in the argument of the running coupling in the
integrand corresponds to neglecting terms of $\order{\as}$
without any enhancements from logarithms of $R$ or $p_{t1}/{Q_0}$.

The results for $d_{\JA,R}$ are
\begin{subequations}\label{eq:area-scaling-passive-coefficient}
  \begin{align}
    d_{\text{anti-}k_t,R} &= 0,\\ 
    d_{k_t,R} &= \bigg(\frac{\sqrt{3}}{8} + \frac{\pi}{3} + \xi\bigg)
    R^2\,\simeq\,
    0.5638 \, \pi R^2\,,
    \\
    d_{\cam,R} &= \bigg(\frac{\sqrt{3}}{8} + \frac{\pi}{3} - 2\xi \bigg) R^2
                 \,\simeq\,
                 0.07918 \,\pi R^2\,,
    \\
    d_{\cone,R}
    &= \bigg(-\frac{\sqrt{3}}{8}+\frac{\pi }{6} - \xi\bigg) R^2 \,\simeq\, 
    -0.06378 \, \pi R^2\,,
  \end{align}
\end{subequations}
where
\begin{equation}
  \label{eq:xi}
  \xi \equiv
  \frac{\psi'(1/6)+\psi'(1/3)-\psi'(2/3)-\psi'(5/6)}{48\sqrt{3}}
  \,\simeq\, 0.507471\,,
\end{equation}
with $\psi'(x) = d^2/dx^2 (\ln \Gamma(x))$. One notes that the
coefficient for the $k_t$ algorithm is non-negligible, given that it
is multiplied by the quantity $2\as C_1/\pi \ln R p_{t1}/{Q_0}$ in
\eq~(\ref{eq:delta-ajf-res}) (or its running coupling analogue), which
can be of order $1$. In contrast the coefficients for Cambridge/Aachen
and SISCone are much smaller and similar (the latter being however of
opposite sign).  Thus $k_t$ areas will depend significantly more on
the jet $p_t$ than will those for the other algorithms.
In contrast, the soft-resilience of the anti-$k_t$ jet algorithm is
manifest here with an area of $\pi R^2$ independently of $p_t$.

The fluctuation of the areas (over the emission $p_2$) can be
calculated in a similar way. Let us define
\begin{equation}
  \label{eq:passive-fluct-decomp}
  \langle \sigma^2_{\JA,R} \rangle = \langle a_{\text{JA},R}^2\rangle
  - \langle a_{\text{JA},R}\rangle^2 = \sigma^2_{\JA,R}(0) + \langle
  \Delta \sigma^2_{\JA,R} \rangle\,,\qquad\quad \sigma^2_{\JA,R}(0) = 0\,,
\end{equation}
where we have introduced $\sigma^2_{\JA,R}(0)$, despite its being null,
so as to facilitate comparison with later results. We then evaluate
\begin{equation}
\label{eq:fluct}
\langle
  \Delta \sigma^2_{\JA,R} \rangle=
\langle \Delta a_{\text{JA},R}^2\rangle -
\langle \Delta a_{\text{JA},R}\rangle^2 \simeq \langle \Delta a_{\text{JA},R}^2\rangle \, ,
\end{equation}
where we neglect $\langle \Delta a_{\text{JA},R}\rangle^2$ since it is of 
$\order{\as^2\ln^2(R p_{t1}/{Q_0})}$.
The calculation of $\langle \Delta a_{\text{JA},R}^2\rangle$ proceeds
much as for $\langle \Delta a_{\text{JA},R}\rangle$ and gives
\begin{equation}
  \label{eq:delta-ajf-res2}
  \langle \Delta a_{\text{JA},R}^2\rangle =
  s_{\text{JA},R}^2\frac{C_1}{\pi b_0}
  \ln \frac{\as({Q_0})}{\as(R p_{t1})}\;, \qquad
  s_{\text{JA},R}^2 = \int_0^{2R} \frac{d\theta}{\theta} ( a_{\text{JA},R}(\theta) -
  \pi R^2)^2\,
\end{equation}
for running coupling.
The results are
\begin{subequations}\label{eq:fluct_passive_coefs}
  \begin{align}
    s_{\text{anti-}k_t,R}^2 &= 0,\\ 
    s_{k_t,R}^2 &=  \bigg( \frac{\sqrt{3}\pi}{4} - \frac{19}{64}
    -\frac{15\zeta(3)}{8} + 2\pi \xi \bigg) R^4 \,\simeq\, 
    (0.4499 \, \pi R^2)^2\,,
    \\
    s_{\cam,R}^2 &=  \bigg( \frac{\sqrt{3}\pi}{6} -\frac{3}{64} -
    \frac{\pi^2}{9} -\frac{13\zeta(3)}{12} +\frac{4\pi}{3} \xi
    \bigg) R^4 \,\simeq\, 
    (0.2438\, \pi R^2)^2\,,
    \\
    s_{\cone,R}^2
    &=  \bigg( \frac{\sqrt{3}\pi}{12} -\frac{15}{64} -\frac{\pi^2 }{18}
    -\frac{13\zeta(3)}{24} +\frac{2\pi}{3} \xi \bigg) R^4 \,\simeq\, 
    (0.09142\, \pi R^2)^2\,.
  \end{align}
\end{subequations}
We have a hierarchy between algorithms that is similar to that
observed for the average area scaling violations, with anti-$k_t$
showing no deviation from $\pi R^2$. The coefficient for
Cambridge/Aachen is now more intermediate between the $k_t$ and
SISCone values.

\subsection{Active areas}
\label{sec:areamed-areaanalytics-active}

\subsubsection{Areas for 1-particle  configurations and for ghost jets}
\label{sec:areamed-areaanalytics-active_1point}

This time, we consider a configuration with a single hard particle and
an ensemble of ghosts, with the ghosts taking active part in the
clustering.

\begin{figure}
\centering
  \includegraphics[width=0.48\textwidth]{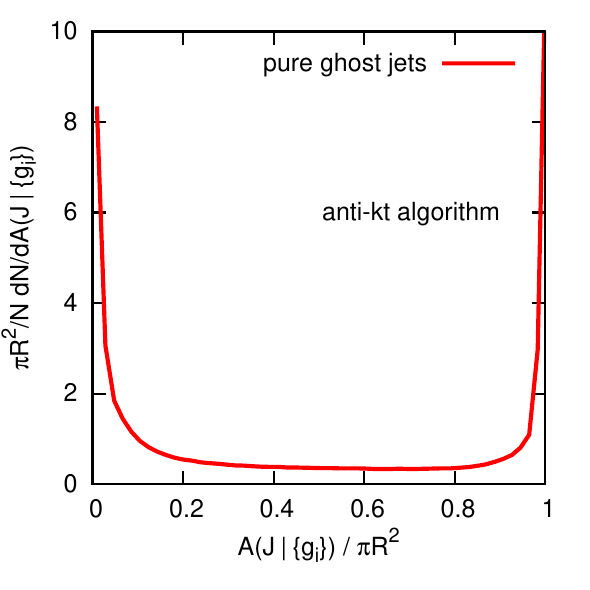}
\caption{Distribution of the active area for pure-ghost jets obtained
  with the anti-$k_t$ algorithm.}\label{fig:antikt-pure-ghost}
\end{figure}

\paragraph{Anti-$\boldsymbol{\kt}$.}
The shortest anti-$k_t$ distance will be between the hard particle and
either a ghost or the beam, depending on the pure geometrical
distance.
The hard particle $p_1$ will therefore cluster with all the ghosts
within a geometrical distance $R$ from it before it becomes a jet (by
clustering with the beam).
Ghost--ghost distances are much larger, enhanced by a factor $p_t^2/g_t^2$.
The active area will therefore be $\pi R^2$ (with no fluctuations),
once again a direct consequence of the soft resilience of the
anti-$k_t$ algorithm:
\begin{subequations}\label{eq:area-hard-particle-antikt}
\begin{align}
    A_{\antikt,R}  (\text{one-particle-jet}) &\,=\, \pi R^2\,,\\
    \Sigma_{\antikt,R} (\text{one-particle-jet}) &\,=\, 0\,.
\end{align}
\end{subequations}

Beside the hard jets, there will also be pure-ghost jets. 
Their distribution is most readily studied numerically, by directly
clustering large numbers of ghost particles, with or without
additional hard particles.
Typically we add ghosts with a density $\nu_g$ of $\sim 100$ per unit
area,\footnote{They are placed on a randomly scattered grid, in order
  to limit the impact of the finite density, \ie one effectively
  carries out quasi Monte Carlo integration of the ghost ensembles, so
  that that finite density effects ought to vanish as $\nu_g^{-3/4}$,
  rather than $\nu_g^{-1/2}$ as would be obtained with completely
  random placement. }
in the rapidity region $|y| < y_{\text{max}} = 6$, and study jets in
the region $|y| < y_{\text{max}}-R$. This leads to about 7500 ghost
particles. Each ghost is given a transverse momentum
$\sim 10^{-100}\GeV$.

The resulting distribution of the anti-$k_t$ active areas
$A(J \,|\, {\{g_i\}})$ is presented in
Fig.~\ref{fig:antikt-pure-ghost} for a large ensemble of sets of
ghosts.\footnote{In this particular case we have used about $10^7$
  separate random ghost sets, in order to obtain a smooth curve for
  the whole distribution. When calculating areas in physical events
  (or even at parton-shower level) the multiple real particles in the
  jet ``fix'' most of the area, and between 1 and 5 sets of ghosts
  particles are usually sufficient to obtain reliable area results
  (this is the case also for the other algorithms studied below).}
This distribution appears to be dominated by jets which are either
almost circular and of area close to $\pi R^2$ and jets of much
smaller area. We observe about 30\% of the jets with an area smaller
than $0.1\pi R^2$ and about 30\% with an area larger than
$0.95\pi R^2$.
This distribution can be understood in the following way: pure-ghost
jets will only be clustered once all the hard jets have been
clustered. That clustering will start aggregating ghosts, giving them
a larger $p_t$ than the other ghosts, seeding a process of further
aggregating the ghosts in the vicinity of that original cluster, in a
way similar to the clustering of the hard jets. This would give a peak
close to $A=\pi R^2$. As this goes on, the space between the jets
becomes smaller and small pure-ghost jets, of area close to 0, fill
the spaces between the jets of area close to $\pi R^2$.

Note that this distribution justifies why anti-$k_t$ jets are not
suited for the area--median estimation of the pileup density
$\rho$. The patches used for pileup estimation would follow a
distribution close to that of the pure-ghost jets --- potentially with
an additional bias coming from pileup fluctuations --- and the large
number of small jets would lead to a large uncertainty on the
estimation of $\rho$.

\begin{figure}
  \centering
  \includegraphics[width=0.48\textwidth]{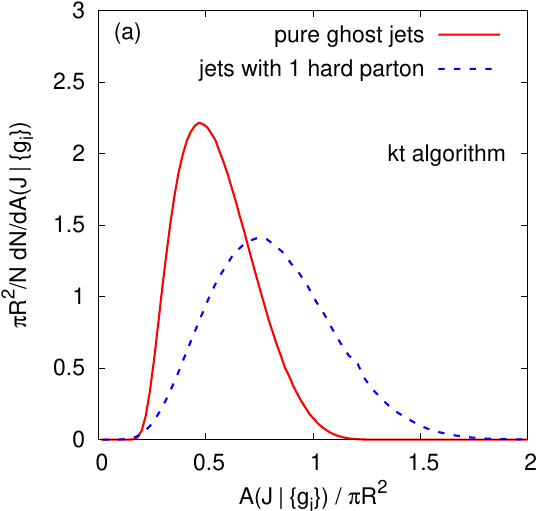}\hfill
  \includegraphics[width=0.48\textwidth]{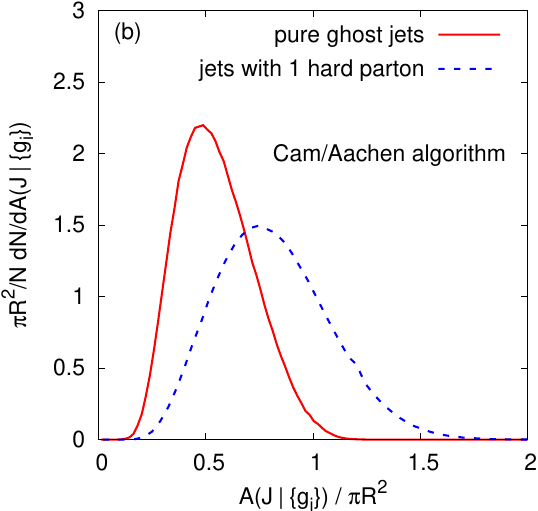}
  \caption{Distribution of active areas for pure ghost jets and
    jets with a single hard particle: (a) $k_t$ algorithm, (b)
    Cambridge/Aachen algorithm. }
  \label{fig:ghost-areas}
\end{figure}

\paragraph{$\boldsymbol{\kt}$ and Cambridge/Aachen.}
%
The active area for the $k_t$ and Cambridge/Aachen algorithms is again
best studied numerically. On top of the ghosts distributed as
discussed above, we added one hard particle with a transverse momentum
of $100\GeV$ that we study.
The results are insensitive to the transverse momentum as long as it
is sufficiently larger than the ghost $p_t$.
We briefly investigate in Appendix~\ref{sec:transition} how the
distribution of the ``1-hard-parton'' jet area gets modified when the
transverse momentum of the parton is progressively reduced below the
scale of a generic set of soft particles.
Figure~\ref{fig:ghost-areas} shows the distribution of values of $A(J
\,|\, {\{g_i\}})$ for pure ghost jets and jets with one hard particle.
Let us concentrate initially on the case with a hard particle. Firstly,
the average active area, \eq~(\ref{eq:act_area}) differs noticeably
from the passive result of $\pi R^2$:
\begin{subequations}
\label{eq:area-hard-particle}
  \begin{align}
    A_{k_t,R}  (\text{one-particle-jet}) &\,\simeq\, 0.812 \,\pi R^2\,, \\
    A_{\cam,R} (\text{one-particle-jet}) &\,\simeq\, 0.814 \,\pi R^2\,.
  \end{align}
\end{subequations}
Secondly, the distributions of the area in Fig.~\ref{fig:ghost-areas}
are rather broad. The randomness in the initial distribution of ghosts
propagates all the way into the shape of the final jet and hence its
area. This occurs because the $k_t$ and Cambridge/Aachen algorithms
flexibly adapt themselves to local structure (a good property when
trying to reconstruct perturbative showering), and once a random
perturbation has formed in the density of ghosts this seeds further
growth of the soft part of the jet.
The standard deviations of the resulting distributions are
\begin{subequations}
  \label{eq:ktcamsigmas}
  \begin{align}
    \Sigma_{k_t,R}  (\text{one-particle-jet}) &\,\simeq\, 0.277 \,\pi R^2\,, \\
    \Sigma_{\cam,R} (\text{one-particle-jet}) &\,\simeq\, 0.261 \,\pi R^2\,.
  \end{align}
\end{subequations}

Figure~\ref{fig:ghost-areas} also shows the distribution of areas for
pure ghost jets. One sees that pure ghost jets typically have a
smaller area than hard-particle jets:\footnote{Obtaining these values
  actually requires going beyond the ghost density and the rapidity
  range previously mentioned. In fact, when going to higher accuracy
  one notices the presence of small edge and finite-density effects,
  ${\cal O}(R/(y_{max}-R))$ and ${\cal O}(1/(\nu_g R^2))$ to 
  some given power.  Choosing
  the ghost area sufficiently small to ensure that finite-density
  effects are limited to the fourth decimal (in practice this
  means $1/(\nu_g R^2) < 0.01$) and extrapolating to infinite
  $y_{max}$ one finds
\begin{subequations}
  \begin{align}
    A_{k_t}  (\text{ghost-jet}) &\,\simeq\, (0.5535 \pm 0.0005) \,\pi R^2\,, \\
    A_{\cam} (\text{ghost-jet}) &\,\simeq\, (0.5505 \pm 0.0005) \,\pi R^2\,,
  \end{align}
\end{subequations}  
with a conservative estimate of the residual uncertainty.
This points to a small but statistically significant 
difference between the two algorithms.
}
\begin{subequations}
\label{eq:area-pure-ghost}
  \begin{align}
    A_{k_t,R}  (\text{ghost-jet}) &\,\simeq\, 0.554 \,\pi R^2\,, \\
    A_{\cam,R} (\text{ghost-jet}) &\,\simeq\, 0.551 \,\pi R^2\,,
  \end{align}
\end{subequations}
and the standard deviations are
\begin{subequations}
\label{eq:area-pure-ghost-stddev}
  \begin{align}
    \Sigma_{k_t,R}  (\text{ghost-jet}) &\,\simeq\, 0.174 \,\pi R^2\,, \\
    \Sigma_{\cam,R} (\text{ghost-jet}) &\,\simeq\, 0.176 \,\pi R^2\,.
  \end{align}
\end{subequations}
The fact that pure ghost jets are smaller than hard jets has an
implication for certain physics studies: one expects jets made of soft
`junk' (minimum bias, pileup, thermal noise in heavy ions) to have
area properties similar to ghost jets; since they are smaller on
average than true hard jets, the hard jets will emerge from the junk
somewhat more clearly than if both had the same area.

\paragraph{SISCone.}
%
The SISCone active area is amenable to analytical treatment, at least
in some cases.

\begin{figure}
  \centering
  \includegraphics[width=\textwidth]{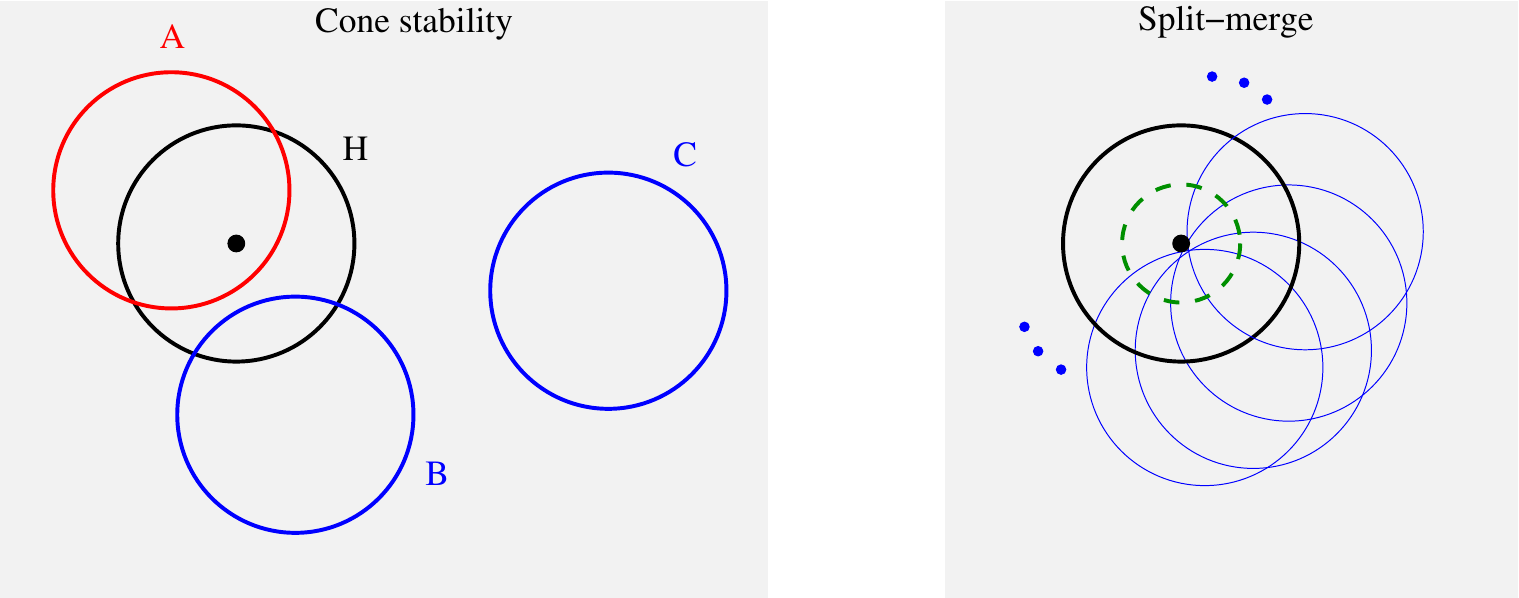}
  \caption{Left: the hard particle with the stable cone (H) centred
    on it, an example of a cone (A) that is unstable because it also
    contains the hard particle, and of two cones (B) and (C) that
    contain just ghost particles and are therefore stable. Right: some
    of the stable ghost cones (thin blue circles) that have the
    largest possible overlap with (H), together with the boundary of
    the hard jet after the split--merge procedure (dashed green line).
    In both diagrams, the grey background represents the uniform
    coverage of ghosts.  }
  \label{fig:cone-active-diag}
\end{figure}

We recall that SISCone starts by finding all stable cones.
A first stable cone is centred on the single hard particle. Additionally,
there will be a large number of other stable cones, of order of the
number of ghost particles added to the event~\cite{siscone}. In the
limit of an infinite number of ghosts, all cones that can be drawn in
the rapidity--azimuth plane and that do not overlap with the hard particle
will be stable. Many of these stable cones will still overlap with the
cone centred on the hard particle, as long as they do not contain
the hard particle itself (see figure~\ref{fig:cone-active-diag}, left).

Next, the SISCone algorithm involves a split--merge procedure. One defines 
$\tilde p_t$ for a jet as the scalar sum of the transverse momenta of its 
constituents. 
During the split--merge step, SISCone finds the stable cone with the
highest $\tilde p_t$, and then the next hardest stable cone that
overlaps with it. To decide whether these two cones (protojets) are to
be merged or split, it determines the fraction of the softer cone's
$\tilde p_t$ that is shared with the harder cone. If this fraction is
smaller than some value $f$ (the overlap parameter of the cone
algorithm), the two protojets are split: particles that are shared
between them are assigned to the protojet to whose
centre\footnote{This centre is given by the sum of momenta in the
  protojet before the split--merge operation.} they are closer.
Otherwise they are merged into a single protojet. This procedure is
repeated until the hard protojet no longer has any overlap with other
protojets. At this point it is called a jet, and the split--merge
procedure continues on the remaining protojets (without affecting the
area of the hard jet).

The maximum possible overlap fraction,\footnote{The fraction of
  momentum coincides with the fraction of area because the ghosts have
  uniform transverse momentum density.} $f_{\text{max}}$, between the
hard protojet and a ghost protojet occurs in the situation depicted in
figure~\ref{fig:cone-active-diag} (right), \ie when the ghost protojet's
centre is just outside the edge of the original hard stable cone (H).
It is given by $f_{\text{max}} =
2-u(1)=\frac{2}{3}-\frac{\sqrt{3}}{2\pi}\approx 0.391$. This means
that for a split--merge parameter $f>f_{\text{max}}$ (commonly used
values are $f = 0.5$ and $f = 0.75$) every overlap between the hard
protojet and a pure-ghost stable cone will lead to a splitting. Since
these pure-ghost stable cones are centred at distances $d > R$
from the hard particle, these splittings will reduce the hard jet to a
circle of radius $R/2$ (the dashed green line in the right hand part
of figure~\ref{fig:cone-active-diag}). The active area of the hard jet
is thus
\begin{equation}\label{eq:active_1point_cone}
A_{\cone,R}(\text{one-particle-jet}) = \frac{\pi R^2}{4} \,.
\end{equation}
This result has been verified numerically using the same technique
employed above for $k_t$ and Cambridge/Aachen.

Note that this area differs considerably from the passive area,
$\pi R^2$. This shows that the cone area is very sensitive to the
structure of the event, and it certainly does not always coincide with
the naive geometrical expectation $\pi R^2$, contrary to assumptions
sometimes made in the literature (see for
example~\cite{Bhatti:2005ai}).

We further note that in contrast to $k_t$ and Cambridge/Aachen
algorithms, the SISCone algorithm always has the same active area for a
single hard particle, independently of fluctuations of an infinitely
dense set of ghosts, \ie
\begin{equation}\label{eq:active_1point_cone_sigma}
  \Sigma_{\cone,R}(\text{one-particle-jet}) = 0 \,,
\end{equation}
a property that SISCone shares with the anti-$k_t$ algorithm.

\begin{figure}
  \centering
  \includegraphics[height=0.47\textwidth]{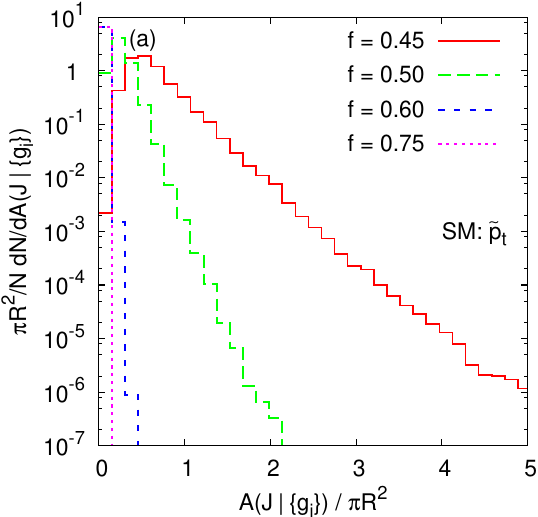}\hfill
  \includegraphics[height=0.47\textwidth]{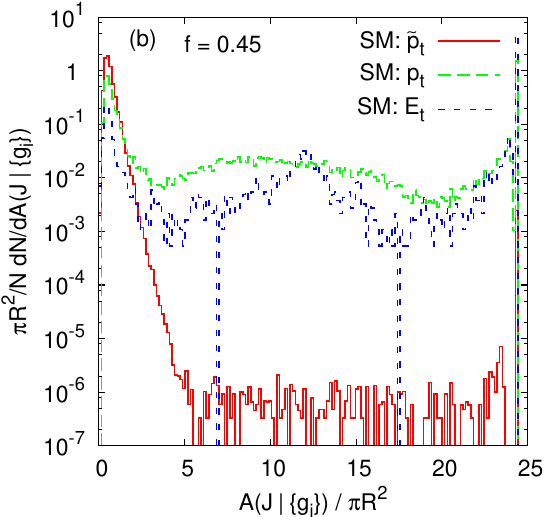}
  \caption{Distribution of pure-ghost jet areas for the SISCone algorithm,
    (a) with different values of the split--merge parameter $f$ and
    (b) different choices for the scale used in the split--merge
    procedure.
    Ghosts are placed on a grid up to $|y|<6$, with an average spacing
    of $0.2\times 0.2$ in $y$, $\phi$, and a random displacement of up
    to $\pm 0.1$ in each direction, and transverse momentum values
    that are uniform modulo a $\pm 5\%$ random rescaling for each
    ghost. We consider all jets with $y < 5$. All jet definitions use
    $R=1$ and multiple passes.}
  \label{fig:siscone-ghost-areas}
\end{figure}

\paragraph{SISCone ghost-jet areas.}
While the area of a hard particle jet could be treated analytically,
this is not the case for the pure-ghost jet area. Furthermore,
numerical investigations reveal that the pure-ghost
area distribution has a much more complicated behaviour than for
$k_t$ or Cambridge/Aachen. One aspect is that the distribution of
pure-ghost jet areas is sensitive to the fine details of how
the ghosts are distributed (density and transverse momentum
fluctuations). Another is that it depends significantly on the details
of the split--merge procedure. Figure~\ref{fig:siscone-ghost-areas}a
shows the distribution of areas of ghost jets for SISCone, for
different values of the split--merge overlap threshold $f$. One sees,
for example,
that for smaller values of $f$ there are occasional rather large ghost
jets, whereas for $f\gtrsim 0.6$ nearly all ghost jets have very small
areas.

One of the characteristics of SISCone that differs from previous cone
algorithms is the specific ordering and comparison variable used to
determine splitting and merging. As explained above, 
the choice made in SISCone was $\tilde
p_t$, the scalar sum of transverse momenta of all particles in a jet.
Previous cone algorithms used either the vector sum of constituent
transverse momenta, $p_t$ (an infrared unsafe choice), or the
transverse energy $E_t = E \sin \theta$ (in a 4-vector recombination
scheme). With both of these choices of variable, split--merge
thresholds $f\lesssim 0.55$ can lead to the formation of `monster'
ghost jets, which can even wrap around the whole cylindrical phase
space.  For $f=0.45$ this is a quite frequent occurrence, as
illustrated in figure~\ref{fig:siscone-ghost-areas}b, where one sees a
substantial number of jets occupying the whole of the phase space (\ie
an area $4\pi y_{\max} \simeq 24 \pi R^2$). Monster jets can be formed
also with the $\tilde p_t$ choice, though it is a somewhat rarer
occurrence --- happening in `only' $\sim 5\%$ of events.\footnote{This
  figure is not immediately deducible from
  Fig.~\ref{fig:siscone-ghost-areas}b, which shows results normalised
  to the total number of ghost jets, rather than to the number of
  events.}

We have observed the formation of such monster jets also from normal
pileup momenta simulated with
Pythia~\cite{Sjostrand:2000wi,Sjostrand:2003wg,Sjostrand:2006za},
indicating that this disturbing characteristic is not merely an
artefact related to our particular choice of ghosts. This indicates
that a proper choice of the split--merge variable and threshold is
critical in high-luminosity environments. The results from
Figure~\ref{fig:siscone-ghost-areas}a, suggest that if one wants to
avoid monster jets, one has to choose a large enough value for
$f$. Our recommendation is to adopt $f=0.75$ as a default value for
the split--merge threshold (together with the use of the $\tilde p_t$
variable, already the default in SISCone, for reasons related to
infrared safety and longitudinal boost invariance).

\subsubsection{Areas for 2-particle configurations}
\label{sec:areamed-areaanalytics-active_2point}

In this section we study the same problem described in
section~\ref{sec:areamed-areaanalytics-passive-area-2particle}, \ie the area of jets
containing two particles, a hard one and a softer (but still
``perturbative'') one, \eq~(\ref{eq:strord}), but now for active
areas. As before, the results will then serve as an input in
understanding the dependence of the active area on the jet's
transverse momentum when accounting for perturbative radiation.

\paragraph{Anti-$\boldsymbol{k_t}$.}
%
With little surprise, the addition of the soft particle $p_2$ will not
affect the area of the hard jet. The minimal anti-$k_t$ distance will
be between particle $p_1$ and the closest of $p_2$ and the (remaining)
ghosts, until that geometric distance reaches $R$ and the jet is
clustered with the beam. The resulting area will be $\pi R^2$,
\begin{equation}\label{eq:active_2points_antikt}
A_{\antikt,R}(\Delta_{12}) = \pi R^2,
\end{equation}
with no fluctuations
\begin{equation}\label{eq:fluct_active_2points_antikt}
\Sigma_{\antikt,R}(\Delta_{12}) = 0,
\end{equation}
independently of the distance $\Delta_{12}$ between particles $p_1$
and $p_2$.

\paragraph{$\boldsymbol{k_t}$ and Cambridge/Aachen.}
%
As was the case for the active area of a jet containing a single hard
particle, we again have to resort to numerical analyses to study that
of jets with two energy-ordered particles. We define
$A_{\JA,R}(\Delta_{12})$ to be the active area for the energy-ordered
two particle configuration already discussed in section
\ref{sec:areamed-areaanalytics-passive-area-2particle}. 

Additionally since we have a distribution of areas for the
single-particle active area case, it becomes of interest to study also
$\Sigma_{\JA,R}(\Delta_{12})$ the standard deviation of the
distribution of areas obtained for the two-particle configuration.

\begin{figure}[t]
  \centering
  \includegraphics[width=0.6\textwidth]{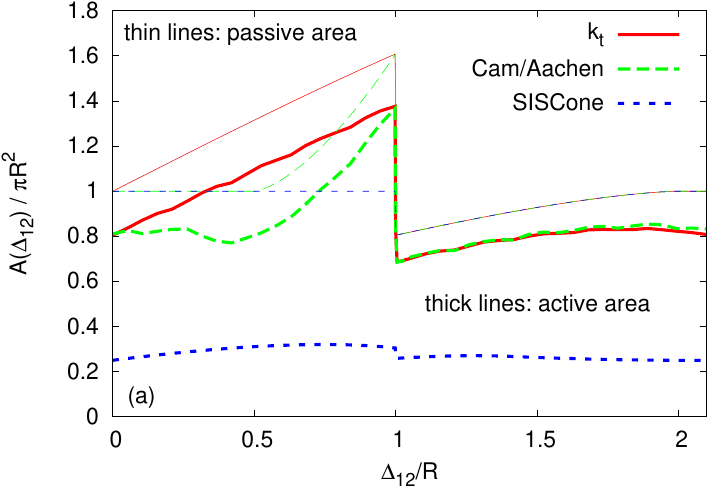}
  \includegraphics[width=0.6\textwidth]{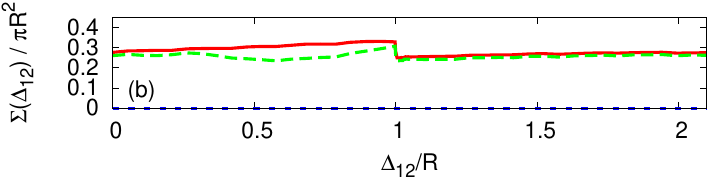}
  \caption{(a) Active areas (divided by $\pi R^2$) for the three jet
    algorithms as a function of the separation between a hard and a
    softer particle. For comparison we also include the passive areas,
    previously shown in Fig.~\ref{fig:2point}. (b) The corresponding
    standard deviations.}
  \label{fig:2point-areas-active-3alg}
\end{figure}

The results are shown in figure~\ref{fig:2point-areas-active-3alg}:
the active areas can be seen to be consistently smaller than the
passive ones, as was the case for the 1-particle area, but retain the
same dependence on the angular separation between the two particles.
Among the various features, one can also observe
that the active area does not quite reach the single-particle value
($\simeq 0.81 \pi R^2$) at $\Delta_{12}=2R$, but only somewhat beyond
$2R$.
This contrasts with the behaviour of the passive area.
The figure also shows results for the cone area, discussed in the
following subsection. 

\paragraph{SISCone}

In the case of the SISCone algorithm it is possible to find an analytical
result for the two-particle active area, in an extension of what was
done for one-particle case.\footnote{This is only possible for
  configurations with strong energy ordering between all particles ---
  as soon as 2 or more particles have commensurate transverse momenta
  then the cone's split--merge procedure will include `merge' steps,
  whose effects on the active area are currently beyond analytical
  understanding.}

The stable-cone search will find one or two ``hard'' stable cones: the
first centred on the hard particle and the second centred on the soft
one, present only for $\Delta_{12}>R$. The pure-ghost stable cones
will be centred at all positions distant by more than $R$ from both
$p_1$ and $p_2$, \ie outside the two circles centred on $p_1$ and
$p_2$.

We shall consider the active area of the jet centred on the hard particle
$p_1$. When $\Delta_{12}>R$, the jet centred on the soft particle has
the same area.

\begin{figure}
\includegraphics[scale=0.50]{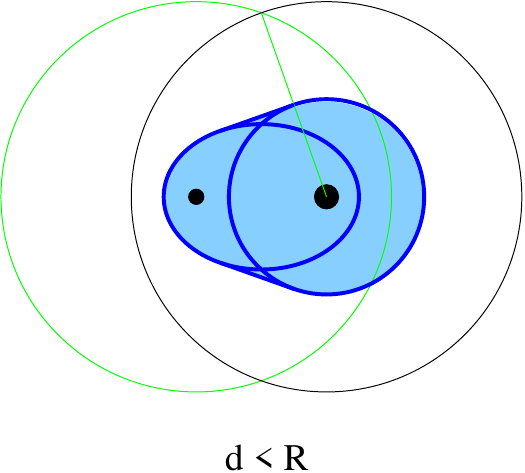}\;\;
\includegraphics[scale=0.50]{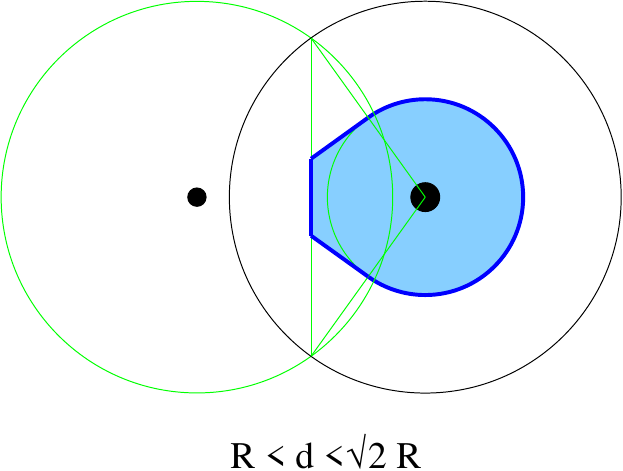}\;\;
\includegraphics[scale=0.50]{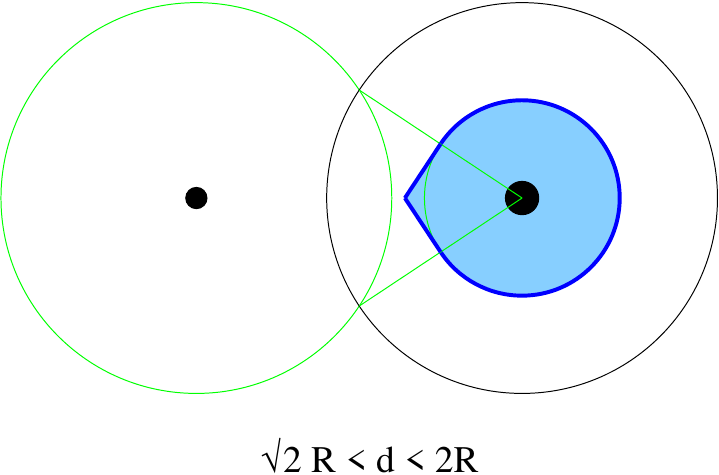}
\caption{Picture of the active jet area for 2-particle configurations
  in the case of the SISCone jet algorithm. The black points represent
  the hard (big dot) and soft (small dot) particles, the black circle
  is the hard stable cone. The final hard jet is represented by the
  shaded area. The left (a) (centre (b), right (c)) plot corresponds
  to $\Delta_{12}<R$ ($R<\Delta_{12}<\sqrt{2}R$,
  $\sqrt{2}R<\Delta_{12}<2R$).}
\label{fig:active_2point_cone_pict}
\end{figure}

As in section~\ref{sec:areamed-areaanalytics-active_1point}, the
split--merge procedure first deals with the pure-ghost protojets
overlapping with the hard stable cone. For $f>f_{\text{max}}$, this
again only leads to splittings. Depending of the value of
$\Delta_{12}$, different situations are found as shown on figure
\ref{fig:active_2point_cone_pict}. If the $y$-$\phi$ coordinates of
the particles are $p_1\equiv(\Delta,0)$ and $p_2\equiv (0,0)$, the
geometrical objects that are present are: the circle centred on $p_1$
with radius $R/2$; the tangents to this circle at $y$-$\phi$
coordinates $(\Delta_{12}/4,\pm\frac12\sqrt{R^2-(\Delta_{12}/2)^2})$;
and, for $\Delta_{12}<R$, the ellipse of eccentricity $\Delta_{12}/R$
whose foci are $p_1$ and $p_2$ (given by the equation,
$\Delta_{a1} + \Delta_{a2} = R$, where $a$ is a point on the
ellipse). For $\Delta_{12}<R$ the area of the jet is given by the
union of the ellipse, the circle and the regions between the ellipse,
the circle and each of the tangents
(Fig.~\ref{fig:active_2point_cone_pict}a). For
$R <\Delta_{12}<\sqrt2 R$ it is given by the circle plus the region
between the circle, the two tangents and the line equidistant from
$p_1$ and $p_2$ (Fig.~\ref{fig:active_2point_cone_pict}b). For
$\sqrt2 R <\Delta_{12}<2 R$ it is given by the circle plus the region
between the circle and two tangents, up to the intersection of the two
tangents (Fig.~\ref{fig:active_2point_cone_pict}c). Finally, for
$\Delta\ge 2R$, the area is $\pi R^2/4$.

\begin{figure}
\centerline{\includegraphics{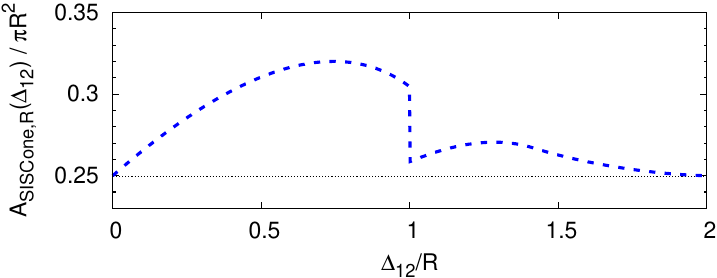}}
\caption{Active area of the hardest jet as a function of the distance
  between the hard and soft particle for the SISCone algorithm, \cf
  \eq~(\ref{eq:active_2points_cone}).}
\label{fig:active_2point_cone}
\end{figure}

An analytic computation of the active area gives
\begin{eqnarray}\label{eq:active_2points_cone}
  \frac{A_{\cone,R}(\Delta_{12})}{\pi R^2}
  & = & \frac{1}{4}
  \left[1-\frac{1}{\pi}\arccos\left(\frac{x}{2}\right)\right] \\
  & + & \begin{cases}
    \frac{x}{2\pi}\sqrt{1-\frac{x^2}{4}}
    +\frac{1}{4\pi}\sqrt{1-x^2}\arccos\left(\frac{x}{2-x^2}\right)
    & \Delta_{12}<R \\
    \frac{x}{2\pi}\sqrt{1-\frac{x^2}{4}}
    -\frac{x}{8\pi}\frac{1}{\sqrt{1-\frac{x^2}{4}}}
    & R<\Delta_{12}\le \sqrt{2}R \\
    \frac{1}{2\pi x}\sqrt{1-\frac{x^2}{4}}
    & \sqrt{2}R < \Delta_{12} \le 2R
  \end{cases}.\nonumber
\end{eqnarray}
with $x=\Delta_{12}/R$, while for $\Delta_{12}>2R$, one recovers the
result $\pi R^2/4$. The SISCone active area is plotted in
figure~\ref{fig:active_2point_cone} and it is compared to the results
for $k_t$ and Cambridge/Aachen in
figure~\ref{fig:2point-areas-active-3alg}. One notes that the SISCone
result is both qualitatively and quantitatively much further from the
passive result than was the case for $k_t$ and Cambridge/Aachen.

The main reason why the 2-point active area is larger than the 1-point
active area (whereas we saw the opposite behaviour for the passive
areas) is that the presence of the 2-particle configuration causes a
number of the pure-ghost cones that were present in the 1-particle case
to now contain the second particle and therefore be unstable. Since
these pure ghost cones are responsible for reducing the jet area
relative to the passive result (during the split--merge step), their
absence causes the active area to be `less-reduced' than in the
1-particle case.


\subsubsection{Area scaling violation}
\label{sec:areamed-areaanalytics-area-scal-viol-active}

We can write the order $\as$ contribution to the average active area
in a manner similar to the passive area case,
\eq~(\ref{eq:ajfr}):
\begin{equation}
  \label{eq:Ajfr-base}
    \langle A_{\text{JA},R}\rangle =  
    A_{\text{JA},R}(0) + \langle \Delta
    A_{\text{JA},R}\rangle\,, 
\end{equation}
where we have used the relation
$A_{\text{JA},R}(\text{one-particle-jet})\equiv A_{\text{JA},R}(0)$
and where
\begin{equation}
  \label{eq:Ajfr}
  \langle \Delta  A_{\text{JA},R}\rangle \simeq \int_0 d\Delta_{12} 
  \int_{{Q_0}/\Delta_{12}}^{p_{t1}} d p_{t2} 
  \frac{dP}{dp_{t2} \, d\Delta_{12}} (
  A_{\text{JA},R}(\Delta_{12}) 
  - A_{\text{JA},R}(0))\,.
\end{equation}
Note that compared to \eq~(\ref{eq:ajfr}) we have removed the explicit
upper limit at $2R$ on the $\Delta_{12}$ integral since, for active
areas of some sequential recombination algorithms,
$(A_{\text{JA},R}(\Delta_{12}) - A_{\text{JA},R}(0))$ may be 
non zero even for $\Delta_{12}>2R$.
Note also the notation for averages: we use $\avg{\cdots}$ to
refer to an average over perturbative emissions, while
$\avg{\cdots}_g$, implicitly contained in $A_{\JA,R}$ (see 
\eq~(\ref{eq:act_area})), refers to an average over ghost ensembles.
We now proceed as for the passive area scaling violations and write
\begin{equation}
  \label{eq:delta-Ajf-res}
  \langle \Delta A_{\text{JA},R}\rangle \simeq
  D_{\text{JA},R} \frac{C_1}{\pi b_0}
  \ln \frac{\as({Q_0})}{\as(R p_{t1})}\;, \qquad
  D_{\text{JA},R} = \int_0 \frac{d\theta}{\theta} ( 
  A_{\text{JA},R}(\theta)
  -  A_{\text{JA},R}(0))\,,
\end{equation}
where for brevity we have given just the running-coupling result.
One observes that the result continues to depend on $Q_0$, an
indication that the active area is an infrared-unsafe quantity, just
like the passive area.
The coefficients for the anomalous dimension of the  active area for
the various algorithms are
\begin{subequations}
  \label{eq:Dalg1}
  \begin{align}
    D_{\antikt,R} &\simeq 0\,,\\
    D_{k_t,R}    &\simeq 0.52 \, \pi R^2\,,\\
    D_{\cam,R}   &\simeq 0.08 \, \pi R^2\,, \\
    D_{\cone,R}  &\simeq 0.1246 \, \pi R^2 \,,
  \end{align}
\end{subequations}
where the anti-$k_t$ and SISCone results has been obtained by
integrating the analytical result,
\eqs~(\ref{eq:active_2points_antikt}) and
(\ref{eq:active_2points_cone}) respectively, while the results for
$k_t$ and Cambridge/Aachen have been obtained both by integrating the
2-point active-area results shown in
Fig.~\ref{fig:2point-areas-active-3alg} and by a direct Monte Carlo
evaluation of \eq~(\ref{eq:delta-Ajf-res}). Note that while the
coefficients for $k_t$ and Cambridge/Aachen are only slightly
different from their passive counterparts, the one for SISCone has the
opposite sign relative to the passive one.

The treatment of higher-order fluctuations of active areas is more
complex than that for the passive ones, where the one-particle area
was a constant. We can separate the fluctuations of active areas into
two components, one (described above) that is the just the
one-particle result and the other, $\langle \Delta \Sigma^2_{\JA,R}
\rangle$, accounting for their modification in the presence of
perturbative radiation:
\begin{equation}
  \label{eq:act-fluct-seq}
  \langle \Sigma^2_{\JA,R} \rangle =
  \Sigma^2_{\JA,R}(0) + 
  \langle \Delta \Sigma^2_{\JA,R} \rangle \,,
\end{equation}
where $\Sigma_{\JA,R}(0)$ is given by \eqs~(\ref{eq:ktcamsigmas}a,b)
for the $k_t$ and the Cambridge/Aachen algorithms respectively, and it
is equal to zero for the anti-$k_t$ and SISCone algorithms.
The perturbative modification
$\langle \Delta \Sigma^2_{\JA,R} \rangle$ is itself now driven by two
mechanisms: the fact that the second particle causes the average area
to change, and that it also causes modifications of the fluctuations
associated with the sampling over many ghost sets. We therefore write
\begin{align}
  \label{eq:sigma2_pt}
  \langle \Delta \Sigma^2_{\text{JA},R}\rangle &\simeq
  S_{\text{JA},R}^2 \frac{C_1}{\pi b_0}
  \ln \frac{\as({Q_0})}{\as(R p_{t1})}
  \;,\\ \qquad
  \label{eq:sigma2_pt_mid}
  S_{\text{JA},R}^2 &= \int_0 \frac{d\theta}{\theta} \big[
  (A_{\text{JA},R}(\theta) -  A_{\text{JA},R}(0))^2 + 
  \Sigma^2_{\text{JA},R}(\theta) -  \Sigma^2_{\text{JA},R}(0)
  \big]\, \\
  &= \int_0 \frac{d\theta}{\theta} 
  (A^2_{\text{JA},R}(\theta) -  A^2_{\text{JA},R}(0))
  -2 A_{\text{JA},R}(0) D_{\text{JA},R}\, ,
  \label{eq:sigma2_pt_end}
\end{align}
where as usual we neglect contributions that are not enhanced by any
logarithm or that are higher-order in $\as$. The details of how to
obtain these results are given in Appendix~\ref{sec:app_fluct}.

The coefficient $S_{\text{JA},R}^2$ can  be determined only numerically for the $k_t$ and
Cambridge/Aachen algorithms, while for the SISCone algorithm the result
can be deduced from \eq~(\ref{eq:active_2points_cone}) together with
the knowledge that $\Sigma_{\cone,R}(\theta) \equiv 0$:
\begin{subequations}
  \label{eq:Sigmaalg1}
  \begin{align}
    {S}_{\antikt,R}^2 &\simeq 0\,,\\ 
    {S}_{\kt,R}^2 &\simeq (0.41 \, \pi R^2)^2\,,\\ 
    {S}_{\cam,R}^2 &\simeq  (0.19 \, \pi R^2)^2\,,\\ 
    {S}_{\cone,R}^2 &\simeq  (0.0738\, \pi R^2)^2\,.
  \end{align}
\end{subequations}
Again, both the values and their ordering are similar to what we have
obtained for the passive areas (see \eq~\eqref{eq:fluct_passive_coefs}).

\subsubsection{$\bs n$-particle properties and  large-$\bs n$ behaviour}
\label{sec:areamed-areaanalytics-n-particle-active}

\paragraph{Anti-$\bs{k_t}$ algorithm.}
%
If we consider a situation with a single hard particle $p_1$, and multiple
soft emissions $p_i$, $i\ge 2$, with $p_{t1}\gg p_{ti}$, the situation
for the anti-$k_t$ jet areas does not change compared to what we have
seen so far, with both the passive and active areas of the jet
containing $p_1$ being simply $\pi R^2$.
This is the main property of the anti-$k_t$ algorithm, namely that the
hard (isolated) jets are circular. Beside the fact that this
soft-resilience makes it a nice algorithm in an experimental
framework, it also has important consequences when studying soft gluon
emissions analytically in QCD. For example, the calculation of
non-global logarithm contributions will be considerably simplified and
the Milan factor will take the ``universal'' value, 1.49 for 3 active
non-perturbative flavours.

In the limit where multiple particles are hard and become close to
each other, the anti-$k_t$ areas can depart from $\pi R^2$. We shall
comment a bit more on this below, when we discuss back-reaction.

\paragraph{$\bs{k_t}$ algorithm.}
%
As for the passive area, the $k_t$ algorithm's active area has the
property that it can be expressed as a sum of individual particle
areas:
\begin{equation}
  \label{eq:particle-areas-active}
  A_{\kt,R}(J) = \sum_{p_i \in J} A_{\kt,R}(p_i)\,.
\end{equation}
This is the case because the presence of the momentum scale $k_t$ in
the distance measure means that all ghosts cluster among themselves
and with the hard particles, before any of the hard particles start
clustering between themselves. However in contrast to the passive-area
situation, there is no known simple geometrical construction for the
individual particle area.

\paragraph{Equivalence of all areas for large $\bs{n}$}

The existence of different area definitions is linked to the ambiguity
in assigning `empty space' to any particular jet. In the presence of a
sufficiently large number of particles $n$, one expects this ambiguity to
vanish because real particles fill up more and more of the empty space
and thus help to sharpen the boundaries between jets. Thus in the
limit of many particles, all sensible area definitions should give
identical results for a given jet.

To quantify this statement, we examine (a bound on) the
scaling law for the relation between the density of particles and the
magnitude of the potential differences between different area
definitions.
We consider the limit of `dense' coverage, defined as follows: take
square tiles of some size and use them to cover the rapidity--azimuth
cylinder (up to some maximal rapidity).  Define $\lambda$ as the
smallest value of tile edge-length such that all tiles contain at
least one particle. In an event consisting of uniformly distributed
particles, $\lambda$ is of the same order of magnitude as the typical
inter-particle distance. The event is considered to be dense if
$\lambda \ll R$.

Now let us define a boundary tile of a jet to be a tile that contains
at least one particle of that jet and also contains a particle from
another jet or has an adjacent tile containing one or more
particle(s) belonging to a different jet.
We expect that the difference between different jet-area definitions
cannot be significantly larger than
the total area of the boundary tiles for a jet. 

The number of boundary tiles for jets produced by a given jet
algorithm (of radius $R$) may scale in a non-trivial manner with the
inter-particle spacing $\lambda$, since the boundary may well have a
fractal structure. We therefore parametrise the average number of
boundary tiles for a jet, $N_{b,\JA,R}$ as
\begin{equation}
  \label{eq:Nb}
  N_{b,\JA,R} \sim \left(\frac{R}{\lambda}\right)^{{\!\daleth}_{\,\JA}},
\end{equation}
where the fractal dimension ${\!\daleth}_{\,\JA}=1$ would correspond to a smooth
boundary. The total area of these boundary tiles gives an upper limit
on the ambiguity of the jet area
\begin{equation}
  \label{eq:area-diff-upper-bound}
   \langle |a_{\JA,R} - A_{\JA,R}| \rangle \lesssim N_{b,\JA,R}\, \lambda^2 \sim 
      R^{{\!\daleth}_{\,\JA}} \lambda^{2-{\!\daleth}_{\,\JA}}\,,
\end{equation}
and similarly for the difference between active or passive and Voronoi
areas.
As long as ${\!\daleth}_{\,\JA} < 2$ the differences between various
area definitions are guaranteed to vanish in the infinitely dense
limit, $\lambda \to 0$.  We note that ${\!\daleth}_{\,\JA} = 2$
corresponds to a situation in which the boundary itself behaves like
an area, \ie occupies the same order of magnitude of space as the jet
itself. This would be visible in plots representing jet active areas
(such as Fig.~\ref{fig:example-active-area}), in the form of finely
intertwined jets. We have seen no evidence for this and therefore
believe that $1 \le {\daleth}_{\,\JA} < 2$ for all three jet
algorithms considered here.

In practice we expect the difference between any two area definitions
to vanish much more rapidly than \eq~(\ref{eq:area-diff-upper-bound}) as $\lambda \to 0$, since the
upper bound will only be saturated if, for every tile, the difference
between two area definitions has the same sign. This seems highly
unlikely. If instead differences in the area for each tile are
uncorrelated (but each of order $\lambda^2$) then one would expect to
see
\begin{equation}
  \label{eq:area-diff-guessed-bound}
   \langle |a_{\JA,R} - A_{\JA,R}| \rangle \sim \sqrt{N_{b,\JA,R}}\, \lambda^2 \sim 
      R^{{\!\daleth}_{\,\JA}/2} \lambda^{2-{\!\daleth}_{\,\JA}/2}\,.
\end{equation}
We have measured the fractal dimension for the $k_t$ and
Cambridge/Aachen algorithms and find ${\daleth}_{\,\kt} \simeq
{\!\daleth}_{\,\cam} \simeq 1.20-1.25$.\footnote{This has been
  measured on pure ghost jets, because their higher multiplicity
  facilitates the extraction of a reliable result, however we strongly
  suspect that it holds also for single-particle jets.}
Note that any measurement of the fractal dimension of jet algorithms
in real data would be severely complicated by additional structure
added to jets by QCD branching, itself also approximately fractal in
nature.

The fact that active and passive (or Voronoi) areas all give the same result
in dense events has practical applications in real-life situations where
an event is populated by a very large number of particles (heavy ion 
collisions being an example). In this case it will be possible to choose
the area type which is fastest to compute (for instance the Voronoi area) 
and  use the results in place of the active or passive one.

\subsection{Back reaction}
\label{sec:areamed-analytic-back-reaction}

So far we have considered how a set of infinitely soft ghosts clusters
with a hard jet, examining also cases where the jet has some
finitely-soft substructure. This infinitely-soft approximation for the
ghosts is not only adequate, but also necessary from the point of view
of properly defining jet areas.  
However if we are to understand the impact of pileup and
underlying-event radiation on jet finding --- the original motivation
for studying areas --- then we should take into account the fact that
these contributions provide a dense background of particles at some
small but \emph{finite} soft scale $\sim \Lambda_{QCD}$.

This has two main consequences. The first (more trivial) one is that
pileup or underlying event can provide an alternative, dynamic
infrared cutoff in the $p_{t2}$ integration in equations such
as~\eq~(\ref{eq:ajfr}): assuming that the density of PU transverse
momentum per unit area is given by $\rho$ then one can expect that
when $p_{t2} \ll \pi R^2 \rho$, the presence of $p_{2}$ will no longer
affect the clustering of the ghosts (\ie the PU particles). In this
case the $p_{t2}$ integral will then acquire an effective infrared
cutoff $\sim \pi R^2 \rho$ and in expressions such as
\eqs~(\ref{eq:delta-ajf-res},\ref{eq:delta-ajf-res-running}), $Q_0$
will be replaced by $\pi R^3
\rho$. 
Note that neither with infinitely nor finitely-soft ghosts do we claim
control over the coefficient in front of this cutoff, though we do
have confidence in the prediction of its $R$ dependence for small
$R$.

The second consequence of the finite softness of the PU contribution
is that the addition of the PU particles can modify the set of non-PU
particles that make it into a jet. This is the effect we called {\em
  back reaction} in our introductory discussion (see
Section~\ref{sec:description-convolution}).

That back reaction should happen is quite intuitive since it
concerns non-PU particles whose softness is commensurate with the PU
scale. However the extent to which it occurs depends significantly on
the jet algorithm. Furthermore for some algorithms it can also occur
(rarely) even for non-PU particles that are much harder than the PU
scale.

As with the studies of areas, there are two limits that can usefully
be examined for back-reaction: the illustrative and mathematically
simpler (but less physical) case of a pointlike minimum-bias
background and the more realistic case with diffuse pileup radiation.

\subsubsection{Back reaction from pointlike minimum-bias}
\label{sec:areamed-areaanalytics-pointlike-back-reaction}

Let us first calculate back reaction in the case of pointlike minimum
bias. We will consider minimum-bias particles with transverse momentum
$p_{tm}$ distributed uniformly on the $y$--$\phi$ cylinder with density $\nu_m \ll
1$. We use a subscript $m$ rather than $g$ to differentiate them from
ghost particles, the key distinction being that $p_{tm}$ is small but
finite, where $p_{tg}$ is infinitesimal.

We thus consider the situation in which a particle $p_1$, with large
transverse momentum, $p_{t1} \gg p_{tm}$, has emitted a soft particle
$p_2$ on a scale commensurate with the minimum-bias particles, $p_{t2} \sim
p_{tm}$. We shall calculate the probability that $p_2$ was part of the
jet in the absence of the minimum-bias particle, but is \emph{lost}
from it when the minimum-bias particle is added. This can be written
\begin{equation}
  \label{eq:loss-passive-master}
  \frac{dP^{(\text{loss})}_{\JA,R}}{d p_{t2}} = 
  \int d\phi_m dy_m \nu_m \int d\Delta_{12} \frac{dP}{dp_{t2} \,
    d\Delta_{12}} H_{\JA,R}(p_2 \in J_1) \, 
  H_{\JA,R}(p_2 \notin J_1 | \,p_m)\,,
\end{equation}
where $H_{\JA,R}(p_2 \in J_1)$ is $1$ ($0$) if, in the absence of
$p_m$, $p_{2}$ is inside (outside) the jet that contains $p_1$.
Similarly, $H_{\JA,R}(p_2 \notin J_1 | p_m)$ is $1$ ($0$) if, in the
presence of $p_m$, $p_{2}$ is outside (inside) the jet that contains
$p_1$. One can also define the probability for $p_2$ to not be part of
the jet in the absence of the minimum-bias particle, but to be
\emph{gained} by the jet when the minimum-bias particle is added,
\begin{equation}
  \label{eq:gain-passive-master}
  \frac{dP^{(\text{gain})}_{\JA,R}}{d p_{t2}} = 
  \int d\phi_m dy_m \nu_m \int d\Delta_{12} \frac{dP}{dp_{t2} \,
    d\Delta_{12}} H_{\JA,R}(p_2 \notin J_1) \, 
  H_{\JA,R}(p_2 \in J_1 | \,p_m)\,.
\end{equation}
It is convenient to factor out the particle production probability as
follows, in the small $R$ limit,
\begin{equation}
  \label{eq:loss-passive-factorize}
  \frac{dP^{(\text{loss,gain})}_{\JA,R}}{d p_{t2}} = \left. \Delta_{12} 
      \frac{dP}{dp_{t2}\, d\Delta_{12}}
  \right|_{\Delta_{12}=R} \nu_m \; b^{(\text{loss,gain})}_{\JA,R}(p_{t2}/p_{tm})\,,
\end{equation}
where $b^{(\text{loss})}_{\JA,R}(p_{t2}/p_{tm})$
($b^{(\text{gain})}_{\JA,R}(p_{t2}/p_{tm})$) can be thought of as
the effective `back-reaction area' over which the minimum-bias
particle causes a loss (gain) of jet contents, given a
$d\Delta_{12}/\Delta_{12}$ angular distribution for the jet
contents:\footnote{Strictly speaking it is the integral over area of
  the probability of causing a loss (or gain) of jet contents.}
\begin{align}
  \label{eq:loss-effective-area}
  b^{(\text{loss})}_{\JA,R}(p_{t2}/p_{tm})& = \int d\phi_m dy_m \int
  \frac{d\Delta_{12}}{\Delta_{12}} 
  H_{\JA,R}(p_2 \in J_1) \, 
  H_{\JA,R}(p_2 \notin J_1 | \,p_m)\,,\\
  \label{eq:gain-effective-area}
  b^{(\text{gain})}_{\JA,R}(p_{t2}/p_{tm}) &= \int d\phi_m dy_m \int
  \frac{d\Delta_{12}}{\Delta_{12}} 
  H_{\JA,R}(p_2 \notin J_1) \, 
  H_{\JA,R}(p_2 \in J_1 | \,p_m)\,.  
\end{align}

Let us first consider the $k_t$ and Cambridge/Aachen
sequential-recombination algorithms for which the $H$ functions in
\eqs~(\ref{eq:loss-effective-area},\ref{eq:gain-effective-area})
translate to a series of $\Theta$-functions, \eg
\begin{multline}
  \label{eq:HL-kt}
  H_{\kt,R}(p_2 \in J_1) H_{\kt,R}(p_2 \notin J_1 |\, p_m) 
    = \Theta(R - \Delta_{12}) \Theta(R-\Delta_{2m}) 
      \Theta(\Delta_{1(2+m)} - R) \times \\
    \times
    \Theta(\Delta_{1m}- \min(1,p_{t2}/p_{tm}) \Delta_{2m}) 
    \Theta(\Delta_{12}- \min(1,p_{tm}/p_{t2}) \Delta_{2m}) \,,
\end{multline}
for the $k_t$ algorithm and 
\begin{multline}
  \label{eq:HL-cam}
  H_{\cam,R}(p_2 \in J_1) H_{\cam,R}(p_2 \notin J_1 |\, p_m) 
    = \Theta(R - \Delta_{12}) \Theta(R-\Delta_{2m}) 
      \Theta(\Delta_{1(2+m)} - R) \times \\
    \times
    \Theta(\Delta_{1m}-  \Delta_{2m}) 
    \Theta(\Delta_{12}-  \Delta_{2m}) \,,
\end{multline}
for Cambridge/Aachen, where $\Delta_{1(2+m)}$ is the distance between
$p_1$ and the recombined $p_{2}+p_{m}$. Evaluating integrals with the
above $\Theta$-functions is rather tedious, but one can usefully
consider the limit $p_{tm} \ll p_{t2} \ll p_{t1}$. This of physical
interest because it relates to the probability that the minimum-bias
particle induces changes in jet momentum that are much larger than
$p_{tm}$, and a number of simplifications occur in this limit. Since
\begin{equation}
  \label{eq:BR-distance-approx-large-pt2}
  \Delta_{1(2+m)} = 
  \left|\vec \Delta_{12} + \frac{p_{tm}}{p_{t2}+p_{tm}}\vec \Delta_{2m}\right|
  = \Delta_{12} + \frac{p_{tm}}{p_{t2}} \frac{\vec \Delta_{12} \cdot \vec
  \Delta_{2m}}{\Delta_{12}} + \order{\frac{p_{tm}^2}{p_{t2}^2} R}\,,
\end{equation}
$p_2$ must be close to the edge of the jet in order for it to be
pulled out by $p_m$, $|\Delta_{12}-R| \ll 1$. Without loss of
generality, we can set 
$y_1=\phi_1=\phi_2=0$, so that $\Delta_{12} = y_2 \simeq R$, and
\begin{equation}\label{eq:delta12m}
  \Delta_{1(2+m)}
  = y_2 + \frac{p_{tm}}{p_{t2}} (y_m - R) +
  \order{\frac{p_{tm}^2}{p_{t2}^2} R}.
\end{equation}
We can then carry out the integrations over $\phi_m$ and $y_2$
straightforwardly, leading to following the result for loss at high
$p_{t2}$,
\begin{equation}
  \label{eq:bLres}
  b^{(\text{loss})}_{\kt,R}(p_{t2}/p_{tm} \gg 1) \simeq 
  b^{(\text{loss})}_{\cam,R}(p_{t2}/p_{tm} \gg 1) \simeq
   \int_{R}^{2R} d y_m \, \frac{(y_m-R)}{R} \frac{p_{tm}}{p_{t2}}\, 
   2\sqrt{R^2-(y_m-R)^2} =
    \frac{2}{3} \frac{p_{tm}}{p_{t2}} R^2\,.
\end{equation}

For the anti-$k_t$ algorithm it is tempting to follow the same path,
\ie to impose that the loss term is obtained by clustering $p_m$ with
$p_2$ in order to pull $p_2$ out of the jet, and write the $H$
functions (for the loss term and assuming $p_{t2}\gg p_{tm}$) as
\begin{equation}
  \label{eq:HL-antikt-subleading}
  \Theta(R - \Delta_{12}) \, \Theta(\Delta_{1(2+m)} - R) \,
    \Theta(\Delta_{1m}- (p_{t1}/p_{t2}) \Delta_{2m}) \,
    \Theta(\Delta_{12}- (p_{t1}/p_{t2}) \Delta_{2m}) \,
    \Theta(R-\Delta_{2m})\,.
\end{equation}
In the limit of interest, $p_{tm}\ll p_{t2}\ll p_{t1}$, the constraint
that $p_2$ must be in the jet and that $p_2+p_m$ must be outside the
jet imposes, as before (see \eqref{eq:delta12m}), that $\Delta_{12}$
must be between $R-\frac{p_{tm}}{p_{t2}}(y_m-R)$ and $R$. 
But now, the condition that $p_2$ clusters with $p_m$ and not $p_1$
(the fourth term in \eqref{eq:HL-antikt-subleading}) imposes that
$p_m$ lies within a half-circle of radius $R p_{t2}/p_{t1}$ centred on
$p_2$. Ultimately, this gives a contribution proportional to
$\frac{p_{t2}^2p_{tm}}{p_{t1}^3} R^2$ which, as we shall show below,
is subleading for the case of interest, $p_{t2}\ll p_{t1}$.

Instead, the dominant contribution comes from the situation where
$p_m$ clusters with $p_1$ and the recoil causes $p_2$ to fall outside
the jet, giving
\begin{multline}
  \label{eq:HL-antikt}
  H_{{\rm anti-}k_t,R}(p_2 \in J_1) H_{{\rm anti-}k_t,R}(p_2 \notin J_1 |\, p_m) 
    = \Theta(R - \Delta_{12}) \Theta(\Delta_{(1+m)2} - R) \times \\
    \times
    \Theta(\Delta_{1m}-\Delta_{12}) 
    \Theta(\Delta_{12}- \min(p_{t1}/p_{t2},p_{t1}/p_{tm}) \Delta_{2m}) 
    \Theta(R-\Delta_{1m})\,.
\end{multline}
In the limit $p_{t1}\gg p_{t2},p_{tm}$, the recoil (along the
direction $\vec{\Delta}_{12}$) is found to be $p_{tm}/p_{t1} y_m$ and this
ultimately leads to
\begin{equation}
  \label{eq:bLres-antikt}
  b^{(\text{loss})}_{\antikt,R} \simeq 
    \frac{2}{3} \frac{p_{tm}}{p_{t1}} R^2\,,
\end{equation}
\ie suppressed by a factor $p_{t2}/p_{t1}$ compared to the result
obtained in (\ref{eq:bLres}) for the $k_t$ and Cambridge/Aachen
algorithms.\footnote{A similar contribution, with the same
  $p_{tm}/p_{t1}$ scaling, would also be present for the $k_t$ and
  Cambridge/Aachen algorithms but would be subleading compared to the
  result of \eqref{eq:bLres}.} Note also that, contrary to the results
for the $k_t$ and Cambridge/Aachen algorithms, \eqref{eq:bLres-antikt}
is valid independently of the ordering between $p_{t2}$ and $p_{tm}$,
as long as they are both small compared to $p_{t1}$.

In a similar manner we obtain for the gain,
\begin{multline}
  \label{eq:bGres}
  b^{(\text{gain})}_{\kt,R}(p_{t2}/p_{tm} \gg 1) \simeq 
  b^{(\text{gain})}_{\cam,R}(p_{t2}/p_{tm} \gg 1) \simeq
   \int_{R/2}^{R} d y_m \, \frac{(R-y_m)}{R} \frac{p_{tm}}{p_{t2}} \, 
   2\sqrt{R^2-(R-y_m)^2} =\\=
    \left(\frac{2}{3} - \frac{\sqrt{3}}{4}\right) \frac{p_{tm}}{p_{t2}} R^2\,.
\end{multline}
and for the anti-$k_t$ algorithm,
\begin{equation}
  b^{(\text{gain})}_{\antikt,R} \simeq 
  b^{(\text{loss})}_{\antikt,R} \simeq 
    \frac{2}{3} \frac{p_{tm}}{p_{t1}} R^2\,,
\end{equation}

\begin{figure}
  \centering
  \includegraphics[width=0.48\textwidth]{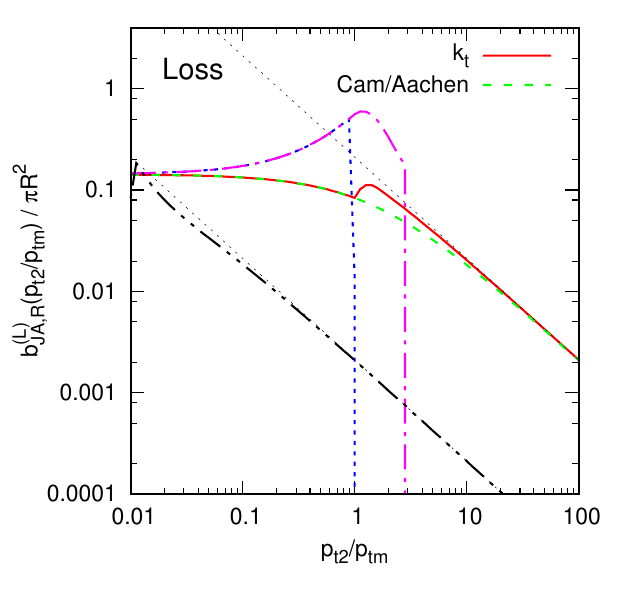}
  \hfill
  \includegraphics[width=0.48\textwidth]{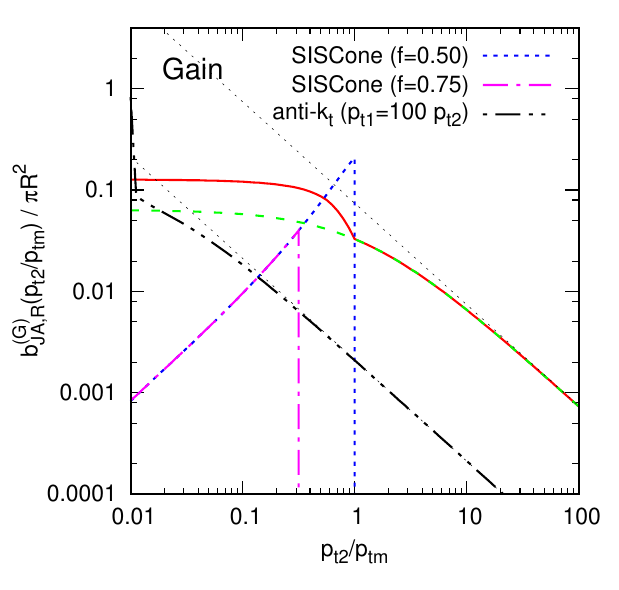}
  \caption{The effective area for back-reaction as a function of the
    ratio of the soft perturbative scale $p_{t2}$ and the point-like
    minimum-bias scale $p_{tm}$, showing separately the loss (left) and
    gain (right) components for four jet definitions. For the
    anti-$k_t$ results, we have set $p_{t1}=100\, p_{t2}$. The thin
    black dotted lines represent our analytic results valid for
    $p_{t2}\gg p_{tm}$.}
  \label{fig:passive-loss-gain}
\end{figure}%

The SISCone results are included and have the property that
\begin{subequations}
\label{eq:cone-GL-cutoff}
  \begin{align}
    b^{(\text{loss})}_{\cone,R}(p_{t2}/p_{tm}) &= 0 \quad \text{for} \quad
    \frac{p_{t2}}{p_{tm}} > \frac{f}{1-f}\,,\\
    b^{(\text{gain})}_{\cone,R}(p_{t2}/p_{tm}) &= 0 \quad \text{for} \quad
    \frac{p_{t2}}{p_{tm}} > \frac{1-f}{f}\,.
  \end{align}
\end{subequations}
\ie at high $p_{t2}$, point-like minimum bias never induces
back-reaction in the cone algorithm.\footnote{For large but finite
  $p_{t1}$, back-reaction can actually occur beyond the above limits,
  but only with probability $\sim p_{tm}/p_{t1}$.}

The results for general $p_{t2}/p_{tm}$, determined numerically, are
shown in figure~\ref{fig:passive-loss-gain}. 
On sees that our analytic estimates work very well in the region where
$p_{t2}\gg p_{tm}$ and that indeed, the anti-$k_t$ algorithm has a
suppressed back reaction and SISCone has zero back-reaction at large
$p_{t2}$ as expected. 
On the other hand, for $p_{t2} \sim p_{tm}$ back-reaction is more
likely with the cone algorithm --- the effective area over which the
MB particle can cause a change in jet contents is $\sim 0.5 \pi R^2$,
to be compared to $\sim 0.1 \pi R^2$ for the $\kt$ and
Cambridge/Aachen algorithms.

One may use the results
\eqs~(\ref{eq:bLres})--(\ref{eq:cone-GL-cutoff}) to determine the
average change in jet-momentum due to back reaction.
Because of the logarithmic spectrum of emissions $dP/(dp_{t2} \,
d\Delta_{12})$, one finds that it receives contributions from the whole
logarithmic region $p_{tm} < p_{t2} < p_{t1}$,
\begin{equation}
  \label{eq:deltapt_back}
  \langle \Delta p_{t,\JA,R}^{(\text{gain}-\text{loss})} \rangle \simeq
  \int_{p_{tm}}^{p_{t1}} dp_{t2}\, p_{t2} \left[\frac{dP^{(\text{gain})}_{\JA,R}}{dp_{t2}} -
    \frac{dP^{(\text{loss})}_{\JA,R}}{dp_{t2}} \right] =
  \beta_{\JA,R} \, 
  \rho  \cdot \frac{C_1}{\pi b_0} \ln
  \frac{\as(p_{tm} R)}{\as(p_{t1} R)}\,,
\end{equation}
(evaluated for running coupling), where $\rho = \nu_m p_{tm}$
corresponds to the average transverse momentum of minimum-bias
radiation per unit area and
\begin{equation}
  \label{eq:passive-beta}
  \beta_{\JA,R} = \lim_{p_{t2} \to \infty} \frac{p_{t2}}{p_{tm}}
  \left(b^{(\text{gain})}_{\JA,R}(p_{t2}/p_{tm}) -
    b^{(\text{loss})}_{\JA,R}(p_{t2}/p_{tm}) \right)\,.
\end{equation}
The structure of the correction in \eq~(\ref{eq:deltapt_back}) is very
similar to that for the actual contamination from minimum bias,
$\rho \langle \Delta a_{\JA,R}\rangle$, with
$ \langle \Delta a_{\JA,R}\rangle$ as determined in
section~\ref{sec:areamed-areaanalytics-area-scal-viol-passive}:
notably, for fixed coupling, the average back-reaction scales with the
logarithm of the jet $p_t$.  The coefficients $\beta_{\JA,R}$,
\begin{subequations}
  \label{eq:passive-beta-res}
  \begin{align}
    \beta_{\kt,R} = \beta_{\cam,R} &= -\frac{\sqrt{3}}{4} R^2 \simeq
    -0.1378 \pi R^2\,,\\
    \beta_{\antikt,R} = \beta_{\cone,R} &= 0\,.
  \end{align}
\end{subequations}
can be directly compared to the results for the $d_{\JA,R}$ there. The
values are relatively small, similar in particular to what one observes
for the Cambridge/Aachen algorithm, though of opposite sign.
Note also that for the SISCone and anti-$k_t$ algorithms, one would
expect a contribution similar to (\ref{eq:deltapt_back}), proportional
to $\rho$, but without the logarithmic enhancement, although, in the
case of the anti-$k_t$ algorithm, further cancellations could happen
since $b^{(\text{gain})}_{\antikt,R} \simeq b^{(\text{loss})}_{\antikt,R}$.

Though the average change in jet momentum,
both from scaling violations of the area and from back-reaction, have
a similar analytical structure,
it is worth bearing in mind that these similar analytical structures
come about quite differently in the two cases.
Regarding area scaling violations, a significant fraction of jets,
$\sim \as \ln p_{t1}/p_{tm}$, are subject to a change in area
$\sim R^2$ (\cf
section~\ref{sec:areamed-areaanalytics-area-scal-viol-passive}), and a
consequent modification of the minimum-bias contamination by a modest
amount $\sim p_{tm}$.
In contrast the average back reaction effect $\sim \as p_{tm} \ln
p_{t1}/p_{tm}$, is due to large modifications of the jet momentum
$\sim p_{t2}$ with $p_{tm} \ll p_{t2} \ll p_{t1}$ occurring rarely,
with a differential probability that depends on $p_{t2}$ as $\sim \as
d p_{t2} p_{tm}/p_{t2}^2$.
One consequence of this is that the mean square change in transverse
momentum due to back-reaction is dominated by very rare modifications
of jets in which $p_{t2} \sim p_{t1}$, giving
\begin{equation}
  \label{eq:BR-dispersion}
  \left\langle \left(\Delta p_{t,\JA,R}^{(\text{gain,loss})}\right)^2
  \right\rangle \sim \as p_{t1} p_{tm} \nu_m\,.
\end{equation}
Note that the coefficient of this dispersion is non-zero even for the
SISCone algorithm, due to the residual probability $\sim p_{tm}/p_{t1}$
that it has, for finite $p_{t1}$, for any change in structure with
$p_{tm} \ll p_{t2} \lesssim p_{t1}$.
For the anti-$k_t$ algorithm, one would also obtain the behaviour
given by \eq~\eqref{eq:BR-dispersion} since, for $p_{t2}\sim p_{t1}$,
the gain and loss probabilities also behave as
$\sim \as d p_{t2} p_{tm}/p_{t2}^2$.

\subsubsection{Back reaction from diffuse pileup}
\label{sec:areamed-areaanalytics-diffuse-back-reaction}

Let us now examine back reaction in the case where the minimum-bias
radiation has a uniform diffuse structure, consisting of a high
density of pileup particles, $\nu_m \gg 1$. The relation between back
reaction in the point-like and diffuse cases is rather similar to the
relation between passive and active areas: key analytical features
remain unaffected by the change in the structure of the soft
background, but certain specific coefficients change.

We define the probability for loss in the presence of diffuse PU as
\begin{equation}
  \label{eq:BR-active-loss-master}
  \frac{dP^{(\text{loss})}_{\JA,R}}{d p_{t2}} = 
  \mathop{\lim_{\nu_m \to \infty}}_{\rho = \nu_m \langle
    p_{tm}\rangle\: \text{fixed}}
  \left\langle 
    \int d\Delta_{12} \frac{dP}{dp_{t2} \,
      d\Delta_{12}} H_{\JA,R}(p_2 \in J_1) \, 
    H_{\JA,R}(p_2 \notin J_1 | \,\rho)
  \right\rangle_{\mathrm{PU}},
\end{equation}
where the average is performed over the ensemble of PU configurations,
and now $H_{\JA,R}(p_2 \notin J_1 | \,\rho)$ is 1 (0) if $p_2$ is
outside (inside) the jet containing $p_1$ in the presence of the
specific pileup instance. A similar equation holds for the gain
probability.

Then, as with \eq~(\ref{eq:loss-passive-factorize}) we factorise this,
\begin{equation}
  \label{eq:loss-active-factorize}
  \frac{dP^{(\text{loss})}_{\JA,R}}{d p_{t2}} = \left. \Delta_{12} \frac{dP}{dp_{t2}
      \, d\Delta_{12}}
  \right|_{\Delta_{12}=R} \; B^{(\text{loss})}_{\JA,R}(p_{t2} / \rho)\,,
\end{equation}
into one piece related to the probability for perturbative emission
and a second piece $B^{(\text{loss})}_{\JA,R}$ that is the diffuse
analogue of the effective back-reaction area
$b^{(\text{loss})}_{\JA,R}$ in the point-like case.  Note however that
it is not so obvious exactly what geometrical area
$B^{(\text{loss})}_{\JA,R}$ actually corresponds to.\footnote{A
  related issue is that the precise choice of normalisation of
  $B^{(\text{loss})}_{\JA,R}$ is somewhat arbitrary --- our specific
  choice is intended to provide a meaningful connection with
  $b^{(\text{loss})}_{\JA,R}$ in the large $p_{t2}$ limit.}
Similarly, we introduce $dP^{(\text{gain})}_{\JA,R}/d p_{t2}$ and
$B^{(\text{gain})}_{JA,R}$ corresponding to the gain in the presence
of a diffuse PU.

As for the point-like case, we can obtain analytical results for
$B^{(\text{loss,gain})}_{\JA,R}(p_{t2} / \rho)$ in the case of the SISCone
algorithm, for which (with $f > f_{\max}\simeq 0.391$)
\begin{equation}
  \label{eq:cone-active-BR-is-zero}
  B^{(\text{loss})}_{\cone,R}(p_{t2} / \rho) 
  = B^{(\text{gain})}_{\cone,R}(p_{t2} / \rho)
  = 0\,.
\end{equation}
This is a consequence of the facts (a) that the addition of a uniform
background of PU particles has no effect on the stability (or
instability) of a specific cone, (b) that for $p_{t2} \ll p_{t1}$ the
split--merge step is immaterial to the jet finding if $p_2$ is within
the cone around $p_1$, and (c) that for $p_{2}$ outside the cone
around $p_1$, the maximal possible overlap is of $p_{2}$'s stable cone
with that of $p_{1}$ is $f_{\max}$ and if $f>f_{\max}$ then the two
cones will always be split, ensuring that $p_{2}$ remains in a jet
distinct from $p_{1}$. We believe that real-life corrections to the
zero in \eq~(\ref{eq:cone-active-BR-is-zero}) are proportional to the
standard deviation, $\sigma$, of PU transverse-momentum density from
point to point in the event.

\begin{figure}
  \centering
  \includegraphics[width=0.48\textwidth]{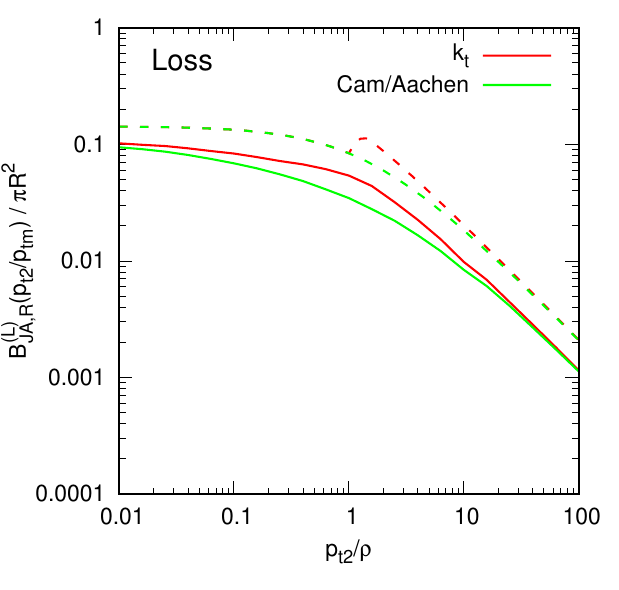}
  \hfill
  \includegraphics[width=0.48\textwidth]{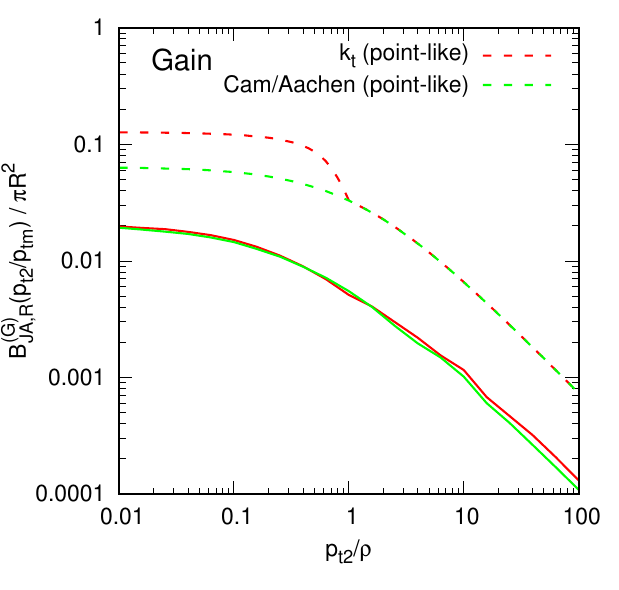}
  \caption{Numerical results for the diffuse effective back-reaction
    `area', $B^{(\text{loss,gain})}_{\JA,R} (p_{t2}/\rho)$, for the
    $k_t$ and Cambridge/Aachen algorithms, with the point-like results
    $b^{(\text{loss,gain})}_{\JA,1} (p_{t2}/p_{tm})$ shown for
    comparison also. Results obtained for $R=1$ and verified also for
    $R=0.7$.}
  \label{fig:active-loss-gain}
\end{figure}%

Numerical results for the
$B^{(\text{loss,gain})}_{\JA,R}(p_{t2} / \rho)$ for the $k_t$ and
Cambridge/Aachen algorithms are given in
figure~\ref{fig:active-loss-gain} and compared to the results in the
point-like case. One sees that the general functional form is rather
similar though the normalisations are somewhat smaller (by a factor of
$2$ for loss, a factor $\sim 10$ for gain). The asymptotic
large-$p_{t2}$ behaviours are observed to be
\begin{subequations}
  \label{eq:BLG-asymptotic}
  \begin{align}
    B^{(\text{loss})}_{\kt,R}(p_{t2} / \rho) \simeq B^{(\text{loss})}_{\cam,R}(p_{t2} / \rho) &\simeq 0.11 \,\pi R^2 \, \frac{\rho}{p_{t2}}\\
    B^{(\text{gain})}_{\kt,R}(p_{t2} / \rho) \simeq B^{(\text{gain})}_{\cam,R}(p_{t2} / \rho) &\simeq 0.013 \,\pi R^2 \,\frac{\rho}{p_{t2}}
  \end{align}
\end{subequations}

The case of the anti-$k_t$ algorithm is slightly more delicate. 
With a uniform background, the momentum imbalance, due to the
background within a distance $R$ of $p_1$ clustered instead with
particle $p_2$, will be of order
$\pi/2\, R^2\rho\, (p_{t2}/p_{t1})^2 $, with a corresponding recoil of
order $\pi/2\, R^3\,(\rho/p_{t1}) (p_{t2}/p_{t1})^2$. This means
that one would obtain
\begin{equation}
B^{(\text{loss})}_{\antikt,R}(p_{t2} / \rho) \simeq B^{(\text{gain})}_{\antikt,R}(p_{t2} / \rho) \simeq \frac{1}{2}\frac{\pi R^2\rho}{p_{t1}}\frac{p_{t2}^2}{p_{t1}^2}.
\end{equation}
One should however also notice that, as for SISCone, one should expect
corrections proportional to the PU fluctuations $\sigma$ which, as for
the point-like case, should be proportional to $\pi
R^2\sigma/p_{t1}$. For $p_{t2}\ll p_{t1}$ the latter is expected to be
the dominant behaviour.

As in the point-like case we can calculate the mean change in jet
transverse momentum due to back reaction and we obtain
\begin{equation}
  \label{eq:deltapt_back_active}
  \langle \Delta p_{t,\JA,R}^{(\text{gain}-\text{loss})} \rangle \simeq
  \int_{p_{tm}}^{p_{t1}} dp_{t2} p_{t2} \left[\frac{dP^{(\text{gain})}_{\JA,R}}{dp_{t2}} -
    \frac{dP^{(\text{loss})}_{\JA,R}}{dp_{t2}} \right] =
  {\mathcal{B}}_{\JA,R} \, 
  \rho  \cdot \frac{C_1}{\pi b_0} \ln
  \frac{\as(\rho R^3)}{\as(p_{t1} R)}\,,
\end{equation}
with 
\begin{equation}
  \label{eq:active-beta}
  {\mathcal{B}}_{\JA,R} = \lim_{p_{t2} \to \infty} \frac{p_{t2}}{\rho}
  \left(B^{(\text{gain})}_{\JA,R}(p_{t2}/\rho) -
    B^{(\text{loss})}_{\JA,R}(p_{t2}/\rho) \right)\,. 
\end{equation}
Even though $b^{(\text{loss,gain})}_{\JA,R}$ had $p_{t2}/p_{tm}$ as its argument
and $B^{(\text{loss,gain})}_{\JA,R}$ has $p_{t2}/\rho$, the final expressions for
the average back-reaction in the point-like and diffuse cases,
\eqs~(\ref{eq:deltapt_back}), (\ref{eq:deltapt_back_active}), have
almost identical forms --- in particular the overall scale appearing
in each is $\rho$ and the only difference appears in the denominator
for the argument of the logarithm. The coefficients are slightly smaller,
\begin{subequations}
  \label{eq:active-beta-res}
  \begin{align}
    {\mathcal{B}}_{\kt,R} = {\mathcal{B}}_{\cam,R} &\simeq
    -0.10 \pi R^2\,,\\
    {\mathcal{B}}_{\cone,R} = {\mathcal{B}}_{\antikt,R} &= 0\,,
  \end{align}
\end{subequations}
and will again translate into modest effects compared to the overall
pileup contamination in the jets.
Note however that in general the scaling with $R$ of
$b^{(\text{loss,gain})}_{\kt,R}(p_{t2} / p_{tm})$ and
$B^{(\text{loss,gain})}_{\kt,R}(p_{t2} / \rho)$ is subtly
different. The former truly behaves like an area, in that
$b^{(\text{loss,gain})}_{\kt,R}(p_{t2} / p_{tm})/ R^2$ is
$R$-independent; the latter instead has the property that it is
$B^{(\text{loss,gain})}_{\kt,R}(R^2 p_{t2} / \rho)$ that is
$R$-independent.

Finally, as for the case of the point-like background, the diffuse PU
will lead to a mean square change of the jet due to back-reaction
which will be dominated by rare modifications with $p_{t2}\sim p_{t1}$
such that
\begin{equation}
  \label{eq:BR-dispersion-diffuse}
  \left\langle \left(\Delta p_{t,\JA,R}^{(\text{gain,loss})}\right)^2
  \right\rangle \sim \as p_{t1}\rho\,.
\end{equation}
This is also true for the anti-$k_t$ algorithm since the extra
suppression factor become finite for $p_{t2}\sim p_{t1}$.

\subsection{Comparison to Monte-Carlo simulations}
\label{sec:areamed-areaanalytics-real-life}

In this section we examine the properties of jet areas in the context
of realistic events, as simulated with
Herwig~\cite{Corcella:2000bw,Corcella:2002jc} and
Pythia~\cite{Sjostrand:2000wi,Sjostrand:2003wg,Sjostrand:2006za}. There
are two purposes to doing so.
Firstly we wish to illustrate the extent to which the
simple single-gluon emission arguments of the previous section hold
once one includes full parton-showering and hadronisation.
Secondly, jet areas play an important role in the estimation and
subtraction of underlying event and pileup contamination. The study of
jet areas in realistic events can help to highlight some of the issues
that arise in such a procedure.

\begin{figure}[tp]
  \centering
  \includegraphics[width=\textwidth]{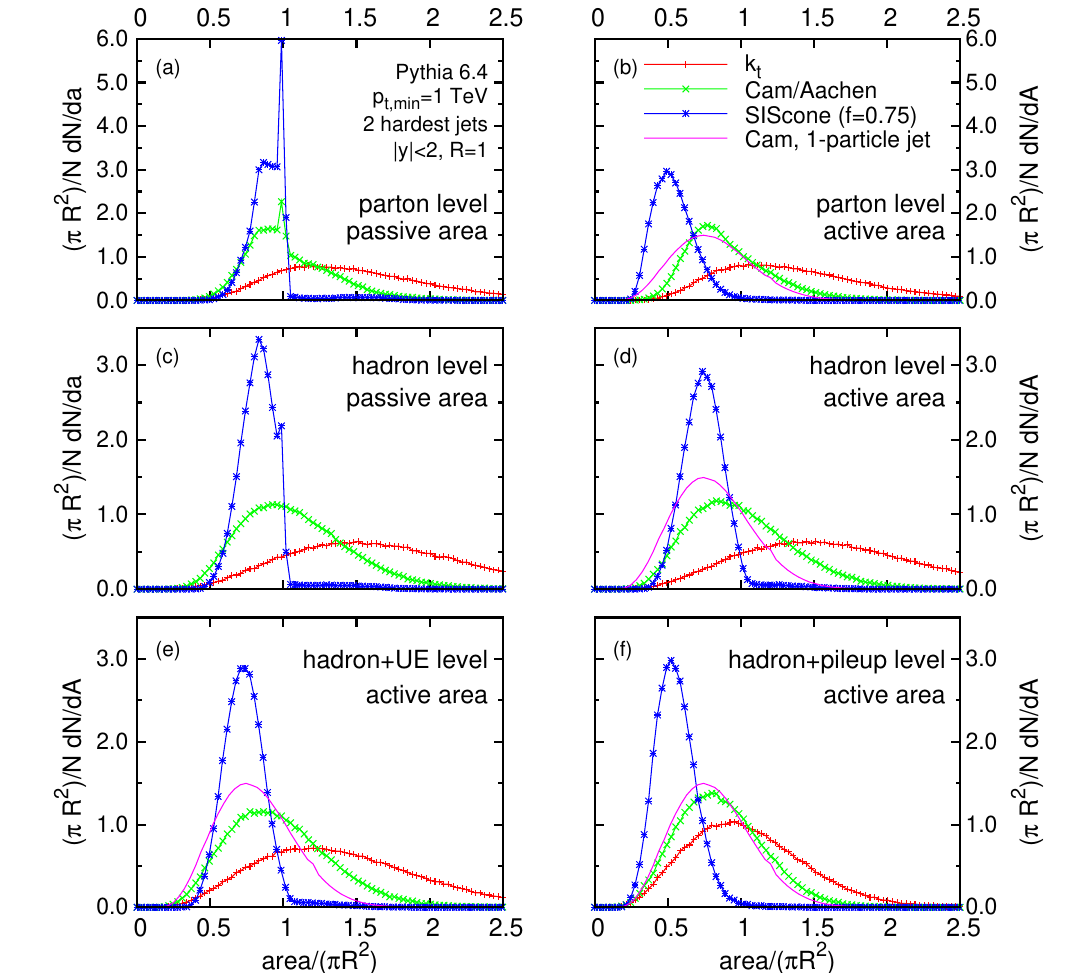}
  \caption{Distribution of active and passive areas of the two hardest
    jets in a range of simulated LHC dijet events, with a minimum
    $p_t$ of $1\TeV$ in the Pythia event generation (v6.4, default
    tune).
    Only jets with $|y|<2$ have been included in the histogram;
    `parton' indicates parton-level, `hadron' indicates hadron-level
    with the UE switched off, `UE' corresponds to hadron-level with UE
    switched on and in the `pileup' case the UE-level event is
    supplemented with additional pileup events corresponding to
    $0.25\mb^{-1}$ per bunch crossing ($\sim 25$ simultaneous
    interactions). For all jet algorithms we use $R=1$.
    A ghost area $\simeq 0.02$ was used throughout except for
    SISCone, where the ghost area was roughly $0.1$ in the active area
    cases.
    Note that ``area'' in these plots corresponds to
    $a_{\JA,R}(J_i)$ and $A_{\JA,R}(J_i)$ respectively for the passive
    and active areas. As such the latter has been averaged over ghost
    ensembles (see \eq~(\ref{eq:act_area})) and the dispersion is a
    consequence of the event and jet structure. 
    We have found that $5$ ghost ensembles were sufficient for $k_t$
    and Cambridge/Aachen algorithms, and $3$ for SISCone (with pileup,
    1 ensemble would actually have been enough).
    In contrast, the Cam/Aachen 1-particle jet result corresponds to
    $\pi R^2/N dN/dA(\text{1-particle-jet}|\{g_i\})$ and serves to
    illustrate how the impact of variability in the event structure in
    real events has consequences rather similar to that of the
    variability of ghost-particle ensembles in the theoretical
    arguments of the previous sections.  }
  \label{fig:area-hist-pythia}
\end{figure}

Let us start with an investigation of the distribution of the areas of
hard jets with $p_t\gtrsim 1\TeV$ in simulated LHC dijet events, and
first concentrate on the $k_t$, Cambridge/Aachen and SISCone
algorithms.
The area distributions are shown on Fig.~\ref{fig:area-hist-pythia}
for various `levels' in the Monte Carlo: just after parton showering,
after hadronisation, both with and without an UE, and finally with a
pileup contribution corresponding to a luminosity of 0.25~mb$^{-1}$
per bunch crossing (\ie $\mu\sim 25$). We examine both passive and
active areas.

The passive areas distributions at parton-shower level,
Fig.~\ref{fig:area-hist-pythia}a, are those that are most amenable to
understanding in terms of our analytical results.  Firstly one notes
that the SISCone and Cambridge/Aachen algorithms have a clear peak at
$a=\pi R^2$. These two algorithms both have the property (cf.\
section~\ref{sec:areamed-areaanalytics-passive-area-2particle}) that
the area is not affected by moderately collinear ($\Delta_{12} < R$
for SISCone, $\Delta_{12} < R/2$ for Cambridge/Aachen) soft particle
emission.  Thus it is possible, even in the presence of parton
showering (which is mostly collinear), for the passive area to remain
$\pi R^2$.
For the cone algorithm, the other main structure is a ``shoulder''
region $0.8 \lesssim a/(\pi R^2) \lesssim 1$, which coincides nicely
with the range of values that can be induced by 2-particle
configurations (cf.\ Fig.~\ref{fig:2point}). A similar shoulder exists
for the Cambridge algorithm, which however additionally has jets with
$a > \pi R^2$ --- again the range of values spanned, up to $a\simeq
1.6 \pi R^2$, is rather similar to what would be expected from the
two-particle calculations. Further broadening of these distributions
at the edges is attributable to parton-level states with more than two
particles.
In contrast the parton-level passive area distribution for the $k_t$
algorithm seems less directly related to the $2$-particle calculations
of
section~\ref{sec:areamed-areaanalytics-passive-area-2particle}. This
can be understood by observing that the $k_t$ passive area is modified
even by rather collinear emissions, and the multiple collinear
emissions present in parton showers add together to cause substantial
broadening of the area distribution.

At parton shower level, there are relatively few particles in the
event and there is no obvious boundary to the jet --- the ghosts that
we add provide a way of assigning that empty area.  It is therefore
not surprising to see significant differences between the two ways of
adding ghosts, \ie the passive and active area distributions. This is
most marked for SISCone, as is to be expected from the results of
section~\ref{sec:areamed-areaanalytics-active_1point}, which showed
that for a 1-particle jet the active area is $\pi R^2/4$.  There is a
trace of this result in Fig.~\ref{fig:area-hist-pythia}b, where one
sees that the SISCone distribution now extends down to $A=\pi
R^2/4$. There is however no peak structure, presumably because even a
highly collinear emission gives a slight modification of the area,
cf.\ Fig.~\ref{fig:active_2point_cone} (the same argument given for
the absence of a peak for the passive $k_t$ area). For the
Cambridge/Aachen and $k_t$ algorithms, there is less difference
between active and passive area distributions, again as expected.

As one moves to events with more particles, for example hadron level,
Fig.~\ref{fig:area-hist-pythia}c and d, the particles themselves start
to give a clearer outline to the jets. Thus the passive and active
distributions are far more similar. This is less so for SISCone than
the others, as is to be expected, for which one still sees a trace of
the peak at $a = \pi R^2$ for the passive area.

For events with many particles, for example with the UE
(Fig.~\ref{fig:area-hist-pythia}e) and then pileup
(Fig.~\ref{fig:area-hist-pythia}f) added, the difference between
passive and active area distributions is so small that we show only
the active area result. These last two plots are the most relevant to
phenomenology. A feature of both is that the dispersion is smallest
for SISCone (but significantly different from zero) and largest for
$k_t$, precisely as one would expect from \eqs~(\ref{eq:act-fluct-seq})
and (\ref{eq:Sigmaalg1}). Another feature is how for Cambridge/Aachen
the dispersion is essentially that associated with $\Sigma(0)$, the
distribution being very similar to the 1-parton active area result,
shown as the solid line. This similarity is strongest when there is
pileup: the logarithmic enhancement that enters in
\eq~(\ref{eq:sigma2_pt}) is reduced because the pileup introduces a
large value for $Q_0 \sim 20 \GeV$. In this case even the $k_t$
algorithm starts to have a distribution of areas that resembles the
$1$-parton active area result, and for SISCone one once again sees
signs of the lower limit at $A=\pi R^2/4$, the one-particle active
area result.

\begin{figure}[tp]
  \centering
  \includegraphics[width=0.7\textwidth]{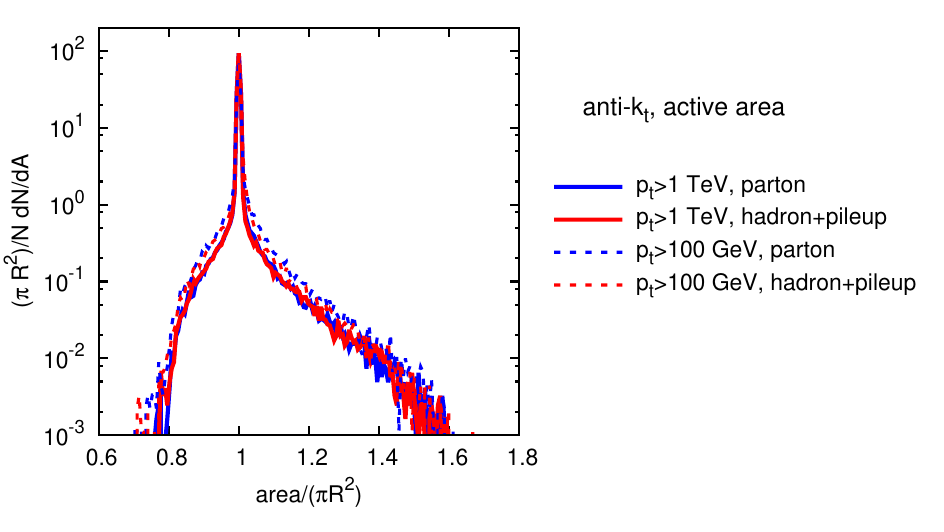}
  \caption{Distributions obtained in the same context as for
    Fig.~\ref{fig:area-hist-pythia}, this time for the anti-$k_t$
    algorithm. We just show the two extreme situations, corresponding
    to parton-level (blue lines) and hadron level with pileup (red
    lines). We show the distributions for two different jet minimum
    transverse momentum: 100~GeV (solid lines) and 1~TeV (dotted
    lines).}\label{figg:area-hist-pythia-antikt}
\end{figure}

Let us now move to area distributions for the anti-$k_t$
algorithm. A few representative distributions are plotted in
Fig.~\ref{figg:area-hist-pythia-antikt}. Only the active area
distribution is plotted for the two extreme particle multiplicities ---
the ``parton'' level and the ``hadron+pileup'' level --- and for two
jet transverse momenta. Most often, the jet area comes out as
$\pi R^2$ as expected from our analytic studies so far. The use of a
logarithmic scale on the vertical axis on
Fig.~\ref{figg:area-hist-pythia-antikt} allows one to highlight the
rare tails away from $\pi R^2$, occurring for about 1\% of the jets.
These features are stable when we add pileup to the event of vary the
jet $p_t$, showing that jet areas are constrained by the hard
structure of the event.

\begin{figure}[p]
  \centering
  \includegraphics[width=0.9\textwidth]{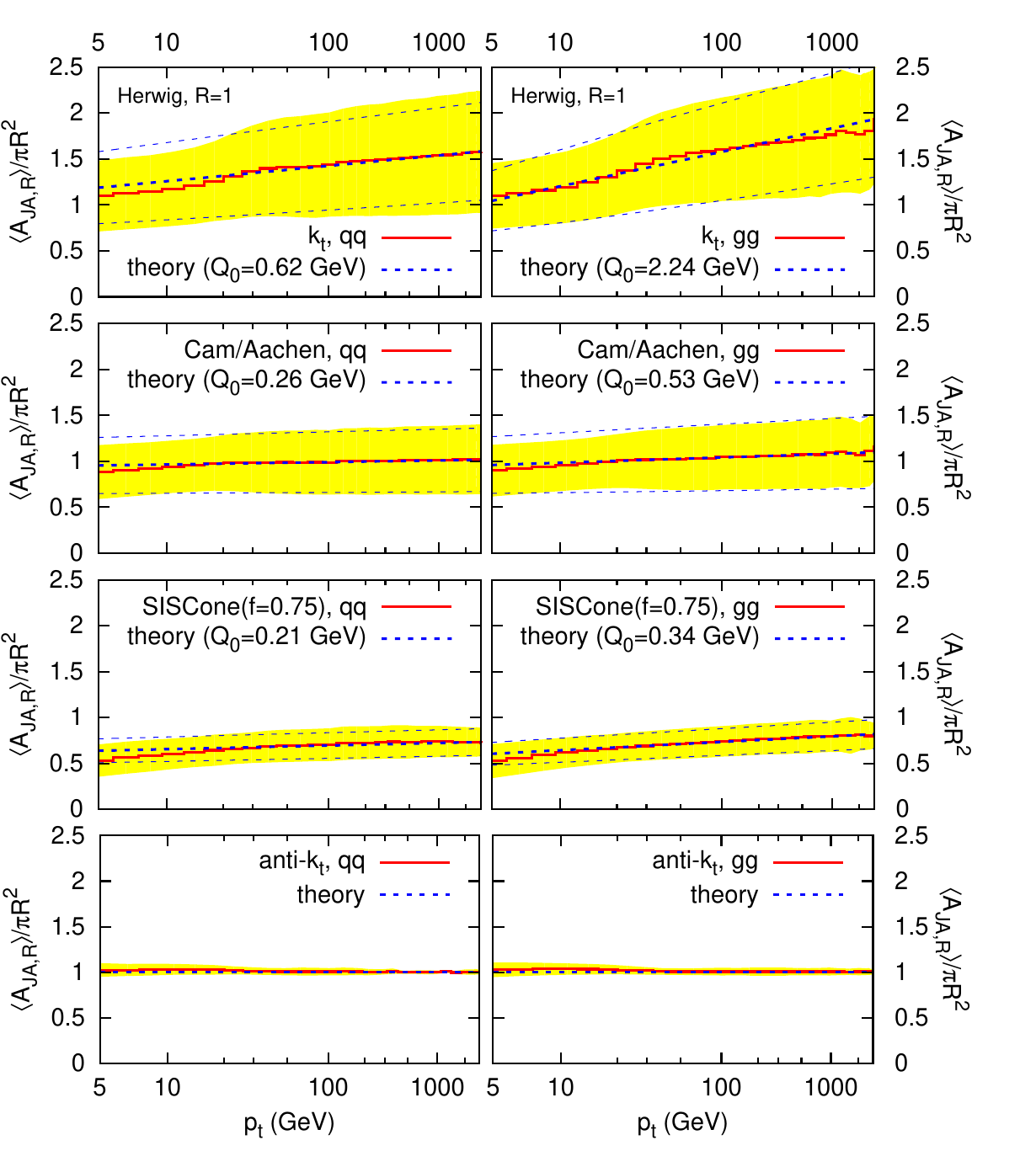}
  \caption{The mean (solid line) and standard deviation (band) of the
    active area for each of the two hardest jets in $qq\to qq$ and
    $gg \to gg$ events simulated with Herwig 6.5, as a function of the
    jet transverse momentum. The theory curves (thick dashed for mean,
    thin dashed for standard deviation) correspond to
    \eqs~(\ref{eq:Ajfr-base})--(\ref{eq:Sigmaalg1}) with a 1-loop
    coupling ($\Lambda_\mathrm{QCD} = 0.2\GeV$, $n_f=5$), and $Q_0$
    fitted (for the mean values).  The areas have been obtained with
    ghosts of size $0.02$ for anti-$k_t$, $k_t$ and Cambridge/Aachen,
    and $0.1$ for SISCone.  For all algorithms, we used $R=1$.
    Note that the horizontal scale is uniform in
    $\ln \ln p_t/\Lambda$. We considered $pp$ collisions with
    $\sqrt{s} = 14\TeV$.  }
  \label{fig:anom_dim_herwig_gg}
\end{figure}

While the plots of Figs.~\ref{fig:area-hist-pythia} and
\ref{figg:area-hist-pythia-antikt} provide an illustration of many of
the features that we have discussed in the earlier sections of this
Chapter, the fact that we have not systematically calculated area
distributions means that the discussion necessarily remains
qualitative. For quantitative checks, one should instead examine the
mean active area and its dispersion and compare them to the results of
section~\ref{sec:areamed-areaanalytics-active}. This is done in
Fig.~\ref{fig:anom_dim_herwig_gg}, separately for $qq\to qq$ and for
$gg \to gg$ scattering, as a function of the jet transverse momentum.
The horizontal scale has been chosen uniform in $\ln \ln p_t/\Lambda$
so that our predictions correspond to straight lines.

The predictions of section~\ref{sec:areamed-areaanalytics-active} set
the slope of those lines for each algorithm, and the agreement is
reasonable in all cases. The infrared cutoff scale $Q_0$ is not
predicted, and may differ both between quark and gluon jets and
between algorithms. The values for $Q_0$ have therefore been fitted,
using the results for the mean, and are consistent with a general
non-perturbative origin (modulo issues in SISCone discussed
below). The standard deviation (indicated by the band for the Monte
Carlo simulation and the thin dashed lines for the theory result) is
then entirely predicted, and also agrees remarkably well. Thus,
overall, our simple analytical calculations provide a surprisingly
successful picture of the mean and dispersions for various algorithms,
over a range of jet transverse momenta. This is all the more striking
considering that the calculation is based on just the first term in a
series $\as^n \ln^n p_{t}$, and is in part based on a small angle
approximation.

Some remarks are due regarding the $Q_0$ values. Clear patterns
emerge: it is largest for $k_t$, smallest for SISCone, and
systematically larger for gluon jets than for quark jets. In the case
of SISCone with quark jets, the value is uncomfortably close to the
value of $\Lambda_{QCD}=0.2\GeV$ used in the one-loop coupling. It may
be that this low value is an artefact whose origin lies in the finite
density both of actual event particles and of ghosts (the latter due
to speed limitations in SISCone): when one has a limited density of
particles and/or ghosts, the measured area may be intermediate between
the passive and ``ideal'' active areas; because SISCone has such a
large difference between passive and active areas (a factor of 4 for
the 1-particle results), the finite density effects can be
significant, and so $Q_0$ may be taking an extreme value in order to
compensate for this.

Compared to $k_t$, Cambridge/Aachen and SISCone, the anti-$k_t$
algorithm shows an area very close to $\pi R^2$ with almost no
dependence on $p_t$ and a very small dispersion.

A final comment concerns the choice of event generator. Here we have
shown results from Herwig (v.~6.5). Pythia with the original shower
(the default choice in v.~6.4) is known to have difficulties
reproducing anomalous dimensions associated with soft large angle
radiation~\cite{SalamWicke,BCD}, whereas the new shower
\cite{PythiaNewShower} mostly resolves these issues \cite{BCD}.
Similar concerns are potentially relevant here too. In our original
studies~\cite{Cacciari:2008gn}, we actually found that both Pythia
showers give results similar to Herwig's (and ours) in all cases
except the $k_t$ algorithm, for which both Pythia showers give a slope
a factor of two smaller than expected. This suggests that an
experimental measure of the $p_t$ dependence of jet areas might
provide some non-trivial constraints on parton-showering dynamics.

\subsection{Summary}\label{sec:area-analytics-summary}

The concept of a jet area is one that initially seems rather
intuitive. Yet, as we have seen both in Chapter~\ref{chap:areamed} (in
particular in Section~\ref{sec:areamed-defareas}) and in this Chapter,
there is a wealth of physics associated with it, both in terms of how
one defines it and as concerns the interplay between a jet's internal
structure and its area. This is reflected in the range of quantities
that can be studied to characterise the behaviour of jet areas,
summarised in table~\ref{tab:all-defs}.

Our guiding principle in defining jet areas has been that they should
provide a measure of a jet's susceptibility to additional underlying
event and pileup radiation, in the limit in which these are infinitely
soft. Two opposite hypotheses for the structure of
such radiation, pointlike or diffuse, lead to two definitions of
the area, respectively passive or active, both calculated in practice
with the help of infinitely soft ``ghost'' particles.
The two definitions can be used with any infrared safe jet algorithm,
and we have studied them both with sequential recombination algorithms
(anti-$k_t$, $k_t$ and Cambridge/Aachen) and with a stable-cone with
split--merge algorithm, SISCone.

The area of a jet may depend on its substructure. For the simplest
case of a jet made of a single hard particle, the passive area
coincides with one's naive expectation of $\pi R^2$ for all jet
algorithms considered here. The active area --- that related to the
more realistic, diffuse picture for UE and pileup --- instead differs
from $\pi R^2$, except for the anti-$k_t$ algorithm. The most
surprising result is that for SISCone, whose active area is
$\pi R^2/4$, a consequence of the split--merge dynamics (the midpoint
algorithm widely used at the Tevatron behaves similarly).
Thus the widespread assumption that cone-based algorithms
automatically have an area of $\pi R^2$ is, in many cases,
unjustified.

Real jets of course consist of more than a single particle. The first
level of substructure involves the addition of a soft particle to the
neighbourhood of the jet. For {\it all} except the anti-$k_t$
algorithm, we have seen that this modifies the jet area, and that the
average jet area then becomes an infrared unsafe quantity.
Even if it looks like a bad feature at first sight, the infrared
unsafety of jet areas is actually desirable and tightly related to the
fact that areas are meant to precisely measure the jet's
susceptibility to soft radiation.
In that respect, the soft resilience of the anti-$k_t$ algorithm
corresponds to an area remaining $\pi R^2$ when we add any number of
soft particles to the jet.
Also, the small pileup contamination often observed with the SISCone
algorithm can be related to its smaller catchment area than that of
recombination algorithms.

From our analytic viewpoint, it emerges that the effects of gluon
emission can usefully be summarised in terms of an anomalous
dimension, which encodes how the jet's average area depends on its
transverse momentum.
We have calculated this to leading order and seen that it agrees
remarkably well with the $p_t$ dependence of measures of the jet areas
in hadron-level Monte Carlo simulations.

\begin{table}
  \centering
  \begin{tabular}{|l|l|p{0.58\textwidth}|}
    \hline
    quantity & discussed in & description \\
    \hline
     && \\[-10pt]
    Passive area & Section \ref{sec:areamed-areaanalytics-passive} & single-ghost area \\
      $\bullet$ average area 
           & 
           & 
	   \\
        \hspace*{0.3cm} $a$(1PJ) 
           & Sect. \ref{sec:areamed-areaanalytics-passive-area-1particle} 
           & \hspace*{0.3cm}passive area for a 1-particle jet\\
        \hspace*{0.3cm} $\langle a \rangle$
           & \eq~(\ref{eq:ajfr-base}) 
           & \hspace*{0.3cm}average passive area of jets with QCD branching: \\ && 
	\hspace*{0.3cm}  $\langle a \rangle = a(\oPJ) + d\, \frac{C_1}{\pi b_0}\ln \frac{\as(Q_0)}{\as(R p_t)} + \ldots$ \\
        \hspace*{0.3cm} $d$
           & \eq~(\ref{eq:area-scaling-passive-coefficient})
           & \hspace*{0.3cm}coefficient of leading scaling violations of $\langle a \rangle$\\
      $\bullet$ area fluctuations
           & 
           & 
	   \\
        \hspace*{0.3cm} $\sigma^2(\oPJ)$
           & 
           & \hspace*{0.3cm}variance of one-particle passive area (0 by definition)\\
        \hspace*{0.3cm} $\langle\sigma^2 \rangle$
           & \eq~(\ref{eq:passive-fluct-decomp})
           & \hspace*{0.3cm}variance of passive area of jets with QCD branching: \\ &&
	   \hspace*{0.3cm} $\langle\sigma^2 \rangle = \sigma^2(\oPJ)  + s^2 \, \frac{C_1}{\pi b_0}\ln \frac{\as(Q_0)}{\as(R p_t)} + \ldots$ \\
        \hspace*{0.3cm} $s^2$
           & \eq~(\ref{eq:fluct_passive_coefs})
           & \hspace*{0.3cm}coefficient of leading scaling violations of $\langle\sigma^2 \rangle$
    \\[5pt]\hline 
     && \\[-10pt]
    Active area & Section \ref{sec:areamed-areaanalytics-active} & many-ghost area (ghosts also cluster among themselves)\\
      $\bullet$ average area
           & 
           & 
	   \\
        \hspace*{0.3cm} $A$(1PJ) 
           & (\ref{eq:area-hard-particle-antikt},\ref{eq:area-hard-particle},\ref{eq:active_1point_cone})
           & \hspace*{0.3cm}active area for a 1-particle jet\\
        \hspace*{0.3cm} $\langle A \rangle$
           & \eq~(\ref{eq:Ajfr-base})
           & \hspace*{0.3cm}average active area of jets with QCD branching: \\ && 
	   \hspace*{0.3cm} $\langle A \rangle
                = A(\oPJ)
                + D \, \frac{C_1}{\pi b_0}\ln\frac{\as(Q_0)}{\as(R p_t)} + \ldots$ \\
        \hspace*{0.3cm} $D$  
           & \eq~(\ref{eq:Dalg1})
           & \hspace*{0.3cm}coefficient of leading scaling violations of
	   $\langle A \rangle$\\
        \hspace*{0.3cm} $A$(GJ)
           & \eq~(\ref{eq:area-pure-ghost})
           & \hspace*{0.3cm}average active area for pure-ghost jets \\[5pt]
      $\bullet$ area fluctuations&
           & 
	   \\
        \hspace*{0.3cm} $\Sigma^2(\oPJ)$
           & (\ref{eq:area-hard-particle-antikt},\ref{eq:ktcamsigmas},\ref{eq:active_1point_cone_sigma})
           & \hspace*{0.3cm}variance of one-particle active area\\
        \hspace*{0.3cm} $\langle\Sigma^2 \rangle$
           & \eq~(\ref{eq:act-fluct-seq})
           & \hspace*{0.3cm}variance of active area of jets with QCD branching: \\ &&
	   $\hspace*{0.3cm}\langle\Sigma^2 \rangle
                = \Sigma^2(\oPJ)
                + S^2 \,  \frac{C_1}{\pi b_0}\ln \frac{\as(Q_0)}{\as(R p_t)} + \ldots$\\
        \hspace*{0.3cm} $S^2$
           & \eq~(\ref{eq:Sigmaalg1})
           & \hspace*{0.3cm}coefficient of leading scaling violations of
	   $\langle\Sigma^2 \rangle$\\
        \hspace*{0.3cm} $\Sigma^2$(GJ)
           & \eq~(\ref{eq:area-pure-ghost-stddev})
           & \hspace*{0.3cm}variance of active area for pure-ghost jets
    \\[5pt]\hline 
     && \\[-10pt]
    Back reaction & Section \ref{sec:areamed-analytic-back-reaction} & action of
    finite-momentum min-bias (MB) on the clustering of the non-MB particles
    \\
      $\bullet$ pointlike MB
           & 
           & 
	   \\
        \hspace*{0.3cm} $dP^{(L,G)}/dp_{t2}$
           & \eqs~(\ref{eq:loss-passive-master}, \ref{eq:gain-passive-master})
           & \hspace*{0.3cm}probability of jet losing (L) or gaining
           (G) $p_{t2}$ worth \\
           & & \hspace*{0.3cm}of non-MB particles\\
        \hspace*{0.3cm} $b^{(L,G)}(p_{t2}/p_{tm})$
           & \eqs~(\ref{eq:loss-effective-area}, \ref{eq:gain-effective-area})
           & \hspace*{0.3cm}eff.\ area for loss, gain (MB
           particle has $\perp$ mom.\ $p_{tm}$) \\
        \hspace*{0.3cm} $\beta$  
           & \eq~(\ref{eq:passive-beta})
           & \hspace*{0.3cm}coeff.~of net high-$p_{t2}$ gain$-$loss: 
           $\lim_{p_{t2}\to\infty} \frac{p_{t2}}{p_{tm}}
           b^{(G-L)}(\frac{p_{t2}}{p_{tm}}\!)\!$
           \\[5pt]
      $\bullet$ diffuse MB&
           & 
	   \\
        \hspace*{0.3cm} $dP^{(L,G)}/dp_{t2}$
           & \eq~(\ref{eq:BR-active-loss-master})
           & \hspace*{0.3cm}probability of jet losing (L) or gaining
           (G) $p_{t2}$ worth \\
           & & \hspace*{0.3cm}of non-MB particles\\
        \hspace*{0.3cm} $B^{(L,G)}(p_{t2}/\rho)$
           & \eq~(\ref{eq:loss-active-factorize})
           & \hspace*{0.3cm}eff.\ area for loss, gain (MB
           has $\perp$-mom.\ density $\rho$) \\
        \hspace*{0.3cm} $\cal B$  
           & \eq~(\ref{eq:active-beta})
           & \hspace*{0.3cm}coeff.~of net high-$p_{t2}$ gain$-$loss: 
           $\lim_{p_{t2}\to\infty} \frac{p_{t2}}{\rho}
           B^{(G-L)}(\frac{p_{t2}}{\rho}\!)\!$
    \\[5pt]\hline
  \end{tabular}
  \caption{A summary of the mathematical quantities defined throughout
    this article, together with descriptions of the associated
    physical concepts.}
  \label{tab:all-defs}
\end{table}

\begin{table}\small
  \newcommand{\D}[1]{\comment{$#1$}}
  \centering   
  \begin{tabular}{r|cc|cc|cc|cc||cc|}
                 & $a$(1PJ) & $A$(1PJ)  & $\sigma$(1PJ) & $\Sigma$(1PJ)  & $d$         &  $D$   & $s$     & $S$      & A(GJ)  & $\Sigma$(GJ) \\\hline
   $k_t$         & $1$      & $0.81$    & $0$           & $0.28$         & $\,\,0.56$  & $0.52$ & $0.45$  & $0.41$   & $0.55$ &  $0.17$      \\ \hline
   Cam/Aachen    & $1$      & $0.81$    & $0$           & $0.26$         & $\,\,0.08$  & $0.08$ & $0.24$  & $0.19$   & $0.55$ &  $0.18$      \\ \hline
   anti-$k_t$    & $1$      & $0$       & $0$           & $0$            &  $\,\,0$    & $0$    & $0$     & $0$      &  ---   &  ---         \\ \hline
   SISCone       & $1$      & $1/4$     & $0$           & $0$            & $\!\!-0.06$ & $0.12$ & $0.09$  & $0.07$   &  ---   &  ---         \\ \hline
  \end{tabular}
  \caption{A summary of the numerical results for the main quantities
    worked out in the article and described in table~\ref{tab:all-defs}.
    All results are normalised to $\pi R^2$, and rounded
    to the second decimal figure.
    Anomalous dimensions multiply powers of $\as^n \ln^n p_t/Q_0$ that
    for typical jet transverse momenta sum to something of order $1$.
    Active-area and anomalous-dimension results hold only in the
    small-$R$ limit, though finite-$R$ corrections are small.
    Back reaction is not easily summarised by a single number, and the
    reader is therefore referred to 
    Figs.~\ref{fig:passive-loss-gain}, \ref{fig:active-loss-gain}.
  }
  \label{tab:summary}
\end{table}

A summary of our main results for active and passive areas, their
fluctuations and their anomalous dimensions is given in
table~\ref{tab:summary}. As is visible also in
figure~\ref{fig:anom_dim_herwig_gg} for a broad range of $p_t$, there
is a hierarchy in the areas of the jet algorithms,
$\langle A_{\cone,R} \rangle \lesssim \langle A_{\antikt,R} \rangle
\sim \langle A_{\cam,R} \rangle \lesssim \langle A_{\kt,R} \rangle$.
A likely consequence is that SISCone jets will be the least affected
by UE contamination, though the extent to which this holds depends on
the precise balance between pointlike and diffuse components of the
UE.
The above hierarchy might also suggest an explanation for the opposite
hierarchy in the size of hadronisation corrections, observed in Monte
Carlo studies for the different jet algorithms in
\cite{Dasgupta:2007wa}. There the $k_t$ algorithm was seen to be least
affected, which now appears natural, since a larger jet is less likely
to lose momentum through hadronisation.

In the context of pileup subtraction, the significant fluctuations of
the area from one jet to the next (for algorithms other than
anti-$k_t$) mean that it is important to take into account the area of
each individual jet, rather than assume some typical mean value. Among
results of relevance to the subtraction procedure, we highlight the
demonstration that all areas (passive, active, Voronoi) are identical
for highly populated events, which is important in ensuring that the
subtraction procedure is free of significant ambiguities. We also
remark on the calculation of back-reaction of pileup on jet structure,
which is found to be a small effect, especially for the anti-$k_t$
algorithm.

\subsection{Future investigations}\label{sec:area-analytics-future}

There are many avenues for potential further study of jet areas. The
calculations presented here have extracted only the leading
logarithmic part of the first non-trivial order in $\as$, and usually
we have concentrated on properties of the mean area. 
The Monte Carlo results in
Section~\ref{sec:areamed-areaanalytics-real-life} also suggest interesting
structures in the distributions of areas, and these merit further
investigation.

In the specific case of the widely used anti-$k_t$ algorithm, studying
the deviations of the jet areas from $\pi R^2$ would also be of
interest, especially in the context of jet calibration at the LHC. In
particular, one would want to study the jet area when two hard jets
become close to one another.
A possibly related question is the one of the reach of the anti-$k_t$
algorithm, where one may ask if there is a maximal area for an
anti-$k_t$ jet, or a maximal distance at which two particles can be
clustered.

Stepping away from the anti-$k_t$ algorithm, another question for
future work is that of the transition between passive and active areas
if one considers a continuous transformation of the ghosts from
pointlike to diffuse.
Since real UE and pileup contamination is neither fully
pointlike, nor fully diffuse, this question is of
particular relevance for the stable-cone type algorithms, for which
there is a large difference between the passive and active areas.

Finally, an understanding of the behaviour of jet areas can play a key
role in the choice of parameters for jet algorithms, as well as in the
design of new algorithms. One example of this concerns SISCone, for
which we saw in section~\ref{sec:areamed-areaanalytics-active} that a
split--merge overlap threshold $f \simeq 0.5$ can lead to the
formation of ``monster jets,'' whereas a choice of $f \simeq 0.75$,
eliminates the problem, suggesting that the latter is a more
appropriate default value.

\section{Estimation of the pileup properties}\label{sec:analytic-pileup}

We show in this Section that it is also possible to obtain analytic
results about the estimation of $\rho$ using from the area--median
approach.
The natural question in this respect is the precision that one should
expect on the estimated value of $\rho$.

Many of the results included in this Section are adapted from
Ref~\cite{Cacciari:2009dp}, where the authors considered several
methods used to estimate event-by-event the Underlying Event
characteristics at the LHC. Here, we basically follow the same steps
but focus on the specific case of the area--median approach and put
the derivation in the context of pileup subtraction.

We shall also discuss several other aspects like estimations of the
impact of the size of the range used to estimate $\rho$, \ie address the
question of the minimal range size required to estimate $\rho$
reliably.

In the following pages, we will concentrate on grid-based
estimates. If, instead, we consider a jet-based determination of
$\rho$, most of the results derived below hold with the replacement of
the grid-cell size $a$ by $\sqrt{c_J}R$ with $R$ the jet radius and
$c_J\approx 0.65\pi \approx 2.04$ (see Ref.~\cite{Cacciari:2009dp}).

\subsection{A toy model for pileup}

To discuss the estimation of $\rho$, it is helpful to start by
introducing a simple toy-model to describe the non-perturbative pileup
particle distributions.

We will assume that pileup is producing, on average, $\nu$ particles
per unit area and that the probability to find $n$ particles in a
patch of area $A$ follows a Poisson distribution, \ie
\begin{equation}\label{eq:toy-multiplicity-distrib}
P_n=\frac{(\nu A)^n}{n!}e^{-\nu A}.
\end{equation}
We further assume that the particles have independent transverse
momenta and that the single-particle transverse-momentum probability
distribution decreases exponentially\footnote{Assuming a power-low
  behaviour would be slightly more physical but it would also
  complicate the calculations without changing significantly the
  features discussed here.}
\begin{equation}\label{eq:toy-one-particle-pt}
\frac{1}{P_1}\frac{dP_1}{dp_t} = \frac{1}{\kappa}e^{-p_t/\kappa}.
\end{equation}
Both the average transverse momentum of a particle and its dispersion
are therefore given by $\kappa$.

If one takes a patch of area $A$, containing $n$ particles, its total
$p_t$ probability distribution is given by
\begin{equation}\label{eq:toy-n-particle-pt}
  \frac{1}{P_n}\frac{dP_n}{dp_t}=\frac{1}{(n-1)!}\frac{p_t^{n-1}}{\kappa^n}e^{-p_t/\kappa}.
\end{equation}
From \eqref{eq:toy-multiplicity-distrib} and
\eqref{eq:toy-n-particle-pt}, it is possible to compute the overall
probability distribution for the transverse momentum in the patch. One
finds
\begin{equation}
\frac{dP}{dp_t}(A) 
 = \sum_{n=0}^\infty\frac{dP_n}{dp_t}
 = \delta(p_t)e^{-\nu A}+e^{-\nu A-p_t/\kappa}\sqrt{\frac{A\nu}{\kappa
     p_t}}I_1\left(2\sqrt{\frac{A\nu p_t}{\kappa}}\right),
\label{eq:toy-patch-pt-distrib}
\end{equation}
with $I_1(x)$ the modified Bessel function of the first kind.
This distribution has a mean of $\nu A\kappa$ and a standard deviation of
$\sqrt{2\nu A}\kappa$, where we recognise the expected scaling in $\kappa$
and $\nu A$.

If we want to match this toy model to our description of pileup in
terms of a deposit $\rho$ per unit area and fluctuations $\sigma$, we
have
\begin{align}
\rho   &= \kappa\nu\\
\sigma &= \sqrt{2\nu}\kappa
\end{align}
so that, in the end, 
\begin{equation}
p_t(A) = \rho A \pm \sigma \sqrt{A}.
\end{equation}
Note that, in the high-density limit, \ie in the large pileup limit
$\nu A\gg 1$, the distribution (\ref{eq:toy-patch-pt-distrib}) will
tend to a Gaussian.

In what follows, we shall consider typical values obtained in
Section~\ref{sec:area-median:pileup-properties}, say
$\rho=\rho_1\nPU$, with $\rho_1\approx 700$~MeV, and
$\sigma=\sigma_1\sqrt{\nPU}$, with $\sigma_1\approx 550$~MeV.
These would correspond to $\nu=2\rho^2/\sigma^2\approx 3.25\nPU$ and
$\kappa=\sigma^2/(2\rho)\approx 215$~MeV.
%
If, instead, we use $1/P_1\,dP_1/dp_t=4p_t\kappa^2 e^{-p_t/\kappa}$, for
which $\sigma/\rho=\sqrt{3/(2\nu)}$, one obtains
$\nu=(3/2)\rho^2/\sigma^2\approx 2.40\nPU$ and
$\kappa=2\sigma^2/(3\rho)\approx 290$~MeV.
In both cases, we should expect a few particles per unit area for each
pileup interaction.

\subsection{Pileup distribution impact on $\rho$ estimation}

A first source of potential bias for the estimation of $\rho$ using
the area--median approach is coming from the fact that there might be
a mismatch between the average and the median of the distribution
(\ref{eq:toy-patch-pt-distrib}).
To estimate this effect, let us consider pure-pileup events, \ie event
which are just the superposition of $\nPU$ minimum bias event without
any hard interaction.
Assuming that we use the grid-based approach with cells of area
$A_{\text{cell}}$, the estimated $\rho_{\text{est}}$ median would
correspond to the solution of
\begin{equation}
  \int_0^{A_{\text{cell}}\rho_{\text{est}}}dp_t\,\frac{dP}{dp_t}(A_{\text{cell}})=\frac{1}{2}.
\end{equation}

The integration above is delicate to handle analytically but one can
obtain an approximate solution that is numerically close to the exact
solution and keeps the right asymptotic behaviour:
\begin{equation}
\rho_{\text{est}}^{\text{(PU)}}
  = \rho\,
    \frac{\nu A_{\text{cell}}-\log(2)}{\nu A_{\text{cell}}-\log(2)+\frac{1}{2}}\,
    \Theta(\nu A_{\text{cell}}>\log(2)).
\end{equation}
This shows a turn-on point for $\nu A_{\text{cell}}=\log(2)$ and has
the property that for asymptotic $\nu A_{\text{cell}}$ one tends to
the expected value for $\rho$ according to
\begin{equation}\label{eq:asymptotic-shift-from-soft}
\frac{\rho_{\text{est}}^{\text{(PU)}}-\rho}{\rho} = -\frac{1}{2\nu A_{\text{cell}}} +
\order{(\nu A_{\text{cell}})^{-2}}.
\end{equation}

This parametric behaviour is actually fairly independent of the
details of the underlying 1-particle transverse momentum
distribution. Although its coefficient changes in
\eqref{eq:asymptotic-shift-from-soft}, other distribution, including
those behaving like $p_t^{-m}e^{-(m+1)p_t/\kappa}$ would all show a
$1/(\nu A_{\text{cell}})$ bias.

It is also helpful to rewrite
\eq~\eqref{eq:asymptotic-shift-from-soft} in terms of the pileup
properties $\rho$ and $\sigma$. If $a$ is the size of the grid cells,
\ie $A_{\text{cell}}=a^2$, we find
\begin{equation}
\frac{\rho_{\text{est}}^{\text{(PU)}}-\rho}{\rho} = -\frac{\sigma^2}{4\rho^2 a^2} +
\order{\frac{\sigma^4}{\rho^4 a^4}}.
\end{equation}

For the alternative model, $1/P_1\,dP_1/dp_t=4p_t\kappa^2 e^{-p_t/\kappa}$,
the above results would become (see \cite{Cacciari:2009dp} for
details)
\begin{equation}
\rho_{\text{est}}^{\text{(PU)}}
  = \rho\,
    \sqrt{\frac{\nu A_{\text{cell}}-\log(2)}{\nu A_{\text{cell}}-\log(2)+\frac{2}{3}}}\,
    \Theta(\nu A_{\text{cell}}>\log(2)).
\end{equation}
and
\begin{equation}
\frac{\rho_{\text{est}}^{\text{(PU)}}-\rho}{\rho} = -\frac{2\sigma^2}{9\rho^2 a^2} +
\order{\frac{\sigma^4}{\rho^4 a^4}}.
\end{equation}

\subsection{Perturbative impact of the hard event}\label{sec:analytic-rhoest-hard}

Let us now take the opposite situation and consider events without
pileup or Underlying Event but with just a hard interaction.
If we assume that we have selected events with at least 2 hard jets,
extra jets can arise from perturbative radiation at higher orders. 
These ``hard'' jets will contribute to the estimation of the median,
potentially biasing the estimated value of $\rho$ towards larger
values.
We estimate these effects below.

To a decent approximation, we can consider that these emissions are
soft, that they are distributed uniformly in azimuth and rapidity,
and with a $p_t$ spectrum proportional to $C_i\alpha_s(p_t)/p_t$, with
$C_i$ a colour factor. This means
\begin{equation}\label{eq:toy-perturbative-rad}
\frac{dn}{dp_t\,dy\,d\phi}\approx\frac{C_i}{\pi^2}\frac{\alpha_s(p_t)}{p_t}.
\end{equation}
We will consider emissions with $p_t$ between an infra-red scale
$Q_0$, that we shall set to 1~GeV from now on, and a hard scale
$Q=p_{t,\text{hard}}/2$. 
Assuming independent emissions, the probability distribution for the
number of emissions will be Poissonian with a mean given by the
integral of \eqref{eq:toy-perturbative-rad} over the whole phasespace
available for radiation.

The average total number of ``hard'' jets $\avg{n_h}$ can be written
as $\avg{n_h}=n_b+\avg{n_p}$ with $n_b$ the number of original,
Born-level, jets and $\avg{n_p}$ the number of perturbative emissions
in the total area $A_{\rm tot}$, given by
\begin{equation}
\avg{n_p}
 \approx \frac{C_i}{\pi^2}\int_{A_{\rm tot}}dy\,d\phi\int_{Q_0}^{Q}\frac{dp_t}{p_t}\alpha_s(p_t)
 = \frac{C_iA_{\rm tot}}{2\pi^2\beta_0}\log\left(\frac{\alpha_s(Q_0)}{\alpha_s(Q)}\right),
\end{equation}
with $\beta_0$ the 1-loop QCD $\beta$ function.

For a given cell of area $A_{\text{cell}}=a^2$, the probability to
have at least an emission in the cell is
\begin{align}
P_{\text{em}}
 &\approx 1-\exp\left[\frac{C_i}{\pi^2}\int_{A_{\text{cell}}}dy\,d\phi
   \int_{Q_0}^{Q}\frac{dp_t}{p_t}\alpha_s(p_t) \right]\\
 &=1-\exp\left[\frac{C_iA_{\text{cell}}}{2\pi^2\beta_0}\log\left(\frac{\alpha_s(Q_0)}{\alpha_s(Q)}\right)\right].\label{eq:estim-Pem}
\end{align}
Then, for the background density estimation to be non-zero, we need at
least half of the cells non-empty. If there is a number
$N=A_{\text{tot}}/A_{\text{cell}}$ cells, this probability is given by
(recalling that we already have $n_b$ initial hard jets)
\begin{equation}
P_{\text{nonzero}}
 =\sum_{k=N/2-n_b}^{N-n_b}\frac{(N-n_b)!}{k!(N-n_b-k)!}P_{\text{em}}^k
 \approx \frac{(N-n_b)!}{(N/2-n_b)!(N/2)!}P_{\text{em}}^{N/2-n_b},
\end{equation}
where we have assumed that in the limit $P_{\text{em}}\ll 1$, the sum
is dominated by the first term.
Given $P_{\text{nonzero}}$, one can estimate $\rho_{\text{est}}$ by
observing that the $(N/2)^{\text{th}}$ jet will be the softest of all
and therefore have $p_t\approx Q_0$, hence
\begin{equation}
\rho_{\text{est}}^{\text{(pert)}} \approx \frac{Q_0}{A_{\text{cell}}}P_{\text{nonzero}}.
\end{equation}

\subsection{Perturbative radiation and soft pileup}

In practice, real events will have both soft radiation coming from
pileup and perturbative radiation. 
The combination of the two effects is more delicate to handle
analytically, unless we work in the dense pileup region, $\nu
A_{\text{cell}}\gg 1$, far from the turn-on point
$\nu A_{\text{cell}}=\log(2)$.
In this region, relevant for most phenomenological considerations,
the pileup distribution in the grid cells can be considered
approximately Gaussian.

Out of the $N=A_{\text{tot}}/A_{\text{cell}}$ cells, an average of
$n_b+NP_{\text{em}}$, with $P_{\text{em}}$ given by
\eq~\eqref{eq:estim-Pem}, will be contaminated by hard perturbative
radiation. The
$N(1-P_{\text{em}})-n_b\approx N-\avg{n_h}+\avg{n_h}^2/(2N)$ remaining
jets will have an approximately Gaussian
$\rho_{\text{cell}}=p_{t,\text{cell}}/A_{\text{cell}}$ distribution of
the form
\begin{equation}
\frac{dn^{\text{PU}}}{d\rho_{\text{cell}}}
 \approx \frac{1}{\sigma}\sqrt{\frac{A_{\text{cell}}}{2\pi}}
         \exp\left[-\frac{A_{\text{cell}}}{2\sigma^2}(\rho_{\text{cell}}-\rho)^2\right]
         (N(1-P_{\text{em}})-n_b).
\end{equation}
Assuming $n_b/N+P_{\text{em}}<1/2$, the median procedure will find
$\rho_{\text{est}}$ such that $N/2$ of the $N(1-P_{\text{em}})-n_b$
remaining Gaussian-distributed pileup jets have
$\rho_{\text{jet}}<\rho_{\text{est}}$. This means that
$\rho_{\text{est}}$ will be determined by
\begin{equation}
\int_{-\infty}^{\rho_{\text{est}}}d\rho_{\text{jet}}\,\frac{dn^{\text{pileup}}}{d\rho_{\text{jet}}}=\frac{N}{2}.
\end{equation}

In the small $\avg{n_h}$ limit, this can be solved analytically and we
find
\begin{align}
\rho_{\text{est}}
 & \approx \rho_{\text{est}}^{\text{(PU)}}+\sqrt{\frac{\pi}{2A_{\text{cell}}}}\sigma
   \left(\frac{\avg{n_h}}{N}+\frac{\avg{n_h}^2}{2N^2}\right),\\
 & \approx \rho-\frac{\sigma^2}{4\rho A_{\text{cell}}}
   +\sqrt{\frac{\pi A_{\text{cell}}}{2}}\sigma
   \left(\frac{\avg{n_h}}{A_{\text{tot}}}+A_{\text{cell}}\frac{\avg{n_h}^2}{2A_{\text{tot}}^2}\right),
\label{eq:rhoest-total-average}
\end{align}
where $\rho_{\text{est}}^{\text{(PU)}}$ in the first equality is the
result obtained in \eq~\eqref{eq:asymptotic-shift-from-soft} in the
pure-pileup case.\footnote{In the second line, we have dropped the
  $\Theta$ function in $\rho_{\text{est}}^{\text{(PU)}}$. In practical
  applications in Section~\ref{sec:rhoest-mc-comparison}, we will
  simply require that the final estimation be positive.} For
large-enough pileup, the scale $Q_0$ in the definition of $\avg{n_h}$
should be replaced by a scale of the order of the pileup fluctuations
$\sigma \sqrt{A_{\text{cell}}}$, so that we have
\begin{equation}
\frac{\avg{n_h}}{A_{\text{tot}}} 
  \approx \frac{n_b}{A_{\text{tot}}} + \frac{C_i}{2\pi^2\beta_0}\log\left(\frac{\alpha_s(\text{max}(Q_0,\sigma\sqrt{A_{\text{cell}}}))}{\alpha_s(Q)}\right).
\label{eq:rhoest-nh-expression}
\end{equation}

At the end of the day, the bias in the estimation of the pileup
density $\rho$ receives two main contributions: the non-Gaussianity
of the pileup distributions results in a negative bias decreasing with
the grid size $a$, the second term in the above equation, and the
contamination from the hard radiation yields a positive bias
increasing with $a$. We should therefore expect that there is an
intermediate range in $a$ where these two biases cancel approximately.

If we ignore the $n_b$ contribution to $\avg{n_h}$ and the term
proportional to $\avg{n_h}^2$, and set the logarithm in $\avg{n_h}$ to
a typical value of 1, we find that the two biases cancel for
\begin{equation}
a 
 \approx \sqrt{\frac{\pi}{2^{1/3}}}\left(\frac{\beta_0}{C_i}\frac{\sigma}{\rho}\right)^{1/3}
 \approx 0.9\left(\frac{C_A}{C_i}\frac{\sigma}{\rho}\right)^{1/3}.
\end{equation}

\subsection{Fluctuations in the estimated $\rho$}

The estimation of $\rho$ will fluctuate depending on the exact
distribution of the pileup particles and the hard radiation.
This is due to the fact that the pileup distribution is only sampled
in a finite area and with a finite number of objects. The pileup
fluctuations therefore also imply some fluctuations in the estimated
value of $\rho$.
In that sense, the results provided in the previous Section are
really averaged values for the estimated $\rho$.
In this Section, we concentrate on the fluctuations of the estimated
$\rho$ around this averaged value.
As for the average estimate of $\rho$ above, we shall first consider
the contribution from the soft pileup itself and then the contribution
from the hard radiation.

\paragraph{Pure-pileup case.} For our purpose, it is sufficient to
work in the dense limit where $\nu A_{\rm cell}\gg 1$ so that we can
approximate the $\rho_{\rm cell}=p_{t,\rm cell}/A_{\rm cell}$
distribution (see \eq~\eqref{eq:toy-patch-pt-distrib}) by a Gaussian
of average $\rho$ and dispersion $\sigma/\sqrt{A_{\rm cell}}$
\begin{equation}
\frac{dP}{d\rho_{\rm cell}}
 = \sqrt{\frac{A_{\rm cell}}{2\pi}}\frac{1}{\sigma}
   \exp\left[-\frac{A_{\rm cell}}{2\sigma^2}(\rho_{\rm cell}-\rho)^2\right].
 \end{equation}
 From that we can easily obtain the cumulative probability
\begin{equation}
  P(<\rho_{\rm cell})
  = \int_{-\infty}^{\rho_{\rm cell}} d\rho_{\rm cell}\,\frac{dP}{d\rho_{\rm cell}}
  = \frac{1}{2}+\frac{1}{2}{\rm erf}\left(
       \sqrt{\frac{A_{\rm cell}}{2}}
       \frac{\rho_{\rm cell}-\rho}{\sigma}\right)
  \approx \frac{1}{2}+\sqrt{\frac{A_{\rm cell}}{2\pi}}
       \frac{\rho_{\rm cell}-\rho}{\sigma},
\end{equation}
where the last expression is valid when
$\sqrt{A_{\text{cell}}}(\rho_{\rm cell}-\rho)/\sigma$ is small.

Now, say we have $N=2m+1$ such cells, where we have taken $N$ odd for
simplicity. In that case, the median will be $\rho_{\rm est}$ if $m$
cells have a smaller $\rho_{\rm cell}$, one cell has
$\rho_{\rm cell}=\rho_{\rm est}$ and the remaining $m$ cells a larger
$\rho_{\rm cell}$. This corresponds to the following probability
distribution
\begin{equation}
\frac{dP_{\rm median}}{d\rho_{\rm est}}
 = (2m+1)\frac{(2m)!}{(m!)^2}\,\frac{dP}{d\rho_{\rm est}}
   [P(<\rho_{\rm est})]^m [1-P(<\rho_{\rm est})]^m.
\end{equation}
For large $m$, we can use the Stirling formula,
$m!\approx\sqrt{2\pi m}(m/e)^m$, so that, working in the small
$\sqrt{A_{\text{cell}}}(\rho_{\rm cell}-\rho)/\sigma$ limit, we get
\begin{equation}
\frac{dP_{\rm median}}{d\rho_{\rm est}}
 \approx \frac{\sqrt{2mA_{\rm cell}}}{\pi\sigma}
   \left[1-\frac{2A_{\rm cell}}{\pi}\frac{(\rho_{\rm cell}-\rho)^2}{\sigma^2}\right]^m
 \approx \frac{\sqrt{A_{\rm tot}}}{\pi\sigma}
   \exp\left[-\frac{A_{\rm tot}}{\pi}\frac{(\rho_{\rm cell}-\rho)^2}{\sigma^2}\right],
\end{equation}
where, to obtain the last equality, we have used $N\approx 2m$,
$NA_{\rm cell}=A_{\rm tot}$ and taken the large $N$ limit.
This means that the dispersion in the estimated $\rho$ will be
\begin{equation}\label{eq:rhoest-purepu-dispersion}
S_{\rm est}^{\rm (PU)} \approx \sqrt{\frac{\pi}{2A_{\rm tot}}}\,\sigma.
\end{equation}
This result has actually already been used in the previous Chapter
when estimating the minimal size of a range required to obtain a good
estimate of $\rho$ (see Appendix~\ref{app:minrange-fluct}).

Note that if we had computed $\rho$ from the average of the
$\rho_{\rm cell}$, we would have obtained a dispersion
$\sigma/\sqrt{NA_{\rm cell}}=\sigma/\sqrt{A_{\rm tot}}$. The median
estimate is larger by a factor $\sqrt{\pi/2}\approx 1.25$. This
moderate increase in resolution is the price to pay to have a method
which is robust against hard perturbative radiation.

\paragraph{Hard perturbative radiation.} To compute this effect, we
take a slightly different approach than the one we have adopted
before to calculate the average effect. 

We start with an event with pileup and investigate the effect of
adding hard perturbative radiation.
For each emission, we then have two possible situations: either the
hard emission contributes to a cell which was already above the
median, leaving the median unchanged, or it contributes to a cell
which was below the median, putting it above.
If there are $k$ emissions belonging to the second category, the
median will be shifted by (still working in the large number of cell
$N$ limit)
\begin{equation}
\avg{\delta\rho_{\rm est}} = k\delta_1,
\qquad\text{with }
\delta_1=\sqrt{\frac{2\pi}{A_{\rm cell}}}\frac{\sigma}{N}.
\end{equation}
For $k=\avg{n_h}/2$, we correctly recover the hard contamination from
\eqref{eq:rhoest-total-average}.

Around that average value, we can have two sources of fluctuations:
the first is related to the fact that, for a given $k$, the exact
shift of the median will depend on how the $\rho_{\rm cell}$ are
distributed, and the second comes from the fluctuations in $k$ 
itself.

Let us start with the first source. For $k=1$, the median will shift
up by one jet, and the distribution of $\delta\rho$ will be simply be
given by the distribution of the difference in $\rho$ cell between two
neighbouring cells in the sorted sequence of cells (at position in the
sequence that is near the median). That distribution is an exponential
distribution with mean $\delta_1$:
\begin{equation}
\frac{dP_{\rm shift}}{d\delta\rho}(k=1)
 =\frac{1}{\delta_1}e^{-\delta\rho/\delta_1}.
\end{equation}
For an arbitrary $k$ shifted jets, this therefore gives
\begin{equation}
\frac{dP_{\rm shift}}{d\delta\rho}(k)
 =\frac{1}{k!}\frac{(\delta\rho)^{k-1}}{\delta_1^k}e^{-\delta\rho/\delta_1},
\end{equation}
which has a mean $k\delta_1$ as expected and standard deviation
$\sqrt{k}\delta_1\simeq\sqrt{\avg{n_h}/2}\delta_1$.

Now, for the second effect, $k$ will follow a Poisson distribution
with mean $\avg{n_h}/2$ and standard deviation
$\sqrt{\avg{n_h}/2}$.\footnote{For simplicity, we have treated Born
  partons and other emissions on equal footing. A more precise
  calculation would only include the hard perturbative emissions in
  the Poisson distribution.}
This corresponds to a fluctuation $\sqrt{\avg{n_h}/2}\delta_1$ in the
estimated $\rho$.

Putting these two results together, we obtain
\begin{equation}
S_{\rm est}^{\rm (hard)}
 = \sqrt{\avg{n_h}}\delta_1
 = \sqrt{2\pi \frac{A_{\rm cell}}{A_{\rm tot}}\frac{\avg{n_h}}{A_{\rm tot}}} \sigma
\end{equation}
with $\avg{n_h}/A_{\rm tot}$ given by \eq~(\ref{eq:rhoest-nh-expression}).

\subsection{Comparison to Monte-Carlo studies}\label{sec:rhoest-mc-comparison}

\begin{figure}
\centering
\includegraphics[width=0.48\textwidth]{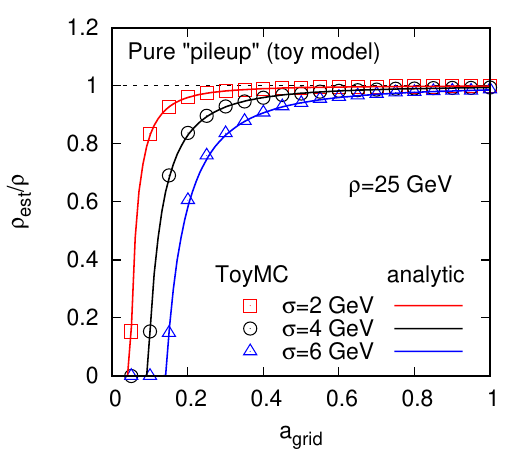}%
\hfill%
\includegraphics[width=0.48\textwidth]{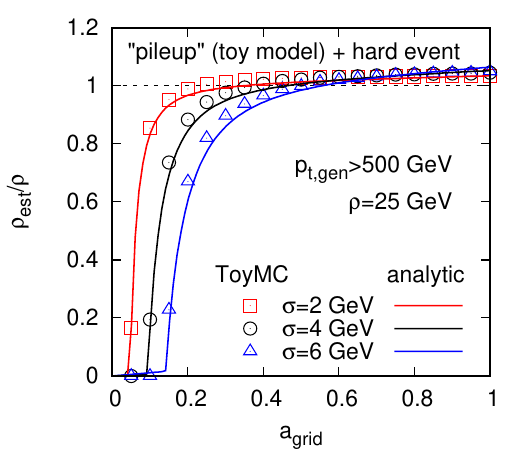}
\caption{Analytic predictions obtained in this Section for the
  estimated value of $\rho$ compared to numerical simulations of our
  toy-model pileup. For the hard events, we have generated dijet
  events with a generator $\hat{p}_t$ of 500
  GeV.}\label{fig:rho-estim-toy-model}
\end{figure}

\begin{figure}
\centering
\includegraphics[width=0.48\textwidth]{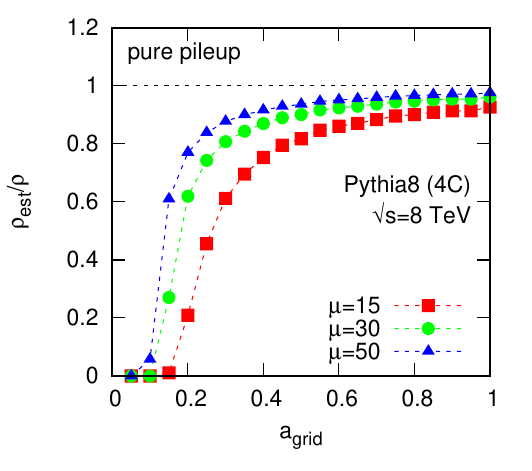}%
\hfill%
\includegraphics[width=0.48\textwidth]{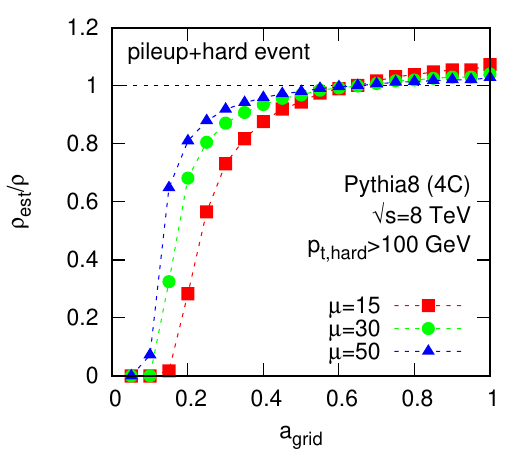}
\caption{Similar results as in Fig.~\ref{fig:rho-estim-toy-model} for
  pure pileup and pileup with an additional hard dijet, now with
  pileup also obtained from Pythia~8 simulations.  For the hard events,
  we have generated dijet events with a generator $\hat{p}_t$ of 100
  GeV.}\label{fig:rho-estim-pythia}
\end{figure}

In order to cross-check our calculations, we implemented the toy pileup
model in a Monte-Carlo event generator and compared results of
numerical simulations to our analytic expectations.

Figure~\ref{fig:rho-estim-toy-model} shows the comparison for
pure-pileup events on the left, and, on the right, for pure-pileup
events in which we have embedded dijet events simulated with Pythia~8
with $p_{t,\text{gen}}\ge 500$~GeV (right).
In all cases, we have kept the particles up to $|y|=5$ and varied the
size of the grid used to estimate $\rho$ between 0.05 and 1 by steps
of 0.05.\footnote{Note that we have used the
  \ttt{GridMedianBackgroundEstimator} class provided with \fastjet. This
  will adapt the grid-size in both rapidity and azimuthal angle to fit
  an integer number of cells in the available interval. This should
  have a negligible influence on our results.}
We have fixed $\rho=25$~GeV, a value in the ballpark of what one could
expect at the LHC and varied $\sigma$ to probe the scaling of our
analytic results.
For the analytic curves, we have used $n_b=2$, $C_i=C_A$ and a
one-loop running coupling with $n_f=5$ and
$\Lambda_{\text{QCD}}=100$~MeV.\footnote{We have checked that varying
  $\Lambda_{\text{QCD}}$ does not lead to any visible changes in our
  results.}
We see that, globally speaking, the analytic expressions do provide a
very good qualitative and quantitative description of the features
observed in the Monte Carlo simulations.

Finally, we also show on Fig.~\ref{fig:rho-estim-pythia} simulations
obtained with Pythia~8 for both the hard events and the
pileup. Although the quantitative details are different, especially
for the transition from $0$ to the region with $\rhoest\sim 1$. We
know however that this region is sensitive to the details of the
pileup model and we should therefore expect that our simplified toy
model should at most capture the qualitative features of the full
Pythia simulations. The plots on Figs.~\ref{fig:rho-estim-toy-model}
and \ref{fig:rho-estim-pythia} show that it is indeed the case. 


\part{Grooming techniques for fat jets}

\chapter{Main methods}\label{chap:grooming-description}

For the second part of this document, we shall switch gear and
consider another approach that can be used to mitigate pileup, namely
{\it grooming techniques}. In this Chapter, we discuss what is meant
by grooming and describe the most common techniques that have recently
been introduced and used at the LHC.
As for the area--median subtraction method discussed in the previous
part, we will then carry on with some Monte-Carlo studies, presented
in Chapter~\ref{chap:grooming-mcstudy}, and finally show that groomers
are amenable to some level of analytic understanding in
Chapter~\ref{chap:grooming-analytic}.

Today, grooming techniques are mostly used for specific applications,
usually in the context of boosted jets and jet substructure. Compared
to the area--median techniques, groomers are not as widely used as
generic pileup subtraction tools at the LHC. We shall therefore remain
a bit more concise in our presentation than what we have done in the
previous part of this document.
However, I believe that grooming techniques have the potential to play
a crucial role in the development of more advanced pileup mitigation
techniques. They are for example an element of the cleansing pileup
mitigation technique \cite{Krohn:2013lba} recently introduced. In the
third part of this document, which discusses new pileup mitigation
techniques, we shall therefore take an excursion into currently
unpublished, preliminary, results trying to use grooming techniques as
a generic pileup subtraction tool (see
Chapter~\ref{chap:beyond-grooming}).

\section{Grooming and tagging fat jets}\label{sec:fatjets}

Let us put aside for a moment the ``standard'' use of jets at the LHC
with relatively small jet radii and work with jets defined with larger
$R$, in the range 0.8-1.5. These {\it fat jets} are severely affected
by pileup and even without any pileup, are still substantially
affected by the Underlying Event.
One therefore often tries to exclude from the jet a fraction of its
constituents which is deemed to contribute essentially to pileup or
the UE. For example, one tends to remove soft contamination away from
the jet centre, a region indeed expected to be dominated by UE or
pileup.
This is typically what {\em groomers} do: given an input jet, they
clean it from soft and large angle radiation and return a groomed jet.

To give a slightly more concrete perspective to this discussion, let
us illustrate our arguments with the specific example of {\em
  filtering}. It works as follows: given a jet, one re-clusters its
constituents into subjets (with a smaller radius) and keeps only a
(pre-decided) fixed number of the hardest subjets (\ie with the
largest $p_t$). The union of the subjets that have been kept is the
filtered jet.
This filtered jet is therefore made of a subset of the original jet's
constituents.
Over a bit less than a decade, many grooming techniques have been
introduced and we shall briefly review them in the next Section.

We should notice that there is an overlap between the grooming
techniques and {\em boosted objects tagging}.
In a nutshell, the latter refers to the fact that massive objects
($W/Z/H$ bosons or top quarks), when produced with a $p_t$ much larger
than their mass and when decaying into hadrons, will be reconstructed
as single jets. To separate these boosted fat jets, originating from
the hadronic decay of a massive object, from the large QCD background,
one also exploits the internal properties of jets, \ie {\em jet
  substructure}. Such techniques are often referred to as {\em
  taggers}.

Taggers and groomers share many similarities, going beyond the simple
observation that they are both used on fat jets.
Physically, one approach to tag boosted jets is to rely on the fact
that massive objects share their $p_t$ across their decay products
more symmetrically than QCD jets which tend to be dominated by soft
radiation. In essence, this approach of finding hard prongs inside a
jet bears resemblance to eliminating soft contamination from the
Underlying Event or pileup. Hence taggers bear resemblance with
groomers.
In practice, many tools introduced as groomers can be used as taggers
and vice versa. Furthermore, combinations of the two are not uncommon,
{\em c.f.}  the use of filtering after applying a mass-drop tagger in
\cite{Butterworth:2008iy}.

In my opinion (see also the discussion in \cite{Dasgupta:2013ihk}),
the main difference between a tagger and a groomer lies in what we
expect it to do: when we use a tagger, we understand that we want to
identify multiple hard prongs in a fat jet with the ultimate goal to
maximise signal-to-background efficiency in boosted objects searches;
when we use a groomer, we mean that we want to eliminate soft
contamination and improve the resolution on the jet kinematics and
properties.
In other words, a tagger either succeeds or fails, typically according
to whether or not it finds hard substructure in the jet, while a
groomer always returns a (groomed) jet.

An additional point which is worth keeping in mind in the context of
pileup mitigation is that taggers and groomers can be seen as acting
on different physical scales (although, in a practical context, the
separation may not be as clear): improving the resolution on a jet,
the work of groomers, is expected to be done by imposing cuts around
the soft UE/pileup scale, while, for taggers, identifying hard prongs
in a jet is likely to involve cuts at the perturbative scale(s) of the
jet.
In the end, this is an area of jet substructure which is still in full
development and we can hope for exciting improvements in the next
years.

Let us now come back to our original purpose of mitigating pileup. One
should understand that grooming technique adopt a different approach
to pileup mitigation than the area--median subtraction method used so
far.
The main conceptual difference is that ``standard'' pileup subtraction
techniques like the area--median method try to correct the jet
contaminated by soft radiation back to a jet as it was without that
contamination, while grooming techniques would most often affect the
hard jets event in the absence of pileup.

This can be understood in the context of our qualitative discussion of
pileup effects in Section~\ref{sec:areamed-idea-characterisation}. The
main goal of the area--median subtraction method is to eliminate the
shift due to pileup independently of the hard process one is
considering. This means, eliminate, on average, the $\rho A_{\rm jet}$
contamination from the jet. In doing so, the smearing effects
are reduced to $\sigma \sqrt{A_{\rm jet}}$.
Grooming techniques, on the other hand, target an improvement in $p_t$
(or mass) resolution of the jet, \ie want to reduce as much as
possible the $\sigma \sqrt{A_{\rm jet}}$ without guaranteeing an
unbiased average contamination. By excluding soft parts of the jet,
grooming typically reduces its area from $A_{\rm jet}$ to, say,
$A_{\rm groomed}<A_{\rm jet}$. This is not bias-free since it comes at
the expense of removing some of the constituents of the original hard
jet (a negative bias) and leaving a positive $\rho A_{\rm groomed}$
contribution from pileup. However, it significantly lowers the
resolution degradation by reducing it to
$\sigma \sqrt{A_{\rm groomed}}$.  Adjusting the free parameters of the
groomer we could try to minimise the bias while keeping the resolution
improvements but this has no guarantee to work on a wide range of
applications and contexts.

Both approaches have their advantage. For a broad and robust
applications of pileup mitigation to jets with not-so-large $R$, it is
crucial to control the average bias and this is what the area--median
subtraction method is good at.
On the other hand, if one considers a search based on fat jets,
appropriate grooming techniques are perfectly suited to obtain a
narrow signal peak on a reduced QCD background.
Note that, in this case, we also wish to supplement the groomer with
area--median subtraction in order to also correct for the positive
bias due to $\rho A_{\rm groomed}$ and further improve the resolution
when averaging over the full jet sample.

Before digging into the grooming techniques themselves, it is
interesting to illustrate some of the arguments above with a concrete
example.
Let us consider the measurement of the mass of boosted tops
reconstructed using fat jets. 
The area--median subtraction alone would, on average reconstruct the
correct average jet mass for both top quarks and pure-QCD jets, with a
potentially large smearing due to pileup fluctuations (and the UE).
A grooming technique, like filtering briefly introduced above where we
would keep the 3 or 4 hardest subjets, would have a different
behaviour.
For signal jets coming from the hadronic decay of top quarks, one
would keep the main structure of the 3-prong decays and hence
reconstruct a mass with little average bias, and, at the same time,
benefit from the grooming to obtain a good resolution on the mass \ie
a narrower mass peak than without filtering. For standard QCD jets,
where most often the mass comes from soft radiation at larger angles,
grooming would strongly reduce their mass, already at the perturbative
level. This can be a large bias which, in the case of our search for
top quarks, plays in our favour since it further suppresses the QCD jet
background.
Combining grooming and area--median subtraction would further improve
the top signal and further reduce the QCD background.
We shall see this explicitly in Chapter~\ref{chap:grooming-mcstudy}.

\section{Mass-drop, filtering, trimming, pruning, soft drop}\label{sec:grooming-list-techniques}

Here, we briefly introduce the main grooming techniques that have been
used over the last decade. Some were initially introduced
either as taggers, or as both taggers and groomers, but at the end of
the day, they all eliminate soft and large-angle radiation and, in
that sense, can be seen as groomers.

\paragraph{Filtering.} Given a fat jet,
filtering~\cite{Butterworth:2008iy} re-clusters its constituents
with the Cambridge--Aachen algorithm with a small radius $R_{\rm
  filt}$. It then only keeps the $n_{\rm filt}$ hardest subjets (in
terms of their $p_t$). The filtered jet is the union of the subjets
that have been kept.
This has two adjustable parameters: $R_{\rm filt}$ and $n_{\rm filt}$.
It is typically used to reduce soft contamination in situations where
we have a prior knowledge of the number of hard prongs in a
jet. For a jet with $n_{\rm prong}$ hard prongs --- $n_{\rm prong}=2$
for a $W/Z/H$ bosons and $n_{\rm prong}=3$ for a top --- we would
typically use $n_{\rm filt}=n_{\rm prong}+1$ which would also keep the
(perturbative) radiation of an extra gluon.
Filtering is implemented directly in \fastjet where it is available as
the \ttt{Filter} tool (using \ttt{SelectorNHardest} for the selection
criterion).

\paragraph{Trimming.} Trimming~\cite{Krohn:2009th} shares some
similarities with filtering. It also starts with re-clustering the jet
with a smaller radius, $R_{\rm trim}$, using either the $k_t$ or the
Cambridge/Aachen algorithm. It then keeps all the subjets with a
transverse momentum larger than a fraction $f_{\rm trim}$ of the
initial jet transverse momentum.
On top of the choice of algorithm, this also has two parameters:
$R_{\rm trim}$ and $f_{\rm trim}$.
It is often used as a generic groomer and tagger in boosted-jet
studies.
Trimming can be obtained from the \ttt{Filter} class in \fastjet
(using \ttt{SelectorPtFractionMin} for the selection criterion).

\paragraph{Pruning.} Pruning~\cite{Ellis:2009su} is similar in spirit
to trimming except that it adopts a bottom-up approach (with trimming
seen as a top-down approach). Given a jet, pruning reclusters its
constituents using a user-specified jet definition (based on pairwise
recombinations) with the additional constraint that at each step of
the clustering, objects $i$ and $j$ are only recombined if they
satisfy at least one of these two criteria: (i) the geometric distance
$\Delta R_{ij}=\sqrt{\Delta y_{ij}^2+\Delta\phi_{ij}^2}$ is smaller
than $R_{\rm prune}=f_{\rm prune}\times 2m_{\rm jet}/p_{t,\rm jet}$,
with $p_{t,\rm jet}$ and $m_{\rm jet}$ the original jet transverse
momentum and jet mass, (ii) the splitting between $i$ and $j$ is
sufficiently symmetric, \ie
${\rm min}(p_{t,i},p_{t,j})\ge z_{\rm prune}p_{t,(i+j)}$. If none of
these criteria are met, only the hardest of $i$ and $j$ (in terms of
their $p_t$) is kept and the other is rejected.
On top of the jet definition used for the re-clustering, which is
usually taken to be either $k_t$ or Cambridge/Aachen with a radius
much larger than the one of original jet, this has two parameters:
$f_{\rm prune}$ and $z_{\rm prune}$. 
$z_{\rm prune}$ plays the same role as $f_{\rm trim}$ for trimming and
$f_{\rm prune}$ plays a role similar to $R_{\rm trim}$. Note however
that the $R_{\rm prune}$ is defined dynamically with the jet
kinematics, compared to $R_{\rm trim}$ which is kept fixed.  This can
have important consequences both analytically and
phenomenologically.\footnote{It would be interesting to study the
  properties of trimming with a dynamical trimming radius.}
Pruning can also be considered as a general-purpose groomer and tagger
and is often used in situations similar to trimming.
It is available in \fastjet as the \ttt{Pruner} class.

\paragraph{$I$ and $Y$-Pruning.} When pruning a jet, there might be
situations where a soft emission at large angle dominates the mass of
the jet but gets pruned away because it does not satisfy the symmetry
condition. The mass of the pruned jet is then determined by radiation
at smaller angle, typically within the pruning radius. This situation
where the jet mass and the pruning radius are determined by different
emissions within the jet would result in a jet with a single prong and
are therefore called I-Pruning. For these situations, the pruning
radius has no relation to the hard substructure of the jet.
More precisely, I-Pruning is defined~\cite{Dasgupta:2013ihk} as the
subclass of pruned jets for which, during the sequential clustering,
there was never a recombination with $\Delta R_{ij}>R_{\rm prune}$ and
${\rm min}(p_{t,i},p_{t,j})> z_{\rm prune}p_{t,(i+j)}$.
The other situation, \ie a pruned jet for which there was at least one
recombination for which $\Delta R_{ij}>R_{\rm prune}$ and
${\rm min}(p_{t,i},p_{t,j})> z_{\rm prune}p_{t,(i+j)}$, corresponds to
a genuine two-prong structure and is called Y-Pruning.
This distinction between I- and Y-Pruning is mostly relevant for
boosted jet tagging. However, it has been shown to have an impact on
the analytical behaviour of Pruning, with Y-Pruning being under better
control and than I-Pruning, the latter adding quite a lot of
complexity to the calculation. If one wishes to reach some level of
analytic control over groomed jets, Y-Pruning appears as a more
natural choice than Pruning which also includes the I-Pruning class.

\paragraph{Mass-drop.} Originally proposed~\cite{Butterworth:2008iy}
as a tool to isolate boosted Higgs bosons, decaying to $b\bar b$
pairs, from the QCD background, the mass-drop tagger assumes one works
with a fat jet clustered with the Cambridge/Aachen algorithm --- if it
is not the case, one is always at liberty to re-cluster the jet. One
then iteratively undoes the last step of the clustering
$p_{i+j}\to p_i+p_j$ until both the following criteria are met: (i)
there is a ``mass drop'' \ie
${\rm max}(m_i.m_j)<\mu_{\rm cut}m_{i+j}$, (ii) the splitting is
sufficiently symmetric \ie
${\rm min}(p_{t,i}^2,p_{t,j}^2)\Delta R_{ij}^2>y_{\rm cut}
m_j^2$.
When both criteria are met, we keep ``$i+j$'' as the result of the
mass-drop tagger, otherwise the procedure is repeated iteratively
keeping only the most massive of $i$ and $j$. If the procedure fails
to find two subjets satisfying the conditions, \ie end up recursing
until it reaches a single constituent which cannot be further
de-clustered, it is considered as having failed and returns an empty
jet.
The mass-drop tagger has two parameters: $\mu_{\rm cut}$, the
mass-drop parameter itself, and $y_{\rm cut}$, the symmetry cut.
The two conditions imposed by the mass-drop tagger make full sense in
the physical situation where we want to tag the decay of 2-pronged
boosted objects: the symmetry cut requires that one indeed finds two
hard prongs and the mass-drop condition imposes that one goes from a
massive boson jet to two jets originated from massless QCD partons.
Although it was originally introduced as a tagger, the mass-drop
tagger also acts as a groomer since, via the Cambridge/Aachen
declustering, it would iteratively remove the soft radiation at the
outskirts of the jet, hence limiting the pileup/UE contamination.
This is available as the \ttt{MassDropTagger} tool in \fastjet.

\paragraph{Modified Mass-drop (mMDT).} It has been shown recently
\cite{Dasgupta:2013ihk} that substructure tools can be understood
analytically. When trying to understand the behaviour of the mass-drop
tagger on QCD jets, it was realised that the fact that the iterative
de-clustering was following the most massive of the two branches could
lead to pathological situations. It was therefore suggested to adapt
the procedure so that it instead follows the hardest branch (in terms
of $p_t$).
This modification has the advantage that it makes the analytical
calculation much easier and more robust without affecting the
performance of the method (even improving it slightly).
The same study also added two more minor modifications. First, it was
realised that the symmetry condition could be replaced by
${\rm min}(p_{t,i},p_{t,j})>z_{\rm cut}(p_{t,i}+p_{t,j})$ and that the
latter would have a slightly reduced sensitivity to non-perturbative
corrections. Second, perturbatively, the mass-drop condition would
only enter as a subleading correction (in the strong coupling constant
$\alpha_s$) compared to the symmetry condition and so could usually be
ignored.\footnote{This argument was made and checked in the context of
  boosted object tagging \cite{Dasgupta:2013ihk}. It would still be
  interesting to see if the $\mu_{\rm cut}$ parameter in the mass-drop
  condition has an impact on grooming properties.}
The \ttt{ModifiedMassDropTagger} is available as part of the
\ttt{RecursiveTools} contribution in \fjcontrib. By default, it uses a
$z_{\rm cut}$ symmetry condition, does not impose a mass-drop
condition and follows the highest-$p_t$ branch, but variants are
allowed in the code.

\paragraph{\SD.} \SD~\cite{Larkoski:2014wba} can be seen a
generalisation of the modified Mass-drop tagger. It also proceeds by
iteratively de-clustering a jet obtained --- either directly or
re-clustered --- with the Cambridge/Aachen algorithm until the \SD
condition is met. If we decluster $p_{i+j}$ into $p_i$ and $p_j$, we
stop the de-clustering when
\begin{equation}\label{eq:soft-drop-condition}
  \frac{{\rm min}(p_{t,i},p_{t,j})}{p_{t,i}+p_{t,j}} 
    > \zcut \left(\frac{\Delta R_{ij}}{R}\right)^\beta,
\end{equation}
with $R$ the jet radius.
In the limit $\beta\to 0$, this condition reduces to the mMDT.  Hence,
\SD can be seen as an extension of the mMDT with the extra parameter
$\beta$ allowing for more freedom in constraining the
phasespace.\footnote{The \SD procedure returns by default a single
  particle if it fails to find two subjets satisfying the \SD
  condition. This ``grooming mode'' is different from the default
  ``tagging mode'' of the mMDT which would fail, \ie return an empty
  jet, if no substructure are found.}
\SD has two parameters. The $\zcut$ parameter plays the same
role as in the (m)MDT of keeping the hard structure and exclude soft
emissions, starting from large angles due to the Cambridge/Aachen
de-clustering. The $\beta$ parameter controls how aggressive the
groomer is. It goes from no grooming at all for $\beta \to \infty$ to
a quite aggressive grooming for $\beta=0$ (mMDT) and carries on with a
2-prong tagging region for $\beta\le 0$.
Used as a groomer, we believe that values of $\beta$ in the range
between 1 and 2 are appropriate as they retain some of the large-angle
structure of the jet, although larger values could also be
used.\footnote{Note that in the context of boosted object tagging,
  when used alone, negative values of $\beta$, typically the mMDT or
  \SD with $\beta=-1$, give perfectly adequate and efficient taggers
  (see the original \SD paper~\cite{Larkoski:2014wba}). However, if
  one wishes to combine \SD with other tagging techniques, like
  $N$-subjettiness or Energy-Correlation Functions, which constrain
  radiation patterns in boosted jets, we wish to retain some of the
  large-angle structure of the jets, where the radiation pattern
  between a colourless $W/Z/H$ and a QCD jet is most different. In
  that case, \SD with positive $\beta$ would be a good
  compromise between retaining some of this structure while excluding
  soft and non-perturbative UE and pileup effects. When combining
  two-prong taggers with radiation constraints, it would then appear
  natural to use in parallel negative, or zero, $\beta$ to identify
  the 2-prong structure and positive $\beta$ with a jet shape, to
  impose a cut on radiation.}
It is also interesting to realise that, for $\beta=2$, the \SD
condition imposes a momentum cut proportional to the square of the
angular separation of the two subjets, \ie proportional to the area
of the subjets. One might therefore expect that the choice $\beta=2$
plays a special role when it comes to discussing pileup
effects. For example, for $\beta=2$, one could expect the pileup
contamination to correspond to a shift in $z_{\rm cut}$.
In practice \SD is made available as a \ttt{SoftDrop} class in
the \ttt{RecursiveTools} contribution to \fjcontrib, where, similarly
to the mMDT, it also offers the option to impose an additional
mass-drop condition.

\paragraph{Recursive \SD.}
One of the potential issues with mMDT and \SD regarding their ability
to groom pileup away is that the declustering procedure stops once the
condition~(\ref{eq:soft-drop-condition}) has been met, leaving the jet
unchanged at smaller angular sizes.
{\em Recursive} \SD~\cite{Dreyer:2018tjj} instead iterates the
declustering procedure, either until a fixed number $N+1$ of hard
prongs have been found, or until the subjets can no longer be
declustered.
As a consequence, the grooming is carried on to smaller angular scales
and one can expect that the resulting jets have a smaller sensitivity
to pileup.
For simplicity, we will only consider \SD in this report and refer the
reader to Ref.~\cite{Dreyer:2018tjj} for pileup-mitigation studies
with Recursive \SD.

\paragraph{Combination with area--median subtraction.} As discussed
earlier, area--median subtraction and groomers target different
aspects of pileup mitigation. In many cases it is however possible to
combine both approaches in order to benefit both from the correction
of the bias implemented in the area--median subtraction and the
resolution improvements brought by grooming. We briefly comment on how
this can be achieved for the various grooming procedures defined
above.

\begin{itemize}
\item For the case of {\bf filtering} and {\bf trimming}, the idea is
  to use the area--median subtraction to correct the momentum of the
  subjets before selecting which ones have to be kept. In the end, the
  groomed jet would be the union of the {\it subtracted} subjets that
  have been kept.
  Note that in the case of trimming there is an ambiguity as to
  whether the condition on the (subtracted) subjet $p_t$ uses the
  subtracted full jet as a reference or the unsubtracted jet. Both
  options are acceptable and have their respective
  advantage. Basically, the former tends to be more aggressive than
  the latter. Depending on the application and the choice of
  parameters, both options show little practical difference.
  In terms of the implementation, the combination of
  filtering/trimming with area--median subtraction can be
  straightforwardly done by passing a \ttt{Subtractor} to the
  \ttt{Filter} object. For trimming, it is the end-user's
  responsibility to decide whether to pass a subtracted or an
  unsubtracted jet.
\item For {\bf pruning}, we could also decide, at least on paper, to
  apply an area--median subtraction at each step of the clustering and
  impose the pruning criteria on the subtracted jets. This has not yet
  been implemented in practice. This is due to a purely technical
  reason: jet areas in \fastjet are assigned to the jets a posteriori,
  hence are not available when one would need them to veto or not a
  recombination during the re-clustering.
\item For the case of the {\bf (modified) Mass-drop Tagger} and {\bf
    \SD}, it is again sufficient to apply the kinematic conditions on
  subtracted jets. The groomed jet at the end of the procedure would
  then be the union of the two subtracted subjets found by the
  iterative de-clustering.
  In practice, a \ttt{Subtractor} can be passed to both the
  \ttt{ModifiedMassDropTagger} and to the \ttt{SoftDrop} classes. Note
  that by default, the original jet is assumed to be unsubtracted and
  is subtracted internally.\footnote{This default behaviour can be
    changed.} Note that the \ttt{MassDropTagger} class in \fastjet has
  no support for internal subtraction so the
  \ttt{ModifiedMassDropTagger} should be preferred, knowing that the
  latter could, if needed, be parametrised so as to reproduce the
  behaviour of the former.
\end{itemize}

\paragraph{Further discussion, similarities and differences.} 
Before carrying on with some Monte-Carlo studies of groomers and an
example of their analytic properties, there are still a few generic
remarks that we find of generic use.
\begin{itemize}
\item all the groomers introduced above cut soft particles (or
  subjets) at large angles. This is either achieved explicitly by
  removing soft subjets (which are automatically away from the hard
  subjets by, at least, the subjet radius), or, when using
  de-clustering, relying on the specific choice of the
  Cambridge/Aachen jet algorithm, which reflects an angular ordering
  and therefore throws away soft radiation at large angles until some
  hard structure is met.
\item Since they all implement mainly the same physical ideas, the
  physical behaviour of all these tools share many similarities,
  something that can be seen both in Monte Carlo studies and analytic
  calculations.
\item At some point, when the jet boost is increased, important
  differences between the various groomers start to appear. For
  example, the fact that the Mass Drop tagger can recurse all the way
  to arbitrary small opening angle, compared to trimming which has a
  built-in subjet radius, or to Pruning which has a dynamical minimum
  angular scale, translates into different behaviour of these three
  tools. These can sometimes be seen in Monte-Carlo simulations but
  are best understood when assessing the analytic properties of the
  groomers.
\item \SD differs from the other groomers in that the symmetry cut is
  angle-dependent. In the end, I believe that this gives \SD some
  extra flexibility which could prove useful in a broad range of
  applications. It makes \SD a powerful groomer with nice
  analytic properties (see also next Chapters).
\item In terms of the contamination from pileup (or the UE) at an
  angular scale $\Delta R$, one would expect an average $p_t$ shift of
  order $\rho\,\pi\Delta R^2$. To some extent, this is reminiscent of
  \SD with $\beta=2$ where the condition would read
  $z>\zcut (\Delta R/R)^2$, suggesting
  $\zcut\gtrsim A_{\rm jet}\rho/p_{t,\rm jet}$.\footnote{A similar
    argument can be made for pileup fluctuations and $\beta=1$ \SD.}
\end{itemize}


\chapter{Monte-Carlo studies}\label{chap:grooming-mcstudy}

So far, most of the practical applications of grooming techniques have
concentrated on boosted jets studies.
We shall therefore investigate here the performance of grooming as a
tool to mitigate pileup effects on boosted jets.
This is mostly based on a study that we initially conducted in the
context of the proceedings of the Boost~2012
workshop~\cite{Altheimer:2013yza}, although the new version presented
here has been updated to include a broader set of groomers, especially
more modern ones.
We will come back to grooming techniques as a way to mitigate pileup
from generic jets in Chapter~\ref{chap:beyond-grooming}.

Compared to Ref.~\cite{Altheimer:2013yza}, this study includes more
recent groomers like the modified Mass Drop Tagger and \SD.
Similar to the initial study, the events are obtained by combining the
hard (signal) events embedded in a series of minimum-bias events to
simulate pileup.
We consider collisions at a centre-of-mass energy of
13~TeV. Minimum-bias events are generated with the Pythia~8 Monte Carlo
(MC)
generator~\cite{Sjostrand:2000wi,Sjostrand:2003wg,Sjostrand:2006za,Pythia8},
with its 4C tune \cite{Corke:2010yf}.
We considered different pileup conditions, Poisson distributed
with $\mu = \{ 30, 60, 100, 140 \}$.
This set of choices covers several situations of interest ranging from
typical pileup conditions for Run I and II of the LHC\footnote{The
  change in centre-of-mass energy should not bring any sizeable
  difference compared to the present study.} up to the exploration of
future high intensity scenarii at the LHC.

We have selected two example signals for the Monte Carlo simulation
studies presented in this Section. The first is the decay of a
possible heavy $Z'$ boson with a chosen $M_{Z'} = 1.5$~TeV to a
(boosted) top quark pair, as included in the proceedings of the Boost
2012 workshop. The top and anti-top quarks then decay fully
hadronically ($t \to W b \to jjb$).
The second also consists of a $Z'$ boson with $M_{Z'} = 1.5$~TeV, now
decaying into $WW$ with both $W$ decaying hadronically. 
These two samples are meant as sources of 3-pronged and 2-pronged
jets, respectively.
The signal samples were also generated using Pythia~8 with tune 4C.  

The analysis uses the tools either directly available in the \fastjet
package or in \fjcontrib for the jet finding and the jet substructure
analysis. The initial jet finding is done with the anti-$k_t$
algorithm with $R = 1$, to ensure that most of the final state $W$ and
top-quark decays can be collected into one jet. For the latter, this
corresponds to top-quarks generated with $p_t \gtrsim 400$~GeV.

In practice, we will apply several grooming procedures, listed in the
next Section, to the original anti-$k_t$ jets. For the studies
presented in this report we require that the jet $p_t$ before grooming
and pileup subtraction is larger than 100~GeV. We keep jets with
$|y|<3$. We only consider the two hardest $p_t$ jets in the event. We
further require that the rapidity difference between the two jets
$|y_1 - y_2|$ is less than one.

\section{Jet grooming configurations} \label{sec:grooming}

Several jet grooming techniques, in use at the LHC, are included in
our study (see Chapter~\ref{chap:grooming-description} for a more
complete description):
\begin{description}
\item {\bf Trimming.} In this study, we used the $k_t$ algorithm
  for the reclustering, with $R_{\mathrm{trim}} = 0.2$, and we used
  $f_{\text{trim}} = 0.03$.
\item {\bf Filtering.} Here, we reclustered the jets with the
  Cambridge-Aachen algorithm with $R_{\mathrm{filt}} = 0.2$, and
  kept the $n_{\mathrm{filt}}$ hardest subjets. $n_{\mathrm{filt}}$
  has been varied according to the process under consideration and we
  found good results using $n_{\mathrm{filt}}=2$ for the $Z'\to WW$ process
  and $n_{\mathrm{filt}}=4$ for the $Z'\to t\bar t$ case.
\item {\bf modified Mass Drop (mMDT)} and {\bf \SD (SD).} We have
  fixed $z_{\mathrm{cut}}=0.1$ and studied $\beta=0$ (mMDT), $1$ and
  $2$ (SD).
  Following the original prescription of \cite{Butterworth:2008iy}, we
  have optionally applied to the result of the mMDT/SD procedure, an
  extra step of filtering, using
  $R_{\rm filt}={\rm max}(R_{\rm SD},0.2)$ with $R_{\rm SD}$ the
  distance between the two prongs found by the mMDT/SD, and
  $n_{\rm filt}$ as for the plain filtering case described above.
  In a nutshell, once the mMDT/SD procedure has found a 2-pronged
  structure, thereby excluding radiation at angles larger than
  $R_{\rm SD}$, this extra filtering step further cleans the
  contamination at smaller angles.
\end{description} 

A specific point we want to study is the interplay between jet
grooming and area-based pileup correction. The subtraction is applied
directly on the 4-momentum of the jet using the area--median approach,
\eq~\eqref{eq:subtraction-with-rhom}, with the active area of the jet
computed using \fastjet. We included the $\rho_m$ term which corrects
for contamination to the total jet mass due to the mass of the pileup
particles.
The estimation of $\rho$ and $\rho_m$ is performed with \fastjet using
a grid-based estimate with a grid of cell-size 0.55 extending to
$|y|=4$. Corrections for the rapidity dependence of the pileup density
$\rho$ are applied using a rapidity rescaling (see
Sections~\ref{sec:areamed-position} and
\ref{sec:area-median:pileup-properties}).
Ghosts are placed up to $y_{\rm max}=4$ and explicit ghosts are
enabled, ensuring the availability of the area information to grooming
tools.
When applying this subtraction procedure, we discard jets with
negative transverse momentum or (squared) mass of the jet.

When we apply this background subtraction together with grooming, the
subtraction is performed directly on the subjets, before applying any
kinematic constraint, \ie before deciding which subjets should be
kept, in the case of trimming and filtering, or before checking if the
symmetry criterion is met in the case of mMDT and \SD. This is meant
to limit the potential effects of pileup on kinematic tests.
Note that in the case of trimming, the unsubtracted jet $p_t$ is used
as the reference to compute the $p_t$ threshold for subjets. It was
found that this gives slightly better results than using the
subtracted jet.

\section{Jet substructure performance}\label{sec:results_grooming}

\begin{figure} [p!]
\begin{center}
\begin{tabular}{cc}
\subfloat[raw, ungroomed jets]{\includegraphics[width=0.4\textwidth]{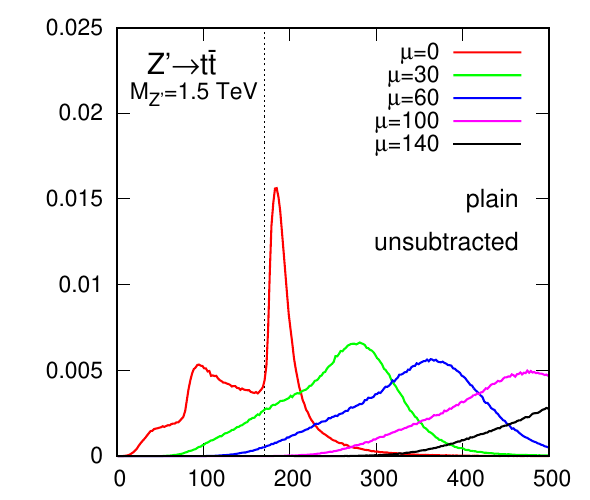}} &
\subfloat[raw jets with pileup subtraction]{\includegraphics[width=0.4\textwidth]{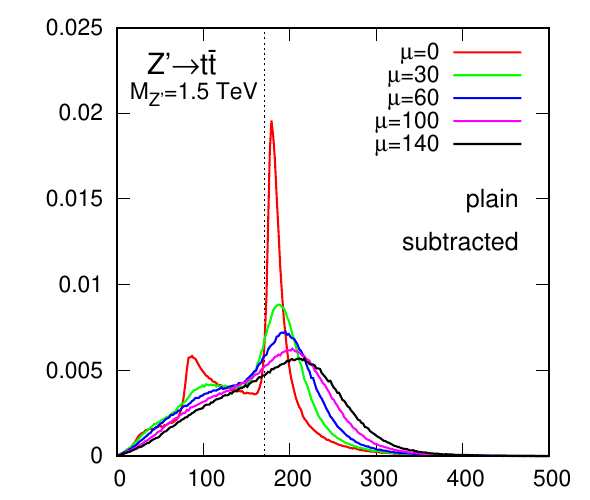}} \\
\subfloat[trimmed jets]{\includegraphics[width=0.4\textwidth]{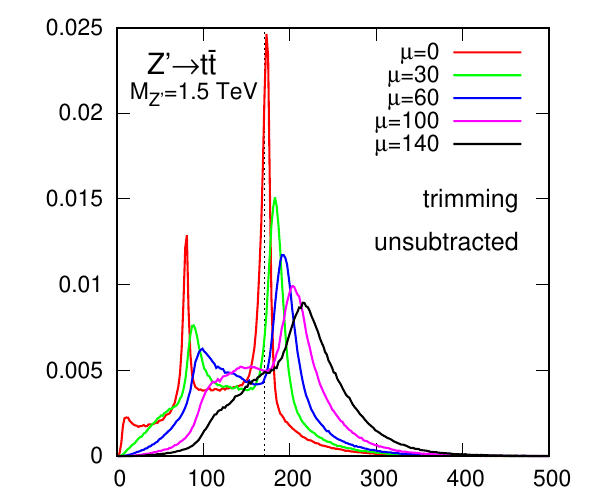}} &
\subfloat[trimmed jet with pileup subtraction]{\includegraphics[width=0.4\textwidth]{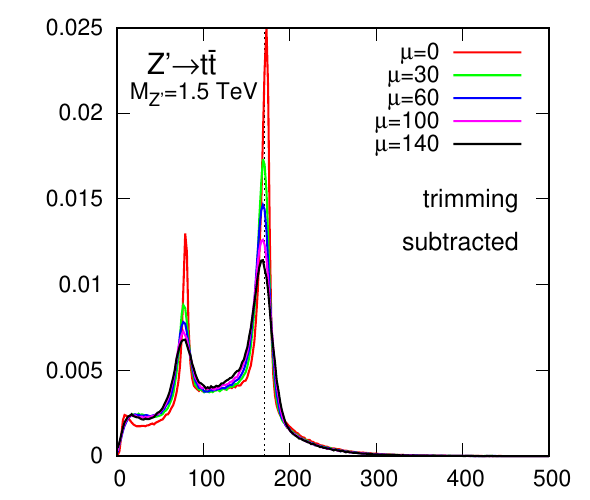}}\\
\subfloat[filtered jets]{\includegraphics[width=0.4\textwidth]{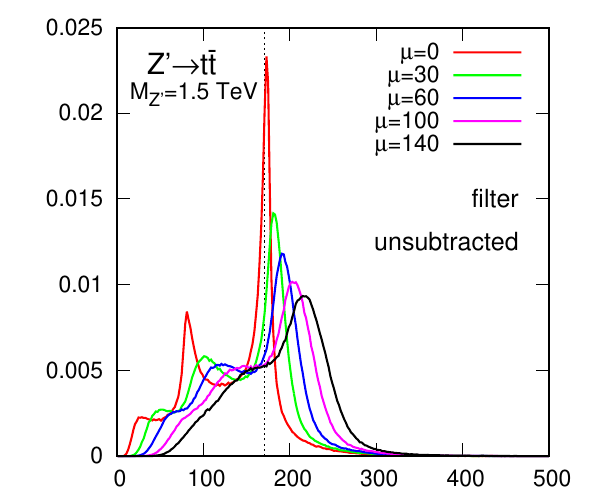}} &
\subfloat[filtered jet with pileup subtraction]{\includegraphics[width=0.4\textwidth]{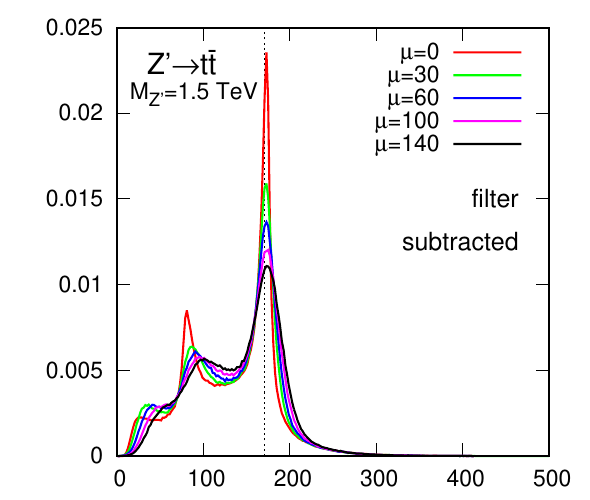}} 
\end{tabular}
\end{center}
\caption{The impact of \nPU on the jet mass distribution. Top row: the
  raw jet mass distribution for $Z' \to t\bar t$ final states with
  $m_{Z'} =$ 1.5~TeV, in the presence of pileup with $\mu = 30$, 60,
  100, and 140, before (left) and after (right) area--median pileup
  subtraction. The second and third rows show the same results after
  trimming (middle row) and filtering (lower row).}\label{fig:mass_spectra}
\end{figure}

\begin{figure} [p!]
\begin{center}
\begin{tabular}{cc}
\subfloat[jets after mMDT]{\includegraphics[width=0.4\textwidth]{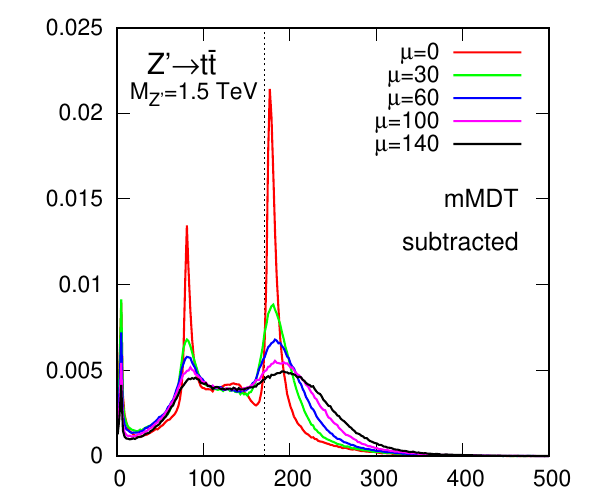}} &
\subfloat[jets after mMDT+filtering]{\includegraphics[width=0.4\textwidth]{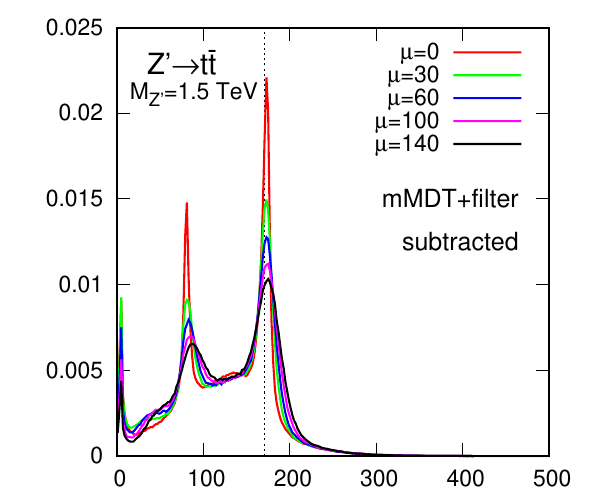}} \\
\subfloat[jets after SoftDrop($\beta=1$)]{\includegraphics[width=0.4\textwidth]{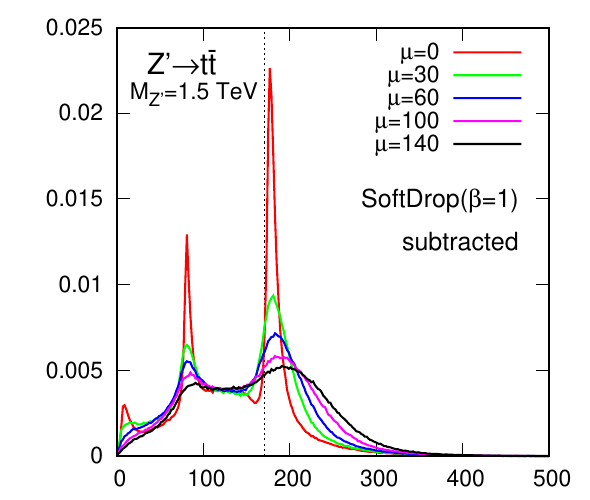}} &
\subfloat[jets after SoftDrop($\beta=1$)+filtering]{\includegraphics[width=0.4\textwidth]{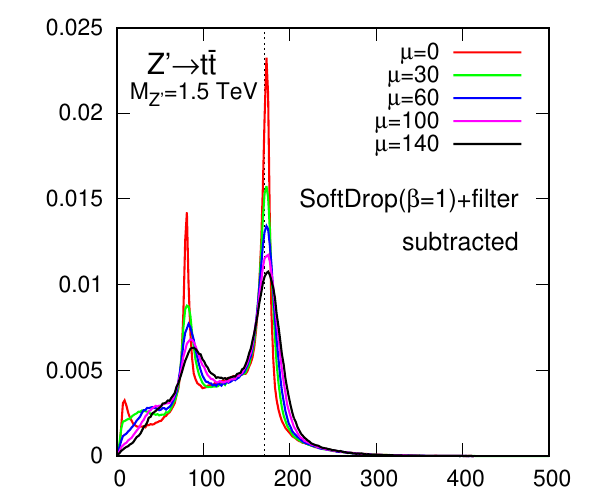}}\\
\subfloat[jets after SoftDrop($\beta=2$)]{\includegraphics[width=0.4\textwidth]{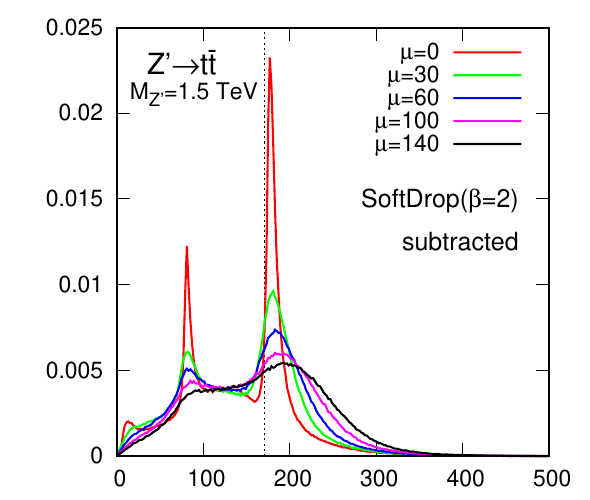}} &
\subfloat[jets after SoftDrop($\beta=2$)+filtering]{\includegraphics[width=0.4\textwidth]{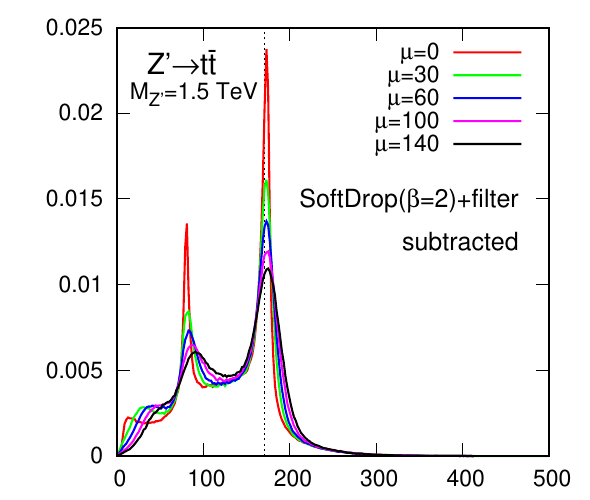}} 
\end{tabular}
\end{center}
\caption{Same as Fig.~\ref{fig:mass_spectra} now showing the jet mass
  distribution after applying the modified Mass Drop (top row) or \SD
  with $\beta=1$ (middle row) or $\beta=2$ (lower row). All
  distributions are shown after area--median pileup subtraction. Plots
  on the left column show the results of the initial grooming
  procedure alone, and plots on the right column show the results with
  an extra filtering step (see text for details).}
\label{fig:mass_spectra_2}
\end{figure}

As a first test, we look at the jet mass distribution obtained for out
$Z'\to t\bar t$ event sample. This is plotted in
Figs.~\ref{fig:mass_spectra} and \ref{fig:mass_spectra_2}.
The immediate expectation for the reconstructed jet mass is the top
mass, \ie $m \approx 171$~GeV, and no residual dependence on the
pileup activity $\mu$, after the pileup subtraction. The two plots in
the upper row of Figure~\ref{fig:mass_spectra} show the distributions
of the reconstructed jet masses: without pileup subtraction on the
left (\ref{fig:mass_spectra}(a)) and with area--median pileup
subtraction on the right (\ref{fig:mass_spectra}(b)). Concentrating on
the sample without any pileup added, we see that a large fraction of
the events are reconstructed close to the top mass and about 30\% are
compatible with the $W$ mass.
The effect of pileup on the mass scale and resolution is clearly
visible, increasingly shifting the peak towards much larger jet masses
and smearing it.

Applying only the pileup subtraction, without applying any grooming to
the jets, already improves the mass reconstruction significantly. The
pileup shift is mostly removed from the jet mass spectrum, as shown in
Fig.~\ref{fig:mass_spectra}(b) and the position of the mass peak is
recovered, independently of $\mu$. With increasing pileup, the mass
peak gets more and more smeared, an effect that we have already seen
in the first part of this document, due to the fact that the pileup is
not perfectly uniform. We have already stressed that these
point-to-point fluctuations in an event lead to a smearing by
$\pm \sigma\sqrt{A}$ in the subtracted jet $p_t$. For very large
pileup, this smearing extends all the way to $m=0$ as seen in
Fig.~\ref{fig:mass_spectra}(b) and we lose the clean separation
between the $W$ and the top mass peaks.\footnote{The remaining
  apparent small shift of the peak position still remaining after
  pileup subtraction is likely due to a selection bias coming from our
  selection of the hardest jets before grooming or subtraction is
  applied. Applying the $p_t$ cut on the subtracted jets would likely
  remove that remaining bias.}

The effect of the grooming techniques on the reconstructed jet mass
distributions is summarised in the remaining plots of
Figs.~\ref{fig:mass_spectra} and \ref{fig:mass_spectra_2}. For
filtering and trimming, Figs.~\ref{fig:mass_spectra}(c)-(f), we show
the mass distributions with and without subtraction, while for the
plot of Fig.~\ref{fig:mass_spectra_2} we show the result of the
mMDT/SD procedure with and without the extra filtering step.
The spectra show that grooming clearly improves the mass
reconstruction. Already without applying area--median pileup
subtraction, one sees that the peak is much less shifted and smeared
than in the ungroomed case. The application of the pileup subtraction
in addition to the grooming procedure further improves the mass
reconstruction performance, fully correcting for the peak position and
further improving the resolution.
From Fig.~\ref{fig:mass_spectra_2}, we also see that applying
filtering on the result of the mMDT/SD grooming further improves the
quality of the reconstructed spectrum.

The fact that the combination of grooming and area--median pileup
subtraction appears as the optimal combination should not be a
surprise. As discussed many times already, pileup has mainly two
effects on the jet: a constant shift proportional to $\rho A$ and a
smearing effect proportional to $\sigma \sqrt{A}$, with $\sigma$ a
measure of the fluctuations of the pileup within an event. In that
language, the subtraction corrects for the shift leaving the smearing
term untouched. Grooming, on the contrary, since it selects only part
of the subjets, acts as if it was reducing the area of the
jet\footnote{Note that grooming techniques do more than reducing the
  catchment area of a jet. Noticeably, the selection of the hardest
  subjets introduces a bias towards including upwards fluctuations of
  the background. This positive bias is balanced by a negative one
  related to the perturbative radiation discarded by the
  grooming. These effects go beyond the generic features explained
  here.}. This reduces both the shift and the dispersion. Combining
grooming with subtraction thus allows to correct for the shift
leftover by grooming and reduce the smearing effects at the same
time. All these effects are observed on the plots.

The effect of the filtering step applied after the mMDT/SD grooming
can also be understood from fairly simple arguments: the declustering
procedure applied by mMDT/SD discards the soft activity at large
angles in the jet until two hard prongs are found. This effectively
reduces the jet area with the immediate consequence of reducing the
shift and smearing of the mass peak. However, this results in a
groomed jet with an effective radius dictated by the hard-prong
structure of the jet. Say we have a 500~GeV $W$ boson, this
corresponds to a grooming radius around $R_{\rm SD}\approx 0.4$ which
gives an area $\pi R_{\rm SD}^2$ still larger that the typical
$2 \pi R_{\rm filt,trim}^2$ obtained with filtering or
trimming. Applying an extra filtering step to the result of the
mMDT/SD, as already suggested in the original work
\cite{Butterworth:2008iy}, would further groom the jet at angular
scales smaller than the original grooming radius, further reducing the
effective jet area and hence further reducing the peak shift and
smearing.

\begin{figure}[ht]
\begin{center}
\includegraphics[width=0.45\textwidth]{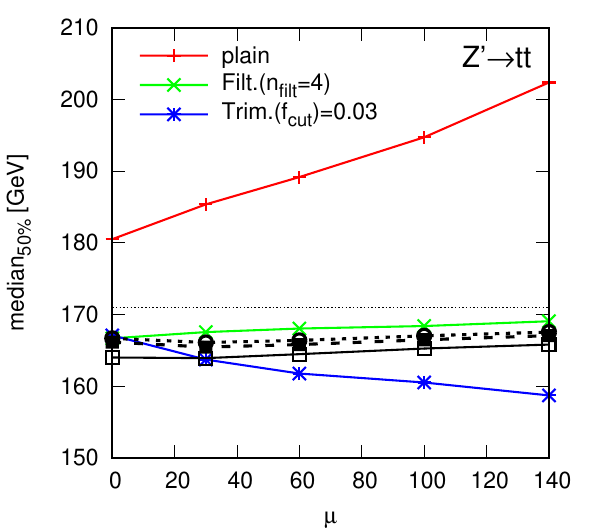}
\includegraphics[width=0.45\textwidth]{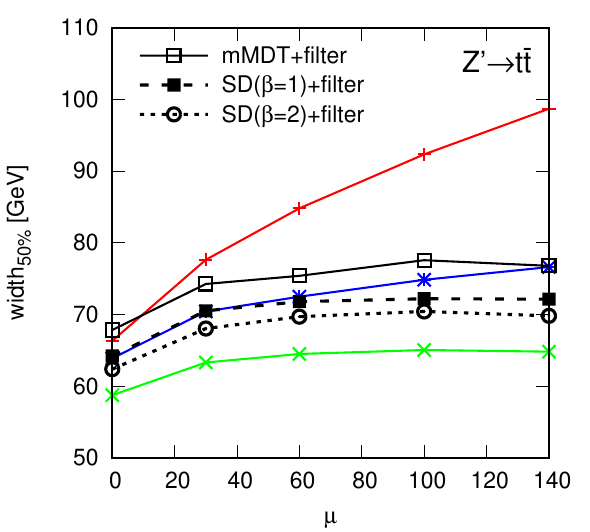}
\includegraphics[width=0.45\textwidth]{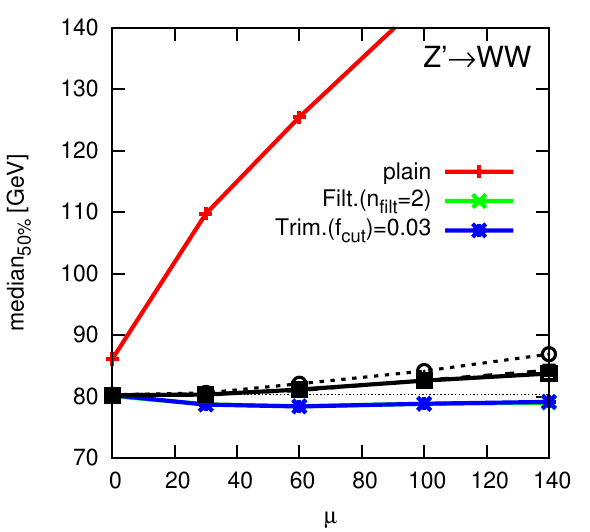}
\includegraphics[width=0.45\textwidth]{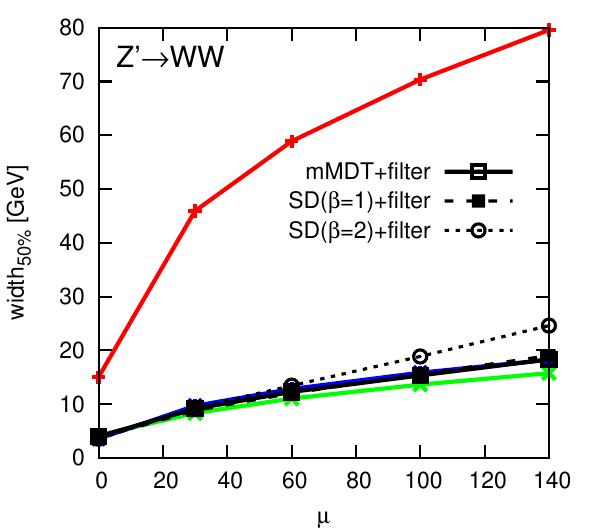}
\caption[]{Position (leftmost figure) and width (rightmost figure) of
  the reconstructed jet mass distribution in $Z' \to t\bar t$ final
  states (top row) and $WW$ events (bottom row), as a function of the
  pileup activity $\mu$, for various jet grooming techniques.
  Area--median pileup subtraction is applied in all cases.}
\label{fig:mass_summary}
\end{center}
\end{figure}

To summarise in a more compact way the above findings and quantify
more precisely the performance of the various grooming techniques
analysed in this study, we have measured the position and width of the
peak in all our different scenarii.
We define this by looking, for each individual mass distribution, at
the narrowest mass window that contains 50\% of the reconstructed
jets.\footnote{The original Boost report used 67\%. In the case of the
  top mass, where a certain fraction of the events are initially
  compatible with a $W$ boson, we find that a lower fraction, more
  often including only the top peak, is a more representative quality
  figure.}
We then define the peak position as the median of that window and the
peak width as the width of the mass window.
The findings from the spectra are quantitatively summarised in
Fig.~\ref{fig:mass_summary} for the peak position (left) and width
(right), for both the top (top) and the $W$ (bottom) samples.
Since we have already seen that the combination of grooming with
area--median subtraction was performing better than grooming alone,
and that mMDT/SD was working best when supplemented with a filtering
step, we focused only on these optimal cases.

We see that all methods succeed at maintaining the jet mass
independent of the pileup conditions and around the expected mass
value of~171~GeV, although the reconstructed average top mass tends
to be slightly underestimated.
This works well for all groomers even at large pileup multiplicity,
although some differences are observed between the various groomers.
All the groomers studied here improve the mass resolution to different
degrees, but in any case perform better than pileup subtraction alone,
as expected.
Combining area-subtraction and filtering yields the least sensitivity
to pileup in terms of the peak position and width, especially for the
$t\bar t$ sample. For the $WW$ sample, all groomers tend to perform
similarly.
The observed differences are anyway beyond the scope of this study and
depend also on the specific choice of parameters for the different
groomers.\footnote{A perhaps surprising feature is that in the top
  case, filtering alone tends to outperform mMDT/SD+filtering in terms
  of dispersion. We suspect that it might be related to situations
  with extra radiation at large angle and that radiation, kept by
  filtering alone but groomed away by the mMDT/SD de-clustering
  procedure. This would agree with the absence of that phenomena in
  the $W$ case but seems to contradict the presumption that
  large-angle radiation in the case of the top should happen before
  the on-shell top decays. In that respect, it would be helpful to
  investigate more carefully the dependence on $R_{\text{filt}}$ and
  $n_{\text{filt}}$.}

\section{Summary}\label{subsec:concl}

A brief Monte Carlo study of the effect of jet grooming techniques on
the jet mass reconstruction in $Z' \to t\bar t$ and $Z' \to WW$ final
states has been conducted. Jet trimming, filtering, the modified
Mass-Drop tagger and \SD are used by themselves, or in combination
with the pileup subtraction using the four-vector area, to reconstruct
the single jet mass and evaluate the stability of the mass scale and
resolution at pileup levels of 30, 60, 100, and 140 extra $pp$
collisions, in addition to the signal event. We find that for these
particular final states all groomers work well for maintaining the
mass scale and resolution, provided they are applied together with
pileup subtraction so as to benefit both from the average shift
correction from subtraction and noise reduction from the grooming. The
results are stable up to large pileup multiplicities. In the case of
the mMDT/SD procedures, we find useful to supplement them with an
extra filtering step as suggested in the original work from
Ref.~\cite{Butterworth:2008iy}.

Finally, note that a more thorough comparison of the performance of
these groomers would require varying their parameters. This goes
beyond the scope of this study.


\chapter{Analytical insight}\label{chap:grooming-analytic}

\section{Foreword}

We have seen that groomers are an efficient way to reduce pileup
contamination from fat jets. Another aspect of groomers is that, even
in the absence of pileup, they affect the structure of the hard jets
in the event. This is usually a desired feature since it implies a
reduction of the QCD background compared to the heavy-object signal
when tagging boosted objects.

In this chapter we illustrate that the perturbative behaviour of
groomers is amenable to an analytic treatment. After the very first
results on the MassDrop Tagger, trimming and pruning in
Refs.~\cite{Dasgupta:2013ihk,Dasgupta:2013via}, we have shown
\cite{Larkoski:2014wba} that an analytic treatment is also possible
for SoftDrop. To keep the discussion relatively contained and to avoid
departing too much from the main theme of this document, we will just
concentrate on a single observable, namely the groomed jet radius
after one applies SoftDrop to a boosted jet. We will show that it can
be computed in QCD in for boosted objects.
This also covers the case of the modified
MassDropTagger~\cite{Dasgupta:2013ihk} which can be obtained from the
SoftDrop results by taking the limit $\beta\to 0$.
We refer directly to~\cite{Dasgupta:2013ihk},~\cite{Larkoski:2014wba}
and \cite{Dasgupta:2015lxh} for additional analytic results for
substructure techniques.

At first sight, this does not seem directly related to pileup
mitigation.
However, we believe that an analytic understanding of common tools in
the field of jet substructure is crucial-enough that we devote a few
pages to that topic.
Furthermore, one might ultimately want to use insight from QCD
first-principle calculations to develop optimal pileup subtraction
techniques, \eg to control their biases and sensitivities to different
dynamical properties of the jet. The analytic control over groomers
described in the next few pages can therefore also be seen as a step
towards this broader goal.
In particular, the groomed jet radius is one observable that is
directly related to the residual sensitivity of the groomed jet to
pileup. Gaining analytic control over it could be helpful in devising
better pileup mitigation techniques.

We will also provide a brief Monte-Carlo study of non-perturbative
effects for SoftDropped jets.
In practice, it is partially possible to address the question of
non-perturbative effects from an analytic perspective. Unfortunately,
this work is still too preliminary to deserve a place here, although we
can say that the underlying idea follows the non-perturbative
calculation done for the modified Mass-Drop Tagger in
Ref.~\cite{Dasgupta:2013ihk}.

\section{Groomed Jet Radius}
\label{sec:pileup}

Because the soft drop procedure is defined through declustering a
Cambridge/Aachen branching tree, there is a well-defined and IRC-safe
meaning to the groomed jet radius.  Concretely, the groomed jet radius
$R_g$ is the angle between the branches that first satisfy the
SoftDrop condition~(\ref{eq:soft-drop-condition}).
This is sensible since for a Cambridge/Aachen tree, all subsequent
branches are separated by an angle less than $R_g$.
From a practical perspective, $R_g$ is particularly interesting, since
the groomed jet area is approximately $\pi R_g^2$.  Thus, $R_g$ serves
as a proxy for the sensitivity of the groomed jet to possible
contamination from pileup.

\subsection{Analytic calculation}
\label{sec:MLLradius}

In what follows, we will provide an analytic calculation of the $R_g$
distribution valid in the small-angle limit, $R_g\ll R \ll 1$,
pertinent \eg for the boosted jet regime, and focus on the leading
logarithmic terms.

At this accuracy it is sufficient to assume that the jet is composed
of the ``leading parton'' accompanied by independent
soft-and-collinear emissions strongly ordered in angle and energy.
Assume we have $n$ such emissions with strongly-ordered angles to the
jet axis, $\theta_1\gg\theta_2\gg\dots\gg\theta_n$. We also denote by
$z_1,\dots,z_n$ their respective transverse momentum fraction.
The grooming radius will be $\theta_i$ if emission $i$ is kept by \SD
and all previous emissions $1,\dots,i-1$ are groomed away.
The grooming radius condition for that set of emissions can thus be
written as
\begin{equation}\label{eq:sdRg-condition}
\delta(R_g(\{\theta_i,z_i\})-R_g)
 = \sum_{i=1}^n
 \Theta_i^{\text{in}}\prod_{j=1}^{i-1}\Theta_j^{\text{out}}\delta(\theta_i-R_g)
 + \prod_{i=1}^{n}\Theta_i^{\text{out}}\delta(R_g),
\end{equation}
where $\Theta_i^{\text{in}}$ indicates that emission $i$ is kept by
\SD while $\Theta_i^{\text{out}}$ indicates that it is groomed away.
The last term in the above expression corresponds to the situation
where all emissions are groomed and where we have decided to set
$R_g=0$ in this case for simplicity.
Recall also that if emission $i$ is kept by grooming, all subsequent
emissions at smaller angles will be kept as well.

We then have to sum over any number of such emissions and obtain the
following expression for the groomed jet radius
\begin{align}\label{eq:sdRg-start}
\frac{R_g}{\sigma}\frac{d\sigma}{dR_g} = \lim_{\epsilon\to 0}\sum_{n=0}^\infty
 \prod_{i=0}^n\int_\epsilon^R\frac{d\theta_i}{\theta_i}\int_\epsilon^1dz_i\,p(z_i)&\,\frac{\alpha_s(z_i\theta_ip_t)}{\pi}\:\Theta(\theta_i>\theta_{i+1})\\
& R_g\delta(R_g(\{\theta_i,z_i\})-R_g)\:\exp\bigg[-\int_\epsilon^R\frac{d\theta}{\theta}\int_\epsilon^1dz\,p(z)\,\frac{\alpha_s(z\theta p_t)}{\pi}\bigg].\nonumber
\end{align}
To write \eqref{eq:sdRg-start}, we have introduced $\epsilon$ to
regulate the soft and collinear divergences. The last, exponential,
term corresponds to virtual corrections and will cancel the soft and
collinear divergences in the end so that one can safely take the limit
$\epsilon\to 0$. The argument of the strong coupling is taken as the
relative transverse momentum of the emitted gluon \wrt the parent
parton. Finally, $p(z)$ is the appropriate splitting function
different for quark-initiated jets and gluon-initiated jets.

To evaluate \eqref{eq:sdRg-start}, we first notice that the
$\delta(R_g)$ in~\eqref{eq:sdRg-condition} will vanish in the limit
$\epsilon\to 0$. Indeed, if all the emissions are outside of the
grooming radius, the sum over real emissions factorises into
\begin{equation}
\sum_{n=0}^\infty
  \prod_{i=0}^n\int_\epsilon^R\frac{d\theta_i}{\theta_i}\int_\epsilon^1dz_i\,p(z_i)\,\frac{\alpha_s(z_i\theta_ip_t)}{\pi}\:\Theta(\theta_i>\theta_{i+1})\Theta_i^{\text{out}}
  = \exp\bigg[\int_\epsilon^R\frac{d\theta}{\theta}\int_\epsilon^1dz\,p(z)\,\frac{\alpha_s(z\theta p_t)}{\pi}\,\Theta^{\text{out}}\bigg].
\end{equation}
Combined with virtual corrections, this gives
\begin{equation}
\exp\bigg[-\int_\epsilon^R\frac{d\theta}{\theta}\int_\epsilon^1dz\,p(z)\,\frac{\alpha_s(z\theta
  p_t)}{\pi}\,\Theta^{\text{in}}\bigg]
\end{equation}
which vanishes in the limit $\varepsilon\to 0$.

Then, for the remaining terms in~\eqref{eq:sdRg-condition}, it is
helpful to reorganise the emissions where we pull out in front the
integration over $\theta_i$ such that $\theta_i=R_g$, and rename it
$\theta_0$. We then separate the remaining emissions in two groups:
those at larger angles for which we have to impose the condition that
they are outside the \SD region (see (\ref{eq:sdRg-condition})), and
those at smaller angles which are automatically kept by \SD. The first
series can set to have $n$ emissions with momenta and angles
$(z_1,\theta_1),\dots,(z_n,\theta_n)$ and the second series $\bar n$
emissions with momenta and angles
$(\bar z_1,\bar \theta_1),\dots,(\bar z_{\bar n},\bar\theta_{\bar
  n})$, where we have to sum over $n$ and $\bar n$. This gives
\begin{align}\label{eq:sdRg-next}
\frac{R_g}{\sigma}\frac{d\sigma}{dR_g} = \lim_{\epsilon\to 0}&
  \int_\epsilon^R\frac{d\theta_0}{\theta_0}
  \int_\epsilon^1dz_0\,p(z_0)\,\frac{\alpha_s(z_0\theta_0p_t)}{\pi}
  R_g\delta(R_g(\theta_0-R_g))\Theta(z_0>\zcut(R_g/R)^\beta)\\
& \sum_{n=0}^\infty \frac{1}{n!}
  \prod_{i=0}^n\int_{\theta_0}^R\frac{d\theta_i}{\theta_i}
  \int_\epsilon^1dz_i\,p(z_i)\,\frac{\alpha_s(z_i\theta_ip_t)}{\pi}
  \Theta(z_i<\zcut(\theta_i/R)^\beta)\nonumber\\
& \sum_{\bar n=0}^\infty \frac{1}{\bar n!}
  \prod_{j=0}^n\int_\epsilon^{\theta_0}\frac{d\bar\theta_j}{\bar\theta_j}
  \int_\epsilon^1d\bar z_j\,p(\bar z_j)\,
  \frac{\alpha_s(\bar z_j\bar\theta_jp_t)}{\pi}
  \:\exp\bigg[-\int_\epsilon^R\frac{d\theta}{\theta}\int_\epsilon^1dz\,p(z)\,\frac{\alpha_s(z\theta p_t)}{\pi}\bigg].\nonumber
\end{align}
Once again, the sums over $n$ and $\bar n$ lead to exponentials:
\begin{align}\label{eq:sdRg-simplifies-n}
 \sum_{n=0}^\infty \frac{1}{n!}
  \prod_{i=0}^n\int_{\theta_0}^R\frac{d\theta_i}{\theta_i}
  \int_\epsilon^1dz_i\,p(z_i)\,&\frac{\alpha_s(z_i\theta_ip_t)}{\pi}
  \Theta(z_i<\zcut(R_g/R)^\beta)\\
 & =
   \exp\bigg[-\int_{\theta_0}^R\frac{d\theta}{\theta}\int_\epsilon^1dz\,p(z)\,\frac{\alpha_s(z\theta p_t)}{\pi}\,\Theta(z<\zcut(\theta/R)^\beta)\bigg],\nonumber
\end{align}
and
\begin{equation}
\sum_{\bar n=0}^\infty \frac{1}{\bar n!}
  \prod_{j=0}^n\int_\epsilon^{\theta_0}\frac{d\bar\theta_j}{\bar\theta_j}
  \int_\epsilon^1d\bar z_j\,p(\bar z_j)\,
  \frac{\alpha_s(\bar z_j\bar\theta_jp_t)}{\pi}
  = \exp\bigg[-\int_\epsilon^{\theta_0}\frac{d\theta}{\theta}\int_\epsilon^1dz\,p(z)\,\frac{\alpha_s(z\theta p_t)}{\pi}\bigg].
\end{equation}
These exponentials combine with the virtual corrections in particular
to cancel the divergence at $\epsilon\to 0$ as expected and give,
after integration over $\theta_0$
\begin{equation}\label{eq:sdRg-final}
\frac{R_g}{\sigma}\frac{d\sigma}{dR_g} = 
  \int_0^1dz_0\,p(z_0)\,\frac{\alpha_s(z_0R_g p_t)}{\pi}
  \Theta\Big(z_0>\zcut\frac{R_g^\beta}{R^\beta}\Big)
\exp\bigg[-\int_{R_g}^R\frac{d\theta}{\theta}\int_0^1dz\,p(z)\,\frac{\alpha_s(z\theta p_t)}{\pi}\,\Theta\Big(z>\zcut\frac{\theta^\beta}{R^\beta}\Big)\bigg].
\end{equation}

The above expression can easily be evaluated for quark and gluon jets,
using a 1-loop running-coupling approximation for $\alpha_s$. This is
what we have done for the comparison to Monte-Carlo simulations
presented in the next Section. 
It is nevertheless illustrative to consider the fixed-coupling
approximation which gives (up to subleading power corrections)
\begin{align}
\int_0^1dz_0\,p(z_0)\,\frac{\alpha_s}{\pi} 
 & = \frac{\alpha_sC_R}{\pi}\bigg[\log\Big(\frac{1}{\zcut R_g^\beta}\Big) + B_i \bigg],\\
\int_{R_g}^R\frac{d\theta}{\theta}\int_0^1dz\,p(z)\,\frac{\alpha_s}{\pi}
  \,\Theta\Big(z>\zcut\frac{\theta^\beta}{R^\beta}\Big)
 & =  \frac{\alpha_sC_R}{\pi}\bigg[\log\Big(\frac{R^{\beta/2}}{\zcut
   R_g^{\beta/2}}\Big) + B_i \bigg]\log\Big(\frac{R}{R_g}\Big),
\end{align}
with $C_R=C_F$ and $B_i=-3/4$ for quark-initiated jets, and $C_R=C_A$
and $B_i=(11C_A-4n_fT_R)/(12C_A)$ for gluon-initiated jets.
This corresponds to a double-logarithmic suppression in $R/R_g$ which
becomes single-logarithmic for $\beta=0$.

\begin{figure}
\centerline{\includegraphics{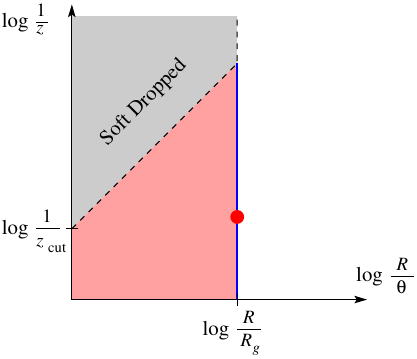}}
\caption{Phase space for emissions relevant for groomed jet radius
  $R_g$ in the $(\log \frac{1}{z},\log \frac{R}{\theta})$ plane.  The
  region groomed away by \SD is grey and the first emission satisfying
  the soft drop criteria is illustrated by the red dot which can be
  anywhere on the thick blue line. The forbidden emission region (the
  Sudakov exponent) is shaded in pink.  }
\label{fig:groomedregions-radius}
\end{figure}

It is also helpful to give a physical interpretation of
(\ref{eq:sdRg-final}). This is illustrated in
Fig.~\ref{fig:groomedregions-radius} where we represent the
phase-space available for emission in the
$(\log \frac{1}{z},\log \frac{R}{\theta})$ plane.
If one wants a given grooming radius $R_g$, we need an emission at an
angle $R_g$ passing the \SD condition. This corresponds to the red dot
in the Figure which can lie anywhere on the thick blue line. It leads
to the prefactor in~(\ref{eq:sdRg-final}) where the integration over
$z_0$ corresponds to the red dot spanning the blue line. We then need
to guarantee that the \SD procedure does not stop at a larger
radius. We should therefore veto all real emissions at larger angles
and that would pass the \SD criterion. This is represented by the pink
shaded area in the figure and to the Sudakov suppression factor in
\eq~(\ref{eq:sdRg-final}). Since real emissions are vetoed in that
region, only virtual corrections remain. In all the rest of the
phasespace, real emissions are allowed and cancel against virtual
corrections.

On a more technical basis, it is worth noting that the groomed jet
radius $R_g$ is also simple in terms of multiple-emission corrections.
Indeed, once one emission satisfies the soft drop criteria, the jet
radius is set, so multiple emissions do not contribute to this
observable. We can also verify that non-global contributions are
suppressed by $R_g$ for $\beta<\infty$. For these reasons, we believe
that the expression in \eq~(\ref{eq:sdRg-final}) is fully accurate to
single-logarithmic level.\footnote{Strictly speaking, NLL accuracy
  would require evaluating the strong coupling at two loops in the CMW
  scheme \cite{Catani:1990rr}, \ie including two extra terms
  proportional to $\beta_1$ and $K$, suppressed by one power of
  $\alpha_s$ compared to the leading $\beta_0$ term.}

\subsection{Comparison to Monte Carlo}
\begin{figure}[]
\centering
\includegraphics[width=0.48\textwidth]{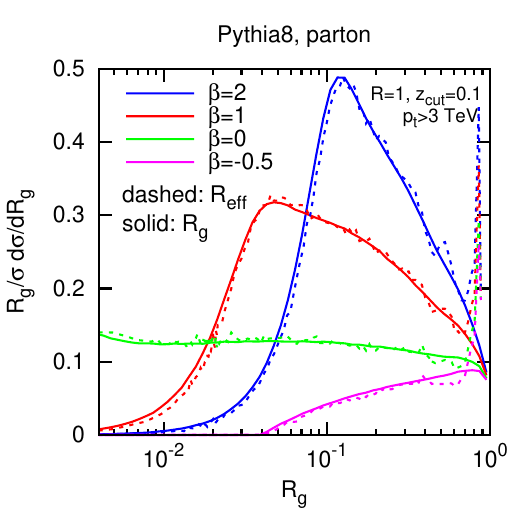}
\hfill
\includegraphics[width=0.48\textwidth]{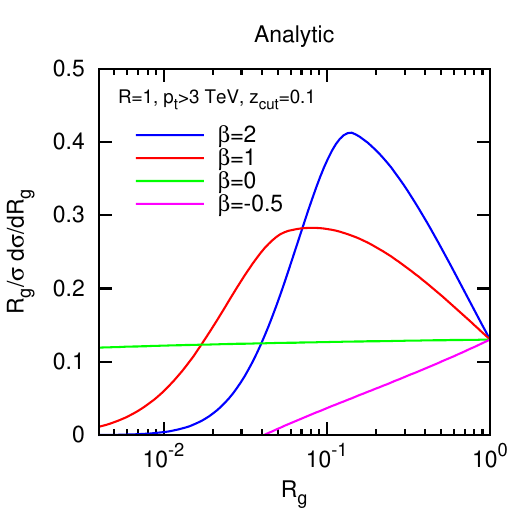}
\caption{ Comparison of the jet radius $R_g$ distribution extracted
  from \pythia{8} (left, solid) and inferred from the active area in
  \pythia{8} using $R_\text{eff} =\sqrt{A_\text{active}/\pi \xi}$
  (left, dashed), and computed to modified leading logarithmic
  accuracy (right). The original jet radius is set to be $R = 1$ and
  the jets have an ungroomed energy of $3$ TeV.  The \SD parameter is
  $z_\text{cut}=0.1$, while $\beta$ is varied.  }
\label{fig:rad_dist}
\end{figure}

There are two different ways one can define the groomed jet radius in
Monte Carlo. The first method is to simply measure the $R_g$ value of
the Cambridge/Aachen branching that satisfies the \SD condition. A
second approach, more directly connected with pileup mitigation, is to
determine the effective radius of the groomed jet from its active area
(see Section~\ref{sec:areamed-defareas-active}). An effective jet
radius $R_\text{eff}$ can then be defined from the groomed jet active
area using:
\begin{equation}\label{eq:Reffdef}
 R_\text{eff} \equiv \left(\frac{A_\text{active}}{\pi \xi }\right)^{1/2} \ ,
\end{equation}
where $A_\text{active}$ is the active jet area, and
$\xi \simeq (1.16)^2$ accounts for the fact that a typical
Cambridge/Aachen jet of radius $R$ has an average active area
$\xi \pi R^2$.\footnote{The numerical value for $\xi$ can be read
  from Fig.~\ref{fig:2point-areas-active-3alg}. Strictly speaking,
  this result is only valid for a jet made of two particles separated
  by $R_g$ with one of them much softer than the other. However, for
  Cambridge/Aachen jets, one expects that this would not vary much for
  more symmetric two-particle configurations (see
  \eg Ref.~\cite{Sapeta:2010uk}).}

To compare $R_g$ and $R_{\text{eff}}$ and to validate our analytic
calculation (\ref{eq:sdRg-final}), we have generated with Pythia~8
(v8.176) 14~TeV $qq\to qq$ events and selected anti-$k_t(R=1)$ jets
with $p_t\ge 3$~TeV. We have applied the \SD procedure with
$\zcut=0.1$ and different values of $\beta$ and computed $R_g$ and
$R_{\text{eff}}$. To obtain $R_\text{eff}$ in practice, we have
computed the groomed jet area using active areas as implemented in
\fastjet~(v3), and we used a ghost area of 0.0005 and 10 repetitions
in order to reach sufficiently small values of $R_{\rm eff}$.  

The resulting distributions are plotted in the left plot of
Fig.~\ref{fig:rad_dist}.
With the $\xi$ offset factor, the two techniques give remarkably
similar results, giving strong evidence that the groomed jet radius
$R_g$ is an effective measure of pileup sensitivity.  The main
difference is the spike at $R_{\rm eff} = 1/\sqrt{\xi}$, corresponding
to cases where the first Cambridge/Aachen branching already satisfies
the soft drop condition, yet typically with $R_g < 1$.  The nice
reduction of the jet area even with mild grooming (\eg~$\beta = 2$)
suggests that \SD should work well for pileup mitigation.

In the right plot of Fig.~\ref{fig:rad_dist}, we show the distribution
obtained from our analytic calculation, \eq~(\ref{eq:sdRg-final}).
There is good qualitative agreement with \pythia{} for a range of
angular exponents $\beta$, suggesting that our modified
leading-logarithmic calculation for $R_g$ captures much of the
relevant physics effects present in the Monte Carlo simulation and
that analytic calculations can be helpful in understanding how
grooming techniques mitigate pileup.

Note that when increasing the jet $p_t$, one would effectively reduce
the coupling $\alpha_s(p_tR)$ arising naturally from
(\ref{eq:sdRg-final}). Both the pre-factor and Sudakov exponent would
then be reduced, shifting the peak in the $R_g$ distribution to
smaller values.

\section{Non-Perturbative Contributions}
\label{sec:NP}

\begin{figure}
\centering
\includegraphics[width=0.48\textwidth]{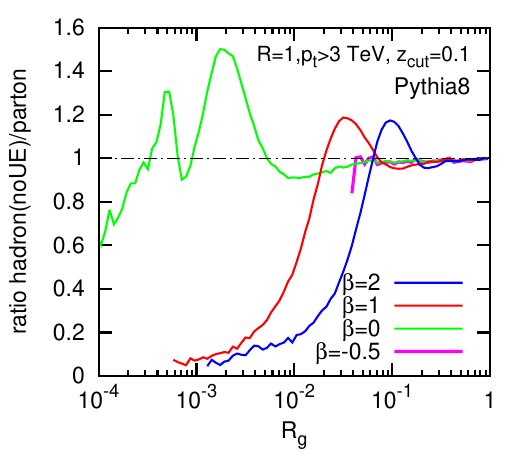}
\hfill
\includegraphics[width=0.48\textwidth]{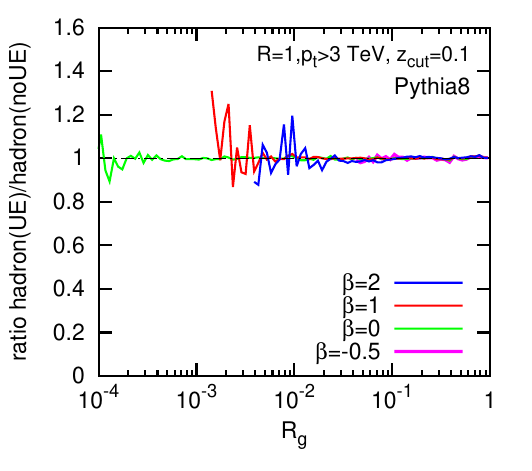}
\caption{Effect of non-perturbative corrections on $R_g$.  The left
  plot shows the ratio between hadron level and parton level
  predictions obtained with \pythia{8} (without UE). The right plot
  instead shows the ratio of hadron-level results with and without
  UE.}
\label{fig:np_corrections}
\end{figure}

In the above analytic calculations and Monte-Carlo simulations, we
only considered the distributions generated at parton level.  In this
section, we present a brief Monte Carlo study to estimate the impact
that non-perturbative effects from hadronisation and UE can have on
the groomed jet radius distribution.

In Fig.~\ref{fig:np_corrections}, we show the effect of hadronisation
(left) and UE (right) on the $R_g$ distribution for different values
of $\beta$.  In the case of hadronisation, we plot the ratio between
the hadronic and partonic distributions obtained from \pythia{8}. In
the case of UE, we plot the ratio between the distributions with and
without UE.  Apart from including non-perturbative effects, the
details of the analysis are the same as for the previous Monte Carlo
studies.

The left plot in Fig.~\ref{fig:np_corrections} shows that \SD
declustering pushes the onset of hadronisation corrections to smaller
values of $R_g$ with decreasing $\beta$.  As shown in the right plot
of Fig.~\ref{fig:np_corrections}, \SD has the remarkable ability
to reduce the UE contribution to almost zero down to very small values
of $R_g$.


\part{Towards more efficient techniques}

\chapter{Challenges beyond the area--median subtraction}\label{chap:beyond-motivation}

\section{Motivation}

Let us put aside the specific case of fat jets and their grooming, and
come back to our original task: providing a generic-purpose pileup
mitigation technique that can be used for standard jet reconstruction
at the LHC. This covers the majority of cases where jets are used at
the LHC today.
In the same kind of spirit, in most of the analysis where jets are
used, the fundamental quantity that we use is the jet transverse
momentum. 
In this last part, we shall therefore focus on pileup mitigation for
``standard'' jet clustering and mainly consider the most widely used
jet characteristic: its transverse momentum. 
We shall only briefly investigate the case of other jet properties
like the jet mass or even other jet shapes, and some applications to
boosted jets. One should however keep in mind that even if these jet
substructure observables are of growing importance, they usually have
to be considered together with grooming techniques for practical
applications.

In the first part of this document we have seen that the area--median
pileup subtraction technique does a very good job at correcting, event
by event, for the transverse momentum shift induced by pileup. A
benefit of the event-by-event nature of the subtraction is that it
also reduces the associated resolution degradation.
The area--median is therefore an efficient pileup-mitigation tool that
has proven perfectly adequate for the Run I of the LHC and at least
the beginning of Run II, with values of $\mu$ slightly above 40 as of
October 2016.
However, as the LHC luminosity increases, so will $\mu$, the average
number of pileup interactions per bunch crossing. One might expect
values of $\mu$ around 60 at maximal luminosity already in Run II, and
in perspectives for the High-Luminosity LHC, one quotes values of
$\mu$ that can reach the $140{-}200$ range. 
While the area--median subtraction method would still be efficient at
correcting for the average $p_t$ shift, with higher pileup
multiplicities, the effects of fluctuations would increase
substantially. From the 6~GeV observed for 30 pileup interactions in
Fig.~\eqref{fig:pusub_resolution}, one should expect a resolution
smearing of about 7.5~GeV for 60 pileup interactions and more than
10~GeV above 100, in good agreement with a scaling proportional to
$\sqrt{\nPU}$.
This increase should not be too dramatic for very high-$p_t$ jets
where detector resolution effects dominate over the pileup smearing
effects, but become an issue for low-$p_t$ jets.

One could simply decide to increase the minimal jet $p_t$ threshold.
However, the feasibility of key measurements at the LHC --- like
double-Higgs production, jet vetoes and the study of the properties
of the Higgs boson --- depend on the experiments' ability to
reconstruct jets down to transverse momenta as low as 20~GeV.
This situation is therefore far from ideal and has lead people to think
about alternative pileup mitigation techniques.

In what follows, we shall discuss new ideas going along two different
main avenues.
The first one follows a logic similar to the area--median subtraction
technique: it focuses on correcting for the right average pileup
contamination but further uses additional information in the
event. The typical example here is the case of charged tracks. With
current technology, the experiments can associate charged tracks to a
vertex of origin and discard those coming from pileup vertices.
One is then left with pileup mitigation on the neutral part of the
event.
Although we do not want to enter into the experimental details, this
is a core idea in the Charged-Hadron Subtraction (CHS) method in CMS.
In an ideal situation, this would correct exactly for the charged part
of pileup. Applying pileup subtraction on the remaining neutral part,
\eg using the area--median subtraction, would mean that, in the end,
the resolution degradation is only affected by the neutral
fluctuations of pileup. Since about 60\% of the energy goes into
charged particles, this would automatically reduce the resolution
degradation by a factor around $\sqrt{0.4}\approx 0.63$. With this
technique, the pileup smearing effects for $\nPU=100$ would only be
slightly worse than what one sees without using charged-track
information at $\nPU=30$.
We shall discuss at length the use of charged tracks to mitigate
pileup in Chapter~\ref{chap:charged_tracks} below, including the
adaptation of the area--median method as well as other alternatives.

The second approach is to explicitly try to reduce the effect of
pileup fluctuations. Thinking about the LHC as a (very fast) camera,
this can be seen as a noise reduction algorithm.\footnote{With the
  important difference that the boundary of the hard event are not as
  well defined as boundaries between objects on a standard picture.}
This is a potentially vast domain of investigation and a few methods
going in that direction have already been proposed. We will discuss a
few such methods below: in Chapter~\ref{chap:beyond-grooming}, we will
discuss potential use of grooming techniques as generic pileup
mitigation tools (\ie beyond their applications to fat jets); in
Chapter~\ref{chap:soft-killer} we shall report on the SoftKiller
method~\cite{Cacciari:2014gra} that we have introduced recently, and
in Chapter~\ref{chap:beyond-prelimn} we will briefly comment on
possible alternatives that have not yet been fully investigated but
are of potential interest for the development of future pileup
mitigation techniques.
Note that around the same time as we introduced the SoftKiller method,
another noise-reduction technique called PUPPI has also been
proposed~\cite{PUPPI} and is currently increasingly used in the CMS
collaboration. We will not discuss the method here but will briefly
come back to it in our final discussion at the end of this document.

\section{Challenges of ``noise-reduction'' techniques}

Before going into the investigation of the techniques themselves, we
would like to have a generic discussion about one faces when designing
pileup-mitigation techniques that try to reduce the effects of
fluctuations.
We will argue that the main issue is that these techniques come with a
high level of fine-tuning needed to keep the method unbiased.

In what follows, we will consider two kinds of methods. The first
family starts by clustering the event and then, for a given jet
decides on a subset of its constituents (or subjets) to keep, similar
to what grooming procedures do. The other approach is to first decide
on a set of particles to keep, deeming the others as pileup and
removing them from the event prior to the clustering. 
Let us therefore consider a generic method that decides for each
particle or subjet whether they are pileup particles or not, keeping
the latter and rejecting the former. 
The discussion below trivially extends to the more generic case where
one instead associates a weight to each particle or subjet.

The procedure of keeping or rejecting particles comes with two sources
of bias.
First some of the particles originally coming from pileup will be
kept, yielding an under-subtraction, \ie a positive bias.
Then, some of the particles from the hard event will be deemed as
pileup and removed, giving a negative bias.
One will therefore only have an overall unbiased method if these two
sources of bias compensate each other on average. This does require
some level of fine-tuning of the free parameters of all the methods
considered below.
In other words, the positive tail of upwards pileup fluctuations, kept
in the event, will have to balance the negative bias coming from,
essentially soft, particles in the hard event that have been rejected
by the pileup mitigation technique.
The resulting fine-tuning has to be contrasted with the area--median
method for which the estimation of the pileup density $\rho$ remains
acceptable in a relatively large range of the grid-size parameter used
for the estimation.
However, if the bias of our new method can be kept small, one can
expect an improvement of the jet resolution if the fluctuations of the
positive and negative biases discussed above are themselves small.
In that respect, it is reasonable that the set of parameters for which
the bias remains small should also roughly correspond to the values of
the parameters for which the resolution degradation is at its minimum.

An important point to be aware of is that the parameters for which the
bias remains small are likely to vary with the process under
consideration, the energy scale of the hard event and the pileup
conditions.
It is therefore important to scan different hard processes, jet $p_t$
and pileup multiplicities when validating any new method.

The last potential issue to be aware of concerns methods applying an
event-wide pileup subtraction prior to the clustering. Pileup is
roughly uniformly spread across the event while the hard jets are
localised, peaked, energy deposits. One must therefore be careful,
when balancing the positive pileup fluctuations left uniformly in the
event, to avoid leaving some of these positive fluctuations at the
expense of over-subtracting the hard jets.
Ideally, an efficient pileup mitigation technique should therefore cut
pileup more aggressively when moving away from the hard jets.
This is somehow encoded in the PUPPI approach \cite{PUPPI} but is not
taken into account in a method like the SoftKiller (see
Chapter~\ref{chap:soft-killer}).
This is not a concern if one wishes to subtract pileup from jets of a
given radius but requires a re-tuning for different radii (see also
Section~\ref{sec:sk-improvements-Rdep}).

In the long term, it would be very interesting to understand if a
first-principles QCD understanding of jet substructure could help
controlling the biases mentioned above, in particular their variation
with respect to the jet $p_t$, the jet radius and the pileup
conditions, as well as how the subtraction should vary when
progressively moving away from the hard jets.


\chapter{Grooming as a generic tool for pileup subtraction}\label{chap:beyond-grooming}

\section{Foreword and motivation}

Given the success of grooming techniques in the context of fat jets
and boosted objects reconstruction observed in
Chapter~\ref{chap:grooming-mcstudy}, one naturally wonders if they
could also be useful for pileup mitigation in a more generic context.
This is what we discuss in this Chapter.
This work was initiated in the context of the ``Les Houches, Physics
at TeV colliders'' workshop in 2013 but never really finalised or, at
least, never pushed to a point where I thought it contained
solid-enough a message to deserve a place in the workshop proceedings
or as a standalone publication.
The study below is therefore unpublished and should really be
considered as preliminary.
I find useful to include it in this review, not only because it
directly pertains to the topic of pileup mitigation, but also because
I want this document to keep an opened door towards future
developments and I believe that grooming techniques can play a key
role in that context.

That said, the study carried over the next few pages is also similar
in spirit to the tests we have carried with the
``Cambridge/Aachen+filtering'' algorithm in our heavy-ion studies in
Chapter~\ref{chap:mcstudy-hi} up to one crucial difference.
In our heavy-ion studies, we considered the
``Cambridge/Aachen+filtering'' for both the hard reference and the
full event including Underlying Event, \ie as a jet algorithm, used
in combination with the area--median subtraction.
Here we only want to consider grooming on the full jet and compare
that to the ungroomed hard jets. In a sense, the idea is really to
consider grooming as a tool for pileup mitigation, which implies that,
in the absence of pileup, we work with ungroomed anti-$k_t$
jets.\footnote{It might also be interesting to consider the case where
  the grooming is included as part of the jet definition, as done in
  Chapter~\ref{chap:mcstudy-hi}.}
Despite this sizeable difference, the reason for which we want to use
grooming remains the same: by effectively reducing the area of the
jet, we expect the groomed jet to be less sensitive to pileup
fluctuations \ie have a better energy resolution.
This obviously comes with a price: grooming also affects the hard
contents of jets and so, compared to the robust area--median approach,
there is no guarantee that the average bias will remain close to 0.
We refer the reader to Chapter~\ref{chap:beyond-motivation} for a
discussion of biases and reduction of fluctuation effects.

\section{Grooming configurations and workflow}

Our study includes the three types of jet grooming techniques listed
below. In each case, the subjets are subtracted using the area--median
method before making the selection of which subjets have to be kept.
\begin{description}
\item {\bf Filtering.} We will consider both $n_{\rm filt}=2$ and
  $n_{\rm filt}=3$ and scan over a range of $R_{\rm filt}$ values,
  with subjets clustered with the Cambridge/Aachen algorithm.
\item {\bf Trimming.} We use the Cambridge/Aachen algorithm with
  $R_{\rm trim}=0.2$ and scanned over a range of $f_{\rm trim}$
  values. The reference $p_t$ is taken to be that of the unsubtracted
  jet.
\item {\bf Area-trimming.} This has not been introduced
  before. Area-trimming starts similarly to trimming by re-clustering
  the jet into subjets using the Cambridge/Aachen algorithm with
  $R_{\rm sub}=0.2$. We then keep the subjets for which the
  (subtracted) $p_t$ is larger than
  $n_{\sigma}\sigma\sqrt{A_{\rm subjet}}$, with $\sigma$ the pileup
  fluctuations as estimated with \fastjet.
  The motivation for this choice is that, after the pileup
  subtraction, one has removed the negative tail of the pileup
  fluctuations but one is left with the positive tail which has
  $\sigma\sqrt{A_{\rm subjet}}$ as a characteristic scale.
  We have tested a range of $n_\sigma$ values in our studies.
\end{description}
Note that in a more complete study it would be interesting to add \SD
to the above list and to study different subjet radii for trimming and
area-trimming.

The rest of the simulation procedure follows the same pattern as what
we have done in Section~\ref{sec:areamedian-mcstudy:jetpt} for our
basic assessment of the area--median approach. We simulate hard events
at $\sqrt{s}=13$~TeV with Pythia~8.186 (tune 4C), with the Underlying
Event switched off. We have only considered dijet events although
tests with different event samples would be desirable.
The hard events will be embedded in pileup, also generated with the
same Pythia~8 version and tune. Following the same terminology as in
Chapter~\ref{sec:areamedian-mcstudy:jetpt}, the latter will be
referred to as the ``full event''.
The estimation of the pileup properties are obtained using a
grid-based area--median estimation with a cell size of 0.55 and
rapidity rescaling.

In practice, we select jets from the hard event with $|y|<4$ and a
given $p_{t,\rm cut}$, then match the jets in the full event to the
hard jets.\footnote{As before, we do that by requiring that the $p_t$
  of the common constituents is at least 50\% of the $p_t$ of the hard
  jet.
}
Once we have a pair of matching jets, we apply one of the grooming
techniques listed above and compute the $p_t$ difference,
$\Delta p_t$, between the original hard jet and the subtracted full
jet. 
As before, we study the average shift $\avg{\Delta p_t}$, which should
be close to 0 for an unbiased method, and the dispersion
$\sigma_{\Delta p_t}$ which measures the residual resolution
degradation.

\section{Results}

\begin{figure}[!t]
\centerline{\includegraphics[width=0.8\textwidth]{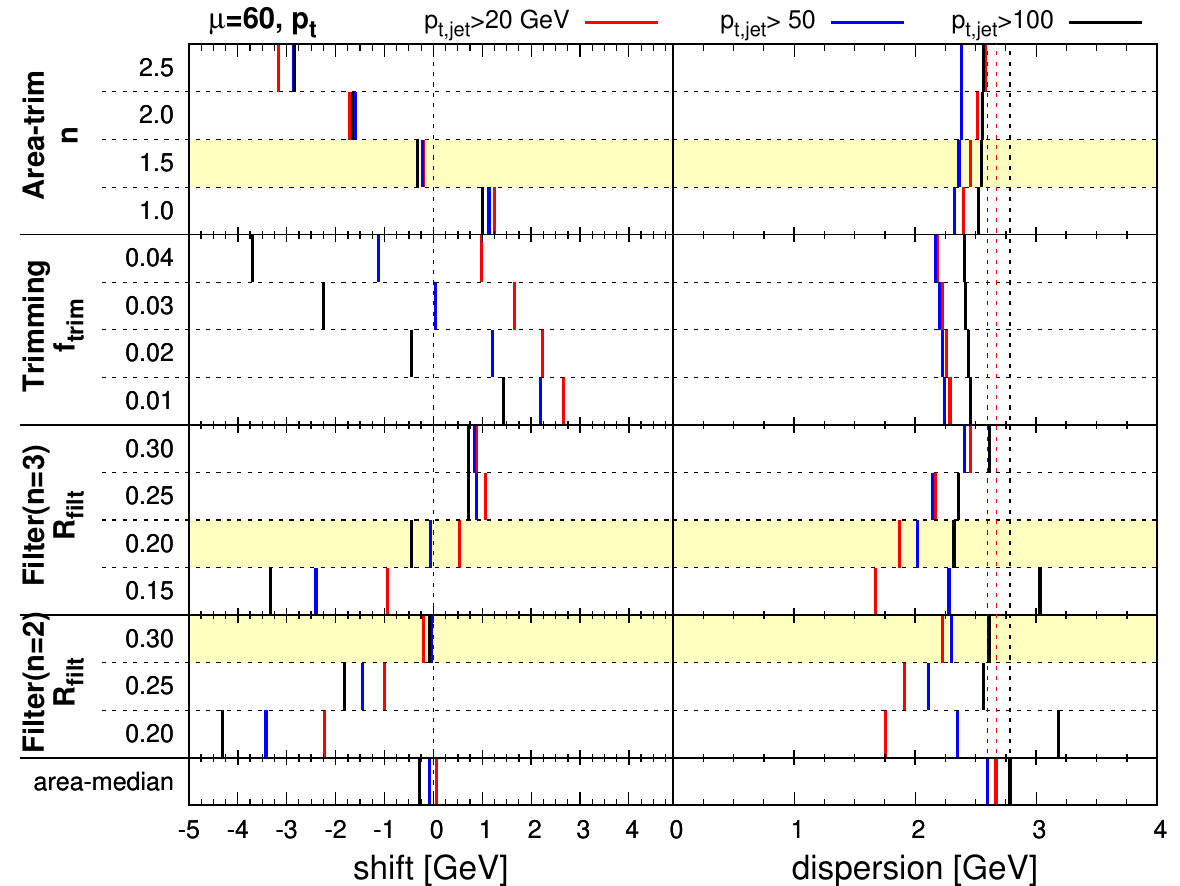}}%
\caption{Summary plot of the average shift and bias for the different
  grooming methods we have tested and different values of their
  respective parameters. We have used $\mu=60$ and different cuts on
  the jet $p_t$: $20$~GeV (red), $50$~GeV(blue) and $100$~GeV
  (black). The ``standard'' area--median results are quoted at the
  bottom of the plot. On the average shift plot (left), the vertical
  dashed line corresponds to a zero bias, and on the dispersion plot
  (right), the vertical dashed lines correspond to the area--median
  results.}\label{fig:groomingsub-summary-v-pt}
\end{figure}

\begin{figure}[!t]
\centerline{\includegraphics[width=0.8\textwidth]{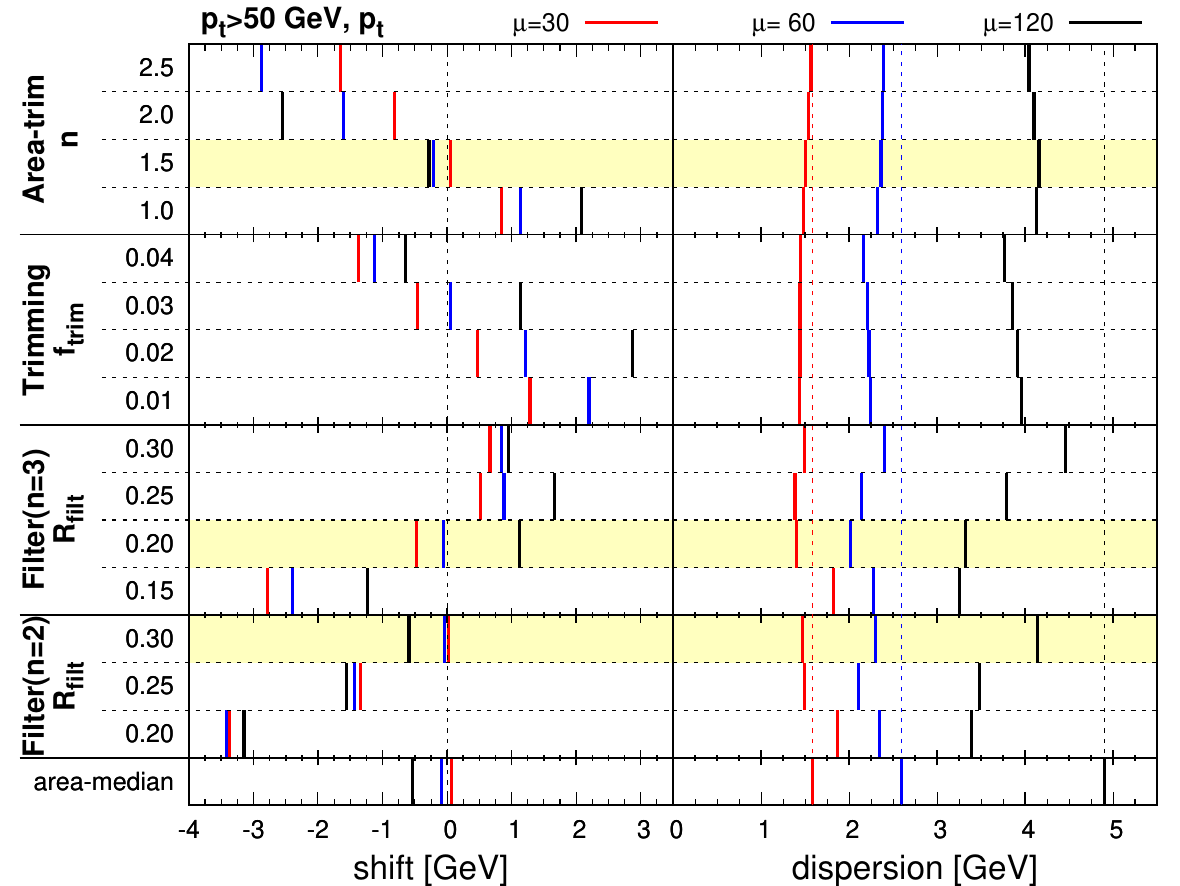}}%
\caption{Same as Fig.~\ref{fig:groomingsub-summary-v-pt} by with a
  fixed cut of $50$~GeV on the jet $p_t$ and, instead, varying the
  pileup average multiplicity $\mu=30$ (red), $60$ (blue) or $120$
  (black).}\label{fig:groomingsub-summary-v-mu}
\end{figure}

A summary of our findings is presented in
Figs.~\ref{fig:groomingsub-summary-v-pt} and
\ref{fig:groomingsub-summary-v-mu}.
In both cases, we show the average shift and the corresponding
dispersion for the different grooming techniques under scrutiny,
varying their free parameter. In
Fig.~\ref{fig:groomingsub-summary-v-pt}, we have fixed the pileup
conditions to $\mu=60$ and varied the cut on the $p_t$ of the jet,
while in fig.~\ref{fig:groomingsub-summary-v-mu} we have worked with a
fixed $p_t$ cut of 50~GeV and varied the pileup conditions.

The first properties we expect from an efficient pileup mitigation
technique is that it remains unbiased and robust, \ie that the average
shift is close to 0 regardless of the pileup conditions and of the
details of the hard process.
In all cases, we see that this requirement is not trivially
satisfied. At the very least, the free parameter of each grooming
technique needs to be carefully tuned for the bias to remain
small. This contrasts with the area--median approach where the average
bias is very close to 0 in a wide variety of cases and where the free
parameter, controlling the size of the patches, can be varied around
the optimal scale without affecting significantly the performance.

Furthermore, we already see differences between the grooming methods
at the level of the average bias. Filtering with $n_{\rm filt}=2$ and
$R_{\rm filt}=0.3$, and area-trimming with $n_\sigma=1.5$ show small
biases, comparable with the area--median approach, for all the cases
we have studied. Then filtering with $n_{\rm filt}=3$ and
$R_{\rm filt}=0.2$ shows a reasonably small bias albeit with a larger
dependence on the pileup multiplicity. For trimming, the smallest bias
is obtained for $f_{\rm trim}=0.03$ but this choice appears less
robust that what we observe with other grooming techniques.
Physically, this seems to indicate that if we want to use a method
that decides which subjets are kept based on a $p_t$ cut, one obtains
better results if the cut is set based on the pileup properties than
based on the jet $p_t$. If some dependence on $p_t$ is to be
included --- \eg to account for the residual biases of the
area-trimming method --- it should at least be milder than
linear.\footnote{This is also a question that one should be able to
  address analytically.}

Turning to the dispersion, we see that the resolution degradation is,
as expected, smaller than what one obtains with the area--median
approach. 
We also see that the gain becomes larger as the pileup increases.

\begin{figure}
\centerline{%
\includegraphics[width=0.48\textwidth]{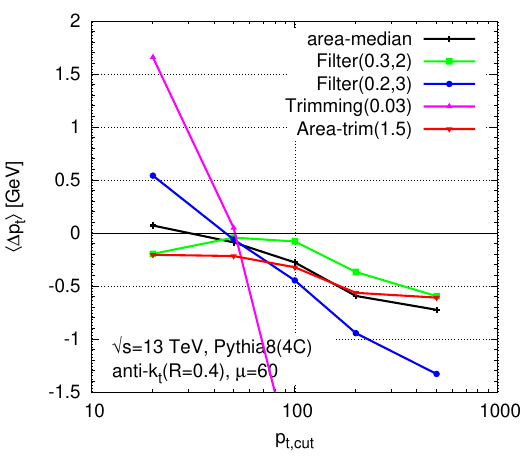}%
\hfill%
\includegraphics[width=0.48\textwidth]{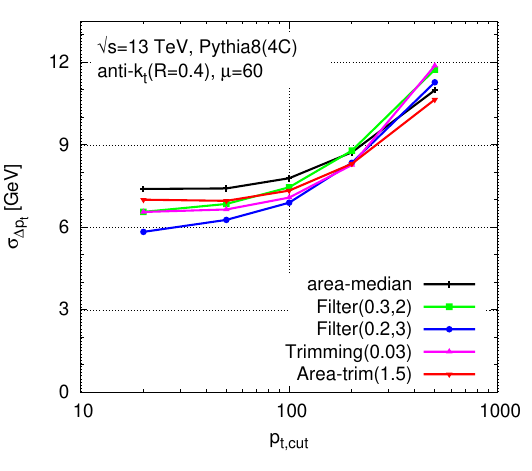}}%
\caption{Plot of the average bias (left) and of the corresponding
  dispersion (right), as a function of the jet $p_t$ cut, for a few
  specific grooming techniques and for $\mu=60$,
}\label{fig:groomingsub-stats-v-pt}
\end{figure}

\begin{figure}
\centerline{%
\includegraphics[width=0.48\textwidth]{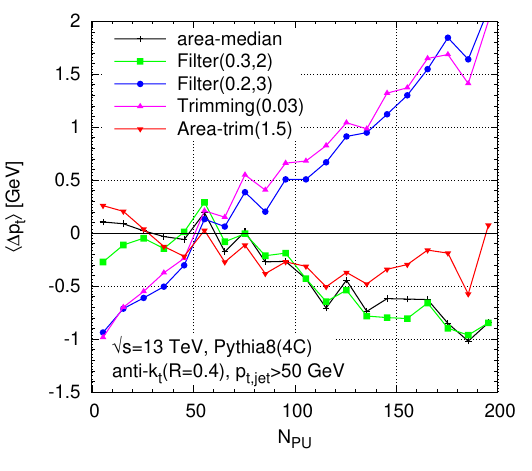}%
\hfill%
\includegraphics[width=0.48\textwidth]{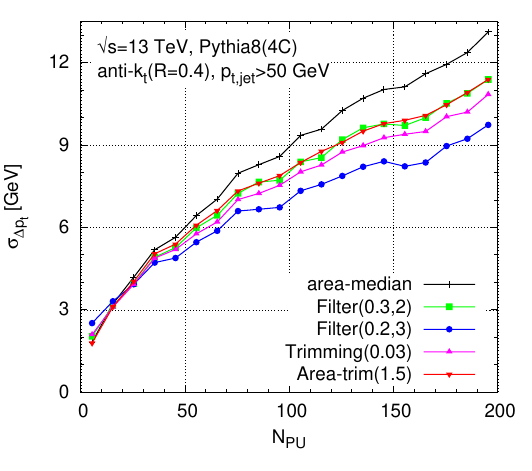}}%
\caption{Plot of the average bias (left) and of the corresponding
  dispersion (right), as a function of the average number of pileup
  interactions, for jets with
  $p_t>50$~GeV.}\label{fig:groomingsub-stats-v-mu}
\end{figure}

To investigate these results a bit further, we have selected for each
of the 4 groomer families the one showing the best performance and we
have looked more closely at the average shift and dispersion as a
function of the jet $p_t$, Fig.~\ref{fig:groomingsub-stats-v-pt}, and
of the pileup conditions, Fig.~\ref{fig:groomingsub-stats-v-mu}.
This shows clearly that filtering with $n_{\rm filt}=2$ and
$R_{\rm filt}=0.3$, and area-trimming with $n_\sigma=1.5$ have a bias
similar to the area--median method, at least for the dijet processes
that we have studied. Except for the filter at large $p_t$, these two
options show an improvement over the area--median approach in terms of
resolution degradation.
At small $p_t$ (and $\mu=60$), we see an improvement around 5\% for
area-trimming and 15\% for filtering.
While this improvement persists across the whole $p_t$ range for the
case of area-trimming, it disappears for $p_t\ge 200$~GeV in the case
of filtering.
This might not be too much of an issue since pileup is most
problematic at low $p_t$.
Nonetheless, it suggests that improvements are possible at large
$p_t$.
At high pileup multiplicities, both methods show a resolution
improvement slightly above 10\%.

If one wishes to tolerate a slightly larger bias, filtering with
$n_{\rm filt}=3$ and $R_{\rm filt}=0.2$ can also be considered and
gives a more sizeable improvement in terms of dispersion, although
once again, this is mostly seen at small $p_t$.

\section{Concluding remarks and future prospects}

The study we have presented here is fairly minimal and definitely
requires more tests if one wishes to draw a firm conclusion.
We see however some patterns emerging: 
\begin{itemize}
\item using grooming techniques as a tool for pileup subtraction
  requires some level of fine-tuning if one wishes to keep the average
  bias small.
\item In the region where the bias is small, one sees, as initially
  foreseen, a reduction of the sensitivity to pileup fluctuations.
\item Filtering with $n_{\rm filt}=2$ and $R_{\rm filt}=0.3$ or with
  $n_{\rm filt}=3$ and $R_{\rm filt}=0.2$ as well as the, new,
  area-trimming with $n_\sigma=1.5$ show overall good
  performances. Trimming seems to be disfavoured, in particular
  because it shows a larger dependence of the bias on the jet $p_t$.
\end{itemize}

The main message to remember from this preliminary pileup-mitigation
study is that grooming can be a valuable tool in the context of pileup
mitigation. It would also be interesting to carry further studies of
the robustness of the methods that are seen to perform well here. This
involves more detailed tests, including the rapidity dependence, a more
complete scan over the jet $p_t$ and \nPU and the study of other
processes, but also tests of detector effects or applications to CHS
events (see also Chapter~\ref{chap:charged_tracks} and our final
discussion in Chapter~\ref{chap:ccl}).

Finally, one may also consider dynamical adjustments of the
parameters, \eg $R_{\rm filt}$ or $n_\sigma$, with the jet $p_t$ and
the pileup conditions. For example, we could groom more aggressively
for low-$p_t$ jets where we see larger gains in terms of resolution
and where these gains are the most important for practical
applications at the LHC. Note that these parameters are expected to
vary relatively slowly, typically only logarithmically, with the jet
$p_t$.


\chapter{Using information from charged tracks}\label{chap:charged_tracks}

In this chapter, we consider the situation where we can use
vertex information to tell whether a charged track comes from the
hard interaction vertex of from a pileup vertex.
We shall assume perfect identification and measurement of charged
tracks and all vertices.
This can be seen as an idealised Charge-Hadron-Subtracted (CHS)
approach.

In this approach, the particles in each event can therefore be split
in three main categories: neutral particles, charged particles from
the hard interaction vertex, and charged particles from any of the
pileup vertices.
In that context, we can perform an exact subtraction of pileup for the
charged part of the event, keeping the charged tracks associated with
the leading vertex and discarding the ones associated with a pileup
vertex.
However, if one wishes to retain information from the charged
particles associated with a pileup vertex, it is always possible to
keep them as ghosts in the event, \ie to keep them with transverse
momenta scaled down to infinitesimal values.
This is what we shall consider for most of our CHS-like events in what
follows.

We are therefore left with the subtraction of the neutral part of the
event.
In this chapter, we investigate several methods that concentrate on
pileup mitigation for CHS-type of events and compare their performance.

\section{Pileup subtraction methods for CHS events}\label{sec:pusub-with-chargedinfo}

\paragraph{Area--median with charged-tracks information.}
The generalisation of the area--median method to CHS events is
straightforward: we simply compute $\rho$ (and $\rho_m$) as described
in Section~\ref{sec:areamed-areamed}, this time using the particles
from the CHS events. The rapidity profile can also be obtained from
the neutral component of minimum bias. 
Note that, in practice, we could either use only the neutral particles
or all CHS particles for the estimation of $\rho$. Since this makes
very little difference in the end, we decided to compute the pileup
density from all particles in the CHS event.

Furthermore, we still use \eqref{eq:subtraction-base} to subtract
pileup from the jets. In the context of CHS events, we can also
slightly improve our positivity constraints for the final jet
transverse momentum and mass. If, after subtraction, the transverse
momentum of the jet is less than the transverse momentum carried by
charged tracks from the leading vertex, we set the subtracted
4-momentum to the sum of the charged tracks from the leading
vertex. If it has a larger momentum but a smaller mass, we set the
transverse momentum and azimuthal angle according to the result of the
pileup subtraction and set the rapidity and mass as the one obtained
from summing the charged tracks from the leading vertex. We shall study
later in this chapter the effect of this ``safe'' subtraction on the
reconstruction of the jet mass.

Note finally that, in \fastjet, one can still use the \ttt{Subtractor}
tool as for the standard usage. One can even use the \ttt{Subtractor}
directly on full events using
\ttt{Subtractor::set\_known\_selectors(...)} which takes two
\ttt{Selector}s as arguments: the first one isolates particles which
can be associated to a known vertex, while the second one selects the
particles from the leading vertex among the particles with a known
vertex association.

If we want to impose positivity constraints on the mass, we can still
use \ttt{Subtractor::set\_safe\_mass}. With a previous call to
\ttt{Subtractor::set\_known\_selectors(...)}, the more constraining
conditions, involving information from the particles associated with
the leading vertex, will be used (see
Section~\ref{sec:areamedian-safesubtraction}).

\paragraph{Neutral-proportional-to-charged (NpC).} An alternative
approach to pileup subtraction in a context where we know for each
charged track whether it comes from the leading vertex or from a
pileup interaction is to deduce the neutral momentum of the jet based
on its charged information. 
The NpC method estimates and subtracts the neutral pileup component by
assuming it to be proportional to the charged pileup component.
At least two variants can be conceived of. 

If the charged pileup particles are kept (untouched) as part of the
jet during clustering, then the corrected jet momentum
is~\cite{ourtalk4cms}
\begin{equation}
  \label{eq:npc-full}
  p_\mu^\text{jet,sub} = p_\mu^\text{jet} -
  \frac{1}{\gamma_0} p_\mu^\text{jet,chg-PU} \,,
\end{equation}
where $p_\mu^\text{jet,chg-PU}$ is the four-momentum of the
charged-pileup particles in the jet and $\gamma_0$ is the
average fraction of pileup transverse momentum that is carried by
charged particles. 
Specifically, one can define
\begin{equation}
  \label{eq:gamma0}
  \gamma_0 \equiv \left\langle 
    \frac{\sum_{i\in \text{charged particles}} \,p_{ti}}%
         {\sum_{i\in \text{all particles}} \,p_{ti}} 
         \right\rangle_\text{\!\!events}\!\!\!\!\!,
\end{equation}
where the sums run over particles in a given event (possibly limited
to some central region with tracking), and the average is carried out
across minimum-bias events. With the Pythia~8 simulations we have used
in this Chapter --- see Section~\ref{sec:npc-mcstudies} below and
Appendix~\ref{app:details-npc} --- we have $\gamma_0\approx 0.612$.

If the charged pileup particles are not directly included in the
clustering (\ie it is the CHS event that is provided to the
clustering), then one needs to make sure that pileup vertices are
still included as ghost tracks, with their momenta rescaled by an
infinitesimal factor $\epsilon\lll 1$.
In this case the correction becomes
\begin{equation} 
  \label{eq:npc-chs}
  p_\mu^\text{jet,sub} = p_\mu^\text{jet,CHS} -
  \frac{1-\gamma_0}{\gamma_0\, \epsilon}
  p_\mu^\text{jet,rescaled-chg-PU}\,,
\end{equation}
where $p_\mu^\text{jet,CHS}$ is the momentum of the jet as obtained
from the CHS event, while $p_\mu^\text{jet,rescaled-chg-PU}$ is the
summed momentum of the rescaled charged-pileup particles that are in
the jet.
When carrying out NpC-style subtraction, this is our preferred
approach because it eliminates any back-reaction associated with the
charged pileup (this is useful also for the area--median subtraction),
while retaining the information about charged pileup tracks.

There are multiple issues that may be of concern for the NpC method.
For example, calorimeter fluctuations can limit the experiments'
ability to accurately remove the charged pileup component as measured
with tracks.
For out-of-time pileup, which contributes to calorimetric energy
deposits, charged-track information may not be available at all.
In any case, charged-track information covers only a limited range of
detector pseudo-rapidities.
Additionally there are subtleties with hadron masses: in effect,
$\gamma_0$ is different for transverse components and for longitudinal
components.
In the following we will avoid this problem by treating all particles
as massless.\footnote{Particle momenta are modified so as to become
  massless while conserving $p_t$, rapidity and azimuth.}
The importance of the above limitations can only be fully evaluated in
an experimental context.

NpC also comes with a few potential (experimental) advantages compared
to the area--median subtraction method.  
First, the area--median method includes questions of non-trivial
rapidity dependence and detector non-linearities (the latter is
relevant also for NpC).
These have, to a reasonable extent, been successfully overcome by the
experiments~\cite{ATLAS-PU-Performance,CMS-PU-Performance}.
One respect in which NpC may have advantages of the area--median
method is that the latter fails to correctly account for the fact
pileup fluctuates from point to point within the event, a feature that
cannot be encoded within the global pileup estimate $\rho$.
Furthermore NpC does not need a separate estimation of the background
density $\rho$, which can have systematics related to the event
structure (\eg $t\bar t$ events v.\ dijet events); and there is no
need to include large numbers of ghosts for determining jet areas, a
procedure that has a non-negligible computational cost.

\paragraph{Jet cleansing.} Jet cleansing \cite{Krohn:2013lba}
essentially puts several ingredients together into a pileup
subtraction technique. It first re-clusters the jet into subjets as
done for filtering and trimming, see
Sec.~\ref{sec:grooming-list-techniques}, (cf.\ also the early work by
Seymour~\cite{Seymour:1993mx}). It then applies an NpC-like technique
to each of the subjets, before summing the resulting 4-momenta into
the final subtracted jet. It also applies an additional constraint
that we decided to call {\it zeroing} and that we shall discuss
separately later in this Section.
Cleansing may also be used in conjunction with trimming-style cuts to
the subtracted subjets, specifically it can remove those whose
corrected transverse momentum is less than some fraction
$f_\text{cut}$ of the overall jet's transverse momentum (as evaluated
before pileup removal).\footnote{They also investigated the use of a
  variable known as the jet vertex fraction, widely used
  experimentally to reject jets from a vertex other than the leading
  one~\cite{ATLAS-PU-Performance,CMS:2013wea,ATL-Pileup-2}.}

To be a little more precise, the implementation of (linear) Jet
Cleansing, available in \fjcontrib, slightly differs from our
definition of NpC. 
According to the description in Ref.~\cite{Krohn:2013lba}, one
additional characteristic of linear cleansing relative to NpC is that
it switches to jet-vertex-fraction (JVF) cleansing when the NpC-style
rescaling would give a negative answer.
In contrast, area-subtraction plus trimming simply sets the (sub)jet
momentum to zero.
In our Monte-Carlo studies presented below in
Section~\ref{sec:npc-mcstudies}, we explicitly tried turning the
switch to JVF-cleansing on and off and found it had a small effect and
did not explain the differences.

Furthermore, \eq~\eqref{eq:npc-chs} for NpC corrects for the jet
4-momentum, while cleansing scales the 4-momentum $p_\mu^{(i)}$ of each
neutral particle $i$ in the jet as follows
\begin{equation}
  \label{eq:cl-rescaling}  
  p_\mu^{(i)} \;\to\; p_\mu^{(i)} \times \left(1- \frac{1-\gamma_0}{\gamma_0 \epsilon}
    \frac{p_t^\text{jet, rescaled-chg-PU}}{p_t^\text{jet, ntr}}\right)\,,
\end{equation}
where the notation is as in \eq~(\ref{eq:npc-chs}) and additionally
$p_t^\text{jet, ntr}$ is the transverse momentum of the neutral part
of the jet (including pileup).
\Eq~(\ref{eq:cl-rescaling}) is correct for the jet $p_t$, but would be
only be correct for the jet mass if the pileup and hard components had
identical angular distributions.
However in practice, the hard component is usually collimated, whilst,
to a first approximation, the pileup component is uniformly
distributed across the jet.
Accordingly, the scaling gives incorrect jet mass
results.
The issue is less severe when cleansing is used as originally
intended, \ie including trimming.

A last small difference is that cleansing assumes that all the
particles in the event are clustered, in a spirit similar to
\eq~\eqref{eq:npc-full}. As already mentioned in our description of
NpC, this can bring an additional source of back-reaction compared to
\eq~\eqref{eq:npc-chs}, where the charged tracks from pileup vertices
are only included as ghosts.

Note finally that besides the ``linear'' version of cleansing
presented above, Ref.~\cite{Krohn:2013lba} also introduces a variant
called {\it Gaussian cleansing} which is particularly interesting in
that it effectively carries out a $\chi^2$ minimisation across
different hypotheses for the ratio of charged to neutral energy flow,
separately for the pileup and the hard event. We shall see below that
its performance is usually marginally better than the much simpler
linear cleansing.

\paragraph{Charged-tracks-based trimming, zeroing and protected
  zeroing.} It turns out that Jet Cleansing imposes an additional
condition to subjets: if a subjet contains no charged particles from
the leading vertex (LV), then its momentum is set to
zero.\footnote{Note that zeroing does not seem to be documented in
  Ref.~\cite{Krohn:2013lba}. It can however be clearly identified
  (cf.~Ref.~\cite{Cacciari:2014jta} on which this Chapter is largely
  based) in the public code for jet cleansing. (Version 1.0.1 from
  \fjcontrib.)}
Since we will be discussing it extensively, it is useful to give it a
name, ``\emph{zeroing}''.
Zeroing can be thought of as an extreme limit of the charged-track
based trimming procedure introduced by ATLAS~\cite{ATL-Pileup-2},
whereby a JVF-style cut is applied to reject subjets whose
charged-momentum fraction from the leading vertex is too
low.
As we will show in the Monte-Carlo studies presented in
Section~\ref{sec:npc-mcstudies} zeroing turns out to be crucial in
explaining the differences observed between CHS area-subtraction (or
with NpC subtraction) with $f_\cut=0$ trimming and cleansing.

This approach may seem a little dangerous since there are rare cases
where a hard jet --- or a subjet carrying a large fraction of the jet
$p_t$ --- can be purely neutral and zeroing would simply discard
it. This can correspond to the loss of subjets with tens of GeV, while
it is very unlikely that a subjet from a pileup collision will be
responsible for such a large energy.
Therefore we introduce a modified procedure that we call
``\emph{protected zeroing}'': one rejects any subjet without LV tracks
\emph{unless} its $p_t$ after subtraction is $n$ times larger than the
largest charged $p_t$ in the subjet from any single pileup vertex (or,
more simply, just above some threshold $p_{t,\text{min}}$; however,
using $n$ times the largest charged subjet $p_t$ could arguably be
better both in cases where one explores a wide range of $N_\PU$ and
for situations involving a hard subjet from a pileup collision).
We have found good results, and improvements compared to plain
zeroing, taking $n=10$ (as we shall use later in our Monte-Carlo
studies) or a fixed $p_{t,\text{min}} = 20\GeV$.

Before starting an in-depth comparison of the methods described above,
we also want to point out that similarly to the combination of a
neutral-proportional-to-charged method with trimming, done by jet
cleansing, several other similar approaches have been proposed and
studied in the past in the context of the area--median
approach\cite{Cacciari:2008gd}. We have already discussed in
Chapter~\ref{chap:mcstudy-hi} (see also Ref.\cite{Cacciari:2010te})
and in Chapter~\ref{chap:grooming-mcstudy} (see also
\cite{Altheimer:2013yza}) the possibility to combine the area--median
subtraction together with grooming techniques and this combination can
trivially be extended to the case of CHS events.
The preliminary studies presented in
Chapter~\ref{chap:beyond-grooming} also relate to similar ideas.

\section{Monte-Carlo studies and physical discussions}\label{sec:npc-mcstudies}

Since our testing framework remains close to the one we have used for
the studies in Chapter~\ref{chap:mcstudy}, modulo a few adaptations
meant to be closer to the setup used in Ref.~\cite{Krohn:2013lba}, we
will directly concentrate on the results of our studies and their
physical interpretation. We refer to Appendix~\ref{app:details-npc} for a
more detailed description of our simulations.

Since the main goal of this Chapter is to discuss the use of charged
tracks in pileup mitigation, the first and most important point we
shall discuss is a comparison between the area--median method and
the NpC approach.
It will show that the former gives a slightly better performance than
the latter and we will explain why this could be expected.
We shall then discuss extensively the comparison of our findings with
the, apparently contradictory, findings from jet cleansing in
Ref.~\cite{Krohn:2013lba}.

\subsection{Performance of NpC v. area--median}\label{sec:npc-v-areamed}

\begin{figure}[t]
  \centering
  \includegraphics[width=0.48\textwidth]{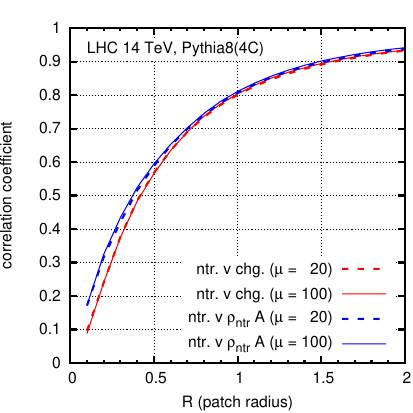}
  \hfill
  \includegraphics[width=0.48\textwidth]{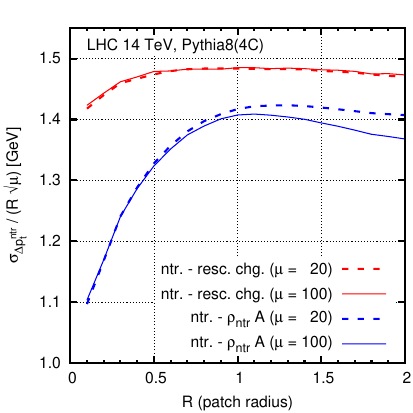}
  \caption{
    Left: the correlation coefficient between the neutral transverse
    momentum in a central patch and either the charged transverse
    momentum in that patch (rescaling this component would not change
    the correlation coefficient) or the prediction using the area--median
    method, \ie $\rho_\text{ntr} A$.
    Right: the standard deviation of the difference between neutral transverse
    momentum in a central patch and either the rescaled charged transverse
    momentum in that patch or the prediction using the area--median
    method, \ie $\rho_\text{ntr} A$.
    The events are composed of superposed zero-bias collisions
    simulated with Pythia~8, tune 4C, and the number of collisions per
    event is Poisson distributed with average $\mu$.}
  \label{fig:correl-central}
\end{figure}

Let us therefore proceed with an investigation of NpC's performance
compared to the now-familiar area--median method, focusing our
attention on particle-level events for simplicity.
The key question is the potential performance gain due to NpC's use of
local information.
To study this quantitatively, we consider a circular patch of radius
$R$ centred at $y = \phi = 0$ and examine the correlation coefficient
of the actual neutral energy flow in the patch with two estimates: (a)
one based on the charged energy flow in the same patch and (b) the
other based on a global energy flow determination from the neutral
particles, $\rho_\text{ntr}$.
Fig.~\ref{fig:correl-central} (left) shows these two correlation
coefficients, ``ntr v.\ chg'' and ``ntr v.\ $\rho_\text{ntr} A$'', as
a function of $R$, for two average pileup multiplicities, $\mu = 20$
and $\mu = 100$.
One sees that the local neutral-charged correlation is slightly \emph{lower},
\ie slightly worse, than the neutral-$\rho_\text{ntr}$ correlation.
Both correlations decrease for small patch radii, as is to be
expected, and the difference between them is larger at small
patch radii.
The correlation is largely independent of the number of pileup events
being considered, which is consistent with our expectations, since all
individual terms in the determination of the correlation coefficient
should have the same scaling with $N_\PU$.

As already pointed out before, quantitative interpretations of
correlation coefficients can sometimes be delicate (see also 
Appendix~\ref{app:correlation-coefs}).
We instead find that it is more robust to investigate 
$\sigma_{\Delta p_t^\text{ntr}}$, defined (similarly to our previous
studies in this document) as the standard deviation of
\begin{equation}
  \label{eq:delta-pt}
  \Delta p_t^\text{ntr} = p_{t}^\text{ntr} - p_{t}^\text{ntr,estimated}\,,
\end{equation}
where the estimate of neutral energy flow,
$p_{t}^\text{ntr,estimated}$, may be either from the rescaled charged
flow or from $\rho_\ntr A$.
The right-hand plot of Fig.~\ref{fig:correl-central} shows
$\sigma_{\Delta p_t^\text{ntr}}$ for the two methods, again as a function of $R$,
for two levels of pileup.
It is normalised to $R \sqrt{\mu}$, to factor out the expected
dependence on both the patch radius and the level of pileup.
A lower value of $\sigma_{\Delta p_t^\text{ntr}}$ implies better performance, and
as with the correlation we reach the conclusion that a global estimate
of $\rho_\ntr$ appears to be slightly more effective at predicting local
neutral energy flow than does the local charged energy flow.

If one hoped to use NpC to improve on the performance of area--median
subtraction, then figure~\ref{fig:correl-central} suggests that
one will be disappointed: the local estimate of the neutral component
of pileup from NpC is marginally worse than that given by the global
$\rho_\ntr$ from the area--median method.

In striving for an understanding of this finding, one should recall
that the ratio of charged-to-neutral energy flow is almost entirely
driven by non-perturbative effects.
Inside an energetic jet, the non-perturbative effects are at scales
$\sim \Lambda_\text{QCD}$ that are tiny compared to the jet transverse
momentum $p_t$.
There are fluctuations in the relative energy carried
by charged and neutral particles, for example because a leading
$u$-quark might pick up a $\bar d$ or a $\bar u$ from the
vacuum. 
However, because $\Lambda_\text{QCD} \ll p_t$, the charged and neutral
energy flow mostly tend to go in the same direction.

The case that we have just seen of an energetic jet gives an intuition
that fluctuations in charged and neutral energy flow are going to be
locally correlated.
It is this intuition that motivates the study of NpC.
We should however examine if this intuition is actually valid for
pileup.
We will examine one step of hadronisation, namely the production of
short-lived hadronic resonances, for example a $\rho^{+}$.
The opening angle between the $\pi^+ \pi^0$ decay products of the
$\rho^{+}$ is of order $2m_\rho / p_{t,\rho}$.
Given that pileup hadrons are produced mostly at low $p_t$, say
$0.5-2\GeV$, and that $m_\rho \simeq 0.77\GeV$, the angle between the
charged and neutral pions ends up being of order $1$ or even larger.
As a result, the correlation in direction between charged and neutral
energy flow is lost, at least in part.
Thus, at low $p_t$, non-perturbative effects specifically tend to wash out the
charged-neutral angular correlation.

\begin{figure}
  \centering
  \begin{minipage}[c]{0.48\linewidth}
    \includegraphics[width=\textwidth]{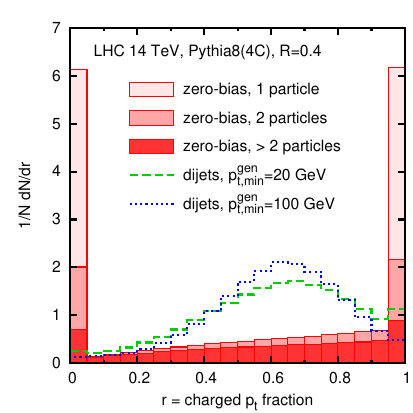}  
  \end{minipage}
  \hfill
  \begin{minipage}[c]{0.48\linewidth}
    \caption{The filled histogram shows the distribution, for simulated
      zero-bias collisions, of the
      fraction, $r$, of the transverse 
      momentum in a central circular patch of radius $R=0.4$ that is due to
      charged particles. 
      It is separated into components according to the multiplicity of
      particles in the patch.
      The dashed and dotted histograms show the corresponding
      charged-fraction distributions for each of the two hardest
      anti-$k_t$, $R=0.4$ jets in simulated dijet events, with two
      choices for the hard generation cut $p_{t,\min}^\text{gen}$.}
    \label{fig:why-NpC-bad}
  \end{minipage}
\end{figure}

This point is illustrated in Fig.~\ref{fig:why-NpC-bad}.
We consider zero-bias events and examine a circular patch of radius
$R=0.4$ centred at $y = \phi = 0$. 
The figure shows the distribution of the charged $p_t$ fraction, $r$,
\begin{equation}
  \label{eq:r-definition}
  r = \frac{p_{t}^\text{chg}}{p_{t}^{\text{chg}+\text{ntr}}}\,,
\end{equation}
in the patch (filled histogram, broken into contributions where the
patch contains $1$, $2$ or more particles).
The same plot also shows the distribution of the charged $p_t$ fraction
in each of the two leading anti-$k_t$, $R=0.4$ jets in dijet events
(dashed and dotted histograms). 
Whereas the charged-to-total ratio for a jet has a distribution peaked
around $0.6$, as one would expect, albeit with a broad distribution,
the result for zero-bias events is striking: in about 60\% of events
the patch is either just charged or just neutral, quite often
consisting of just a single particle.
This is probably part of the reason why charged information provides
only limited local information about neutral energy flow in pileup
events.

These considerations are confirmed by an analysis of the actual
performance of NpC and area--median subtraction.
We reconstruct jets using the anti-$k_t$ algorithm, as implemented in
FastJet, with a jet radius parameter of $R=0.4$.
We study dijet and pileup events generated with Pythia~8.176, in tune
4C; we assume idealised CHS, treating the charged pileup particles as
ghosts.
In the dijet (``hard'') event alone, \ie without pileup, we run the
jet algorithm and identify jets with absolute rapidity $|y|<2.5$ and
transverse momentum $p_t> 150\GeV$.
Then in the event with superposed pileup (the ``full'' event) we rerun
the jet algorithm and identify the jets that match those selected in
the hard event\footnote{\label{footnote:matching}For the matching, we
  introduce a quantity
  $p_t^\text{shared}(j_i^\text{hard}, j_j^\text{full})$, the scalar
  sum of the constituents that are common to a given pair $i,j$ of
  hard and full jets.
  For a hard jet $i$, the matched jet in the full event is the one that
  has the largest $p_t^\text{shared}(j_i^\text{hard},
  j_j^\text{full})$. 
  In principle, one full jet can match two hard jets, \eg if two
  nearby hard jets end up merged into a single full jet due to
  back-reaction effects. However this is exceedingly rare, since we
  consider only the two hardest jets in the hard event, which are nearly always
  back-to-back.}
and subtract them using either NpC, \eq~(\ref{eq:npc-chs}), or the
area--median method, \eq~\eqref{eq:subtraction-base}, with $\rho$
estimated from the CHS event.
The hard events are generated with the underlying event turned off so
as to avoid subtleties related to the simultaneous subtraction of the
Underlying Event.

\begin{figure}
  \centering
  \includegraphics[width=0.48\textwidth]{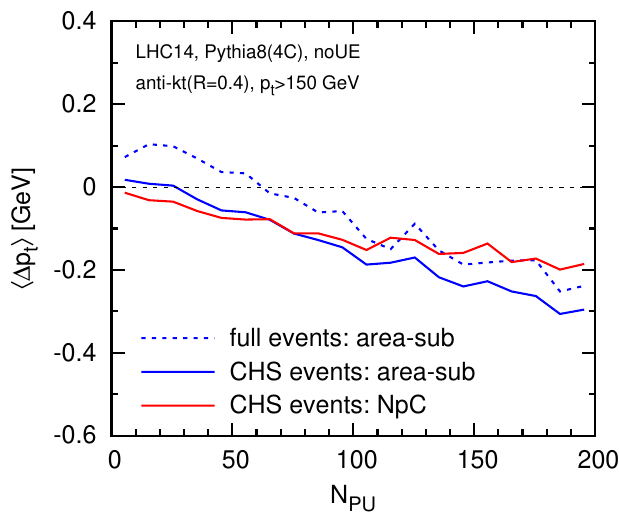}
  \hfill
  \includegraphics[width=0.48\textwidth]{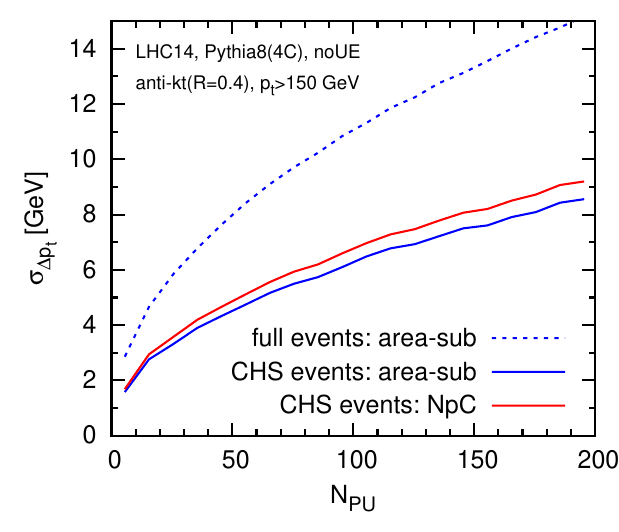}
  \caption{A comparison of the performance of the NpC and area--median 
    subtraction methods. 
    The left-hand plot shows, as a function of the
    number of pileup vertices $N_\text{PU}$, the average
    difference in $p_t$
    between a jet after pileup addition and subtraction and the
    corresponding matched jet in the hard sample, $\Delta p_t=p_t^{\rm
      jet,sub}-p_t^{\rm jet,hard}$. 
    The right-hand plot shows the standard deviation of $\Delta p_t$
    (lower values are better).
    NpC is shown only for CHS events, while area--median subtraction
    is shown both for events with CHS and for events without it
    (``full''). }
  \label{fig:npc-performance}
\end{figure}

Figure~\ref{fig:npc-performance} provides the resulting comparison of
the performance of the NpC and area--median subtraction methods (the
latter in CHS and in full events).
The left-hand plot shows the average difference between the subtracted
jet $p_t$ and the $p_t$ of the corresponding matched hard jet, as a
function of the number of pileup interactions.
Both methods clearly perform well here, with the average difference
systematically well below $1\GeV$ even for very high pileup levels.
The right-hand plot shows the standard deviation of the difference
between the hard and subtracted full jet $p_t$. 
A lower value indicates better performance, and one sees that in CHS
events the area--median method indeed appears to have a small, but
consistent advantage over NpC.
Comparing area--median subtraction in CHS and full events, one
observes a significant degradation in resolution when one fails to use
the available information about charged particles in correcting the
charged component of pileup, as is to be expected for a particle-level
study.

The conclusion of this section is that the NpC method fails to give a
superior performance to the area--median method in CHS events. This is
because the local correlations of neutral and charged energy flow are
no greater than the correlations between local neutral energy flow and
the global energy flow.
We believe that part of the reason for this is that the hadronisation
process for low $p_t$ particles intrinsically tends to produce hadrons
separated by large angles, as illustrated concretely in the case of
$\rho^\pm$ resonance decay.

\subsection{Performance of cleansing v. area--median}
\label{sec:appraisal}

\begin{figure}[pt]
  \centering
  \begin{minipage}{0.48\linewidth}
    \includegraphics[width=\textwidth]{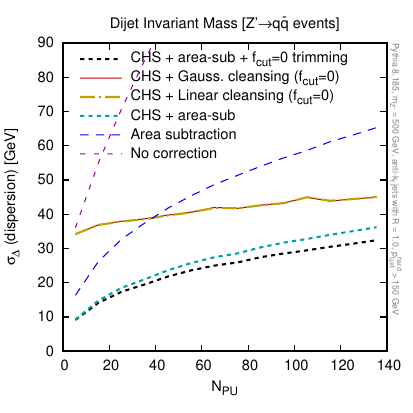}
  \end{minipage}
  \hfill
  \begin{minipage}{0.48\linewidth}
    \includegraphics[width=\textwidth]{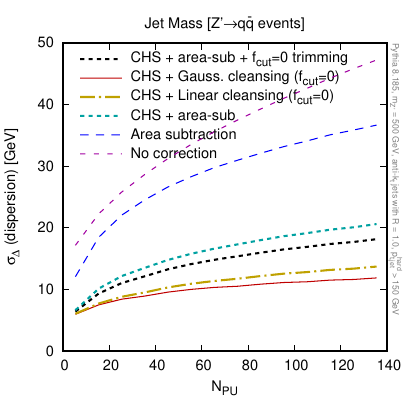}%
  \end{minipage}%
  \\
  \begin{minipage}{0.48\linewidth}
    \includegraphics[width=\textwidth]{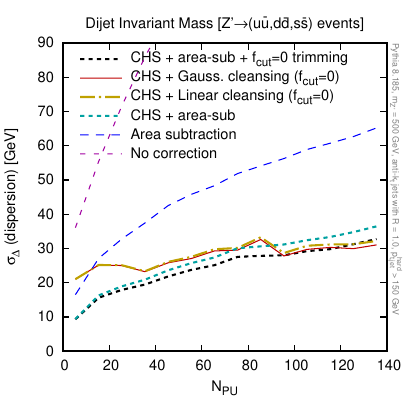}
  \end{minipage}
  \hfill
  \begin{minipage}{0.48\linewidth}
    \caption{ %
      Upper left plot: the dispersion of
      $\Delta m_{jj}=m_{jj}^{\text{sub}}-m_{jj}^{\text{hard}}$ after
      adding pileup and applying various pileup-removal methods; shown
      as a function of the number of pileup events, $N_\text{PU}$.
      Upper right plot: similarly for the single jet mass.
      Both plots are for a hadronically decaying $Z'$ sample with $m_{Z'} =
      500\GeV$.
      Decays are included to all flavours except $t\bar t$ and
      $B$-hadrons are taken stable. 
      Lower-left plot: the dispersion of $\Delta m_{jj}$, as
      in the upper-left plot, but with a sample of $Z'$ bosons that decay
      only to $u$, $d$ and $s$ quarks.
      Jets are reconstructed, as in Ref.~\cite{Krohn:2013lba}, with
      the anti-$k_t$ algorithm with $R=1$. 
      For both trimming and cleansing, subjets are reconstructed with
      the $k_t$ algorithm with $R_\text{sub}=0.3$ and the
      $f_\cut$ value that is applied is $f_\cut=0$.
      \label{fig:f00}
    }
  \end{minipage}
\end{figure}

We now turn to a broader discussion, comparing our discussion about the
NpC and area--median approaches with the results found using Jet
Cleansing.

The top left-hand plot of Fig.~\ref{fig:f00} shows the dispersion
$\Delta m_{jj}=m_{jj}^{\text{sub}}-m_{jj}^{\text{hard}}$ for the dijet
mass for several pileup mitigation methods.
%
The results are shown as a function of $N_\PU$.
The pileup mitigation methods include two forms of cleansing (with
$f_\cut=0$), area--median subtraction, CHS+area subtraction, and
CHS+area subtraction in conjunction with trimming (also with
$f_\cut=0$).
The top right-hand plots shows the corresponding results for the jet mass.
For the dijet mass we see that linear (and Gaussian) cleansing
performs worse than area subtraction, while in the right-hand plot,
for the jet mass, we
see linear (and Gaussian)
cleansing performing better than area subtraction, albeit not to the
extent found in Ref.~\cite{Krohn:2013lba}.
These (and, unless explicitly stated, our other $Z'$ results) have
been generated with the $Z'$ decaying to all flavours except
$t\bar t$, and $B$-hadrons have been kept stable.\footnote{%
  We often find this to be useful for particle-level $b$-tagging
  studies.
  Experimentally, in the future, one might even imagine an
  ``idealised'' form of particle flow that attempts to reconstruct
  $B$-hadrons (or at least their charged part) from displaced tracks
  before jet clustering. }
The lower plot shows the dijet mass for a different $Z'$ sample, one
that decays only to $u$, $d$ and $s$ quarks, but not $c$ and $b$
quarks. 
Most of the results are essentially unchanged. 
The exception is cleansing, which turns out to be very sensitive to
the sample choice.
Without stable $B$-hadrons in the sample, its performance improves
noticeably and at high pileup becomes comparable to that of
area-subtraction.\footnote{Note that %
Both of the left-hand plots in Fig.~\ref{fig:f00} differ
noticeably from the conclusion drawn from Fig.~6 (top) of
Ref.~\cite{Krohn:2013lba} (arXiv v2). In particular the former do not
show the same performance improvement that the latter for
the dijet mass with cleansing relative to area+CHS subtraction.
Fig.~\ref{fig:f00} also shows that, as one might have naively expected,
CHS  significantly reduces the impact of pileup.}

In light of our results on NpC in section~\ref{sec:npc-v-areamed}, the
difference between the performance of area-subtraction plus trimming
versus that of cleansing are puzzling: our expectation is that their
performances should be similar.\footnote{Note that our observations
  for area--median (with and without CHS) are consistent with a
  dispersion increasing like $\sqrt{\nPU}$, with the switch from full
  events to CHS events having the effect of reducing the coefficient in
  front of $\sqrt{N_\PU}$.}
The strong sample-dependence of the cleansing performance also calls
for an explanation.
We thus pursue with an in-depth study of the question.

As mentioned in our description of cleansing in
Section~\ref{sec:pusub-with-chargedinfo}, one additional
characteristic of linear cleansing relative to area-subtraction is
that it switches to jet-vertex-fraction (JVF) cleansing when the
NpC-style rescaling would give a negative answer.
In contrast, area-subtraction plus trimming simply sets the (sub)jet
momentum to zero.
We explicitly tried turning the switch to JVF-cleansing on and off and
found it had a small effect and did not explain the differences
observed in Fig.~\ref{fig:f00}.

The next difference that we will discuss at length is the zeroing step
that we introduced in Section~\ref{sec:pusub-with-chargedinfo}.
Zeroing turns out to be crucial: if we use it in conjunction with CHS
area-subtraction (or with NpC subtraction) and $f_\cut=0$ trimming, we
obtain results that are very similar to those from cleansing.
Conversely, if we turn this step off in linear-cleansing, its results come into
accord with those from (CHS) area-subtraction or NpC-subtraction with
$f_\cut=0$ trimming.

\begin{figure}[tp]
  \centering
  \includegraphics[width=0.48\textwidth]{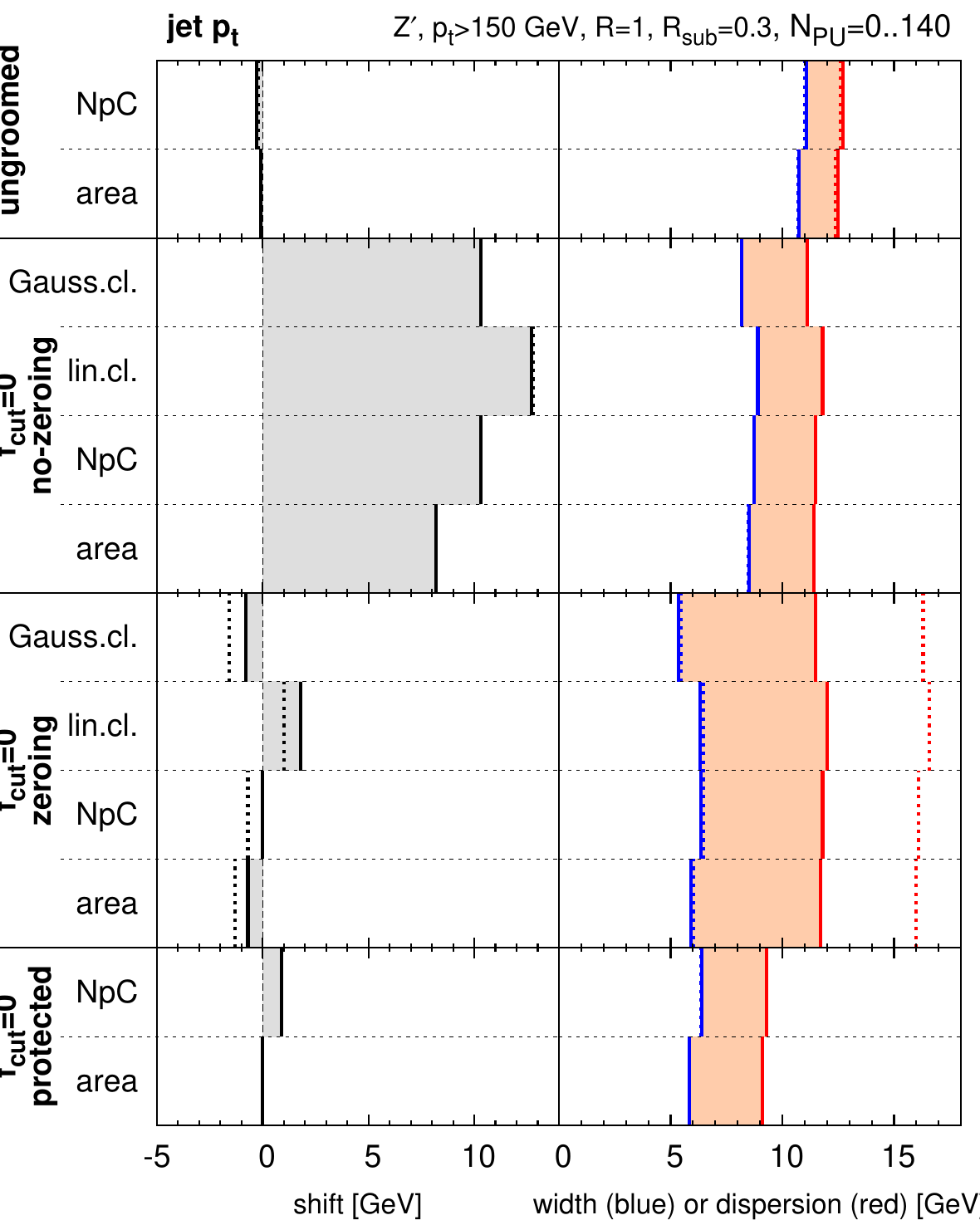}
  \hfill
  \includegraphics[width=0.48\textwidth]{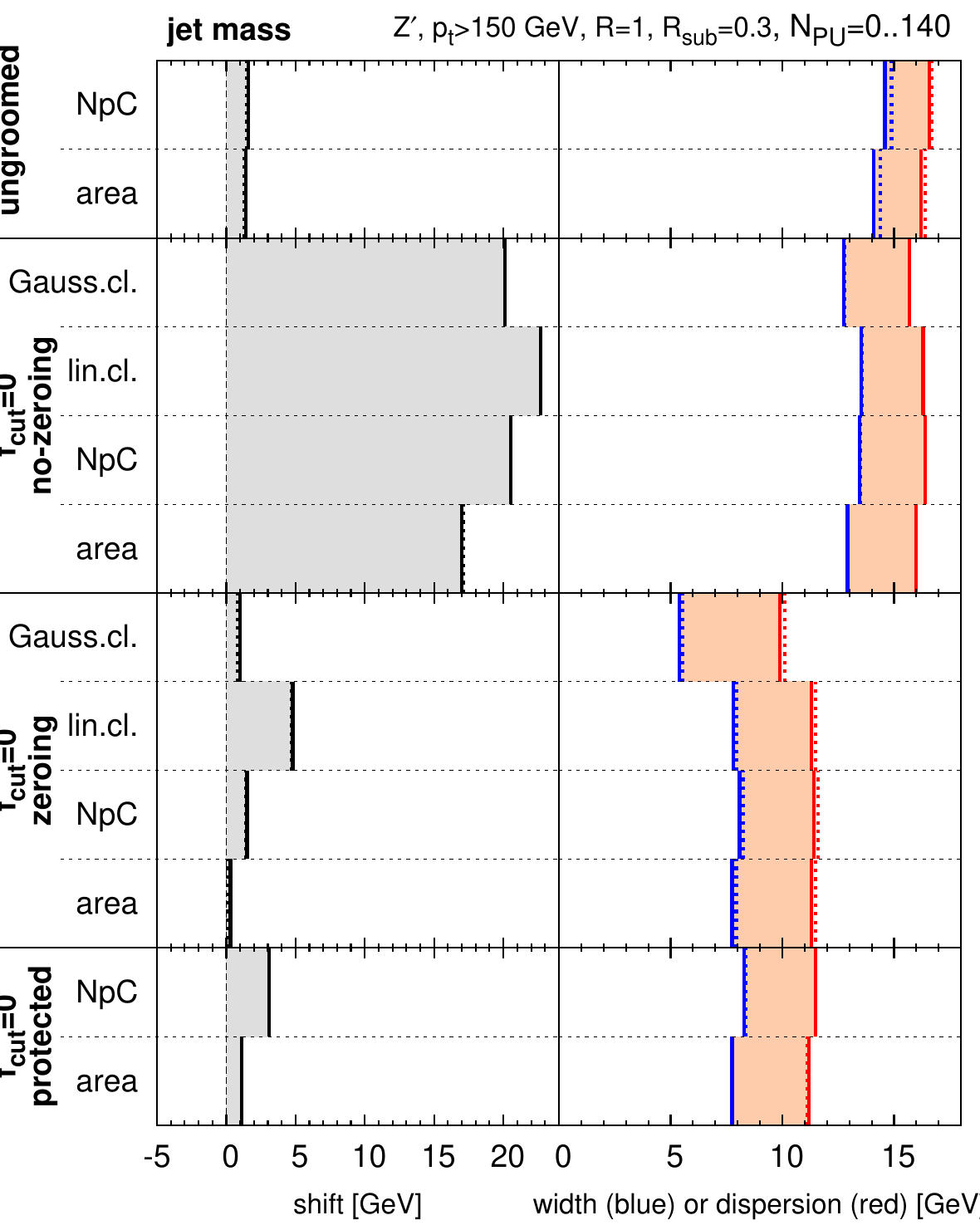}
  \caption{The left-hand plot illustrates various characteristics of the
    change ($\Delta p_t$) in the jet $p_t$ after addition of pileup and removal
    by a range of methods.
    It shows the average shift $\langle \Delta p_t
    \rangle$ (in black) and the peak width (in blue) and dispersion
    (in red) of the $\Delta p_t$ distribution. 
    The peak width is defined as the smallest window of $\Delta p_t$ that
    contains 90\% of the $\Delta p_t$ distribution, scaled by a
    factor $\simeq 0.304$
    such that in the case of a Gaussian distribution the result agrees
    with the dispersion.
    The right-hand plot shows the same set of results for the jet mass.
    The results are obtained in a sample of events with the number of
    pileup vertices distributed uniformly between $0$ and $140$.
    The hard events consist of hadronic $Z'$ decays: for the solid
    vertical lines the sample is $Z'\to d\bar d, u\bar u, s\bar s$,
    while for the dotted lines (sometimes not visible because directly
    over the solid lines), the sample
    additionally includes $Z'\to
    c\bar c, b\bar b$ with $B$ hadrons kept stable.
    The $Z'$ mass is $m_{Z'}=500\GeV$
    and jets are reconstructed with the anti-$k_t$ algorithm with $R=1$.
    All results in this figure include charged-hadron subtraction by default.
    The default form of cleansing, as used \eg in Fig.~\ref{fig:f00},
    is ``$f_\cut=0$ zeroing''.  
  }
  \label{fig:shifts-dispersions-R1}
\end{figure}

To help illustrate this, Fig.~\ref{fig:shifts-dispersions-R1} shows a
``fingerprint'' for each of several pileup-removal methods, for both
the jet $p_t$ (left) and mass (right).
The fingerprint includes the average shift
($\langle \Delta p_t\rangle$ or $\langle \Delta m\rangle$) of the
observable after pileup removal, shown in black.
It also includes two measures of the width of the $\Delta p_t$ and
$\Delta m$ distributions: the dispersion (\ie standard deviation) in
red and an alternative peak-width measure in blue. 
The latter is defined as follows: one determines the width of the
smallest window that contains $90\%$ of the entries and then scales
this width by a factor $0.304$.
For a Gaussian distribution, the rescaling ensures that the resulting
peak-width measure is equal to the dispersion.
For a non-Gaussian distribution the two measures usually differ and
the orange shaded region quantifies the extent of this difference. 
The solid black, blue and red lines have been obtained from samples in
which the $Z'$ decays just to light quarks; the dotted lines are for a
sample including $c\bar c$ and $b\bar b$ decays (with stable
$B$-hadrons), providing an indication of the sample dependence; in
many cases they are indistinguishable from the solid lines.

Comparing $f_\cut=0$ grooming for NpC, area (without zeroing) and cleansing
with zeroing manually disabled, all have very similar fingerprints.
Turning on zeroing in the different methods leads to a significant
change in the fingerprints, but again NpC, area and cleansing are very
similar.\footnote{One exception is that for the jet mass, Gaussian
  cleansing does differ in the $f_\cut=0$ case with zeroing, and shows
  an advantage from its combinations of different constraints on
  subjet momenta.  
  Unfortunately, as we shall see this later, this advantage does not
  seem to carry over to jet masses with $f_\cut \neq 0$ trimming, which are
  phenomenologically more relevant than the full jet mass.
}

When used with $f_\cut=0$ trimming, and when examining quality
measures such as the dispersion (in red, or the closely related correlation
coefficient, cf.\ Appendix~\ref{app:correlation-coefs}), subjet
zeroing appears to be advantageous for the jet mass, but potentially
problematic for the jet $p_t$ and the dijet mass.
However, the dispersion quality measure does not tell the full story
regarding the impact of zeroing.
Examining simultaneously the peak-width measure (in blue) makes it
easier to disentangle two different effects of zeroing.
On one hand we find that zeroing correctly rejects subjets that are
entirely due to fluctuations of the pileup.
This narrows the peak of the $\Delta p_t$ or $\Delta m$ distribution,
substantially reducing the (blue) peak-width measures in
Fig.~\ref{fig:shifts-dispersions-R1}.
On the other hand, zeroing sometimes incorrectly rejects subjets that
have no charged tracks from the LV but do have significant
neutral energy flow from the LV.
This can lead to long tails for the $\Delta p_t$ or $\Delta m$
distributions, adversely affecting the dispersion.\footnote{A
  discrepancy between dispersion and peak-width measures is to be seen
  in Fig.~15 of Ref.~\cite{CMS:2014ata} for jet masses. Our
  ``fingerprint'' plot is in part inspired by the representation
  provided there, though our choice of peak-width measure differs.}
It is the interplay between the narrower peak and the longer tails
that affects whether overall the dispersion goes up or down with
zeroing.
In particular the tails appear to matter more for the jet $p_t$ and
dijet mass than they do for the single-jet mass.
Note that accurate Monte Carlo simulation of such tails may be quite
challenging: they appear to be associated with configurations where a
subjet contains an unusually small number of energetic neutral
particles. 
Such configurations are similar to those that give rise to fake
isolated photons or leptons and that are widely known to be difficult
to simulate correctly.

We commented earlier that the cleansing performance has a significant
sample dependence.
This is directly related to the zeroing: indeed
Fig.~\ref{fig:shifts-dispersions-R1} shows that for cleansing without
zeroing, the sample dependence (dashed versus solid lines) vanishes,
while it is substantial with zeroing.
Our understanding of this feature is that the lower multiplicity of
jets with undecayed $B$-hadrons (and related hard fragmentation of the
$B$-hadron) results in a higher likelihood that a subjet will contain
neutral but no charged particles from the LV, thus enhancing the
impact of zeroing on the tail of the $\Delta p_t$ or $\Delta m_{jj}$
sample.

The long tails produced by the zeroing can be avoided by using the
alternative {\em protected zeroing} described in
Section~\ref{sec:pusub-with-chargedinfo}.
Taking $n=10$ (or a fixed $p_{t,\text{min}} = 20\GeV$) we have found
reduced tails and, consequently, noticeable improvements in the jet
$p_t$ and dijet mass dispersion (with little effect for the jet mass).
This is visible for area and NpC subtraction in
Fig.~\ref{fig:shifts-dispersions-R1}.
Protected zeroing also eliminates the sample dependence.%
\footnote{%
  Note that for the identification of pileup v.\ non-pileup full jets
  in ATLAS, some form of protection is also used, in that JVF-type
  conditions are not applied if $p_t > 50 \GeV$.  We thank David
  Miller for exchanges on this point.
}

Several additional comments can be made about $f_\cut=0$ trimming combined with
zeroing.
Firstly, $f_\cut=0$ trimming alone introduces a bias in the jet
$p_t$, which is clearly visible in the $f_\cut=0$ no-zeroing shifts in
Fig.~\ref{fig:shifts-dispersions-R1}. 
This is because the trimming removes negative fluctuations of the
pileup, but keeps the positive fluctuations.
Zeroing then counteracts that bias by removing some of the positive
fluctuations, those that happened not to have any charged tracks from
the LV.
It also introduces further negative fluctuations for subjets that
happened to have some neutral energy flow but no charged tracks.
Overall, one sees that the final net bias comes out to be relatively
small.
This kind of cancellation between different biases is common in
noise-reducing pileup-reduction
approaches~\cite{Kodolova:1998laa,Gavrilov:2003pza,Kodolova:2007hd,Cacciari:2014gra,Bertolini:2014bba}
as we have already seen in Chapter~\ref{chap:beyond-grooming} and will
see again in the next two Chapters of this review.

\begin{figure}[tp]
  \centering
  \begin{minipage}{0.48\linewidth}
    \includegraphics[width=\textwidth]{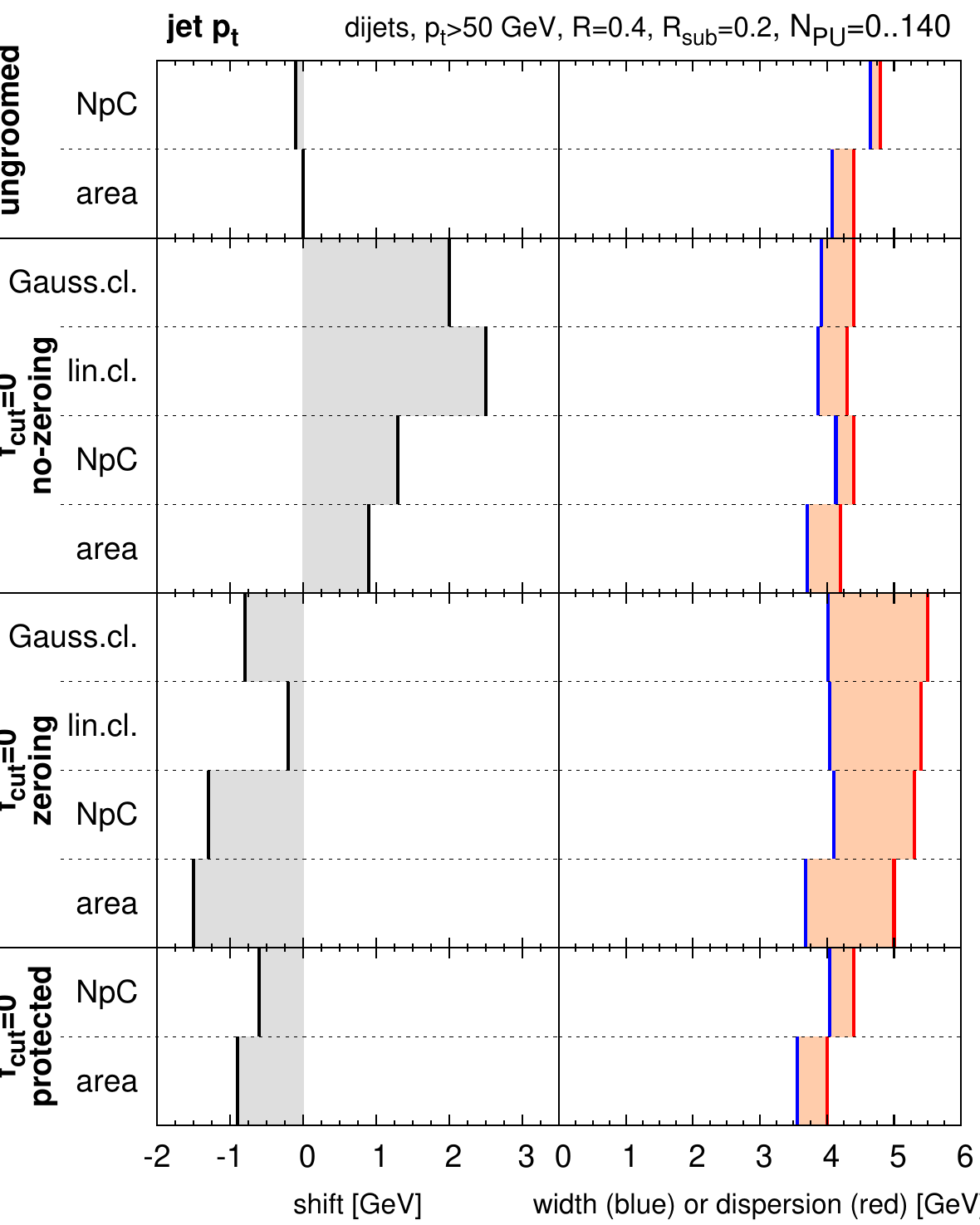}
  \end{minipage}
  \hfill
  \begin{minipage}{0.48\linewidth}
    \includegraphics[width=\textwidth]{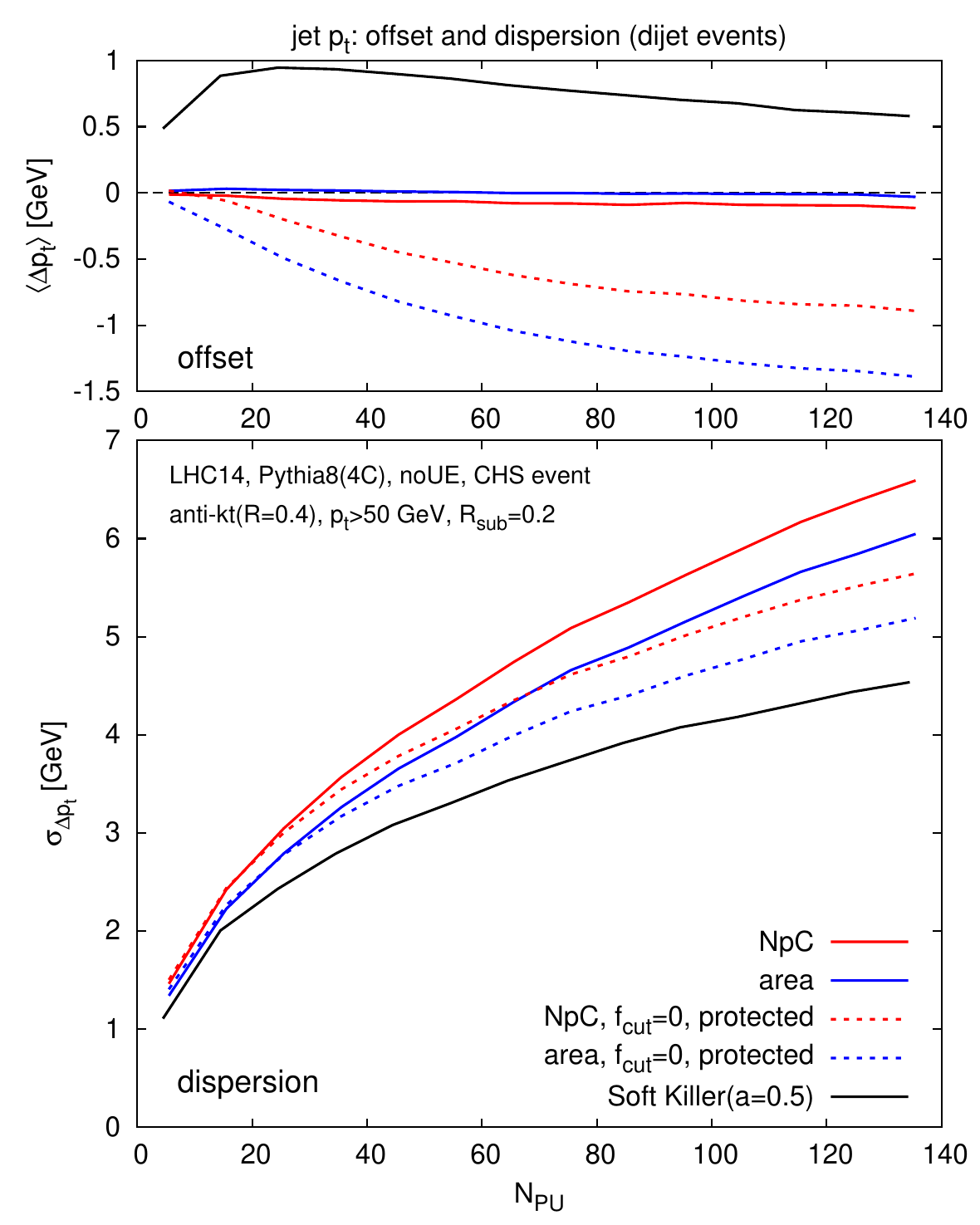}
  \end{minipage}
  \caption{
    Left, Analogue of Fig.~\ref{fig:shifts-dispersions-R1} (left) for a
    jet radius of $R=0.4$, subjet radius (where relevant) of
    $R_\text{sub}=0.2$ and a QCD continuum dijet sample generated with Pythia~8.
    The underlying event is turned off in the sample and $B$ hadrons
    decay. 
    We consider only jets that in the hard sample have $p_t > 50\GeV$
    and $|y|< 2.5$.
    Right: the dispersions for a subset of the methods, shown as a function
    of the number of pileup events.
    \label{fig:shifts-dispersions-dijet-R0.4}
  }
\end{figure}

\begin{figure}[tp]
  \centering
  \begin{minipage}{0.48\linewidth}
    \includegraphics[width=\textwidth]{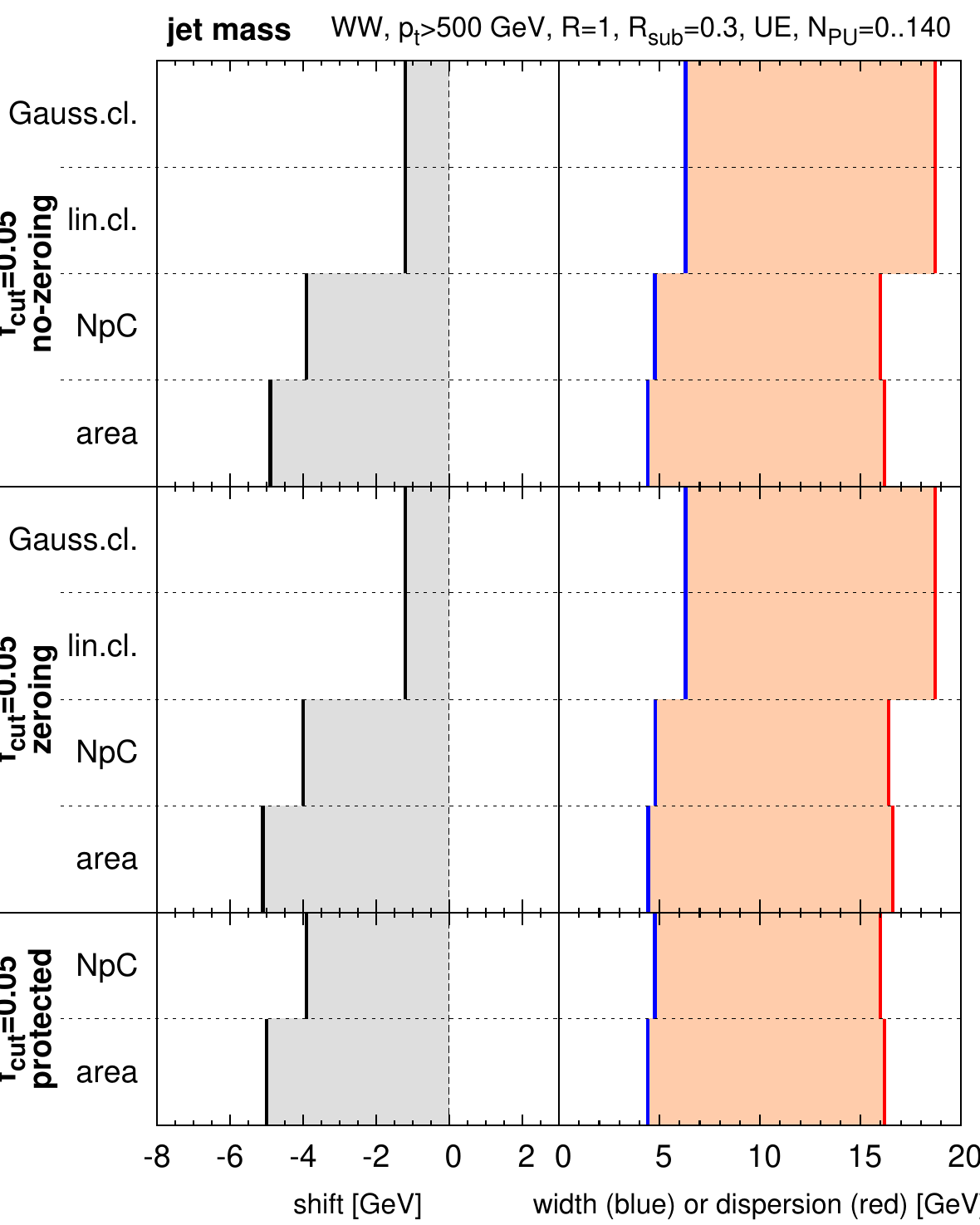}
  \end{minipage}
  \hfill
  \begin{minipage}{0.48\linewidth}
    \caption{
      Analogue of Fig.~\ref{fig:shifts-dispersions-R1} (right),
      showing the performance for the jet mass, but
      now with $f_\cut=0.05$ applied to both trimming and cleansing and in
      a sample of hadronically-decaying boosted $W$ bosons ($pp \to
      W^+W^-$).
      The jets reconstructed after addition and subtraction of pileup
      are compared to trimmed hard jets.
      Jets are reconstructed with a jet radius of $R=1$ and a subjet
      radius of $R_\text{sub} = 0.3$. 
      Only hard jets with $p_t > 500\GeV$ and $|y| < 2.5$ (before
      trimming) are considered and we let $B$ hadrons decay.
      \label{fig:shifts-dispersions-WW500-R10}
    }
  \end{minipage}
\end{figure}

Most of the studies so far in this section have been carried out with a setup
that is similar to that of Ref.~\cite{Krohn:2013lba}, \ie $R=1$ jets
in a $Z'$ sample with $f_\cut=0$ trimming.
This is not a common setup for most practical applications.
For most uses of jets, $R=0.4$ is a standard choice and pileup is
at its most severe at low to moderate $p_t$.
Accordingly, in Fig.~\ref{fig:shifts-dispersions-dijet-R0.4} (left) we show
the analogue of Fig.~\ref{fig:shifts-dispersions-R1}'s summary for the
jet $p_t$, but now for $R=0.4$, with $R_\text{sub}=0.2$ in a QCD dijet
sample, considering jets that in the hard event had $p_t > 50\GeV$.
We see that qualitatively the pattern is quite similar to that in
Fig.~\ref{fig:shifts-dispersions-R1}, with the area--median approach
performing slightly better than NpC-based techniques as expected from
our discussion in Section~\ref{sec:npc-v-areamed}.\footnote{Note,
  however, that at even lower jet $p_t$'s, the difference between
  zeroing and protected zeroing might be expected to disappear. This
  is because the long negative tails are suppressed by the low jet
  $p_t$ itself.}
Quantitatively, the difference between the various choices is much
smaller, with about a $10\%$ reduction in dispersion (or width) in going
from ungroomed CHS area-subtraction to the $f_\cut=0$ protected
subjet-zeroing case.
One should be aware that this study is only for a single $p_t$, across
a broad range of pileup.
The dispersions for a subset of the methods are shown as a function of
the number of pileup vertices in the right-hand plot of
Fig.~\ref{fig:shifts-dispersions-dijet-R0.4}.
That plot also includes results from the SoftKiller method discussed
in Chapter~\ref{chap:soft-killer} (see also~\cite{Cacciari:2014gra})
and illustrates that the benefit from protected zeroing (comparing the
solid and dashed blue curves) is about half of the benefit that is
brought from SoftKiller (comparing solid blue and black curves).
These plots show that protected zeroing is potentially of interest for
jet $p_t$ determinations in realistic conditions. 
Thus it would probably benefit from further study: one should, for
example, check its behaviour across a range of transverse momenta,
determine optimal choices for the protection of the zeroing and
investigate also how best to combine it with particle-level
subtraction methods such as SoftKiller.\footnote{An interesting
  feature of protected zeroing, SoftKiller and another
  recently introduced method, PUPPI~\cite{Bertolini:2014bba}, is that
  the residual degradation in resolution from pileup appears to scale
  more slowly than the $\sqrt{N_\PU}$ pattern that is observed for
  area and NpC subtraction alone.} We will come back to this
combination in Section~\ref{sec:sk-improvements}.

Turning now to jet masses, the use of $R=1$ is a not uncommon choice,
however most applications use a groomed jet mass with a non-zero
$f_\cut$ (or its equivalent): this improves mass resolution in the
hard event even without pileup, and it also reduces backgrounds,
changing the perturbative structure of the
jet~\cite{Dasgupta:2013ihk,Dasgupta:2013via} even in the absence of
pileup.\footnote{In contrast, for $f_\cut=0$ trimming, the jet
  structure is unchanged in the absence of pileup.}
Accordingly in Fig.~\ref{fig:shifts-dispersions-WW500-R10} we show
$f_\cut=0.05$ results (with shifts and widths computed relative to
$f_\cut=0.05$ trimmed hard jets) for a hard $WW$ sample where the hard
fat jets are required to have $p_t > 500\GeV$. 
Zeroing, whether protected or not, appears to have little impact. 
One potential explanation for this fact is as follows: zeroing's
benefit comes primarily because it rejects fairly low-$p_t$ pileup
subjets that happen to have no charged particles from the leading
vertex.
However for a pileup subjet to pass the $f_\cut=0.05$ filtering
criterion in our sample, it would have to have $p_t > 25\GeV$.
This is quite rare.
Thus filtering is already removing the pileup subjets, with little
further to be gained from the charged-based zeroing.
As in the plain jet-mass summary plot, protection of zeroing appears
to have little impact for the trimmed jet mass.\footnote{%
  \label{footnote:trimming-ref}%
  Cleansing
  appears to perform slightly worse than trimming with NpC or area
  subtraction. 
  One difference in behaviour that might explain this is that the
  $p_t$ threshold for cleansing's trimming step is
  $f_\text{cut} p_t^\text{full,no-CHS}$ (even in the CHS-like
  \texttt{input\_nc\_separate} mode that we use).
  In contrast, for the area and NpC-based results, it is
  $f_\text{cut} p_t^\text{full,CHS}$.
  In both cases the threshold, which is applied to subtracted subjets, is
  increased in the presence of pileup, but this increase is more
  substantial in the cleansing case.
  This could conceivably worsen the correspondence between trimming in
  the hard and full samples.
  For the area and NpC cases, we investigated the option of using
  $f_\text{cut} p_t^\text{area-sub,CHS}$ or
  $f_\text{cut} p_t^\text{NpC-sub,CHS}$ and found that this brings a
  small additional benefit. 
}
Does that mean that (protected) zeroing has no scope for improving the
trimmed-jet mass?
The answer is ``not necessarily'': one could for example imagine
first applying protected zeroing to subjets on some angular scale
$R_\text{zero}$ in order to eliminate low-$p_t$ contamination;
then reclustering the remaining constituents on a scale $R_\text{trim}
\gtrsim R_\text{zero}$, subtracting according to the area or NpC
methods, and finally applying the trimming momentum cut (while also
keeping in mind the considerations of footnote
\ref{footnote:trimming-ref}).
In that context, it would also be interesting to consider the
combination of zeroing with the grooming techniques suggested as
pileup mitigation methods in Chapter~\ref{chap:beyond-grooming}.

Further systematic comparisons of the different methods, with and
without zeroing can be found in the Appendices of Ref.~\cite{Cacciari:2014jta}.

We close this section with a summary of our findings.
Based on its description in Ref.~\cite{Krohn:2013lba} and our findings
about NpC v.\ area subtraction, cleansing with $f_\cut=0$ would be
expected to have a performance very similar to that of CHS+area
subtraction with $f_\cut=0$ trimming.\footnote{We note that
  Ref.~\cite{Krohn:2013lba} reports large improvements for the
  correlation coefficients of the dijet mass and the single jet mass
  using $R=1$ jets, while only an improvement for the jet mass is
  seen in the results presented here.}
The differences in behaviour between (linear) cleansing and trimmed
CHS+area-subtraction therefore call for an explanation, and appear to
be due to a step in the cleansing code that was undocumented in
Ref.~\cite{Krohn:2013lba} and that we dubbed ``zeroing'': if a subjet
contains no charged tracks from the leading vertex it is discarded.
Zeroing is an extreme form of a procedure described in
Ref.~\cite{ATL-Pileup-2}.
In can be used also with area or NpC subtraction, and we find that it
brings a benefit for the peak of the $\Delta p_t$ and $\Delta m$
distributions, but appears to introduce long tails in $\Delta p_t$.
A variant, ``protected zeroing'', can avoid the long tails by still
accepting subjets without leading-vertex tracks, if their $p_t$ is
above some threshold, which may be chosen dynamically based on the
properties of the pileup.
In our opinion, a phenomenologically realistic estimate of the benefit
of zeroing (protected or not) requires study not of $R=1$ plain jets,
but instead of $R=0.4$ jets (for the jet $p_t$) or larger-$R$ trimmed
jets with a non-zero $f_\cut$ (for the jet mass).
In a first investigation, there appear to be some phenomenological
benefits from protected zeroing for the $R=0.4$ jet $p_t$, whereas to
obtain benefits for large-$R$ trimmed jets would probably require
further adaptation of the procedure.
In any case, additional study is required for a full evaluation of
protected zeroing and related procedures.

\section{Digression: combining different methods}

\begin{figure}[t]
  \centering
  \includegraphics[width=0.48\textwidth]{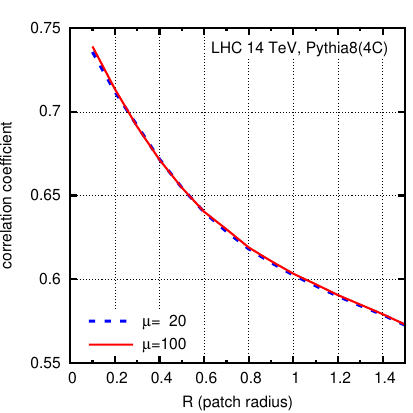}
  \hfill
  \includegraphics[width=0.48\textwidth]{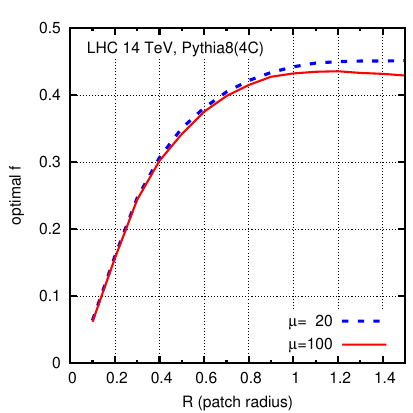}
  \caption{Left: correlation between $p_t^\ntr - \hat
    p_\mu^\text{ntr(NpC)}$ and $p_t^\ntr - \hat p_\mu^{\text{ntr(${\rho}
        A$)}}$, shown as a function of $R$. 
    Right: optimal weight $f$ for combining NpC and area pileup
    subtraction, \eq~(\ref{eq:best f}), as a function of $R$.
  }
  \label{fig:correl-mistakes}
\end{figure}

\begin{figure}
  \centering
  \includegraphics[width=0.48\textwidth]{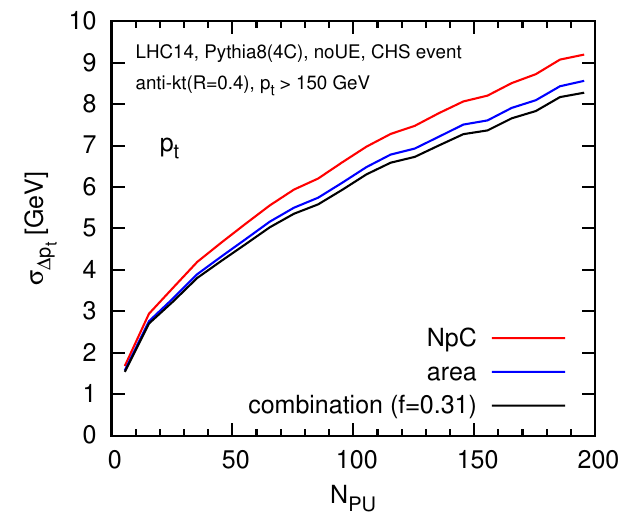}
  \hfill
  \includegraphics[width=0.48\textwidth]{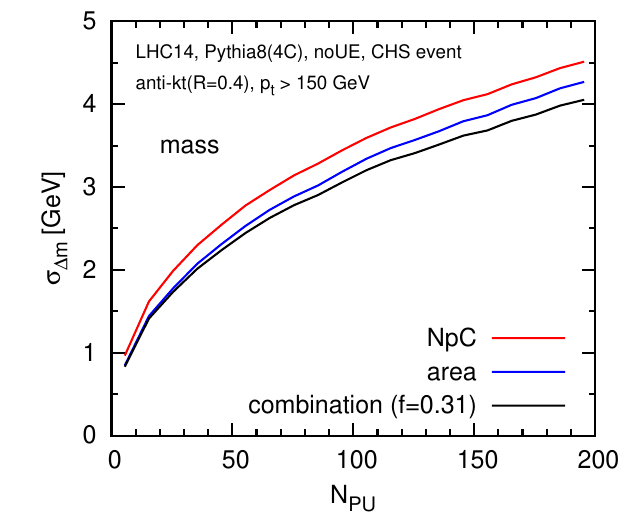}
  \caption{Comparison of the performance of NpC, area--median and
    combined subtraction, as a function of the number of pileup
    vertices. The left-hand plot is for the jet $p_t$ and the
    right-hand one for the jet mass.}
  \label{fig:chg+rhoA-improvement}
\end{figure}

It is interesting to further probe the relation between NpC and the
area--median method and to establish whether there might be a benefit
from combining them:
the area--median method makes a mistake in predicting local energy flow
mainly because local energy flow fluctuates from region to region;
NpC makes a mistake because charged and neutral energy flow are not
$100\%$ locally correlated.
The key question is whether, for a given jet, NpC and the area--median
method generally make the same mistake, or if instead they are making
uncorrelated mistakes.
In the latter case it should be possible to combine the information
from the two methods to obtain an improvement in subtraction
performance.

Let $p_t^\ntr$ be the actual neutral pileup component flowing into a
jet, while
\begin{equation}
  \label{eq:pthat}
  \hat p_\mu^\text{ntr(NpC)} = \frac{1-\gamma_0}{\gamma_0 \epsilon}
  p_\mu^\text{jet,rescaled-chg-PU} \,, 
  \qquad 
  \hat p_\mu^{\text{ntr(${\rho} A$)}} = {\rho} A\,,
\end{equation}
are, respectively, the estimates for the neutral pileup based on the
local charged $p_t$ flow and on ${\rho} A$. 
Note that we for our analysis in this section we use CHS events and in
particular ${\rho}$ is as determined from the CHS event, \ie 
$\rho\equiv\rho_\text{CHS}$.
This means that it is essentially equal to $\rho_\ntr$, with small
differences arising principally due to the presence of the charged
component of the LV's underlying event contribution.
One could of course also use the actual $\rho_\ntr$ instead of
$\rho_\text{CHS}$ to estimate $\hat p_\mu^{\text{ntr(${\rho} A$)}}$,
and this would in some respects be a more coherent approach, though
the numerical differences will be small.

Concentrating on the transverse components, the extent to which the
two estimates in \eq~(\ref{eq:pthat}) provide complementary
information can be quantified in terms of (one minus) the correlation
coefficient, $r$, between $p_t^\ntr - \hat p_\mu^\text{ntr(NpC)}$ and
$p_t^\ntr - \hat p_\mu^{\text{ntr(${\rho} A$)}}$.
That correlation is shown as a function of $R$ in
Fig.~\ref{fig:correl-mistakes} (left), and it is quite high, in the range
$0.6$--$0.7$ for commonly used $R$ choices. 
It is largely independent of the number of pileup vertices.

Let us now quantify the gain to be had from a linear combination of
the two prediction methods, \ie using an estimate
\begin{equation}
  \label{eq:NpC+rhoA}
  \hat p_{\mu}^\text{ntr} = f \hat p_\mu^\text{ntr(NpC)} + (1-f) \hat p_\mu^{\text{ntr(${\rho} A$)}}\,,
\end{equation}
where $f$ is to be chosen to as to minimise the dispersion of
$p_t^\ntr - \hat p_t^\text{ntr}$.
Given dispersions $\sigma_\text{NpC}$ and $
\sigma_{{\rho} A}$
respectively for $p_t^\ntr - \hat p_\mu^\text{ntr(NpC)}$ and $p_t^\ntr - \hat
p_\mu^{\text{ntr(${\rho} A$)}}$, we can show that the optimal $f$ is
\begin{equation}
  \label{eq:best f}
  f = \frac{\sigma_{{\rho} A}^2 - r \, \sigma_{\NpC} \, \sigma_{{\rho}
      A}}{\sigma_{\NpC}^2 +  \sigma_{{\rho} A}^2 - 2 r \,\sigma_{\NpC} \, \sigma_{{\rho} A}}\,,
\end{equation}
which is plotted as a function of $R$ in
Fig.~\ref{fig:correl-mistakes} (right). The resulting squared
dispersion for $p_t^\ntr - \hat p_t^\text{ntr}$ is
\begin{equation}
  \label{eq:combined-dispersion}
  \sigma^2
  =
  \frac{(1-r^2) \, \sigma_{\NpC}^2\, \sigma_{{\rho}
      A}^2}{\sigma_{\NpC}^2 + \sigma_{\rho A}^2 - 2 r\, \sigma_{\NpC}\, \sigma_{{\rho} A}}\,.
\end{equation}
Reading $r = 0.67$ from Fig.~\ref{fig:correl-mistakes} (left) for
$R=0.4$, and $ \sigma_{\NpC} \simeq 1.14\, \sigma_{{\rho} A}$ from
Fig.~\ref{fig:correl-central} (right), one finds $f\simeq 0.31$ and
$\sigma \simeq 0.96 \, \sigma_{{\rho} A}$.
Because of the substantial correlation between the two methods, ones
expects only a modest gain from their linear combination.

In Fig.~\ref{fig:chg+rhoA-improvement} we compare the performance of
pileup subtraction from the combination of the NpC and the
area--median methods, using the optimal value $f=0.31$ that can be
read from Fig.~\ref{fig:correl-mistakes} (right) for $R=0.4$, both for
the jet $p_t$ and the jet mass.
The expected small gain is indeed observed for the jet $p_t$, and it
is slightly larger for the jet mass.\footnote{We also examined
  results with other choices for $f$: we found that the true optimum
  value of $f$ in the Monte Carlo studies is slightly different from
  that predicted by \eq~(\ref{eq:best f}). However the dependence on
  $f$ around its minimum is very weak, rendering the details of its
  exact choice somewhat immaterial. }
Given the modest size of the gain, one may wonder how
phenomenologically relevant it is likely to be. 
Nevertheless, one might still consider investigating whether the gain
carries over also to a realistic experimental environment with full
detector effects.

Note finally that the approach described above can be applied to
combine a different pair of pileup mitigation techniques. It can also
easily be extended to more than two methods.

\section{Concluding remarks and discussions}\label{sec:npc-conclusions}

One natural approach to pileup subtraction is to use the charged
pileup particles in a given jet to estimate the amount of neutral
pileup that needs to be removed from that same jet.
In this chapter, with the help of particle-level simulations, we have
studied such a method (NpC) and found that it has a performance that
is similar to, though slightly worse than the existing, widely used
area--median method.
This can be related to the observation that the correlations between
local charged and neutral energy flow are no larger than those between
global and local energy flow.
Tentatively, we believe that this is in part because the non-perturbative
effects that characterise typical inelastic proton-proton collisions
act to destroy local charged-neutral correlation.

When studying cleansing, we have seen an improved performance for the
jet mass.\footnote{Note however that this study presented does not
  reproduce the large improvement for both the jet mass and the dijet
  mass seen in Ref.~\cite{Krohn:2013lba}.}
This improvement may come as a surprise since cleansing is largely
based on Npc.
We trace a key difference in the behaviour of cleansing and area
subtraction (or pure NpC) to the use in the cleansing code of a step
that was not documented in Ref.~\cite{Krohn:2013lba} and that discards
subjets that contain no tracks from the leading vertex.
This ``zeroing'' step, similar to the charged-track based trimming
introduced by ATLAS~\cite{ATL-Pileup-2}, can indeed be of benefit. 
It has a drawback of introducing tails in some distributions due to
subjets with a substantial neutral $p_t$ from the leading vertex, but
no charged tracks.
As a result, different quality measures (and different event samples)
lead to different conclusions as to the benefits of zeroing.
The tails can be alleviated by a variant of zeroing that we introduce
in Ref.~\cite{Cacciari:2014jta}, ``protected zeroing'', whereby subjets
without LV charged tracks are rejected only if their $p_t$ is below
some (possibly pileup-dependent) threshold.
Protected zeroing does in some cases appear to have phenomenological
benefits, which are observed across all quality
measures. 

Code for the above implementation of area subtraction with
positive-definite mass is available as part of FastJet versions 3.1.0
and higher (see Section~\ref{sec:areamed-implementation} and, in
particular, \ref{sec:BGE-masses}).
Public code and samples for carrying out a subset of the comparisons
with cleansing described in section~\ref{sec:appraisal}, including
also the NpC subtraction tools, are available from
Ref.~\cite{public-code}.

To conclude this Chapter, we want to point out that NpC and cleansing
are not the only ways to use charged-track information in the context
of jet clustering and pileup mitigation.
For example, one could simply imagine clustering jets using only the
charged tracks in the event, for which an almost perfect separation
between the leading vertex and pileup interactions can be
performed. One could then scale the result to mimic the inclusion of
the neutral contribution to the jets. In this context where we are
dominated by the high-$p_t$ dynamics of jets, we expect a
much larger correlation between charged and neutral particles than
what was seen in Section~\ref{sec:npc-v-areamed} for soft pileup
particles. Furthermore, for idealised CHS events for which tracks from
the leading vertex can be exactly identified, such a rescaling would be
independent of the pileup conditions.
However, this would result in fluctuations which scale with the $p_t$
of the jet rather than the, much softer, $\sigma\sqrt{A_\text{jet}}$
scale driving fluctuations in the area--median subtraction. Unless
$p_t$ is very small and \nPU very large this is therefore not
practical.\footnote{A quick estimate using the framework we will use
  for the summary study in Chapter~\ref{chap:beyond-prelimn} shows a
  dispersion of order 10~GeV jets with $p_t\ge 20$~GeV and of order
  175~GeV for jets above 500~GeV. This is (much) larger than the
  corresponding 6.5~GeV and 11~GeV found for the CHS+area--median
  subtraction in Chapter~\ref{chap:beyond-prelimn} for $\nPU=140$.}

Charged tracks could be used when designing more elaborate pileup
mitigation techniques which rely on additional kinematic properties of
the jets and of the event in general. If these kinematic properties
are the same (or very similar) for charged and neutral particles, one
could then obtain them directly from the data. The determination of
the ``$\alpha$ distribution'' used in the PUPPI method is an
illustration of this (see Fig.~1 of Ref.~\cite{PUPPI}).
Similarly, one could imagine determining the SoftKiller grid-size
parameter (see the next Chapter) based on an estimation using charged
tracks.\footnote{We thank the referee for this
  suggestion.}\footnote{Note that this is not totally trivial since,
  as we shall see in the next Chapter
  (Section~\ref{sec:adapt-calor-towers}), the SoftKiller grid-size
  parameter increases if we correct CHS events compared to full
  events. A determination based on charged tracks would then
  presumably have to be somehow corrected.}


\chapter{Noise reduction with the SoftKiller}\label{chap:soft-killer}

In this Chapter, we describe a new method that we introduced in
Ref.~\cite{Cacciari:2014gra} and that we called SoftKiller (SK).
This method acts as a noise-reduction method which has shown nice
resolution improvements compared to the area--median method in
Monte-Carlo simulations.

We will first describe the method itself in Section
\ref{sec:area-median-recall} followed by some validation in
Monte-Carlo studies (Section~\ref{sec:performance}) as
well as possible adaptations to specific situations like CHS events
and detector effects (Section~\ref{sec:adapt-calor-towers}).

One of the interesting features of the SoftKiller method is that it
shows remarkable performance in terms of computing time. We will
discuss that in Section~\ref{sec:computing-time}.

The work presented in this Chapter is a minor adaptation of the
original SoftKiller paper ~\cite{Cacciari:2014gra}. We will discuss in
the next Chapter, devoted to preliminary ideas and results, some
possible ways to further improve the SoftKiller method.

\section{The SoftKiller method}
\label{sec:area-median-recall}

The SoftKiller method involves eliminating particles below some $p_t$
cutoff, $\ptcut$, chosen to be the minimal value that ensures that $\rho$ is
zero.
Here, $\rho$ is the event-wide estimate of transverse-momentum flow
density in the area--median approach, \ie given by
\eq~(\ref{eq:rhoest-base}). In practice we will use a grid-base
estimate for $\rho$.\footnote{Note also that the estimation of $\rho$
  does not use any rapidity rescaling. Such a rapidity rescaling would
  actually have no effect on the determination of the SoftKiller $p_t$
  cut. This is further discussed at the end of Section~\ref{sec:performance}.}

\begin{figure}[t]
  \centering
  \includegraphics[width=0.48\textwidth]{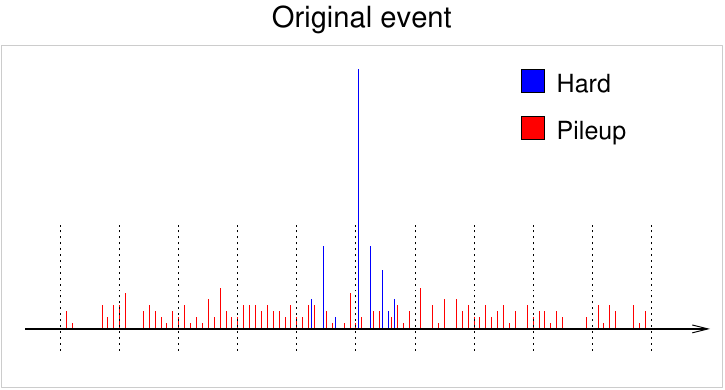}%
  \hfill
  \includegraphics[width=0.48\textwidth]{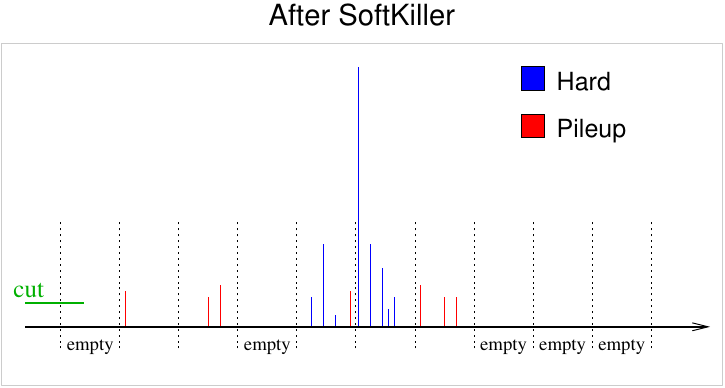}%
  \caption{Illustration of the SoftKiller method. The left plot
    depicts particles in an event, with the hard event particles shown in blue
    and the pileup particles shown in red. On the right, the same
    event after applying the SoftKiller. The vertical dotted lines
    represent the edges of the patches used to estimate the pileup
    density $\rho$.} 
  \label{fig:SK-illustration}
\end{figure}

Choosing the minimal transverse momentum threshold, $p_t^\cut$,
that results in $\rho = 0$ is equivalent to gradually raising the
$p_t$ threshold until exactly half of the patches contain no
particles, which ensures that the median is zero.
This is illustrated in Fig.~\ref{fig:SK-illustration}.
Computationally, $p_t^\cut$ is straightforward to evaluate: one
determines, for each patch $i$, the $p_t$ of the hardest particle in
that patch, $p_{ti}^{\max}$ and then $p_t^\text{cut}$ is given by the 
median of $p_{ti}^{\max}$ values:
\begin{equation}
  \label{eq:ptmin}
  p_t^\cut = \underset{i \in \text{patches}}{\text{median}} \left\{ p_{ti}^{\max} \right\}\,.
\end{equation}
With this choice, half the patches will contain only particles that
have $p_t < \ptcut$. 
These patches will be empty after application of the $p_t$ threshold,
leading to a zero result for $\rho$ as defined in
\eq~(\ref{eq:rhoest-base}).\footnote{Applying a $p_t$ threshold to
  individual particles is not collinear safe; in the specific context
  of pileup removal, we believe that this is not a significant issue,
  as we discuss in more detail in
  Appendix~\ref{app:sk-collinear-safety}. }
The computational time to evaluate $\ptcut$ as in \eq~(\ref{eq:ptmin})
scales linearly in the number of particles and the method should be
amenable to parallel implementation.

Imposing a cut on particles' transverse momenta eliminates most of the
pileup particles, and so might reduce the fluctuations in residual pileup
contamination from one point to the next within the event.
However, as with other event-wide noise-reducing pileup and
underlying-event mitigation approaches, notably the CMS heavy-ion
method~\cite{Kodolova:1998laa,Gavrilov:2003pza,Kodolova:2007hd} (cf.\
the analysis in Appendix A.4 of Ref.~\cite{Cacciari:2011tm}; see also
our introductory discussion in Chapter~\ref{chap:beyond-motivation}),
the price that one pays for noise reduction is the introduction of
biases.
Specifically, some particles from pileup will be above
$p_t^\cut$ and so remain to contaminate the jets, inducing a
net positive bias in the jet momenta.
Furthermore some particles in genuine hard jets will be lost, because
they are below the $p_t^\cut$, inducing a negative bias in the
jet momenta.
The jet energy scale will only be correctly reproduced if these two
kinds of bias are of similar size,\footnote{For patch areas that are
  similar to the typical jet area, this can be expected to happen
  because half the patches will contain residual pileup of order
  $\ptcut$, and since jets tend to have only a few low-$p_t$
  particles from the hard scatter, the loss will also be order of
  $\ptcut$.} so that they largely cancel.
There will be an improvement in the jet resolution if the fluctuations in
these biases are modest.

\begin{figure}
  \includegraphics[width=0.48\textwidth]{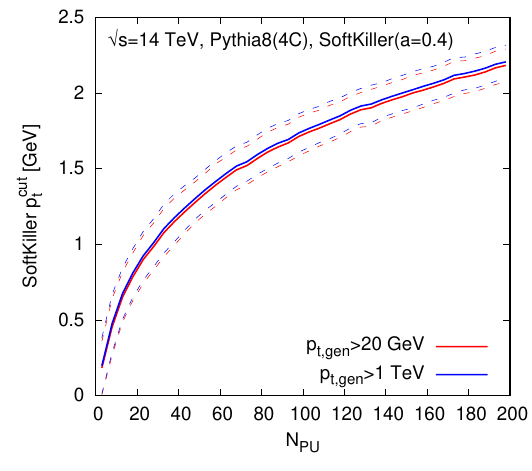}
  \hfill
  \includegraphics[width=0.48\textwidth]{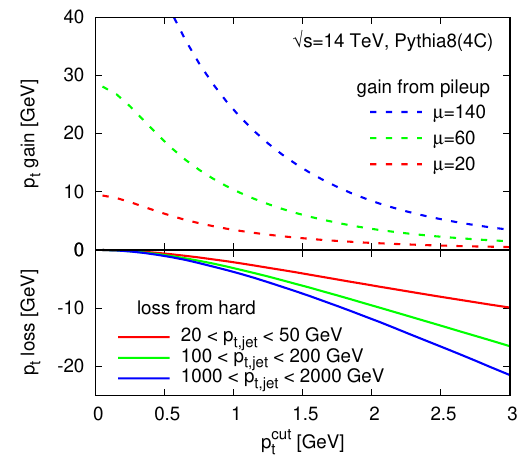}
  \caption{Left: value of the $p_t$ cut applied by the SoftKiller,
    displayed as a function of the number of pileup events. We show
    results for two
    different values of the generator minimal $p_t$ for the hard
    event, $p_{t,\rm gen}$. The solid line is the average $\ptcut$ value while
    the dashed lines indicate the one-standard-deviation band.
    Right: plot of the $p_t$ that is lost when applying a given
    $p_t$ cut (the $x$ axis) to the constituents of jets
    clustered (anti-$k_t$, $R=0.4$) from the hard event (solid lines)
    and the residual pileup $p_t$ that remains after applying that
    same cut to the constituents of circular patches of radius $0.4$ in
    pure-pileup events (dashed lines).}
  \label{fig:killer-cut}
\end{figure}

Figure~\ref{fig:killer-cut} shows, on the left, the average
$p_t^\cut$ value, together with its standard deviation (dashed
lines), as a function of the number of pileup interactions, $\nPU$.
The event sample consists of a superposition of $\nPU$ zero-bias on
one hard dijet event, in 14~TeV proton--proton collisions, all
simulated with Pythia~8 (tune 4C)~\cite{Pythia8}.\footnote{For a study
  of the tune dependence, see Appendix~C of
  Ref.~\cite{Cacciari:2014gra} where we briefly examined the
  Pythia~6~\cite{Sjostrand:2000wi,Sjostrand:2003wg,Sjostrand:2006za}
  Z2 tune~\cite{Field:2011iq}, and find very similar results.}
The underlying event in the hard event has been switched off, and all
particles have been made massless, maintaining their $p_t$, rapidity
and azimuth.\footnote{If one keeps the underlying event in the hard
  event, much of it (about $1\GeV$ for both the area--median approach
  and the SoftKiller) is subtracted together with the pileup
  correction, affecting slightly the observed shifts. Keeping massive
  particles does not affect the SK performance but requires an extra
  correction for the area--median subtraction, see
  Section~\ref{sec:areamed-particle-masses}. We therefore use massless
  particles for simplicity.}
These are our default choices throughout this Chapter.
The grid used to determine $p_t^\cut$ has a spacing of $a \simeq 0.4$
and extends up to $|y| < 5$. 
One sees that $p_t^\cut$ remains moderate, below $2\GeV$, even
for pileup at the level foreseen for the high-luminosity upgrade of
the LHC (HL-LHC), which is expected to reach an average
(Poisson-distributed) number of 
pileup interactions of $\mu \simeq 140$.
The right-hand plot shows the two sources of bias: the lower (solid)
curves, illustrate the bias on the hard jets induced by the loss of
genuine hard-event particles below $p_t^\cut$.
Jet clustering is performed as usual with the anti-$k_t$ jet algorithm
with $R=0.4$ (as implemented in FastJet~3.1).
The three line colours correspond to different jet $p_t$ ranges. 
The loss has some dependence on the jet $p_t$ itself, notably for
higher values of $p_t^\cut$.%
\footnote{In a local parton--hadron duality type approach to calculate
  hadron spectra, the spectrum of very low $p_t$ particles in a jet of
  a given flavour
  is actually independent of the jet's $p_t$~\cite{KLO}.}
In particular it grows in absolute terms for larger jet $p_t$'s,
though it decreases relative to the jet $p_t$.
The positive bias from residual pileup particles (calculated in
circular patches of radius $0.4$ at rapidity $y=0$) is shown as dashed
curves, for three different pileup levels.
To estimate the net bias, one should choose a value for $\nPU$,
read the average $p_t^\cut$ from the left-hand plot, and for that
$p_t^\cut$ compare the solid curve with the dashed curve that
corresponds to the given $\nPU$.
Performing this exercise reveals that there is indeed a reasonable
degree of cancellation between the positive and negative biases.
Based on this observation, we can move forward with a more detailed
study of the performance of the method.\footnote{A study of fixed
  $p_t$ cutoffs, rather than dynamically determined ones, is performed
  in Appendix~D of Ref.\cite{Cacciari:2014gra}.}

\section{SoftKiller performance}
\label{sec:performance}

\paragraph{Choice of grid size.}
For a detailed study of the SoftKiller method, the first step is to
choose the grid spacing $a$ so as to break the event into patches.
The spacing $a$ is the one main free parameter of the method. 
The patch-size parameter is present also for area--median pileup
subtraction.
There the exact choice of this parameter is not too critical because
the median is quite stable when pileup levels are high: all grid cells
are filled, and nearly all are dominated by pileup.
However the SoftKiller method chooses the $p_t^\cut$ so as
to obtain a nearly empty event. 
In this limit, the median operation becomes somewhat more sensitive to
the grid spacing (see also the discussion in
Section~\ref{sec:analytic-pileup}).

\begin{figure}
  \includegraphics[width=\textwidth]{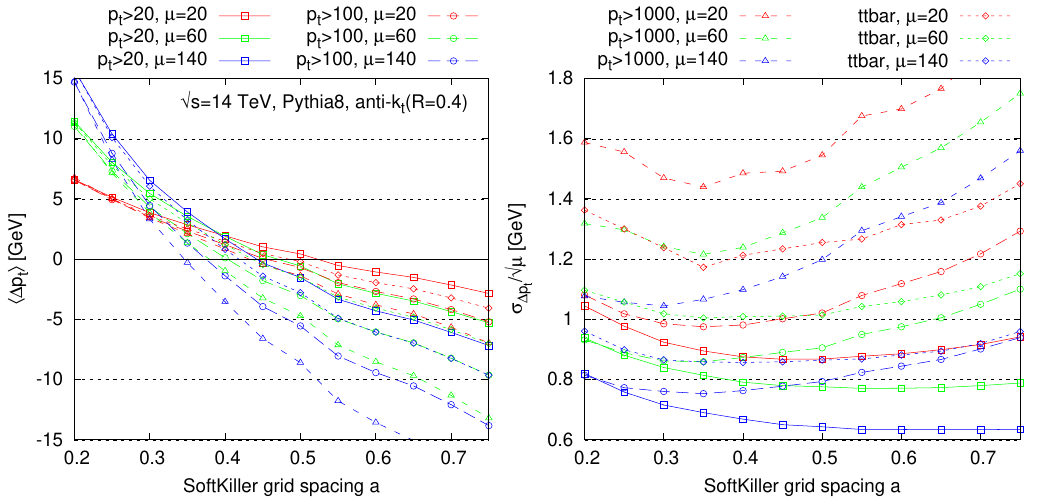}
  \caption{Scan of the SoftKiller performances as a function the
    grid-spacing parameter $a$ for different hard-event samples and 
    three different pileup levels (Poisson-distributed with average
    pileup event multiplicities of $\mu=20,60,140$).
    Left: average $p_t$ shift; right: $p_t$ shift dispersion, normalised
    to $\sqrt{\mu}$ for better readability.
    %
    Results are given for a variety of hard processes
    and pileup conditions to illustrate robustness. 
    Curves labelled $p_t > X$ correspond to dijet events, in which one
    studies only those jets that in the hard event have a transverse
    momentum greater than $X$.
    For the $t\bar t$ sample, restricted to fully hadronic decays of
    the top quarks, the study is limited to jets that have $p_t > 50
    \GeV$ in the hard event.  }
  \label{fig:killer-scan}
\end{figure}

Fig.~\ref{fig:killer-scan} considers a range of hard event samples
(different line styles) and pileup levels (different colours). 
For each, as a function of the grid spacing $a$, the left-hand plot
shows the average, $\langle \Delta p_t\rangle$, of the net shift in
the jet transverse momentum, defined as before\footnote{The matching
  between a hard jet and a ``corrected'' jet is done essentially as
  before: for each hard jet passing our selection cut, we select the
  jet in the subtracted event (\ie after applying the SoftKiller) for
  which the sum of the transverse momentum of the constituents it
  has in common with the original hard jet is maximal.}
\begin{equation}
  \label{eq:Delta-pt}
  \Delta p_t = p_t^\text{corrected} -
  p_t^\text{hard} \,,
\end{equation}
while the right-hand plot shows the dispersion, $\sigma_{\Delta p_t}$,
of that shift from one jet to the next, here normalised to
$\sqrt{\mu}$ (right).

One sees that the average jet $p_t$ shift has significant dependence on
the grid spacing $a$. 
However, there exists a grid spacing, in this case $a\simeq 0.4$,
for which the shift is not too far from zero and not too dependent
either on the hard sample choice or on the level of pileup. 
In most cases the absolute value of the shift is within about $2\GeV$,
the only exception being for the $p_t > 1000\GeV$ dijet sample, for which
the bias can reach up to $4\GeV$ for $\mu=140$.
This shift is, however, still less than the typical best experimental
systematic error on the jet energy scale, today of the order of
$1\%$ or slightly better~\cite{Kirschenmann:2012toa,Aad:2014bia}.

It is not trivial that there should be a single grid spacing that is
effective across all samples and pileup levels: the fact that there is
can be considered phenomenologically fortuitous.
The value of the grid spacing $a$ that minimises the typical shifts is
also close to the value that minimises the dispersion in the
shifts.\footnote{In a context where the net shift is the sum of two
  opposite-sign sources of bias, this is perhaps not too surprising:
  the two contributions to the dispersion are each likely to be of the
  same order of magnitude as the individual biases, and their sum
  probably minimised when neither bias is too large.}
That optimal value of $a$ is not identical across event samples, and
can also depend on the level of pileup.
However the dispersion at $a = 0.4$ is always close to the actual
minimal attainable dispersion for a given sample.
Accordingly, for most of the rest of this article, we will work with a grid
spacing of $a=0.4$.\footnote{A single value of $a$ is adequate as long
  as jet finding is carried out mostly with jets of radius $R \simeq
  0.4$. Later in this section we will supplement our $R=0.4$ studies with
  a discussion of larger jet radii.}

\begin{figure}
  \centering
  \begin{minipage}[c]{0.48\linewidth}
  \includegraphics[width=\textwidth]{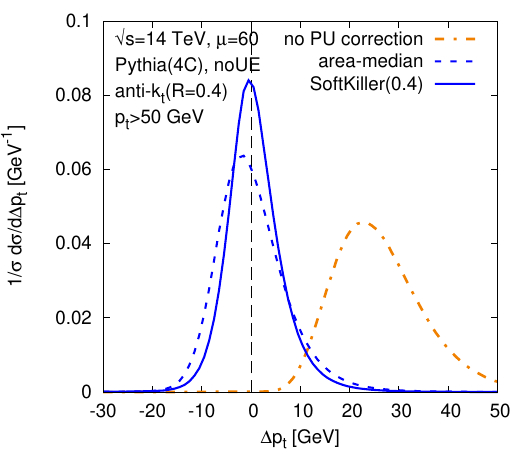}
  \end{minipage}
  \hfill
  \begin{minipage}[c]{0.48\linewidth}
  \caption{Performance of SoftKiller for 50 GeV jets and $\mu=60$
    Poisson-distributed pileup events. We plot the distribution of the
    shift $\Delta p_t$ between the jet $p_t$ after pileup removal and
    the jet $p_t$ in the hard event alone. Results are compared
    between the area--median approach and the SoftKiller. For
    comparison, the (orange) dash-dotted line corresponds to the situation
    where no pileup corrections are applied.}
  \label{fig:killer-pt-hdiff}
  \end{minipage}
\end{figure}

\begin{figure}
  \includegraphics[width=0.48\textwidth]{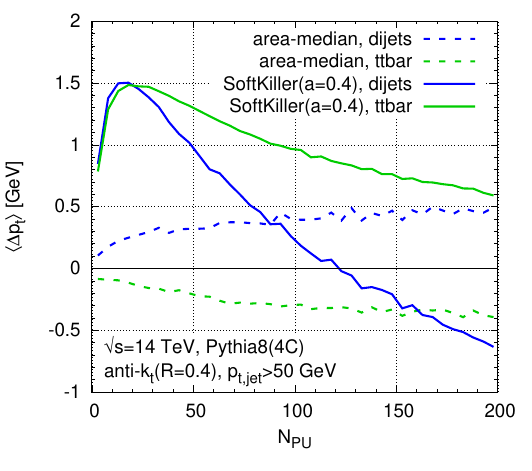}
  \hfill
  \includegraphics[width=0.48\textwidth]{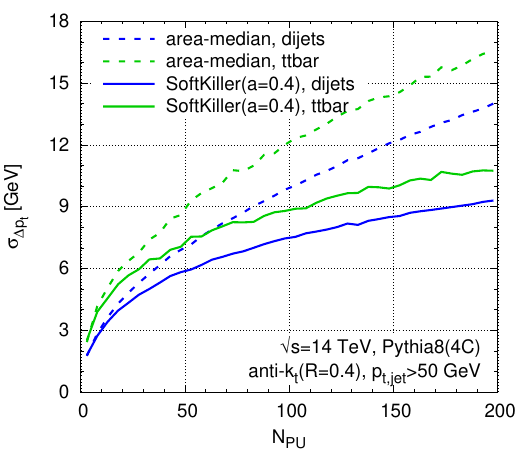}
  \caption{Performance of SoftKiller shown as a function of the pileup
    multiplicity and compared to the area--median subtraction method.
    Left: the average $p_t$ shift after subtraction, compared to the
    original jets in the hard event. 
    Right: the corresponding dispersion.}
  \label{fig:killer-v-npu}
\end{figure}
 
\paragraph{Performance compared to the area--median method.}
We now compare the performance of the SoftKiller to that of
area--median subtraction.
Figure~\ref{fig:killer-pt-hdiff} shows the distribution of shift in
$p_t$, for (hard) jets with $p_t > 50\GeV$ in a dijet sample.
The average number of pileup events is $\mu = 60$, with a Poisson
distribution.
One sees that in the SoftKiller approach, the peak is about $30\%$
higher than what is obtained with the area--median approach and the
distribution correspondingly narrower.
The peak, in this specific case, is well centred on $\Delta p_t = 0$.

Figure~\ref{fig:killer-v-npu} shows the shift (left) and dispersion
(right) as a function of $\nPU$ for two different samples: the $p_t >
50\GeV$ dijet sample (in blue), as used in
Fig.~\ref{fig:killer-pt-hdiff}, and a hadronic $t\bar t$ sample,
with a $50\GeV$ $p_t$ cut on jets (in green). 
Again, the figure compares the area--median (dashed) and SoftKiller
results (solid).
One immediately sees that the area--median approach gives a bias that
is more stable as a function of $\nPU$.
Nevertheless, the bias in the SoftKiller approach remains between
about $-0.5$ and $1.5\GeV$, which is still reasonable when one
considers that, experimentally, some degree of recalibration is anyway
needed after area--median subtraction.
As concerns sample dependence of the shift, comparing $t\bar t$ v.\
dijet, the area--median and SoftKiller methods appear to have similar
systematic differences.
In the case of SoftKiller, there are two main causes for the sample
dependence: firstly the higher multiplicity of jets has a small effect
on the choice of $\ptcut$ and secondly the dijet sample is mostly composed
of gluon-induced jets, whereas the $t\bar t$ sample is mostly composed of
quark-induced jets (which have fewer soft particles and so lose less
momentum when imposing a particle $p_t$ threshold).
Turning to the right-hand plot, with the dispersions, one sees that
the SoftKiller brings a significant improvement compared to
area--median subtraction for $\nPU \gtrsim 20$.
The relative improvement is greatest at high pileup levels, where
there is a reduction in dispersion of $30-35\%$, beating the
$\sqrt{\nPU}$ scaling that is characteristic of the area--median
method. 
While the actual values of the dispersion depend a little on the
sample, the benefit of the SoftKiller approach is clearly visible for
both.

\begin{figure}
  \includegraphics[width=0.48\textwidth]{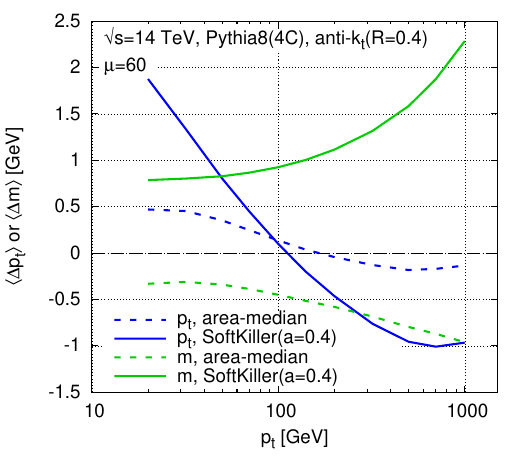}
  \hfill
  \includegraphics[width=0.48\textwidth]{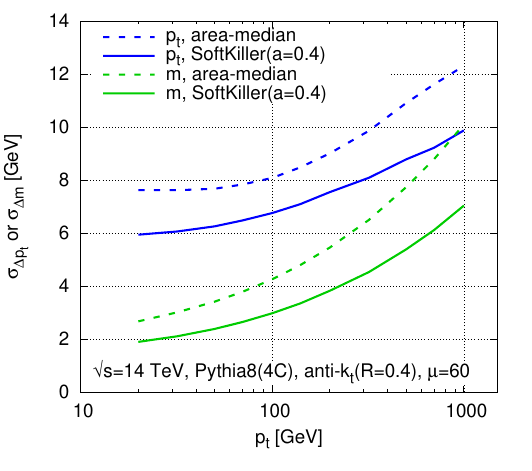}
  \caption{Performance of SoftKiller shown as a function of the hard
    jet $p_t$ (solid lines), with a comparison to the area--median
    subtraction method (dashed lines).
    Left: the average shifts for the jet $p_t$ (blue) and mass (green),
    after subtraction.  
    Right: the corresponding dispersions.  }
  \label{fig:killer-v-pt}
\end{figure}

Figure~\ref{fig:killer-v-pt} shows the shift (left) and dispersion
(right) for jet $p_t$'s and jet masses, now as a
function of the hard jet minimum $p_t$.
Again, dashed curves correspond to area--median subtraction, while solid
ones correspond to the SoftKiller results.
All curves correspond to an average of $60$ pileup interactions.
For the jet $p_t$ (blue curves) one sees that the area--median shift
ranges from $0.5$ to $0 \GeV$ as $p_t$ increases from $20\GeV$ to
$1\TeV$, while for SK the dependence is stronger, from about $2$ to
$-1\GeV$, but still reasonable.
For the jet mass (green curves), again the area--median
method\footnote{using a ``safe'' subtraction procedure that replaces
  negative-mass jets with zero-mass jets, as described in
  Section~\ref{sec:areamedian-safesubtraction}.} is more stable than
SK, but overall the biases are under control, at the $1$ to $2\GeV$
level.
Considering the dispersions (right), one sees that SK gives
a systematic improvement, across the whole range of jet $p_t$'s. 
In relative terms, the improvement is somewhat larger for the jet mass
($\sim 30\%$) than for the jet $p_t$ ($\sim 20\%$).

\begin{figure}
  \includegraphics[width=0.48\textwidth]{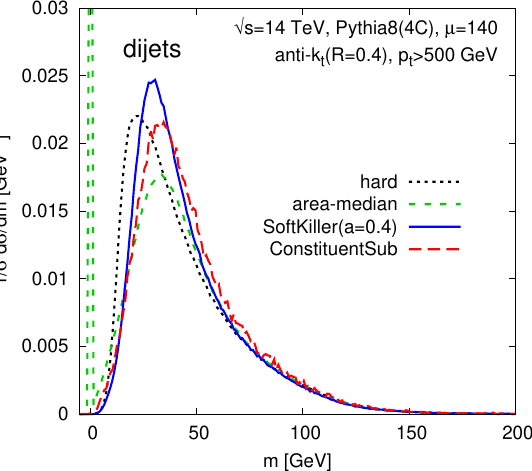}
  \hfill
  \includegraphics[width=0.48\textwidth]{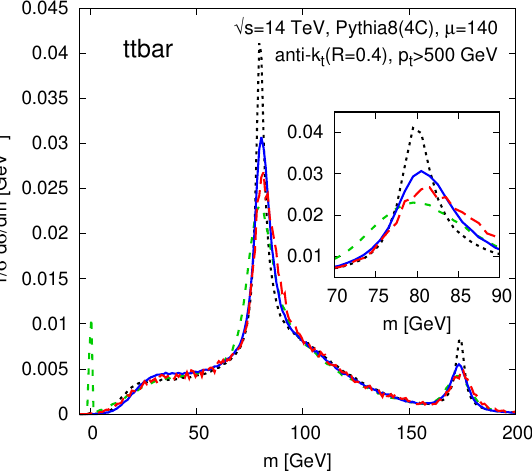}
  \caption{Performance of SoftKiller on jet mass reconstruction. We
    show the jet mass distribution after applying different pileup
    mitigation techniques. Left: dijet events, right:
    hadronically-decaying $t\bar t$ events.
}
  \label{fig:mass-spectra}
\end{figure}

Fig.~\ref{fig:mass-spectra} shows the actual mass spectra of $R=0.4$
jets, for two samples: a QCD dijet sample and a boosted $t\bar t$
sample.
For both samples, we only considered jets with $p_t > 500\GeV$ in the
hard event.
One sees that SK gives slightly improved mass peaks relative to the
area--median method and also avoids area--median's spurious peak at
$m=0$, which is due to events in which the squared jet mass came out
negative after four-vector area-subtraction and so was reset to zero.
The plot also shows results from the recently proposed Constituent
Subtractor method~\cite{Berta:2014eza}, using v.~1.0.0 of the
corresponding code from FastJet Contrib~\cite{fjcontrib}.
It too performs better than area--median subtraction for the jet mass,
though the improvement is not quite as large as for
SK.\footnote{A further option is to use an ``intrajet killer'' that
  removes soft particles inside a given jet until a total $p_t$ of
  $\rho A_{\rm jet}$ has been subtracted. This shows performance
  similar to that of the Constituent Subtractor.\label{ft:intrajet}}

\begin{figure}
  \includegraphics[width=0.48\textwidth]{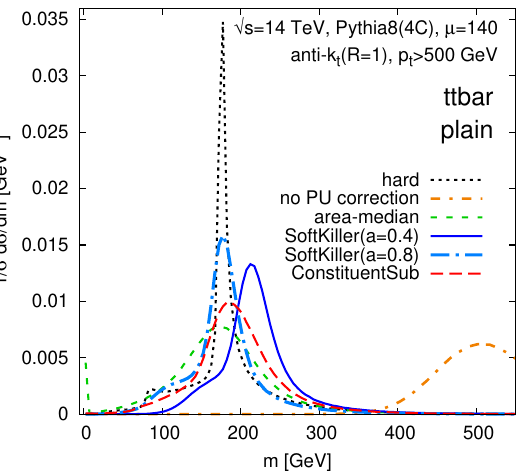}
  \hfill
  \includegraphics[width=0.48\textwidth]{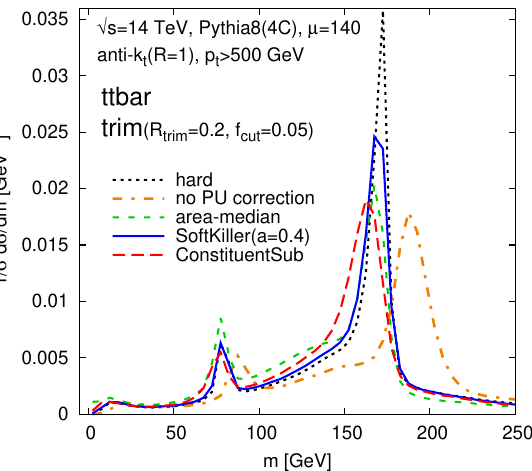}
  \caption{The results of the SoftKiller and other methods applied to
    large-$R$ jets. The left-hand plot is without grooming, the
    right-hand plot with grooming (trimming with $R_\text{trim}=0.2$
    and $f_\cut = 0.05$).
  }
  \label{fig:large-R}
\end{figure}

\paragraph{Large-$R$ jets and grooming.}
One might ask why we concentrated on $R=0.4$ jets here, given that
jet-mass studies often use large-$R$ jets.
The reason is that large-$R$ jets are nearly always used in
conjunction with some form of grooming, for example trimming, pruning
or filtering (see also Chapter~\ref{chap:grooming-description}).
Grooming reduces the large-radius jet to a collection of small-radius
jets and so the large-radius groomed-jet mass is effectively a
combination of the $p_t$'s and masses of one or more small-radius
jets.

For the sake of completeness, let us briefly also study the SoftKiller
performance for large-$R$ jets.
Figure~\ref{fig:large-R} shows jet-mass results for the same $t\bar t$
sample as in Fig.~\ref{fig:mass-spectra} (right), now clustered with
the anti-$k_t$ algorithm with $R=1$.
The left-hand plot is without grooming: one sees that SK with our
default spacing of $a=0.4$ gives a jet mass that has better resolution
than area--median subtraction (or the ConstituentSubtractor), but a
noticeable shift, albeit one that is small compared to the
effect of uncorrected pileup.
That shift is associated with some residual contamination from pileup
particles: in an $R=0.4$ jet, there are typically a handful of
particles left from pileup, which compensate low-$p_t$ particles lost
from near the core of the jet.
If one substantially increases the jet radius without applying
grooming, then that balance is upset, with substantially more pileup
entering the jet, while there is only slight further loss of genuine
jet $p_t$.
To some extent this can be addressed by using the SoftKiller with a
larger grid spacing (cf.\ the $a=0.8$ result), which effectively
increases the particle $\ptcut$. 
This comes at the expense of performance on small-$R$ jets (cf.\
Fig.~\ref{fig:killer-scan}).
An interesting, open problem is to find a simple way to remove pileup
from an event such that, for a single configuration of the pileup
removal procedure, one simultaneously obtains good performance on
small-$R$ and large-$R$ jets. We will come back to candidate solutions
to this open question in Chapter~\ref{chap:beyond-prelimn}
(Section~\ref{sec:sk-improvements-Rdep}).

As we said above, however, large-$R$ jet masses are nearly always used
in conjunction with some form of grooming.
Fig.~\ref{fig:large-R} (right) shows that when used together with
trimming, SoftKiller with our default $a=0.4$ choice performs well
both in terms of resolution and shift.

\begin{figure}
  \includegraphics[width=\textwidth]{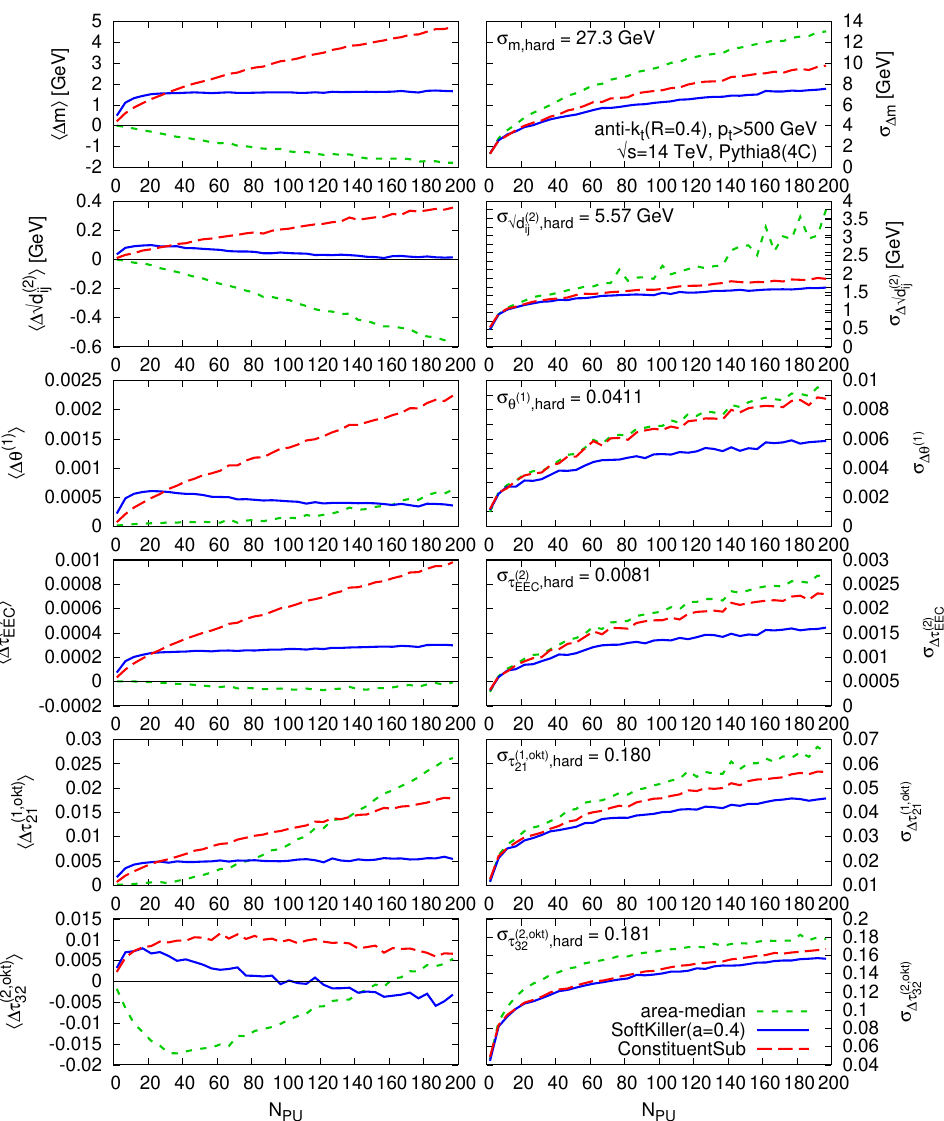}
  \caption{Performance of the SoftKiller on jet shapes, compared to
    area--median subtraction and the recently proposed Constituent Subtractor method
     \cite{Berta:2014eza}. All
    results are shown for dijet events with a 500 GeV $p_t$ cut on
    anti-$k_t$, $R=0.4$ jets.
    For comparison of the subtraction performance we also quote, for 
    each observable $X$,  $\sigma_{X,\rm hard}$, the dispersion of the
    distribution of the observable in the hard event.
    For $\tau_{21}$ there is the additional requirement (in the
    hard event) that the jet mass is above 30 GeV and for $\tau_{32}$
    we further impose that $\tau_{21}\ge 0.1$ (again in the hard
    event), so as to ensure infrared safety.
  \label{fig:shape-stats}
}
\end{figure}

\paragraph{Jet shapes.}
Returning to $R=0.4$ jets, the final figure of this section,
Fig.~\ref{fig:shape-stats}, shows average shifts (left) and
dispersions (right) as a function of $\nPU$ for several different jet
``shapes'': jet masses, $k_t$ clustering
scales~\cite{Catani:1993hr,Ellis:1993tq}, the jet width (or broadening
or girth~\cite{Catani:1992jc,Berger:2003iw,Almeida:2008yp}), an
energy--energy correlation moment~\cite{Larkoski:2013eya} and the
$\tau_{21}^{(\beta=1)}$ and $\tau_{32}^{(\beta=2)}$ N-subjettiness
ratios~\cite{Thaler:2010tr,Thaler:2011gf}, using the exclusive $k_t$
axes with one-pass of minimisation.
Except in the case of the jet mass (which uses ``safe'' area
subtraction, as mentioned above), the area--median results have been
obtained using the shape subtraction technique introduced in
Section~\ref{sec:areamed-shapes}, as implemented in v.~1.2.0 of the
GenericSubtractor in FastJet Contrib.

As regards the shifts, the SK approach is sometimes the best, other
times second best.
Which method fares worst depends on the precise observable. 
In all cases, when considering the dispersions, it is the SK that
performs best, though the extent of the improvement relative to other
methods depends strongly on the particular observable.
Overall this figure gives us confidence that one can use the SoftKiller
approach for a range of properties of small-radius jets.

\begin{figure}
  \centering
  \includegraphics[width=0.48\textwidth]{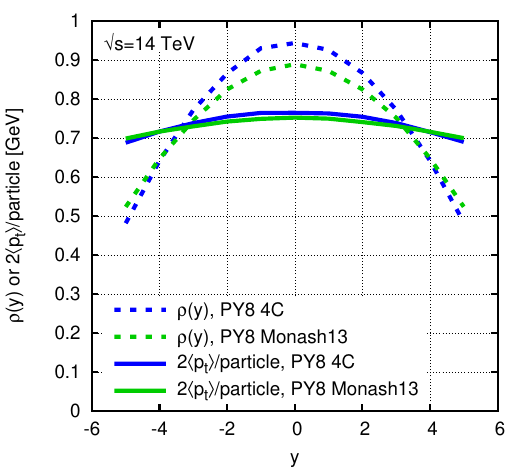}
  \hfill
  \includegraphics[width=0.48\textwidth]{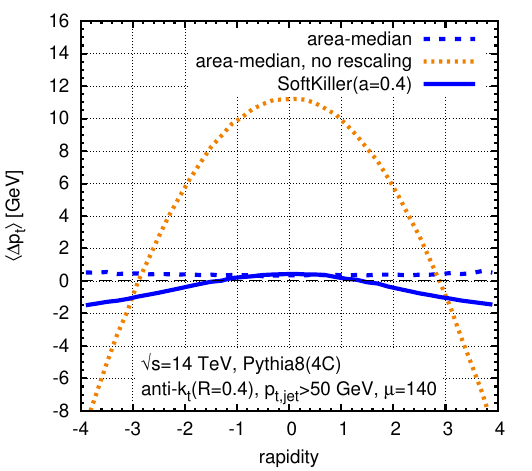}
  \caption{Left: rapidity dependence, in simulated zero-bias events,
    of the $p_t$ density per
    unit area ($\rho(y)$), and of the average $p_t$ per particle (scaled
    by a factor of $2$ for readability), comparing Pythia~8's 4C and
    Monash~2013~\cite{Skands:2014pea} tunes.
    Right: the average shift in jet $p_t$ as a function of rapidity
    for the area--median method with rapidity rescaling (our default),
    without rapidity rescaling, and for the SoftKiller.}
  \label{fig:rapidity-dep}
\end{figure}

\paragraph{Rapidity dependence.}
Given that the SoftKiller imposes a fixed $p_t$ cut, one might wonder
if it correctly account for the rapidity dependence of pileup.
In the case of the area--median subtraction, that is handled by the
rapidity rescaling used in Eq.~(\ref{eq:rhoest-rescaled}) which we
have used throughout this Chapter.\footnote{For our 14~TeV Pythia~8(4C)
  simulations, we use
  $f(y)=1.1685397-0.0246807 \, y^2+5.94119\cdot10^{-5}\, y^4$.}
The rapidity dependence of $\rho$ is shown as the dashed lines in
Fig.~\ref{fig:rapidity-dep} (left) for two different tunes (4C and
Monash~2013), illustrating the fact that they have somewhat different
rapidity dependence.

The SoftKiller method acts not on the average energy flow, but instead
on the particle $p_t$'s. 
The solid lines in Fig.~\ref{fig:rapidity-dep} (left) show that the
average particle $p_t$ is nearly independent of rapidity.
This suggests that there may not be a need to explicitly account for
rapidity in the SK method, at least at particle level (detector
effects introduce further non-trivial rapidity dependence).

This is confirmed in the right-hand plot of
Fig.~\ref{fig:rapidity-dep}, which shows the rapidity-dependence of
the shift in the jet $p_t$ with the area--median and SK methods.
Our default area--median curve, which includes rapidity rescaling,
leads to a nearly rapidity-independent shift. 
However, without the rapidity rescaling, there are instead
rapidity-dependent shifts of up to $10\GeV$ at high pileup.
In contrast, the SK method, which in our implementation does not
involve any parametrisation of rapidity dependence, automatically
gives a jet $p_t$ shift that is fairly independent of rapidity, to
within about $2\GeV$.
We interpret this as a consequence of the fact (cf.\ the left-hand
plot of Fig.~\ref{fig:rapidity-dep}) that the average particle $p_t$
appears to be far less dependent on rapidity than the average $p_t$
flow.
In a similar spirit to~(\ref{eq:rhoest-rescaled}), one could also
imagine introducing a rapidity-dependent rescaling of the particle
$p_t$'s before applying SoftKiller, and then inverting the rescaling
afterwards. Our initial tests of this approach suggest that it does
largely correct for the residual SK rapidity dependence.

\section{Adaptation to CHS events and calorimetric events}
\label{sec:adapt-calor-towers}

It is important to verify that the SoftKiller (or any other new pileup
mitigation method with noise reduction) works not just at particle
level, but also at detector level, essentially to make sure that the
balance between the positive and negative biases discussed in
Section~\ref{sec:area-median-recall} is not spoiled.
There are numerous subtleties in carrying out detector-level
simulation, from the difficulty of correctly treating the detector
response to low-$p_t$ particles, to the reproduction of actual detector
reconstruction methods and calibrations, and even the determination of which
observables to use as performance indicators.
Here we will consider two cases: idealised charged-hadron-subtraction,
which simply examines the effect of discarding charged pileup
particles; and simple calorimeter towers.

\paragraph{Performance on CHS events.}
As in the previous Chapter, for events with particle flow~\cite{pflow}
and charged-hadron subtraction (CHS), we imagine a situation in which
all charged particles can be unambiguously associated either with the
leading vertex or with a pileup vertex.
We then apply the SK exclusively to the neutral particles, which we
assume to have been measured exactly.
This is almost certainly a crude approximation, however it helps to
illustrate some general features.

\begin{figure}
  \centering
  \includegraphics[width=0.48\textwidth]{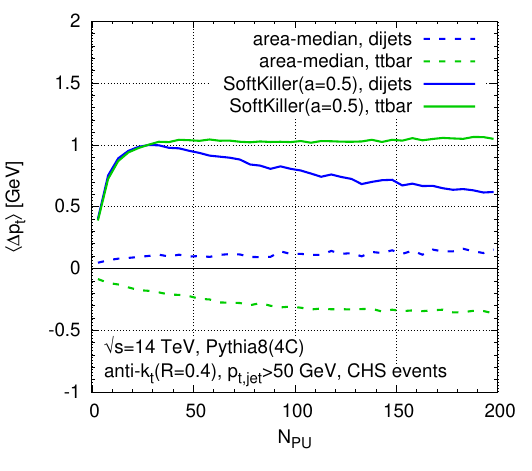}
  \hfill
  \includegraphics[width=0.48\textwidth]{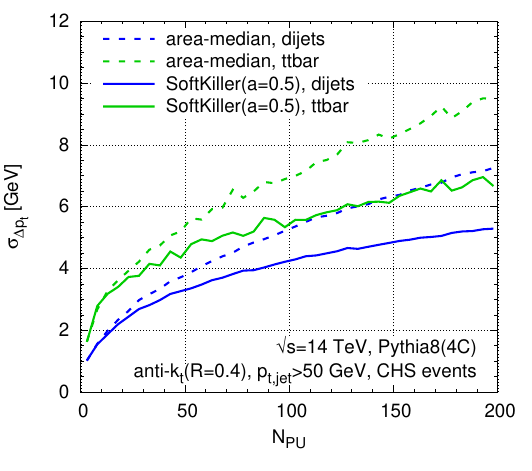}
  \caption{Same as Fig.~\ref{fig:killer-v-npu}, for events with
    charged-hadron subtraction (CHS). Note that the grid size used for
    the SoftKiller curves has been set to $a=0.5$.}
  \label{fig:chs-performance}
\end{figure}

One important change that arises from applying SK just to the neutral
particles is that there is a reduced contribution of low-$p_t$
hard-event particles. 
This means that for a given actual amount of pileup contamination (in
terms of visible transverse momentum per unit area), one can afford to
cut more aggressively, \ie raise the $p_t^\cut$ as compared to the
full particle-level case, because for a given $p_t^\cut$ there will be a
reduced loss of hard-event particles.
This can be achieved through a moderate increase in the grid spacing,
to $a=0.5$.
Figure~\ref{fig:chs-performance} shows the results, with the shift
(left) and dispersion (right) for the jet $p_t$ in dijet and $t\bar t$
samples. 
The SK method continues to bring an improvement, though that
improvement is slightly more limited than in the
particle-level case. 
We attribute this reduced improvement to the fact that SK's greatest
impact is at very high pileup, and for a given
$\nPU$, SK with CHS is effectively operating at lower pileup
levels than without CHS.
In Ref.~\cite{Cacciari:2014gra}, Appendix~E, we also carried on a
further study with our toy CHS simulation concerning lepton isolation.

\paragraph{Performance with calorimetric events.}
Next let us turn to events where the particles enter calorimeter
towers.
Here we encounter the issue (discussed also in
Appendix~\ref{app:sk-collinear-safety}) that SK is not collinear safe.
While we argue there that this is not a fundamental drawback from the point
of view of particle-level studies, there are issues at calorimeter
level: on one hand a single particle may be divided between two
calorimeter towers (we will not attempt to simulate this, as it is very
sensitive to detector details); on the other,
within a given tower (say $0.1\times0.1$) it is quite likely
that for high pileup the tower may receive contributions from multiple
particles.
In particular, if a tower receives contributions from a hard particle
with a substantial $p_t$ and additionally from pileup particles, the
tower will always be above threshold, and the pileup contribution will
never be removed.
There are also related effects due to the fact that two pileup
particles may enter the same tower.
To account for the fact that towers have finite area, we therefore
adapt the SK as follows.
In a first step we subtract each tower:
\begin{equation}
  \label{eq:tower-correction}
  p_t^\text{tower,sub} = \max\left(0,\; p_t^\text{tower} - \rho A^\text{tower}\right)\,,
\end{equation}
where $\rho$ is as determined on the event prior to any
correction.\footnote{We use our standard choices for determining
  $\rho$, namely the grid version of the area--median method, with a
  grid spacing of $0.55$ and rapidity scaling as discussed in
  Section~\ref{sec:areamedian-mcstudy:jetpt}. One could equally well
  use the same grid spacing for the $\rho$ determination as for the
  SoftKiller.}
This in itself eliminates a significant fraction of pileup, but there
remains a residual contribution from the roughly $50\%$ of towers
whose $p_t$ was larger than  $\rho A^\text{tower}$. 
We then apply the SoftKiller to the subtracted towers, \ie with a cut
given by
\begin{equation}
  \label{eq:ptcut-subtowers}
  p_t^{\cut,\text{sub}} = \underset{i \in \text{patches}}{\text{median}} \left\{ p_{ti}^{\text{tower,sub,}\max} \right\}\,,
\end{equation}
where $p_{ti}^{\text{tower,sub,}\max}$ is the $p_t$, after
subtraction, of the hardest tower in patch $i$, in analogy with
\eq~(\ref{eq:ptmin}).
In the limit of infinite granularity, a limit similar to particle
level, $A^\text{tower}=0$, the step in \eq~\eqref{eq:tower-correction}
then has no effect and one recovers the standard SoftKiller procedure
applied to particle level.

\begin{figure}
  \centering
  \includegraphics[width=0.48\textwidth]{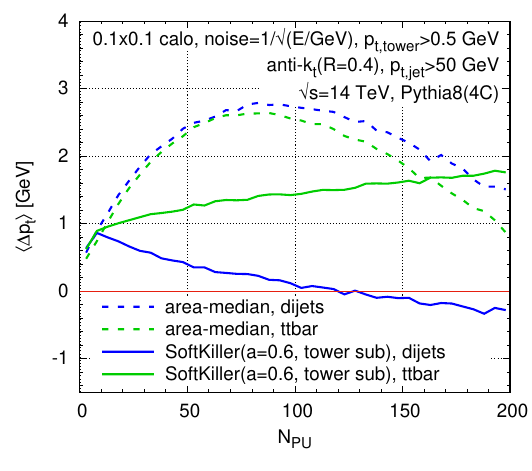}
  \hfill
  \includegraphics[width=0.48\textwidth]{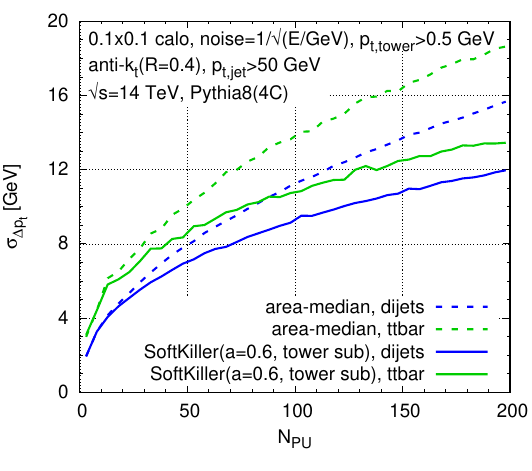}
  \caption{Same as Fig.~\ref{fig:killer-v-npu} for events with a
    simple calorimeter simulation. The SoftKiller
    was used here with a grid spacing of $a=0.6$ and includes the tower
    subtraction of \eq~(\ref{eq:tower-correction}).}
  \label{fig:calo-performance}
\end{figure}

Results are shown in Fig.~\ref{fig:calo-performance}.
The energy $E$ in each $0.1\times 0.1$ tower is taken to have Gaussian
fluctuations with relative standard deviation $1/\sqrt{E/\!\GeV}$.
A $p_t$ threshold of $0.5\GeV$ is applied to each tower after fluctuations.
The SK grid spacing is set to $a = 0.6$. 
Interestingly, with a calorimeter, the area--median method starts to
have significant biases, of a couple of GeV, which can be attributed
to the calorimeter's non-linear response to soft energy.
The SK biases are similar in magnitude to those in
Fig.~\ref{fig:killer-v-npu} at particle level (note, however, the need
for a different choice of grid spacing $a$).
The presence of a calorimeter worsens the resolution both for
area--median subtraction and SK, however SK continues to perform
better, even if the improvement relative to area--median subtraction
is slightly smaller than for the particle-level results.

We have also investigated a direct application of the particle-level
SoftKiller approach to calorimeter towers, \ie without the
subtraction in \eq~(\ref{eq:tower-correction}). 
We find that the biases were larger but still under some degree of
control with an appropriate tuning of $a$, while the performance on
dispersion tends to be intermediate between that of area--median subtraction
and the version of SoftKiller with tower subtraction.

The above results are not intended to provide an exhaustive study of
detector effects. 
For example, particle flow and CHS are affected by detector
fluctuations, which we have ignored; purely calorimetric jet
measurements are affected by the fact that calorimeter towers are of
different sizes in different regions of the detector and furthermore
may be combined non-trivially through topoclustering.
Nevertheless, our results help illustrate that it is at least
plausible that the SoftKiller approach could be adapted to a full
detector environment while retaining much of its performance advantage
relative to the area--median method.

\section{Computing time}
\label{sec:computing-time}

The computation time for the SoftKiller procedure has two components:
the assignment of particles to patches, which is $\order{N}$, \ie
linear in the total number of particles $N$ and the determination of
the median, which is $\order{P \ln P}$ where $P$ is the number of
patches.
The subsequent clustering is performed with a reduced number of
particles, $M$, which, at high pileup is almost independent of the
number of pileup particles in the original event.
In this limit, the procedure is therefore expected to be dominated by
the time to assign particles to patches, which is linear in $N$.
This assignment is almost certainly amenable to being parallelised.

In studying the timing, we restrict our attention to particle-level
events for simplicity.
We believe that calorimeter-type extensions as described in
section~\ref{sec:adapt-calor-towers} can be coded in such a way as to
obtain similar (or perhaps even better) performance.

\begin{figure}[t]
  \centering
  \includegraphics[width=0.48\textwidth]{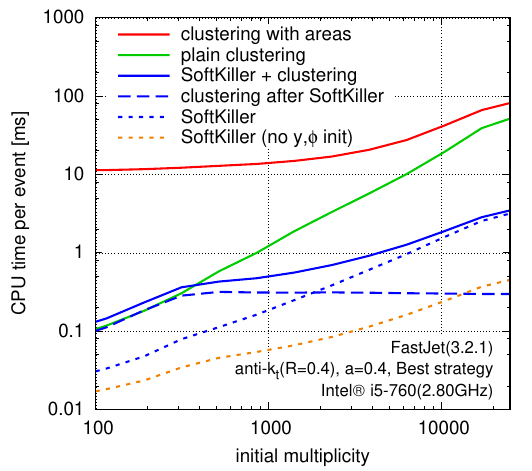}
  \hfill
  \includegraphics[width=0.48\textwidth]{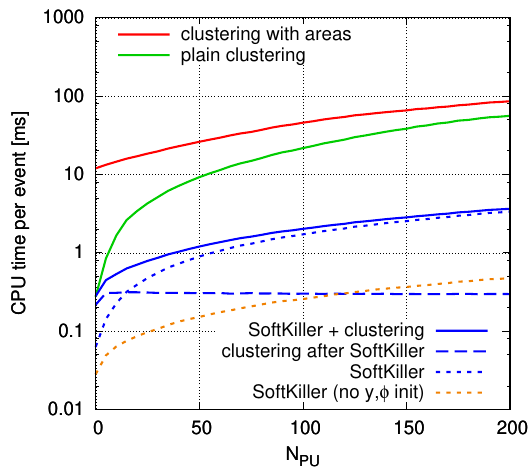}  
  \caption{Timings of the SoftKiller compared to standard clustering
    shown as a function of the number of particles in the event (left)
    or as a function of the number of pileup vertices (right). We
    compare the SoftKiller timings to the time to cluster the full
    event with (red) or without (green) jet area
    information. For the SoftKiller timings (in blue), we show
    individually the time spent to apply the SoftKiller to the event
    (dotted line), the time spent to cluster the resulting event
    (dashed line) and their sum (solid line).
    The orange dotted line corresponds to the SoftKiller timing when the
    particle's rapidity and azimuth have been precomputed.
  }
  \label{fig:timing}
\end{figure}

Timings are shown in Fig.~\ref{fig:timing} versus initial multiplicity
(left) and versus the number of pileup vertices (right).\footnote{
  These timings have been obtained on an Intel processor, i5-760
  (2.80\,GHz), using FastJet 3.2.1, with the ``Best'' clustering
  strategy.
  Note that this figure has been updated compared to what was
  originally published in~\cite{Cacciari:2014gra}, so as to include
  the speed improvements brought with the release of Fastjet 3.1.}
Each plot shows the time needed to cluster the full event and the time to
cluster the full event together with ghosts (as needed for area-based
subtraction).
It also shows the time to run the SoftKiller procedure, the time to
cluster the resulting event, and the total time for SK plus
clustering.

Overall, one sees more than one order of magnitude improvement in
speed from the SK procedure, with run times per event ranging from
$0.15$ to $4\,$ms as compared to $10$ to $90\,$ms for clustering with
area information.
At low multiplicities, the time to run SK is small
compared to that needed for the subsequent clustering.
As the event multiplicity increases, SK has the effect of limiting the
event multiplicity to about $300$ particles, nearly
independently of the level of pileup and so the clustering time
saturates.
However the time to run SK grows and comes to dominate over the
clustering time.
Asymptotically, the total event processing time then grows linearly
with the level of pileup.
A significant part of that time (about $130\,\text{ns}$ per particle,
75\% of the run-time at high multiplicity) is taken by the
determination of the particles' rapidity and azimuth in order to
assign them to a grid cell. If the particles' rapidity and azimuth are
known before applying the SoftKiller to an event (as it would be the
case \eg for calorimeter towers), the computing time to apply the
SoftKiller and cluster the resulting event would be yet faster,
staying below $0.5\,$ms, as indicated by the dotted orange line on
Fig.~\ref{fig:timing}.

Because of its large speed improvement, the SoftKiller method has
significant potential for pileup removal at the trigger level. Since
SoftKiller returns an event with fewer particles, it will have a speed
performance edge also in situations where little or no time is spent
in jet-area calculations (either because Voronoi areas or fast
approximate implementations are used). This can be seen in
Fig.~\ref{fig:timing} by comparing the green and the solid blue curves
which stays well below $1\,\text{ms}$ even at high multiplicities.

\section{Discussion and conclusions}
\label{sec:concl}

The SoftKiller method appears to bring significant improvements in
pileup mitigation performance, in particular as concerns the jet
energy resolution, whose degradation due to pileup is reduced by
$20-30\%$ relative to the area--median based methods.
As an example, the performance that is obtained with area--median
subtraction for 70 pileup events can be extended to 140 pileup events
when using SoftKiller.
This sometimes comes at the price of an increase in the biases on the
jet $p_t$, however these biases still remain under control.

Since the method acts directly on an event's particles, it
automatically provides a correction for jet masses and jet shapes, and
in all cases that we have studied brings a non-negligible improvement
in resolution relative to the shape subtraction method, and also (albeit
to a lesser extent) relative to the recently proposed Constituent
Subtractor approach.

The method is also extremely fast, bringing nearly two orders of
magnitude speed improvement over the area--median method for jet
$p_t$'s.
This can be advantageous both in time-critical applications, for
example at trigger level, and in the context of fast detector
simulations.

There remain a number of open questions.
It would be of interest to understand, more quantitatively, why such a
simple method works so well and what dictates the optimal choice of
the underlying grid spacing.
This might also bring insight into how to further improve the method. 
In particular, the method is known to have deficiencies when applied to
large-$R$ ungroomed jets, which would benefit from additional study
(see Section~\ref{sec:sk-improvements-Rdep} for a possible solution).
Finally, we have illustrated that in simple detector simulations it is
possible to reproduce much of the performance improvement seen at
particle level, albeit at the price of a slight adaption of the method
to take into account the finite angular resolution of calorimeters.
These simple studies should merely be taken as indicative, and we
look forward to proper validation (and possible further adaptation)
taking into account full detector effects.
It would also be interesting to see a performance comparison of the
SoftKiller with the recently-proposed PUPPI pileup mitigation
technique~\cite{PUPPI}.
We will come back to this point in our conclusions.


\chapter{A (rich) forest of ideas to explore}\label{chap:beyond-prelimn}

\section{Alternative noise-reduction methods based on particle areas}\label{sec:beyond-prelim-tests}

Besides the SoftKiller, we have considered a series of alternative
simple methods.
These have only been put through a very minimal set of tests and
should be considered as preliminary. 
Nonetheless, we find helpful to document our findings since some of
the corresponding ideas can prove useful for future pileup mitigation
techniques.
Note that, since we were after a simple pileup-mitigation technique,
all the methods presented in this Section have essentially a single
free parameter. Introducing additional parameters could, at the
expense of adding complexity, bring some extra flexibility and help
finding improvements.\footnote{Like with the area--median subtraction
  (Section~\ref{sec:analytic-pileup}) and the substructure-based
  methods discussed in Chapter~\ref{chap:beyond-grooming}, it would be
  very interesting to see if analytic QCD calculations could help
  constraining the free parameters.}

Since our goal is to provide potential alternatives for generic
pileup-mitigation techniques, we shall focus mainly on the
performance for the reconstruction of the jet transverse momentum.
We will also consider briefly performance for the reconstruction of
the jet mass.
As we have already argued in the last two Chapters, jet properties
like the jet mass or jet shapes are often used in the context of
boosted jets, where one starts from a large-$R$ jets and applies a
grooming technique.
The study of the jet mass presented below should therefore merely be
seen as a first basic check that our methods do not spoil critically
measurements of the internal properties of jets.
Additional studies like jet shapes (the equivalent of
Fig.~\ref{fig:shape-stats} in our SoftKiller studies) or boosted jet
reconstruction (the equivalent of Fig.~\ref{fig:large-R} in our
SoftKiller studies) are left for future studies.

One avenue that we have investigated is to try to extend our
area-based approach beyond jets and to apply it directly to the
particles in the event. This first requires that we extend the concept
of jet area to individual particles.
The definition we have adopted can be seen as a {\it particle Voronoi
  area}: we build the Voronoi graph from all the particles in the
event, withing a given rapidity acceptance, and the area associated to
each particle is simply the area of its Voronoi cell, restricted to
the particle's rapidity acceptance so as to avoid infinite cells at
the edges of the acceptance.
The 4-vector area is then defined as a the massless 4-vector with
transverse momentum given by the Voronoi (scalar) area, and rapidity
and azimuthal angle of the particle.

With that definition at hand, there is an apparent straightforward way
to apply the area--median subtraction directly at the level of the
particles: if $A_{{\rm particle}}^\mu$ is the area of a given particle,
one subtracts $\rho A_{{\rm particle}}^\mu$ from its momentum,
discarding the particle altogether if its transverse momentum $p_t$ is smaller
than $\rho A_{{\rm particle}}$. 
This is unfortunately over-simplified since many soft particles,
corresponding to local downwards fluctuations of the pileup background
would be discarded (\ie set to a zero $p_t$ instead of a negative
one), cutting the negative tail of pileup fluctuations while keeping
the positive tail, ultimately leading to an under-subtraction of
pileup.

In our preliminary studies, we have considered several particle-based
pileup mitigation methods trying to avoid this under-subtraction
issue. We have compared them to the area--median approach as well as
to the SoftKiller approach presented in the previous Chapter.

\subsection{List of methods}\label{sec:prelim-list}

Here is the list of methods we shall discuss in this Section:
\begin{itemize}
\item The {\bf Area--median} method serves as a robust unbiased
  baseline for comparison. Following our recommendations from
  Section~\ref{sec:areamed-practical} we have used a grid-based
  estimation of $\rho$ with a grid-size parameter of 0.55 and rapidity
  rescaling.
  We have imposed positivity constraints on the jet $p_t$ and mass.
  %
  %
  %
%
\item The {\bf SoftKiller} method discussed at length in the previous
  Chapter, that removes the soft particles in the event until the
  estimated $\rho$ vanishes.
\item A somewhat similar idea, that we shall call the {\bf
    Soft-Removal} method in what follows, is to remove the soft
  particles in the event until the total (scalar) $p_t$ that has been
  eliminated reaches $f_{\text{SR}}\rho A_{\text{tot}}$, where $\rho$
  is the pileup density (if rescaling is active, it is estimated at
  $y=0$), $A_{\text{tot}}$ is the total detector area and
  $f_{\text{SR}}$ an overall factor, taken as a free parameter. Taking
  $f_{\text{SR}}$ to 1, we should expect to over-subtract the
  transverse momentum of the hard jets since the unsubtracted pileup
  fluctuations left across the whole event would be partially
  compensated by the subtraction of particles from the hard jets. We
  shall therefore consider values of $f_{\text{SR}}$ smaller than 1 in
  what follows.
%
\item A {\bf Voronoi Killer} method which works, in a similar spirit
  as the SoftKiller, by removing the softest particles in the event
  until, a fraction $f_{\text VK}$ of the total event area, computed
  from the particle Voronoi area, is empty.
%
\item Let us come back to the idea of subtracting
  $\rho A_{{\rm particle}}$ from each particle, setting to 0 the
  particles with $p_t<\rho A_{{\rm particle}}$. As argued above, this
  would leave a potentially large positive bias (\ie an
  under-subtraction). This bias is expected to be proportional to the
  internal fluctuations of the pileup in the event, \ie to $\sigma$.
  A possible solution would be to subtract, on top of
  $\rho A_{{\rm particle}}$ from each particle, an additional amount
  proportional to $\sigma\sqrt{A_{{\rm particle}}}$. We have tried 3
  approaches based on this idea:
  \begin{itemize}
  \item {\bf Voronoi Subtraction}: we apply a subtraction to each of
    the original particle. For a particle with original transverse
    momentum $p_t$ and Voronoi area $A$, we set its subtracted
    transverse momentum to
    $\max(p_t-\rho A-f_{\text{VS}}\sigma\sqrt{A},0)$ where
    $f_{\text{VS}}$ is an adjustable parameter.
  \item {\bf Voronoi Cut}: the subtracted event is built by keeping
    the particles with $p_t>\rho A-f_{\text{VC}}\sigma\sqrt{A}$
    (untouched) and discarding the others. $f_{\text{VC}}$ is again an
    adjustable parameter.
  \item {\bf Voronoi Subtract\&Cut}: we first subtract $\rho\sqrt{A}$
    from each particle in the event, discarding particles with
    $p_t<\rho\sqrt{A}$. We then remove all the particles for which the
    subtracted $p_t$ is below $f_{\text{VSC}}\sigma\sqrt{A}$, with
    $f_{\text{VSC}}$ an adjustable parameter.
    Note that this last option is somewhat reminiscent of the
    adaptation of the SoftKiller method to calorimeters --- see
    Section~\ref{sec:adapt-calor-towers} --- and can therefore likely
    be applied straightforwardly in an experimental context.
  \end{itemize}
\end{itemize}

There are also two other methods that we briefly investigated but were
relatively rapidly abandoned since our initial findings did not give
very promising results. For the sake of completeness, we introduce
them here, discuss why we believe they are not efficient and quote
possible options to improve them.
Both methods subtract from the event a total transverse momentum of
$\rho A_{\rm tot}$, with $A_{\rm tot}$ the total area of the
detector. They both first subtract $\rho A_{{\rm particle}}$ from each
particle, discarding the particles for which this would give a
negative transverse momentum.
Since we have put some of the particles to $0$ transverse momentum,
the total subtracted scalar $p_t$ is less than $\rho A_{\rm tot}$ and
an extra subtraction is needed. We then consider two possible options:
\begin{itemize}
\item the excess, between $\rho A_{\rm tot}$ and what has been
  subtracted from the initial step is subtracted uniformly over the
  remaining particles. We again subtract an amount proportional to
  their area, setting to 0 the particles for which this would give a
  negative $p_t$. The operation is repeated until a total of $\rho
  A_{\rm tot}$ has been subtracted. We dubbed this method {\it
    Voronoi subtraction with Uniform Balance}.
\item alternatively, after the initial subtraction step, we
  successively remove the softest particle in the event until a total
  $p_t$ of $\rho A_{\rm tot}$ has been subtracted. We call this the
  {\it Voronoi subtraction with Soft Balance}.
\end{itemize}
In both cases, the idea here is to balance the negative part of the
fluctuations, cut by the initial subtraction, with the positive part,
hopefully discarded in subsequent subtractions or particle removal.

In both cases, this actually lead to a rather larger negative bias on
the reconstruction of the jet $p_t$. We believe this is due to a
reason similar to our argument in to use $f_{\text{SR}}<1$ for the
Soft-Removal method. After the initial subtraction, we are essentially
left with two kinds of objects: hard jets located around a few
directions in the event and pileup upwards fluctuations distributed
uniformly across the event.
Ideally, only the pileup component need to be subtracted but in
practice, both these components will be affected by the subsequent
steps of the subtraction. This will lead to some upwards pileup
fluctuations left uniformly across the event and an oversubtraction of
the hard jets to compensate for that effect.
Similarly to the Soft-Removal method, a tentative solution to that
issue would be to adapt the method and only subtract a fraction $f$ of
the total $\rho A_{\rm tot}$, with $f$ an adjustable parameter. We
have not (yet) investigated that option.

Note also that we could have considered a third option where the
amount left to subtract after the initial subtraction is balanced
using the particle's neighbours (there are actually several ways to
achieve that). In practice, the hard particles in the jet would be
used to compensate the negative fluctuations in the jet's vicinity,
yielding once more an over-subtraction, possibly even larger than with
the other two approaches.\footnote{The Constituent Subtractor has
  likely a similar issue when applied to the whole event.}

Finally, in a more complete study, it would be interesting to include
a comparison to the ConstituentSubtractor, as well as other intra-jet
methods, like an {\it intrajet killer} which would iteratively remove
the softest particle in a given jet until a total $p_t$ of
$\rho A_{\text{jet}}$ has been subtracted (see
footnote~\ref{ft:intrajet} in Chapter~\ref{chap:soft-killer}).
These methods are expected to give results very close to the
area--median for the jet transverse momentum and a small improvement
for the jet mass.

\subsection{Monte-Carlo studies}\label{sec:prelim-mcstudies}

\paragraph{Description.} As usual, we work by embedding hard events in
a superposition of minimum bias events to simulate pileup. Both the
signal events and the minimum bias events have been generated with
Pythia~8186 with the 4C tune and, for simplicity, we have set the mass
of the initial particles to zero, preserving their transverse momenta,
rapidities and azimuthal angles.\footnote{Compared to the studies done
  in the previous Chapter, we have kept the $\pi^0$ stable. As we
  shall see, this slightly affects the optimal parameter for the
  SoftKiller.}
We keep particles up to $|y|=5$.
We cluster the jets using the anti-$k_t$ algorithm with $R=0.4$. In
the hard event, we select the jets passing a given $p_t$ cut that we
shall vary to probe the robustness of the pileup mitigation methods,
and with $|y|<4$.

We apply the various event-wide pileup subtraction techniques
introduced in the previous Section to the full event --- including the
hard event and pileup --- and for each of the original hard jets, find
the most overlapping jet in that subtracted event.
We then focus on the shift $\avg{\Delta p_t}$ and dispersion
$\sigma_{\Delta p_t}$ quality figures already used several times so
far, in order to assess the performance of the methods introduced
above.
As an additional check, we also look at the reconstruction of the jet
mass by studying $\avg{\Delta m}$ and $\sigma_{\Delta m}$.

The parameters of our candidate pileup mitigation techniques are
fixed/varied as follows.
For the area--median subtraction, acting as our baseline. we use a
grid-based estimation of $\rho$ with the grid-size parameter set to
0.55 and rescaling to handle the rapidity dependence.
The SoftKiller size parameter is varied between 0.3 and
0.7.\footnote{This was already studied in the previous Chapter. We
  included it here in order to provide an easier comparison between
  the different pileup mitigation methods.}
For the Soft~Removal technique, we use the pileup density $\rho$
coming from the area--median method, calculated at $y=0$ and vary
$f_{\text{SR}}$ between 0.66 and 0.76.
Moving to the methods based on Voronoi areas, we calculate
the particle areas up to $|y|=5$. For the Voronoi~Killer, we vary
$f_{\text{VK}}$ between 0.94 and 0.98.
Finally, for the last three Voronoi-based methods, we compute $\rho$
and $\sigma$ independently for each particle, using the area--median
technique, including the rapidity dependence. We vary
$f_{\text{VC}}$ between 1.5 and 3.5, $f_{\text{VSC}}$ between 0.5 and
3, and $f_{\text{VS}}$ between 0.2 and 0.7.

Since a key element common to all these new techniques is their
potentially uncontrolled biases, we want to study different processes
and pileup conditions to assess the robustness of the methods.
We do this by selecting a few working points. First, we work with
fixed pileup conditions, taking $\mu=60$ as a default, and study a few
representative cuts on the jet $p_t$, namely $20$, $100$ and
$500$~GeV. Conversely, for a fixed jet $p_t$ cut of $50$~GeV, we vary
the pileup conditions, choosing $\mu=30$, $60$ or $140$.
This should be sufficient to give a first overview of the performance
of our candidate methods when varying the two main scales in the
problem: the scale of the hard jet and that of pileup.

\begin{figure}[!p]
\centerline{\includegraphics[width=0.82\textwidth]{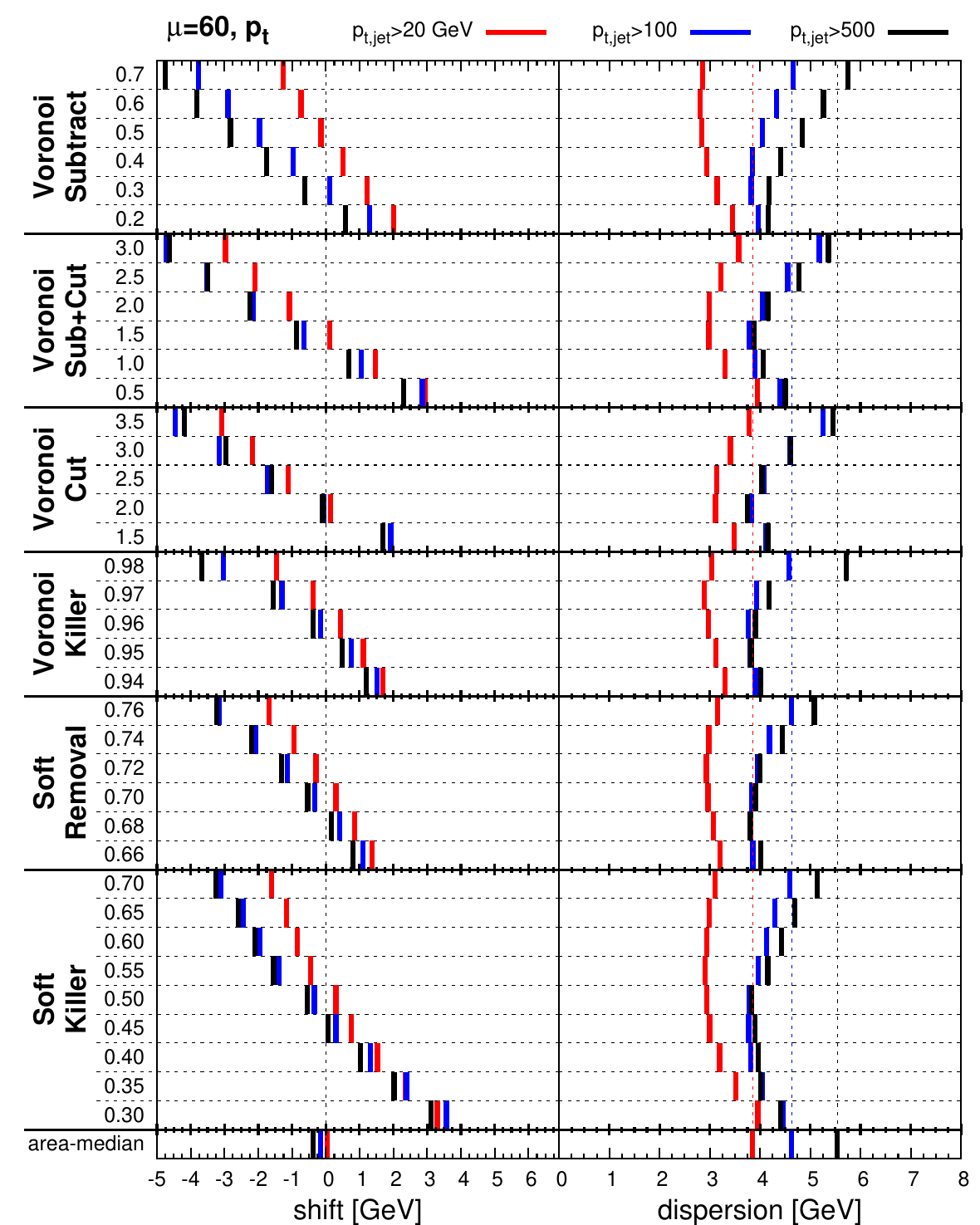}}
\caption{Summary of the quality measures for various pileup mitigation
  methods. The average $p_t$ shift $\avg{\Delta p_t}$ (left panel) and
  the associated dispersion $\sigma_{\Delta p_t}$ (right panel) are
  plotted for $\mu=60$ and different jet $p_t$ cuts (20, 100 and
  500~GeV respectively in red, blue and black).
  From bottom to top, we show the results for the area--median 
  method, for the SoftKiller for various grid size $a$, for the
  SoftRemoval for different $f_{\text{SR}}$, for the Voronoi
  Cut, the Voronoi Subtract\&Cut and the Voronoi Subtraction methods,
  for different $f_{\text{VC}}$, $f_{\text{VSC}}$ and
  $f_{\text{VS}}$, respectively.
  For the average shift, the dashed vertical line corresponds to the
  ideal situation of a zero bias. For the dispersion plot, the
  vertical lines correspond to the baseline obtained with the
  area--median subtraction method.}\label{figs:prelim-summary-v-pt-pt}
\end{figure}

\begin{figure}[!p]
\centerline{\includegraphics[width=0.82\textwidth]{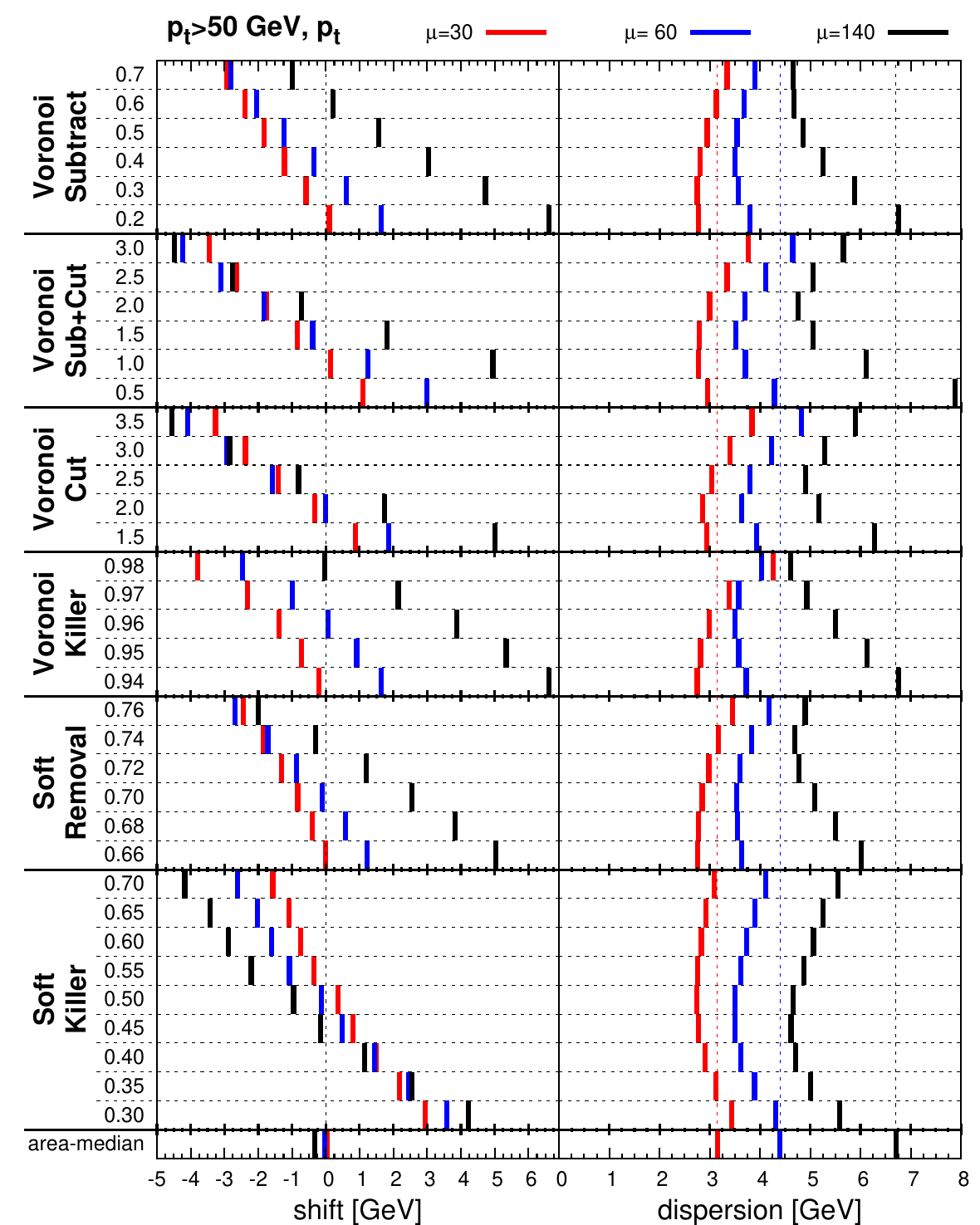}}
\caption{Same as Fig.~\ref{figs:prelim-summary-v-pt-pt} but this time
  for a sample of hard jets with $p_t\ge 50$~GeV and studying the
  sensitivity of our results \wrt to variations of the pileup
  multiplicity. Different curves correspond to $\mu=30$, $60$ and
  $140$ (respectively in red, blue and
  black).}\label{figs:prelim-summary-v-npu-pt}
\end{figure}

\paragraph{Results.} Summary plots are presented in
Fig.~\ref{figs:prelim-summary-v-pt-pt} for the $p_t$ variations at
fixed $\mu$ and in Fig.~\ref{figs:prelim-summary-v-npu-pt} for the
$\mu$ dependence at fixed $p_t$.
Overall, the message to take from these figures is that, as foreseen,
one can obtain a much better resolution than with the area--median
approach but at the expense of potentially large biases, strongly
dependent on the adjustable parameter of the method.\footnote{It would
  be interesting to introduce a measure of the fine-tuning associated
  with these free parameters.}
Below, we make more precise comments individually for each method.

\begin{figure}[!p]
\centerline{\includegraphics[width=0.82\textwidth]{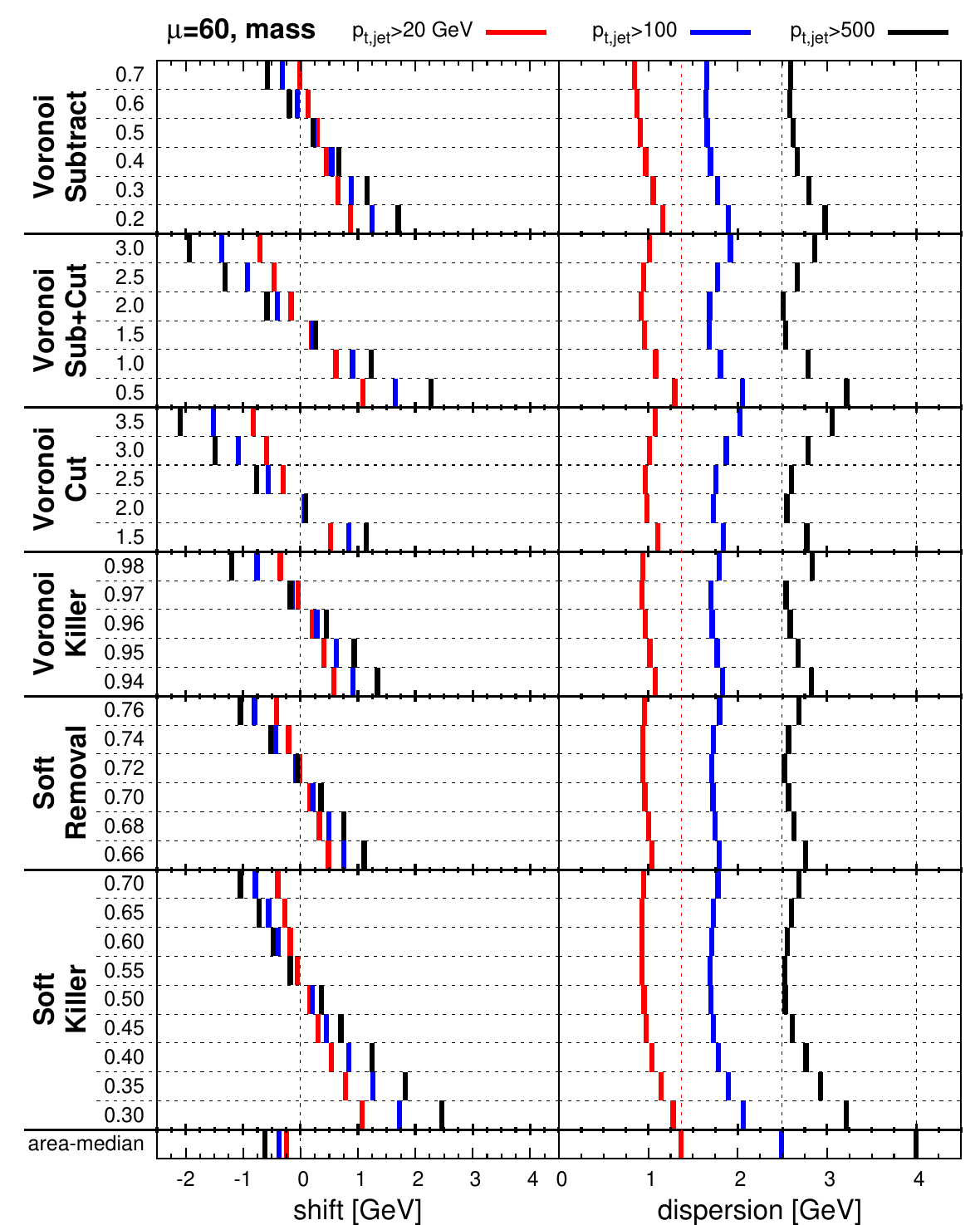}}
\caption{Same as Fig.~\ref{figs:prelim-summary-v-pt-pt} but this time
  for the jet mass rather than for the jet $p_t$. The left plot shows
  the average shift and the right plot shows the
  dispersion.}\label{figs:prelim-summary-v-pt-m}
\end{figure}

\begin{figure}[!p]
\centerline{\includegraphics[width=0.82\textwidth]{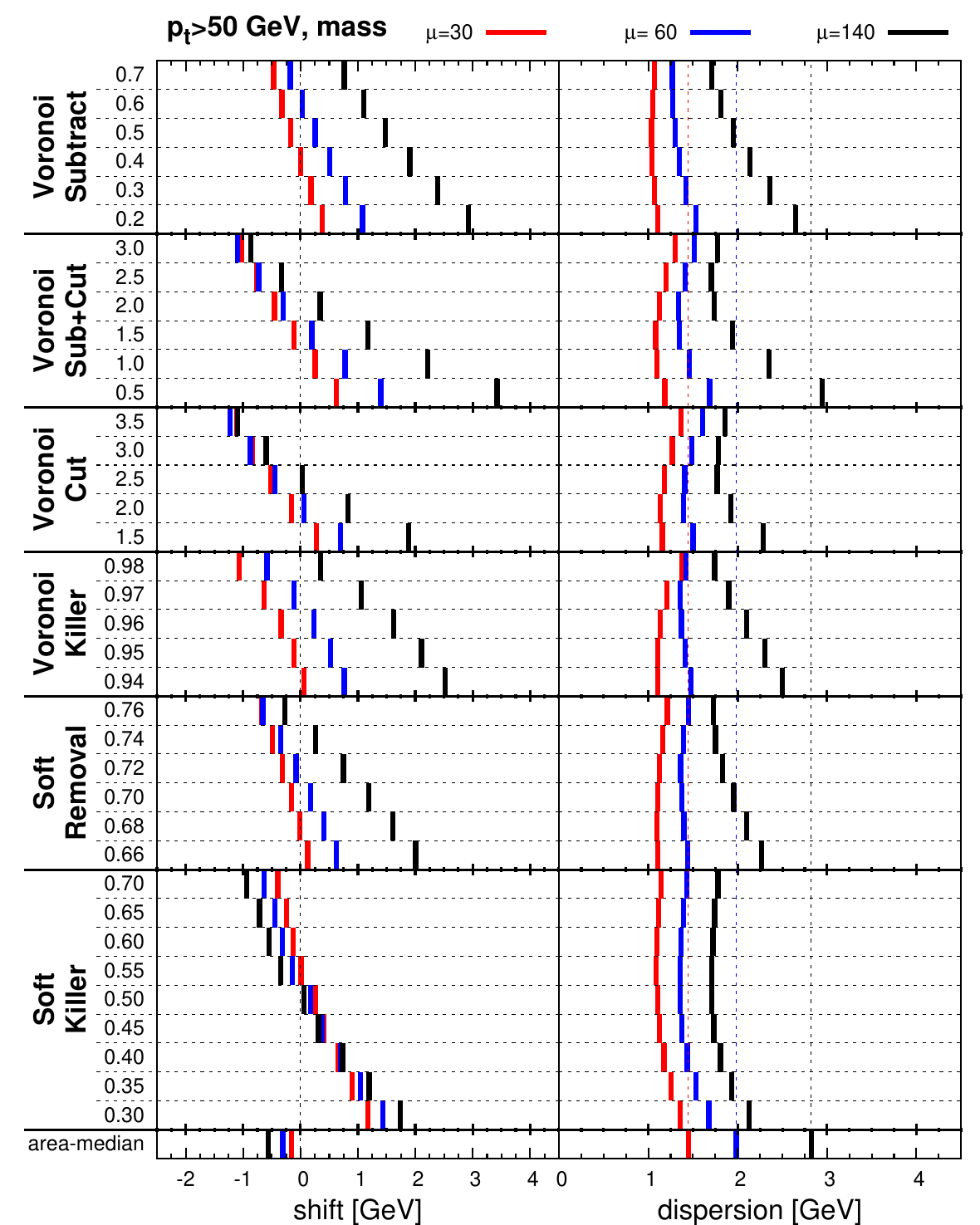}}
\caption{Same as Fig.~\ref{figs:prelim-summary-v-npu-pt} but this time
  for the jet mass rather than for the jet $p_t$. The left plot shows
  the average shift and the right plot shows the
  dispersion.}\label{figs:prelim-summary-v-npu-m}
\end{figure}

Note that while the discussion focuses on the reconstruction of the
most fundamental property of jets, their $p_t$, similar conclusions
can be reached for the jet mass for which summary plots are presented
in Figs.~\ref{figs:prelim-summary-v-pt-m} and
\ref{figs:prelim-summary-v-npu-m}.

\paragraph{SoftKiller.} The SoftKiller with $a\approx 0.4-0.5$ shows
a nice stability against both $p_t$ and $\mu$ variations, as well as a
significant resolution improvement compared to the area--median
approach.
Here, we recover the results presented in details in the previous
Chapter.\footnote{Up to a small variation of the dependence on the
  SoftKiller parameter $a$ related to our different event simulation
  with stable $\pi^0$.}

\paragraph{Soft Removal and Voronoi Killer.} These two methods show
similar characteristics. For fixed pileup conditions, the subtraction
bias is reasonably stable against changes in the jet $p_t$ with a
region where it remains close to zero, but a fixed value of the
adjustable parameter is unable to stay unbiased when varying the
pileup conditions. 

This could actually have been anticipated. If one takes, for example,
the case of the Voronoi Killer, as pileup increases, the area occupied
by pileup particles (resp. hard particles) increases
(resp. decreases). Removing a fixed area from the event will then
result in a bias increasing with pileup.
A possible improvement of the method would be to adjust
$f_{\text{SR}}$ or $f_{\text{VK}}$ with the pileup
conditions.\footnote{For example, a behaviour like
  $f_{\text{VK}}=\nPU/(a+\nPU)$ --- with $a\sim 2$ according to
  Fig.~\ref{figs:prelim-summary-v-npu-pt} --- would probably give the
  right asymptotic behaviour at large pileup and nicely go to 0 in the
  absence of pileup.}

It is also interesting to note that, while values of $f_{\text{VK}}$
are close to one, $f_{\text{SR}}$ is much smaller. This is likely due
to the fact that we have used the value of $\rho$ estimated at
$y=0$. If we take into account the rapidity profile of $\rho$,
introduced in Chapter~\ref{chap:mcstudy}, we find that, for
$\sqrt{s}=13$~TeV, the averaged $\rho$ for $|y|<5$ is about 83\% of
the $\rho(y=0)$, a value only slightly above our range for
$f_{\text{SR}}$.

That said, if we consider a value of their parameter where the bias
remains small, both the Soft Removal and the Voronoi Killer methods
show a resolution similar to that obtained with the SoftKiller. This
means that, modulo the dependence of $f_{\text{SR}}$ and
$f_{\text{VK}}$ on the pileup multiplicity, which might likely be
parametrised easily, these methods could give results similar to that
of the SoftKiller. 

\paragraph{Voronoi Cut, Subtract\&Cut and Subtract.} Comparing the
average bias observed with these three methods, one sees that, in the
region where the bias is small, the Voronoi Subtract Method shows a
larger dependence on $p_t$ and $\mu$ that the Voronoi Cut, with the
mixed Voronoi Subtract\&Cut method being intermediate. 
This indicates that a method that either keeps a particle untouched or
discards it shows more stability than a method that applies a more
democratic subtraction.
The Voronoi Cut technique actually shows a behaviour comparable to the
Soft Removal and Voronoi Killer methods discussed above with an even
slightly smaller dependence on the jet $p_t$ and $\mu$.
As long as we do not go to very large pileup multiplicities, the
Voronoi Cut method with $f_{\text{VC}}\approx 2$ shows similar
performance and robustness than the SoftKiller, both in terms of the
average bias than in terms of dispersion, with a small advantage for
the latter in terms of dispersion.
Going to larger pileup multiplicities would probably necessitate a
small increase of $f_{\text{VC}}$ to maintain a similar level of
performance.

\paragraph{Conclusions.} In the end, all the methods show relatively
similar potential resolution gains compared to the original
area--median approach. Amongst the list of methods we have considered,
the SoftKiller is the one showing the best stability when varying
$p_t$ and $\mu$ and so is the one we ultimately decided to select for
a deeper study.\footnote{Speed arguments also play in favour of the
  SoftKiller, although we have not put any effort in optimising the
  speed of the other methods.}
It is nevertheless interesting to notice that the Soft Removal, the
Voronoi Killer and, especially, the Voronoi Cut methods, for fixed
pileup conditions, all have a working point for which the average bias
remains small independently of the jet $p_t$.\footnote{This test
  should also be carried with busier events like hadronically decaying
  $t\bar t$ events.} Adjusting their respective free parameter with
the pileup conditions would deserve further study and can potentially
give performances similar to what is reached with the SoftKiller.
Since the studies here remain fairly basic, \eg they do not discuss
CHS events or detector effects, the alternative approaches presented
above, and the ideas they are based on, should be kept in mind for
future pileup mitigation studies.

\section{SoftKiller improvements}\label{sec:sk-improvements}

Studies in the previous Chapter and in the previous Section have shown
that the SoftKiller is a very promising noise-reduction pileup
mitigation technique. In this Section, we investigate possible
improvements of the method.
All the results presented here are to be considered as preliminary. We
nevertheless believe that they carry useful information and might prove
useful in the longer run.

In practice, we want to study two specific points. The first one is
the question of the angular dependence: as we have seen in the
previous Chapter, the parameter $a$ of the SoftKiller tends to depend
on the jet radius $R$ used to cluster the final jets and we want to
propose a possible workaround that would work independently of
$R$. 
Note that fat jets are only one possible application of this
study. The other application, potentially of broader impact, could be
an improvement of the SoftKiller already at $R=0.4$ through an
improved description of the angular dependence at smaller $R$.

The second question is related to our observation in
Chapter~\ref{chap:charged_tracks} that ``(protected) zeroing'' can be
useful in reducing the impact of pileup fluctuations at the expense of
potentially biasing the reconstructed averaged jet properties. We want
to see if zeroing can be combined with the SoftKiller for further
performance gains.

\subsection[Improved \boldmath $R$ dependence]{Improved \boldmath $R$
  dependence}\label{sec:sk-improvements-Rdep}

\paragraph{Motivation and description.} One piece of physical
information that is not encoded in the SoftKiller technique is the
fact that the QCD radiation from a hard jet is collinear to the jet,
peaking at small angles, while pileup tends to be mostly uniform in
rapidity and azimuthal angle.

In practice, this means that if one goes further away from the jet axis,
particles (of a given $p_t$) are more likely to come from pileup.
In the context of the SoftKiller, the consequence is that if we want
to reconstruct jets with a larger radius, one has to increase the
$p_t$ cut on the particles which is done in practice by increasing the
size parameter $a$ of the SoftKiller.

In that context, it is natural to wonder if one can encode some form
of angular dependence directly in the SoftKiller recipe to
automatically allow for jet clustering with any radius.
A side benefit of that approach might also be that we could also
obtain a performance gain at $R=0.4$.

Implementing the physical idea that hard jets have collinear
radiation while pileup is more uniform has one major difficulty: it
requires some form of knowledge of where the hard jets are.

The approach we have tested is the following. For a given event, we
first apply the SoftKiller with an initial grid-size parameter $a$.
This corresponds to a $p_t$ cut $p_{t,0}$.
We then cluster the event with a jet radius $R_0$ for which we
expect the initial SoftKiller to give good performances.
In our studies, we chose the anti-$k_t$ algorithm but this choice
should not matter too much. 
We then keep the jets above a certain $p_{t,\rm cut}$ as our
``reference hard jets''.
Finally, for each particle in the event, we apply the following cut:
we compute its distance to the nearest reference hard jet, $\Delta$,
and discard that particle if its transverse momentum is below
\begin{equation}\label{eq:rdep-killer}
p_{t,0}\left(\frac{{\rm max}(\Delta,R_0)}{R_0}\right)^\gamma,
\end{equation}
where $\gamma$ is a parameter controlling the angular dependence of
the cut.

Compared to the plain SoftKiller, this has three extra parameters:
$R_0$ and $p_{t,\rm cut}$ defining the reference jets, and $\gamma$.
Adding parameters allows for much more fine tuning but, at the same
time, leads to a much larger phase-space to explore.
A complete and thorough exploration of this phase-space goes beyond
the reach of this preliminary analysis. In what follows, we shall just
present a few results based on a first, very basic, exploration of the
ideas we just introduced.

\paragraph{Testing framework.} In practice, we have used the same
framework as for our SoftKiller studies. We have worked with full
events (\ie not assuming anything about charged track or neutral
particles), and let the $\pi^0$ decay.
As a first attempt to test a $R$-dependent version of the SoftKiller,
we have focused on a single hard process. We have considered dijet
events with a $p_t$ cut of 50~GeV. 
We have assumed $\mu=60$ in all our studies.
In order to explore the parameter phasespace of the method, we have
varied $a$ between 0.2 and 0.6 by steps of 0.1, $R_0$ between 0.1 and
0.4 by steps of 0.1, tested $\gamma=0.5,\:1$ or $2$, and $p_{t,\rm
  cut}=5,\:10$ or $20$~GeV.
Since our main goal is to test the $R$ dependence of the method, we
have studied independently $R=0.4,\:0.7$ and $1$.
With no surprise, we have looked at the average shift $\Delta p_t$ and
the associated dispersion $\sigma_{\Delta p_t}$ with the same matching
procedure as the one used for our SoftKiller studies.

\begin{figure}
\centerline{\includegraphics[width=0.95\textwidth]{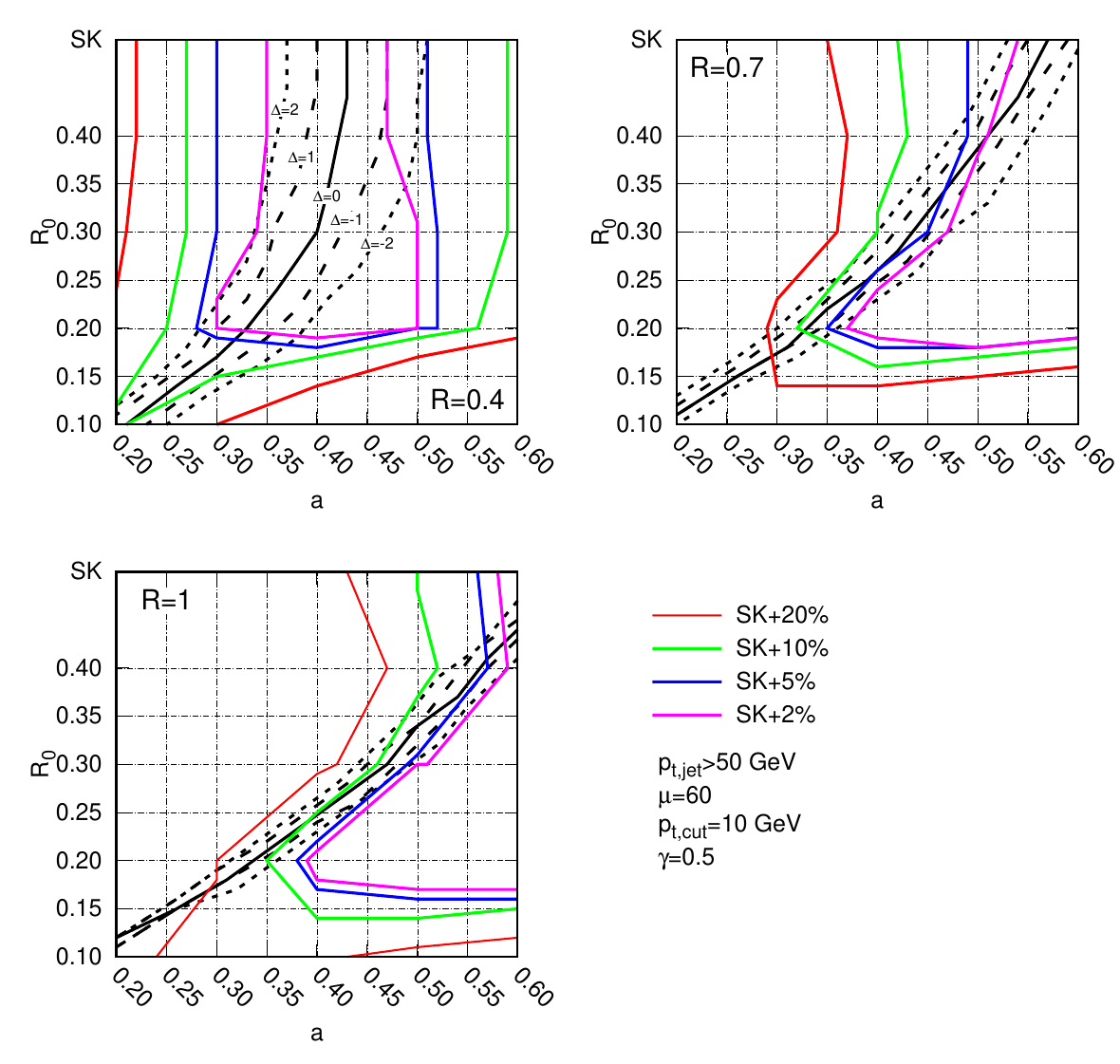}}
\caption{Summary of our results for the $R$-dependent extension of the
  SoftKiller. Each plot corresponds to different radius used to
  cluster jets: $R=0.4$ on the top left plot, $0.7$ on the top right
  plot and $R=1$ on the bottom plot.
  On each plot, we have plotted contours of equal shift and dispersion
  in the plane of the $a$ and $R_0$ parameters of the method. The top
  of the plot corresponds to the SoftKiller which only depends on
  $a$.
  The black lines indicate contours where the average bias
  $\Delta\equiv\avg{\Delta p_t}$ is $0,\pm 1$ or $\pm 2$~GeV.
  The other lines indicate contours of equal dispersion
  $\sigma_{\Delta p_t}$ where we have used the optimal dispersion for
  the plain SoftKiller (with $0.2\le 1\le 0.6$) as a reference.
}\label{fig:rdep-soft-killer}
\end{figure}

\paragraph{Results.} We present a summary of our findings in
Fig.~\ref{fig:rdep-soft-killer}. 
For simplicity we show only results for $p_{t,\rm cut}=10$~GeV and
$\gamma=0.5$. Smaller values of $p_{t,\rm cut}$ tend to give worse
results and larger values only give a small improvement and could be
risky for the reconstruction of jets of lower $p_t$. Larger $\gamma$
values show similar performance but a slightly larger dependence on
the jet radius $R$.\footnote{One should however also keep in mind that
  a too small value for the $p_t$ cut could increase sensitivity to
  detector effects as well as to pileup jets.}

In order to visualise the performance in a compact form, we have shown
on each plot both the average shift and the dispersion, as a function
of the remaining parameters, $a$ and $R_0$.
The black lines indicate contours of equal $\avg{\Delta p_t}$ and we
show them for $\avg{\Delta p_t}=0, \pm 1$ or $\pm 2$~GeV (respectively
the solid, dashed and dotted lines).  Ideally, we want to stay as
close as possible to the solid black line which corresponds to an
unbiased method.
The other lines indicate contours of equal $\sigma_{\Delta p_t}$. For
those we have first computed the optimal $\sigma_{\Delta p_t}$ for the
plain SoftKiller (with $a$ kept within the values mentioned above,
namely $0.2-0.6$) and show contours where the dispersion is larger
than this minimum by $2,5,10$ or $20\%$.
The results for the SoftKiller alone correspond to the values reported
at the top of the plot, as a function of the parameter $a$ only.

In the end, it seems that a choice $a\approx 0.35$, $R_0\approx
0.2-0.25$ shows good performance across the whole range of $R$
values. However, this comes with a series of remarks and caveats.

First, we have only tested one process, one jet $p_t$ and one value of
$\mu$. Checking the robustness of our findings for a range of
processes and pileup conditions is mandatory before claiming any
solid conclusion.
In a similar spirit, it would be interesting to repeat the study with
CHS events and, optionally, with (protected) zeroing (see the next
Section). Performance for other observables like missing $E_T$ would
also be interesting to study.

Moreover, it appears that the SoftKiller alone, with $a\approx R$ seems to
give small biases and a dispersion comparable or even better than a
fixed ``$R$-dependent Killer''. Since larger values of $R$ are mostly
used for specific purpose, one might therefore also consider applying
the SoftKiller as described in Chapter~\ref{chap:soft-killer},
guaranteeing good performance for $R=0.4$ and then apply (or re-apply)
the SoftKiller with a larger $a$ whenever one needs a larger $R$.
In that context, it is important to note that the use of grooming
techniques, inherent to the majority of the analyses involving larger
$R$ values, are likely to affect the picture. This was already
discussed in Section~\ref{sec:performance} where we noticed that a
value of $a$ valid for $R=0.4$ gives good results for the groomed mass
distribution of large-$R$ jets.

Finally, focusing on the results for $R=0.4$, there is a small region
around $a\approx 0.35-0.4$, $R_0\approx 0.2-0.25$,\footnote{That
  region will also depend on the precise value chosen for
  $p_{t,\rm cut}$ and $\gamma$.} where the bias is small and the
dispersion marginally better than the SoftKiller alone.
This means that this approach has a potential to bring a marginal
improvement over the SoftKiller method for the standard $R=0.4$
use-case.
A more definitive statement would involve a more complete study of the
other parameters of the model and, in particular, an in-depth study
of the robustness of the method across a range of pileup conditions
and jet $p_t$.

\paragraph{Conclusions.} In the end, we find that the $R$-dependent
extension of the SoftKiller method proposed here brings at best minor
improvements compared to the SoftKiller alone. However, it indicates
that there might be a region of our parameter space where a single
working point could allow clustering with any jet radius, where a
plain SoftKiller would require adapting the size parameter $a$.

There are also some signs indicating that a small improvement could be
waiting around the corner, although this requires more systematic
tests.
In particular, we have found improvements of a few percents for
$R=0.4$, compared to the SoftKiller alone. 
Further tests could also involve or suggest modifications of our
simple central formula, \eq~(\ref{eq:rdep-killer}). This points again
to the need for studies with different processes and different pileup
conditions as well as for performance tests with CHS events.
Also, given similarities between the performance obtained with the
SoftKiller and PUPPI (see also the summary study presented in our
conclusions, Chapter~\ref{chap:ccl}), with the latter involving some
form of angular dependence, one might wonder about the size of the
gains that could be obtained from such an approach.

Finally, it would be interesting to investigate whether a toy-model
approach like the one discussed in Section~\ref{sec:analytic-pileup}
could help understanding from an analytic perspective the angular
dependence of the SoftKiller or its extensions.


\subsection{Combination with (protected) zeroing}\label{sec:sk-improvements-zeroing}

The next extension of the SoftKiller approach that we want to
investigate is its combination with (protected) zeroing.

\paragraph{Description.} 
We have tested this using the following method: given a CHS event, we
first apply the SoftKiller, with a size parameter $a$, on it; we then
apply protected zeroing i.e. discard each of the remaining neutral
particle which have no charged tracks coming from the leading vertex
within a distance $R_{\rm zero}$, unless they have a transverse
momentum larger than $p_{t, \rm prot}$.

We have followed the same testing scheme as for the SoftKiller in the
previous Chapter and so refer to Section~\ref{sec:performance}
for details of our simulations.
The approach described above has three parameters: the grid size $a$
for the SoftKiller that we varied between 0.2 and 0.6, the
zeroing radius $R_{\rm zero}$ that we varied between 0.1 and 0.3, and
the protection cut $p_{t, \rm prot}$ that we fixed to 10~GeV.

\begin{figure}
\centerline{
\includegraphics[width=0.48\textwidth]{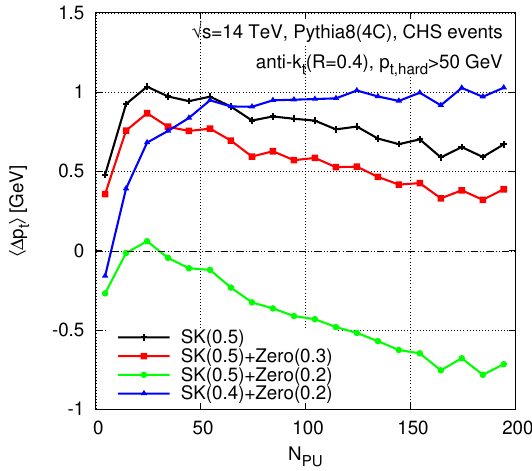}%
\hfill
\includegraphics[width=0.48\textwidth]{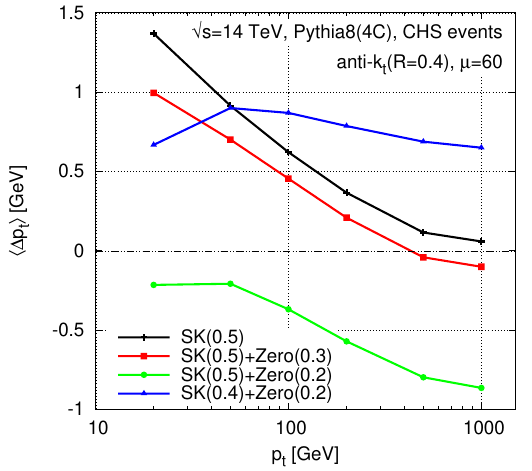}}
\caption{Average shift, $\avg{\Delta p_t}$, obtained after applying
  the SoftKiller either alone (black line) or followed by a zeroing
  step. The shift is shown as a function of the number of pileup
  vertices for hard jets above 50~GeV on the left plot and as a
  function of the hard jet $p_t$ for $\mu=60$ Poisson-distributed
  pileup events on the right plot.}\label{fig:zerokill-pt-shift}
\end{figure}

\begin{figure}
\centerline{
\includegraphics[width=0.48\textwidth]{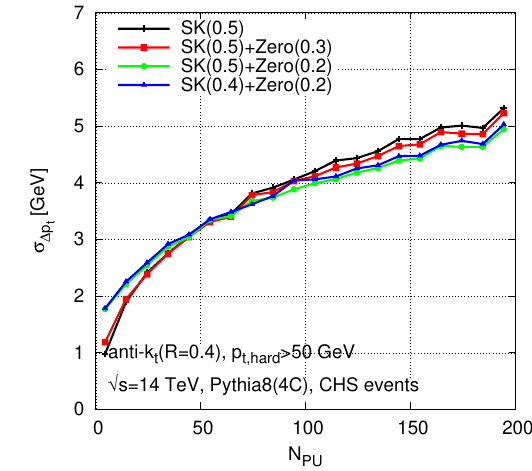}%
\hfill
\includegraphics[width=0.48\textwidth]{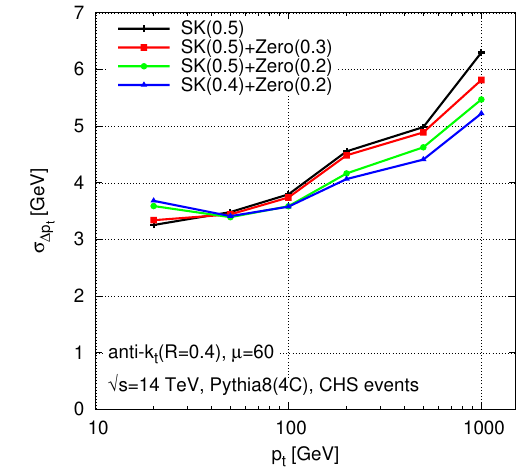}}
\caption{Dispersion $\sigma_{\Delta p_t}$ corresponding to the bias
  from Fig.~\ref{fig:zerokill-pt-shift}.}\label{fig:zerokill-pt-dispersion}
\end{figure}

\paragraph{\boldmath Results for the jet $p_t$.} On
Figs.~\ref{fig:zerokill-pt-shift} and \ref{fig:zerokill-pt-dispersion}
we show the usual performance indicators: the average bias
$\avg{\Delta p_t}$ of the reconstructed jets after pileup mitigation
compared to the original hard jets, Fig.~\ref{fig:zerokill-pt-shift},
and the associated dispersion $\sigma_{\Delta p_t}$,
Fig.~\ref{fig:zerokill-pt-dispersion}. In both cases, the results are
shown as a function of the $p_t$ of the hard jet and as a function of
the pileup multiplicity.
We have selected a few representative values of $a$ and $R_{\rm zero}$
which show good performance. As expected, lowering the zeroing radius
shifts the bias downwards and a first inspection seems to indicate
that values of $R_{\rm zero}$ in the 0.2-0.3 range give the optimal
performance.
In terms of the average bias, all the combinations shown in the Figure
give results of similar quality to, or marginally better than, the
SoftKiller alone.

Let us now turn to the dispersion. We see that zeroing gives an
improvement in extreme situations: for $a=0.4$ or $0.5$ and
$R_{\rm zero}=0.2$, we gain about 15\% in dispersion at large jet $p_t$
and around 8\% at large pileup multiplicities.
Unfortunately, this improvement is not present uniformly, especially
at low $p_t$ or low pileup multiplicity where our results indicate a
preference for a larger zeroing radius and only bring marginal
improvement, if any, compared to the SoftKiller alone.
A more careful tuning of the parameters is probably needed to reach a
performance gain over the whole kinematic spectrum.

\begin{figure}[t]
\centerline{
\includegraphics[width=0.48\textwidth]{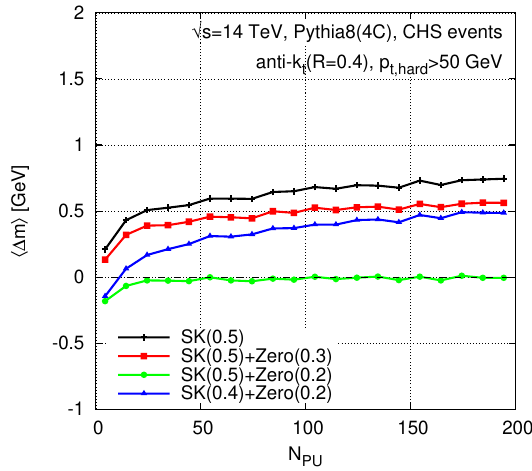}%
\hfill
\includegraphics[width=0.48\textwidth]{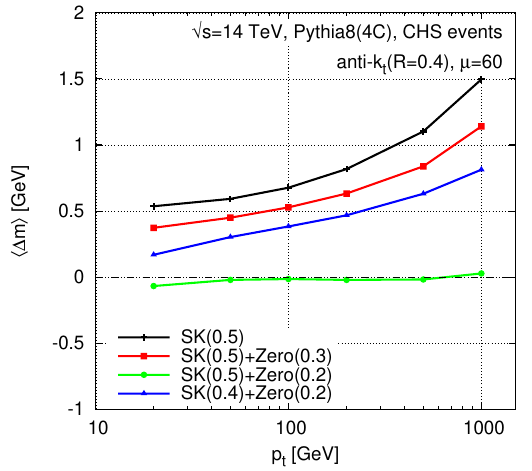}}
\centerline{
\includegraphics[width=0.48\textwidth]{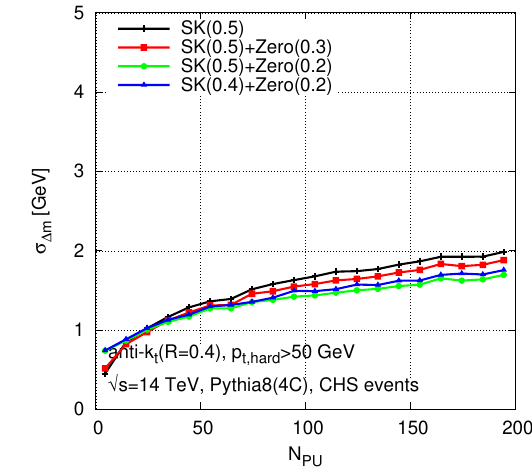}%
\hfill
\includegraphics[width=0.48\textwidth]{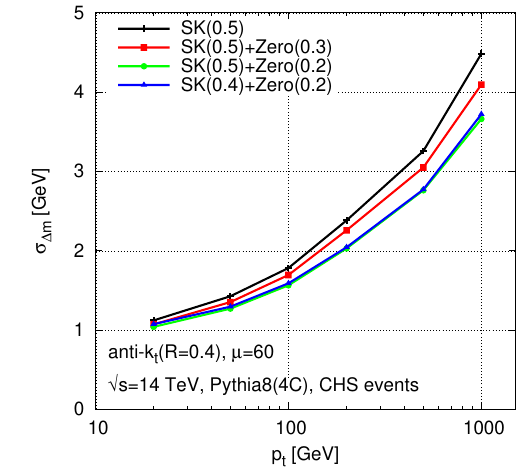}}
\caption{Same as Figs.~\ref{fig:zerokill-pt-shift} and
  \ref{fig:zerokill-pt-dispersion} now for the jet mass instead of
  the jet $p_t$.}\label{fig:zerokill-mass}
\end{figure}

\paragraph{Results for the jet mass.} If one now considers the
reconstruction of the jet mass instead of the jet transverse momentum,
see Fig.~\ref{fig:zerokill-mass}, one sees a more interesting
behaviour. First, the performance gain brought by the extra zeroing
step is more manifest. Then, the combination of the SoftKiller with
$a=0.5$ and zeroing with $R_{\rm zero}=0.2$ shows remarkable
performance: on average the mass is reconstructed with very little
bias over the whole kinematic range we have tested, and the dispersion
comes out 15-20\% lower than with the SoftKiller alone, without a sign
of degradation at low $p_t$.
This is coherent with our earlier observations from
Chapter~\ref{chap:charged_tracks}.

\paragraph{Conclusions.} In the end, this preliminary study indicates
that (protected) zeroing can indeed improve the performance of the
SoftKiller but that some degree of tuning is needed to fully optimise
the method.
For example, our studies show that a better performance would be
obtained with a larger zeroing radius at low pileup multiplicity and
low jet $p_t$.
After that tuning is carefully done, one could expect a performance
gain over the SoftKiller alone that reaches up to 15\% in terms of
dispersion with no significant impact on the average bias. 
Finally, if one focuses on the reconstruction of the jet mass instead
of the jet transverse momentum, the combination of the SoftKiller with
$a=0.5$ and zeroing with $R_{\rm zero}=0.2$ shows remarkable
performance with nearly negligible bias and nice resolution gains.
This would therefore certainly deserve more extensive studies,
including tests with more busy environments like $t\bar t$ events and
tests with a more realistic, experimental, setup.


\chapter{Conclusion and perspectives}\label{chap:ccl}

Pileup, the superposition of simultaneous proton-proton collision,
contaminates almost all measurements at the LHC. Being hadronic
objects, jets are particularly affected by pileup and one needs to
design procedures to subtract/mitigate pileup effects if one wants to
reach optimal precise for measurements involving jets.
This is the topic of this document.
The main quantity of concerns is the widely used jet transverse
momentum but we have also discussed other jet properties.

In this concluding Chapter, we will first give a brief overview of the
methods that have been covered with their respective advantages and
potential issues.
To put that overview into perspective, we will provide a benchmark
study comparing their performance in a couple of summary plots.
We will then spend the last few paragraphs discussing future
challenges and perspectives.

A significant fraction (the first part) of this review has
concentrated on the area--median pileup subtraction method which has
been used in Run~I of the LHC and is still extensively in use during
Run~II.
In Chapters~\ref{chap:areamed}{-}\ref{chap:analytics}, we have
described how the method can be applied to a wide range of jet
observables: the jet transverse momentum, the jet mass, jet shapes and
even the collinear-unsafe jet fragmentation function; we have then
tested its performance and studied some of its analytical properties.
A summary of our recommendations for the usage of the area--median
method at the LHC can be found in Section~\ref{sec:areamed-practical}.
The main benchmark of the area--median method is that it provides a
robust pileup subtraction, with a very small residual bias --- of
order of a few 100~MeV --- and a large reduction of the pileup
resolution degradation effects.
By ``robust'', we mean that the method depends essentially on a single
free parameter --- the patch size used for the estimation of the
pileup density --- which can be chosen over a relatively wide range
without affecting the overall performance of the method, independently
of the process and transverse momentum scale under consideration.

One aspect that we have not discussed at all in this document is
detector effects which will certainly require additional
jet-calibration corrections.\footnote{One specific issue we have not
  discussed is the non-linearity coming from the fact that pileup
  contamination can push a calorimeter tower from below to above the
  noise threshold.}
However, the stability of the area--median approach \wrt the process
and kinematic dependence and \wrt the choice of its free parameter,
indicates that the residual detector effects should also be small.

Furthermore, as we have shown in Chapter~\ref{chap:charged_tracks},
the area--median method can be straightforwardly extended to
Charged-Hadron-Subtracted (CHS) events, and is expected to perform
slightly better than the Neutral-proportional-to-Charged (NpC) method
which uses the (known) charged pileup contamination to the jet to
infer their neutral contamination.
This comparison could however be affected by several detector effects
like the effectiveness of the CHS event reconstruction and out-of-time
pileup.

In the end, the area--median pileup subtraction method can be seen as a
robust and efficient reference method for pileup subtraction.

The main limitation of the area--median method is that it leaves some
residual resolution degradation after the subtraction has been applied.
This smearing remains small with current pileup conditions
($\mu\sim 30-50$) --- see below for more quantitative information --- but
will increase significantly in future runs of the LHC ($\mu\sim
60-140$).
This suggests that one investigates alternative pileup mitigation
techniques which are less sensitive to pileup fluctuations between
different points within an event.
This has been investigated in the third part of our review. 
We have proposed a series of new methods following two main directions
of though: jet grooming techniques and event-wide pileup subtraction.

Jet grooming techniques are substructure tools extensively used in the
context of boosted fat jets at the LHC, and we have briefly studied
this in the second part of this document.
Our preliminary results for the use of jet grooming as a generic
pileup mitigation tool have been presented in Chapter
\ref{chap:beyond-grooming} and a discussion of the jet cleansing
technique, which uses jet substructure as well, is included in Chapter
\ref{chap:charged_tracks}.
As far as event-wide techniques are concerned, we have made an
extensive study of the SoftKiller in
Chapter~\ref{chap:beyond-prelimn}.
We have also provided a brief Monte-Carlo study of a series of
alternative (and still preliminary) event-wide pileup-mitigation
methods, in Chapter~\ref{chap:beyond-prelimn}, together with a few
possible (also preliminary) extensions of the SoftKiller method.

These substructure-based or event-wide techniques all share similar
generic patterns. As expected, they come with sometimes sizeable gains
in jet resolution, especially at large pileup multiplicities. In
particular, they often show a dependence on $\nPU$ which grows slower
that the naive $\sqrt{\nPU}$ statistical behaviour.
However, this gain in resolution does not come for free: in order to
keep the average transverse momentum bias small, one usually needs to
fine-tune the free parameters of the method, sometimes including
variations of these parameters with the pileup conditions or jet
energy. This contrasts with the robustness of the area--median method.
Among the methods we have studies, the most stable is the SoftKiller
for which a single value of the one free parameter gives acceptable
biases independently of the pileup conditions, process and jet energy
under consideration. The remaining average bias is around 1-2~GeV,
larger than what we obtain from the area--median method but of the
order of, or smaller than, the jet-energy-scale corrections applied in
an experimental context.

To summarise our main findings and the many pileup mitigation
techniques introduced throughout this document, we have selected a few
representative methods (see below) and studied them with the framework
developed for the {\em ``PileUp WorkShop''} held at CERN in May 2014
\cite{PUWS}.
Since the goal of the workshop was precisely to discuss and compare
different pileup mitigation techniques using a common open-source
software framework, this is perfectly suited for our summary.
The code used to obtain the results presented hereafter is publicly
available in the \ttt{comparisons/review} folder of the
\url{https://github.com/PileupWorkshop/2014PileupWorkshop}
GitHub~\cite{github} repository.

To avoid a proliferation of curves, we have limited the study
presented here to the following representative methods:
\begin{itemize}\setlength\itemsep{0.1cm}
\item {\em Area--median}: this is the main method described in this
  review (see Part~I) and should be considered as a baseline for
  comparisons.
\item {\em Filtering}: this is meant to illustrate how jet grooming
  techniques can be used as generic pileup subtraction methods (see
  Chapter~\ref{chap:beyond-grooming} for more details and other
  groomer options).
\item {\em Linear cleansing}: this is the technique introduced in
  Ref.~\cite{Krohn:2013lba} and discussed in
  Chapter~\ref{chap:charged_tracks}.
  It also illustrates the use of the Neutral-proportional-to-Charged
  method, combined with subjet techniques (and zeroing).
\item {\em SoftKiller}: this method discussed in
  Chapter~\ref{chap:soft-killer} is an event-wide, noise-reduction,
  pileup mitigation method.
\item {\em SoftKiller+Zeroing}: this is the extension of the
  SoftKiller method proposed in
  Section~\ref{sec:sk-improvements-zeroing}, supplemented with
  (protected) zeroing.
\item {\em PUPPI}: this is the recent method proposed in
  Ref.~\cite{PUPPI} and used by the CMS collaboration in Run~II of the
  LHC.
\end{itemize}
The first 3 methods are applied individually on each jet in the full
event, the last three methods are applied on the full event prior to
the clustering. For all cases, we have assumed massless particles and
perfect CHS events where the charged particles can be exactly
reconstructed and associated either to the leading vertex or to a
pileup vertex.
Note that the software available from the GitHub repository includes
additional options --- typically, filtering with alternative
parameters, area--trimming, the ConstituentSubtractor and the
subtractions based on the Voronoi particle area introduced in
Section~\ref{sec:beyond-prelim-tests} --- which we will not discuss
here.

The details of how the simulation and analysis are carried on are
described in Appendix~\ref{app:ccl-setup}, together with the
parameters adopted for each of the pileup mitigation techniques.
In the end, we study our usual quality measures: the average bias for
the jet transverse momentum and the associated resolution. 
We also consider the jet mass but this should merely be considered as
a cross-check that the pileup mitigation does not severely affects jet
substructure measurements.

\begin{figure}[t]
\centerline{\includegraphics[width=0.85\textwidth]{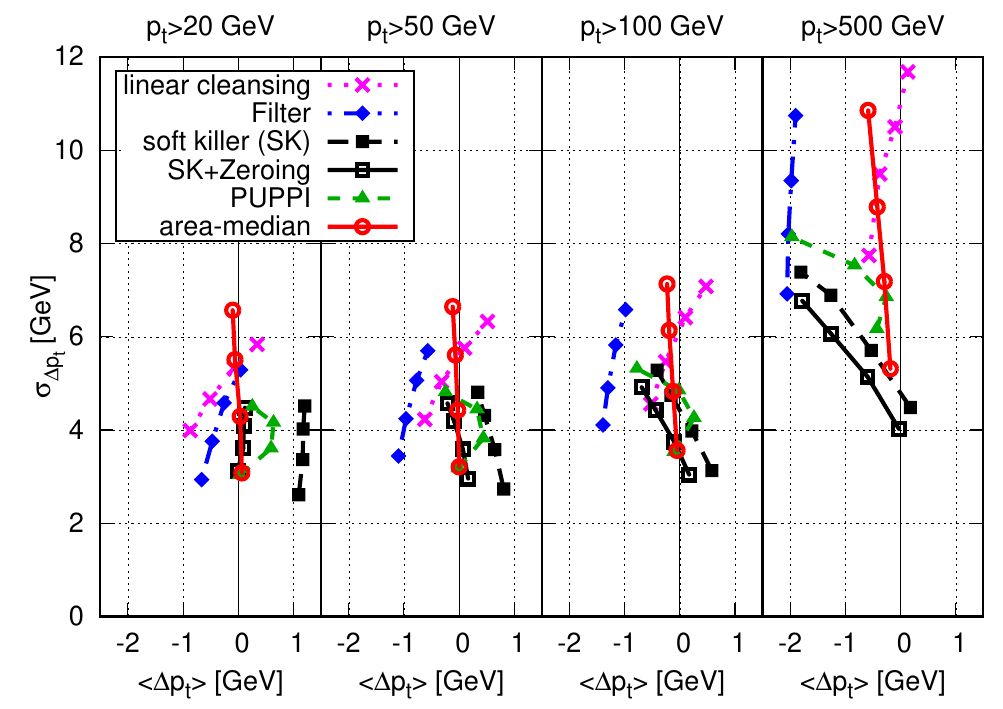}}
\caption{Performance summary for 6 representative pileup mitigation
  techniques. The $p_t$ resolution degradation due to pileup is
  plotted vs.~the average $p_t$ bias for cuts on the jet $p_t$ ranging
  from 20~GeV (left panel) to 500~GeV (right panel). The 4 points on
  each curve corresponds to $\nPU=30$, 60, 100 and 140 from bottom to
  top. All methods are applied on idealistic CHS
  events.}\label{fig:ccl-summary-pt}
\end{figure}

\begin{figure}[t]
\centerline{\includegraphics[width=0.85\textwidth]{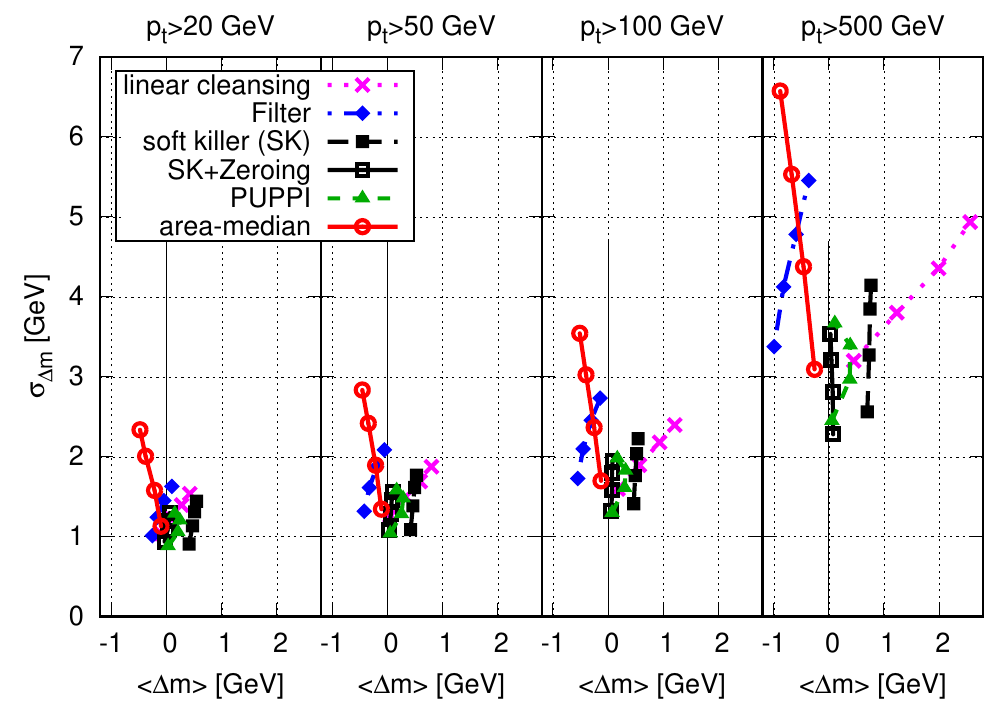}}
\caption{Same as Fig.~\ref{fig:ccl-summary-pt}, now for the jet mass
  instead of the jet $p_t$.}\label{fig:ccl-summary-m}
\end{figure}

Results are plotted in Fig.~\ref{fig:ccl-summary-pt} for the jet $p_t$
and in Fig.~\ref{fig:ccl-summary-m} for the jet mass.
We have used the format advocated during the workshop which shows the
dispersion versus the average bias. Each curve corresponds to a
different method with the 4 points corresponding to the 4 pileup
multiplicities, $\nPU=30$, 60, 100 and 140 from bottom to top.
Each panel of the plot corresponds to a different cut on the jet $p_t$
in the initial sample without pileup.
We first discuss the case of the jet $p_t$ and move to the jet mass
later on.

First and above all, we clearly see the stability of the area--median
subtraction method (the solid red curves with empty circles): the
$p_t$ bias remains well below 1~GeV over a wide range of pileup
conditions and jet $p_t$.
We also observe many of the other features highlighted in the first
part of this document, namely that the dispersion increases like
$\sqrt{\nPU}$ (we have, roughly,
$\sigma_{\Delta p_t}\approx 0.55\sqrt{\nPU}$~GeV) and that the
dispersion starts increasing at large $p_t$ (an effect associated with
back-reaction).

Let us now examine to the two other jet-by-jet techniques: filtering,
combined with the area--median subtraction for the subjets (the
dot-dot-dashed blue line with diamond symbols), and jet cleansing,
using the neutral-proportional-to-charged approach (the dotted magenta
line with crosses).
We see that they typically bring gains in resolution but come with a
few caveats: the average bias increases and depends on either the
pileup multiplicity (for cleansing) or the jet $p_t$ (for filtering).
Furthermore, while the gain in resolution is clear at small transverse
momentum and large pileup, arguably where it matters the most, the
situation is less clear for large $p_t$ where the more robust
area--median approach also gives a better resolution, or at smaller
pileup multiplicity and small $p_t$, where only filtering shows a good
performance.
This illustrates the fact that, compared to the area--median method,
these new techniques require an additional fine-tuning to keep the
average bias under control.
Some level of fine-tuning (\eg of the subjet radius used in both
filtering and jet cleansing) with $\nPU$ and $p_t$ could also lead to
a more systematic improvement in jet resolution.
Note also that it would be interesting to study a combination of
filtering with (protected) zeroing since this has shown performance
gains with cleansing.

Finally, let us consider the 3 event-wide techniques we have included
in our study: the SoftKiller alone (the long-dashed black line with
filled squares) or supplemented with (protected) zeroing (the solid
black line with empty squares) and PUPPI (the short-dashed green line
with triangles).
In terms of the average bias, we see the same trend as with the subjet
techniques: a fine-tuning of the free parameters of the method is
necessary to keep the bias under reasonable control, reasonable here
meaning around 1-2~GeV with some residual dependence on the jet
transverse momentum and the pileup conditions.
Such a bias can be considered as acceptable since it remains
comparable to jet energy scale corrections one would anyway have to
apply in an experimental context.
Furthermore, the three event-wide approaches show substantial
resolution gains, especially at large pileup multiplicities, with only
small differences between the methods.
At low $p_t$, PUPPI and SoftKiller combined with zeroing behave very
similarly with the SoftKiller alone showing a larger bias but a
slightly better resolution for $\nPU\sim 30-60$. Moving to
intermediate and large $p_t$, the SoftKiller supplemented with zeroing
seems to perform slightly better than the other two methods.
It would be definitely interesting to see how this comparison behaves
in a more realistic experimental context.

Note that it is worth comparing our numbers obtained for the
resolution, to the CMS detector performance for the jet energy
resolution~\cite{CMS:2016JER} using particle flow~\cite{pflow} and CHS
events.
With almost no pileup ($\mu<10$), CMS reports a jet energy resolution
around 3.2~GeV for 20~GeV jets, growing to $\sim 5.5$~GeV for 50~GeV
jets and $\sim 10$~GeV for 100~GeV jets.
This indicates, as expected, that pileup affects more low and
intermediate-$p_t$ jets than high-$p_t$ jets. 
For 20~GeV jets, a pileup resolution degradation of 4 GeV would give
about a 50\% degradation in jet energy resolution, which is about what
is obtained with current (end of 2016) pileup conditions.
For larger pileup multiplicities, fig.~\ref{fig:ccl-summary-pt}
clearly points in the direction of using the new event-wide mitigation
techniques, like the SoftKiller and its possible improvements, or
PUPPI. For those techniques the resolution degradation is much less
dramatic.
Whether this can be further improved or not remains however an open
question.

Turning briefly to the jet mass, Fig.~\ref{fig:ccl-summary-m}, we see
roughly the same pattern as for the jet $p_t$, although the resolution
gain is larger in the case of the jet mass. In particular, it can
already be seen for filtering and cleansing which use subjets, and the
combination of SoftKiller and protected zeroing shows a very good
performance in terms of resolution with a remarkably small average
bias.

This summary study illustrates the main points developed in this
document: we see that the area--median does an excellent job at
providing a robust method with very small bias and process dependence,
and new techniques, especially those based on event-wide subtractions,
have a potential for significant resolution improvements in the coming
runs of the LHC, at the expense of requiring a fine-tuning of their
free parameters and a potentially more involved jet
calibration.

We will now conclude with a few last generic remarks and open
questions for future investigations.

First, a brief comment about speed: jet clustering (including pileup
mitigation) is one of the most computer-expensive tasks in the
reconstruction of an event, together with the reconstruction of charged
tracks.
The fast implementation of sequential algorithms provided by FastJet
was clearly a step forwards in that context and recent releases of
FastJet brought valuable additional gains in speed. 
With the increasing LHC luminosity, one might wonder if further gains
can be found. In that context, one has to highlight that the
SoftKiller can be implement so at to run extremely fast (see
Section~\ref{sec:computing-time}). 
Given the results of our summary study above, one might wonder if the
extension of SoftKiller with protected zeroing is also amenable to
such a fast implementation?

Additionally, the many discussions throughout this report has focused
essentially on the impact of pileup on jets. This is the most common
application but is far from the only one: pileup affects the
reconstruction of all objects in a hadronic collider. The typical
other examples are lepton/photon isolation and missing transverse
energy (MET) reconstruction.
These observables are however more dependent on the details of the
detectors and hence more delicate to address using Monte Carlo
studies.
A few elements in this document point in the direction that a good
reconstruction of MET should be more aggressive that what is used for
jets. For example, in the case of an area--median-like approach, one
should typically consider an extra pileup subtraction to remove the
potentially large positive bias from positive fluctuations of the
pileup density. Similarly, our SoftKiller studies require a larger
grid-size parameter $a$ (and hence a larger associated $p_t$ cut)
when the jet radius increases.
However, other indices suggest a less aggressive pileup mitigation
for MET reconstruction. For example, with an ideal, fully hermetic
detector, energy-momentum conservation would guarantee a good
reconstruction of MET even without any pileup mitigation.

More generally, we have (almost) not addressed detector effects and
have instead tried to concentrate on more analytic aspects of pileup
mitigation (see the remarks below).
Most definitely, any pileup mitigation technique has to be validated
in an experimental context but I believe that it is preferable to
leave that delicate task directly to members of the experimental
collaborations which are more aware of the associated subtle technical
details.
In that viewpoint, tests based on Monte Carlo simulations should serve
as a solid indication that a method is worth of further
investigations.
Active discussions between the LHC phenomenologists and the
experimental collaborations can only be beneficial to address
potential issues or technical details and to ultimately aim at
efficient pileup mitigation techniques.

Let me finish this final Chapter with a few remarks on how analytic
insight and calculations have been, and can still be, useful in the
context of pileup mitigation, a topic that might seem purely
experimental at first sight.
First, we have seen repeatedly in this document that simple analytic
descriptions of the pileup energy deposit often allow one to understand
the main features observed in Monte Carlo simulations. Already using
a Gaussian approximation is sufficient to grasp many aspects of how
pileup affects jets and how the area--median subtraction behaves. It
is also interesting to point out that, as pileup multiplicity
increases, the Gaussian approximation becomes valid over a
increasingly wide range of scales.
This fundamental understanding is key to gain confidence in a pileup
mitigation method. 
Identifying physical properties of existing pileup mitigation
techniques through a simple analytic understanding can also be of
potential help towards developing new, more efficient, methods.

On a different level, analytic developments presented in this review
are intimately connected to other aspects of perturbative QCD. The
area--median method introduces a new property of jets, namely their
area, and we have seen in Chapter~\ref{chap:analytics} that many
analytic results can be derived for that quantity.\footnote{Or,
  to be more precise, ``quantities'', since there are several ways to
  define the area of a jet.}
Our results show rich structures which can be of potential interest
for tests of Monte Carlo event generators. 
Since jet areas are intrinsically non-perturbative quantities,
sensitive to soft particles, this is naturally true for the tuning of
non-perturbative models of the Underlying Event.
Additionally, we have seen that the dependence of the jet area on the
jet $p_t$ can be understood from perturbative QCD. This could provide
constraints on parton showers, and we have indeed seen some
differences between different generators and shower models included in
our studies.
Even on a more technical ground, jet areas show fundamental properties
that are still to be understood. This includes for example the
distribution of pure-ghost jets (see Fig.~\ref{fig:ghost-areas}) or
the appearance of fractal jet boundaries (see
Section~\ref{sec:areamed-areaanalytics-n-particle-active}).

Another situation where analytic calculations are important is the
case of jet grooming, discussed briefly in
Chapter~\ref{chap:grooming-analytic}. 
This is intimately related to the understanding of jet substructure
from first-principles in QCD, an area that is still fairly young and
has received quite a lot of interest over the past couple of years.

More generally, I believe that a combination of perturbative QCD
understanding of jets and modelling of the properties of pileup ---
either using a simple Gaussian approximation, more refined toy
models like those described in Section~\ref{sec:analytic-pileup}, or
directly measurements from data --- is one of the most promising way
towards developing and controlling new pileup mitigation techniques.

As for jet areas and jet substructure, a first-principles understanding
comes with two advantages.
First, it can provide a simple picture to explain the main
features of a given pileup mitigation method.
More importantly, it could help understanding from basic principles
how the average bias and resolution behaves when varying pileup
conditions, the jet transverse momentum, the jet radius, or the
parameters of a given pileup mitigation technique.
This is somehow similar to the calculation of the anomalous dimension
for jet areas and to our analytic discussions of jet grooming,
In a context where the tendency is to develop event-wide pileup
mitigation methods with noise reduction, such a first-principles
understanding might be the key to keep under control the potentially
large biases inherent to these methods.
It could therefore largely reduce the need for fine-tuning and improve
both the effectiveness and the robustness of future pileup mitigation
techniques.


\begin{acknowledgements}

  The body of work presented here is the result of many scientific
  collaborations and human interactions. Elements of this document
  date back from the early days of 2007 when I started to be involved
  in jet physics. Trying to look back, it seems to me like the path
  between then and now has been paved with a pile of positive
  encounters which have left a trace on who I am today.
  
  First and above all, a very special thanks goes to Matteo Cacciari
  and Gavin Salam for a decade worth of fruitful team work. I hardly
  see how any of this could have existed without them. It has been,
  and still is, a great adventure during which I have learned a lot
  about physics and human aspects of a longstanding collaboration. I
  can only look forwards to more of this.
  
  There are many other scientists who contributed to various degrees
  in some of the results presented here. This includes direct
  collaborators like Jihun Kim, Paloma Quiroga-Arias, Juan Rojo, Jesse
  Thaler, Simone Marzani, Andrew Larkoski and Souvik Dutta. They each
  carry a part of my memories associated with the work detailed below.

  This list can be extended to a long series of people I have
  interacted with and talked to over the same period. This obviously
  includes colleagues at the IPhT, LPTHE, BNL and CERN but also
  countless people met in conferences, seminars, visits and schools.
  This includes in particular Nestor Armesto, Mrinal Dasgupta, David
  Krohn, Matthew Low, Michelangelo Mangano, Guilherme Milhano, Carlos
  Salgado, Sebastian Sapeta, Matthew Schwartz, Liantao Wang and Urs
  Wiedemann.

  Next, even though the title of this document might suggest
  otherwise, I have greatly benefited from many interactions with the
  experimental colleagues: Helen Caines, David d'Enterria, Guenther
  Dissertori, Phil Harris, Peter Jacobs, David Miller, Filip Moortgat,
  Matt Nguyen, Matteusz Ploskon, Sal Rappoccio, Guenther Roland,
  Christophe Royon, Sevil Salur, Ariel Schwartzman, and, last but not
  least, Peter Loch from whom I have learned a great deal of
  experimental details in lively discussions during the Les-Houches
  workshop series and beyond.

  There are 6 other scientists who deserve a special thanks, namely
  Fawzi Boudjema, Jon Butterworth, Abdelhak Djouadi, Marumi Kado,
  Samuel Wallon and Bryan Webber, who have accepted to review this
  document, initially written as a French habilitation thesis, as
  members of the jury.
  I know it is a heavy task and I hope they nevertheless managed to
  get something out of this lengthy document.
  In a similar spirit, I wish to thank the anonymous Physics Reports
  referee for the careful reading of the manuscript and the helpful
  comments.

  As I said above this is both a scientific and a human story. All
  this work has been built on a firm ground of family and
  friends. Their continuous presence is the crucial part which has
  allowed me to enjoy life at least as much as I have enjoyed the
  science. 
  This includes a special thought for a few close family members who
  have sadly passed away during the past ten (or fifteen) years.
  On a much brighter side, I am infinitely grateful (and many of you
  should probably be as well) to Emilie who has lived with me through
  all of this journey, for example supporting my work at odd times and
  in odd places.
  My final thanks go to Thiago, Neil and Alana, together with
  apologies for having more than once traded playing games with them
  for research. I have no doubt that the years to come will bring many
  moments of great fun with all of them.
  
  Over the years, this work has been supported by many institutions
  (the LPTHE at the University Pierre et Marie Curie, the Brookhaven
  National Laboratory, CERN, the French CNRS and the IPhT at the CEA
  Saclay) and several grants (Contract No. DE-AC02-98CH10886 with the
  U.S. Department of Energy, the French ANR grant Jets4LHC under
  number ANR-10-CEXC-009-01, the ERC advanced grant Higgs@LHC and the
  French ANR grant OptimalJets under number ANR-15-CE31-0016). I thank
  them all for their support.
\end{acknowledgements}

\appendix

\chapter{Jet algorithms and definitions}\label{app:jetdefs}

Throughout this document we have mostly considered four jet
algorithms: anti-$k_t$~\cite{Cacciari:2008gp},
$k_t$~\cite{Catani:1993hr,Ellis:1993tq},
Cambridge/Aachen~\cite{Dokshitzer:1997in,Wobisch:1998wt} and SISCone
\cite{siscone}.
We provide here a very brief description of how they work.

\section{Recombination algorithms}

The anti-$k_t$, $k_t$, and Cambridge/Aachen algorithms
cluster the jets via successive recombinations. Then can all be seen
as special cases of the generalised-$k_t$
algorithm~\cite{Cacciari:2008gp}.

One first introduces a distance $d_{ij}$ between each pair of
particles and a beam distance $d_{iB}$ for each particle:
\begin{equation}
  \label{eq:dij-genkt}
  d_{ij} = \text{min}(k_{ti}^{2p},k_{tj}^{2p})
  \frac{\Delta y_{ij}^2 + \Delta \phi_{ij}^2}{R^2}\,,\qquad
  d_{iB} = k_{ti}^{2p}\,.
\end{equation}
The free parameters are $R$, the jet radius, and $p$. The anti-$k_t$,
$k_t$, and Cambridge/Aachen algorithms respectively correspond to
$p=-1$, $1$ and $0$.

One then iteratively repeat the following procedure until no particles
are left: at each step, the smallest distance is computed. If it
involves a pair of particles, they are recombined using a $E$-scheme
sum of four-momenta (\ie direct addition of the four-momenta),
otherwise the particle is ``clustered with the beam'', \ie called a
jet. 

\section{The SISCone algorithm}

The SISCone jet algorithm is an infrared- and collinear-safe
implementation of a modern cone algorithm. It first finds all stable
cones of radius $R$, a stable cone being a circle in the $(y,\phi)$
plane such that the $E$-scheme sum of the momenta inside the cone
points in the same direction as the centre of the cone. It then runs a
Tevatron run~II type~\cite{Blazey} split--merge procedure to deal with
overlapping stable cones. The stable cones are ordered in $\tilde
p_t$, the scalar sum of the $p_t$ of a cone's constituents, to produce
the initial list of protojets. One takes the hardest protojet and finds the
next hardest one that overlaps with it (its scaled transverse momentum
being $\tilde p_{t,j}$).  If the shared $\tilde p_t$ is larger than
$f\tilde p_{t,j}$ ($f$ is the split--merge overlap threshold parameter
for the algorithm), they are merged, otherwise, the common particles
are attributed to the protojet with the closer centre. If no overlaps
are found, the protojet is added to the list of jets and removed from the
list of protojets.
This is repeated until no protojets remain.



\chapter{Representations of the jet fragmentation }
\label{app:fragmentation-function-representations}

A jet fragmentation function can be defined as the distribution
$dN_{h}/dz$ of the momentum fraction
$z = p_{t,h}/(\sum_{i\in \text{jet}} p_{t,i})$,
of hadrons in the jet, where $p_{t,h}$ is the transverse momentum of
the hadron.

\begin{figure}[t]
\centerline{\includegraphics[width=0.9\textwidth]{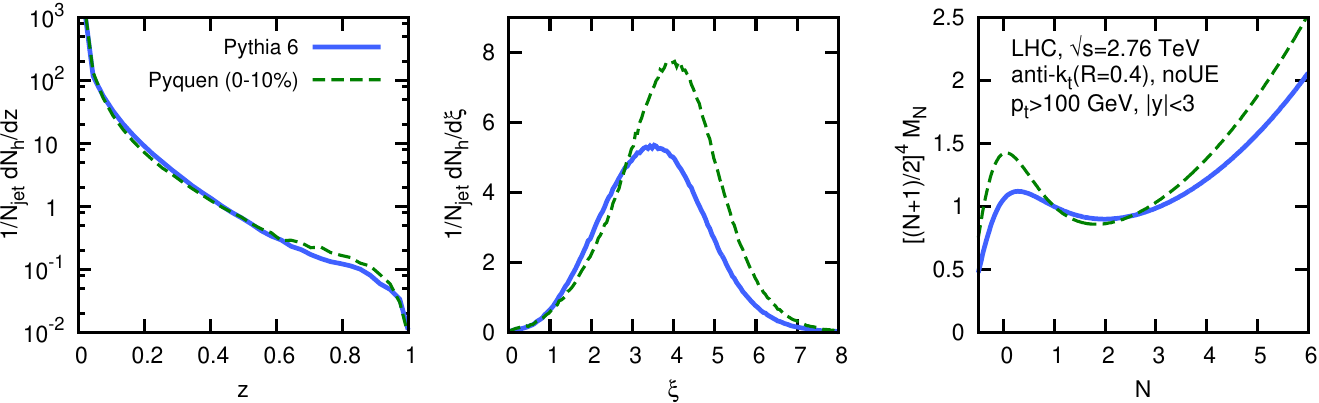}}
\caption{Jet fragmentation functions (versus $z$ and $\xi$) and moments
  (versus $N$) in $pp$ collisions at the 
  LHC ($\sqrt{s_{NN}} = 2.76$~TeV), obtained without quenching
  (Pythia~6), and with quenching (Pyquen).
}
\label{fig:hard}
\end{figure}

The left-most plot of Figure~\ref{fig:hard} shows the typical shape of
the $z$ distribution (normalised to the total number of jets
$N_{\text{jet}}$), given for anti-$k_t$ ($R=0.4$) jets with
$p_t > 100\GeV$ in $pp$ with $\sqrt{s}=2.76\TeV$.
The middle plot shows identical information, represented as a
differential distribution in $\xi = \ln 1/z$.
The $z$ representation helps visualise the hard region of the FF, while
$\xi$ devotes more space to the soft part.
Two curves are shown: the solid blue, labelled ``Pythia 6'', is a $pp$
reference curve obtained from Pythia 6.425 with its virtuality-ordered
shower;
the dashed green was obtained with the Pyquen program~\cite{pyquen,
  pyquen_tune}, which modifies Pythia showering so as to simulate
quenching. We have used it with settings corresponding to $0-10\%$
centrality.
Its effect on the FF is of the same order of magnitude as the effects
seen experimentally~\cite{CMS:2012wxa,ATLAS:2012ina}.
Independently of any question of whether it correctly models the
underlying physics, Pyquen provides therefore a useful reference when
establishing whether FFs are being reconstructed with sufficient
accuracy by some given procedure.

The right-most plot of Fig.~\ref{fig:hard} shows moments of the
fragmentation functions.
The $N^\text{th}$ moment, $M_N$, of the fragmentation function is
given by the integral
\begin{equation}
  M_N = \frac{1}{N_{\text{jet}}}\int_0^1 z^N\, \frac{dN_{h}}{dz}\, dz
      = \frac{1}{N_{\text{jet}}}\int_{0}^{\infty} e^{-N \xi}\, \frac{dN_{h}}{d\xi}\, d\xi\,.
\end{equation}
In practice, the moments for a single jet can be calculated as
\begin{equation}
  \label{eq:MN-direct}
  M_N^{\text{jet}} =\frac{\sum_{i\in{\text{jet}}}
    p_{t,i}^N}{(\sum_{i\in \text{jet}} p_{t,i})^N} \, ,
\end{equation}
where the sum runs over all the jet's constituents.\footnote{We have
  chosen to normalise the moments using the scalar sum of the
  transverse momenta of the constituents, so that $M_1=1$. 
  Using instead the transverse component of the jet momentum rather
  than the scalar sum leads to small violations (of the order of a
  fraction of one percent, for the jet radius $R=0.4$ used in this
  work) of the $M_1=1$ relation.} The results can then be averaged
over many jets, so that
$M_N = \left\langle
  \smash{M_N^{\text{jet}}}\right\rangle_{\text{jets}}$.
Obviously, $M_0$ represents the average particle multiplicity in a
jet, and $M_1$ is equal to one by virtue of momentum conservation,
provided one measures all hadrons, as we assume here (taking $\pi_0$'s
to be stable).
If instead only charged tracks are used in the numerator, then it is
clear that $M_1$ will be significantly below $1$.
There is another value of $N$ that is of special interest: given a jet
spectrum $d\sigma_\text{jet}/dp_t$ that falls as $p_t^{-n}$, the ratio
of the inclusive hadron spectrum and inclusive jet spectrum is given
by $M_{n-1}$.
Thus, $M_{n-1}^\text{AA}/M_{n-1}^\text{pp}$ corresponds to the ratio
of (charged-)hadron and jet $R_\text{AA}$ values (in the approximation that
$n$ is exactly independent of $p_t$),
\begin{equation}
\frac{M_{n-1}^{\text{AA}}}{M_{n-1}^{\text{pp}}}= \frac{R_{\text{AA}}^{\text{h}}}{R_{\text{AA}}^{\text{jet}}}\, .
\end{equation}
For $p_{t}$ in the range $100-200\GeV$, at $\sqrt{s}_{NN}=2.76\TeV$,
the relevant $n$ value has some dependence on $p_t$ and is in the
range $n=6-7$, corresponding to $N=5-6$.

\begin{figure}[t]
\centerline{\includegraphics[width=0.35\textwidth]{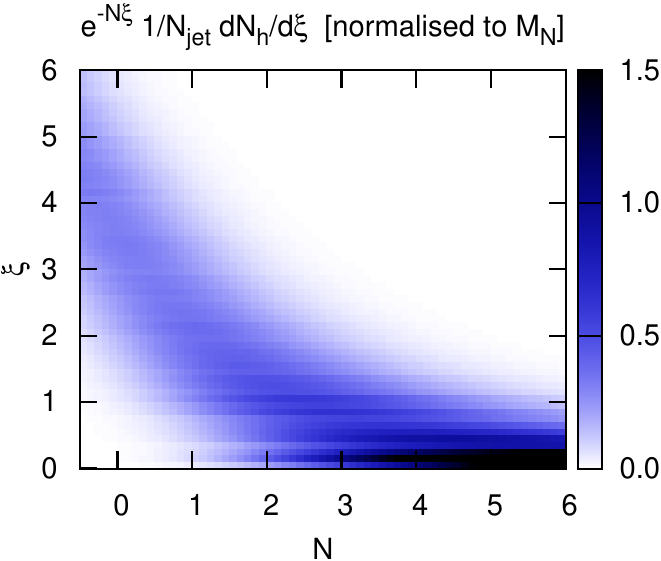}}
\caption{Representation of the $\xi$ values that contribute dominantly
  to the $M_N$ integral for a given $N$, shown as a function of
  $N$. Shown for the Pythia~6 results and cuts of
  Fig.~\ref{fig:hard}.}
\label{fig:ximap}
\end{figure}

In representing the moments in figure~\ref{fig:hard}(right), we
include a factor $((N+1)/2)^4$, which allows a broad range of $N$
values to be shown on a linear vertical scale.
The same features visible in the plots versus $z$ and $\xi$ are
visible versus $N$ too, for example that Pyquen leads to higher
multiplicities than Pythia at small $N$ (corresponding to $z < 0.05$
or $\xi > 3$)
and large $N$ ($z>0.5$, $\xi < 0.7$), and a slightly reduced multiplicity at
intermediate $N$ ($z\sim 0.2$, $\xi \sim 1.5$).

To help understand the quantitative relationship between $N$ and
$\xi$, one may examine figure~\ref{fig:ximap}, a colour-map that shows
as a function of $N$ the contribution to the $M_N$ moment from each
$\xi$ value.
It shows clearly how large $\xi$ values dominate for low $N$ (and vice-versa).
This $\xi, N$ relationship depends to some extent on the shape of the
fragmentation function and it is given for the same Pythia~6
fragmentation function that was used in Fig.~\ref{fig:hard}.


\chapter{The correlation coefficient as a quality
  measure}\label{app:correlation-coefs}

In this appendix, we discuss some characteristics of correlation
coefficients that affect their appropriateness as generic quality
measures for pileup studies.

Suppose we have an observable $v$.
Define 
\begin{equation}
  \Delta v = v^\text{sub} - v^\text{hard}\,,
\end{equation}
to be the difference, in a given event, between the pileup-subtracted
observable and the original ``hard'' value without pileup.
Two widely used
quality measures for the performance of pileup subtraction are the
average offset of $v$, $\langle \Delta v\rangle$ and the standard
deviation of $\Delta v$, which we write as $\sigma_{\Delta v}$.
One might think there is a drawback in keeping track of two measures,
in part because it is not clear which of the two is more important.
It is our view that the two measures provide complementary
information: if one aims to reduce systematic errors in a precision
measurement then a near-zero average offset may be the most important
requirement, so as not to be plagued by issues related to the
systematic error on the offset.
In a search for a resonance peak, then one aims for the narrowest
peak, and so the smallest possible standard deviation.\footnote{This
  statement assumes the absence of tails in the $\Delta v$
  distribution. 
  For some methods the long tails can affect the relevance of the
  standard-deviation quality measure.
  Other measures of the peak width that are less affected by long
  tails can be considered, see \eg the discussion in
  Section~\ref{sec:appraisal}.}

Another quality measure, advocated for example in
\cite{Krohn:2013lba}, is the correlation coefficient between
$v^\text{sub}$ and $v^\text{hard}$.
This has the apparent simplifying advantage of providing just a single
quality measure. 
However, it comes at the expense of masking potentially important
information: for example, a method with a large offset and one with no
offset will give identical correlation coefficients, because the
correlation coefficient is simply insensitive to (constant)
offsets.

\begin{figure}[t]
  \centering
  \includegraphics[width=0.48\textwidth]{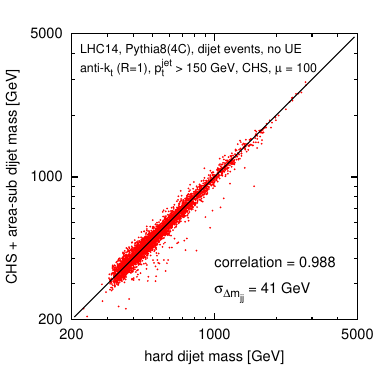}\hfill
  \includegraphics[width=0.48\textwidth]{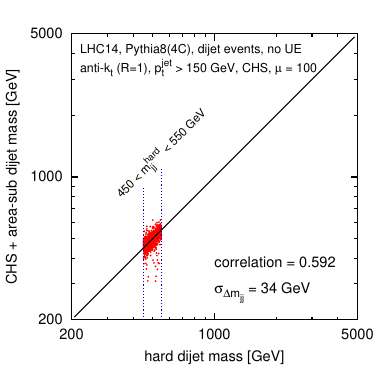}
  \caption{Left: scatter plot of the dijet mass after addition of an
    average of $100$ pileup 
    events and area--median subtraction (in CHS events) versus the dijet mass in the original
    hard event. 
    The hard dijet sample and the analysis are as described in
    appendix~\ref{app:details-npc}, with a jet radius of $R=1$.
    The right-hand plot is identical except for the following
    additional condition on the hard event: $450 < m_{jj} < 550
    \GeV$. Note the lower correlation coefficient, even though the
    lower $\sigma_{\Delta m_{jj}}$ suggests better typical subtraction in
    this specific mass bin. }
  \label{fig:correls-are-bad}
\end{figure}

The correlation coefficient has a second, more fundamental flaw, as
illustrated in Fig.~\ref{fig:correls-are-bad}. On the left, one has a
scatter plot of the dijet mass in PU-subtracted events versus the
dijet mass in the corresponding hard events, as obtained in an
inclusive jet sample.
There is a broad spread of dijet masses, much wider than the standard
deviation of $\Delta m_{jj}$, and so the correlation coefficient
comes out very high, $c= 0.988$.
Now suppose we are interested in reconstructing resonances with a mass
near $500\GeV$, and so consider only hard events in which $450 <
m_{jj} < 550\GeV$ (right-hand plot).
Now the correlation coefficient is $0.59$, \ie much
worse.
This does not reflect a much worse subtraction: actually,
$\sigma_{\Delta m_{jj}}$ is better (lower) in the sample with a limited
$m_{jj}$ window, $\sigma_{\Delta m_{jj}}=34\GeV$, than in the full sample,
$\sigma_{\Delta m_{jj}}=41\GeV$.
The reason for the puzzling decrease in the correlation coefficient is
that the dispersion of $m_{jj}$ is much smaller than before, and so
the dispersion of $\Delta m_{jj}$ is now comparable to that of
$m_{jj}$: it is this, and not an actual degradation of performance,
that leads to a small correlation.

This can be understood quantitatively in a simple model with two
variables: let $v^\text{hard}$ have a standard deviation of
$\sigma_{v,\text{hard}}$, and
for a given $v^\text{hard}$ let $v^\text{sub}$ be distributed with a mean value
equal to $v^\text{hard}$ (plus an optional constant offset) and a
standard deviation of $\sigma_{\Delta v}$ (independent of $v^\text{hard}$).
Then the correlation coefficient of $v^\text{hard}$ and
$v^\text{sub}$ is
\begin{equation}
  \label{eq:correl-model}
  c = \frac{\sigma_{v,\text{hard}}}{\sqrt{(\sigma_{v,\text{hard}})^2 +
      \sigma_{\Delta v}^2}}\,,
\end{equation}
\ie it tends to zero for $\sigma_{v,\text{hard}} \ll \sigma_{\Delta
  v}$ and to $1$ for large $\sigma_{v,\text{hard}} \gg \sigma_{\Delta
  v}$, in accord with the qualitative behaviour seen in
Fig.~\ref{fig:correls-are-bad}.
The discussion becomes more involved if $\langle v^\text{sub}\rangle$
has a more complicated dependence on $v^\text{hard}$ or if
$\sigma_{\Delta v}$ itself depends on $v^\text{hard}$ (as is actually
the case for several practical applications, like dijet mass studies).

The main conclusion from this appendix is that correlation
coefficients mix together information about the quality of pileup
mitigation and information about the hard event sample being studied.
It is then highly non-trivial to extract just the information about
the pileup subtraction.
This can lead to considerable confusion, for example, when evaluating
the robustness of a method against the choice of hard sample.
Overall therefore, it is our recommendation that one consider direct
measures of the dispersion introduced by the pileup and subtraction
and not correlation coefficients.
In cases with severely non-Gaussian tails in the $\Delta v$
distributions it can additionally be useful to consider quality
measures more directly related to the peak structure of the $\Delta v$
distribution.


\chapter{More analytics results for jet areas}\label{app:area-extra-details}

\section{Transition from one-particle jet to soft jet.} \label{sec:transition}

Having observed in Chapter~\ref{chap:analytics} the different
properties of the area of pure ghost jets and jets containing a hard
particle, one may wonder what happens to the area of a jet containing
a ``trigger'' particle whose transverse momentum $p_t$ is
progressively reduced, until it becomes much smaller (ultrasoft) than
the momentum scale of a soft background,
$\pi R^2 \rho$.\footnote{Imagine the trigger particle being a $B$
  meson, which you could tag through a secondary vertex --- you could
  then recognise its presence in any jet, regardless of its transverse
  momentum relative to the background.}

\begin{figure}
  \centering
  \includegraphics[width=0.5\textwidth]{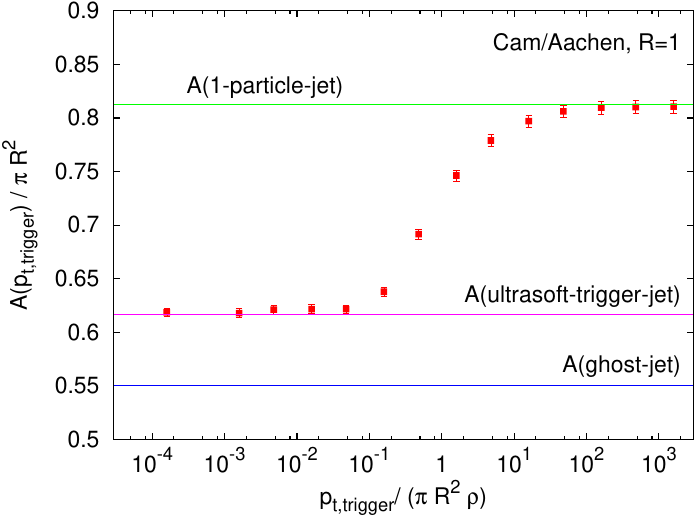}
  \caption{Area of the jet containing a trigger particle when it is
    immersed in a bath of soft particles with transverse momentum
    momentum density $\rho$; shown as a function of the trigger
    particle transverse momentum.}
  \label{fig:soft-hard}
\end{figure}

Figure \ref{fig:soft-hard} show our findings for the areas obtained
with the Cambridge/Aachen algorithm. As one expects, that there are
two asymptotic regions. At large $p_t$, the area tends to that of a
jet with an infinitely hard particle,
\eq~(\ref{eq:area-hard-particle}).  At small $p_t$, it tends not to
the pure ghost jet area, \eq~(\ref{eq:area-pure-ghost}), as one might
naively expect, but rather to a value which can be predicted as
\begin{equation}
  \label{eq:area-ghost-biased}
  A(\text{ultrasoft-trigger-jet}) = { { \int dA \, A^2\, dN/dA}\over {\int
      dA\, A\, dN/dA }} \, ,
\end{equation}
where $dN/dA$ is the distribution of the number of pure ghost jets
with a given area, and corresponds to the solid curves depicted in
figure~\ref{fig:ghost-areas}. This equation can be understood in the
following way: when the momentum of the trigger particle becomes
negligible compared to that of the soft background, it does not
influence the size of its own jet. However, the likelihood of the
trigger particle being found in a soft-background jet is proportional
to that soft-background jet's area and one gets an extra factor of $A$
in the integrand of the numerator of \eq~(\ref{eq:area-ghost-biased}).
One is able to use the pure-ghost-jet area distribution in
\eq~(\ref{eq:area-ghost-biased}), because it coincides with that of
soft background jets.

Figure~\ref{fig:soft-hard} helps to illustrate the point that a
diffuse background of particles such as pileup (represented here by
the ghosts) provides an effective cutoff scale, below which a particle
will not have any effect on the jet's area. Numerically, the effective
cutoff does indeed coincide with the soft-transverse momentum scale
$\pi R^2 \rho$.

\section{Fluctuations of the active area}\label{sec:app_fluct}

Finally, we derive the results
\eqs~(\ref{eq:sigma2_pt})--(\ref{eq:sigma2_pt_end}) for the
fluctuation coefficient $S^2_{\JA,R}$ in the case of active areas.
This is slightly more technical than for passive areas as we also have
to deal with averages over ghosts distributions. Let us briefly recall
our notation: $\avg{\cdots}$ represents an average over perturbative
emission, while $\avg{\cdots}_g$ is an average over ghosts ensembles.
When a quantity is evaluated for a specific ghost ensemble $\{g_i\}$,
we will explicitly state so with notation of the form
$A_{\JA,R}(\cdots\!\gghosts)$. Otherwise, we have the implicit
notation that a quantity specified without any mention of ghosts, such
as $A_{\JA,R}(0)$ is already averaged over ghost ensembles.

At order $\alpha_s$, the mean active area can then be written (using
$A_{\JA,R}(\Delta=0)=A_{\JA,R}(\text{one particle})$)
\begin{eqnarray}
  \label{eq:def_avg}
\avg{A_{\JA,R}} &\equiv& \avg{\avg{A_{\JA,R}(\cdots\!\gghosts)}}_g
\\
 & \simeq & \avg{A_{\JA,R}(0\gghosts)
     + \int dP\, [A_{\JA,R}(\Delta\gghosts)-A_{\JA,R}(0\gghosts)]}_{\!g}\\
 & \simeq & A_{\JA,R}(0) + D_{\JA,R} \frac{C_1}{\pi b_0}
       \log\left(\frac{\alpha_s(Q_0)}{\alpha_s(Rp_{t,1})}\right),
\end{eqnarray}
with
\begin{equation}
D_{\JA,R} = \int \frac{d\Delta}{\Delta}\,
          (\avg{A_{\JA,R}(\Delta\gghosts)}_g-\avg{A_{\JA,R}(0\gghosts)}_g).
\end{equation}
Here, we have taken into account both the corrections due to the
radiation of a soft particle and the average over the distribution of
the ghosts. Note that we have used the shorthand $dP$ to represent \eq~(\ref{eq:softcoll}),
$dP/(dp_{t2} d\Delta)$, times $dp_{t2} d\Delta$, and that ``$\cdots$'' in
\eq~\eqref{eq:def_avg} represents all possible perturbative states.

For the corresponding fluctuations, the derivation goes along the same
line
\begin{eqnarray}
\avg{\Sigma^2_{\JA,R}}
 & = & \avg{\avg{A^2_{\JA,R}(\cdots\!\gghosts)}}_g - \avg{A_{\JA,R}}^2 \\
 & = & \avg{A^2_{\JA,R}(0\gghosts)+\int dP\,
       [A^2_{\JA,R}(\Delta\gghosts)-A^2_{\JA,R}(0\gghosts)]}_{\!g} \nonumber \\
 & & - \avg{A_{\JA,R}(0\gghosts)+\int dP\,[A_{\JA,R}(\Delta\gghosts)-A_{\JA,R}(0\gghosts)]}_{\!g}^{\!2}.
\end{eqnarray}
Neglecting the terms proportional to $\alpha_s^2$, we can write
$\avg{\Sigma^2_{\JA,R}} = \Sigma^2_{\JA,R}(0) + \avg{\Delta\Sigma^2_{\JA,R}}$,
where
\begin{equation}
 \Sigma^2_{\JA,R}(\Delta) = \avg{A^2_{\JA,R}(\Delta \gghosts)}_g -
 \avg{A_{\JA,R}(\Delta \gghosts)}_g^2\,,
\end{equation}
which for $\Delta=0$ is the leading order result. We can also write
\begin{equation}
\avg{\Delta\Sigma^2_{\JA,R}} = S^2_{\JA,R} \frac{C_1}{\pi b_0}
\log\left(\frac{\alpha_s(Q_0)}{\alpha_s(Rp_{t,1})}\right).
\end{equation}
Using straightforward algebra one obtains
\begin{eqnarray}
S^2_{\JA,R}
  & = & \int \frac{d\Delta}{\Delta}\,\left\{
        \avg{[A^2_{\JA,R}(\Delta\gghosts)-A^2_{\JA,R}(0\gghosts)]}_g 
        \right.\nonumber \\
        & & \qquad\qquad\qquad\left.
      - 2\avg{A_{\JA,R}(0\gghosts)}_g
      \avg{[A_{\JA,R}(\Delta\gghosts)-A_{\JA,R}(0\gghosts)]}_g 
      \right\}\\
  & = & \int \frac{d\Delta}{\Delta}\,
        \avg{[A^2_{\JA,R}(\Delta\gghosts)-A^2_{\JA,R}(0\gghosts)]}_g
      - 2\avg{A_{\JA,R}(0\gghosts)}_g D_{\JA,R} \\
  & = & \int \frac{d\Delta}{\Delta}\,\left[
        \avg{A_{\JA,R}(\Delta\gghosts)-A_{\JA,R}(0\gghosts)}_g^2
      + \Sigma^2_{\JA,R}(\Delta)-\Sigma^2_{\JA,R}(0)\right].
\end{eqnarray}
The second equality is a direct rewriting of the first. One gets to
the last line by rearranging the different terms of the first one.
The last two lines correspond exactly to \eqs~(\ref{eq:sigma2_pt_end})
and (\ref{eq:sigma2_pt_mid}), respectively.


\chapter{Details of our NpC and cleansing study}\label{app:details-npc}

Let us first fully specify what we have done in our study and then
comment on (possible) differences relative to the KLSW study from
Ref.~\cite{Krohn:2013lba}.

Our hard event sample consists of dijet events from $pp$ collisions at
$\sqrt{s}=14\TeV$, simulated with Pythia~8.176~\cite{Pythia8}, tune
4C, with a minimum $p_t$ in the $2\to2$ scattering of $135\GeV$ and
with the underlying event turned off, except for the plots presented
in Figs.~\ref{fig:f00} and \ref{fig:shifts-dispersions-R1},
where we 
use $Z'$ events with $m_{Z'}=500\GeV$.
Jets are reconstructed with the anti-$k_t$
algorithm~\cite{Cacciari:2008gp} after making all particles massless
(preserving their rapidity) and keeping only particles with
$|y|<4$. We have $R=0.4$, except for the some of the results presented
in Section~\ref{sec:appraisal}, where we use $R=1$ as in
Ref.~\cite{Krohn:2013lba}.

Given a hard event, we select all the jets with $p_t > 150\GeV$ and
$|y| < 2.5$. We then add pileup and cluster the resulting full event
without imposing any $p_t$ or rapidity cut on the resulting jets. For
each jet selected in the hard event,
we find the jet in the full event that overlaps the most with it.
Here, the overlap is defined as the scalar $p_t$ sum of all the common
jet constituents.
Given a pair of jets, one in the hard
event and the matching one in the full event, we can apply
subtraction/grooming/cleansing to the latter and study the quality of
the jet $p_t$ or jet mass reconstruction.
For studies involving the dijet mass (cf. Fig.~\ref{fig:f00})
we require that at least two jets pass the jet selection in the
hard event and use those two hardest jets, and the corresponding
matched ones in the full event, to reconstruct the dijet
mass.\footnote{In the case of the $Z'$ events used for
  Fig.~\ref{fig:f00}, this does not exactly reflect how we
  would have chosen to perform a dijet (resonance) study ourselves.
  One crucial aspect is that searches for dijet resonances always
  impose a rapidity cut between the two leading jets, such as $|\Delta
  y| < 1.2$~\cite{ATLASDijet,CMSDijet}. This ensures that high
  dijet-mass events are not dominated by low $p_t$ forward--backward
  jet pairs, which are usually enhanced in QCD v.\ resonance
  production.
  Those forward--backward pairs can affect conclusions about pileup,
  because for a given dijet mass the jet $p_t$'s in a forward--backward
  pair are lower than in a central--central pair, and so relatively
  more sensitive to pileup.
  Also the experiments do not use $R=1$ for their dijet studies: ATLAS
  uses $R=0.6$~\cite{ATLASDijet}, while CMS uses $R=0.5$ with a form of
  radiation recovery based on the inclusion of any additional jets
  with $p_t > 30 \GeV$ and within $\Delta R = 1.1$ of either of the
  two leading jets (``wide jets'')~\cite{CMSDijet}.
  This too can affect conclusions about pileup.
}
This approach avoids having to additionally consider the impact of
pileup on the efficiency for jet selection, which cannot
straightforwardly be folded into our quality measures.\footnote{One
  alternative would have been to impose the cuts on the jets in the
  full event (after subtraction/grooming/cleansing) and
  consider as the ``hard jet'', the subset of the particles in the
  full-event jet that come from the leading vertex (\ie the
  hard event).
  We understand that this is close to the choice made in
  Ref.~\cite{Krohn:2013lba}. 
  This can give overly optimistic results because it neglects
  backreaction.
  %
}

Most of the studies shown in this paper use idealised particle-level
CHS events. 
In these events, we scale all charged pileup hadrons by a factor
$\epsilon=10^{-60}$ before clustering, to ensure that they do not
induce any backreaction (see
Section~\ref{sec:areamed-analytic-back-reaction}). The jet selection
and matching procedures are independent of the use of CHS or full
events.
%
%
As throughout this document, results plotted as a function of $\nPU$
use a fixed number of zero-bias events, while results shown for a
fixed value of $\mu$ use Poisson-distributed zero-bias events with an
average $\mu$.
Clustering and area determination are performed with a development
version of FastJet 3.1 and with FastJet 3.1.0 and 3.1.1 for
the results in section~\ref{sec:appraisal}. They all behave
identically to the current 3.2 series for the features used here.

Details of how the area--median subtraction is performed could
conceivably matter.
Jet areas are obtained using active area with explicit ghosts placed
up to $|y|=4$ and with a default ghost area of 0.01. We use FastJet's
\texttt{GridMedianBackgroundEstimator} with a grid spacing of 0.55 to
estimate the event background density $\rho$, using the particles (up
to $|y| = 4$) from the full or the CHS event as appropriate.
When subtracting pileup from jets, we account for the rapidity
dependence of $\rho$ using a rescaling as discussed in
Section~\ref{sec:areamed-position}.
We carry out 4-vector subtraction, Eq.~\eqref{eq:subtraction-base}.

We additionally avoid unphysical situations as described in
Sections~\ref{sec:areamedian-safesubtraction},
\ref{sec:areamedian-CHSsubtraction} and \ref{sec:BGE-masses}, with
{\em known selectors} specified to identify the charged particles and
their vertex of origin. 
In particular, we enabled the \ttt{safe\_mass} option for both the
area--median and NpC subtractions.

One difference between our study and KLSW's is that we carry out a
particle-level study, whereas they project their event onto a toy
detector with a $0.1\times0.1$ tower granularity, removing charged
particles with $p_t < 0.5\GeV$ and placing a $1\GeV$ threshold on
towers.
In our original (v1) studies with $f_\cut=0.05$ we tried including a
simple detector simulation along these lines and did not find any
significant modification to the pattern of our results, though
CHS+area subtraction is marginally closer to the cleansing curves in
this case.\footnote{We choose to show particle-level results here
  because of the difficulty of correctly simulating a full detector,
  especially given the non-linearities of responses to pileup and the
  subtleties of particle flow reconstruction in the presence of real
  detector fluctuations.}

Cleansing has two options: one can give it jets clustered from the
full event, and then it uses an analogue of \eq~(\ref{eq:npc-full}):
this effectively subtracts the exact charged part and the NpC estimate
of the neutrals.
Or one can give it jets clustered from CHS events, and it then applies
the analogue of \eq~(\ref{eq:npc-chs}), which assumes that there is no
charged pileup left in the jet and uses just the knowledge of the
actual charged pileup to estimate (and subtract) the neutral pileup.
These two approaches differ by contributions related to back-reaction.
Our understanding is that KLSW took the former approach, while we used
the latter. 
Specifically, our charged-pileup hadrons, which are scaled down in the CHS event,
are scaled back up to their original $p_t$ before passing them to the
cleansing code, in its \texttt{input\_nc\_separate} mode.
If we use cleansing with full events, we find that its performance
worsens, as is to be expected given the additional backreaction
induced when clustering the full event. Were it not for backreaction,
cleansing applied to full or CHS events should essentially be
identical. 

Regarding the NpC and cleansing parameters, our value of $\gamma_0 =
0.612$ differs slightly from that of KLSW's $\gamma_0=0.55$, and
corresponds to the actual fraction of charged pileup in our simulated
events.
In our tests with a detector simulation, we adjusted $\gamma_0$ to its
appropriate (slightly lower) value.

Finally, for trimming we use $R_{\rm trim}=0.3$ and the
reference $p_t$ is taken unsubtracted, while the subjets are
subtracted before performing the trimming cut, which removes subjets
with $p_t$ below a fraction $f_\text{cut}$ times
the reference $p_t$.
Compared to using the subtracted $p_t$ as the reference for trimming,
this effectively places a somewhat harder cut as pileup is
increased.\footnote{And is the default behaviour in FastJet if one
  passes an unsubtracted jet to a trimmer with subtraction, \eg a
  \texttt{Filter(Rtrim,SelectorPtFractionMin(fcut),$\rho$)}. 
  One may of course choose to pass a subtracted jet to the trimmer, in
  which case the reference $p_t$ will be the subtracted one.
}
For comparisons with cleansing we generally use $f_\cut = 0$ unless
explicitly indicated otherwise.


\chapter{SoftKiller collinear safety issued}
\label{app:sk-collinear-safety}

Collinear safety is normally essential in order to get reliable
results from perturbation theory.
One reaction to the SoftKiller proposal is that it is not
collinear safe, because it relies only on information about individual
particles' transverse momenta.
There are at least two perspectives on why this is not a severe issue.

The first relates to the intrinsic low-$p_t$ nature of the $p_t^\cut$,
which is typically of order $1-2\GeV$.
At these scales, non-perturbative dynamics effectively regulates the
collinear divergence.
Consider one element of the hadronisation process, namely resonance
decay, specifically $\rho \to \pi \pi$: if the $\rho$ has a $p_t$ of
order $2\GeV$, the rapidity--azimuth separation of the two pions is of
the order of $0.7-1$ (see \eg Section~\ref{sec:npc-v-areamed}).
Alternatively, consider the emission from a high-energy parton of a
gluon with a $p_t$ of the order of $1\GeV$: this gluon can only be
considered perturbative if its transverse momentum relative to the
emitter is at least of order a GeV, \ie if it has an angle relative
to the emitted of order $1$.
Both these examples illustrate that the collinear divergence that is of
concern at parton level is smeared by non-perturbative
effects when considering low-$p_t$ particles.
Furthermore, the impact of these effects on the jet $p_t$ will remain of the
order of $\ptcut$, \ie power-suppressed with respect to the scale of
hard physics.

The second perspective is from the strict point of view of
perturbative calculations.
One would not normally apply a pileup reduction mechanism in such a
context.
But it is conceivable that one might wish to \emph{define} the final
state such that it always includes a pileup and underlying event (UE)
removal procedure.\footnote{For example, so as to reduce prediction
  and reconstruction uncertainties related to the modelling of the UE
  (we are grateful to Leif L\"onnblad for discussions on this
  subject). This might, just, be feasible with area--median
  subtraction, with its small biases, but for the larger biases of SK
  does not seem phenomenologically compelling. Still, it is
  interesting to explore the principle of the question. }
Then one should understand the consequences of applying the
method at parton level.
Considering patches of size $0.5 \times \pi/6$ and particles with
$|y|<2.5$, there are a total of 120 patches; only when the
perturbative calculation
has at least $60$ particles, \ie attains order $\alpha_s^{60}$, can
$p_t^\cut$ be non-zero; so the collinear safety issue would
enter at an inconceivably high order, and all practical fixed-order
parton-level calculations would give results that are unaffected by
the procedure.

Collinear safety, as well as being important from a
theoretical point of view, also has experimental relevance: for
example, depending on its exact position, a particle may shower
predominantly into one calorimeter tower or into two. Collinear
safety helps ensure that results are independent of these details. 
While we carried out basic detector simulations in
section~\ref{sec:adapt-calor-towers}, a complete study of the
impact of this type of effect would require full simulation and
actual experimental reconstruction methods (\eg particle flow or
topoclustering).

\chapter{Details of the summary simulations}\label{app:ccl-setup}

We give here the details of the analysis presented as a summary study
in Chapter~\ref{chap:ccl}.

\paragraph{Framework.}
All studies have been preformed using the framework developed for the
Pileup Workshop~\cite{PUWS} held at CERN in May
2014. The framework is publicly available from the GitHub repository\\
\centerline{\url{https://github.com/PileupWorkshop/2014PileupWorkshop}}\\
which includes the code used to obtain the results presented in this
document in the \ttt{comparisons/review} folder (as of December 15
2016).

\paragraph{Hard event samples.} 
We have used the ``\ttt{dijetsel}'' set of event samples
available from \\
\centerline{\url{http://puws2014.web.cern.ch/puws2014/events},}\\
corresponding to dijet events with at least one reconstructed
anti-$k_t$($R=0.4$) jet satisfying $p_t\ge 20,\: 50,\: 100$ or
$500$~GeV. Each event sample contains 100$\,$000 events.
The events have been simulated with Pythia~8~(v8.185) with tune 4C. $B$
hadrons have been kept stable and the Underlying Event has been
switched off. The last point avoids bringing into the discussion extra
complications related to the subtraction of the Underlying Event.

\paragraph{Pileup simulation.}
Pileup is simulated here as the superposition of a fixed number $\nPU$ of
minimum bias events simulated using the Pythia~8 event generator
(v8.185) with tune 4C, at $\sqrt{s}=14$~TeV.
In practice, we have used the sample of 10 million zero-bias events
also available from\\
\centerline{\url{http://puws2014.web.cern.ch/puws2014/events}.}

We have studied four pileup multiplicities: $\nPU=30$, 60, 100 or
140, ranging between conditions obtained at the end of Run~I and the
beginning of Run~II, to multiplicities expected for HL-LHC.

\paragraph{Event and jet selections.}
To avoid potential difficulties related to the inclusion of
the $\rho_m$ term, we take particles to be massless, preserving
rapidity and azimuthal angle,
We also work with idealised CHS events, where charged tracks can be
exactly reconstructed and their vertex of origin exactly identified.
Charged tracks from pileup vertices are kept in the analysed event,
with their transverse momentum scaled down by a factor $10^{-60}$.
We consider the hard event alone, as well as ``full'' events which
include both the hard event and pileup. In both cases, we keep
particles with $|y|<4$.

We then build a set of reference jets by clustering the hard event
using the anti-$k_t$ algorithm with $R=0.4$ and selecting jets with
$p_t\ge p_{t,\rm min}$ and $|y|<2.5$.\footnote{We could also apply the
  selection cuts on the jets obtained from the full event, after
  pileup mitigation. This would however come with additional
  complications related to the selection bias inherent to this
  approach.}
The $p_{t,\rm min}$ cut is taken to match that of the initial jet
sample being studied, \ie 20, 50, 100 or 500~GeV. 
For each pileup mitigation technique, we then reconstruct and subtract
the jets: for methods which work on each individual jets, we first
cluster the jets and then subtract each jet individually, while for
event-wide techniques, we first subtract the whole event and then
cluster the resulting subtracted event.
In both cases, the clustering is also done using the anti-$k_t$
algorithm with $R=0.4$ and no selection cuts are applied.

For each jet in the full event, we then search for the closest
selected jet in the hard event, requiring that both are separated by
$\Delta R=\sqrt{\Delta y^2+\Delta\phi^2}<0.3$.
We have checked that the resulting matching efficiencies are close to
one, generally well above 99.9\%, with some methods showing a small
decrease to $\sim$99.5\% for $p_{t,\min}=20$~GeV and $\nPU=140$.
For matching pairs of jets, $(j_{\text{hard}}, j_{\text{full,sub}})$, we
measure the bias in transverse momentum,
$\Delta p_t=p_{t,\text{full,sub}}-p_{t,\text{hard}}$ and mass,
$\Delta m=p_{t,\text{full,sub}}-p_{t,\text{hard}}$.
We study the our usual quality measures, namely the average bias,
$\avg{\Delta p_t}$ and the associated dispersion, $\sigma_{\Delta
  p_t}$, computed over the full jets sample.\footnote{In practice, the
  code also computes the correlation coefficient between $p_t$ (or the
  mass) of the hard and full jets. Since we have argued in
  Chapter~\ref{chap:charged_tracks} and
  Appendix~\ref{app:correlation-coefs} that correlation coefficients
  are more delicate to interpret, we will not use them in the results
  presented here.}

Finally, unless explicitly mentioned otherwise, all jet manipulations
are done using FastJet~(v3.2.1).

\paragraph{Pileup mitigation methods.}
For the methods under consideration in this study, we use the
following configurations and parameters:
\begin{itemize}
\item {\em Area--median}: we use a grid-based estimation of $\rho$
  with a grid-size parameter of 0.55 and rapidity rescaling according
  to the profile
  $f(y)=0.9084412-0.0189777\,y^2+4.05981\,10^{-5}\,y^4$. We have
  enabled the protection against negative masses after subtraction.
  We use the full CHS event for the estimation of $\rho$.
  For the computation of the jet area, the ghosts are placed up
  to the edge of the particle rapidity acceptance, $|y|<4$, with a
  ghost area of 0.01.
\item {\em Filtering}: for each jet in the full event, we recluster
  the jet constituents into subjets using the Cambridge/Aachen
  algorithm with $R_{\rm filt}=0.2$. We then apply the area--median
  subtraction described above and keep the $n_{\rm filt}=3$ hardest
  resulting subjets.
\item {\em Linear cleansing}: we use the linear cleansing variant
  with the \ttt{input\_nc\_separate} mode. We take $\gamma_0=0.612$
  and applied trimming with $R_{\text{trim}}=0.2$ and
  $f_{\text{trim}}=0$, as suggested by the authors of
  Ref.~\cite{Krohn:2013lba} during the PileUp
  WorkShop~\cite{PUWS}.
\item {\em SoftKiller}: we have used a grid-size parameter of 0.5,
  with the grid extending to the particle acceptance, $|y|<4$. We
  apply the SoftKiller only on the neutral particles in the event,
  leaving the charged particles from the leading vertex untouched.
\item {\em SoftKiller+Zeroing}: here, we first apply a SoftKiller (as
  described above but now with a grid-size parameter of 0.45) to the
  neutral particles in the event. We then apply protected zeroing: we
  remove neutral particles with $p_t< 10$~GeV for which there are no
  charged tracks within a radius $R_{\rm zero}=0.2$ of the neutral
  particle.
\item {\em PUPPI}: we have used the implementation provided by the
  PUPPI authors in the context of the PileUp WorkShop~\cite{PUWS}. The
  code has initially been taken from revision 158 of the GitHub repository
  (\ttt{puppiContainer} from the \ttt{example-puppi} directory) with
  the following adaptation meant to retain the user-defined
  information of the jet constituents without affecting any of the
  PUPPI results: for a particle \ttt{PseudoJet p} with 4-momentum
  $(p_x,p_y,p_z,E)$ and PUPPI weight $w$, we set the 
  resulting subtracted \ttt{PseudoJet} to \ttt{$w$*p} instead of $(w
  p_x,w p_y,w p_z,w E)$.
\end{itemize}


\bibliographystyle{JHEP}
\bibliography{refs}

\end{document}